\documentstyle[11pt,epsf,epsfig]{article}
\def\smill{\small\baselineskip 0.45cm}
\def\raya{\vskip0.4cm\hrule width5.73cm height1pt}
\def\rayan{\vskip0.4cm\hrule width9.27cm height1pt}
\def\hatch{/\hskip-0.5em}

\def\Acal{\cal A}
\def\Bcal{\cal B}
\def\Ccal{\cal C}
\def\Rcal{\cal R}
\def\Ncal{\cal N}
\def\Mcal{\cal M}
\def\Zcal{\cal Z}
\def\Qcal{\cal Q}
\def\Ocal{\cal O}

\def\Fcal{\cal F}
\def\Scal{\cal S}
\def\Pcal{\cal P}

\def\Hcal{\cal H}

\def\alphabold{\mbox{\boldmath ${\alpha}$}}
\def\betabold{\mbox{\boldmath ${\beta}$}}
\def\thetabold{\mbox{\boldmath ${\theta}$}}
\def\Thetabold{\mbox{\boldmath ${\Theta}$}}
\def\mubold{\mbox{\boldmath ${\mu}$}}
\def\pibold{\mbox{\boldmath ${\pi}$}}
\def\lambdabold{\mbox{\boldmath ${\lambda}$}}
\def\phibold{\mbox{\boldmath ${\phi}$}}
\def\taubold{\mbox{\boldmath ${\tau}$}}

\def\Sigmabold{\mbox{\boldmath ${\Sigma}$}}
\def\sigmabold{\mbox{\boldmath ${\sigma}$}}
\def\abold{\mbox{\boldmath ${a}$}}
\def\xbold{\mbox{\boldmath ${x}$}}
\def\rbold{\mbox{\boldmath ${r}$}}
\def\vbold{\mbox{\boldmath ${v}$}}
\def\wbold{\mbox{\boldmath ${w}$}}
\def\Xbold{\mbox{\boldmath ${X}$}}
\def\Ybold{\mbox{\boldmath ${X}$}}

\def\ybold{\mbox{\boldmath ${y}$}}
\def\zbold{\mbox{\boldmath ${z}$}}

\def\tbold{\mbox{\boldmath ${t}$}}
\def\Ibold{\mbox{\boldmath ${I}$}}
\def\Abold{\mbox{\boldmath ${A}$}}
\def\Tbold{\mbox{\boldmath ${T}$}}
\def\Mbold{\mbox{\boldmath ${M}$}}
\def\Sbold{\mbox{\boldmath ${S}$}}
\def\Zbold{\mbox{\boldmath ${Z}$}}
\def\Vbold{\mbox{\boldmath ${V}$}}
\def\Wbold{\mbox{\boldmath ${W}$}}
\def\Ybold{\mbox{\boldmath ${Y}$}}
\def\Rbold{\mbox{\boldmath ${R}$}}

\begin{document}

\setcounter{page}{0}
\pagenumbering{gobble}
 
\vspace*{2.cm}
\oddsidemargin 1.5cm

\rm

\begin{center}
{\Huge \bf Three Lectures on}
\end{center}
\begin{center}
{\Huge \bf Probability and Statistics}
\end{center}

\vspace{2.0cm}
\begin{center}
{\Large Carlos Mana}\\ 
\vspace{0.2cm}
{\Large {\sl Departamento de Investigaci\'on B\'asica}} \\
\vspace{0.2cm}
{\Large {\sl CIEMAT, E-28040 Madrid, Spain}}
\end{center}

\newpage\null\thispagestyle{empty}\newpage

\pagenumbering{roman}
\noindent

\footnotesize
\begin{tabular}{p{2.9cm}p{11.0cm}}
& "They say that understanding ought to work by the rules of right reason.
 These rules are, or ought to be, contained  in Logic; but the actual 
 science of logic is conversant at present only with things either certain,
 impossible, or entirely doubtful, none of which (fortunately) we have to 
 reason on. Therefore the true logic of this world is the calculus of 
 Probabilities, which takes account of the magnitude of the probability 
 which is, or ought to be, in a reasonable man's mind" 
\end{tabular}
\begin{flushright}
\emph{J.C. Maxwell}
\end{flushright}
\vspace{1.0cm}

\Large
\noindent
{\bf Introduction}
\footnotesize
\vspace*{0.5cm}

These notes, based on a one-semester-course on
Probability and Statistics given in the former
{\sl Doctoral Program} of the {\sl Department of Theoretical Physics} at the 
{\sl Universidad Complutense} in Madrid, are a more elaborated version of
three lectures on statistics given at different places
to advanced graduate and PhD students. A tailored version, more suited to
gaduate students was prepared for the
{\sl CERN Latin-American School of Physics} to be held in Mexico, 2017.
They contain a humble
overview of the basic concepts and ideas one should have 
in mind before getting involved in data analysis and I belive they will
be a useful reference for both students and researchers.  

I feel, maybe wrongly, that there is a recent tendency in a subset of the
Particle Physics community to consider statistics as a collection of 
prescriptions {\sl written in some holy references}  that are used
blindly with the only arguments that either 
{\sl "everybody does it that way"} 
or that {\sl "it has always been done this way"}. 
In the lectures I have tried to demystify the {\sl ``how to''} 
recipes not because they are not useful but because, on the one hand,
they are applicable under some conditions that tend to be
forgotten and, on the other, because if the concepts are clear 
so will be the way to proceed ("at least formally") for 
the problems that come across in Particle Physics.
At the end, 
the quote from Laplace given at the beginning of the first lecture 
is what it is all about.

There is a countable set of books on probability and statistics and a 
sizable subset of them are very good out of which 
I would recommend the following ones (a personal choice function).
Chapter 1 deals with probability and this is just a measure, a finite
non-negative measure, so it will be very useful to read some sections of 
{\sl Measure Theory} by V.I. Bogachev ([Bo06]); in particular the
chapters 1 and 2 of the first volume. 
A large fraction of the 
material presented in this lecture can be found in more depth, together with 
other interesting subjects, in the 
book {\sl Probability: A Graduate Course} by A. Gut [Gu13].
Chapter 2 is about statistical inference, Bayesian Inference in fact, 
and a must for this topic is the 
{\sl Bayesian Theory} of J.M. Bernardo and A.F.M. Smith [Be94] that
contains also an enlightening discussion about the Bayesian and 
frequentist approaches in the appendix B. It is beyond question that 
in any worthwile course on statistics the ubiquitous frequentist methodology 
has to be taught as well. This is not explained in this notes because
there are already excellent references on the subject. 
Students are encouraged to look for instance at 
{\sl Statistical Methods in Experimental Physics} by F. James [Ja06],
{\sl Statistics for Nuclear and Particle Physicists} by L.Lyons [Ly89] or
{\sl Statistical Data Analysis} by G. Cowan [Co97].
Last, Chapter 3 is devoted to Monte Carlo simulation, an essential tool in
Statistics and Particle Physics and, 
like for the first chapters, there are interesting references along the 
text.

{\sl ``Time is short, my strength is limited,...·''}, Kafka dixit, so many
interesting subjects that deserve a whole lecture by themselves
are left aside. To mention some: an historical development of probability
and statistics, Information Theory and its connection with Geometry,
Bayesian Networks, Generalized Distributions 
(a different approach to probability distributions), 
Decision Theory (Games Theory), ...
and Markov Chains for which we shall state only the relevant properties 
without further explanation.
I encourage you take look at these subjects elsewhere although, 
if you wish, I can provide some notes or slides on them as additional material.

\newpage\null\thispagestyle{empty}\newpage


\Large
\noindent
{\bf Content}
\footnotesize
\vspace*{0.5cm}

\rm
\small
\noindent
{\bf Lecture 1: Probability}
\vspace{0.3cm}

\footnotesize
\noindent
{\bf 1 The Elements of Probability $(\Omega,{\Bcal},{\mu})$} (1):
Events and Sample Space (1);
$\sigma$-algebras and Measurable Spaces (3);
Set Functions and Measure Space (5); Probability Measure (6);
What is {\sl probability}? (7); Random Quantities (8).
\noindent
{\bf 2 Conditional Probability and Bayes Theorem} (11):
Statistically Independent Events (12);
Theorem of Total Probability (14), Bayes Theorem (15).
\noindent
{\bf 3 Distribution Function} (18): 
Discrete and Continuous Distribution Functions (19); 
Radon-Nikodym Theorem (19);
Distributions in more dimensions, Marginal and Conditional 
Distributions (23).
\noindent
{\bf 4 Stochastic Characteristics} (27):
Mathematical Expectation (27);
Moments of a Distribution (27); Position parameters (29);
Dispersion parameters (30); Asymmetry and Peakiness (30);
Correlation Coefficient (32); Information as a measure of independence (33); 
The "Error Propagation Expression" (34).
\noindent
{\bf 5 Integral Transforms} (34):
The Fourier Transform (34) [Inversion Theorem (37),
Changes of variable (37),
Sum of random quantities (38), Moments of a distribution (39)];
The Mellin Transform (40) [Inversion (40),
Useful properties (41), Some useful examples (42), 
Distributions with support in $\Rcal$ (46)].
\noindent
{\bf 6 Ordered Samples} (47). 
\noindent
{\bf 7 Limit Theorems and Convergence} (51):
Chebyshev's Theorem (51);
Convergence in Probability-Weak Law of Large Numbers (52);
Almost Sure Convergence-Strong Law of Large Numbers (53);
Convergence in Distribution-Central Limit Theorem (53);
Convergence in $L_p$ Norm (56);
Uniform Convergence and Glivenko-Cantelli Theorem (56); 
Divergence of measures (62).

\vspace*{0.5cm}
\small
\noindent
{\bf Lecture 2: Bayesian Inference}
\vspace{0.3cm}

\footnotesize
\noindent
{\bf 1 Elements of Parametric Inference} (64).
\noindent
{\bf 2 Exchangeable sequences} (65).
\noindent
{\bf 3 Predictive Inference} (67).
\noindent
{\bf 4 Sufficient Statistics} (68).
\noindent
{\bf 5 Exponential Family} (69).
\noindent
{\bf 6 Prior functions} (70):
Principle of Insufficient Reason (70);
Parameters of position and scale (71);
Covariance under reparameterizations (75);
The Fisher's Matrix (76);
Invariance under a Group of Transformations (81);
Conjugated Distributions (85);
Probability Matching Priors (88);
Reference Analysis (92).
\noindent
{\bf 7 Hierarchical Structures} (97).
\noindent
{\bf 8 Priors for discrete parameters} (99).
\noindent
{\bf 9 Constrains on parameters and priors} (100).
\noindent
{\bf 10 Decision Problems} (101): 
Loss and Risk functions (101); Hypothesis testing, Bayes Factor and BIC (102); 
Point estimation (106).
\noindent
{\bf 11 Credible Regions} (107).
\noindent
{\bf 12 Bayesian vs Classical Philosophy} (108): Confidence Intervals,
Neyman Construction, Feldman-Cousins Confidence Belts.
\noindent
{\bf 13 Some worked examples}: Regression (113); 
Characterization of a Possible Source of 
Events (116); Anisotropies of Cosmic Rays (118).

\vspace*{0.5cm}
\small
\noindent
{\bf Lecture 3: Monte Carlo Methods}
\vspace{0.3cm}

\footnotesize
\noindent
{\bf 1 Pseudo-Random Sequences} (125).
\noindent
{\bf 2 Basic Algorithms} (127): Inverse Transform (127); Simulation of a 
Photomultiplier Tube (130); Bootstrap (132),
Acceptance-Rejection (132);
Incorrect estimation of $\max_x\{p(x|\cdot)\}$ (134);
Weighted events (135);
Importance Sampling (136);
Stratified Sampling (136);
Decomposition of the probability density (137).
\noindent
{\bf 3 Everything at work} (138): The Compton Scattering (138);
An incoming flux of particles (143);
Sampling some continuous distributions of interest (145)
(Beta, Cauchy, Chi-squared, Dirichlet, Generalized Dirichlet,
Exponential, Gamma, Laplace, Logistic, Normal,
Pareto, Snedecor, Student, Uniform, Weibull).
\noindent
{\bf 4 Markov Chain Monte Carlo} (148): Metropolis-Hastings (150);
Path Integrals in Quantum Mechanics (154);
Sampling from Conditionals and Gibbs Sampling (160).
\noindent
{\bf 5 Evaluation of definite integrals} (162).

\vspace*{1.cm}
\small
\noindent
{\bf References} (165).

\newpage\null\thispagestyle{empty}\newpage

\setcounter{page}{1} 
\pagenumbering{arabic}
\rm

\footnotesize
\begin{tabular}{p{6.0cm}p{8.0cm}}
   & ``The Theory of Probabilities is basically nothing else but common
       sense reduced to calculus''
\end{tabular}
\begin{flushright}
\emph{P.S. Laplace}
\end{flushright}

\vspace*{1.cm}
\huge
\noindent
{\bf Lecture 1}

\vspace{0.5cm}
\noindent
\Huge
{\bf Probability}
\vspace{0.8cm}
\smill

\section{\LARGE \bf The Elements of Probability: $(\Omega,{\Bcal},{\mu})$}
The axiomatic definition of probability was introduced by
A.N. Kolmogorov in 1933 and starts with the concepts of
{\sl sample space} $(\Omega)$ and {\sl space of events}
$({\Bcal}_{\Omega})$ with structure of ${\sigma}$-algebra. When the pair 
$(\Omega,{\Bcal}_{\Omega})$ is equipped with a {\sl measure}
$\mu$ we have a {\sl measure space} $(E,{\Bcal},{\mu})$ and, if the
measure is a {\sl probability measure} $P$
we talk about a {\sl probability space}
$(\Omega,{\Bcal}_{\Omega},P)$. Lets discuss all these elements.

\subsection{Events and Sample Space: $(\Omega)$}

To learn about the state of nature, we do experiments and observations of the
natural world and ask ourselves questions about the outcomes.
In a general way, the {\sl object} of questions we may ask about the 
result of an experiment such that the possible answers are 
{\sl it occurs} or {\sl it does not occur} are called {\bf events}.
There are different kinds of events and among them we have the 
{\bf elementary events}; that is, those results of the
random experiment that {\bf can not} be
decomposed in others of lesser entity. 
The {\bf sample space} $(\Omega)$ is the set of {\bf all} the possible
{\bf elementary outcomes (events)} of a random experiment and they
have to be:
\begin{itemize}
   \item[i)]   {\bf exhaustive}: any possible outcome of the experiment
               has to be included in $\Omega$;
   \item[ii)]  {\bf exclusive}: there is no overlap of elementary results.
\end{itemize}
To study random phenomena we start by specifying the
{\sl sample space} and, therefore, we have to have a clear idea of what are the
possible results of the experiment. To center the ideas, consider the simple
experiment of rolling a die with 6 faces numbered from 1 to 6. We consider as
{\sl elementary events} 
 \begin{eqnarray*}
      e_i\,=\,\{ {\rm get\,\,the\,\,number\,\,} i {\rm \,\,on\,\,the\,\,
        upper\,\,face}\}\,\,;\,\,\,\,\,i=1,{\ldots},6
        \nonumber  
   \end{eqnarray*}
so ${\Omega}=\{e_1,{\ldots},e_6\}$.
Note that any possible outcome of the roll is included in
$\Omega$ and we can not have two or more elementary results simultaneously.
But there are other types of events besides the elementary ones.
We may be interested for instance in the parity of the number 
so we would like to consider also the possible results
\footnote{Given two sets $A,B{\subset}{\Omega}$, we shall denote
          by $A^c$ the {\sl complement} of $A$ (that is, the set of all 
          elements of $\Omega$ that are not in $A$) and by
          $A{\setminus}B\equiv A{\cap}B^c$ the {\sl set difference} 
          or {\sl relative complement
          of $B$ in $A$} (that is, the set of elements that are in $A$ but not
          in $B$). It is clear that $A^c=\Omega{\setminus}A$.}
\begin{eqnarray*}
      A=\{ {\rm get\,\,an\,\,even\,\,number} \} 
    \hspace{1.cm}{\rm and}\hspace{1.cm}
      {A}^c=
     \{ {\rm get\,\,an\,\,odd\,\,number} \} 
\end{eqnarray*}
They are not {\sl elementary} since the result $A=\{e_2,e_4,e_6\}$ is
equivalent to get $e_2$, $e_4$ or $e_6$ and ${A}^c=\Omega\setminus A$ to get 
$e_1$, $e_3$ or $e_5$. In general, 
an {\bf event} is any subset
\footnote{This is not completely true
if the sample space is non-denumerable since
there are subsets that can not be considered as events.
It is however true for the subsets
of ${\Rcal}^n$ we shall be interested in. We shall talk about that in
section 1.2.2.}
of the sample space and we shall distinguish between:

\begin{tabular}{p{3.5cm}p{7.7cm}}
                 & \\
 {\bf elementary events}:    &
          any element of the {\sl sample space} $\Omega$;
                                     \\ & \\
\end{tabular}

\begin{tabular}{p{3.5cm}p{7.7cm}}
 {\bf events}:             &
          any subset of the {\sl sample space};
                                     \\ & \\
\end{tabular}

\noindent
and two extreme events:

\begin{tabular}{p{3.5cm}p{7.7cm}}
                 & \\
 {\bf sure events}:    &
          $S_S=\{ {\rm get\,\,any\,\,result\,\,contained\,\,in}\,\,\Omega \}
          \equiv \Omega$
 \\ & \\
\end{tabular}

\begin{tabular}{p{3.5cm}p{7.7cm}}
 {\bf impossible events}:             &
      $S_I=\{ {\rm get\,\,any\,\,result\,\,not\,\,contained\,\,in}
             \,\,\Omega \}
          \equiv \emptyset$
 \\ & \\
\end{tabular}

\noindent
Any event that is neither {\sl sure} nor {\sl impossible} is called
{\bf random event}. Going back to the rolling of the die, sure events are
\begin{eqnarray*}
      S_S\,&=&\,\{ {\rm get\,\,a\,\,number\,\,} n\,| \, 1{\le}n{\le}6 \}\,=\,
      \Omega\hspace{1.cm}{\rm or} \\
      S_S\,&=&\,\{ {\rm get\,\,a\,\,number\,\,that\,\,is\,\,even\,\,or\,\,
      odd} \}\,=\,\Omega
\end{eqnarray*}
impossible events are
\begin{eqnarray*}
   S_I\,&=&\,\{ {\rm get\,\,an\,\,odd\,\,number\,\,that\,\,is\,\,not\,\,
                    prime} \}\,=\,\emptyset
      \hspace{1.cm}{\rm or} \\
  S_I\,&=&\,\{ {\rm get\,\,the\,\,number}\,\,7 \}\,=\,\emptyset
\end{eqnarray*}
and random events are any of the $e_i$ or, for instance,
\begin{eqnarray*}
      S_r\,=\,\{ {\rm get\,\,an\,\,even\,\,number
      } \}\,=\,\{e_2,\,e_4,\,e_6 \}
\end{eqnarray*}

Depending on the number of possible outcomes of the experiment, the 
the sample space can be:

\begin{tabular}{p{3.2cm}p{8.cm}}
                 & \\
 {\bf finite}:   &
                 if the number of elementary events is finite; \\
                 &  \\
                 &
                 \footnotesize
                 {\bf Example:}  In the rolling of a die, 
                 $\Omega=\{ e_i ;\,i=1,\ldots,6\}$ so
                 ${\rm dim}({\Omega})=6$.
                 \smill
                                     \\ & \\
\end{tabular}

\begin{tabular}{p{3.2cm}p{8.0cm}}
 {\bf countable}:    &
                 when there is a one-to-one correspondence between
                 the elements of $\Omega$ and ${\Ncal}$;\\
                 &  \\
                 &
                 \footnotesize
                 {\bf Example:} Consider the experiment of flipping a coin and 
                 stopping when we get $H$. Then
                 $\Omega=\{ H,\,TH,\,TTH,\,TTTH,\,{\ldots} \}$.
                 \smill
                                     \\ & \\
\end{tabular}

\begin{tabular}{p{3.2cm}p{8.cm}}
 {\bf non-denumerable}:    &
                 if it is neither of the previous; \\
                 &  \\
                 &
                 \footnotesize
                 {\bf Example:} For the decay time of an unstable particle 
                 $\Omega=\{t \in R | t {\ge} 0 \}=[0,{\infty})$ and for
                 the production polar angle of a particle
                 $\Omega=\{\theta \in R | 0 {\le} \theta {\le} \pi
                    \}=[0,{\pi}]$.
                 \smill
                                     \\ & \\
\end{tabular}
 
It is important to note that the {\sl events} are not necessarily 
numerical entities. We could have for instance the die with colored faces
instead of numbers. We shall deal with that when discussing {\sl random
quantities}. Last, given a sample space $\Omega$ we shall talk quite 
frequently about a {\sl partition} (or a {\sl complete system of events}); 
that is, a sequence $\{S_i\}$ of events, finite or countable, such that
\begin{eqnarray*}
    \Omega=\bigcup_{i}\,S_i \,\,\,\,{\rm (complete\,\,system)}
    \hspace{0.8cm}{\rm and}\hspace{0.8cm}
    S_i\,\bigcap_{\forall i,j}\,S_j\,=\,{\emptyset}
     \,\,;\,\,\,i{\neq}j\,\,\,\,
             {\rm (disjoint\,\,events)}
\end{eqnarray*}


\subsection{$\sigma$-algebras $({\Bcal}_{\Omega})$ and 
            Measurable Spaces $(\Omega,{\Bcal}_{\Omega})$} 

As we have mentioned, in most cases we are interested in events other than
the elementary ones. It is therefore interesting to consider a class of
events that contains all the possible results of the experiment we are
interested in such that when we ask about the union, intersection and
complements of events we obtain elements that belong the same class. 
A non-empty family ${\Bcal}_{\Omega}=\{S_i\}_{i=1}^n$ 
of subsets of the sample space $\Omega$ that is {\sl closed} (or stable)
under the operations of {\sl union} and {\sl complement}; that is
\begin{eqnarray*}
     S_i\, \cup \,S_j \, \in \, {\Bcal}\,\,;\,\,\,\,\,
      \forall S_i,S_j\, \in \, {\Bcal}
\hspace{1.cm}{\rm and} \hspace{1.cm}
     {S_i}^c\, \in \, {\Bcal}\,\,;\,\,\,\,\,
      \forall S_i\, \in \, {\Bcal} 
\end{eqnarray*}
is an {\bf algebra} ({\sl Boole algebra}) if $\Omega$ is finite. 
It is easy to see that if it is closed under
unions and complements it is also closed under intersections and the
following properties hold for all $S_i,S_j{\in}{\Bcal}_{\Omega}$:

\begin{tabular}{p{5.0cm}p{5.0cm}p{5.0cm}}
     \\ & \\
$\Omega \in {\Bcal}_{\Omega} $     &
     ${\emptyset}\in{\Bcal}_{\Omega} $  &
     $S_i \cap S_j \in {\Bcal}_{\Omega}$ \\ & & \\
     ${S_i}^c \cup  {S_j}^c \in {\Bcal}_{\Omega}$ &  
$({S_i}^c \cup {S_j}^c)^c \in \,{\Bcal}_{\Omega}$ & 
     $S_i \setminus  S_j \in {\Bcal}_{\Omega}$  
        \\ & & \\    
$\cup_{i=1}^{m}S_i \in {\Bcal}_{\Omega}$ & 
     $\cap_{i=1}^{m} S_i \in {\Bcal}_{\Omega}$ &
 \\ & & 
\end{tabular}

Given a sample space $\Omega$ we can construct different Boole algebras
depending on the events of interest. The smaller one is
${\Bcal}_m\,=\,\{ {\emptyset},\, \Omega \}$, the minimum algebra that
contains the event $A \subset \Omega$ has 4 elements:
${\Bcal}=\{ {\emptyset}, \Omega, A, A^c \}$
and the largest one,
${\Bcal}_M\,=\,\{ {\emptyset},\, \Omega,\,{\rm all\,\,possible\,\,subsets\,\, 
     of\,\,} \Omega \}$ will have $2^{{\rm dim}({\Omega})}$ elements.
From ${\Bcal}_M$ we can engender any other algebra by a finite number
of unions and intersections of its elements.


\subsubsection{$\sigma$-algebras}

If the sample space is countable, we have to generalize the Boole algebra
such that the unions and intersections can be done a countable number
of times getting always events that belong to the same class; that is: 
\begin{eqnarray*}
    \bigcup_{i=1}^{\infty}\, \,S_i \, \in \, {\Bcal} \hspace{1.cm}{\rm and}
    \hspace{1.cm}\bigcap_{i=1}^{\infty}\, \,S_i \, \in \, {\Bcal}
\end{eqnarray*}
with $\{S_i\}^{\infty}_{i=1} \in {\Bcal}$. These algebras are called
{\bf $\sigma$-algebras}. Not all the Boole algebras satisfy these properties
but the {\sl $\sigma$-algebras} are always Boole algebras (closed under finite 
union).

\vspace{0.5cm}
\noindent
{\rayan}                   
\vspace{0.35cm}
\footnotesize
\noindent

   Consider for instance a finite set $E$ and the class
   ${\cal A}$ of subsets of $E$ that are either finite
   or have finite complements. The finite union of subsets of  
   ${\cal A}$ belongs to ${\cal A}$ because the finite union of finite
   sets is a finite set and the finite union of sets that have finite
   complements has finite complement. However, the countable union of
   finite sets is countable and its complement will be an infinite set so
   it does not belong to ${\cal A}$. Thus, 
   ${\cal A}$ is a Boole algebra but not a $\sigma$-algebra.

   Let now $E$ be any infinite set and
   ${\cal B}$ the class of subsets of $E$ that are either countable
   or have countable complements. The finite or countable union of
   countable sets is countable and therefore belongs to ${\cal B}$.
   The finite or countable union of 
   sets whose complement is countable has a countable complement
   and also belongs to ${\cal B}$. Thus, ${\cal B}$ is a Boole algebra and
   $\sigma$-algebra.

{\rayan}                   
\vspace{1.0cm}
\smill

\subsubsection{Borel $\sigma$-algebras}

Eventually, we are going to assign a probability to the events of interest
that belong to the algebra and, anticipating concepts, probability is just 
a bounded measure so we need a class of {\sl measurable sets} with
structure of a {\sl $\sigma$-algebra}.
Now, it turns out that when the sample space $\Omega$  
is a non-denumerable topological space there exist non-measurable
subsets that obviously can not be considered as events
\footnote{Is not difficult to show the existence
of Lebesgue non-measurable sets in $\Rcal$.
One simple example is the {\sl Vitali set} constructed by G. Vitali in 1905 
although there are other interesting examples (Hausdorff, Banach-Tarsky) and
they all assume the Axiom of Choice. In fact, the work of
R.M. Solovay around the 70's shows that one can not prove the
existence of Lebesgue non-measurable sets without it. However, one can
not specify the choice function so one can prove their existence but
can not make an explicit construction in the sense Set Theorists would like.
In Probability Theory, we are interested only in Lebesgue measurable 
sets so those which are not have nothing to do in 
this business and Borel's algebra contains only measurable sets.}. 
We are particularly interested in ${\Rcal}$ (or, in general, in ${\Rcal}^n$)
so we have to construct a family ${\Bcal}_{\Rcal}$ of measurable 
subsets of ${\Rcal}$ that is
\begin{itemize}
    \item[i)] closed under countable number of intersections:
    $ \{ B_i\}^{\infty}_{i=1} \in {\Bcal}_{\Rcal} \longrightarrow
     \cap^{\infty}_{i=1}\,B_i \in {\Bcal}_{\Rcal}$
    \item[ii)] closed under complements:
    $ B  \in {\Bcal}_{\Rcal} \rightarrow
     {B}^c = {\Rcal}{\setminus}B \in {\Bcal}_{\Rcal}$
\end{itemize}
Observe that, for instance, the family
of all subsets of  ${\Rcal}$ satisfies the conditions $i)$ and $ii)$
and the intersection of any collection of families that satisfy them is
a family that also fulfills this conditions but not all are measurable.
Measurably is the key condition.
Let's start identifying what could be considered 
the equivalent of an {\sl elementary event}. The sample space
${\Rcal}$ is a linear set of points and, among it subsets, we have the
{\bf intervals}. In particular, if $a{\leq}b$ are any two points of
${\Rcal}$ we have: 
\begin{itemize}
    \item[$\bullet$] open intervals:
    $ (a,b)\, =\, \{ x \in {\Rcal}\, | \, a < x < b \} $
    \item[$\bullet$] closed intervals:
    $ [a,b]\, =\, \{ x \in {\Rcal}\, | \, a \leq x \leq b \} $
    \item[$\bullet$] half-open intervals on the right:
    $ [a,b)\, =\, \{ x \in {\Rcal}\, | \, a \leq x < b \} $
    \item[$\bullet$] half-open intervals on the left:
    $ (a,b]\, =\, \{ x \in {\Rcal}\, | \, a < x \leq b \} $
\end{itemize}
When $a=b$ the closed interval reduces to a point
$\{x=a\}$ ({\sl degenerated interval}) and the other three to the null set and,
when $a{\rightarrow}-{\infty}$ or $b{\rightarrow}{\infty}$ 
we have the {\sl infinite intervals}
$(-{\infty},b)$, $(-{\infty},b]$, $(a,{\infty})$ and $[a,{\infty})$.
The whole space ${\Rcal}$ can be considered as the interval
$(-{\infty},{\infty})$ and any interval will be a subset of ${\Rcal}$.
Now, consider the class of all intervals of 
${\Rcal}$ of any of the aforementioned types.
It is clear that the intersection of a finite or countable number of
intervals is an interval but the union is not necessarily an interval; for
instance $[a_1,b_1]\cup [a_2,b_2]$ with $a_2>b_1$ is not
an interval. Thus, this class is not additive and therefore not a closed family.
However, it is possible to construct an additive class including, along
with the intervals, other measurable sets so that any set formed by
countably many operations of unions, intersections
and complements of intervals is included in the family. 
Suppose, for instance, that we take the half-open intervals on the right 
$[a,b)$, $b>a$ as the initial class of sets
\footnote{The same algebra is obtained if 
one starts with $(a,b), (a,b]$ or $[a,b]$.}
to generate the algebra ${\Bcal}_{\Rcal}$
so they are in the bag to start with.
The open, close and degenerate intervals are
\begin{eqnarray*}
     (a,b)=\bigcup^{\infty}_{n=1}[a-1/n,b) \,\,;\hspace{0.5cm}
     [a,b]=\bigcap^{\infty}_{n=1}[a,b+1/n)
      \hspace{0.5cm}{\rm and}\hspace{0.5cm}
       {a}=\{x \in {\Rcal}|x = a\}=[a,a]
\end{eqnarray*}
so they go also to the bag as well as the
half-open intervals $(a,b]=(a,b)\cup [b,b]$ and the
countable union of unitary sets and their complements. Thus, 
countable sets like 
${\Ncal},{\Zcal}$ or ${\Qcal}$ are in the bag too. Those are the sets we shall
deal with.

The smallest family 
${\Bcal}_{\Rcal}$ (or simply $\Bcal$) of measurable subsets of ${\Rcal}$ 
that contains all intervals and is
closed under complements and countable number of intersections 
has the structure of a {\sl $\sigma$-algebra}, is called
{\bf Borel's algebra} and its elements are generically called 
{\bf Borel's sets} or {\sl borelians}
Last, recall that half-open
sets are Lebesgue measurable (${\lambda}((a,b])=b-a$) 
and so is any set built up from
a countable number of unions, intersections and complements so
all Borel sets are Lebesgue measurable and every Lebesgue measurable set
differs from a Borel set by at most a set of measure zero.
Whatever has been said about $\Rcal$ is applicable to the n-dimensional
euclidean space ${\Rcal}^n$.

The pair $(\Omega,{\Bcal}_{\Omega})$ is called {\bf measurable space} and
in the next section it will be equipped with a {\sl measure} and "upgraded" to
a {\sl measure space} and eventually to a {\sl probability space}. 

\subsection{Set Functions and Measure Space: $(\Omega,{\Bcal}_{\Omega},{\mu})$}

A function
$f:A\,\in\,{\Bcal}_{\Omega}\,\longrightarrow {\Rcal}$
that assigns to each set $A \in {\Bcal}_{\Omega}$ one, and only one real number,
finite or not, is called a {\bf set function}. Given a sequence
$\{A_i\}_{i=1}^n$ of subset of ${\Bcal}_{\Omega}$ pair-wise disjoint,
($A_i\cap A_j={\emptyset};\,\,i,j=1,{\ldots},n\,\,;\,\,i{\neq}j$) we say that
the {\sl set function} is {\bf additive} ({\sl finitely additive}) if:
\begin{eqnarray*}
  f\,\left( \bigcup^{n}_{i=1}\,A_i \right) \,=\,
  \sum^{n}_{i=1}\,f(A_i)
\end{eqnarray*}
or {\bf $\sigma$-additive} if, for
a countable the sequence  $\{A_i\}_{i=1}^{\infty}$ of pair-wise disjoint
sets of ${\Bcal}$, 
\begin{eqnarray*}
  f\,\left( \bigcup^{\infty}_{i=1}\,A_i \right) \,=\,
  \sum^{\infty}_{i=1}\,f(A_i)
\end{eqnarray*}
It is clear that any 
$\sigma$-additive set function is additive but the converse is not true.
A {\sl countably additive set function} is a {\bf measure on} 
the algebra ${\Bcal}_{\Omega}$, a {\sl signed measure} in fact. If
the $\sigma$-additive set function is 
$\mu :A\,\in\,{\Bcal}_{\Omega}\,\longrightarrow [0,\infty)$ (i.e.,
$\mu(A){\geq}0\,$) for all $ A\,\in\,{\Bcal}_{\Omega}$ it is a 
{\bf non-negative measure}. In what follows, 
whenever we talk about {\sl measures} ${\mu},{\nu},..$ on a $\sigma$-algebra
we shall assume that they are always non-negative measures without further 
specification.
If $\mu (A)=0$ we say that $A$ is a set of {\sl zero measure}.

The {\sl ``trio''} $(\Omega,{\Bcal}_{\Omega},\mu )$, 
with $\Omega$ a non-empty set,
${\Bcal}_{\Omega}$ a $\sigma$-algebra of the sets of $\Omega$
and $\mu$ a measure over ${\Bcal}_{\Omega}$ is called
{\bf measure space} and the elements of ${\Bcal}_{\Omega}$
{\em measurable sets}.

In the particular case of the n-dimensional euclidean space
$\Omega={\Rcal}^n$, the $\sigma$-algebra is the Borel algebra and
 {\bf all} the Borel sets are measurable. Thus, the intervals
$I$ of any kind are {\sl measurable sets} and satisfy that
\begin{itemize}
    \item[i)] If $I \in {\Rcal}$ is measurable 
              $\longrightarrow\,\,{I}^c={\Rcal}-I$ is measurable;
    \item[ii)] If $\{I\}^{\infty}_{i=1} \in {\Rcal}$ are measurable
              $\longrightarrow\,\,\cup ^{\infty}_{i=1}\, I_i$
              is measurable;
\end{itemize}
{\sl Countable sets} are Borel sets of zero measure for, if
$\mu$ is the Lebesgue measure, we have that
$\mu\,([a,b))\,=\,b-a$ and therefore:
\begin{eqnarray*}
  \mu\,(\{a\})\,=\,\lim_{n{\rightarrow}\infty}\,\mu \left(
     [a,a+1/n) \right)\,=\,\lim_{n{\rightarrow}\infty}\,
     \frac{\textstyle 1}{\textstyle n}\,=\,0
\end{eqnarray*}
Thus, {\sl any point} is a Borel set with {\em zero Lebesgue measure} and,
being ${\mu}$ a $\sigma$-additive function, any countable set has zero measure.
The converse is not true since there are borelians with zero measure
that are not countable (i.e. Cantor's ternary set).

In general, a measure $\mu$ over {\Bcal} satisfies that, for any
$A,B \in {\Bcal}$ not necessarily disjoint:

\begin{tabular}{p{0.2cm}p{12.5cm}}
&\begin{itemize}
\item[m.1)] $\, \mu (A \cup B)=\mu (A)+\mu (B{\setminus}A) $ 
\item[m.2)] $\, \mu (A \cup B)=\mu (A)+\mu (B) - \mu (A \cap B)$ 
          $\hspace{2.cm}(\mu (A \cup B) \le \mu (A)+\mu (B))$  
\item[m.3)] If $A\subseteq B$, then $\, \mu (B{\setminus}A)=\mu (B)-\mu (A)$ 
          $\hspace{1.4cm}(\ge 0$ since $\mu (B) \ge \mu (A))$ 
\item[m.4)] $\mu ({\emptyset})=0 $ 
\end{itemize}
\end{tabular}

\vspace{0.5cm}
\noindent
{\rayan}                   
\vspace{0.35cm}
\footnotesize
\noindent
\begin{itemize}
\item[m.1)] $A \cup B$ is the union of two disjoint sets
                 $A$ and $B{\setminus}A$ and the measure is an additive 
                 set function;

\item[m.2)]      $A\cap {B}^c$ and $B$ are disjoint and
                 its union is $A \cup B$ so
                 $\mu(A \cup B)=\mu (A\cap {B}^c)+\mu (B)$.
                 On the other hand $A\cap {B}^c$ and
                 $A\cap B$ are disjoint at its union is A so
                 $\mu (A\cap {B}^c)+
                 \mu (A\cap B) = \mu (A)$. It is enough to substitute
                 $\mu (A\cap {B}^c)$ in the previous expression;
\item[m.3)]      from m.1) and considering that, if
                 $A\subseteq B$, then $A\cup B=B$
\item[m.4)]      from m.3) with $B=A$. 
\end{itemize}

{\rayan}                   
\vspace{1.0cm}
\smill

A measure ${\mu}$ over a measurable space $(\Omega,{\Bcal}_{\Omega})$ is 
{\bf finite} if ${\mu}(\Omega)<{\infty}$ and
${\sigma}${\bf -finite} if 
$\Omega=\cup_{i=1}^{\infty}A_i$, with $A_i{\in}{\Bcal}_{\Omega}$ and 
$\mu(A_i)<\infty$.
Clearly, any finite measure is ${\sigma}$-finite but the converse is not 
necessarily true. For instance,  
the Lebesgue measure
${\lambda}$ in $({\Rcal}^n,{\Bcal}_{{\Rcal}^n})$ is not finite because
${\lambda}({\Rcal}^n)={\infty}$ but is ${\sigma}$-finite because
\begin{eqnarray}
  {\Rcal}^n\,=\,\bigcup_{k{\in}{\Ncal}}\,{[-k,k]}^n
                    \nonumber
\end{eqnarray}
and ${\lambda}({[-k,k]}^n)=(2k)^n$ is finite. As we shall see in lecture 2,
in some circumstances we shall be interested in the limiting behaviour of
${\sigma}$-finite measures over a sequence of compact sets.
As a second example, consider
the measurable space $({\Rcal},{\Bcal})$ and ${\mu}$ such that for
$A{\subset}{\Bcal}$ is ${\mu}(A)={\rm card}(A)$ if $A$ is finite and 
${\infty}$ otherwise.
Since ${\Rcal}$ is an uncountable union of finite sets,
${\mu}$ is not ${\sigma}$-finite in ${\Rcal}$. 
However, it is ${\sigma}$-finite in $({\Ncal},{\Bcal}_{\Ncal})$.

\subsubsection{Probability Measure}

Let $(\Omega,{\Bcal}_{\Omega})$ be a measurable space. A measure $P$ over
${\Bcal}_{\Omega}$ (that is, with domain in ${\Bcal}_{\Omega}$), image in
the closed interval $[0,1] \in {\Rcal}$ and such that
$P(\Omega)=1$ (finite) is called a {\bf probability measure} and 
its properties a just those of finite (non-negative) measures.
Expliciting the axioms,
a {\em probability measure} is a {\em set function} with 
{\em domain} in ${\Bcal}_{\Omega}$ and
{\em image} in the closed interval $[0,1] \in {\Rcal}$ that satisfies
three {\sl axioms}:

\begin{itemize}
    \item[i)]   {\bf additive:} \hspace{3.0cm}is an additive set function;
    \item[ii)]  {\bf no negativity:} \hspace{2.1cm}is a measure;
    \item[iii)] {\bf certainty:}\hspace{3.0cm}$P(\Omega)=1$.
\end{itemize}

These properties coincide obviously with those of the 
{\sl frequency and combinatorial probability} (see Note 1). 
All probability measures are finite
$(P(\Omega)=1)$ and any bounded measure can be converted in a
{\sl probability measure} by proper normalization.
The {\sl measurable space} $(\Omega,{\Bcal}_{\Omega})$ provided with
and probability measure $P$ is called the {\bf probability space}
$(\Omega,{\Bcal}_{\Omega},P)$. It is straight forward to see that
if $A,B \in {\Bcal}$, then:

\begin{tabular}{p{0.2cm}p{12.5cm}}
& \begin{itemize}
  \item[p.1)] $\, P({A}^c) = 1 - P(A)$
  \item[p.2)] $\, P({\emptyset})=0$
  \item[p.3)] $\, P(A \cup B)=P(A)+P(B{\setminus}A)=
                  P(A)+P(B)-P(A \cap B)\le P(A)+P(B))$
  \end{itemize}
\end{tabular}

\noindent
The property p.3 can be extended by recurrence to an arbitrary number of
events $\{A_i\}_{i=1}^n \in {\Bcal}$ for if $S_k={\cup}_{j=1}^kA_j$, then
$S_k=A_k{\cup}S_{k-1}$ and $P(S_n)=P(A_n)+P(S_{n-1})-P(A_n\cap S_{n-1})$.

Last, note that in the probability space 
$({\Rcal},{\Bcal},P)$ (or in $({\Rcal}^n,{\Bcal}_n,P)$), the set of points
$W\,=\,\{ \forall x \in {\Rcal}\,|\,P(x) > 0 \}$
is countable. Consider the partition 
\begin{eqnarray*}
  W\,=\,\bigcup_{k=1}^{\infty}\,W_k\hspace{1.cm}{\rm where}\hspace{1.cm}
    W_k\,=\,\{ \forall x \in {\Rcal}\,|\,1/(k+1) < P(x) \le 1/k \}
\end{eqnarray*}
If $x \in W$ then it belongs to one $W_k$ and, conversely,
if $x$ belongs to one $W_k$ the it belongs to $W$.
Each set $W_k$ has at most 
$k$ points for otherwise the sum of 
probabilities of its elements is $P(W_k) > 1$. Thus, the sets $W_k$ are finite
and since $W$ is a countable union of finite sets is a countable set.
In consequence, {\sl we can assign finite probabilities on at most a countable
subset of} ${\Rcal}$.
\vspace{0.5cm}
\noindent
{\rayan}                   
\vspace{0.35cm}
\footnotesize
\noindent
{\bf NOTE 1: What is {\sl {\bf probability}}?} 

It is very interesting to see how along the 500 years of history of 
probability many people 
(Galileo, Fermat, Pascal, Huygens, Bernoulli,
Gauss, De Moivre, Poisson, ...) have approached different problems and
developed concepts and theorems 
(Laws of Large Numbers, Central Limit, Expectation, Conditional Probability,...)
and a proper definition of probability has been so elusive.
Certainly there is a before and after Kolmogorov's 
{\sl "General Theory of Measure and Probability Theory"} and
{\sl "Grundbegriffe der Wahrscheinlichkeitsrechnung"} so from the 
mathematical point of view the question is clear after 1930's.
But, as  Poincare said in 1912: 
{\sl "It is very difficult to give a satisfactory definition of Probability'"}. 
Intuitively, What is probability?

The first {\sl ``definition''} 
of probability was the {\sl Combinatorial Probability} (${\sim}1650$). 
This is an objective concept (i.e., independent
of the individual) and is based on Bernoulli's {\sl Principle of Symmetry or
Insufficient Reason}: all the possible outcomes of the
experiment equally likely. For its evaluation we have to
know the cardinal (${\nu}({\cdot})$) of all possible results of the 
experiment $({\nu}(\Omega))$
and the probability for an event 
$A{\subset}\Omega$ is {\sl ``defined''} by the Laplace's rule:
$P(A)={\nu}(A)/{\nu}(\Omega)$
This concept of probability, implicitly admitted by Pascal and Fermat and 
explicitly stated by Laplace, is an {\sl a priory probability} in the sense 
that can be evaluated before or even without doing the experiment.
It is however 
meaningless if $\Omega$ is a countable set (${\nu}(\Omega)={\infty}$)
and one has to justify the validity of the Principle of Symmetry that not 
always holds originating some interesting debates.
For instance, in a problem attributed to D'Alembert, 
a player $A$ tosses a coin twice and wins if $H$ appears in at least one toss. 
According to Fermat, one can get $\{(TT),(TH),(HT),(HH)\}$ and $A$ will loose
only in the first case so being the four cases equally likely, the probability
for $A$ to win is $P=3/4$.
Pascal gave the same result. However, for Roberval one should consider only
$\{(TT),(TH),(H{\cdot})\}$ because if $A$ has won already if $H$ appears 
at the first toss so $P=2/3$. Obviously, Fermat and Pascal were right because,
in this last case, the three possibilities are not all equally likely and
the {\sl Principle of Symmetry} does not apply.

The second interpretation of probability is the {\sl Frequentist Probability},
and is based on the
idea of {\sl frequency of occurrence} of an event. If we repeat the experiment
$n$ times and a particular event $A_i$ appears $n_i$ times, the relative
frequency of occurrence is $f(A_i)=n_i/n$. As $n$ grows, it is observed
(experimental fact) that this number stabilizes around a certain value
and in consequence the probability of occurrence of $A_i$ is defined as
$P(A_i){\equiv}{\lim}_{n \rightarrow \infty}^{exp}f(A_i)$.
This is an objective concept inasmuch it is independent of the observer 
and is {\sl a posteriori} since it is based on what has been observed after the
experiment has been done through an
{\sl experimental limit} that obviously is not attainable. 
In this sense, it is more a practical rule than a definition.
It was also implicitly assumed by Pascal and Fermat (letters of de Mere to
Pascal: {\sl I have observed in my die games...}), by Bernoulli in his
{\sl Ars Conjectandi} of 1705 ({\sl Law of Large Numbers}) and finally
was clearly explicited
at the beginning of the XX'th century (Fisher and Von Mises).

Both interpretations of probability are restricted to observable quantities.
What happen for instance if they are not directly observable?, What if
we can not repeat the experiment a large number of times and/or under the same
conditions?
Suppose that you jump from the third floor down to ground
(imaginary experiment). Certainly, we can
talk about the probability that you break your leg but, how many times can 
we repeat the experiment under the same conditions?

During the XX'th century several people tried to pin down the concept of
probability. Pierce and, mainly, Popper argumented that probability represents
the {\sl propensity} of Nature to give a particular result in a {\bf single}
trial without any need to appeal at {\sl ``large numbers''}. This assumes
that the {\sl propensity}, and therefore the probability, exists in an
{\sl objective} way even though the {\sl causes} may be difficult to
understand. Others, like Knight, proposed that randomness is not a 
measurable property but just a problem of knowledge.
If we toss a coin and know precisely its shape, mass, acting forces,
environmental conditions,... we should be able to determine with certainty
if the result will be head or tail but since we lack the necessary
information we can not predict the outcome with certainty so we are lead to
consider that as a random process and use the Theory of Probability.
Physics suggests that it is not only a question of knowledge but randomness
is deeply in the way Nature behaves.

The idea that probability {\sl is a quantification of the degree
of belief that we have in the occurrence of an event} 
was used, in a more inductive manner, by Bayes
and, as we shall see, Bayes's theorem and the idea of information play 
an essential role in its axiomatization. To quote again Poincare, 
{\sl'' ... the probability of the causes, the most important from the point
of view of scientific applications.''}. It was still an open the question 
whether this quantification is {\sl subjective} or not. In the 20's,
Keynes argumented that it is not because, if we know all the elements
and factors of the experiment, what is likely to occur or not is 
determined in an objective sense regardless what is our opinion. 
On the contrary, Ramsey and de Finetti argued that the probability that is to
be assigned to a particular event depends on the degree of knowledge we have
({\sl personal beliefs}) and those do not have to
be shared by everybody so it is {\sl subjective}.
Furthermore they started the way towards a mathematical formulation of
this concept of probability consistent with Kolmogorov's axiomatic theory.
Thus, within the Bayesian spirit,  it is logical and natural 
to consider that 
{\bf probability is a measure of the degree of belief we have in the
occurrence of an event} that characterizes the random phenomena and 
we shall assign probabilities to events based on the prior knowledge we have.
In fact, to some extent, all statistical procedures used for the analysis of
natural phenomena are subjective inasmuch they all are based on 
a mathematical idealizations of Nature and all require a priory judgments
and hypothesis that have to be assumed.

{\rayan}                   
\vspace{1.0cm}
\smill


\subsection{Random Quantities}

In many circumstances, the possible outcomes of the experiments are not
numeric (a die with colored faces, a person may be sick or healthy, a
particle may decay in different modes,...) and, even in the case they are, 
the possible outcomes of the experiment may form a non-denumerable set.
Ultimately, we would like to deal with numeric values and 
benefit from the algebraic structures of the real numbers and the 
theory behind measurable functions and
for this, given a measurable space $(\Omega,{\Bcal}_{\Omega})$,
we define a function $X(w):w{\in}\Omega{\longrightarrow}{\Rcal}$ 
that assigns to each event $w$ of the sample space $\Omega$ 
{\sl one and only one real number}. 

In a more formal way, consider two measurable spaces 
$(\Omega,{\Bcal}_{\Omega})$ and $(\Omega',{\Bcal}'_{\Omega})$ and a function  
\begin{eqnarray*}
X(w):\,w{\in}\Omega\,{\longrightarrow}\,X(w){\in}\Omega^{'}
\end{eqnarray*}
Obviously, since we are interested in
the events that conform the $\sigma$-algebra
${\Bcal}_\Omega$, the same structure has to be maintained in 
$(\Omega',{\Bcal}'_{\Omega})$ by the application $X(w)$ for otherwise we wont
be able to answer the questions of interest.
Therefore, we require the function $X(w)$ to be
{\sl Lebesgue measurable with respect to the $\sigma$-algebra 
${\Bcal}_\Omega$}; i.e.:
\begin{eqnarray*}
 {X}^{-1}(B^{'})\,=\,B\,{\subseteq}\,{\Bcal}_\Omega\hspace{1.cm}
 {\forall}\,\,\,B^{'}\,\in\,{\Bcal}'_{\Omega}
  \nonumber
\end{eqnarray*}
so we can ultimately identify $P(B^{'})$ with $P(B)$.
Usually, we are interested in the case that 
${\Omega}'={\Rcal}$ (or ${\Rcal}^n$) so ${\Bcal}'_{\Omega}$
is the Borel $\sigma$-algebra and, since we have generated the Borel algebra
${\Bcal}$ from half-open intervals on the left
$I_x=(-{\infty},x]$ with $x{\in}{\Rcal}$, we have that 
$X(w)$ will be a Lebesgue measurable function over the Borel algebra
({\sl Borel measurable}) if, and only if:
\begin{eqnarray*}
 {X}^{-1}(I_x)\,=\,\{w{\in}\Omega\,|\,X(w){\leq}x\}
\,\in\,{\Bcal}_{\Omega}\,\,\,\,\,\,\,\,\,\,\,\,\,\,\,
 {\forall}\,\,\,x\,\in\,{\Rcal}
\end{eqnarray*}
We could have generated as well the Borel algebra from 
open, closed or half-open intervals
on the right so any of the following relations, all equivalent, 
serve to define a Borel measurable function $X(w)$:
\begin{itemize}
\item[1)] $\{w|X(w)>c\}{\in}{\Bcal}_{\Omega}$  ${\forall}c{\in}{\Rcal}$; 
\item[3)] $\{w|X(w){\geq}c\}{\in}{\Bcal}_{\Omega}$  ${\forall}c{\in}{\Rcal}$;
\item[4)] $\{w|X(w)<c\}{\in}{\Bcal}_{\Omega}$  ${\forall}c{\in}{\Rcal}$;
\item[5)] $\{w|X(w){\leq}c\}{\in}{\Bcal}_{\Omega}$  ${\forall}c{\in}{\Rcal}$
\end{itemize} 

\noindent
To summarize:
\begin{itemize}
\item[$\bullet$]
  Given a probability space
$(\Omega,{\Bcal}_{\Omega},Q)$, a {\bf random variable} is a function
$X(w):\Omega{\rightarrow}{\Rcal}$, Borel measurable over the
$\sigma$-algebra ${\Bcal}_{\Omega}$, that allows us to work with the 
induced probability space $(\Rcal,{\Bcal},P)$ 
\footnote{ It is important to note that a random variable
$X(w):\Omega{\longrightarrow}\Rcal$ is measurable
{\bf with respect to} the $\sigma$-algebra ${\Bcal}_\Omega$.}.
\end{itemize} 

Form this definition, it is clear that the name {\sl ``random variable''} is
quite unfortunate inasmuch it is a univoque function, neither random nor 
variable. Thus, at least to get rid of {\sl variable}, the 
term {\bf ``random quantity''} it is frequently used to 
design a numerical entity associated to the outcome of an experiment; outcome 
that is uncertain before we actually do the experiment and observe the result, 
and distinguish between the random quantity $X(w)$, 
that we shall write in upper cases and usually as $X$ assuming understood the 
$w$ dependence, and the value $x$ (lower case) taken in a particular 
realization of the experiment.
If the function $X$ takes values in ${\Omega_X}{\subseteq}{\Rcal}$ it will be a 
{\sl one dimensional random quantity} and, if the image is
${\Omega_X}{\subseteq}{\Rcal}^n$, it will be an ordered n-tuple of real numbers
$(X_1,\,X_2,\,{\ldots},\,X_n)$. Furthermore, attending to the cardinality
of ${\Omega}_X$, we shall talk about {\sl discrete random quantities}
if it is finite or countable and about {\sl continuous random quantities} 
if it is uncountable. This will be explained in more depth in section 3.1.
Last, if for each $w \in \Omega$ is $|X(w)|<k$ with $k$ finite, 
we shall talk about a {\sl bounded random quantity}.

The properties of random quantities are those of the measurable functions.
In particular, if $X(w):{\Omega}{\rightarrow}{\Omega}'$ is measurable with
respect to ${\Bcal}_\Omega$ and
$Y(x):{\Omega}'{\rightarrow}{\Omega}{''}$ is measurable with respect to
${\Bcal}_{{\Omega}{'}}$, the function 
$Y(X(w)):\Omega{\rightarrow}\Omega{''}$ is measurable with respect to
${\Bcal}_\Omega$ and therefore is a random quantity. We have then that
\begin{eqnarray*}
P(Y\,{\leq}\,y)\,=\,
P(Y(X)\,{\leq}\,y)\,=\,P(X\,{\in}\, Y^{-1}(I_X)) 
\end{eqnarray*}
where $Y^{-1}(I_X)$ is the set $\{x|x{\in}\Omega'\}$ such that
$Y(x){\leq}y$. 

\vspace{0.5cm}
\noindent
{\rayan}                   
\vspace{0.35cm}
\footnotesize
\noindent
{\bf NOTE 2: Indicator Function}
\vspace{0.35cm}

This is one of the most useful functions in maths. Given subset
$A{\subset}{\Omega}$ we define the
{\sl Indicator Function} ${\mathbf{1}}_{A}(x)$
for all elements $x{\in}{\Omega}$ as:
\begin{eqnarray}
  {\mathbf{1}}_{A}(x)\,=\,
   \left\{
     \begin{array}{l}
       1 \hspace{1.cm}{\rm if}\,\,x{\in}A  \\
       0 \hspace{1.cm}{\rm if}\,\,x{\notin}A 
     \end{array}
   \right.
  \nonumber
\end{eqnarray}
Given two sets $A,B{\in}{\Omega}$, the following relations are obvious:
\begin{eqnarray}
  {\mathbf{1}}_{A{\cap}B}(x)\,&=&\,{\rm min}\{{\mathbf{1}}_{A}(x),
                                           {\mathbf{1}}_{B}(x)\}\,=\,
           	{\mathbf{1}}_{A}(x)\,{\mathbf{1}}_{B}(x)	
      \nonumber \\
  {\mathbf{1}}_{A{\cup}B}(x)\,&=&\,{\rm max}\{{\mathbf{1}}_{A}(x),
                                           {\mathbf{1}}_{B}(x)\}\,=\,
	     {\mathbf{1}}_{A}(x)+{\mathbf{1}}_{B}(x)-	
         	{\mathbf{1}}_{A}(x)\,{\mathbf{1}}_{B}(x)	
      \nonumber \\
    {\mathbf{1}}_{A^c}(x)\,&=&\,1\,-\,{\mathbf{1}}_{A}(x)
      \nonumber
\end{eqnarray}
It is also called {\sl "Characteristic Function"} but in Probability Theory
we reserve this name for the Fourier Transform.
{\rayan}                   
\vspace{1.0cm}
\smill


\vspace{0.5cm}
\noindent
{\raya}                   
\vspace{0.35cm}
\footnotesize

\noindent
{\bf Example 1.1:} Consider the measurable space $(\Omega,{\Bcal}_\Omega)$ and
$X(w):\,\Omega\,{\rightarrow}\,{\Rcal}$. Then: 

\begin{itemize}
\item[$\bullet$] $X(w)\,=\,k$, constant in $\Rcal$. Denoting by 
$A=\{w{\in}\Omega|X(w)>c\}$ we have that if $c{\geq}k$ then
$A={\emptyset}$ and if $c<k$ then $A=\Omega$. Since
$\{{\emptyset},E\}{\in}{\Bcal}_\Omega$ we conclude that
$X(w)$ is a measurable function. In fact, it is left as an exercise to show
that for the minimal algebra 
${\Bcal}^{min}_{\Omega}=\{{\emptyset},\Omega\}$, the only functions that are
measurable are $X(w)={\rm constant}$.

\item[$\bullet$] Let $G{\in}{\Bcal}_\Omega$ and $X(w)={\mathbf{1}}_{G}(w)$
(see Note 2).
We have that if $I_a=(-{\infty},a]$ with $a{\in}{\Rcal}$, then
$a{\in}(-\infty,0) \rightarrow X^{-1}(I_a)=\emptyset$,
$a{\in}[0,1) \rightarrow X^{-1}(I_a)=G^c$,
and $a{\in}[1,\infty) \rightarrow X^{-1}(I_a)=\Omega$
so $X(w)$ is a measurable function with respect to ${\Bcal}_\Omega$.
A simple function
\begin{eqnarray*}
 X(w)\,=\,\sum_{k=1}^n\,a_k\,{\mathbf{1}}_{A_k}(w)
\end{eqnarray*}
where $a_k{\in}{\Rcal}$ and $\{A_k\}_{k=1}^n$ is a partition of $\Omega$
is Borel measurable and any random quantity that takes a finite number of
values can be expressed in this way.

\item[$\bullet$] Let $\Omega=[0,1]$. It is obvious that if $G$ is a 
non-measurable Lebesgue subset of $[0,1]$,  the function 
$X(w)={\mathbf{1}}_{G^c}(w)$ is not measurable over ${\Bcal}_{[0,1]}$ because 
$a{\in}[0,1) \rightarrow X^{-1}(I_a)=G{\notin}{\Bcal}_{[0,1]}$.

\item[$\bullet$] 
Consider a coin tossing, the elementary events
\begin{eqnarray*}
   e_1\,=\,\{H\},\hspace{0.5cm}{\rm and}\hspace{0.5cm}
   e_2\,=\,\{T\}\hspace{1.0cm}{\longrightarrow}\hspace{1.0cm}
    \Omega\,=\,\{ e_1,\,e_2\} 
\end{eqnarray*}
the algebra
${\Bcal}_\Omega\,=\,\{{\emptyset},\Omega, \{e_1\},\,\{e_2\}\}$ and the function 
$X:\Omega{\longrightarrow}{\Rcal}$ that denotes the number of heads
\begin{eqnarray*}
   X(e_1)\,=\,1 \hspace{0.5cm}{\rm and}\hspace{0.5cm}
   X(e_2)\,=\,0  
\end{eqnarray*}
Then, for $I_a=(-{\infty},a]$ with $a{\in}{\Rcal}$ we have that:
\begin{eqnarray*}
a{\in}(-\infty,0)\,\,\,{\longrightarrow}&&\,\,\,
X^{-1}(I_a)={\emptyset}{\in}{\Bcal}_\Omega \\
a{\in}[0,1)\,\,\,{\longrightarrow}&&\,\,\,
X^{-1}(I_a)={e_2}{\in}{\Bcal}_\Omega \\
a{\in}[1,\infty)\,\,\,{\longrightarrow}&&\,\,\,
X^{-1}(I_a)=\{e_1,e_2\}\,=\,\Omega{\in}{\Bcal}_E 
\end{eqnarray*}
so $X(w)$ is measurable in $(\Omega,{\Bcal}_\Omega,P)$ and therefore an 
admissible random quantity with $P(X=1)=P(e_1)$ and $P(X=0)=P(e_2)$. 
It will not be an admissible random quantity for the trivial minimum
algebra ${\Bcal}_\Omega^{min}\,=\,\{{\emptyset},\Omega\}$ since 
$e_2{\notin}{\Bcal}_\Omega^{min}$.
\end{itemize}

\vspace{0.35cm}

\noindent
{\bf Example 1.2:} 
Let $\Omega=[0,1]$ and consider the sequence of functions
$X_n(w)=2^n\,\mbox{\boldmath $1$}_{{\Omega}_n}(w)$ 
where
$w{\in}\Omega$, ${\Omega}_n=[1/2^n,1/2^{n-1}]$ and $n{\in}{\Ncal}$.
Is each $X_n(w)$ is measurable iff
${\forall}r{\in}{\Rcal}$,
$A=\{w{\in}\Omega\,|\,X_n(w)>r\}$ is a Borel set of
${\Bcal}_\Omega$. Then:
\begin{itemize}
\item[1)] $r{\in}(2^{n},\infty){\longrightarrow}A\,=\,
          {\emptyset}{\in}{\Bcal}_{\Omega}$ with ${\lambda}(A)=0$; 
\item[2)] $r{\in}[0,2^{n}]{\longrightarrow}
          A=[1/2^n,1/2^{n-1}]{\in}{\Bcal}_{\Omega}$
          with ${\lambda}(A)=2/2^n\,-\,1/2^n\,=\,1/2^n$.
\item[3)] $r{\in}(-\infty,0){\longrightarrow}
          A\,=\,[0,1]\,=\,\Omega$ with
          with ${\lambda}(\Omega)=1$.
\end{itemize}
Thus, each $X_n(w)$ is a measurable function.

\vspace{0.35cm}

\noindent
{\bf Problem 1.1:} Consider the experiment of tossing two coins, the
elementary events
\begin{eqnarray*}
e_1=\{H,H\}\,\, , \hspace{0.5cm}
e_2=\{H,T\}\,\, , \hspace{0.5cm}
e_3=\{T,H\}\,\, , \hspace{0.5cm}
e_4=\{T,T\}
\end{eqnarray*}
the sample space $\Omega=\{e_1,\,e_2,\,e_3,\,e_4\}$ and the two algebras
\begin{eqnarray*}
{\Bcal}_1\,&=&\,\{{\emptyset},\Omega,\{e_1\},\{e_4\},
                 \{e_1,e_2,e_3\},\{e_2,e_3,e_4\},
                 \{e_1,e_4\},\{e_2,e_3\}\} \\
{\Bcal}_2\,&=&\,\{{\emptyset},\Omega,\{e_1,e_2\},\{e_3,e_4\}\}
\end{eqnarray*}
The functions $X(w):\Omega{\longrightarrow}{\Rcal}$ such that
$X(e_1)=2;\,X(e_2)=X(e_3)=1;\,X(e_4)\,=\,0$
(number of heads) and $Y(w):\Omega{\longrightarrow}{\Rcal}$ such that
$Y(e_1)=Y(e_2)=1;\,Y(e_3)=Y(e_4)=0$,
with respect to which algebras are admissible random quantities?
(sol.:$X$ wrt ${\Bcal}_1$; $Y$ wrt ${\Bcal}_2$)

\vspace{0.35cm}

\noindent
{\bf Problem 1.2:}
Let $X_i(w):\Rcal{\longrightarrow}{\Rcal}$ with $i=1,\ldots,n$
be random quantities. Show that
\begin{eqnarray*}
Y={\rm max}\{X_1,X_2\}\,,\hspace{0.5cm}
Y={\rm min}\{X_1,X_2\}\,,\hspace{0.5cm}
Y={\rm sup}\{X_k\}_{k=1}^n \hspace{0.5cm}{\rm and}\hspace{0.5cm}
Y={\rm inf}\{X_k\}_{k=1}^n 
\end{eqnarray*}
are admissible random quantities. 
Observe that 
\begin{eqnarray*}
\{w|{\rm max}\{X_1,X_2\}\leq x\}&=&\{w|X_1(w)\leq x\}{\cap}
        \{w|X_2(w)\leq x\}{\in}\Bcal \\
\{w|{\rm min}\{X_1,X_2\}\leq x\}&=&\{w|X_1(w)\leq x\}{\cup}
        \{w|X_2(w)\leq x\}{\in}\Bcal \\
\{w|{\rm sup}_nX_n(w)\leq x\}&=&\cap_n\{w|X_n(w)\leq x\}{\in}\Bcal \\
\{w|{\rm inf}_nX_n(w)< x\}&=&\cup_n\{w|X_n(w)< x\}{\in}\Bcal
\end{eqnarray*}
\smill
{\raya}                   
\vspace{1.0cm}

\section{\LARGE \bf Conditional Probability and Bayes Theorem}

Suppose and experiment that consists on rolling a die
with faces numbered from one to six and the event
$e_2=$\{{\sl get the number two on the upper face}\}.  
If the die is fair, based on the Principle of Insufficient Reason 
you and your friend
would consider reasonable to assign equal chances to any of the possible 
outcomes and therefore a probability of $P_1(e_2)=1/6$.
Now, if I look at the die and tell you, and only you, that the outcome
of the roll is an even number, 
you will change your beliefs on the occurrence of event $e_2$ 
and assign the new value $P_2(e_2)=1/3$. Both of
you assign different probabilities because you do not share 
the same knowledge so it may be a truism but it is clear
that {\sl the probability we assign to an event
is subjective and is conditioned by the information we have about the 
random process}. In one way or another, probabilities are always conditional 
degrees of belief since there is always some state of information (even
before we do the experiment we know that whatever number we shall get is not 
less than one and not greater than six)
and we always assume some hypothesis (the die is fair
so we can rely on the Principle of Symmetry).

Consider a probability space $(\Omega,B_{\Omega},P)$ and  
two events $A,B \subset B_{\Omega}$ that are not disjoint so
$A\cap B \neq \hatch{0}$. The probability for both 
$A$ {\bf and} $B$ to happen is $P(A\cap B){\equiv}P(A,B)$. 
Since $\Omega=B \cup {B}^c$ and $B{\cap}{B}^c=\hatch{0}$
we have that:
\begin{eqnarray*}
 P(A)\,\equiv \,P(A\bigcap \Omega)\,=\,
  \underbrace{\begin{array}{c}
   \,\,\,\,\,\,\,\,\,\,\,\,\, P(A\bigcap B)
   \,\,\,\,\,\,\,\,\,\,\,\,\,
              \end{array}}_{\begin{array}{l}
                     {\rm probability\,\,for\,\,A} \\
                     {\rm {\bf and}\,\,B\,\,to\,\,occur}
                    \end{array}
                           }
   \,\,+\,\,
  \underbrace{\begin{array}{c}
   \,\,\,\,\,\,\,\,\,\,\,\,\, P(A\bigcap {B}^c)=P(A{\setminus}B)
   \,\,\,\,\,\,\,\,\,\,\,\,\,
              \end{array}}_{\begin{array}{l}
                 P(A{\setminus}B): 
                    {\rm probability\,\,for} \\
                     {\rm A\,\,to\,\,happen\,\,and\,\,{\bf not}\,\,B}
               \end{array}
                           }
\end{eqnarray*}
What is the probability for $A$ to happen if we know that $B$ has
occurred? The probability of $A$  {\sl conditioned}
to the occurrence of $B$ 
is called {\bf conditional probability of} $A$ {\bf given} $B$ and
is expressed as $P(A|B)$. This is equivalent to calculate the probability
for $A$ to happen in the probability space $(\Omega',B_{\Omega}',P')$ 
with $\Omega'$ the reduced sample space where $B$ has already
occurred and $B_{\Omega}'$ the corresponding sub-algebra
that does not contain ${B}^c$. We can set
$ P(A|B)\propto P(A\cap B)$ and define
(Kolmogorov) the conditional probability for $A$ 
to happen once $B$ has occurred as:
\begin{eqnarray*}
    P(A|B)\,\stackrel{def.}{=}\,\frac{\textstyle  P(A\, \bigcap \, B)}
                    {\textstyle P(B) }\,=\,
                \frac{\textstyle  P(A,B)}
                    {\textstyle P(B) }
\end{eqnarray*}
provided that $P(B){\neq}0$ for otherwise the conditional probability
is not defined. This normalization factor ensures that
$P(B|B)=P(B\cap B)/P(B)=1$.
Conditional probabilities 
satisfy the basic axioms of probability:
\begin{itemize}
\item[i)]   {\bf non-negative} since
            $(A\,\bigcap\,B)\,\subset\,B\,\rightarrow\,
             0\,\leq\,P(A|B)\,\leq\,1$
\item[   ]  
\item[ii)]  {\bf unit measure} ({\sl certainty}) since
            $
             P(\Omega|B)\,=\,\frac{\textstyle P(\Omega\, \bigcap \, B)}
                             {\textstyle P(B) } \,=\,
                        \frac{\textstyle P(B)}
                             {\textstyle P(B) } \,=\,1
            $

\item[iii)]  {\bf $\sigmabold$-additive}: For a countable sequence of 
       disjoint set 
             $\{A_i\}_{i=1}^{\infty}$
             \begin{eqnarray*}
               P\left(\bigcup^{\infty}_{i=1}\,A_i|B\right)=
               \frac{\textstyle
                 P\left((\bigcup^{\infty}_{i=1}\,A_i)\bigcap B \right)}
                    {\textstyle P(B)}=
                    \frac{\textstyle \displaystyle{\sum^{\infty}_{i=1}}
                      P(A_i\,\bigcap \, B)}
                         {\textstyle P(B)}=
                         \sum^{\infty}_{i=1}\,P(A_i|B)
               \end{eqnarray*}
\end{itemize}
Generalizing, for $n$ events $\{A_i\}_{i=1}^n$ we have, 
with $j=0,\ldots,n-1$ that
\begin{eqnarray*}
P(A_1,\dots,A_n) &=&\,P(A_n,\ldots,A_{n-j}|A_j, \dots ,A_1)P(A_j,\ldots,A_1)=\\
    &=&\,P(A_n|A_1, \ldots ,A_{n-1})P(A_3|A_2,A_1)P(A_2|A_1)P(A_1)
\end{eqnarray*}

\vspace*{0.3cm}
\subsection{Statistically Independent Events}
\vspace*{0.1cm}

Two events $A,B{\in}B_{\Omega}$ are {\bf statistically independent} when
the occurrence of one does not give any information about the occurrence of 
the other
\footnote{
In fact for the events $A,B{\in}B_{\Omega}$ 
we should talk about {\sl conditional independence} for it
is true that if $C{\in}B_{\Omega}$, it may happen that $P(A,B)=P(A)P(B)$ but
conditioned on $C$, $P(A,B|C){\neq}P(A|C)P(B|C)$ so $A$ and $B$ are related
through the event $C$. On the other hand, that $P(A|B){\neq}P(A)$ does not
imply that $B$ has a {\sl "direct"} effect on $A$. Whether this is the case
or not has to be determined by reasoning on the
process and/or additional evidences. Bernard Shaw said that we all should buy
an umbrella because there is statistical evidence that doing so
you have a higher life expectancy. And this is certainly true. However,
it is more reasonable to suppose that instead of the 
umbrellas having any mysterious influence on our health, in London, at the 
beginning of the ${\rm XX}^{\rm th}$ century, if you can afford to buy an 
umbrella you
have most likely a well-off status, healthy living conditions, access to
medical care,... 
}; that is, when
\begin{eqnarray*}
  P(A,B)\,=\,P(A)P(B)   
\end{eqnarray*}
A necessary and sufficient condition for $A$ and $B$ to be
independent is that $P(A|B)=P(A)$  (which implies $P(B|A)=P(B)$). Necessary
because 
\begin{eqnarray*}
P(A,B)=P(A)P(B)\hspace{0.5cm}{\longrightarrow}\hspace{0.5cm}
  P(A|B)\,=\,\frac{\textstyle  P(A,B)}
                  {\textstyle P(B) } \,=\,
             \frac{\textstyle  P(A)P(B)}
                  {\textstyle P(B) } \,=\,P(A)
\end{eqnarray*}
Sufficient because
\begin{eqnarray*}
P(A|B)=P(A)\hspace{0.5cm}{\longrightarrow}\hspace{0.5cm}
   P(A,B)\,=\,P(A|B)P(B)\,=\,P(A)P(B)
\end{eqnarray*}
If this is not the case, we say that they are {\sl statistically dependent} or
{\bf correlated}. In general, we have that:

\begin{tabular}{p{3.2cm}p{10.0cm}}
                 & \\
 $ P(A|B)\,>\, P(A)\,\rightarrow$&
                 the events $A$ and $B$ are 
                 {\bf positively correlated}; that is, that $B$ has already
                 occurred increases the chances for $A$ to happen; \\
    & \\
 $ P(A|B)\,<\, P(A)\,\rightarrow$&
                 the events $A$ and $B$ are
                 {\bf negatively correlated}; that is,
                 that $B$ has already
                 occurred reduces the chances for $A$ to happen; \\
& \\
 $ P(A|B)\,=\, P(A)\,\rightarrow$&
                 the events $A$ and $B$ are {\bf not correlated} so
                 the occurrence of $B$ does not modify
                 the chances for $A$ to happen.
                                     \\ & \\
\end{tabular}

Given a  finite collection of events ${\cal{A}}=\{A_i\}_{i=1}^n$ with
$A_{\forall i}\subset B_{\Omega}$, they are statistically independent if
\begin{eqnarray*}
  P(A_1, {\ldots},A_m)\,=\,P(A_1)\,{\cdots}\,P(A_m)  
\end{eqnarray*}
for any finite subsequence $\{A_k\}_{k=j}^m$; $1\leq j<m \leq n$
of events. Thus, for instance, for a sequence of 3 events 
$\{A_1,A_2,A_3\}$ the condition of independence requires that:
\begin{eqnarray*}
  P(A_1,A_2)=P(A_1)P(A_2)\hspace{0.2cm};\hspace{0.4cm}
  P(A_1,A_3)=P(A_1)P(A_3)\hspace{0.2cm};\hspace{0.4cm}
  P(A_2,A_3)&=&P(A_2)P(A_3)\\
 {\rm {\bf and}}\hspace{0.5cm}
  P(A_1,A_2,A_3)=P(A_1)P(A_2)P(A_3)
\end{eqnarray*}
so the events $\{A_1,A_2,A_3\}$ may be statistically dependent and
be pairwise independent. 


\vspace{0.5cm}
\noindent
{\raya}                   
\vspace{0.35cm}
\footnotesize

\noindent
{\bf Example 1.3:} In four cards ($C_1,C_2,C_3$ and $C_4$)
   we write the numbers 1 $(C_1)$, 2 $(C_2)$, 3 $(C_3)$ and
   123 $(C_4)$ and make a fair random extraction. 
   Let be the events
   \begin{eqnarray*}
      A_i\,&=&\,\{ {\rm the\,\,chosen\,\,card\,\,has\,\,
                        the\,\,number\,\,} i \} 
   \end{eqnarray*}
   with $i=1,2,3$. Since the extraction is fair we have that:
   \begin{eqnarray*}
        P(A_i)\,=\,P(C_i)\,+\,P(C_4)\,=\,1/2 
   \end{eqnarray*}
 Now, I look at the card and tell you that it has number $j$.
Since you know that $A_j$ has happened, you know that
the extracted card was either $C_j$ or $C_4$ and the only possibility to
have $A_i{\neq}A_j$ is that the extracted card was $C_4$ so the
conditional probabilities are
   \begin{eqnarray*}
        P(A_i|A_j)\,=\,1/2\,\,;\,\,\,\,i,j\,=\,1,2,3
        \,\,;\,\,\,\,i{\neq}j  
      \end{eqnarray*}
The, since
   \begin{eqnarray*}
        P(A_i|A_j)\,=\,P(A_i)\,\,;\,\,\,\,i,j\,=\,1,2,3
        \,\,;\,\,\,\,i{\neq}j        
   \end{eqnarray*}
any two events $(A_i,A_j)$ are (pairwise) independent. However:
   \begin{eqnarray*}
        P(A_1,A_2,A_3)\,=\,
        P(A_1|A_2,A_3)\,P(A_2|A_3)\,P(A_3)
   \end{eqnarray*}
and if I tell you that events
$A_2$ and $A_3$ have occurred then you are certain that chosen 
card is $C_4$ and therefore $A_1$ has happened too so
$P(A_1|A_2,A_3)=1$. But
\begin{eqnarray*}
  P(A_1,A_2,A_3)\,=\,1\,\frac{1}{2}\,\frac{1}{2}\,\neq\,
  P(A_1)P(A_2)P(A_3)\,=\,\frac{1}{8}
\end{eqnarray*}
so the events $\{ A_1,A_2,A_3 \}$ are {\bf not} independent even though
they are pairwise independent.

\vspace{0.35cm}

\noindent
{\bf Example 1.4:} {\bf Bonferroni's Inequality}.
Given a finite collection 
$A=\{A_1,{\ldots},A_n\} \subset {\Bcal}$ of events, Bonferroni's inequality 
states that:
\begin{eqnarray*}
  P(A_1\bigcap {\cdots}\bigcap A_n)\,\equiv\,
  P(A_1, {\ldots}, A_n)\,{\geq}\,
  P(A_1)\,+\,{\ldots}\,+\,P(A_n) \,-\,(n-1)   
\end{eqnarray*}
and gives a lover bound for the joint probability $P(A_1, {\ldots}, A_n)$.
For $n=1$ it is trivially true since $ P(A_1)\,{\geq}\,P(A_1)$. 
For $n=2$ we have that
\begin{eqnarray*}
  P(A_1\bigcup A_2)=P(A_1)+P(A_2)-P(A_1\bigcap A_2)\leq 1
  \hspace{0.5cm}{\longrightarrow}\hspace{0.5cm}
  P(A_1\bigcap A_2){\geq}P(A_1)+P(A_2)-1
\end{eqnarray*}
Proceed then by induction.
Assume the statement is true for $n-1$ and see if it is so for $n$.
If $B_{n-1}=A_1\cap {\ldots} \cap A_{n-1}$ and apply the result we got for 
$n=2$ we have that
\begin{eqnarray*}
  P(A_1\bigcap {\cdots} \bigcap A_n)\,=\,
  P(B_{n-1}\bigcap A_n)\,{\geq}\,P(B_{n-1})\,+\,P(A_n)\,-\,1
\end{eqnarray*}
but
\begin{eqnarray*}
  P(B_{n-1})\,=\,
  P(B_{n-2}\bigcap A_{n-1})\,{\geq}\,P(B_{n-2})\,+\,P(A_{n-1})\,-\,1
\end{eqnarray*}
so
\begin{eqnarray}
  P(A_1\bigcap {\ldots} \bigcap A_n)\,{\geq}\,
  P(B_{n-2})\,+\,P(A_{n-1})\,+\,P(A_n)\,-\,2
  \nonumber
\end{eqnarray}
and therefore the inequality is demonstrated.

\smill
{\raya}                   
\vspace{1.0cm}

\subsection{Theorem of Total Probability}

Consider a probability space $(\Omega,B_{\Omega},P)$ and a partition 
${\Scal}=\{S_i\}_{i=1}^n$ of the sample space. Then, 
for any event $A\,{\in}\,B_{\Omega}$ we have that
$A=A \bigcap {\Omega}=A \bigcap ({\bigcup}^{n}_{i=1}\,S_i)$ and therefore:
\begin{eqnarray*}
  P(A)=
       P\left(A \bigcap \left[{\bigcup}^{n}_{i=1}\,S_i\right]\right)=
       P\left({\bigcup}^{n}_{i=1} \left[A \bigcap S_i\right]\right)=
       \sum^{n}_{i=1}\,P(A \bigcap S_i) 
=\displaystyle{\sum^{n}_{i=1}}\,P(A|S_i){\cdot}P(S_i)
\end{eqnarray*}
Consider now a second
different partition of the sample space $\{B_k\}_{k=1}^m$. Then, for
each set $B_k$ we have
\begin{eqnarray*}
  P(B_k)\,=\,\sum^{n}_{i=1}\,P(B_k|S_i)P(S_i)
    \,\,;\,\,\,\, k=1,\ldots,m 
\end{eqnarray*}
and
\begin{eqnarray*}
  \sum_{k=1}^m P(B_k)\,=\,
  \sum^{n}_{i=1}\,P(S_i)\left[\sum_{k=1}^n\,P(B_k|S_i)\right]\,=\,
  \sum^{n}_{i=1}\,P(S_i)\,=1
\end{eqnarray*}
Last, a similar expression can be written for conditional probabilities.
Since
\begin{eqnarray*}
  P(A,B,{\Scal})\,=\,P(A|B,{\Scal})P(B,{\Scal})\,=\,
  P(A|B,{\Scal})P({\Scal}|B)P(B)
\end{eqnarray*}
and
\begin{eqnarray*}
  P(A,B)\,=\,\sum_{i=1}^nP(A,B,S_i)
\end{eqnarray*}
we have that
\begin{eqnarray*}
  P(A|B)=\frac{P(A,B)}{P(B)}=
 \frac{1}{P(B)}\sum_{i=1}^nP(A,B,S_i)=
 \sum_{i=1}^nP(A|B,S_i)P(S_i|B)
\end{eqnarray*}

\vspace{0.5cm}
\noindent
{\raya}                   
\vspace{0.35cm}
\footnotesize

\noindent
{\bf Example 1.5:} We have two indistinguishable urns:
   $U_1$ with three white and two black balls
   and $U_2$ with two white balls and three black ones.
   What is the probability that in a random extraction we get a white ball?

   Consider the events:
   \begin{eqnarray*}
     A_1\,=\,\{ {\rm choose\,\,urn\,\,} U_1 \}\,; 
     \hspace{1.cm}
     A_2\,=\,\{ {\rm choose\,\,urn\,\,} U_2 \} 
     \hspace{1.cm}{\rm and}\hspace{1.cm}
     B\,=\,\{ {\rm get\,\,a\,\,white\,\,ball} \}
   \end{eqnarray*}
   It is clear that $A_1 \cap A_2 = \hatch{0}$ and that
   $A_1 \cup A_2 = \Omega$. Now:
   \begin{eqnarray*}
     P(B|A_1)\,=\,\frac{3}{5}\,;
     \hspace{1.cm}
     P(B|A_2)\,=\,\frac{2}{5}
     \hspace{1.cm}{\rm and}\hspace{1.cm}
     P(A_1)=P(A_2)=\frac{1}{2}
   \end{eqnarray*}
   so we have that
   \begin{eqnarray*}
     P(B)\,=\,\sum_{i=1}^{2}\,P(B|A_i){\cdot}P(A_i)\,=\,
     \frac{3}{5}\frac{1}{2}\,+\,\frac{2}{5}\frac{1}{2}\,=\,\frac{1}{2}
   \end{eqnarray*}
   as expected since out of 10 balls, 5 are white.

\smill
{\raya}                   
\vspace{1.0cm}

\subsection{Bayes Theorem}

Given a probability space $(\Omega,B_{\Omega},P)$ we have seen that
the joint probability for for two events $A,B{\in}B_{\Omega}$ can be expressed
in terms of conditional probabilities as:
\begin{eqnarray*}
  P(A,B)\,=\,P(A|B)P(B)\,=\,P(B|A)P(A)
\end{eqnarray*}
The Bayes Theorem 
(Bayes $\sim$1770's and independently Laplace ${\sim}$1770's)
states that if $P(B){\neq}0$, then
\begin{eqnarray*}
  P(A|B)\,=\,\frac{\textstyle P(B|A)P(A)}{\textstyle P(B)}
\end{eqnarray*}
apparently a trivial statement but with profound consequences. Let's see other
expressions of the theorem. If
${\Hcal}=\{H_i\}_{i=1}^n$ is a partition of the sample space then
\begin{eqnarray*}
  P(A,H_i)\,=\,P(A|H_i)P(H_i)\,=\,P(H_i|A)P(A)
\end{eqnarray*}
and from the Total Probability Theorem
\begin{eqnarray*}
 P(A)\,=\,\sum^{n}_{k=1}\,P(A|H_k)P(H_k)
\end{eqnarray*}
so we have a different expression for Bayes's Theorem:  
\begin{eqnarray*}
  P(H_i|A)\,=\,\frac{\textstyle P(A|H_i)P(H_i)}{\textstyle P(A)}\,=\,
\frac{\textstyle P(A|H_i)P(H_i)}{\textstyle\sum^{n}_{k=1}\,P(A|H_k)P(H_k)}
\end{eqnarray*}
Let's summarize the meaning of these terms
\footnote{Although is usually the case, the terms {\sl prior} 
                 and {\sl posterior} do not necessarily imply a temporal 
                 ordering.}:

\begin{tabular}{p{2.cm}p{11.0cm}}
                 & \\
  $P(H_i):$& 
                is the probability of occurrence of the event $H_i$ 
                before we know if event $A$ has happened or not; that is,
                the degree of confidence we have in the occurrence of the
                event $H_i$ before we do the experiment so it is called
                {\sl {\bf prior probability}};  \\ & \\
$P(A|H_i):$&
                 is the probability for event $A$ to happen given that
                 event $H_i$ has occurred. This may be different 
                 depending on $i=1,2,\ldots,n$ and when considered as
                 function of $H_i$ is usually called {\bf likelihood}; 
                 \\ & \\
$P(H_i|A)\,:$&   is the degree of confidence we have in the occurrence of
                 event $H_i$ given that the event $A$ has happened. 
                 The knowledge that the event $A$ has occurred 
                 provides information about the random process and modifies the
                 beliefs we had in $H_i$ before
                 the experiment was done (expressed by $P(H_i)$)
                 so it is called {\bf posterior probability}; 
                 \\ & \\
$P(A)\,:$ &      is simply the normalizing factor. \\
                         & \\
\end{tabular}

\noindent
Clearly, if the events $A$ and $H_i$ are independent,
the occurrence of $A$ does not provide any information on the chances 
for $H_i$ to happen. Whether it has occurred or not does not modify our
beliefs about $H_i$ and therefore $P(H_i|A)=P(H_i)$.

In first place, it is interesting to note that the occurrence of $A$ restricts 
the sample space for $\Hcal$ and modifies the prior
chances $P(H_i)$ for $H_i$ in the same proportion as the occurrence of 
$H_i$ modifies the probability for $A$ because
\begin{eqnarray*}
  P(A|H_i)P(H_i)\,=\,P(H_i|A)P(A)  
 \hspace{1.0cm}{\longrightarrow}\hspace{1.0cm}
   \frac{\textstyle  P(H_i|A)}
        {\textstyle  P(H_i)}\,=\,
   \frac{\textstyle  P(A|H_i)}
        {\textstyle  P(A)}
\end{eqnarray*}
Second, from Bayes Theorem we can obtain {\sl relative posterior probabilities}
(in the case, for instance, that $P(A)$ is unknown) because 
\begin{eqnarray*}
   \frac{\textstyle  P(H_i|A)}
        {\textstyle  P(H_j|A)}\,=\,
   \frac{\textstyle  P(A|H_i)}
        {\textstyle  P(A|H_j)}\,
   \frac{\textstyle  P(H_i)}
        {\textstyle  P(H_j)}
\end{eqnarray*}
Last, conditioning all the probabilities to $H_0$
(maybe some conditions that are assumed) we get a
third expression of Bayes Theorem
\begin{eqnarray*}
        P(H_i|A,H_0)\,=\,
        \frac{\textstyle  P(A|H_i,H_0)P(H_i|H_0)}
       {\textstyle P(A|H_0)}=
             \frac{\textstyle  P(A|H_i,H_0)P(H_i|H_0)}
             {\textstyle \sum_{k=1}^{n}P(A|H_k,H_0)P(H_k|H_0)}
\end{eqnarray*}
where $H_0$ represents to some initial state of information or some conditions
that are assumed.
The {\sl posterior degree of credibility} we have
on $H_i$ is certainly meaningful when we have an initial degree of information
and therefore is relative to our {\sl prior} beliefs. And those are subjective
inasmuch different people may assign a different prior degree of
credibility based on their previous knowledge and experiences.
Think for instance in soccer pools. Different people will assign
different prior probabilities to one or other team depending on what they
know before the match and this information may not be shared by all of them. 
However, to the extent that they share common prior knowledge they will 
arrive to the same conclusions.  

Bayes's rule provides a natural way to include new information and update
our beliefs in a sequential way. After the event (data) $D_1$ has been 
observed, we have
\begin{eqnarray*}
  P(H_i)\hspace{1.cm}{\longrightarrow}\hspace{1.cm}
  P(H_i|D_1)\,=\,
  \frac{\textstyle P(D_1|H_i)}{\textstyle P(D_1)}\,P(H_i)\propto
  P(D_1|H_i)P(H_i)
      \nonumber
\end{eqnarray*}
Now, if we get additional information provided by the observation of $D_2$
(new data) we {\sl''update''} or beliefs on $H_i$ as:
\begin{eqnarray*}
   P(H_i|D_1){\longrightarrow}
  P(H_i|D_2,D_1)\,=\,
  \frac{\textstyle P(D_2|H_i,D_1)}{\textstyle P(D_2|D_1)}\,
       P(H_i|D_1)=
 \frac{\textstyle P(D_2|H_i,D_1) P(D_1|H_i) P(H_i)}{\textstyle P(D_2,D_1)}
\end{eqnarray*}
and so on with further evidences.

\vspace{0.5cm}
\noindent
{\raya}                   
\vspace{0.35cm}
\footnotesize

\noindent
{\bf Example 1.6:} 
An important interpretation of Bayes Theorem is that based on the
relation {\bf cause-effect}. Suppose that the event $A$ ({\sl effect})
has been produced by a certain {\sl cause} $H_i$. We consider all
possible causes (so $\Hcal$ is a complete set) and among them we have
interest in those that seem more plausible to explain the 
observation of the event $A$. 
Under this scope, we interpret the terms appearing in Bayes's formula as:

\begin{tabular}{p{2.cm}p{11.5cm}}
                 & \\
$P(A|H_i,H_0)\,:$&
                 is the probability that the {\bf effect}
                 $A$ is produced by the {\bf cause} (or hypothesis)
                 $H_i$; \\ & \\
$P(H_i,H_0)\,:$&
                 is the prior degree of credibility we assign to the 
                 {\bf cause} $H_i$ before we know that $A$ has occurred;
                 \\ & \\
$P(H_i|A,H_0)\,:$&
                 is the posterior probability we have for $H_i$ being 
                 the cause of the event (effect) $A$ that has already been 
                 observed.
                 \\ & \\
\end{tabular}

\noindent
Let's see an example of a clinical diagnosis just because the problem 
is general enough and conclusions may be more disturbing. 
If you want, replace {\sl individuals} by {\sl events} and for instance
({\sl sick,healthy}) by ({\sl signal,background}).
Now, the incidence of certain rare disease is of 1 every 10,000 people and
there is an efficient diagnostic test such that:
\begin{itemize}
\item[1)] If a person is sick, the tests gives positive in $99\%$ of the cases;
\item[2)] If a person is healthy, the tests may fail and give positive (false
          positive) in $0.5\%$ of the cases;
\end{itemize}
In this case, the {\bf effect} is to give 
positive $(T)$ in the test
and the exclusive and exhaustive hypothesis for the {\bf cause} are:
\begin{eqnarray*}
     H_1\,:\,\,\,
         \,\,{\rm be\,\,sick} 
         \hspace{1.5cm}{\rm and}\hspace{1.5cm}
     H_2\,:\,\,\,
         \,\,{\rm be\,\,healthy}
   \end{eqnarray*}
with $H_2={H_1}^c$.
A person, say you, is chosen {\bf randomly} ($H_0$) 
among the population 
to go under the test and give positive. Then you are scared when
they tell you: {\sl "The probability of giving positive being 
healthy is $0.5\%$, very small"} (p-value). 
There is nothing wrong with the statement but it has to 
be correctly interpreted and usually it is not. 
It means no more and no less than what the expression
$P(T|H_2)$ says: {\sl  ``under the assumption that you are healthy ($H_2$)
the chances of giving positive are $0.5\%$''} and this
is nothing else but a feature of the test. It doesn't say anything about 
$P(H_1|T)$, the chances you have to be sick giving positive in the test 
that, in the end, is what you are really interested in.
The two probabilities are related by an additional piece of information 
that appears in Bayes's formula: $P(H_1|H_0)$; that is,
{\sl under the hypothesis that you have been chosen at random ($H_0$), 
What are the prior chances to be sick?}. 
From the prior knowledge we have, the degree of credibility we assign to
both hypothesis is
\begin{eqnarray*}
P(H_1|H_0)\,=\,\frac{1}{10000}
\hspace{1.5cm}{\rm and}\hspace{1.5cm}
P(H_2|H_0)\,=\,1\,-\,P(H_1)\,=\,\frac{9999}{10000} 
\end{eqnarray*}
On the other hand, if $T$ denotes the event {\sl give positive in the test}
we know that:
\begin{eqnarray*}
  P(T|H_1)\,=\,\frac{99}{100}   
\hspace{1.5cm}{\rm and}\hspace{1.5cm}
  P(T|H_2)\,=\,\frac{5}{1000}  
\end{eqnarray*}
Therefore, Bayes's Theorem tells that the probability to be sick 
positive in the test is
\begin{eqnarray}
    P(H_1|T)\,=\,\frac{\textstyle  P(T|H_1){\cdot}P(H_1|H_0)}
        {\textstyle \sum_{i=1}^{2}\,P(T|H_i){\cdot}P(H_i|H_0) }  \,=\,
        \frac{\textstyle  \frac{99}{100} \frac{1}{10000}}
        {\textstyle \frac{99}{100} \frac{1}{10000} + 
        \frac{5}{1000} \frac{9999}{10000}}  \,\simeq\,0.02
      \nonumber 
\end{eqnarray}
Thus,  even if the test looks very efficient
and you gave positive, the fact that you were chosen at random 
and that the incidence of the disease in the population is very small, 
reduces dramatically the degree of belief you assign to be sick. 
Clearly, if you were not chosen randomly but because there is a suspicion 
from to other symptoms that you are sic,
prior probabilities change.
 
\smill
{\raya}                   
\vspace{1.0cm}

\section{\LARGE \bf Distribution Function} 
\smill

A one-dimensional Distribution Function is a real
function $F:{\Rcal}{\rightarrow}{\Rcal}$ that:
\begin{itemize}
    \item[p.1)]   is monotonous non-decreasing: 
                $F(x_1)\,\leq\,F(x_2)\,\,\,\,\forall x_1<x_2{\in}\Rcal$
    \item[p.2)]  is everywhere continuous on the right:
                $\lim_{{\epsilon}\rightarrow 0^+}F(x+\epsilon)=F(x)$
                $\forall x{\in}\Rcal$
    \item[p.3)] $F({-\infty})\equiv\lim_{x\rightarrow -\infty}F(x)=0$ and 
                $F({\infty})\equiv\lim_{x\rightarrow +\infty}F(x)=1$. 
\end{itemize}
and there is ([Bo06]) a unique Borel measure $\mu$ on $\Rcal$ that satisfies
$\mu((-\infty,x])=F(x)$ for all $x{\in}{\Rcal}$.
In the Theory of Probability, we define the Probability Distribution Function
\footnote{The condition $P(X\leq x)$ is due to the requirement that
  $F(x)$ be continuous on the right. This is not essential in the sense
  that any non-decreasing function $G(x)$, defined on $\Rcal$, bounded
  between 0 and 1 and continuous on the left
  $(G(x)=\lim_{{\epsilon}{\rightarrow}0^+}G(x-{\epsilon}))$
  determines a distribution function defined as $F(x)$ for all $x$ where
  $G(x)$ is continuous and as $F(x+{\epsilon})$ where $G(x)$ is 
  discontinuous. In fact, in the general theory of measure it is more 
  common to consider continuity on the left.} 
of the random quantity $X(w):\Omega{\rightarrow}\Rcal$ as:
\begin{eqnarray*}
  F(x)\,\stackrel{def.}{=}\,P(X\,{\leq}\,x)\,=\,
  P\left({X}\,{\in}\,(-{\infty},x] \right)\,\,;\,\,\,\,\,
    \forall x\,\in\,{\Rcal} 
\end{eqnarray*}
Note that the Distribution Function $F(x)$ is defined
for all $x{\in}\Rcal$ so if ${\rm supp}\{P(X)\}=[a,b]$, then
$F(x)=0$ ${\forall}x<a$ and $F(x)=1$ ${\forall}x{\geq}b$.
From the definition, it is easy to show the following important properties:
\begin{itemize}
    \item[a)]   $\forall x \in {\Rcal}$ we have that:
         \begin{itemize}
           \item[a.1)]  $P({X}{\leq}x) \,\stackrel{def.}{=}\,F(x)$
           \item[a.2)]  $P({X}{<}x) \,=\,F(x-{\epsilon})\,\,$;
           \item[a.3)]  $P({X}{>}x) \,=\,1-P({X}{\leq}x)\,=\,
                         1-F(x)\,\,$;
           \item[a.4)]  $P({X}{\geq}x)\,=\,
                         1-P({X}{<}x)\,=\,
                         1-F(x-{\epsilon})\,\,$;
         \end{itemize}
    \item[b)]  $\forall x_1<x_2 \in {\Rcal}$ we have that:
         \begin{itemize}
           \item[b.1)]  $P(x_1{<}{X}{\leq}x_2) \,=\,
                         P({X}\in (x_1,x_2])\,=\,
                         F(x_2)-F(x_1)\,\,$;
           \item[b.2)]  $P(x_1{\leq}{X}{\leq}x_2) \,=\,
                         P({X}\in [x_1,x_2])\,=\,
                         F(x_2)-F(x_1-{\epsilon})\,\,$ 
           \item[    ]   (thus, if $x_1=x_2$ then 
                          $P(X=x_1)=F(x_1)-F(x_1-{\epsilon})$); 
           \item[b.3)]  $P(x_1{<}{X}{<}x_2) \,=\,
                         P({X}\in (x_1,x_2))\,=\,
                         F(x_2-{\epsilon})-F(x_1)\,=\,$ \\
                         \hspace*{2.17cm}
                         $=\,
                         F(x_2)-F(x_1)-P({X}=x_2)\,\,$;
           \item[b.4)]  $P(x_1{\leq}{X}{<}x_2) \,=\,
                         P({X}\in [x_1,x_2))\,=\,
                         F(x_2-{\epsilon})-F(x_1-{\epsilon})\,=\,$ \\
                         \hspace*{2.17cm}
                         $=\,F(x_2)-F(x_1)-P({X}=x_2)+
                         P({X}=x_1)
                         \,\,$.
         \end{itemize}
\end{itemize}

The Distribution Function is discontinuous at all $x\in \Rcal$ where
$F(x-{\epsilon}){\neq}F(x+{\epsilon})$. Let $D$ be the set of all points of 
discontinuity. If $x\in D$, then $F(x-{\epsilon}){<}F(x+{\epsilon})$ since
it is monotonous non-decreasing. Thus, we can associate to each $x\in D$ a
rational number $r(x)\in \Qcal$ such that
$F(x-{\epsilon})\,<\,r(x)\,<\,F(x+{\epsilon})$ and all will be different 
because if $x_1<x_2 \in D$ then
$F(x_1+{\epsilon}){\leq}F(x_2-{\epsilon})$. Then, since $\Qcal$ is a countable
set, we have that the set of points of discontinuity of $F(x)$ is either 
{\bf finite} or {\bf countable}. At each of them the distribution function 
has a ``jump'' of amplitude (property b.2):
\begin{eqnarray}
    F(x)\,-\,F(x-{\epsilon})\,=\,P({X}=x)  \nonumber
\end{eqnarray}
and will be continuous on the right (condition p.2).

Last, for each Distribution Function there is a {\bf unique} probability
measure $P$ defined over the Borel sets of $\Rcal$ that assigns the
probability $F(x_2)-F(x_1)$ to each half-open interval $(x_1,x_2]$ and,
conversely, to any probability measure defined on a measurable space
$({\Rcal},{\Bcal})$ corresponds one Distribution Function. Thus, the
Distribution Function of a random quantity contains all 
the information needed to describe the properties of the random process.

\subsection{Discrete and Continuous Distribution Functions}

Consider the probability space $({\Omega},{\Fcal},Q)$, the
random quantity
$X(w)\,:\,w{\in}{\Omega}\,{\rightarrow}\,X(w)\,{\in}\,{\Rcal}$ 
and the induced probability space $({\Rcal},{\Bcal},P)$. 
The function $X(w)$ is a {\bf discrete random quantity} if its range (image)
$D=\{x_1,{\ldots},x_i,{\ldots}\}$, with
$x_i \in {\Rcal}\,$, $i=1,2,{\ldots}$ is a finite or countable set; that is,
if $\{A_k;k=1,2,{\ldots}\}$ is a finite or countable partition of $\Omega$,
the function $X(w)$ is either:
\begin{eqnarray*}
{\rm {\bf simple:}}\hspace{0.5cm}
 X(w)\,=\,\sum_{k=1}^{n}x_k\,{\mathbf{1}}_{A_k}(w)
\hspace{0.5cm}{\rm or\,\,\,\, {\bf elementary:}}\hspace{0.5cm}
 X(w)\,=\,\sum_{k=1}^{\infty}x_k\,{\mathbf{1}}_{A_k}(w)
\end{eqnarray*}
Then, $P(X=x_k)=Q(A_k)$ 
and the corresponding Distribution Function, defined for all $x {\in} \Rcal$,
will be
\begin{eqnarray*}
  F(x)\,=\,P({X}{\leq}x)\,=\,\sum_{{\forall}x_k{\in}D} 
           P({X}=x_k)\,
 {\mathbf{1}}_{A}(x_k)\,=\,
\sum_{{\forall}x_k{\leq}x}P({X}=x_k) 
\nonumber
\end{eqnarray*}
with $A=(-{\infty},x]\,{\cap}\,D$ and satisfies:
\begin{itemize}
\item[i)] $F(-{\infty})=0$ and $F(+{\infty})=1$;
\item[ii)] is a monotonous non decreasing step-function;
\item[iii)] continuous on the right $(F(x+{\epsilon})=F(x))$
            and therefore constant but on the finite or countable set of 
            points of discontinuity $D=\{x_1,{\ldots}\}$ where
             \begin{eqnarray*}
                 F(x_k)\,-\,F(x_k-{\epsilon})\,=\,P(X=x_k)
            \end{eqnarray*}
\end{itemize}
Familiar examples of discrete Distribution Functions are Poisson, Binomial,
Multinomial,...

The random quantity $X(w):{\Omega}{\longrightarrow}{\Rcal}$ is
{\bf continuous} if its range is a non-de\-nu\-me\-ra\-ble set; that is, if for
all $x \in {\Rcal}$ we have that $P(X=x)=0$. In this case, the Distribution
Function   $F(x)\,=\,P(X\,{\leq}\,x)$ is continuous for all
$x \in {\Rcal}$ because
\begin{itemize}
\item[i)] from condition (p.2): $F(x+{\epsilon})=F(x)$;
\item[ii)] from property (b.2): $F(x-{\epsilon})=F(x)-P({X}=x)=F(x)$
\end{itemize}

Now, consider the measure space $(\Omega,{\Bcal}_{\Omega},\mu)$ with 
$\mu$ countably additive. If $f:{\Omega}{\rightarrow}[0,\infty)$ is 
integrable with respect to ${\mu}$, it is clear that $\nu(A)=\int_Afd\mu$ for
$A{\in}{\Bcal}_{\Omega}$ is also a non-negative countably additive set function.
More generally, we have:

\vspace{0.5cm}
\noindent
$\bullet$ {\bf Radon-Nikodym Theorem} (Radon(1913), Nikodym(1930)){\bf :}
If  $\nu$ and $\mu$ are  two $\sigma$-additive measures on 
the measurable space 
$(\Omega,{\Bcal}_{\Omega})$ such that $\nu$ is {\bf absolutely continuous} with
respect to $\mu$ (${\nu}<<{\mu}$; that is, for every set 
$A{\in}{\Bcal}_{\Omega}$ for which $\mu(A)=0$ it is $\nu(A)=0$), then
there exists a $\mu$-integrable function $p(x)$ such that
\begin{eqnarray*}
  \nu(A)\,=\, \int_A\,d\nu(w)\,=\,
 \int_A\,\frac{\textstyle d\nu(w)}{\textstyle d\mu(w)}\,d{\mu}(w)\,=\,
 \int_A\,p(w)d\mu(w)
\end{eqnarray*}
and, conversely, if such a function exists then ${\nu}<<{\mu}$. 
\vspace{0.5cm}

\noindent
The function $p(w)=d{\nu}(w)/d{\mu}(w)$ is called {\sl Radon density} and is
unique up to at most a set of measure zero; that is, if
\begin{eqnarray*}
  \nu(A)\,=\,\int_A\,p(w)d\mu(w)\,=\,\int_A\,f(w)d\mu(w)
\end{eqnarray*}
then then $\mu\{x|p(x)\neq f(x)\}=0$. 
Furthermore, if ${\nu}$ and ${\mu}$ are {\sl equivalent} (${\nu}{\sim}{\mu}$;
${\mu}<<{\nu}$ and  ${\nu}<<{\mu}$) then $d{\nu}/d{\mu}>0$ almost everywhere.
In consequence,
if we have a probability space $({\Rcal},{\Bcal},P)$ with $P$ equivalent
to the Lebesgue measure,
there exists a non-negative Lebesgue integrable function
$p:{\Rcal}\longrightarrow[0,\infty)$, unique a.e., such that
\begin{eqnarray*}
  P(A)\,\equiv\, P(X{\in}A)\,=\,\int_A\,p(x)\,dx
\hspace{0.5cm};\hspace{1.cm}\forall A{\in}{\Bcal}
\end{eqnarray*}
The function $p(x)$ 
is called {\bf probability density function} and satisfies:
\begin{itemize}
    \item[i)]   $p(x)\geq 0$ $\forall x \in {\Rcal}$;
    \item[ii)]  at any bounded interval of $\Rcal$, $p(x)$ is bounded and
                is Riemann-integrable;
   \item[iii)] ${\int_{-\infty}^{+\infty}\,p(x)\,dx}=1$.
\end{itemize}
Thus, for an {\bf absolutely continuous random quantity} $X$, the Distribution
Function $F(x)$ can be expressed as
\begin{eqnarray*}
  F(x)\,=\,P(X{\leq}x)\,=
         \displaystyle {\int_{-\infty}^{x}\,
                       p(w)\,dw}\,\,
\end{eqnarray*}
Usually we shall be interested in random quantities that take values in a 
subset $D\subset{\Rcal}$. It will then be understood that $p(x)$ is
$p(x){\mathbf{1}}_D(x)$ so it is defined for all $x \in {\Rcal}$.
Thus, for instance, if ${\rm supp}\{p(x)\}=[a,b]$ then
\begin{eqnarray*}
  \int_{-\infty}^{+\infty}\,p(x)\,dx\,\equiv\,
  \int_{-\infty}^{+\infty}\,p(x)\,{\mathbf{1}}_{[a,b]}(x)\,dx\,=\,
  \int_{a}^{b}\,p(x)\,dx\,=\,1
\end{eqnarray*}
and therefore
\begin{eqnarray*}
F(x)\,=\,P({X}{\leq}x)\,=\,
{\mathbf{1}}_{(a,\infty)}(x){\mathbf{1}}_{[b,\infty)}(x)\,+\,
{\mathbf{1}}_{[a,b)}(x)\int_{a}^{x}\,p(u)\,du
\end{eqnarray*}

Note that from the previous considerations, the value of the integral will not 
be affected if we modify the integrand
on a countable set of points. In fact, what we actually integrate is an
equivalence class of functions that differ only in a set of measure zero.
Therefore, a probability density function
$p(x)$ has to be continuous for all $x{\in}{\Rcal}$ but, at most, 
on a countable set of points. 
If $F(x)$ is not differentiable at a 
particular point, $p(x)$ is not defined on it but the set of those points 
is of zero measure. However, if $p(x)$ is continuous in $\Rcal$ then
$F'(x)=p(x)$ and the value of $p(x)$ is univocally determined by $F(x)$.
We also have that
\begin{eqnarray*}
  P({X}{\leq}x)=F(x)=\int_{-\infty}^{x}p(w)dw
\hspace{0.5cm}{\longrightarrow}\hspace{0.5cm}
  P({X}>x)=1-F(x)=\int_{x}^{+\infty} p(w)dw
\end{eqnarray*}
and therefore:
\begin{eqnarray*}
  P(x_1<{X}{\leq}x_2)\,=\,
  F(x_2)\,-\,F(x_1)\,=\,
         \displaystyle {\int_{x_1}^{x_2}\,
                       p(w)\,dw}
\end{eqnarray*}
Thus, since $F(x)$ is continuous at all $x{\in}\Rcal$:
\begin{eqnarray*}
  P(x_1<{X}{\leq}x_2)\,=\, P(x_1<{X}<x_2)\,=\,
  P(x_1{\leq}{X}<x_2)\,=\,P(x_1{\leq}{X}{\leq}x_2)
\end{eqnarray*}
and therefore $P({X}=x)=0$ $\forall x \in {\Rcal}$ 
($\lambda([x])=0$) even though $X=x$ is a possible outcome of the 
experiment so, in this sense, unlike discrete random quantities
{\sl ``probability''} 0 does not correspond necessarily to
impossible events.
Well known examples absolutely continuous Distribution Functions are the 
Normal, Gamma, Beta, Student, Dirichlet, Pareto, ...

Last,
if the continuous probability measure $P$ is not absolutely continuous with 
respect to the Lebesgue measure $\lambda$ in $\Rcal$, then the 
probability density function does not exist.  Those are called {\bf singular}
random quantities for which $F(x)$ is continuous but
$F'(x)=0$ almost everywhere.  A well known example is the Dirac's singular
measure ${\delta}_{x_0}(A)={\mathbf{1}}_A(x_0)$
that assigns a measure 1 to a set $A{\in}{\Bcal}$ if $x_0{\in}A$ and 0 
otherwise.
As we shall see in the examples 1.9 and 1.20,
dealing with these cases is no problem
because the Distribution Function always exists. The Lebesgue's
{\sl General Decomposition
Theorem} establishes that any Distribution Function can
be expressed as a convex combination:
\begin{eqnarray*}
 F(x)\,=\,\sum_{i=1}^{N_d}\,a_{i}\,F_d(x)\,+\,
           \sum_{j=1}^{N_{ac}}\,b_{j}\,F_{ac}(x)\,+\, 
           \sum_{k=1}^{N_s}\,c_{k}\,F_s(x) 
\end{eqnarray*}
of a discrete Distribution Functions ($F_d(x)$),
absolutely continuous ones 
($F_{ac}(x)$ with derivative at every point so $F'(x)=p(x)$) and singular
ones ($F_s(x)$). For the cases we shall deal with, $c_k=0$.

\vspace{0.5cm}
\noindent
{\rayan}                   
\vspace{0.35cm}
\footnotesize

\noindent
{\bf Example 1.7:} Consider a real parameter ${\mu}>0$ and a discrete random 
  quantity $X$ that can take values $\{0,1,2,\ldots\}$ with 
a Poisson probability law:
\begin{eqnarray*}
   P(X=k|\mu)\,=\,e^{-\mu}\,\frac{\mu^k}{\Gamma(k+1)}\,\,;
   \hspace{1.cm}k=0,1,2,\ldots
   \end{eqnarray*}
The Distribution Function will be
\begin{eqnarray*}
  F(x|\mu)\,=\,P({X}{\leq}x|\mu)\,=\,
  e^{-\mu}\,\sum_{k=0}^{m=[x]}\frac{\mu^k}{\Gamma(k+1)}
\end{eqnarray*}
where $m=[x]$ is the largest integer less or equal to $x$. Clearly,
for $\epsilon\rightarrow 0^+$:
\begin{eqnarray*}
F(x+\epsilon|\mu)\,=\,F([x+\epsilon]|\mu)\,=\,F([x]|\mu)\,=\,F(x|\mu) 
\end{eqnarray*}
so it is continuous on the right and for $k=0,1,2,\ldots$ 
\begin{eqnarray*}
  F(k|\mu)\,-\,F(k-1|\mu)\,=\,P(X=k|\mu)\,=\,e^{-\mu}\,\frac{\mu^k}{\Gamma(k+1)}
\end{eqnarray*}
Therefore, for reals $x_2>x_1>0$ such that $x_2-x_1<1$,
$P(x_1<{X}{\leq}x_2)=F(x_2)-F(x_1)=0$.
\vspace{0.35cm}

\noindent
{\bf Example 1.8:} Consider the function $g(x)=e^{-ax}$ with $a>0$ real and
   support in $(0,{\infty})$. It is non-negative and Riemann integrable in 
   $\Rcal^+$ so we can define a probability density
   \begin{eqnarray*}
   p(x|a)\,=\,\frac{e^{-ax}}
                   {\int_{0}^{\infty}\,e^{-{\alpha}x}\,dx}\,
                   {\mathbf{1}}_{(0,\infty)}(x)\,=\,
                  a\,e^{-ax}\, {\mathbf{1}}_{(0,\infty)}(x)
   \end{eqnarray*}
and the Distribution Function
\begin{eqnarray}
  F(x)\,=\,P({X}{\leq}x)\,=\,\int_{-\infty}^{x}p(u|a)du\,=\,
   \left\{
     \begin{array}{ll}
       0 & x < 0 \\
       1\,-\,e^{-ax} & x \geq 0
     \end{array}
   \right.
  \nonumber
\end{eqnarray}
Clearly, $F(-\infty)=0$ and $F(+\infty)=1$. Thus, for an absolutely 
continuous random quantity $X{\sim}p(x|a)$ we have that
for reals $x_2>x_1>0$:
 \begin{eqnarray*}
  P({X}{\leq}x_1)\,\,\,\,\,\,\,\,\,\,\,&=&\,
   F(x_1)\,\,\,\,\,\,\,\,\,\,\,\,\,\,\,\,\,\,\,\,\,\,
   =\,1\,-\,e^{-ax_1} \\
  P({X}{>}x_1)\,\,\,\,\,\,\,\,\,\,\,&=&\,
   1\,-\,F(x_1)\,\,\,\,\,\,\,\,\,=\,e^{-ax_1} \\
  P(x_1<{X}{\leq}x_2)\,&=&\,F(x_2)\,-\,F(x_1)\,=\,
  e^{-ax_1}\,-\,e^{-ax_2} 
 \end{eqnarray*}
\vspace{0.35cm}

\begin{figure}[h]
\begin{center}
\mbox{\epsfig{file=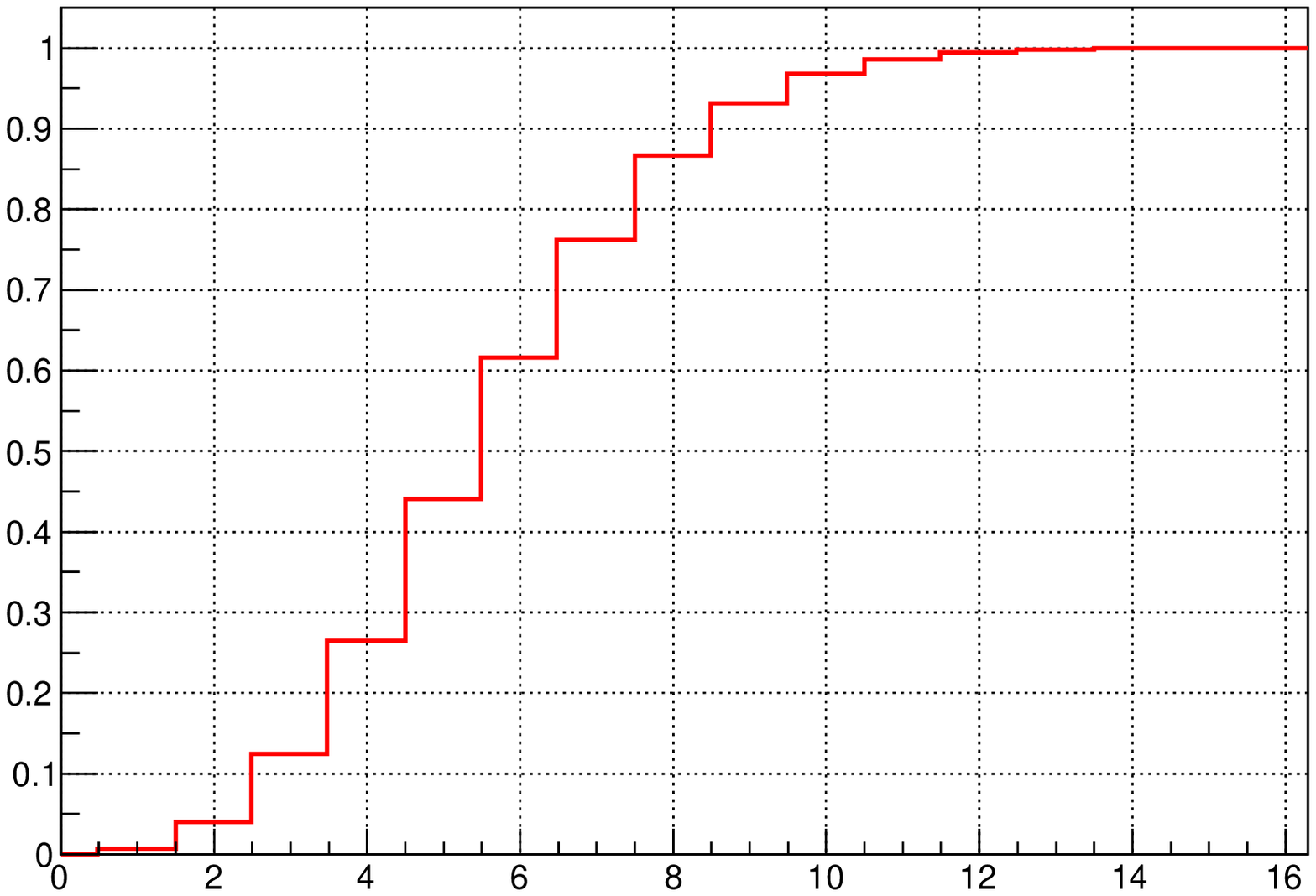,height=6.0cm,width=6.0cm}
      \epsfig{file=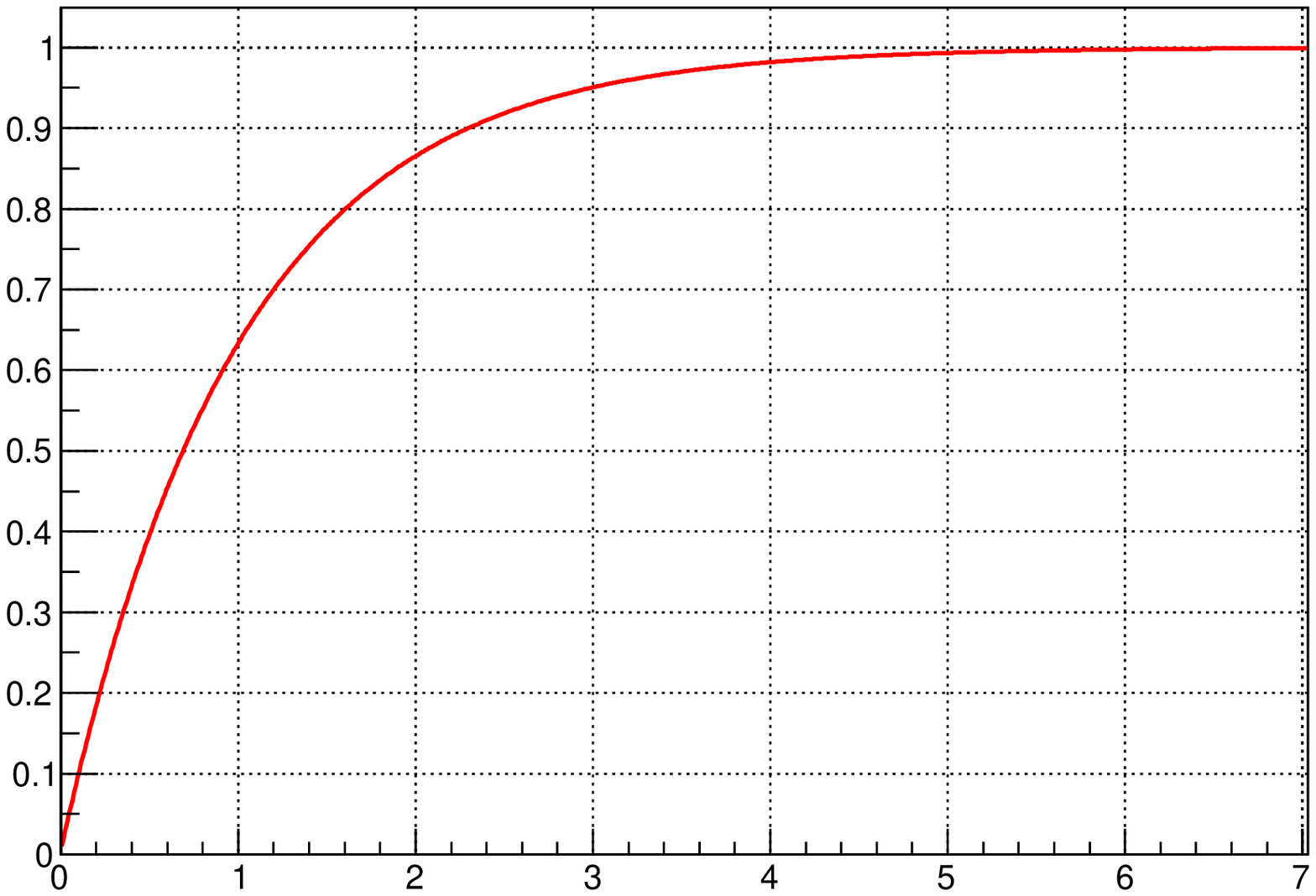,height=6.0cm,width=6.0cm}}
\mbox{\epsfig{file=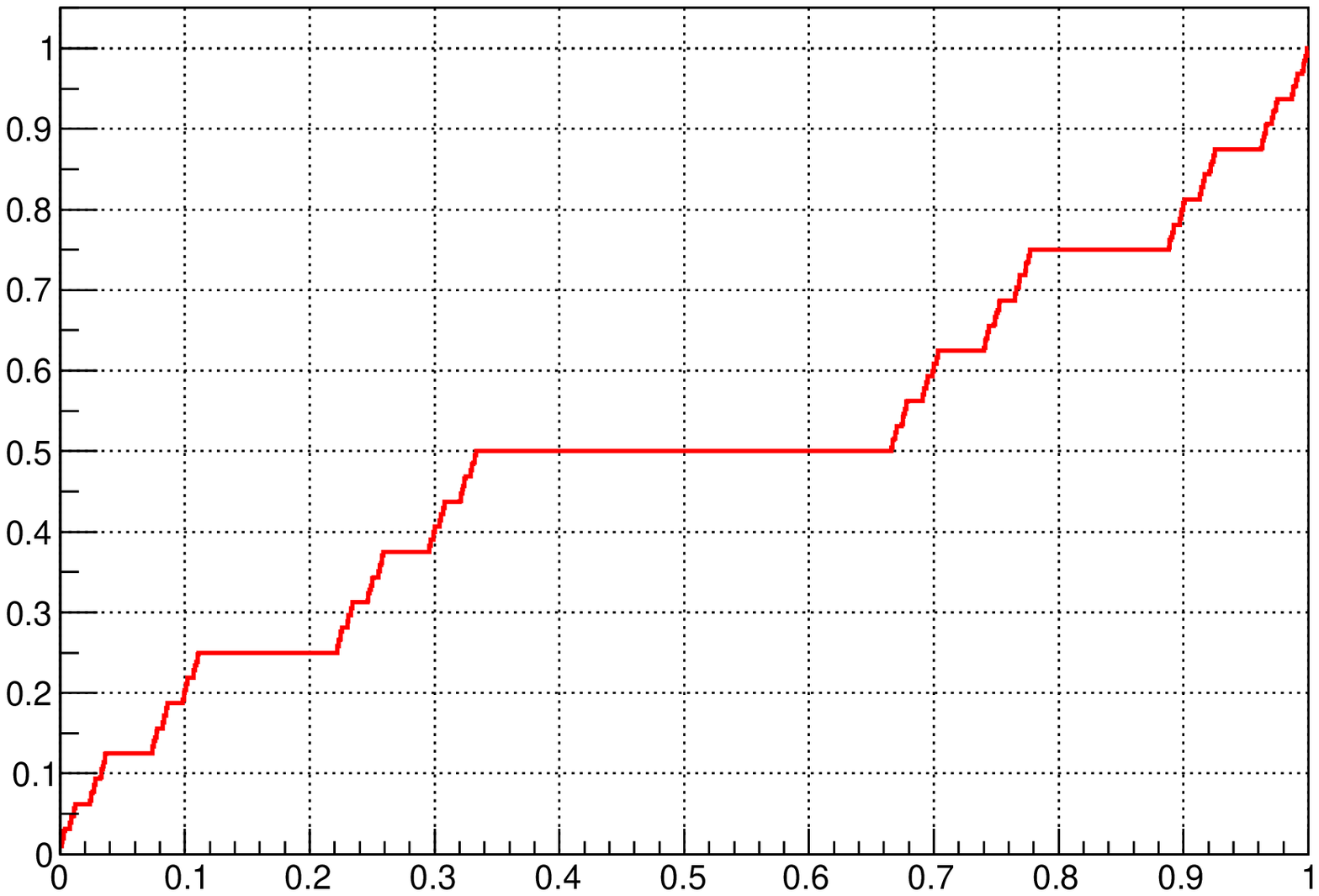,height=6.0cm,width=7.0cm}}

\footnotesize
{\bf Figure 1.1}.- Empiric Distribution Functions (ordinate)
form a Monte Carlo
sampling ($10^6$ events) of the Poisson $Po(x|5)$ (discrete; upper left), 
Exponential $Ex(x|1)$ (absolute continuous; upper right) 
and Cantor (singular; bottom) Distributions.

\end{center}
\end{figure}

\noindent
{\bf Example 1.9:} The ternary Cantor Set $Cs(0,1)$ is constructed iteratively.
Starting with the interval $Cs_0=[0,1]$, at each step one removes the open
middle third of each of the remaining segments. That is; at step one the
interval $(1/3,2/3)$ is removed so $Cs_1=[0,1/3]{\cup}[2/3,1]$ and so on.
If we denote by $D_n$ the union of the $2^{n-1}$  disjoint open intervals 
removed at step $n$, each of length $1/3^n$, the Cantor set is defined as 
$Cs(0,1)=[0,1]{\setminus}\cup_{n=1}^{\infty}D_n$. It is easy to check that
any element $X$ of the Cantor Set can be expressed as
\begin{eqnarray*}
  X\,=\,\sum_{n=1}^{\infty}\frac{X_n}{3^n}
\end{eqnarray*}
with ${\rm supp}\{X_n\}=\{0,2\}$ 
\footnote{Note that the representation of a real number $r{\in}[0,1]$
as $(a_1,a_2,...): \sum_{n=1}^{\infty}a_n3^{-n}$
with $a_i=\{0,1,2\}$ is not unique. In fact $x=1/3{\in}Cs(0,1)$ and can
be represented by $(1,0,0,0,..)$ or $(0,2,2,2,...)$.}
and that $Cs(0,1)$ is a closed set, uncountable,
nowhere dense in $[0,1]$ and with zero measure.
The Cantor Distribution, whose
support is the Cantor Set, is defined assigning
a probability $P(X_n=0)=P(X_n=2)=1/2$. Thus, $X$ is a continuous random
quantity with support on a non-denumerable set of measure zero and
can not be described by a probability density function. 
The Distribution Function $F(x)=P(X{\leq}x)$ (Cantor Function; fig 1.1) is 
an example of singular Distribution.
\vspace{0.35cm}

\smill
{\raya}                   
\vspace{1.0cm}

\subsection{Distributions in more dimensions}

The previous considerations can be extended to random quantities in more
dimensions but with some care. Let's consider the 
the two-dimensional case:
${\Xbold}=(X_1,X_2)$. The Distribution Function will be defined as:
\begin{eqnarray*}
F(x_1,x_2)\,=\,P({X}_1{\leq}x_1,
                              {X}_2{\leq}x_2)\,\,;\,\,\,\,\,
     \forall (x_1,x_2) \in {\Rcal}^2
  \nonumber 
\end{eqnarray*}
and satisfies:
\begin{itemize}
    \item[i)]   monotonous no-decreasing in both variables; that is, if        
                $(x_1,x_2),(x'_1,x'_2) \in {\Rcal}^2$:
                 \begin{eqnarray*}
                x_1 {\leq} x'_1 \,\,\longrightarrow\,\,
                        F(x_1,x_2) {\leq} F(x'_1,x_2)
                \hspace{0.5cm}{\rm and}\hspace{0.5cm}
                x_2 {\leq} x'_2 \,\,\longrightarrow\,\,
                        F(x_1,x_2) {\leq} F(x_1,x'_2)
                \end{eqnarray*}
    \item[ii)]  continuous on the right at $(x_1,x_2) \in {\Rcal}^2$:
                \begin{eqnarray*}
                F(x_1+{\epsilon},x_2)\,=\,
                F(x_1,x_2+{\epsilon})\,=\,F(x_1,x_2) 
                \end{eqnarray*}
    \item[iii)] $F(-\infty ,x_2)\,=\,F(x_1,-\infty)=0\,$ and
                $F(+\infty ,+\infty)=1$.
\end{itemize}

\noindent
Now, if $(x_1,x_2),(x'_1,x'_2) \in {\Rcal}^2$ with $x_1<x'_1$ and
$x_2<x'_2$ we have that:
\begin{eqnarray*}
P(x_1<{X}_1{\leq}x'_1,x_2<{X}_2{\leq}x'_2)= 
 F(x'_1,x'_2)-F(x_1,x'_2)-F(x'_1,x_2)+F(x_1,x_2)\,{\geq}\,0
\end{eqnarray*}
and
\begin{eqnarray*}
P(x_1{\leq}{X}_1{\leq}x'_1,x_2{\leq}{X}_2{\leq}x'_2)=
 F(x'_1,x'_2)-F(x_1-{\epsilon},x'_2)-F(x'_1,x_2-{\epsilon})+
 F(x_1-{\epsilon},x_2-{\epsilon})\,{\geq}\,0
\end{eqnarray*}
so, for discrete random quantities, if $x_1=x'_1$ and $x_2=x'_2$:
\begin{eqnarray*}
P({X}_1=x_1,{X}_2=x_2)=
 F(x_1,x_2)-F(x_1-{\epsilon}_1,x_2)- F(x_1,x_2-{\epsilon})+
 F(x_1-{\epsilon},x_2-{\epsilon})\,{\geq}\,0
\end{eqnarray*}
will give the amplitude of the jump of the Distribution Function at
the points of discontinuity.

As for the one-dimensional case, for absolutely continuous random quantities 
we can introduce a two-dimensional probability density function
$p(\xbold):{\Rcal}^2\longrightarrow{\Rcal}$:
\begin{itemize}
    \item[i)]   $p(\xbold) \geq 0
    \,\,;\,\,\,\,\,\forall \xbold \in {\Rcal}^2$;
    \item[ii)]  At every bounded interval of ${\Rcal}^2$,
                $p(\xbold)$ is bounded and Riemann integrable;
    \item[iii)] $ \int_{{\Rcal}^2}p(\xbold)d{\xbold}\,=\,1$
\end{itemize}
such that:
\begin{eqnarray*}
 F(x_1,x_2)=\int_{-\infty}^{x_1}du_1\int_{-\infty}^{x_2}du_2 \,p(u_1,u_2)
 \hspace{0.5cm}{\longleftrightarrow}\hspace{0.5cm}
    p(x_1,x_2)\,=\,\frac{\textstyle {\partial}^2}
                        {\textstyle {\partial}x_1\,{\partial}x_2}\,F(x_1,x_2)
\end{eqnarray*}

\subsubsection{Marginal and Conditional Distributions}

It may happen that we are interested only in one of the two random 
quantities say, for instance, $X_1$. Then we ignore all aspects concerning
$X_2$ and obtain the one-dimensional Distribution Function 
\begin{eqnarray*}
 F_1(x_1)\,=\,
P({X}_1{\leq}x_1)\,=\,
P({X}_1{\leq}x_1,{X}_2{\leq}+\infty)\,=\,
 F(x_1,+\infty)
\end{eqnarray*}
that is called the {\bf Marginal Distribution Function} of the
random quantity $X_1$. In the same manner, we have
$F_2(x_2)=F(+\infty,x_2)$ for the random quantity $X_2$.
For absolutely continuous random quantities, 
\begin{eqnarray*}
 F_1(x_1)\,=\,
 F(x_1,+\infty)\,=\,
       \displaystyle {\int_{-\infty}^{x_1}}\,du_1\,
       \displaystyle {\int_{-\infty}^{+\infty}}\,
    p(u_1,u_2)\,du_2\,=\,\displaystyle {\int_{-\infty}^{x_1}}\,p(u_1)\,du_1
\end{eqnarray*}
with $p(x_1)$ the {\bf marginal probability density function} 
\footnote{It is habitual to avoid the indices and write $p(x)$ 
meaning {\sl ``the probability density function
of the variable $x$''} since the distinctive features are 
clear within the context.} 
of the random quantity $X_1$:
\begin{eqnarray}
 p_1(x_1)\,=\,
    \frac{\textstyle {\partial}}
         {\textstyle {\partial}x_1}\,F_1(x_1)\,=\,
       \displaystyle {\int_{-\infty}^{+\infty}}\,
       p(x_1,u_2)\,du_2
  \nonumber 
\end{eqnarray}
In the same manner, we have for $X_2$
\begin{eqnarray*}
 p_2(x_2)\,=\,
    \frac{\textstyle {\partial}}
         {\textstyle {\partial}x_2}\,F_2(x_2)\,=\,
       \displaystyle {\int_{-\infty}^{+\infty}}\,
       p(u_1,x_2)\,du_1
  \nonumber 
\end{eqnarray*}

As we have seen, given a probability space $(\Omega,B_{\Omega},P)$, for any
two sets $A,B{\in}B_{\Omega}$ the conditional probability for $A$ given $B$
was {\sl defined} as
\begin{eqnarray*}
  P(A|B)\,\stackrel{def.}{=}\,
     \frac{\textstyle  P(A\, \cap \, B)}{\textstyle P(B) }\,\equiv\,  
     \frac{\textstyle  P(A,B)}{\textstyle P(B) }
      \nonumber
\end{eqnarray*}
provided $P(B)\neq 0$. Intimately related to this definition is the 
Bayes' rule:
\begin{eqnarray*}
  P(A,B)\,=\,P(A|B)\,P(B)\,=\,P(B|A)\,P(A)
\end{eqnarray*}

Consider now the discrete random quantity  ${\Xbold}=({X}_1,{X}_2)$ 
with values on $\Omega{\subset}{\Rcal}^2$. It is then natural to define 
\begin{eqnarray*}
   P({X}_1 = x_1|{X}_2 = x_2)\,\stackrel{def.}{=}\,
  \frac{P({X}_1 = x_1,{X}_2 = x_2)}
       {P({X}_2 = x_2)}
\end{eqnarray*}
and therefore
\begin{eqnarray}
  F(x_1|x_2)\,=\, 
  \frac{\textstyle
  P({X}_1{\leq}x_1,{X}_2 = x_2)}
       {\textstyle P({X}_2 = x_2)}
      \nonumber
\end{eqnarray}
whenever $P({X}_2=x_2){\neq}0$. For absolutely continuous random quantities
we can express the probability density as
\begin{eqnarray*}
  p(x_1,x_2)\,=\,p(x_1|x_2)\,p(x_2)\,=\,p(x_2|x_1)\,p(x_1)
\end{eqnarray*}
and define the {\bf conditional probability density function} as
\begin{eqnarray*}
  p(x_1|x_2)\,\stackrel{def.}{=}\,\frac{\textstyle p(x_1,x_2)}{p(x_2)}
  \hspace{2.cm}=\,
  \frac{\textstyle \partial}{\textstyle \partial x_1}
         \,F(x_1|x_2)
\end{eqnarray*}
provided again that $p_2(x_2) \neq 0$.
This is certainly is an admissible density 
\footnote{Recall that for continuous random quantities 
$P(X_2=x_2)=P(X_1=x_1)=0)$. 
One can {\sl justify} this expression with
kind of heuristic arguments; essentially 
considering $X_1{\in}{\Lambda}_1=(-\infty,x_1]$,
$X_2{\in}{\Delta}_{\epsilon}(x_2)=[x_2,x_2+{\epsilon}]$ and taking the limit
$\epsilon\rightarrow 0^+$ of
\begin{eqnarray*}
P({X}_1{\leq}x_1|{X}_2{\in}{\Delta}_{\epsilon}(x_2))=
\frac{P({X}_1{\leq}x_1,{X}_2{\in}{\Delta}_{\epsilon}(x_2))}
     {P({X}_2{\in}{\Delta}_{\epsilon}(x_2))}=  
\frac{F(x_1,x_2+{\epsilon})-F(x_1,x_2)}{F_2(x_2+{\epsilon})-F_2(x_2)}
\end{eqnarray*}
See however [Bo06]; Vol 2; Chap. 10. for the Radon-Nikodym density with 
conditional measures.}. 
since $p(x_1|x_2){\geq}0$ $\forall (x_1,x_2){\in}{\Rcal}^2$ and
$\int_{\Rcal}p(x_1|x_2)dx_1=1$.

As stated already, two events $A,B \in B_{\Omega}$ are 
statistically independent iff:
\begin{eqnarray*}
  P(A,B)\,\equiv\,
  P(A \cap B)\,=\,P(A)\,\cdot\,P(B)
      \nonumber
\end{eqnarray*}
Then, we shall say that two discrete random quantities $X_1$ and $X_2$ are
{\bf statistically independent} if $F(x_1,x_2)=F_1(x_1)F_2(x_2)$; that is,
if
\begin{eqnarray*}
  P({X}_1=x_1,{X}_2 =x_2)=
  P({X}_1 =x_1)\,P({X}_2 = x_2)
\end{eqnarray*}
for discrete random quantities and
\begin{eqnarray*}
  p(x_1,x_2)\,=\,
  \frac{\textstyle {\partial}^2}{\textstyle \partial x_1\,\partial x_2}
         \,F(x_1)F(x_2)\,=\,
  p(x_1)\,p(x_2)
  \hspace{0.5cm}{\longleftrightarrow}\hspace{0.5cm}
  p(x_1|x_2)=p(x_1)
\end{eqnarray*}
and for absolutely continuous random quantities.

\vspace{0.5cm}
\noindent
{\raya}                   
\vspace{0.35cm}
\footnotesize

\noindent
{\bf Example 1.10:} 
Consider the probability space $({\Omega},{\Bcal}_{\Omega},{\lambda})$ with
${\Omega}=[0,1]$ and ${\lambda}$ the Lebesgue measure. 
If $F$ is an arbitrary Distribution Function,
$X:w{\in}[0,1]{\longrightarrow}F^{-1}(w){\in}{\Rcal}$ is a random quantity
and is distributed as $F(w)$. Take the Borel set
$I=(-{\infty},r]$ with $r{\in}{\Rcal}$. 
Since $F$ is a Distribution Function is monotonous and non-decreasing we have
that:
\begin{eqnarray*}
X^{-1}(I)=\{w{\in}{\Omega}\,|\,X(w){\leq}r\}\,=\,
          \{w{\in}[0,1]\,|\,F^{-1}(w){\leq}r\}\,=\,
          \{w{\in}{\Omega}\,|\,w{\leq}F(r)\}\,=\,=[0,F(r)]{\in}{\Bcal}_{\Omega}
\end{eqnarray*}
and therefore  $X(w)=F^{-1}(w)$ is measurable over ${\Bcal}_{\Rcal}$ and 
is distributed as
\begin{eqnarray*}
P(X(w){\leq}x\}\,=\,P(F^{-1}(w){\leq}x\}\,=\,
P(w{\leq}F(x)\}\,=\,
\int_{0}^{F(x)}d{\lambda}\,=\,F(x)
\end{eqnarray*}

\vspace{0.35cm}

\noindent
{\bf Example 1.11:} Consider the  probability space 
$({\Rcal},{\Bcal},{\mu})$ with ${\mu}$ the probability measure
\begin{eqnarray*}
{\mu}(A)\,=\,\int_{A{\in}{\Bcal}}\,dF
\end{eqnarray*}
The function
$X:w{\in}{\Rcal}{\longrightarrow}F(w){\in}[0,1]$ is measurable on $\Bcal$. Take
$I=[a,b){\in}{\Bcal}_{[0,1]}$. Then
\begin{eqnarray*}
X^{-1}(I)=\{w{\in}{\Rcal}\,|\,a{\leq}F(w)<b\}\,=\,
          \{w{\in}{\Rcal}\,|\,F^{-1}(a){\leq}w<F^{-1}(b)\}\,=\,[w_a,w_b){\in}
          \Bcal_{\Rcal}
\end{eqnarray*}
It is distributed as $X{\sim}Un(x|0,1)$:
\begin{eqnarray*}
P(X(w){\leq}x\}\,=\,P(F(w){\leq}x\}\,=\,
P(w{\leq}F^{-1}(x)\}\,=\,
\int_{-\infty}^{F^{-1}(x)}dF\,=\,x
\end{eqnarray*}
This is the basis of the Inverse Transform sampling method that we shall see
in Chapter 3 on Monte Carlo techniques.

\vspace{0.35cm}

\noindent

\noindent
{\bf Example 1.12:} Suppose that the number of eggs a particular insect may lay 
($X_1$) follows a Poisson distribution $X_1\sim Po(x_1|\mu)$:
\begin{eqnarray*}
      P({X}_1\,=\,x_1|\mu)\,=\,e^{-\mu}
      \frac{\textstyle {\mu}^{x_1}}
           {\Gamma(x_1+1)}\,\,;\,\,\,\,\,x_1=0,1,2,...
\end{eqnarray*}
Now, if the probability for an egg to hatch is $\theta$  
and $X_2$ represent the number of off springs, 
given $x_1$ eggs the probability to have $x_2$ descendants follows a 
Binomial law $X_2{\sim}Bi(x_2|x_1,\theta)$:
   \begin{eqnarray}
      P({X}_2\,=\,x_2|x_1,\theta)\,=\,
      \left(
        \begin{array}{c}
        x_1 \\
        x_2
        \end{array}
      \right)\,\theta^{x_2}\,(1-\theta)^{x_1-x_2}\,\,;\,\,\,\,\,
                       0\,{\leq}\,x_2\,{\leq}\,x_1
     \nonumber
  \end{eqnarray}
In consequence
\begin{eqnarray*}
      P(X_1\,=\,x_1,X_2\,=\,x_2|\mu,\theta)\,&=&\,
      P(X_2\,=\,x_2|X_1=x_1,\theta)\,P(X_1=x_1|\mu)\,= \\
      &=&\, 
      \left(
        \begin{array}{c}
        x_1 \\
        x_2
        \end{array}
      \right)\,\theta^{x_2}\,(1-\theta)^{x_1-x_2}\,
      e^{-\mu}
      \frac{\textstyle {\mu}^{x_1}}
           {\textstyle  \Gamma(x_1+1)}
      \,\,;\,\,\,\,\,0\, {\leq}\, x_2\, {\leq}\, x_1
   \end{eqnarray*}

Suppose that we have not observed the number of eggs that were laid.
What is the distribution of the number of off springs? This is given
by the marginal probability 
\begin{eqnarray*}
      P({X}_2=x_2|\theta,\mu)\,=\,
      \sum^{\infty}_{x_1=x_2}P({X}_1=x_1,{X}_2=x_2)\,=\,
       e^{-{\mu}\theta}
     \frac{\textstyle ({\mu}\theta)^{x_2}}
          {\textstyle \Gamma(x_2+1)}\,=\,Po(x_2|\mu\theta)  
\end{eqnarray*}
Now, suppose that we have found $x_2$ new insects. What is the distribution
of the number of eggs laid? This will be the conditional probability
$P({X}_1=x_1|{X}_2=x_2,\theta,\mu)$ and, since
 $P(X_1=x_1,X_2=x_2)=P(X_1=x_1|X_2=x_2)P(X_2=x_2)$
we have that:
\begin{eqnarray*}
 P({X}_1\,=\,x_1|{X}_2\,=\,x_2,\mu,\theta)\,=\,
 \frac{\textstyle P({X}_1=x_1,{X}_2=x_2)}
      {\textstyle P({X}_2=x_2)}\,=\,
 \frac{\textstyle 1}
      {\textstyle (x_1-x_2) !}\,\left({\mu}(1-\theta)\right)^{x_1-x_2}\,
      e^{-{\mu}(1-\theta)}
\end{eqnarray*}
with $0\,{\leq}\,x_2\,{\leq}\,x_1$; that is, again a Poisson with parameter
${\mu}(1-\theta)$.

\vspace{0.35cm}

\noindent
{\bf Example 1.13:} Let ${X}_1$ and ${X}_2$ two independent Poisson
distributed random quantities with parameters $\mu_1$ and $\mu_2$.
How is $Y=X_1+X_2$ distributed? Since they are independent:
\begin{eqnarray*}
      P({X}_1=x_1,{X}_2=x_2|\mu_1,\mu_2)\,=\,
      \,e^{-({\mu}_1+{\mu}_2)}\,
      \frac{\textstyle {\mu}_1^{x_1}}
           {\textstyle \Gamma(x_1+1)}\,
      \frac{\textstyle {\mu}_2^{x_2}}
           {\textstyle \Gamma(x_2+1)}
\end{eqnarray*}
Then, since $X_2=Y-X_1$:
\begin{eqnarray*}
  P(X_1=x,Y=y)=
  P(X_1=x,X_2=y-x)= 
  e^{-({\mu}_1+{\mu}_2)}\,
      \frac{\textstyle {\mu}_1^x}
           {\textstyle \Gamma(x+1)}\,
      \frac{\textstyle {\mu}_2^{(y-x)}}
           {\textstyle \Gamma(y-x+1)}
\end{eqnarray*}
Being ${X}_2=y-x{\geq}0$ we have the condition $y{\geq}x$ so the marginal
probability for $Y$ will be
\begin{eqnarray*}
  P(Y\,=\,y)\,=\,
     e^{-({\mu}_1+{\mu}_2)}\,
     \displaystyle{ \sum_{x=0}^{y} }\,
      \frac{\textstyle {\mu}_1^x}
           {\textstyle \Gamma(x+1)}\,
      \frac{\textstyle {\mu}_2^{(y-x)}}
           {\textstyle \Gamma(y-x+1)} \,=\,
     e^{-({\mu}_1+{\mu}_2)}\,
      \frac{\textstyle ({\mu}_1+{\mu}_2)^y}
           {\textstyle \Gamma(y+1)} 
\end{eqnarray*}
that is, $Po(y|\mu_1+\mu_2)$.

\vspace{0.35cm}

\noindent
{\bf Example 1.14:} Consider a two-dimensional random quantity
${X}=({X}_1,{X}_2)$ that takes values in ${\Rcal}^2$ with the 
probability density function 
$N(x_1,x_2|{\mubold}={\mathbf{0}},\sigmabold={\mathbf{1}},\rho)$:
   \begin{eqnarray}
     p(x_1,x_2|\rho)\,=\,\frac{\textstyle 1}
                         {\textstyle 2 \pi}\,
                    \frac{\textstyle 1}
                         {\textstyle \sqrt{1-{\rho}^2}}\,\,
              e^{- \frac{\textstyle 1}
                         {\textstyle 2 (1-{\rho}^2)}\,
                     \textstyle{(x_1^2 - 2{\rho} x_1 x_2 + x_2^2)}}
         \nonumber
   \end{eqnarray}
being ${\rho}\in (-1,1)$. The marginal densities are:
\begin{eqnarray*}
   X_1{\sim}p(x_1)\,&=&\,
   \displaystyle{ \int_{-\infty}^{+\infty}}\,
       p(x_1,u_2)\,du_2\,=\,
   \frac{\textstyle 1}
        {\textstyle \sqrt{2 \pi}}\,
        e^{- \frac{\textstyle 1}{2}\,\textstyle{x_1^2}}
       \\
   X_2{\sim}p(x_2)\,&=&\,
   \displaystyle{ \int_{-\infty}^{+\infty}}\,
       p(u_1,x_2)\,du_1\,=\,
   \frac{\textstyle 1}
        {\textstyle \sqrt{2 \pi}}\,
        e^{- \frac{\textstyle 1}{2}\,\textstyle{x_2^2}}
\end{eqnarray*}
and since
\begin{eqnarray*}
     p(x_1,x_2|\rho)\,=\,p(x_1)\,p(x_2)\,
         \frac{\textstyle 1}
              {\textstyle \sqrt{1-{\rho}^2}}\,\,
         e^{- \frac{\textstyle \rho}
                   {\textstyle 2 (1-{\rho}^2)}\,
      \textstyle{(x_1^2{\rho} - 2x_1 x_2 + x_2^2{\rho})}}
         \nonumber
   \end{eqnarray*}
both quantities will be independent iff
${\rho}=0$. The conditional densities are
   \begin{eqnarray*}
     p(x_1|x_2,\rho)\,&=&\,\frac{\textstyle p(x_1,x_2)}
                         {\textstyle p(x_2)}\,=\,
   \frac{\textstyle 1}
        {\textstyle \sqrt{2 \pi}}\,
         \frac{\textstyle 1}
              {\textstyle \sqrt{1-{\rho}^2}}\,\,
         e^{- \frac{\textstyle 1}
                   {\textstyle 2 (1-{\rho}^2)}\,
      \textstyle{(x_1 - x_2{\rho})^2}}
         \\
     p(x_2|x_1,\rho)\,&=&\,\frac{\textstyle f(x_1,x_2)}
                         {\textstyle f_1(x_1)}\,=\,
   \frac{\textstyle 1}
        {\textstyle \sqrt{2 \pi}}\,
         \frac{\textstyle 1}
              {\textstyle \sqrt{1-{\rho}^2}}\,\,
         e^{- \frac{\textstyle 1}
                   {\textstyle 2 (1-{\rho}^2)}\,
      \textstyle{(x_2 - x_1{\rho})^2}}
   \end{eqnarray*}
and when ${\rho}=0$ (thus independent)
$p(x_1|x_2)=p(x_1)$ and $p(x_2|x_1)=p(x_2)$. Last, it is clear that
\begin{eqnarray*}
   p(x_1,x_2|\rho)\,=\,p(x_2|x_1,\rho)\,p(x_1)\,=\,
     p(x_1|x_2,\rho)\,p(x_2) 
\end{eqnarray*}

\smill
{\raya}                   
\vspace{1.0cm}
\section{\LARGE \bf Stochastic Characteristics}
\subsection{Mathematical Expectation}

Consider a random quantity $X(w):\Omega\rightarrow\Rcal$
that can be either discrete
\begin{eqnarray}
  X(w)\,=\,
   \left\{
     \begin{array}{l}
   \displaystyle {X(w)=\sum_{k=1}^{n}x_k{\mbox{\boldmath $1$}_{A_k}(w)}} \\
                   \\
   \displaystyle {X(w)=\sum_{k=1}^{\infty}x_k{\mbox{\boldmath $1$}_{A_k}(w)}} \\
     \end{array}
   \right. \longrightarrow P(X=x_k)=P(A_k)=\int_{\Rcal}
         {\mbox{\boldmath $1$}_{A_k}(w)}\,dP(w) 
  \nonumber
\end{eqnarray}
or absolutely continuous for which
\begin{eqnarray}
  P(X(w){\in}A)\,=\,
   \int_{\Rcal}{\mbox{\boldmath $1$}_{A}(w)}\,dP(w)\,=\, 
   \int_{A}\,dP(w)\,=\, \int_{A}\,p(w)dw
  \nonumber
\end{eqnarray}

The {\bf mathematical expectation} of a n-dimensional random quantity 
$\Ybold=g(\Xbold)$ is defined as
\footnote{
In what follows we consider the Stieltjes-Lebesgue integral so
$\int\rightarrow \sum$ for discrete random quantities and in consequence:
\begin{eqnarray*}
\int_{-\infty}^{\infty}g(x)\,dP(x)= 
\int_{-\infty}^{\infty}g(x)\,p(x)\,dx
\hspace{0.5cm}{\longrightarrow}\hspace{0.5cm}
\sum_{\forall x_k} g(x_k)\,P(X=x_k) 
\end{eqnarray*}
}: 
\begin{eqnarray*}
  E[\Ybold]\,=\,E[g(\Xbold)]\,\stackrel{def.}{=}
         \int_{{\Rcal}^n}g(\xbold)\,dP(\xbold)= 
         \int_{{\Rcal}^n}g(\xbold)\,p(\xbold)\,d\xbold
\end{eqnarray*}
In general, the function $g(x)$ will be unbounded on ${\rm supp}\{X\}$
so both the sum and the integral have to be {\bf absolutely convergent}
for the {\sl mathematical expectation} to exist.

In a similar way, we define the {\sl conditional expectation}. If
$\Xbold=(X_1,\ldots,X_m\ldots,X_n)$, $\Wbold=(X_1\ldots,X_m)$ and
$\Zbold=(X_{m+1}\ldots,X_n)$ we have for
$\Ybold=g(\Wbold)$ that 
\begin{eqnarray*}
  E[\Ybold|\Zbold_0]\,=\,
  \int_{{\Rcal}^m}\,g(\wbold)\,
                 p(\wbold|\zbold_0)\,d\wbold\,=\,
  \int_{{\Rcal}^m}\,g(\wbold)\,
   \frac{\textstyle p(\wbold,\zbold_0)}{p(\zbold_0)}\,d\wbold
\end{eqnarray*}

\subsection{Moments of a Distribution}

Given a random quantity $X{\sim}p(x)$, we define the 
{\sl moment or order} $n$ 
$({\alpha}_n)$ as:
\begin{eqnarray*}
  {\alpha}_n\,{=}\,E[{X}^n]\,\stackrel{def.}{=}\,
             \int_{-\infty}^{\infty}\,x^n\,p(x)\,dx
\end{eqnarray*}
Obviously, they exist if $x^np(x){\in}L_1(\Rcal)$ so it may happen that
a particular probability distribution has only a finite number of moments.
It is also clear that if the moment of order $n$ exists, so do the moments
of lower order and, if it does not, neither those of higher order. In 
particular, the moment of order $0$ always exists 
(that, due to the normalization condition, is $\alpha_0=1$) 
and those of even order, if exist, are non-negative.
A specially important moment is that order $1$: the {\bf mean} ({\sl mean
value}) ${\mu}=E[X]$ that has two important properties:
\begin{itemize}
\item[$\bullet$] It is a {\bf linear operator} since
$ X=c_0+\sum_{i=1}^{n} c_iX_i
       \hspace{0.5cm}{\longrightarrow}\hspace{0.5cm}
       E[X]\,=\,c_0\,+\,\sum_{i=1}^{n} c_i\,E[X_i]$
\item[$\bullet$] If $ X=\prod_{i=1}^{n}c_iX_i$ with
$\{X_i\}_{i=1}^n$ independent random quantities, then
       $E[X]\,=\,\prod_{i=1}^{n} c_i\,E[X_i]$.
\end{itemize}
We can define as well the moments $({\beta}_n)$ with respect to any point 
$c{\in}\Rcal$ as:
\begin{eqnarray*}
  {\beta}_n\,=\,E[({X}-c)^n]\,\stackrel{def.}{=}\,
        \int_{-\infty}^{\infty}\,(x-c)^n\,p(x)\,dx
\end{eqnarray*}
so ${\alpha}_n$ are also called
{\sl central moments or moments with respect to the origin}. 
It is easy to see that the non-central moment of second order, 
${\beta}_2\,=\,E[({X}-c)^2]$, is minimal for
$c={\mu}=E[X]$. Thus, of special relevance are the {\sl moments or order n
with respect to the mean}
\begin{eqnarray*}
  {\mu}_n\,{\equiv}\,E[({X}-{\mu})^n]\,=\,
             \int_{-\infty}^{\infty}\,(x-{\mu})^n\,p(x)\,dx
\end{eqnarray*}
and, among them, the moment of order 2: the {\bf variance}
${\mu}_2=V[{X}]={\sigma}^2$. 
It is clear that ${\mu}_0=1$ and, if exists, ${\mu}_1=0$. 
Note that:
\begin{itemize}
\item[$\bullet$] $V[{X}]={\sigma}^2=E[({X}-{\mu})^2]>0$
\item[$\bullet$] It is {\bf not a linear operator} since
$       X=c_0+c_1X_1
       \hspace{0.3cm}{\longrightarrow}\hspace{0.3cm}
       V[X]\,=\,{\sigma}_X^2\,=\,c_1^2\,V[X_1]\,=\,c_1^2{\sigma}_{X_1}^2$
\item[$\bullet$] If $X=\sum_{i=1}^{n}c_iX_i$ and
$\{X_i\}_{i=1}^n$ are independent random quantities,
  $V[X]\,=\,\sum_{i=1}^{n} c_i^2\,V[X_i]$.
\end{itemize}

Usually, is less tedious to calculate the moments
with respect to the origin and evidently they are related so, from the binomial
expansion
 \begin{eqnarray*}
  ({X}-{\mu})^n\,=\,
    \sum^{n}_{k=0}\,
       \left(  \begin{array}{c}
                 n \\ k
               \end{array}
     \right)\,{X}^k\,(-{\mu})^{n-k}
       \hspace{0.5cm}{\longrightarrow}\hspace{0.5cm}              
  {\mu}_n\,=\,
    \sum^{n}_{k=0}\,
       \left(  \begin{array}{c}
                 n \\ k
               \end{array}
     \right)\,{\alpha}_k\,(-{\mu})^{n-k}
              \nonumber
 \end{eqnarray*}

The previous definitions are trivially extended to n-dimensional random 
quantities.
In particular, for 2 dimensions, $\Xbold=(X_1,X_2)$, we have the
moments of order $(n,m)$ with respect to the origin:
 \begin{eqnarray*}
  {\alpha}_{nm}\,=\,
  E[{X}_1^n\,{X}_2^m]\,=\,
        \int_{{\Rcal}^2}\,x_1^n\,x_2^m\,p(x_1,x_2)\,dx_1\,
         dx_2
 \end{eqnarray*}
so that ${\alpha}_{01}={\mu}_1$  and ${\alpha}_{02}={\mu}_2$, and
the moments order $(n,m)$ with respect to the mean:
\begin{eqnarray*}
  {\mu}_{nm}\,=\,
  E[({X}_1-{\mu}_1)^n\,({X}_2-{\mu}_2)^m]\,=\,
 \int_{{\Rcal}^2}\,(x_1-{\mu}_1)^n\,(x_2-{\mu}_2)^m\,p(x_1,x_2)\,dx_1\,
         dx_2
 \end{eqnarray*}
for which
 \begin{eqnarray}
  {\mu}_{20}\,=\,E[({X}_1-{\mu}_1)^2]\,=\,V[{X}_1]\,=\,{\sigma}_1^2
\hspace{0.5cm}{\rm and}\hspace{0.5cm}
  {\mu}_{02}\,=\,E[({X}_2-{\mu}_2)^2]\,=\,V[{X}_2]\,=\,{\sigma}_2^2
    \nonumber 
 \end{eqnarray}
The moment
 \begin{eqnarray}
  {\mu}_{11}\,=\,E[({X}_1-{\mu}_1)\,({X}_2-{\mu})]\,=\,
  {\alpha}_{11}\,-\,{\alpha}_{10}\,{\alpha}_{01}\,=\,
    V[{X}_1,{X}_2]\,=\,
    V[{X}_2,{X}_1]\nonumber
 \end{eqnarray}
is called {\bf covariance} between the random quantities
${X}_1$ and ${X}_2$ and, if they are independent, ${\mu}_{11}=0$. 
The second order moments with respect to the mean can be condensed in
matrix form, the {\bf covariance matrix} defined as: 
 \begin{eqnarray}
     \Vbold[{\Xbold}]\,=\,  \left(  \begin{array}{cc}
                 {\mu}_{20} & {\mu}_{11} \\ 
                 {\mu}_{11} & {\mu}_{02} 
               \end{array}
             \right)\,=\,
             \left(  \begin{array}{cc}
                 V[{X}_1,{X}_1] & V[{X}_1,{X}_2] \\
                 V[{X}_1,{X}_2] & V[{X}_2,{X}_2]
               \end{array}
             \right)            \nonumber
 \end{eqnarray}
Similarly, for $\Xbold=(X_1,X_2,{\ldots},X_n)$ we have the moments with 
respect to the origin
 \begin{eqnarray}
  {\alpha}_{k_1,k_2,{\ldots},k_n}\,=\,
  E[{X}_1^{k_1}\,{X}_2^{k_2}{\cdots}
      {X}_n^{k_n}]\,\,;
              \nonumber
 \end{eqnarray}
the moments with respect to the mean
 \begin{eqnarray}
  {\mu}_{k_1,k_2,{\ldots},k_n}\,=\,
  E[({X}_1-{\mu}_1)^{k_1}\,({X}_2-{\mu}_2)^{k_2}{\cdots}
      ({X}_n-{\mu}_n)^{k_n}]
              \nonumber
 \end{eqnarray}
and the covariance matrix:
 \begin{eqnarray}
   \Vbold[{\Xbold}]\,=\,  \left(  \begin{array}{cccc}
 {\mu}_{20{\ldots}0} & {\mu}_{11{\ldots}0} & {\cdots} & {\mu}_{10{\ldots}1}\\
 {\mu}_{11{\ldots}0} & {\mu}_{02{\ldots}0} & {\cdots} & {\mu}_{01{\ldots}1}\\
 {\vdots}            & {\vdots}            & {\cdots} & {\vdots}           \\
 {\mu}_{10{\ldots}1} & {\mu}_{01{\ldots}1} & {\cdots} & {\mu}_{00{\ldots}2}
               \end{array}
             \right)\,=\,
             \left(  \begin{array}{cccc}
 V[{X}_1,{X}_1] &V[{X}_1,{X}_2]&{\cdots}&
               V[{X}_1,{X}_n]\\
 V[{X}_1,{X}_2] & V[{X}_2,{X}_2]&{\cdots}&
               V[{X}_2,{X}_n]\\
 {\vdots}            & {\vdots}          & {\cdots} & {\vdots}       \\
 V[{X}_1,{X}_n]&V[{X}_2,{X}_n]&{\cdots}&V[{X}_n,{X}_n]
               \end{array}
             \right)   \nonumber
 \end{eqnarray}

The covariance matrix 
$\Vbold[\Xbold]=E[(\Xbold-\mubold)(\Xbold-\mubold)^T]$
has the following properties that are easy to prove from basic 
matrix algebra relations:
\begin{itemize}
 \item[1)] It is a {\bf symmetric} matrix 
           ($\Vbold=\Vbold^T$) with {\bf positive diagonal
           elements} ($\Vbold_{ii}{\ge}0$); 
 \item[2)] It is {\bf positive defined}
           ($\xbold^T \Vbold \xbold {\geq}0$;
            $\,\,{\forall}{\xbold}{\in}{\Rcal}^n$
            with the equality when $\forall i$ $x_i=0$);
 \item[3)] Being $\Vbold$ symmetric, all the eigenvalues are real and
           the corresponding eigenvectors orthogonal. Furthermore, since
           it is positive defined all eigenvalues are positive;
 \item[4)] If ${\mathbf{J}}$ is a diagonal matrix
           whose elements are the eigenvalues of $\Vbold$ and
           ${\mathbf{H}}$ a matrix whose columns are the corresponding
           eigenvectors, then
           ${\mathbf{V}}={\mathbf{H}}{\mathbf{J}}{\mathbf{H}}^{-1}$
           (Jordan dixit);
 \item[5)] Since $\Vbold$ is symmetric, there is an orthogonal matrix
           ${\mathbf{C}}$ (${\mathbf{C}}^T={\mathbf{C}}^{-1}$)
           such that ${\mathbf{C}}{\mathbf{V}}{\mathbf{C}}^T={\mathbf{D}}$
           with ${\mathbf{D}}$ a diagonal matrix whose elements are the
           eigenvalues of $\Vbold$;
 \item[6)] Since ${\mathbf{V}}$ is symmetric and positive defined,
           there is a non-singular matrix ${\mathbf{C}}$ such that
           ${\mathbf{V}}={\mathbf{C}}{\mathbf{C}}^T$;
 \item[7)] Since ${\mathbf{V}}$ is symmetric and positive defined,
           the inverse ${\mathbf{V}}^{-1}$ is also symmetric and positive
           defined;
 \item[8)] (Cholesky Factorization) 
           Since ${\mathbf{V}}$ is symmetric and positive defined,
           there exists a unique lower triangular matrix
           ${\mathbf{C}}$ 
           (${\mathbf{C}}_{ij}=0;\,\,{\forall}i<j$) with positive
           diagonal elements such that
           ${\mathbf{V}}={\mathbf{C}}{\mathbf{C}}^T$ (more about this
           in lecture 3).
\end{itemize}

Among other things to be discussed later,
the moments of the distribution are interesting because 
they give an idea of the shape and location of the
probability distribution and, in many cases, the distribution parameters 
are expressed in terms of the moments.

\subsubsection{Position parameters}
Let $X{\sim}p(x)$ with support in $\Omega{\subset}{\Rcal}$. The position
parameters {\sl choose} a {\sl characteristic} value of $X$ and indicate 
more or less where the distribution is located. Among them we have the
{\bf mean value}
\begin{eqnarray*}
 {\mu}\,=\,{\alpha}_1\,=\,E[{X}]\,=\,
                               \int_{-\infty}^{\infty}\,x\,p(x)\,dx
\end{eqnarray*}
The mean is bounded by the minimum and maximum values the random
quantity can take but, clearly, if $\Omega{\subset}{\Rcal}$ it may happen
that ${\mu}{\notin}\Omega$. If, for instance, 
$\Omega=\Omega_1{\cup}\Omega_2$
is the union of two disconnected regions, $\mu$ may lay in between and 
therefore ${\mu}{\notin}\Omega$. On the other hand, as has been mentioned the
integral has to be absolute convergent and there are some probability
distributions for which there is no mean value.
There are however other interesting location quantities. The {\bf mode}
is the value $x_0$ of $X$ for which the distribution is maximum; that is,
\begin{eqnarray*}
  x_0\,=\,{\rm sup}_{x{\in}{\Omega}}p(x)
\end{eqnarray*}
Nevertheless, it may happen that there are
several relative maximums so we talk about uni-modal, bi-modal,... 
distributions. The {\bf median} is the value $x_m$
such that
\begin{eqnarray*}
F(x_m)\,=\,P(X{\leq}x_m)\,=\,1/2
       \hspace{0.5cm}{\longrightarrow}\hspace{0.5cm}
       \int_{-\infty}^{x_m}p(x)dx\,=\,
       \int_{x_m}^{\infty}p(x)dx\,=\,P(X>x_m)\,=\,1/2
\end{eqnarray*}
For discrete random quantities, the distribution function is either a finite
or countable combination of indicator functions
${\mbox{\boldmath $1$}}_{A_k}(x)$ with $\{A_k\}_{k=1}^{n,\infty}$ a 
partition of ${\Omega}$ so it may happen that
$F(x)=1/2$  $\forall x{\in}A_k$. Then, any value of the interval $A_k$ can
be considered the median. 
Last, we may consider the {\bf quantiles} $\alpha$ 
defined as the value $q_{\alpha}$ of the random quantity such that 
$F(q_{\alpha})=P({X}\,{\leq}\,q_{\alpha})\,=\,{\alpha}$
so the {\sl median} is the {\sl quantile} $q_{1/2}$.

\subsubsection{Dispersion parameters}
 
There are many ways to give an idea of how {\sl dispersed} are the values 
the random quantity may take. Usually they are based on the mathematical
expectation of a function that depends on the difference between $X$ and
some characteristic value it may take; for instance
$E[|{X}-{\mu}|]$. By far, the most usual and important one is the
already defined {\bf variance}
\begin{eqnarray*}
V[{X}]={\sigma}^2=E[(X-E[X])^2]\,=\,\int_{\Rcal}(x-\mu)^2p(x)dx
\end{eqnarray*}
provided it exists.
Note that if the random quantity $X$ has dimension $D[{X}]=d_{X}$, the
variance has dimension $D[{\sigma}^2]=d_{X}^2$ so to have a quantity
that gives an idea of the {\sl dispersion} and has the same dimension one
defines the {\bf standard deviation} 
${\sigma}=+\sqrt{V[X]}=+\sqrt{{\sigma}^2}$ and, if both the mean value 
$(\mu)$ and
the variance exist, the {\bf standardized} random quantity
\begin{eqnarray*}
       {Y}\,=\,\frac{\textstyle X\,-\,{\mu}}
                       {\sigma}                       
\end{eqnarray*}
for which $E[Y]=0$ and $V[Y]={\sigma}_Y^2=1$.

\subsubsection{Asymmetry and Peakiness parameters}
\vspace*{0.1cm}

Related to higher order non-central moments, there
are two dimensionless quantities of interest: the skewness and the kurtosis.
The first non-trivial odd moment with respect to the mean is that of 
order 3: ${\mu}_3$. Since it has dimension
$D[{\mu}_3]=d_{X}^3$ we define the {\bf skewness} $({\gamma}_1)$ 
as the dimensionless quantity
\begin{eqnarray*}
           {\gamma}_1\,\stackrel{def}{=}\,\frac{\mu_3}{\mu_2^{3/2}}\,=\,
              \frac{\textstyle {\mu}_3}
                           {\textstyle {\sigma}^3}\,=\,
                          \frac{\textstyle E[({X}-{\mu})^3]}
                           {\textstyle {\sigma}^3}\,\,\,\,\,
\end{eqnarray*}
The skewness is ${\gamma}_1=0$
for distributions that are symmetric with respect to the
mean, ${\gamma}_1>0$ if the probability content is more concentrated on
the right of the mean and ${\gamma}_1<0$ if it is to the left of the mean.
Note however that there are many asymmetric distributions for which
${\mu}_3=0$ and therefore ${\gamma}_1=0$.  For unimodal distributions,
it is easy to see that

\begin{center}
\begin{tabular}{p{1.5cm}p{5.7cm}}
  ${\gamma}_1=0$    &    
  {\sl mode = median = mean}    \\   & \\
  ${\gamma}_1>0$    &    
  {\sl mode $<$ median $<$ mean}   \\   & \\
  ${\gamma}_1<0$    &    
  {\em mode $>$ median $>$ mean}
\end{tabular}
\end{center}

The {\bf kurtosis} is defined, again for dimensional considerations, as
\begin{eqnarray*}
           {\gamma}_2\,=\,\frac{\mu_4}{\mu_2^2}\,=\,
                 \frac{\textstyle {\mu}_4}
                           {\textstyle {\sigma}^4}\,=\,
                          \frac{\textstyle E[({X}-{\mu})^4]}
                           {\textstyle {\sigma}^4}
\end{eqnarray*}
and gives an idea of how {\sl peaked} is the distribution.
For the Normal distribution ${\gamma}_2=3$ so in order to have a reference 
one defines the  {\sl extended kurtosis} as ${\gamma}_2^{ext}={\gamma}_2-3$. 
Thus, ${\gamma}_2^{ext}>0$ $(<0)$ indicates that the distribution is  
{\sl more (less) peaked} than the Normal.
Again, ${\gamma}_2^{ext}=0$ for the Normal density and for any other 
distribution for which ${\mu}_4=3\,{\sigma}^4$. Last
you can check that $\forall a,b{\in}{\Rcal}$
$E[(X-\mu-a)^2(X-\mu-b)^2]>0$ so, for instance, 
defining $u=a+b$, $w=ab$ and taking derivatives, 
${\gamma}_2{\geq}1+{\gamma}_1^2$.

\vspace{0.5cm}
\noindent
{\raya}                   
\vspace{0.35cm}
\footnotesize

\noindent
{\bf Example 1.15:} Consider the discrete random quantity 
$X{\sim}Pn(k|{\lambda})$ with
 \begin{eqnarray*}
 P({X}=k)\,{\equiv}\,Pn(k|{\lambda})\,=\,e^{-{\lambda}}\,
  \frac{\textstyle {\lambda}^k}{\textstyle \Gamma(k+1)}
\hspace{0.5cm};\hspace{1.0cm}
{\lambda}{\in}{\Rcal}^+\,;\,\,k=0,1,2,...
 \end{eqnarray*}
The moments with respect to the origin are
\begin{eqnarray*}
  {\alpha}_n(\lambda)\,{\equiv}\,E[{X}^n]\,=\,e^{-{\lambda}}\,
        \sum_{k=0}^{\infty}\,k^n\,
        \frac{\textstyle {\lambda}^k}{\textstyle k!}
\end{eqnarray*}
If $a_k$ denotes the $k^{th}$ term of the sum, then
\begin{eqnarray*}
 a_{k+1}\,=\,
  \frac{\textstyle {\lambda}}{\textstyle k+1}\,
  \left(1\,+\,\frac{\textstyle 1}{k}\right)^n\,a_k
  \hspace{0.5cm}{\longrightarrow}\hspace{0.5cm}
 {\rm lim}_{k{\rightarrow}{\infty}} \left|
  \frac{\textstyle a_{k+1}}{\textstyle a_k}\right|\,
  \rightarrow\,0
  \nonumber
 \end{eqnarray*}
so being the series absolute convergent all order moments exist.
Taking the derivative of ${\alpha}_n({\lambda})$ with respect to
${\lambda}$ one gets the recurrence relation
\begin{eqnarray*}
  {\alpha}_{n+1}({\lambda})\,=\,{\lambda}\, \left(
        {\alpha}_{n}({\lambda})\,+\,
   \frac{\textstyle d{\alpha}_{n}({\lambda})}{\textstyle d{\lambda}}\right)
\hspace{0.5cm};\hspace{1.0cm}
  {\alpha}_0({\lambda})\,=\,1
\end{eqnarray*}
so we can easily get
\begin{eqnarray*}
  {\alpha}_0\,=\,1 \,;\hspace{0.5cm} 
  {\alpha}_1\,=\,\lambda \,;\hspace{0.5cm} 
  {\alpha}_2\,=\,\lambda(\lambda+1) \,;\hspace{0.5cm} 
  {\alpha}_3\,=\,{\lambda}({\lambda}^2+3{\lambda}+1) \,;\hspace{0.5cm} 
  {\alpha}_4\,=\, {\lambda}({\lambda}^3+6{\lambda}^2+7{\lambda}+1)
\end{eqnarray*}
and from them
\begin{eqnarray*}
     {\mu}_0\,=\,1\,;\hspace{0.5cm}   
     {\mu}_1\,=\,0   \,;\hspace{0.5cm} 
     {\mu}_2\,=\,\lambda \,;\hspace{0.5cm} 
     {\mu}_3\,=\,\lambda \,;\hspace{0.5cm} 
     {\mu}_4\,=\, \lambda(3\lambda+1)
\end{eqnarray*}

\noindent
Thus, for the Poisson distribution $Po(n|\lambda)$ we have that: 
 \begin{eqnarray*}
  E[X]={\lambda}\hspace{0.5cm};\hspace{1.0cm}
  V[X]={\lambda}\hspace{0.5cm};\hspace{1.0cm}
  {\gamma}_1={\lambda}^{-1/2}\hspace{0.5cm};\hspace{1.0cm}
  {\gamma}_2=3+{\lambda}^{-1}
 \end{eqnarray*}

\vspace{0.35cm}

\noindent
{\bf Example 1.16:} Consider $X{\sim}Ga(x|a,b)$ with:
 \begin{eqnarray*}
 p(x)\,=\,
  \frac{\textstyle a^b}{\textstyle {\Gamma}(b)}\,e^{-ax}\,x^{b-1}
{\mbox{\boldmath $1$}}_{(0,\infty)}(x)
{\lambda}\hspace{0.5cm};\hspace{1.0cm} a,b{\in}{\Rcal}^+
 \end{eqnarray*}
The moments with respect to the origin are
 \begin{eqnarray}
 {\alpha}_n\,=\,E[{X}^n]\,=\,
  \frac{\textstyle a^b}{\textstyle {\Gamma}(b)}\,
  \displaystyle {
               \int_{0}^{\infty}\,
  e^{-a x}\,x^{b+n-1}\,dx}\,=\,
  \frac{\textstyle {\Gamma}(b+n)}{\textstyle {\Gamma}(b)}\,a^{-n}
  \nonumber
 \end{eqnarray}
being the integral absolute convergent. Thus we have:
\begin{eqnarray*}
  {\mu}_n\,=\,\frac{1}{a^n\Gamma(b)}
    \sum^{n}_{k=0}\,
       \left(  \begin{array}{c}
                 n \\ k
               \end{array}
     \right)\,(-b)^{n-k}\,\Gamma(b+k)
              \nonumber
 \end{eqnarray*}
and in consequence
 \begin{eqnarray*}
  E[X]=\frac{b}{a}\hspace{0.5cm};\hspace{1.0cm}
  V[X]=\frac{b}{a^2}\hspace{0.5cm};\hspace{1.0cm}
  {\gamma}_1=\frac{2}{\sqrt{b}}\hspace{0.5cm};\hspace{1.0cm}
  {\gamma}_2^{ext.}=\frac{6}{b}
 \end{eqnarray*}

\vspace{0.35cm}

\noindent
{\bf Example 1.17:} For the Cauchy distribution $X{\sim}Ca(x|1,1)$,
 \begin{eqnarray*}
 p(x)\,=\,=\,
  \frac{\textstyle 1}{\textstyle {\pi}}\,
  \frac{\textstyle 1}{\textstyle 1\,+\,x^2}
{\mbox{\boldmath $1$}}_{(-\infty,\infty)}(x)
 \end{eqnarray*}
we have that 
\begin{eqnarray*}
 {\alpha}_n\,=\,E[{X}^n]\,=\,\frac{\textstyle 1}{\textstyle {\pi}}\,
  \displaystyle {
               \int_{-\infty}^{\infty}\,
  \frac{\textstyle x^n}{\textstyle 1\,+\,x^2}\,dx}
\end{eqnarray*}
and clearly the integral diverges for
$n>1$ so there are no moments but the trivial one $\alpha_0$.
Even for $n=1$, the integral
\begin{eqnarray*}
  \displaystyle {
               \int_{-\infty}^{\infty}\,
  \frac{\textstyle |x|}{\textstyle (1\,+\,x^2)}\,dx}\,=\,
 2\, \displaystyle {
               \int_{0}^{\infty}\,
  \frac{\textstyle x}{\textstyle (1\,+\,x^2)}\,dx}\,=\,
 {\rm lim}_{a{\rightarrow}{\infty}}\,{\rm ln}\,(1+a^2)
\end{eqnarray*}
is not absolute convergent so, in strict sense, there is no mean value.
However, the mode and the median are $x_0=x_m=0$,
the distribution is symmetric about $x=0$ and for $n=1$ there exists the
Cauchy's Principal Value and is equal to 0.
Had we introduced the Probability Distributions as a subset of Generalized 
Distributions, the Principal Value is an admissible
distribution.
It is left as an exercise to show that for: 
\begin{itemize}
\item[$\bullet$]{\bf Pareto}: $X{\sim}Pa(x|\nu,x_m)$ with
$p(x|x_m,\nu)\propto x^{-(\nu+1)}
{\mbox{\boldmath $1$}}_{[x_m,\infty)}(x)
\hspace{0.2cm};\hspace{0.5cm} x_m,\nu{\in}{\Rcal}^+$
\item[$\bullet$]{\bf Student}: $X{\sim}St(x|\nu)$ with
$p(x|\nu)\propto
\left(1+x^2/{\nu}\right)^{-(\nu+1)/2}
{\mbox{\boldmath $1$}}_{(-\infty,\infty)}(x)
\hspace{0.2cm};\hspace{0.5cm} \nu{\in}{\Rcal}^+$
\end{itemize}
the moments ${\alpha}_n=E[X^n]$ exist iff $n<\nu$.

\noindent
Another distribution of interest in physics is the Landau Distribution
that describes the energy lost
by a particle when traversing a material under certain conditions.
The probability density, given as the inverse Laplace Transform, is:
\begin{eqnarray*}
 p(x)\,=\,\frac{1}{2\pi i}\int_{c-i\infty}^{c+i\infty}
           e^{s\log s+xs}ds
\end{eqnarray*}
with $c\in \Rcal^+$ and 
closing the contour on the left along a counterclockwise semicircle
with a branch-cut along the negative real axis it has a real representation
\begin{eqnarray*}
 p(x)\,=\,\frac{1}{\pi}\int_{0}^{\infty}
           e^{-(r\log r+xr)}\,\sin(\pi r)\,dr
\end{eqnarray*}
The actual expression of the distribution of the energy loss 
is quite involved and some simplifying assumptions have been made; 
among other things, that the energy transfer in the collisions is unbounded 
(no kinematic constraint). But nothing is for free and
the price to pay is that the Landau Distribution has no moments other
than the trivial of order zero. This is why instead of mean and variance 
one talks about the {\sl most probable energy loss} and the
{\sl full-width-half-maximum}.

\vspace{0.35cm}

\noindent
{\bf Problem 1.3:} Show that if $X_{1,2}{\sim}Un(x|0,1)$, then for
$X=X_1^{X_2}$ we have that $p(x)=-x^{-1}{\rm Ei}(\ln x)$, with ${\rm Ei}(z)$
the exponential integral, and $E[X^m]=m^{-1}{\ln}(1+m)$.

\smill
{\raya}                   
\vspace{1.0cm}
\subsubsection{Correlation Coefficient}

The {\bf covariance} 
between the random quantities ${X}_i$ and ${X}_j$ was defined as:
 \begin{eqnarray*}
   V[{X}_i,{X}_j]\,=\,    V[{X}_j,{X}_i]\,=\,
   E[({X}_i-{\mu}_i)\,({X}_i-{\mu}_j)]\,=\,
   E[X_iX_j]-E[X_i]E[X_j]
 \end{eqnarray*}
If $X_i$ and $X_j$ are independent, then $E[X_iX_j]=E[X_i]E[X_j]$ and 
$V[{X}_i,{X}_j]=0$. Conversely, if $V[{X}_i,{X}_j]\neq 0$ then
$E[{X}_i{X}_j]{\neq}E[{X}_i]E[{X}_j]$ and in consequence $X_i$ and $X_j$ 
are not statistically independent. Thus, the {\sl covariance}
$V[{X}_i,{X}_j]$ serves to quantify, to some extent,
the degree of {\sl statistical dependence}
between the random quantities $X_i$ and $X_j$. Again, for dimensional
considerations one defines the {\sl correlation coefficient}
\begin{eqnarray*}
   {\rho}_{ij}\,=\,
\frac{\textstyle V[{X}_i,{X}_j]}
                  {\textstyle \sqrt{V[{X}_i]\,V[{X}_j]}}\,=\,
      \frac{\textstyle E[{X}_i\,{X}_j]\,-\,E[X_i]E[X_j]}
                  {\textstyle {\sigma}_i\,{\sigma}_j}
\end{eqnarray*}

Since $p(x_i,x_j)$ is a non-negative function we can write
\begin{eqnarray*}
V[{X}_i,{X}_j]=\int_{{\Rcal}^2}\left\{(x_i-\mu_i)\sqrt{p(x_i,x_j)}\right\}
\left\{(x_j-\mu_j)\sqrt{p(x_i,x_j)}\right\}dx_idx_j
\end{eqnarray*}
and from the Cauchy-Schwarz inequality:
\begin{eqnarray*}
   -1\,{\leq}\,{\rho}_{ij}\,\leq\,1
\end{eqnarray*}
The extreme values $(+1,-1)$ will be taken when
$E[{X}_i{X}_j]=E[X_i]E[X_j]{\pm}{\sigma}_i{\sigma}_j$
and ${\rho}_{ij}=0$ when $E[{X}_i{X}_j]=E[X_i]E[X_j]$. In particular,
it is immediate to see that if here is a linear relation between both
random quantities; that is,
${X}_i=a{X}_j+b$, then ${\rho}_{ij}={\pm}1$.
Therefore, it is a {\bf linear correlation coefficient}.
Note however that:
\begin{itemize}
    \item[$\bullet$]
    If $X_i$ and $X_j$ are linearly related,
    ${\rho}_{ij}={\pm}1$, but ${\rho}_{ij}={\pm}1$ {\bf does not}
    imply necessarily a linear relation;
    \item[$\bullet$]
    If $X_i$ and $X_j$ are statistically independent, then
    ${\rho}_{ij}=0$ but ${\rho}_{ij}=0$ {\bf does not}
    imply necessarily statistical independence as the following
    example shows.
\end{itemize}

\vspace{0.5cm}
\noindent
{\raya}                   
\vspace{0.35cm}
\footnotesize

\noindent
{\bf Example 1.18:} Let $X_1\sim p(x_1)$ and define a random quantity $X_2$ as
\begin{eqnarray*}
  {X_2}\,=\,g({X_1})\,=\,a\,+\,b{X_1}\,+\,c{X_1}^2    
\end{eqnarray*}
Obviously, $X_1$ and $X_2$ are not statistically independent for there is
a clear parabolic relation. However
 \begin{eqnarray*}
  V[{X_1},{X_2}]\,=\,E[{X_1}{X_2}]\,-\,E[{X_1}]\,E[{X_2}]\,=\,
  b{\sigma}^2\,+\,c({\alpha}_3-{\mu}^3-{\mu}{\sigma}^2)
 \end{eqnarray*}
with $\mu$, ${\sigma}^2$ and ${\alpha}_3$ respectively the mean, variance
and moment of order 3 with respect to the origin of $X_1$ and, if we take
$b=c{\sigma}^{-2}({\mu}^3+{\mu}{\sigma}^2-{\alpha}_3)$
then $V[{Y},{X}]=0$ and so is the (linear) correlation coefficient.

\vspace{0.5cm}
\noindent
{\rayan}                   
\vspace{0.35cm}
\footnotesize
\noindent
{\bf NOTE 3: Information as a measure of independence.} The {\sl Mutual 
Information} (see Chapter 4.4) serves also to quantify 
the degree of statistical dependence
between random quantities. Consider for instance the two-dimensional
random quantity $\Xbold=(X_1,X_2){\sim}p(x_1,x_2)$. Then:
\begin{eqnarray*}
          I(X_1:X_2)\,=\,
            \int_{\Xbold}\,dx_1\,dx_2
            p(x_1,x_2)\,{\rm ln}\,
            \left( \frac{\textstyle   p(x_1,x_2)}
                        {\textstyle   p(x_1)\,p(x_2)} \right)
\end{eqnarray*}
and $I(X_1:X_2){\geq}0$ with equality iff $p(x_1,x_2)=p(x_1)p(x_2)$.
Let's look as an example to the bi-variate normal distribution:
$N(\xbold|\mubold,\Sigmabold)$: 
\begin{eqnarray*}
    p({\xbold}|{\phibold})\,=\,(2{\pi})^{-1}\,|{\rm det}[{\Sigmabold}]|^{-1/2}\,
       {\rm exp}\left\{ 
     -\,\frac{1}{2}\left((\xbold-\mubold)^T\,{\Sigmabold}^{-1}\,
       (\xbold-\mubold)\right)
                \right\}
\end{eqnarray*}
with covariance matrix
\begin{eqnarray*}
   {\Sigmabold}\,=\,
   \left(\begin{array}{cc}
     {\sigma}_1^{2} & {\rho}{\sigma}_1{\sigma}_2 \\ 
     {\rho}{\sigma}_1{\sigma}_2 & {\sigma}_2^{2} 
\end{array}\right)
\hspace{0.3cm}{\rm and}\hspace{0.3cm}
{\det}[{\Sigmabold}]={\sigma}_1^{2} {\sigma}_2^{2}
(1-{\rho})^2
 \end{eqnarray*}
Since $X_i{\sim}N(x_i|\mu_i,\sigma_i);\,i=1,2$ we have that:
\begin{eqnarray*}
 I(X_1:X_2)\,=\,{\int}_{\Xbold}\,dx_1\,dx_2\,p(\xbold|\mubold,\Sigmabold)\,
{\rm ln}\left( 
               \frac{\textstyle p(\xbold|\mubold,\Sigmabold)}
                    {\textstyle p(x_1|\mu_1,\sigma_1)\,
                                p(x_2|\mu2,\sigma_2)}
          \right)\,=\,-\,\frac{\textstyle 1}{\textstyle 2}\,
       {\rm ln}\,(1\,-\,{\rho}^2)
 \end{eqnarray*}
Thus, if $X_1$ and $X_2$ are independent (${\rho}=0$), 
$I(X_1:X_2)=0$ and when ${\rho}{\rightarrow}{\pm}1$,
$I(X_1:X_2){\rightarrow}{\infty}$.

{\rayan}                   
\vspace{1.0cm}
\smill

\vspace*{0.3cm}
\subsection{The "Error Propagation Expression"}
\vspace*{0.1cm}

Consider a n-dimensional random quantity $\Xbold=(X_1,\ldots,X_n)$
with $E[X_i]={\mu_i}$ and the random quantity $Y=g(\Xbold)$ with $g(x)$ an
infinitely differentiable function. If we make
a Taylor expansion of $g(\Xbold)$ around $E[\Xbold]=\mubold$ we have
\begin{eqnarray*}
Y=g(\Xbold)=g(\mubold)+
 \sum_{i=1}^n
 \left( \frac{\partial g(\xbold)}{\partial x_i}\right)_{\mubold}Z_i
+\frac{1}{2!}\sum_{i=1}^n\sum_{j=1}^n
 \left( \frac{\partial^2 g(\xbold)}{\partial x_i \partial x_j}\right)_{\mubold}
 Z_iZ_j+R
\end{eqnarray*}
where $Z_i=X_i-\mu_i$.
Now, $E[Z_i]=0$ and $E[Z_iZ_j]=V[X_i,X_j]=V_{ij}$ so
\begin{eqnarray*}
E[Y]\,=\,E[g(\Xbold)]\,=\,g(\mubold)\,+\,
+\frac{1}{2!}\sum_{i=1}^n\sum_{j=1}^n
 \left( \frac{\partial^2 g(\xbold)}{\partial x_i \partial x_j}\right)_{\mubold}
 V_{ij}+...
\end{eqnarray*}
and therefore
\begin{eqnarray*}
Y-E[Y]=
 \sum_{i=1}^n
 \left( \frac{\partial g(\xbold)}{\partial x_i}\right)_{\mubold}Z_i
+\frac{1}{2!}\sum_{i=1}^n\sum_{j=1}^n
 \left( \frac{\partial^2 g(\xbold)}{\partial x_i \partial x_j}\right)_{\mubold}
 (Z_iZ_j-V_{ij})+...
\end{eqnarray*}
Neglecting all but the first term
\begin{eqnarray*}
V[Y]\,=\,E[(Y-E[Y])^2]\,=\,\sum_{i=1}^n\sum_{j=1}^n
 \left( \frac{\partial g(x)}{\partial x_i}\right)_{\mubold}\,
 \left( \frac{\partial g(x)}{\partial x_j}\right)_{\mubold}\,V[X_i,X_j]
  \,+\,...
\end{eqnarray*}
This is the first order approximation to $V[Y]$ and usually is reasonable
but has to be used with care. On the one hand, we have assumed that higher
order terms are negligible and this is not always the case so further
terms in the expansion may have to be considered. Take for instance the simple
case $Y=X_1X_2$ with $X_1$ and $X_2$ independent random quantities. The
first order expansion gives 
$V[Y]{\simeq}\mu_1^2\sigma_2^2+\mu_2^2\sigma_1^2$ and including second order
terms (there are no more)
$V[Y]=\mu_1^2\sigma_2^2+\mu_2^2\sigma_1^2+\sigma_1^2\sigma_2^2$; the correct 
result. On the other hand, all this is obviously meaningless if the random
quantity $Y$ has no variance. This is for
instance the case for $Y=X_1X_2^{-1}$ when $X_{1,2}$ are Normal distributed.

\section{\LARGE \bf Integral Transforms}

The Integral Transforms of Fourier, Laplace and Mellin are a very useful
tool to study the properties of the random quantities and their distribution
functions. In particular, they will allow us to obtain the distribution
of the sum, product and ratio of random quantities, the moments
of the distributions and to study the convergence of a sequence 
$\left\{F_k(x)\right\}_{k=1}^{\infty}$ of distribution functions to $F(x)$. 

\subsection{The Fourier Transform}

Let $f:{\Rcal}{\rightarrow}{\Ccal}$ be a complex and 
integrable function $(f{\in}L_1({\Rcal}))$.
The {\sl Fourier Transform} ${\Fcal}(t)$ with $t{\in}{\Rcal}$ of $f(x)$ 
is defined as:
\begin{eqnarray*}
{\Fcal}(t)\,=\,\int_{-\infty}^{\infty}\,f(x)\,e^{ixt}\,dx
\end{eqnarray*}
The class of functions for which the Fourier Transform exists is certainly
much wider than the 
probability density functions $p(x){\in}L_1({\Rcal})$ (normalized 
real functions of real argument) we are interested in for which the
transform always exists. If $X{\sim}p(x)$, the Fourier Transform is
nothing else but the mathematical expectation 
\begin{eqnarray*}
 {\Fcal}(t)\,=\,E [e^{\textstyle i t X}]\,\,;\,\,\,\,\,
     t\,\in\,{\Rcal}  
\end{eqnarray*}
and it is 
called {\sl Characteristic Function} ${\Phi}(t)$. 
Thus, depending on the character of the random quantity $X$, we shall have:

\begin{itemize}
\item[$\bullet$] if $X$ is {\bf discrete}: 
         $
         {\Phi}(t)\,=\,
         \displaystyle{ \sum_{x_k}}\,
         e^{\textstyle i t x_k}\,P(X = x_k)
         $
\item[$\bullet$] if $X$ is {\bf continuous}: 
         $
         {\Phi}(t)\,=\,
         \displaystyle{ \int_{-\infty}^{+\infty}}\,
         e^{\textstyle i t x}\,dP(x)\,
         =\,\displaystyle{ \int_{-\infty}^{+\infty}}\,
         e^{\textstyle i t x}\,p(x)\,dx$
\end{itemize}

Attending to its definition, the Characteristic Function 
${\Phi}(t)$, with $t\in {\Rcal}$, is a complex function and
has the following properties:

\begin{itemize}
\item[1)] $\,{\Phi}(0)\,=\,1$; 

\item[2)] ${\Phi}(t)$ is bounded: $|{\Phi}(t)|\,{\leq}1$;

\item[3)] ${\Phi}(t)$ has schwarzian symmetry:
          ${\Phi}(-t)\,=\,\overline{\Phi}(t)$; 

\item[4)] ${\Phi}(t)$ is uniformly continuous in $\Rcal$.
\end{itemize}

\vspace{0.5cm}
\noindent
{\rayan}                   
\vspace{0.35cm}
\footnotesize
\noindent
The first three properties are obvious. For the fourth one, observe that
for any $\epsilon > 0$ there exists a  $\delta > 0$ such that
$|{\Phi}(t_1)-{\Phi}(t_2)| < {\epsilon}$ when
$|t_1-t_2| < \delta$ with $t_1$ and $t_2$ arbitrary in ${\Rcal}$ because
\begin{eqnarray*} 
           |{\Phi}(t+{\delta})-{\Phi}(t)|{\leq}
           \int_{-\infty}^{+\infty}
           |1-e^{\textstyle -i \delta x}|dP(x)=
           2\,\displaystyle{ \int_{-\infty}^{+\infty}}\,
           |\sin\, {\delta}x/2|\,dP(x) 
           \end{eqnarray*}   
and this integral can be made arbitrarily small taking a
sufficiently small $\delta$. 

{\rayan}                   
\vspace{1.0cm}
\smill

\noindent
These properties, that obviously hold also for a discrete random quantity,
are {\bf necessary} but {\bf not sufficient} for a
function ${\Phi}(t)$ to be the Characteristic Function of a 
distribution $P(x)$ (see example 1.19). Generalizing 
for a n-dimensional random quantity ${\bf X}=(X_1,\ldots ,X_n)$:
\begin{eqnarray*}
       {\Phi}(t_1,\ldots ,t_n)\,=\,
       E[e^{\textstyle i {\bf t}{\bf X}}]\,=\,
       E[e^{\textstyle i (t_1X_1+\ldots +
       t_nX_n)}] 
\end{eqnarray*}
so, for the discrete case:
 \begin{eqnarray*}
         {\Phi}(t_1,\ldots ,t_n)\,=\,
         \displaystyle{ \sum_{x_1}}\,\ldots
         \displaystyle{ \sum_{x_n}}\,
         e^{\textstyle i (t_1 x_1+\ldots +t_n x_n)}\,
         P(X_1 = x_1,\ldots ,
           X_n = x_n)  
 \end{eqnarray*}
and for the continuous case:
 \begin{eqnarray*}
         {\Phi}(t_1,\ldots ,t_n)\,=\,
         \displaystyle{ \int_{-\infty}^{+\infty}}\,dx_1\,\ldots \,
         \displaystyle{ \int_{-\infty}^{+\infty}}\,dx_n\,
         e^{\textstyle i (t_1 x_1+\ldots +t_n x_n)}\,
         p(x_1,\ldots ,x_n)     
 \end{eqnarray*}
The n-dimensional Characteristic Function is such that:
\begin{itemize}
  \item[1)] $\,{\Phi}(0,\ldots ,0)\,=\,1$
  \item[2)] $\,|{\Phi}(t_1,\ldots ,t_n)|\,\leq\,1$
  \item[3)] $\,{\Phi}(-t_1,\ldots ,-t_n)\,=\,
             \overline{\Phi}(t_1,\ldots ,t_n)$
\end{itemize}

\vspace{0.5cm}
\noindent
{\rayan}                   
\vspace{0.2cm}
\footnotesize
\noindent
{\bf Laplace Transform}:
For a function $f(x):{\Rcal}^+{\rightarrow}{\Ccal}$ defined as 
$f(x)=0$ for $x<0$, we may consider also the {\em Laplace Transform} 
defined as 
\begin{eqnarray*}
L(s)\,=\,
\int_{0}^{\infty}\,e^{-sx}\,f(x)\,dx
\end{eqnarray*}
with $s{\in}{\Ccal}$ provided it exists.
For a non-negative random quantity $X{\sim}p(x)$ 
this is just the mathematical expectation $E[e^{-sx}]$ and is named
{\em Moment Generating Function} since the derivatives give the
moments of the distribution (see 5.1.4).
While the Fourier Transform exists for $f(x){\in}L_1({\Rcal})$,
the Laplace Transform exists if $e^{-sx}f(x){\in}L_1({\Rcal}^+)$ and 
thus, for a wider class of functions and
although it is formally defined for functions with 
non-negative support, it may be possible to extend the limits of integration to 
the whole real line ({\sl Bilateral Laplace Transform}).
However, for the functions we shall be interested in
(probability density functions), both Fourier and Laplace Transforms exist
and usually there is no major advantage in using one or the other.  
\vspace{0.2cm}
{\rayan}                   
\smill

\noindent
\noindent

\vspace{1.cm}
\footnotesize
\noindent
{\bf Example 1.19:} There are several criteria (Bochner, Kintchine, 
Cram\`er,...) 
specifying sufficient and necessary conditions for a function
${\Phi}(t)$, that satisfies the four aforementioned conditions, 
to be the Characteristic Function of a random quantity
$X{\sim}F(x)$. 
However, it is easy to find simple functions like
\begin{eqnarray}
g_1(t)\,=\,e^{\textstyle - t^4}\,\,\,\,\,\,\,\,\,\,\,\,\,\,\,{\rm and}
                             \,\,\,\,\,\,\,\,\,\,\,\,\,\,\,
g_2(t)\,=\,\frac{\textstyle 1}{\textstyle 1\,+\,t^4}\,\,; \nonumber
\end{eqnarray}
that satisfy four stated conditions and that
can not be Characteristic Functions associated to any distribution.
Let's calculate the moments of order one with respect to the origin
and the central one of order two.
In both cases (see section 5.1.4) we have that:
\begin{eqnarray*}
{\alpha}_1\,=\,{\mu}\,=\,E[{X}]\,=\,0 \hspace{1.cm}{\rm and}\hspace{1.cm}
{\mu}_2\,=\,{\sigma}^2\,=\,E[({X}-{\mu})^2]\,=\,0 
\end{eqnarray*}
that is, the mean value and the variance are zero so the distribution function
is zero almost everywhere but for 
${X}=0$ where $P({X}=0)=1$... but this is the 
{\em Singular Distribution} ${\rm Sn}(x|0)$ that takes the value 1 if
$X=0$ and 0 otherwise whose Characteristic Function is ${\Phi}(t)=1$. 
In general, any function ${\Phi}(t)$ that in a boundary of $t=0$ 
behaves as
${\Phi}(t)=1 + O(t^{2+ \epsilon})$ with ${\epsilon}>0$ can not be
the Characteristic Function associated to a distribution
$F(x)$ unless ${\Phi}(t)=1$ for all $t{\in}{\Rcal}$.

\vspace{0.35cm}

\noindent
{\bf Example 1.20:} The elements of the Cantor Set $C_S(0,1)$
can be represented in base 3 as:
 \begin{eqnarray}
           X\,=\,\sum_{n=1}^{\infty} \frac{\textstyle X_n}
                         {\textstyle 3^n}
\nonumber
 \end{eqnarray}
with $X_n{\in}\{0,2\}$. This set is non-denumerable and has zero Lebesgue
measure so any distribution with support on it is {\sl singular} and, in
consequence, has no pdf.
The Uniform Distribution on $C_S(0,1)$ is defined assigning a probability
$P(X_n=0)=P(X_n=2)=1/2$ (Geometric Distribution). 
Then, for the random quantity $X_n$ we have that
 \begin{eqnarray}
            {\Phi}_{X_n}(t)\,=\,
            E[e^{\textstyle i t X}]\,=\,
            \frac{\textstyle 1}
                 {\textstyle 2}\left(1\,+\,
              e^{\textstyle 2i t }\right)
  \nonumber
 \end{eqnarray}
and for $Y_n=X_n/3^n$:
\begin{eqnarray}
            {\Phi}_{Y_n}(t)\,=\,{\Phi}_{X_n}(t/3^n)\,=\,
            \frac{\textstyle 1}
                 {\textstyle 2}\left(1\,+\,
              e^{\textstyle 2i t/3^n }\right)
  \nonumber
 \end{eqnarray}
Being all $X_n$ statistically independent, we have that
\begin{eqnarray}
            {\Phi}_{X}(t)\,=\,
\prod_{n=1}^{\infty}\,
            \frac{\textstyle 1}
                 {\textstyle 2}\left(1\,+\,
              e^{\textstyle 2i t/3^n }\right)\,=\,
\prod_{n=1}^{\infty}\,
            \frac{\textstyle 1}
                 {\textstyle 2}
              e^{\textstyle i t/3^n }\cos(t/3^n)\,=\,
              e^{\textstyle i t/2}
\prod_{n=1}^{\infty}\,\cos(t/3^n)
  \nonumber
 \end{eqnarray}
and, from the derivatives (section 5.1.4)
it is straight forward to calculate the moments of the 
distribution. In particular:
\begin{eqnarray*}
      {\Phi}^{1)}_{X}(0)\,=\,
      \frac{\textstyle i}{\textstyle 2} 
      \,\,\,\,\,\longrightarrow \,\,\,\,\,
      E[X]\,\,\,\,=\,\frac{\textstyle 1}{\textstyle 2} 
\hspace{1.cm}{\rm and}\hspace{1.cm}
 {\Phi}^{2)}_{X}(0)\,=\,
      -\frac{\textstyle 3}{\textstyle 8} 
      \,\,\,\,\,\longrightarrow \,\,\,\,\,
      E[X^2]\,=\,\frac{\textstyle 3}{\textstyle 8} \nonumber \\
 \end{eqnarray*}
so $V[X]=1/8$.

\smill
{\raya}                   
\vspace{1.0cm}

\subsubsection{Inversion Theorem \rm (L\'evy, 1925)}

The Inverse Fourier Transform allows us to obtain the
distribution function of a random quantity from the Characteristic
Function.
If $X$ is a continuous random quantity and ${\Phi}(t)$ its Characteristic 
Function, then the pdf $p(x)$ will be given by
\begin{eqnarray*}
         p(x)\,=\,
         \frac{\textstyle 1}{\textstyle 2{\pi}}\,
         \displaystyle{ \int_{-\infty}^{+\infty}}\,
         e^{\textstyle -i t x}\,{\Phi}(t)\,dt 
\end{eqnarray*}
provided that $p(x)$ is continuous at $x$ and, if $X$ is discrete:
\begin{eqnarray*}
         P(X=x_k)\,=\,
         \frac{\textstyle 1}{\textstyle 2{\pi}}\,
         \displaystyle{ \int_{-\infty}^{+\infty}}\,
         e^{\textstyle -i t x_k}\,{\Phi}(t)\,dt 
\end{eqnarray*}
In particular, if the discrete distribution is {\em reticular}
(that is, all the possible values that the random quantity $X$ may take
can be expressed as $a+b\,n$ with $a,b \in {\Rcal}$;
$b{\neq}0$ and $n$ integer) we have that:
\begin{eqnarray*}
         P(X=x_k)\,=\,
         \frac{\textstyle b}{\textstyle 2{\pi}}\,
         \displaystyle{ \int_{-{\pi}/b}^{{\pi}/b}}\,
         e^{\textstyle -i t x_k}\,{\Phi}(t)\,dt 
\end{eqnarray*}

\noindent
From this expressions, we can obtain also the relation between the 
Characteristic Function and the Distribution Function.
For discrete random quantities we shall have:
\begin{eqnarray*}
  F(x_k)\,=\,\displaystyle{ \sum_{x{\leq}x_k} }\,
         P(X=x_k)\,=\,
         \frac{\textstyle 1}{\textstyle 2{\pi}}\,
         \displaystyle{ \int_{-\infty}^{+\infty}}\,
         \displaystyle{ \sum_{x{\leq}x_k} }\,
         e^{\textstyle -i t x}\,{\Phi}(t)\,dt 
\end{eqnarray*}
and, in the continuous case, for $x_1<x_2 \in {\Rcal}$ we have that:
\begin{eqnarray*}
  F(x_2)\,-\,F(x_1)\,=\,
         \displaystyle{ \int_{x_1}^{x_2}}\,p(x)\,dx\,=\,
         \frac{\textstyle 1}{\textstyle 2{\pi}i}\,
         \displaystyle{ \int_{-\infty}^{+\infty}}\,
         {\Phi}(t)\,
         \frac{\textstyle 1}{\textstyle t}\,
         (e^{\textstyle -i t x_1}-e^{\textstyle -i t x_2})
         \,dt 
\end{eqnarray*}
so taking $x_1=0$ we have that (L\'evy, 1925):
\begin{eqnarray*}
  F(x)\,=\,F(0)\,+\,
         \frac{\textstyle 1}{\textstyle 2{\pi}i}\,
         \int_{-\infty}^{+\infty}\,
         {\Phi}(t)\,
         \frac{\textstyle 1}{\textstyle t}\,
         (1\,-\,e^{\textstyle -i t x})
         \,dt 
\end{eqnarray*}

The {\bf Inversion Theorem} states that there is a one-to-one
correspondence between a distribution function and 
its Characteristic Function so to each Characteristic Function 
corresponds {\bf one} and {\bf only one} distribution function  that can 
be either discrete or continuous but not a combination of both.
Therefore, two distribution functions with the same Characteristic Function
may differ, at most, on their points of discontinuity that, as we have
seen, are a set of zero measure.
In consequence, if we have two random quantities $X_1$ and $X_2$ with
distribution functions $P_1(x)$ and $P_2(x)$, a
{\bf necessary} and {\bf sufficient} condition for
$P_1(x)=P_2(x)$ a.e. is that
${\Phi}_1(t)={\Phi}_2(t)$ for all $t \in {\Rcal}$.

\subsubsection{Changes of variable}

Let $X{\sim}P(x)$ be a random quantity with Characteristic Function
${\Phi}_X(t)$
and $g(X)$ a one-to-one finite real function defined for
all real values of $X$. The Characteristic Function of the random quantity
$Y=g(X)$ will be given by:
\begin{eqnarray*}
   {\Phi}_Y(t)\,=\,
        E_Y[e^{\textstyle i t Y}]\,=\,
        E_X[e^{\textstyle i t g(X)}] 
\end{eqnarray*}
that is:
\begin{eqnarray*}
         {\Phi}_Y(t)\,=\,
         \displaystyle{ \int_{-\infty}^{+\infty}}\,
         e^{\textstyle i t g(x)}\,dP(x)  
\hspace{1.cm}{\rm or}\hspace{1.cm}
         {\Phi}_Y(t)\,=\,
         \displaystyle{ \sum_{x_k}}\,
         e^{\textstyle i t g(x_k)}\,P(X = x_k)
\end{eqnarray*}
depending on whether $X$ is continuous or discrete.
In the particular case of a linear transformation
$Y=a X + b$ with $a$ and $b$ real constants, we have that:
\begin{eqnarray*}
         {\Phi}_Y(t)\,=\,
        E_X[e^{\textstyle i t (aX+b)}]\,=\,
        e^{\textstyle i t b}\,
        {\Phi}_X(at) 
\end{eqnarray*}

\subsubsection{Sum of random quantities}

The Characteristic Function is particularly useful to obtain the
Distribution Function of a random quantity defined as the sum
of independent random quantities.
If ${X}_1,\ldots ,{X}_n$ are
$n$ independent random quantities with Characteristic Functions
${\Phi}_1(t_1),\ldots ,{\Phi}_n(t_n)$, then the
Characteristic Function of 
${X}={X}_1+\ldots +{X}_n$ will be:
 \begin{eqnarray*}
       {\Phi}_X(t)\,=\,
        E[e^{\textstyle i t X}]\,=\,
        E[e^{\textstyle i t ({X}_1+\ldots +{X}_n)}]\,=\,
        {\Phi}_1(t)\,\cdots\,{\Phi}_n(t)  
\end{eqnarray*}
that is, the product of the Characteristic Functions of each one,
a {\bf necessary} but {\bf not sufficient} condition for the random quantities 
${X}_1,\ldots ,{X}_n$ to be independent. 
In a similar way, we have that if $X=X_1-X_2$ with $X_1$ and $X_2$
independent random quantities, then
\begin{eqnarray*}
       {\Phi}_X(t)\,=\,
        E[e^{\textstyle i t ({X}_1-{X}_2)}]\,=\,
        {\Phi}_1(t)\,{\Phi}_2(-t)\,=\,  {\Phi}_1(t)\,\overline{\Phi}_2(t) 
\end{eqnarray*}
Form these considerations, it is left as an exercise to show that:
\begin{itemize}
\item[$\bullet$] 
 {\bf Poisson:} The sum of $n$ independent random quantities,
each distributed as $Po(x_k|{\mu}_k)$ with 
$k=1, \ldots ,n$ is Poisson distributed with
parameter  ${\mu}_s={\mu}_1+ \cdots +{\mu}_n$.

\item[$\bullet$] 
 {\bf Normal:} The sum of $n$ independent random quantities,
each distributed as $N(x_k|{\mu}_k,{\sigma}_k)$ with 
$k=1, \ldots ,n$ is Normal distributed with mean
 ${\mu}_s={\mu}_1+ \cdots +{\mu}_n$ and variance
 ${\sigma}^2_s={\sigma}^2_1+ \cdots +{\sigma}^2_n$.

\item[$\bullet$] 
 {\bf Cauchy:}  The sum of $n$ independent random quantities,
each Cauchy distributed $Ca(x_k|{\alpha}_k,{\beta}_k)$ with 
$k=1, \ldots ,n$ is 
is Cauchy distributed with parameters
 ${\alpha}_s={\alpha}_1+ \cdots +{\alpha}_n$ and
 ${\beta}_s={\beta}_1+ \cdots +{\beta}_n$.

\item[$\bullet$] 
 {\bf Gamma:} The sum of $n$ independent random quantities,
each distributed as $Ga(x_k|{\alpha},{\beta}_k)$ with 
$k=1, \ldots ,n$ is Gamma distributed with
parameters $({\alpha},{\beta}_1+ \cdots +{\beta}_n)$.

\end{itemize}
\vspace{0.5cm}
\noindent
{\raya}                   
\vspace{0.35cm}
\footnotesize

\noindent
{\bf Example 1.21:} {\bf Difference of Poisson distributed random quantities}.

\noindent 
Consider two independent random quantities
$X_1{\sim}Po(X_1|{\mu}_1)$ and $X_2{\sim}Po(X_2|{\mu}_2)$ and 
let us find the distribution of
$X=X_1-X_2$. Since for the Poisson distribution:
\begin{eqnarray}
  X_i{\sim}Po({\mu}_i)\,\,\,\,\,{\longrightarrow}\,\,\,\,\,
       {\Phi}_i(t)\,=\,
      e^{\textstyle -{\mu}_i(1-e^{\textstyle it})}
        \nonumber
\end{eqnarray}
we have that
\begin{eqnarray}
       {\Phi}_X(t)\,=\,{\Phi}_1(t)\,\overline{\Phi}_2(t)\,=\, 
      e^{\textstyle -({\mu}_1+{\mu}_2)}
      e^{\textstyle ({\mu}_1e^{\textstyle it}+{\mu}_2e^{\textstyle -it})}
        \nonumber
\end{eqnarray}
Obviously, $X$ is a discrete random quantity with integer
support ${\Omega}_X=\{{\ldots},-2,-1,0,1,2,{\ldots}\}$; that is, a
{\sl reticular} random quantity with $a=0$ and $b=1$. Then
\begin{eqnarray}
         P(X=n)\,=\,
         \frac{\textstyle 1}{\textstyle 2{\pi}}\,
      e^{\textstyle -{\mu}_S}\,
         \displaystyle{ \int_{-{\pi}}^{{\pi}}}\,
         e^{\textstyle -i t n}\,
	 e^{\textstyle ({\mu}_1e^{\textstyle it}+{\mu}_2e^{\textstyle -it})}
         \,dt \nonumber
\end{eqnarray}
being ${\mu}_S={\mu}_1+{\mu}_2$. If we take:
\begin{eqnarray}
         z\,=\,\sqrt{\frac{\textstyle {\mu}_1}{\textstyle {\mu}_2}}\,
      e^{\textstyle it}
	 \nonumber
\end{eqnarray}
we have
\begin{eqnarray}
         P(X=n)\,=\,
\left(\frac{\textstyle {\mu}_1}{\textstyle {\mu}_2}\right)^{n/2}\,
      e^{\textstyle -{\mu}_S}\,
         \frac{\textstyle 1}{\textstyle 2{\pi}i}\,
\displaystyle{{\bigcirc}{\hskip-1.0em}{\int_{C}}}\,
         z^{-n-1}\,
	 e^{\textstyle \frac{w}{2}\,\textstyle{(z+1/z)}}
         \,dz \nonumber
\end{eqnarray}
with $w=2\sqrt{{\mu_1}{\mu}_2}$ and $C$ the circle 
$|z|=\sqrt{{\mu_1}/{\mu}_2}$ around the origin. From the definition of
the Modified Bessel Function of first kind
\begin{eqnarray}
         I_n(z)\,=\,
         \frac{\textstyle 1}{\textstyle 2{\pi}i}\,
\displaystyle{{\bigcirc}{\hskip-1.0em}{\int_{C}}}\,
         t^{-n-1}\,
	 e^{\textstyle \frac{z}{2}\,\textstyle{(t+1/t)}}
         \,dt \nonumber
\end{eqnarray}
with $C$ a circle enclosing the origin anticlockwise
and considering that $I_{-n}(z)=I_n(z)$ we have finally:
\begin{eqnarray*}
         P(X=n)\,=\,
\left(\frac{\textstyle {\mu}_1}{\textstyle {\mu}_2}\right)^{n/2}\,
      e^{\textstyle -({\mu}_1+{\mu}_2)}\,
      I_{|n|}(2\sqrt{{\mu_1}{\mu}_2})
\end{eqnarray*}

\vspace{0.35cm}
\smill
{\raya}                   
\vspace{1.0cm}


\vspace*{0.3cm}
\subsubsection{Moments of a distribution}
\vspace*{0.1cm}

Consider a continuous random quantity $X{\sim}P(x)$
and Characteristic Function
\begin{eqnarray*}
         {\Phi}(t)\,=\,E[e^{\textstyle i t {X}}]\,=\,
         \displaystyle{ \int_{-\infty}^{+\infty}}\,
         e^{\textstyle i t x}\,dP(x) 
\end{eqnarray*}
and let us assume that there exists the moment of order $k$.
Then, upon derivation of the Characteristic Function 
$k$ times with respect to $t$ we have:
\begin{eqnarray*}
  \frac{\textstyle {\partial}^k}
       {\textstyle {\partial}^kt}\,{\Phi}(t)\,=\,
         E[i^k\,{X}^k\,e^{\textstyle i t {X}}]\,=\,
         \displaystyle{ \int_{-\infty}^{+\infty}}\,
         (ix)^k\,e^{\textstyle i t x}\,dP(x) 
\end{eqnarray*}
and taking $t=0$ we get the {\em moments with respect to the origin}:

\begin{eqnarray*}
E[{X}^k]\,=\,
     \frac{\textstyle 1}
          {\textstyle i^k}\, \left(
     \frac{\textstyle {\partial}^k}
       {\textstyle {\partial}^kt}\,{\Phi}(t) \right)_{t=0}
\end{eqnarray*}

Consider now the Characteristic Function referred to an arbitrary point
$a \in {\Rcal}$; that is:
\begin{eqnarray*}
         {\Phi}(t,a)\,=\,E[e^{\textstyle i t ({X}-a)}]\,=\,
         \displaystyle{ \int_{-\infty}^{+\infty}}\,
         e^{\textstyle i t (x-a)}\,dP(x)\,=\,
         e^{\textstyle -i t a}\,{\Phi}(t) 
          \nonumber
\end{eqnarray*}
In a similar way, upon $k$ times derivation 
with respect to $t$ we get the moments with respect to an arbitrary point $a$:
\begin{eqnarray*}
         E[({X}-a)^k]\,=\,
  \frac{1}{i^k}
  \left(  \frac{\textstyle {\partial}^k}
       {\textstyle {\partial}^kt}\,{\Phi}(t,a)\right)_{t=0}
\end{eqnarray*}
and the {\sl central moments} if 
$a=E({X})=\mu$.
The extension to $n$ dimensions immediate: for a
$n$ dimensional random quantity ${\bf X}$ we shall have the
for the moment ${\alpha}_{k_1 \ldots k_n}$ with respect to the origin
that
\begin{eqnarray*}
   {\alpha}_{k_1 \ldots k_n}\,=\,
   E[{X}_1^{k_1} \cdots {X}_n^{k_n}]\,=\,
     \frac{\textstyle 1}
          {\textstyle i^{k_1+\ldots +k_n}}\, \left(
     \frac{\textstyle {\partial}^{k_1+\ldots +k_n}}
       {\textstyle {\partial}^{k_1}{t_1} \cdots 
                   {\partial}^{k_n}{t_n}}\,
       {\Phi}(t_1,\ldots ,t_n) \right)_{t_1=\ldots =t_n=0}
\end{eqnarray*}

\vspace{0.5cm}
\noindent
{\raya}                   
\vspace{0.35cm}
\footnotesize

\noindent
{\bf Example 1.22:} For the difference of Poisson distributed random quantities
analyzed in the previous example, one can easily derive the moments from
the derivatives of the Characteristic Function. Since
\begin{eqnarray*}
       {\log}{\Phi}_X(t)\,=\,
      -({\mu}_1+{\mu}_2)\,+\,
       ({\mu}_1e^{\textstyle it}+{\mu}_2e^{\textstyle -it})
\end{eqnarray*}
we have that
\begin{eqnarray*}
       {\Phi}'_X(0)\,&=&\,i\,({\mu}_1-{\mu}_2)\hspace{2.cm}
       \longrightarrow \hspace{0.5cm}E[X]={\mu}_1-{\mu}_2 \\
       {\Phi}''_X(0)\,&=&\,({\Phi}'_X(0))^2\,-\,({\mu}_1+{\mu}_2)\hspace{0.4cm}
       \longrightarrow \hspace{0.5cm}V[X]={\mu}_1+{\mu}_2 
\end{eqnarray*}
and so on.

\vspace{0.35cm}
\noindent
{\bf Problem 1.4:} The Moyal Distribution, with density 
\begin{eqnarray*}
       p(x)\,=\,\frac{1}{\sqrt{2\pi}}\,\exp\left\{
     -\frac{1}{2}\left(x+e^{-x}\right)\right\}\,
{\mbox{\boldmath $1$}_{(-\infty,\infty)}(x)}
\end{eqnarray*}
is sometimes used as an approximation to the Landau Distribution. Obtain the
Characteristic Function 
${\Phi}(t)={\pi}^{-1/2}2^{-it}{\Gamma}(1/2-it)$ and show that
$E[x]={\gamma}_E+\ln 2$ and $V[X]={\pi}^2/2$. 

\vspace{0.35cm}
\smill
{\raya}                   
\vspace{1.0cm}


\smill
\subsection{The Mellin Transform}
\vspace*{0.1cm}

Let $f:{\Rcal}^+{\rightarrow}{\Ccal}$ be a complex and integrable function
with support on the real positive axis.
The {\sl Mellin Transform} is defined as:
\begin{eqnarray*}
M(f;s)\,=\,M_f(s)\,=\,\int_{0}^{\infty}\,f(x)\,x^{s-1}\,dx
\end{eqnarray*}
with $s{\in}{\Ccal}$, provided that the integral exists.
In general, we shall be interested in continuous probability density functions 
$f(x)$ such that
\begin{eqnarray*}
{\lim}_{x{\rightarrow}0+}f(x)\,=\,O(x^{\alpha}) 
\hspace{1.cm}{\rm and}\hspace{1.cm}
{\lim}_{x{\rightarrow}{\infty}}f(x)\,=\,O(x^{\beta}) 
\end{eqnarray*}
and therefore
\begin{eqnarray*}
|M(f;s)|&{\leq}&\int_{0}^{\infty}|f(x)|x^{Re(s)-1}dx=
\int_{0}^{1}|f(x)|x^{Re(s)-1}dx+
\int_{1}^{\infty}|f(x)|x^{Re(s)-1}dx\,{\leq} \\
&{\leq}&
C_1\,\int_{0}^{1}\,x^{Re(s)-1+{\alpha}}\,dx\,+\,
C_2\,\int_{1}^{\infty}\,x^{Re(s)-1+{\beta}}\,dx
\end{eqnarray*}
The first integral converges for $-{\alpha}<Re(s)$ and the second for
$Re(s)<-{\beta}$ so the Mellin Transform 
exists and is holomorphic on the band 
$-{\alpha}<Re(s)<-{\beta}$,
parallel to the imaginary axis $\Im(s)$ and determined by the conditions of 
convergence of the integral.
We shall denote the holomorphy band 
(that can be a half of the complex plane or the whole complex plane)
by $S_f=<-{\alpha},-{\beta}>$. Last, to simplify the notation when dealing 
with several random quantities, we shall write for $X_n{\sim}p_n(x)$ 
$M_n(s)$ of $M_X(s)$ instead of $M(p_n;s)$.

\smill
\subsubsection{Inversion}
\vspace*{0.1cm}
For a given function $f(t)$ we have that
\begin{eqnarray*}
M(f;s)\,=\,\int_{0}^{\infty}\,f(t)\,t^{s-1}\,dt\,=\,
\int_{0}^{\infty}\,f(t)\,e^{(s-1){\ln}t}\,dt\,=\,
\int_{-{\infty}}^{\infty}\,f(e^u)\,e^{{s}u}\,du
\end{eqnarray*}
assuming that the integral exists.
Since $s{\in}{\Ccal}$, we can write $s=x+iy$ so:
the Mellin Transform of $f(t)$ is the Fourier Transform of
$g(u)=f(e^u)e^{xu}$. Setting now $t=e^u$ we have that
\begin{eqnarray*}
f(t)\,=\,
\frac{\textstyle 1}{\textstyle 2{\pi}}\,
\int_{-{\infty}}^{\infty}\,M(f;s=x+iy)\,t^{-(x+iy)}\,dy\,=\,
\frac{\textstyle 1}{\textstyle 2{\pi}i}\,
\int_{{\sigma}-i{\infty}}^{{\sigma}+i{\infty}}\,M(f;s)\,t^{-s}\,ds
\end{eqnarray*}
where, due to Chauchy's Theorem, ${\sigma}$ lies anywhere within the
holomorphy band. The uniqueness of the result holds with respect to this
strip so, in fact, the Mellin Transform consists on the pair $M(s)$ together
with the band $<a,b>$.

\vspace{0.5cm}
\noindent
{\raya}                   
\vspace{0.35cm}
\footnotesize

\noindent
{\bf Example 1.23:}
It is clear that to determine the function $f(x)$ from the transform $F(s)$
we have to specify the strip of analyticity 
for, otherwise, we do not know which poles should be included.
Let's see as an example $f_1(x)=e^{-x}$. We have that
\begin{eqnarray}
M_1(z)\,=\,\int_{0}^{\infty}\,e^{-x}\,x^{z-1}\,dx\,=\,{\Gamma}(z)
\nonumber
\end{eqnarray}
holomorphic in the band $<0,{\infty}>$
so, for the inverse transform, we shall include the poles $z=0,-1,-2,{\ldots}$.
For $f_2(x)=e^{-x}-1$ we get $M_2(s)={\Gamma}(s)$, the same function, but 
\begin{eqnarray*}
{\rm lim}_{x{\rightarrow}0+}f(x){\simeq}\,O(x^1)
{\longrightarrow}{\alpha}=1 \hspace{1.cm}{\rm and}\hspace{1.cm}
{\rm lim}_{x{\rightarrow}{\infty}}f(x){\simeq}\,O(x^0)
{\longrightarrow}{\beta}=0
\end{eqnarray*} 
Thus, the holomorphy strip is $<-1,0>$ and
for the inverse transform we shall include the poles $z=-1,-2,{\ldots}$.
For $f_3(x)=e^{-x}-1+x$ we get $M_3(s)={\Gamma}(s)$, again the same function,
but 
\begin{eqnarray*}
{\rm lim}_{x{\rightarrow}0+}f(x){\simeq}\,O(x^2)
{\longrightarrow}{\alpha}=2 \hspace{1.cm}{\rm and}\hspace{1.cm}
{\rm lim}_{x{\rightarrow}{\infty}}f(x){\simeq}\,O(x^1)
{\longrightarrow}{\beta}=1
\nonumber
\end{eqnarray*} 
Thus, the holomorphy strip is$<-2,-1>$
and for the inverse transform we include the poles $z=-2,-3,{\ldots}$.
\raya
\vspace{0.5cm}

\smill
\subsubsection{Useful properties}
\vspace*{0.1cm}
Consider a positive random quantity $X$ with continuous
density $p(x)$ and $x{\in}[0,{\infty})$, the Mellin Transform
 $M_X(s)$ (defined only for $x{\geq}0$) 
\begin{eqnarray*}
M(p;s)\,=\,\int_{0}^{\infty}\,x^{s-1}\,p(x)\,dx\,=\,E[X^{s-1}]
\end{eqnarray*}
and the Inverse Transform 
\begin{eqnarray*}
p(x)\,=\,\frac{\textstyle 1}{\textstyle 2{\pi}i}
\int_{c-i{\infty}}^{c+i{\infty}}\,x^{-s}\,M(p;s)\,ds
\end{eqnarray*}
defined for all $x$ where $p(x)$ is continuous with the line of integration
contained in the strip of analyticity of $M(p;s)$. Then:

\vspace{0.5cm}
\noindent
$\bullet$ {\bf Moments}: $E[X^n]=M_X(n+1)$;

\vspace{0.5cm}
\noindent
$\bullet$ For the positive random quantity 
$Z=aX^{\textstyle b}$ ($a,b{\in}{\Rcal}$ and $a>0$) we have that
\begin{eqnarray*}
M_Z(s)\,&=&\,\int_{0}^{\infty}\,z^{s-1}\,f(z)\,dz\,=\,
           \int_{0}^{\infty}\,a^{s-1}\,x^{b(s-1)}\,p(x)\,dx\,=\,
a^{s-1}\,M_X(bs-b+1) \\
2{\pi}i\,p(z)\,&=&\,\int_{c-i{\infty}}^{c+i{\infty}}\,z^{-s}\,M_X(bs-b+1)\,ds
\end{eqnarray*}
In particular, for $Z=1/X$ ($a=1$ and $b=-1$) we have that
\begin{eqnarray*}
M_{Z=1/X}(s)\,=\,M_X(2-s)
\end{eqnarray*}

\vspace{0.5cm}
\noindent
$\bullet$ If $Z=X_1X_2{\cdots}X_n$ with
$\{X_i\}_{i=1}^n$ $n$ independent positive defined random
quantities, each distributed as $p_i(x_i)$, we have that
\begin{eqnarray*}
M_Z(s)&=&\int_{0}^{\infty}\,z^{s-1}\,p(z)\,dz\,=\,
\prod_{i=1}^n  \int_{0}^{\infty}\,x_i^{s-1}p_i(x_i)\,dx_i\,=\,
\prod_{i=1}^n  E[X_i^{s-1}]\,=\,
\prod_{i=1}^n  M_i(s) \\
2{\pi}i\,p(z)\,&=&\,
\int_{c-i{\infty}}^{c+i{\infty}}\,z^{-s}\,M_1(s){\cdots}M_n(s)\,ds
\end{eqnarray*}
In particular, for $n=2$, $X=X_1X_2$ it is easy to check that
\begin{eqnarray*}
p(x)\,=\,
\int_{0}^{{\infty}}\,p_1(w)\,p_2(x/w)\,dw/w\,=\,
\frac{1}{2{\pi}i}\int_{c-i{\infty}}^{c+i{\infty}}\,x^{-s}\,M_1(s)M_2(s)\,ds
\end{eqnarray*}
Obviously, the strip of holomorphy is $S_1{\cap}S_2$.

\vspace{0.5cm}
\noindent
$\bullet$ For $X=X_1/X_2$, with both $X_1$ and $X_2$ positive defined 
and independent, we have that
\begin{eqnarray*}
M_{X}(s)\,=\,M_1(s)\,M_2(2-s)
\end{eqnarray*}
and therefore
\begin{eqnarray*}
p(x)\,=\,\int_{0}^{{\infty}}\,p_1(wx)\,p_2(w)\,w\,dw\,=\,
\frac{\textstyle 1}{\textstyle 2{\pi}i}
\int_{c-i{\infty}}^{c+i{\infty}}\,x^{-s}\,M_1(s)\,M_2(2-s)\,ds
\end{eqnarray*}

\vspace{0.5cm}
\noindent
$\bullet$ Consider the distribution function $F(x)=\int_{0}^{x}p(u)du$
of the random quantity $X$. Since $dF(x)=p(x)dx$ we have that
\begin{eqnarray*}
M(p(x);s)= \int_{0}^{{\infty}}\,x^{s-1}\,dF(x)=
	   \left[x^{s-1}F(x)\right]_{0}^{\infty}\,-\,
           (s-1)\int_{0}^{{\infty}}\,x^{s-2}\,F(x)\,dx
\end{eqnarray*}
and therefore, if ${\lim}_{x{\rightarrow}0+}[x^{s-1}F(x)]=0$ and
${\lim}_{x{\rightarrow}{\infty}}[x^{s-1}F(x)]=0$ 
we have, shifting $s{\rightarrow}s-1$, that
\begin{eqnarray*}
M(F(x);s)\,=\,
M(\int_{0}^{x}p(u)\,du;s)\,=\,-\frac{\textstyle 1}{\textstyle s}
\,M(p(x);s+1)
\end{eqnarray*}

\subsubsection{Some useful examples} 

\vspace{0.5cm}
\noindent
$\bullet$ {\bf Ratio and product of two independent 
Exponential distributed random quantities}

Consider $X_1{\sim}Ex(x_1|a_1)$ and $X_2{\sim}Ex(x_2|a_2)$.
The Mellin transform of $X{\sim}Ex(x|a)$ is
\begin{eqnarray*}
M_{X}(s)\,=\,
\int_{0}^{{\infty}}\,x^{s-1}p(x|a)\,dx\,=\,
a\int_{0}^{{\infty}}\,x^{s-1}e^{-ax}\,dx\,=\,
\frac{\textstyle {\Gamma}(s)}{\textstyle a^{s-1}}
\end{eqnarray*}
and therefore, for $Z=1/X$:
\begin{eqnarray*}
M_Z(s)\,=\,M_{X}(2-s)\,=\,
\frac{\textstyle {\Gamma}(2-s)}{\textstyle a^{1-s}}
\end{eqnarray*}
In consequence, we have that
 
\vspace{0.5cm}
\noindent
$\bullet$ $X=X_1X_2
          \hspace{0.5cm}{\longrightarrow}\hspace{0.5cm}
          M_{X}(z)\,=\,M_1(z)M_2(z)\,=\,
          \frac{\textstyle {\Gamma}(z)^2}{\textstyle (a_1\,a_2)^{z-1}}
          $ 
\begin{eqnarray*}
p(x)\,=\,\frac{\textstyle a_1\,a_2}{\textstyle 2{\pi}i}
\int_{c-i{\infty}}^{c+i{\infty}}\,
(a_1a_2x)^{-z}\,{\Gamma}(z)^2\,dz
\end{eqnarray*}
The poles of the integrand are at $z_n=-n$ and the residuals
\footnote{In the following examples, $-{\pi}{\leq}arg(z)<{\pi}$.} are
\begin{eqnarray*}
Res(f(z),z_n)\,=\,
\frac{\textstyle (a_1a_2x)^{n} }{\textstyle (n!)^2}\,\left(
2{\psi}(n+1)\,-\,\ln(a_1a_2x)
\right)
\end{eqnarray*}
and therefore
\begin{eqnarray*}
p(x)\,=\,a_1\,a_2\,
\sum_{n=0}^{\infty}
\frac{\textstyle (a_1a_2x)^{n} }{\textstyle (n!)^2}\,\left(
2{\psi}(n+1)\,-\,\ln(a_1a_2x)
\right)
\end{eqnarray*}
If we define $w=2\sqrt{a_1a_2x}$
\begin{eqnarray*}
p(x)\,=\,2\,a_1\,a_2\,K_0(2\sqrt{a_1a_2x}){\mbox{\boldmath $1$}_{(0,\infty)}(x)}
\end{eqnarray*}
from the Neumann Series expansion the Modified Bessel Function
$K_0(w)$.

\vspace{0.5cm}
\noindent
$\bullet$ $Y=X_1X_2^{-1}
     \hspace{0.5cm}{\longrightarrow}\hspace{0.5cm}
     M_{Y}(z)\,=\,M_1(z)M_2(2-z)\,=\,
     \left(\frac{\textstyle a_2}{\textstyle a_1}\right)^{z-1}\,
     \frac{\textstyle {\pi}(1-z)}{\textstyle \sin(z{\pi})}
     $
\begin{eqnarray*}
p(x)\,=\,\frac{\textstyle a_1\,a_2^{-1}}{\textstyle 2i}
\int_{c-i{\infty}}^{c+i{\infty}}\,
(a_1a_2^{-1}x)^{-z}\,\frac{\textstyle 1-z}{\textstyle \sin(z{\pi})}dz
\end{eqnarray*}
Considering again the poles of $M_Y(z)$ at $z_n=-n$ we get the residuals
\begin{eqnarray*}
Res(f(z),z_n)\,=\,(1+n)\,(-1)^n\,
\left(\frac{\textstyle a_1}{\textstyle a_2}\right)^{n+1}\,x^n
\end{eqnarray*}
and therefore: 
\begin{eqnarray*}
p(x)\,=\,\frac{\textstyle a_1}{\textstyle a_2}\,
\sum_{n=0}^{\infty}(1+n)\,(-1)^n\,
\left(\frac{\textstyle a_1\,x}{\textstyle a_2}\right)^{n}\,=\,
\frac{\textstyle a_1\,a_2}{\textstyle (a_2+a_1x)^2}
{\mbox{\boldmath $1$}_{(1,\infty)}(x)}
\end{eqnarray*}
To summarize, if $X_1{\sim}Ex(x_1|a_1)$ and $X_2{\sim}Ex(x_2|a_2)$
are independent random quantities:
\begin{eqnarray*}
X=X_1X_2\,&{\sim}&\,2\,a_1\,a_2\,K_0(2\sqrt{a_1a_2x})
{\mbox{\boldmath $1$}_{(1,\infty)}(x)} \nonumber \\
Y=X_1/X_2\,&{\sim}&\,\frac{\textstyle a_1\,a_2}{\textstyle (a_2+a_1x)^2}
{\mbox{\boldmath $1$}_{(0,\infty)}(x)}
\end{eqnarray*}

\vspace{0.5cm}
\noindent
$\bullet$
{\bf Ratio and product of two independent 
Gamma distributed random quantities}

Consider $Y{\sim}Ga(x|a,b)$. Then $X=aY{\sim}Ga(x|1,b)$ with Mellin Transform
\begin{eqnarray*}
M_{X}(s)\,=\,
\frac{\textstyle {\Gamma}(b+s-1)}{\textstyle {\Gamma}(b)}
\end{eqnarray*} The, if
$X_1{\sim}Ga(x_1|1,b_1)$ and $X_2{\sim}Ga(x_2|1,b_2)$; $b_1{\neq}b_2$:

\vspace{0.5cm}
\noindent
$\bullet$ $X=X_1X_2^{-1}
     \hspace{0.5cm}{\longrightarrow}\hspace{0.5cm}
     M_{X}(z)=M_{1}(z)M_{2}(z)=
     \frac{\textstyle {\Gamma}(b_1-1+z)}{\textstyle {\Gamma}(b_1)}
     \frac{\textstyle {\Gamma}(b_2+1-z)}{\textstyle {\Gamma}(b_2)}
     $
\begin{eqnarray*}
2{\pi}i\,{\Gamma}(b_1)\,{\Gamma}(b_2)
p(x)\,=\,
\int_{c-i{\infty}}^{c+i{\infty}}\,x^{-z}\,{\Gamma}(b_1-1+z)\,
{\Gamma}(b_2+1-z)\,dz
\end{eqnarray*}
Closing the contour on the left of the line $Re(z)=c$ contained in the strip
of holomorphy $<0,{\infty}>$ we have poles of order one at
$b_i-1+z_n=-n$ with $n=0,1,2,{\ldots}$, that is, at $z_n=1-b_i-n$. Expansion
around $z=z_n+{\epsilon}$ gives the residuals 
\begin{eqnarray*}
Res(f(z),z_n)\,=\,
\frac{\textstyle (-1)^n}{\textstyle n!}\,
{\Gamma}(b_1+b_2+n)\,x^{n+b_1-1}
\end{eqnarray*}
and therefore the quantity
$X=X_1/X_2$ is distributed as
\begin{eqnarray*}
p(x)=\frac{\textstyle x^{b_1-1}}{\textstyle {\Gamma}(b_1)\,{\Gamma}(b_2)}\,
\sum_{n=0}^{\infty}
\frac{\textstyle (-1)^n}{\textstyle n!}\,
{\Gamma}(b_1+b_2+n)\,x^{n}\,
=\,\frac{\textstyle \Gamma(b_1+b_2)}{\textstyle \Gamma(b_1)\Gamma(b_2)}\,
\frac{\textstyle x^{b_1-1}}{\textstyle (1+x)^{b_1+b_2}}
{\mbox{\boldmath $1$}_{(0,\infty)}(x)}
\end{eqnarray*}

\vspace{0.5cm}
\noindent
$\bullet$ $X=X_1X_2
     \hspace{0.5cm}{\longrightarrow}\hspace{0.5cm}
     M_{X}(z)\,=\,M_1(z)M_2(z)\,=\,
     \frac{\textstyle \Gamma(b_1-1+z)}{\textstyle \Gamma(b_1)} 
     \frac{\textstyle \Gamma(b_2-1+z)}{\textstyle \Gamma(b_2)} 
     $

\noindent
Without loss of generality, we may assume that $b_2>b_1$ so the
strip of holomorphy is $<1-b_1,{\infty}>$. Then, with $c>1-b_1$ real
\begin{eqnarray*}
{\Gamma}(b_1)\,{\Gamma}(b_2)
p(x)\,=\,\frac{1}{2{\pi}i}
\int_{c-i{\infty}}^{c+i{\infty}}\,x^{-z}\,{\Gamma}(b_1-1+z)\,
{\Gamma}(b_2-1+z)\,dz
\end{eqnarray*}
Considering the definition of the Modified Bessel Functions
\begin{eqnarray*}
I_{\nu}(x)\,=\,\sum_{n=0}^{\infty}
\frac{\textstyle 1}{\textstyle n!\,{\Gamma}(1+n+{\nu})}\,
\left(\frac{\textstyle x}{\textstyle 2}\right)^{2n+{\nu}}
     \hspace{0.5cm}{\rm and}\hspace{0.5cm}
K_{\nu}(x)\,=\,
\frac{\textstyle {\pi}}{\textstyle 2}\,
\frac{\textstyle I_{-{\nu}}(x)-I_{{\nu}}(x)}
{\textstyle {\sin}({\nu}{\pi})}\,
\end{eqnarray*}
we get that 
\begin{eqnarray*}
p(x)\,=\,\frac{\textstyle 2}
              {\textstyle {\Gamma}(b_1)\,{\Gamma}(b_2)}\,
              x^{(b_1+b_2)/2-1}\,K_{\nu}(2\sqrt{x})
{\mbox{\boldmath $1$}_{(0,\infty)}(x)}
\end{eqnarray*}
with ${\nu}=b_2-b_1>0$.

To summarize, if
$X_1{\sim}Ga(x_1|a_1,b_1)$ and $X_2{\sim}Ga(x_2|a_2,b_2)$
are two independent random quantities and ${\nu}=b_2-b_1>0$ we have that
\begin{eqnarray*}
X=X_1X_2\,&{\sim}&\,
\frac{\textstyle 2a_1^{b_1}a_2^{b_2}}{\textstyle {\Gamma}(b_1)\,{\Gamma}(b_2)}\,
\left( \frac{a_2}{a_1}\right)^{\nu/2}
x^{(b_1+b_2)/2-1}\,K_{\nu}(2\sqrt{a_1a_2x})
{\mbox{\boldmath $1$}_{(0,\infty)}(x)}
\nonumber \\
X=X_1/X_2\,&{\sim}&\,
\frac{\textstyle {\Gamma}(b_1+b_2)}{\textstyle {\Gamma}(b_1){\Gamma}(b_2)}\,
\frac{\textstyle a_1^{b_1}a_2^{b_2}x^{b_1-1}}{\textstyle (a_2+a_1x)^{b_1+b_2}}
{\mbox{\boldmath $1$}_{(0,\infty)}(x)}
\end{eqnarray*} 

\vspace{0.5cm}
\noindent
$\bullet$
{\bf Ratio and product of two independent 
Uniform distributed random quantities}

Consider $X{\sim}Un(x|0,1)$. Then $M_X(z)=1/z$ with 
with $S=<0,{\infty}>$. For $X=X_1{\cdots}X_n$ we have
\begin{eqnarray*}
p(x)\,=\,\frac{\textstyle 1}{\textstyle 2{\pi}i}
\int_{c-i{\infty}}^{c+i{\infty}}\,e^{-z{\ln}x}\,z^{-n}\,dz\,=\,
\frac{\textstyle (-{\ln}x)^{n-1}}{\textstyle {\Gamma}(n)}
{\mbox{\boldmath $1$}_{(0,1]}(x)}
\end{eqnarray*}
being $z=0$ the only pole or order $n$.

For $X=X_1/X_2$ one has to be careful when defining the contours. In principle,
\begin{eqnarray*}
M_X(s)\,=\,M_1(s)M_2(2-s)\,=\,\frac{\textstyle 1}{\textstyle s}\,
\frac{\textstyle 1}{\textstyle 2-s}
\end{eqnarray*}
so the strip of holomorphy is $S=<0,2>$ and there are two poles, at $s=0$ and
$s=2$. If ${\rm ln}x<0{\rightarrow}x<1$ we shall close the Bromwich
the contour on the left enclosing the pole at $s=0$ and if
${\rm ln}x>0{\rightarrow}x>1$ we shall close the contour on the right
enclosing the pole at $s=2$ so the integrals converge. Then it is easy to get
that
\begin{eqnarray*}
p(x)\,=\,\frac{\textstyle 1}{\textstyle 2}\,\left[
{\mbox{\boldmath $1$}_{(0,1]}(x)}\,+\,x^{-2}\,
{\mbox{\boldmath $1$}_{(1,\infty)}(x)}\right]\,=\,Un(x|0,1)\,+\,Pa(x|1,1)
\end{eqnarray*}
Note that
\begin{eqnarray*}
E[X^n]\,=\,
M_X(n+1)\,=\,\frac{\textstyle 1}{\textstyle n+1}\,
\frac{\textstyle 1}{\textstyle 1-n}
\end{eqnarray*}
and therefore there are no moments for $n{\geq}1$.

\vspace{0.5cm}
\noindent
{\raya}                   
\vspace{0.35cm}
\footnotesize

\noindent
{\bf Example 1.24:}
Show that if $X_i{\sim}Be(x_i|a_i,b_i)$ with $a_i,b_i>0$, then 
\begin{eqnarray*}
M_i(s)\,=\,\frac{\textstyle \Gamma(a_i+b_i)}{\textstyle \Gamma(a_i)}\,
\frac{\textstyle \Gamma(s+a_i-1)}{\textstyle \Gamma(s+a_i+b_i-1)}\,
\end{eqnarray*}
with $S=<1-a_i,\infty>$ and therefore:

\vspace{0.5cm}
\noindent
$\bullet$ $X=X_1X_2$
\begin{eqnarray*}
p(x)\,=\,N_p\,x^{a_1-1}\,(1-x)^{b_1+b_2-1}\,
      F(a_1-a_2+b_1,b_2,b_1+b_2,1-x)
{\mbox{\boldmath $1$}_{(0,1)}(x)}
\end{eqnarray*}
with
\begin{eqnarray*}
N_p\,=\,\frac{\textstyle \Gamma(a_1+b_1)\Gamma(a_2+b_2)}
             {\textstyle \Gamma(a_1)\Gamma(a_2)\Gamma(b_1+b_2)}
\end{eqnarray*}

\vspace{0.5cm}
\noindent
$\bullet$ $X=X_1/X_2$
\begin{eqnarray*}
p(x)\,=\,&N_1&\,x^{-(a_2+1)}\,F(1-b_2,a_1+a_2,b_1+a_1+a_2,x^{-1})
         {\mbox{\boldmath $1$}_{(1,\infty)}(x)}\,+ \\
      +\,&N_2&\,x^{a_1-1}\,F(1-b_1,a_1+a_2,b_2+a_1+a_2,x)
{\mbox{\boldmath $1$}_{(0,1)}(x)}
\end{eqnarray*}
with
\begin{eqnarray*}
N_k\,=\,\frac{\textstyle B(a_1+a_2,b_k)}
             {\textstyle B(a_1+b_1)B(a_2+b_2)}
\end{eqnarray*}

\vspace{0.2cm}
\noindent
{\bf Example 1.25:}
Consider a random quantity
\begin{eqnarray}
X\,{\sim}\,p(x|a,b)\,=\,
\frac{\textstyle 2\,a^{(b+1)/2}}{\textstyle {\Gamma}(b/2+1/2)}\,
e^{\textstyle -ax^2}\,x^{\textstyle b}
\nonumber 
\end{eqnarray}
with $a,b>0$ and $x{\in}[0,{\infty})$. Show that
\begin{eqnarray}
M(s)\,{\sim}\,p(x|a,b)\,=\,
\frac{\textstyle {\Gamma}(b/2+s/2)}{\textstyle {\Gamma}(b/2+1/2)}\,
a^{\textstyle -(s-1)/2}
\nonumber 
\end{eqnarray}
with $S=<-b,{\infty}>$ and, from this, derive that the probability
density function of $X=X_1X_2$, with 
$X_1{\sim}p(x_1|a_1,b_1)$ and $X_2{\sim}p(x_2|a_2,b_2)$
independent, is
given by:
\begin{eqnarray}
p(x)\,=\,
\frac{\textstyle 4\,\sqrt{a_1a_2}}
{\textstyle {\Gamma}(b_1/2+1/2)\,{\Gamma}(b_2/2+1/2)}\,
(\sqrt{a_1a_2}x)^{(b_1+b_2)/2}\,
K_{|{\nu}|}\left(2\sqrt{a_1a_2}x\right)
\nonumber 
\end{eqnarray}
with ${\nu}=(b_2-b_1)/2$ and for $X=X_1/X_2$ by
\begin{eqnarray}
p(x)\,=\,
\frac{\textstyle 2\,{\Gamma}(b+1)}
{\textstyle {\Gamma}(b_1/2+1/2)\,{\Gamma}(b_2/2+1/2)}\,
a^{1/2}\,
\frac{\textstyle (a\,x^2)^{b_1/2}}
{\textstyle (1\,+\,a\,x^2)^{b+1}}
\nonumber 
\end{eqnarray}
with $a=a_1/a_2$ and $b=(b_1+b_2)/2$.
 
\raya
\vspace{0.5cm}
\smill

\subsubsection{Distributions with support in $\Rcal$}

The Mellin Transform is defined for integrable functions with non-negative
support. To deal with the more general case $X{\sim}p(x)$ with 
${\rm supp}\{X\}=\Omega_{x{\geq}0}+\Omega_{x<0}{\subseteq}{\Rcal}$ we have to
\begin{itemize}
\item[$1)$] Express the density as 
$p(x)\,=\,
\underbrace{p(x)\,{\mbox{\boldmath $1$}_{x{\geq}0}(x)}}_{p^+(x)}\,+\,
\underbrace{p(x)\,{\mbox{\boldmath $1$}_{x<0}(x)}}_{p^-(x)}$;
\item[$2)$] Define $Y_1=X$ when $x{\geq}0$ and
            $Y_2=-X$ when $x<0$ so ${\rm supp}\{Y_2\}$ is positive
            and find $M_{Y_1}(s)$ and $M_{Y_2}(s)$; 
\item[$3)$] Get from the inverse transform the corresponding densities 
            $p_1(z)$ for the quantity of interest $Z_1=Z(Y_1,X_2,...)$
            with $M_{Y_1}(s)$ 
            and $p_2(z)$ for $Z_2=Z(Y_2,X_2,...)$ with $M_{Y_2}(s)$ 
            and at the end for  $p_2(z)$ make the corresponding change
            for $X{\rightarrow}-X$. 
\end{itemize}

This is usually quite messy and for most cases of interest it is far easier 
to find the distribution for the product and ratio of random quantities with
a simple change of variables.

\vspace{0.5cm}
\noindent
$\bullet$
{\bf Ratio of Normal and $\chi^2$ distributed random quantities}
Let's study the random quantity $X=X_1(X_2/n)^{-1/2}$ where
$X_1{\sim}N(x_1|0,1)$ with ${\rm sup}\{X_1\}=\Rcal$
and $X_2{\sim}{\chi}^2(x_2|n)$ with ${\rm sup}\{X_2\}=\Rcal^+$. 
Then, for $X_1$ we have
\begin{eqnarray*}
p(x_1)\,=\,
\underbrace{p(x_1)\,{\mbox{\boldmath $1$}_{[0,\infty)}(x_1)}}_{p^+(x_1)}\,+\,
\underbrace{p(x_1)\,{\mbox{\boldmath $1$}_{(-\infty,0)}(x_1)}}_{p^-(x_1)}
\end{eqnarray*}
and therefore for $X$  
\begin{eqnarray*}
X{\sim}p(x)\,=\,
p(x)\,{\mbox{\boldmath $1$}_{[0,\infty)}(x)}\,+\,
p(x)\,{\mbox{\boldmath $1$}_{(-\infty,0)}(x)}\,=\,
p^+(x)+p^-(x)
\end{eqnarray*}
Since
\begin{eqnarray*}
M_2(s)\,=\,
\frac{\textstyle 2^{s-1}\,\Gamma(n/2+s-1)}
     {\textstyle \Gamma(n/2)}
\end{eqnarray*}
we have for $Z=(X_2/n)^{-1/2}$ that
\begin{eqnarray*}
M_Z(s)\,=\,n^{(s-1)/2}\,M_2((3-s)/2)\,=\,
\left(\frac{\textstyle n}{\textstyle 2}\right)^{(s-1)/2}\,
\frac{\textstyle \Gamma((n+1-s)/2)}
     {\textstyle \Gamma(n/2)}
\end{eqnarray*}
for $0<\Re (s)<n+1$. For $X_1{\in}[0,\infty)$ we have that
\begin{eqnarray*}
M_1^+(s)\,=\,
\frac{\textstyle 2^{s/2}\Gamma(s/2)}
     {\textstyle 2\sqrt{2\pi}}\,;\hspace{1.cm}0<\Re (s)
\end{eqnarray*}
and therefore
\begin{eqnarray*}
M_X^+(s)\,=\,M_1^+(s)\,M_Z(s)\,=\,
\frac{n^{s/2}\,\Gamma(s/2)\,\Gamma((n+1-s)/2)}
     {2\sqrt{n\pi}}
\end{eqnarray*}
with holomorphy stripe $0< \Re (s) <n+1$. There are poles at
$s_m=-2m$ with $m=0,1,2,\ldots$ on the negative real axis and
$s_k=n+1+2k$ with $k=0,1,2,\ldots$ on the positive real axis. Closing the
contour on the left we include only $s_m$ so
\begin{eqnarray*}
p^+(x)\,&=&\,
\frac{1}
     {\sqrt{n\pi}\Gamma(n/2)} \sum_{m=0}^{\infty}
\frac{(-1)^m}{\Gamma(m+1)}\,
\left(\frac{x^2}{n}\right)^m\,\Gamma\left(m+\frac{n+1}{2}\right)\,= \\
&=&\,\frac{\Gamma((n+1)/2)}{\sqrt{n\pi}\Gamma(n/2)}\,
\left(1+\frac{x^2}{n}\right)^{-(n+1)/2}
{\mbox{\boldmath $1$}_{[0,\infty)}(x)}
\end{eqnarray*}

For $X_1{\in}(-\infty,0)$ we should in principle define $Y=-X_1$ with
support in $(0,\infty)$, find $M_Y(s)$, obtain the density for
$X'=Y/Z$ and then obtain the corresponding one for $X=-X'$. However,
in this case it is clear by symmetry that $p^+(x)=p^-(x)$ and therefore
\begin{eqnarray*}
X{\sim}p(x)\,=\,
\frac{\Gamma((n+1)/2)}{\sqrt{n\pi}\Gamma(n/2)}\,
\left(1+\frac{x^2}{n}\right)^{-(n+1)/2}
{\mbox{\boldmath $1$}_{(-\infty,\infty)}(x)}\,=\,St(x|n)
\end{eqnarray*}

\vspace{0.5cm}
\noindent
$\bullet$ {\bf Ratio and product of Normal distributed random quantities}
Consider $X_1{\sim}N(x_1|{\mu}_1,{\sigma}_1)$ and
$X_2{\sim}N(x_2|{\mu}_2,{\sigma}_2)$. The Mellin Transform is 
\begin{eqnarray*}
M_Y(s)\,=\,
\frac{\textstyle e^{-{\mu}^2/4{\sigma}^2}}{\sqrt{2\pi}}\,
{\sigma}^{s-1}\,\Gamma(s)\,D_{-s}(\mp{\mu}/{\sigma})
\end{eqnarray*}
with $D_a(x)$ the Whittaker Parabolic Cylinder Functions.
The upper sign $(-)$ of the argument corresponds to $X{\in}[0,\infty)$ 
and the lower one $(+)$ to the quantity  $Y=-X{\in}(0,\infty)$.
Again, the problem is considerably simplified if ${\mu}_1={\mu_2}=0$ because
\begin{eqnarray*}
M_Y(z)\,=\,
\frac{\textstyle 2^{z/2}}{2\sqrt{2\pi}}\,{\sigma}^{z-1}\,\Gamma(z/2)
\end{eqnarray*}
with $S=<0,\infty>$ and, due to symmetry, all contributions are the same. Thus,
summing over the poles at $z_n=-2n$ for $n=0,1,2,\ldots$ we have that
for $X=X_1X_2$ and $a^{-1}=4\sigma_1^2\sigma_2^2$:
\begin{eqnarray*}
p(x)\,=\,\frac{2\sqrt{a}}{\pi}
\sum_{n=0}^{\infty}
\frac{\textstyle (\sqrt{a}|x|)^{2n}}{\textstyle {\Gamma}(n+1)^2}\,
\left(2{\Psi}(1+n)\,-\,\ln(\sqrt{a}|x|)\right)\,=
\frac{2\sqrt{a}}{\pi}K_0(2\sqrt{a}|x|)
\end{eqnarray*}

Dealing with the general case of ${\mu}_i{\neq}0$ it is much more messy to
get compact expressions and life is easier with a simple change of variables.
Thus, for instance for $X=X_1/X_2$ we have that
\begin{eqnarray*}
p(x)\,=\,\frac{\textstyle \sqrt{a_1a_2}}{\textstyle \pi}\,
\int_{-\infty}^{{\infty}}\,
e^{\textstyle -\{a_1(xw-{\mu}_1)^2+a_2(w-{\mu}_2)^2\}}\,|w|\,dw
\end{eqnarray*}
where $a_i=1/(2{\sigma}_i^2)$ and if we define:
\begin{eqnarray*}
w_0=a_2\,+\,a_1x^2\,; \hspace{0.5cm}
w_1=a_1a_2\,(x{\mu}_2-{\mu}_1)^2\hspace{0.5cm}{\rm and}\hspace{0.5cm}
w_2=(a_1{\mu}_1x+a_2{\mu}_2)/\sqrt{w_0} 
\end{eqnarray*}
one has:
\begin{eqnarray*}
p(x)\,=\,\frac{\textstyle \sqrt{a_1a_2}}{\textstyle \pi}\,
         \frac{\textstyle 1}{\textstyle w_0}\,
         e^{\textstyle -w_1/w_0}\,
         \left(
	        e^{\textstyle -w_2^2}\,+\,\sqrt{\pi}\,w_2\,{\rm erf}(w_2)
	 \right)
{\mbox{\boldmath $1$}_{(-\infty,\infty)}(x)}
\end{eqnarray*}

\section{\LARGE \bf Ordered Samples} 

Let $X{\sim}p(x|\thetabold)$ be a one-dimensional random quantity 
and the experiment $e(n)$ that consists on $n$ independent observations 
and results in the exchangeable sequence $\{x_1,x_2,{\ldots},x_n\}$ are 
equivalent to an observation of the n-dimensional random
quantity $\Xbold{\sim}p(\xbold|\thetabold)$ where
\begin{eqnarray*}
p(\xbold|\thetabold)\,=\,p(x_1,x_2,{\ldots},x_n|\thetabold)\,=\,
\prod_{i=1}^np(x_i|\thetabold)
\end{eqnarray*}
Consider now a monotonic non-decreasing ordering of the observations
\begin{eqnarray}
\underbrace{\begin{array}{c}
                x_1\,{\leq}\,x_2\,{\leq}\,{\ldots}\,x_{k-1}  \\
                                                             \\
             \end{array}
           }_{k-1}
       \begin{array}{c}
            \,{\leq}\,x_k\,{\leq}\,        \\
                                           \\
       \end{array}
\underbrace{\begin{array}{c}
                 x_{k+1}\,{\leq}\,{\ldots}\,{\leq}\,x_{n-1}\,{\leq}\,x_n \\
                                                              \\
            \end{array}
           }_{n-k}
              \nonumber
\end{eqnarray}
and the {\sl Statistic of Order $k$}; that is, the random quantity 
$X_{(k)}$ associated with the $k^{th}$ observation $(1{\leq}k{\leq}n)$ of the 
{\sl ordered sample} such that there are $k-1$ observations smaller than
$x_k$ and $n-k$ above $x_k$. Since
\begin{eqnarray*}
P(X{\leq}x_k|\thetabold)\,=\,
\int_{-\infty}^{x_k}p(x|\thetabold)dx=F(x_k|\thetabold)
\hspace{1.cm}{\rm and}\hspace{1.cm}
P(X>x_k|\thetabold)\,=\,1-F(x_k|\thetabold)
\end{eqnarray*}
we have that
\begin{eqnarray*}
X_{(k)}{\sim}p(x_k|\thetabold,n,k)\,
 &=&\,C_{n,k}\,p(x_k|\thetabold)\,\left[F(x_k|\thetabold)\right]^{k-1}\,
                      \left[1\,-\,F(x_k|\thetabold)\right]^{n-k} \\ \\ 
&=&\,C_{n,k}\,p(x_k|\thetabold)\,\,\,\,\,
 \underbrace{
  \left[\int_{-\infty}^{x_k}\,p(x|\thetabold)\,dx\right]^{k-1}}_
  {\left[P(X{\leq}x_k)\right]^{k-1}}
  \,\,\,\,\,
 \underbrace{
  \left[\int^{\infty}_{x_k}\,p(x|\thetabold)\,dx\right]^{n-k}}_
  {\left[P(X{>}x_k)\right]^{n-k}}
\end{eqnarray*}
The normalization factor 
\begin{eqnarray*}
C_{n,k}\,=\,
     k\,\left(\begin{array}{c}
                             n \\
                             k
                           \end{array}\right)
\end{eqnarray*}
is is given by combinatorial analysis although in general it is easier
to get by normalization of the final density. With a similar reasoning
we have that the density function of the two dimensional random quantity 
$X_{(ij)}=(X_i,X_j)$; $j>i$, associated to the observations $x_i$ and $x_j$ 
({\sl Statistic of Order} $i,j;i<j$)
\footnote{If the random quantities $X_i$ are not identically distributed
the idea is the same but one hast to deal with permutations and the expressions
are more involved}
will be:
\begin{eqnarray*}
 X_{(ij)}{\sim}p(x_i,x_j|\thetabold,i,j,n)\,&=&\,C_{n,i,j}\,
    \underbrace{ \left[ \int_{-\infty}^{x_i}p(x|\thetabold)dx \right]^{i-1} }_
               { [P(X<x_i)]^{i-1}}\,\,\,p(x_i|\thetabold)\,\,\,
  \underbrace{ 
         \left[ \int_{x_i}^{x_j}p(x|\thetabold)dx
\right]^{j-i-1} }_
               { [P(x_i<X{\leq}<x_j)]^{j-i-1}}\\
&&p(x_j|\thetabold)\,\,\,
    \underbrace{ \left[ \int_{x_j}^{\infty}p(x|\thetabold)dx \right]^{n-j} }_
               { [P(x_j<X)]^{n-j}}
         \nonumber
\end{eqnarray*}
where
$(x_i,x_j){\in}({-\infty},x_j]{\times}({-\infty},{\infty})$ or
$(x_i,x_j){\in}({-\infty},{\infty}){\times}[x_i,{\infty})$.
Again by combinatorial analysis or integration we have that
\begin{eqnarray*}
      C_{n,i,j}\,=\,\frac{\textstyle n!}
      {\textstyle (i-1)!\,(j-i-1)!\,(n-j)!}
         \nonumber
\end{eqnarray*}
The main {\sl Order Statistics} we are usually interested in are
\begin{itemize}
\item[$\bullet$]{\bf Maximum} $X_{(n)}={\rm max}\{X_1,X_2,{\ldots},X_n\}$:
\begin{eqnarray*}
p(x_n|\cdot)\,=\,
 n\,p(x_n|\thetabold)\,
  \left[\int_{-\infty}^{x_n}\,p(x|\thetabold)\,dx\right]^{n-1}
\end{eqnarray*}
\item[$\bullet$]{\bf Minimum} $X_{(1)}={\rm min}\{X_1,X_2,{\ldots},X_n\}$:
\begin{eqnarray*}
p(x_1|\cdot)\,=\,
 n\,p(x_1|\thetabold)\,
  \left[\int_{x_1}^{\infty}\,p(x|\thetabold)\,dx\right]^{n-1}
\end{eqnarray*}
\item[$\bullet$]{\bf Range} $R=X_{(n)}-X_{(1)}$
\begin{eqnarray*}
p(x_1,x_n|\cdot)\,=\,
 n(n-1)\,p(x_1|\thetabold)\,p(x_n|\thetabold)\,
  \left[\int_{x_1}^{x_n}\,p(x|\thetabold)\,dx\right]^{n-2}
\end{eqnarray*}
If ${\rm supp}(X)=[a,b]$, then $R{\in}(0,b-a)$ and
\begin{eqnarray*}
p(r)\,=\,
 n(n-1)\,\left\{\int_{a}^{b-r}p(w+r)\,p(w)\,
    \left[F(w+r)-F(w) \right]^{n-2}dw
\right\}
\end{eqnarray*}
There is no explicit form unless we specify the Distribution Function 
$F(x|\theta)$.
\item[$\bullet$]{\bf Difference} $S=X_{(i+1)}-X_{(i)}$.
If ${\rm supp}(X)=[a,b]$, then $S{\in}(0,b-a)$ and
\begin{eqnarray*}
p(s)\,=\,\frac{\Gamma(n+1)}{\Gamma(i)\Gamma(n-i)}
 \left\{\int_{a}^{b-s}p(w+s)\,p(w)\,
    \left[F(w) \right]^{i-1}
    \left[1-F(w+s)\right]^{n-i-1}dw
\right\}
\end{eqnarray*}

\end{itemize}

In the case of discrete random quantities, the idea is the same but a bit
more messy because one has to watch for the discontinuities of the 
Distribution Function. Thus, for instance:
\begin{itemize}
\item[$\bullet$]{\bf Maximum} $X_{(n)}={\rm max}\{X_1,X_2,{\ldots},X_n\}$:

\noindent
$X_{(n)}{\leq}x$ iff {\bf all} $x_i$ are {\bf less or equal}
          $x$ and this happens with probability
          \begin{eqnarray*}
            P(x_n{\leq}x)\,=\,\left[F(x)\right]^n
          \end{eqnarray*}

\noindent
$X_{(n)}<x$ iff {\bf all} $x_i$ are {\bf less than} 
          $x$ and this happens with probability
          \begin{eqnarray*}
            P(x_n<x)\,=\,\left[F(x-1)\right]^n
          \end{eqnarray*}
Therefore
\begin{eqnarray*}
            P(x_n=x)\,=\,P(x_n{\leq}x)-P(x_n<x)\,=\,
            \left[F(x)\right]^n-\left[F(x-1)\right]^n
          \end{eqnarray*}

\item[$\bullet$]{\bf Minimum} $X_{(1)}={\rm min}\{X_1,X_2,{\ldots},X_n\}$:

\noindent
$X_{(1)}{\geq}x$ iff {\bf all} $x_i$ are {\bf grater or equal}
          $x$ and this happens with probability
          \begin{eqnarray*}
            P(x_1{\geq}x)\,=\,1-P(x_1<x)\,=\,\left[1-F(x-1)\right]^n
          \end{eqnarray*}

\noindent
$X_{(1)}>x$ iff {\bf all} $x_i$ are {\bf greater than} 
          $x$ and this happens with probability
          \begin{eqnarray*}
            P(x_1>x)\,=\,1-P(x_1{\leq}x)\,=\,\left[1-F(x)\right]^n
          \end{eqnarray*}
Therefore
\begin{eqnarray*}
            P(x_1=x)\,&=&\,P(x_1{\leq}x)-P(x_1<x)\,=\,
            \left[1-P(x_1>x)\right]-\left[1-P(x_1{\geq}x)\right]\,= \\
            &=&\,\left[1-F(x-1)\right]^n-\left[1-F(x)\right]^n
          \end{eqnarray*}
\end{itemize}

\vspace{0.5cm}
\noindent
{\raya}                   
\vspace{0.35cm}
\footnotesize

\noindent
{\bf Example 1.26:} Let $X{\sim}Un(x|a,b)$ and an iid sample of size $n$. Then,
if $L=b-a$:

\vspace{0.35cm}
$\bullet$
{\sl Maximum:}
$p(x_n)\,=\,
         n\,\frac{\textstyle (x_n\,-\,a)^{n-1}}
                 {\textstyle (b\,-\,a)^{n}}
{\mbox{\boldmath $1$}}_{(a,b)}(x_n)$

\vspace{0.35cm}
$\bullet$
{\sl Minimum:}
    $  p(x_1)\,=\,
         n\,\frac{\textstyle (b\,-\,x_1)^{n-1}}
                 {\textstyle (b\,-\,a)^{n}}
{\mbox{\boldmath $1$}}_{(a,b)}(x_1) $

\vspace{0.35cm}
$\bullet$
{\sl Range:} $R=X_{(n)}-X_{(1)}:\,\,$
    $  p(r)\,=\,\frac{n(n-1)}{L}\,
    \left(\frac{r}{L}\right)^{n-2}\,
    \left(1-\frac{r}{L}\right)
 {\mbox{\boldmath $1$}}_{(0,L)}(r) $

\vspace{0.35cm}
$\bullet$
{\sl Difference:} $S=X_{(k+1)}-X_{(k)}:\,\,$
$p(s)\,=\,\frac{n}{L}\left( 1-\frac{s}{L}\right)^{n-1}
{\mbox{\boldmath $1$}}_{(0,L)}(s)$

\vspace{0.35cm}

\noindent
{\bf Example 1.27:}
Let's look at the Uniform distribution in more detail.
Consider a random quantity $X{\sim}Un(x|a,b)$, the experiment $e(n)$
that provides a sample of $n$ independent events and the ordered sample
\begin{eqnarray*}
  {\mathbf{X}}_{n}\,=\,
\{x_1\,{\leq}\,x_2\,{\leq}\,{\ldots}\,{\leq}\,x_k\,{\leq}\,{\ldots}\,
                  {\leq}\,x_{k+p}\,{\leq}\,{\ldots}\,  
{\leq}\,x_{n-1}\,{\leq}\,x_n\}
\end{eqnarray*}
Then, for the ordered 
statistics $X_k$, $X_{k+p}$ 
and $X_{k+p+1}$ with $k,p{\in}{\Ncal}$, $1{\leq}k{\leq}n-1$ and 
$p{\leq}n-k-1$ we have that

\vspace{1.0cm}
{\hspace*{2.7cm}} $k-1$ {\hspace*{1.30cm}} $p-1$ {\hspace*{1.80cm}} $n-(k+p+1)$

\hspace*{2.00cm} {\rule[-0.50mm]{0.50mm}{2.00mm}}
                 \hspace*{-1.50mm}{\rule{1.00cm}{0.50mm}}
                 \hspace*{-1.50mm}{\rule{1.50cm}{0.50mm}}
                 {\hspace*{-1.50mm}\rule[-0.50mm]{0.50mm}{2.00mm}}
                 \hspace*{-1.50mm}{\rule{2.00cm}{0.50mm}}
                 {\hspace*{-1.50mm}\rule[-0.50mm]{0.50mm}{2.00mm}}
                 \hspace*{-1.50mm}{\rule{0.80cm}{0.50mm}}
                 {\hspace*{-1.1mm}\rule[-0.50mm]{0.50mm}{2.00mm}}
                 \hspace*{-1.20mm}{\rule{3.50cm}{0.50mm}}
                 {\hspace*{-1.50mm}\rule[-0.50mm]{0.50mm}{2.00mm}}

{\hspace*{1.95cm}} $a$ {\hspace*{0.80cm}}. . . {\hspace*{0.45cm}} $x_k$
{\hspace*{0.12cm}} . . . {\hspace*{0.12cm}}
$x_{k+p}$ {\hspace*{0.05cm}} 
$x_{k+p+1}$ {\hspace*{0.40cm}} . . . {\hspace*{1.10cm}}$b$
\vspace{0.5cm}

\begin{eqnarray*}
  p(x_k,x_{k+p}|a,b,n,p)\,&{\propto}&\,
\left[\int_a^{x_k}ds_1\right]^{k-1}\,
\left[\int_{x_k}^{x_{k+p}}ds_2\right]^{p-1}\,
\left[\int_{x_{k+p+1}}^{b}ds_3\right]^{n-(k+p+1)}
\end{eqnarray*}
Let's think for instance that those are the arrival times of $n$ events 
collected with a detector in a time window $[a=0,b=T]$.
If we define $w_1=x_{k+p}-x_k$ and $w_2=x_{k+p+1}-x_k$ we have that
\begin{eqnarray*}
  p(x_k,w_1,w_2|T,n,p)=
          x_k^{k-1}w_1^{p-1}(T-x_k-w_2)^{n-k-p-1}
{\mathbf{1}}_{[0,T-w_2]}(x_k)
{\mathbf{1}}_{[0,w_2]}(w_1){\mathbf{1}}_{[0,T]}(w_2)
\end{eqnarray*}
and, after integration of $x_k$:
\begin{eqnarray*}
  p(w_1,w_2|T,n,p)\,=\, \left(  \begin{array}{c}
                 n \\ p
               \end{array}
     \right)
       \frac{\textstyle p(n-p)}{\textstyle T^n}\,
          w_1^{p-1}\,(T\,-\,w_2)^{n-p-1}
{\mathbf{1}}_{[0,w_2]}(w_1){\mathbf{1}}_{[0,T]}(w_2)
\end{eqnarray*}
Observe that the support can be expressed also as
${\mathbf{1}}_{[0,T]}(w_1){\mathbf{1}}_{[w_1,T]}(w_2)$
and that the distribution of $(W_1,W_2)$ does not depend on $k$.
The marginal densities are given by:
\begin{eqnarray}
  p(w_1|T,n,p)\,&=&\,
        \left(  \begin{array}{c}
                 n \\ p
               \end{array}
     \right)
       \frac{\textstyle p}{\textstyle T^n}\,
          w_1^{p-1}\,(T\,-\,w_1)^{n-p}
	  \,\,{\mathbf{1}}_{[0,T]}(w_1)
                 \nonumber \\
   p(w_2|T,n,p)\,&=&\,
       \left(  \begin{array}{c}
                 n \\ p
               \end{array}
     \right)
       \frac{\textstyle n-p}{\textstyle T^n}\,
          w_2^{p}\,(T\,-\,w_2)^{n-p-1}
	  \,\,{\mathbf{1}}_{[0,T]}(w_2)
                 \nonumber 
\end{eqnarray}
and if we take the limit $T{\rightarrow}{\infty}$ and 
$n{\rightarrow}{\infty}$ keeping the the rate ${\lambda}=n/T$ constant we have
\begin{eqnarray*}
\lim_{T,n\rightarrow\infty}\,
  p(w_1,w_2|T,n,p)\,=\,
p(w_1,w_2|{\lambda},p)\,=\,
       \frac{\textstyle {\lambda}^{p+1}}{\textstyle {\Gamma}(p)}\,
          e^{-{\lambda}w_2}\,w_1^{p-1}
\,{\mathbf{1}}_{[0,w_2)}(w_1)\,{\mathbf{1}}_{[0,\infty)}(w_2)
\end{eqnarray*}
and
\begin{eqnarray*}
  p(w_1|{\lambda},p)\,=\,
       \frac{\textstyle {\lambda}^{p}}{\textstyle {\Gamma}(p)}\,
          e^{-{\lambda}w_1}\,w_1^{p-1}
	  \,\,{\mathbf{1}}_{[0,{\infty})}(w_1)
\end{eqnarray*}
In consequence, under the stated conditions
the time difference between two consecutive events $(p=1)$ tends
to an exponential distribution. Let's  consider for simplicity
this limiting behaviour in what follows and leave as an exercise
the more involved case of finite time window $T$.

Suppose now that after having observed one event, say $x_k$, we have a 
dead-time of size $a$ in the detector during which we can not process any 
data. All the events that fall in $(x_k,x_k+a)$ are lost (unless we play with
buffers).
If the next observed event is at time $x_{k+p+1}$, we have lost $p$ events
and the probability for this to happen is
\begin{eqnarray}
  {\Pcal}(w_1{\leq}a,w_2{\geq}a|{\lambda},p)\,=\,
       e^{-{\lambda}a}\,
       \frac{\textstyle ({\lambda}a)^{p}}{\textstyle {\Gamma}(p+1)}\,
                 \nonumber 
\end{eqnarray}
that is, $N_{\rm lost}{\sim}Po(p|{\lambda}a)$ regardless the 
position of the last recorded time ($x_k$) in the ordered sequence. 
As one could easily have intuited, the expected number of events
lost for each observed one is $E[N_{\rm lost}]={\lambda}a$. 
Last, it is clear that the density for the time difference
between two consecutive observed events when $p$ are lost due to the
dead-time is
\begin{eqnarray}
  p(w_2|w_1{\leq}a,{\lambda},p)\,=\,
     {\lambda}\,e^{-{\lambda}(w_2-a)}
	  \,\,{\mathbf{1}}_{[a,{\infty})}(w_2)
                 \nonumber 
\end{eqnarray}
Note that it depends on the dead-time window $a$ and not on the number of 
events lost.

\vspace{0.35cm}

\noindent
{\bf Example 1.28:} Let $X{\sim}Ex(x|{\lambda})$ and an iid sample of size
$n$. Then:
\vspace{0.35cm}

$\bullet$
{\sl Maximum:}
$      p(x_n)\,=\,
         n\,{\lambda}\,e^{-{\lambda}x_n}\,
         (1\,-\,e^{-{\lambda}x_n})^{n-1}
         {\mbox{\boldmath $1$}}_{(0,\infty)}(x_n)$
\vspace{0.35cm}

$\bullet$
{\sl Minimum:}
$      p(x_1)\,=\,         n\,{\lambda}\,e^{-{\lambda}nx_1}
          {\mbox{\boldmath $1$}}_{(0,\infty)}(x_1)$
\vspace{0.35cm}

$\bullet$
{\sl Range:} $R=X_{(n)}-X_{(1)}:$
$p(r)=(n-1){\lambda}^{n-1}e^{-{\lambda}r}
  \left[1-e^{-{\lambda}r}\right]^{n-2}
     {\mbox{\boldmath $1$}}_{(0,\infty)}(r)$
\vspace{0.35cm}

$\bullet$
{\sl Difference:} $S=X_{(k+1)}-X_{(k)}:$
$p(s)\,=\,(n-k)\,{\lambda}e^{-{\lambda}x(n-k)}
{\mbox{\boldmath $1$}}_{(0,\infty)}(s)$

\smill
{\raya}                   
\vspace{1.0cm}

\section{\LARGE \bf Limit Theorems and Convergence}

In Probability, the Limit Theorems are statements that, under the conditions
of applicability, describe the behavior of a sequence of random 
quantities or of Distribution Functions. In principle, whenever we can
define a distance (or at least a positive defined set function) we can
establish a convergence criteria and, obviously, some will be
stronger than others so, for instance, a sequence of random quantities 
$\{X_i\}_{i=1}^{\infty}$ may converge according to one criteria and not to 
other. The most usual types of convergence, their relation and the
Theorems derived from them are:

\begin{tabular}{p{4.cm}p{8.0cm}}
                 & \\
  Distribution & $\Longrightarrow\,$ Central Limit Theorem \\
  $\hspace{0.5cm}\Uparrow$ & $\Longrightarrow\,$ Glivenko-Cantelly Theorem 
                             (weak form) \\ 
            $\hspace{0.5cm}\Uparrow$     & \\
  Probability & $\Longrightarrow\,$ Weak Law of Large Numbers \\ 
              $\hspace{0.5cm}\Uparrow$ $\hspace{0.5cm}\Uparrow$     & \\
  $\hspace{0.5cm}\Uparrow$ \hspace{0.3cm} Almost Sure & 
                 $\Longrightarrow\,$ Strong Law of Large Numbers \\
          $\hspace{0.5cm}\Uparrow$     & \\
  $L_p(\Rcal)$ Norm & $\Longrightarrow\,$ Convergence in Quadratic Mean \\
                 & \\
  Uniform & $\Longrightarrow\,$ Glivenko-Cantelly Theorem \\
                 & \\
\end{tabular}

\noindent
so {\sl Convergence in Distribution} is the weakest of all since does not 
imply any of the others. In principle, there will be no
explicit mention to statistical independence of the random quantities of 
the sequence nor to an specific Distribution Function. In most cases
we shall just state the different criteria for convergence and refer to
the literature, for instance [Gu13], for further details and demonstrations. 
Let's start with the very useful Chebyshev's Theorem.

\subsection{Chebyshev's Theorem}
\vspace*{0.1cm}

Let $X$ be a random quantity that takes values in $\Omega{\subset}{\Rcal}$
with Distribution Function $F(x)$ and consider the random quantity
$Y=g(X)$ with $g(X)$ a non-negative single valued function for all 
$X{\in}\Omega$. Then, for $\alpha{\in}{\Rcal}^+$ 
\begin{eqnarray*}
        P\left(g(X) \geq \alpha \right) \,\leq\,
        \frac{\textstyle E[g(X)]}{\textstyle \alpha}
\end{eqnarray*}

\noindent
In fact, given a measure space $(\Omega,{\Bcal}_{\Omega},\mu)$,
for any $\mu$-integrable function $f(x)$ and $c>0$ we have for
$A=\{x:|f(x)|{\geq}c\}$ that 
$c{\mbox{\boldmath $1$}}_{A}(x){\leq}|f(x)|$ for all $x$ and therefore
\begin{eqnarray*}
c{\mu}(A)=\int c{\mbox{\boldmath $1$}}_{A}(x)d{\mu}{\leq}\int |f(x)|d{\mu}
\end{eqnarray*}

Let's see two particular cases. First, consider
$g(X)=(X-{\mu})^{2n}$ where ${\mu}=E[X]$ and $n$ a positive integer such that
$g(X) \geq 0$ $\forall X \in\Omega$. Applying Chebishev's Theorem:
\begin{eqnarray*}
   P(({X}-{\mu})^{2n} \geq \alpha)\,=\,
   P(|{X}-{\mu}| \geq {\alpha}^{1/2n})\,\leq\,
        \frac{\textstyle E[({X}-{\mu})^{2n}]}
             {\textstyle \alpha }\,=\,
        \frac{\textstyle {\mu}_{2n}}
             {\textstyle \alpha }   
\end{eqnarray*}
For $n=1$, if we take ${\alpha}=k^{2} {\sigma}^2$
we get the {\sl Bienaym\'e-Chebishev}'s inequality
\begin{eqnarray*}
        P(|{X}-{\mu}| \geq k {\sigma}) \,\leq\,
        1/k^2
\end{eqnarray*}
that is, whatever the Distribution Function of the random quantity $X$ is,
the probability that $X$ differs from its expected value 
$\mu$ more than $k$ times its standard deviation is less or equal than
$1/k^2$.
As a second case, assume $X$ takes only positive real values and
has a first order moment $E[X]={\mu}$. Then (Markov's inequality):
\begin{eqnarray*}
   P({X} \geq {\alpha} )\,\leq\,
        \frac{\textstyle {\mu}}
             {\textstyle {\alpha} }
\hspace{1.cm}\stackrel{{\alpha}=k {\mu}}{\longrightarrow}\hspace{1.cm}
        P(X \geq k {\mu}) \,\leq\,1/k
\end{eqnarray*}

The Markov and Bienaym\'e-Chebishev's inequalities 
provide upper bounds for the probability knowing just
mean value and the variance although they are usually very 
conservative. They can be considerably improved if we have
more information about the Distribution Function but, as we shall see,
the main interest of Chebishev's inequality lies on its importance to prove
Limit Theorems.

\subsection{Convergence in Probability}
The sequence of random quantities  $\{X_n(w)\}_{n=1}^{\infty}$
{\sl converges in probability} to $X(w)$ iff:
     \begin{eqnarray*}
     \lim_{n{\rightarrow}\infty}\,
     P(|X_n(w)-X(w)|\, \geq\, {\epsilon})\,=\,
     0 \,\,\,\,\, ;\,\,\, \forall {\epsilon}>0 \,\,;
     \end{eqnarray*}
or, equivalently, iff:
     \begin{eqnarray*}
     \lim_{n{\rightarrow}\infty}\,
     P(|X_n(w)-X(w)|\, <\, {\epsilon})\,=\,
     1 \,\,\,\,\, \forall {\epsilon}>0 \,\,;
     \end{eqnarray*}
Note that $P(|X_n(w)-X(w)| \geq{\epsilon})$ is a a real number so this is
is the usual limit for a sequence of real numbers and, in consequence,
for all ${\epsilon}>0$ and ${\delta}>0$
${\exists}\,n_0({\epsilon},{\delta})$ such that for all
$n>n_0({\epsilon},{\delta})$ it holds that
$ P(|X_n(w)-X(w)|{\geq}{\epsilon})<{\delta}$. 
For a sequence of n-dimensional random quantities, this can be generalized to
$\lim_{n{\rightarrow}\infty}P({\parallel}X_n(w),X(w){\parallel})$ and,
as said earlier, Convergence in Probability implies Convergence in 
Distribution but the converse is not true.
An important consequence of the Convergence in Probability is the

\vspace{0.5cm}
\noindent
$\bullet$ {\bf Weak Law of Large Numbers:}
Consider a sequence of independent
random quantities $\{X_i(w)\}_{i=1}^{\infty}$, all with the same Distribution
Function and first order moment $E[X_i(w)]={\mu}$, and define a new random
quantity
\begin{eqnarray*}
  Z_n(w)\,=\, \frac{\textstyle 1}{\textstyle n} \sum_{i=i}^{n}X_i(w)
\end{eqnarray*}
The, the sequence $\{Z_n(w)\}_{n=1}^{\infty}$ converges in probability to $\mu$;
that is:
\begin{eqnarray*}
     \lim_{n{\rightarrow}\infty}\,
     P(|Z_n(w)-{\mu}|\, \geq\, {\epsilon})\,=\,
     0 \,\,\,\,\, ;\,\,\, \forall {\epsilon}>0 \,\,;
     \end{eqnarray*}
\vspace{0.1cm}
\noindent
{\rayan}                   
\vspace{0.35cm}
\footnotesize

\noindent
The Law of Large Numbers was stated first by J. Bernoulli in 1713 for
the Binomial Distribution, generalized 
(and named {\sl Law of Large Numbers}) by S.D. Poisson and shown in the
general case by A. Khinchin in 1929. 
In the case $X_i(w)$ have variance
$V(X_i)=\sigma^2$ it is straight forward from Chebishev's inequality:
\begin{eqnarray*}
  P\left(|Z_n-\mu|{\geq}\epsilon \right)=
  P\left((Z_n-\mu)^2{\geq}\epsilon^2 \right){\leq}
  \frac{E[(Z_n-\mu)^2]}{\epsilon^2}=
  \frac{\sigma^2}{n\epsilon^2}
\end{eqnarray*}

\smill
{\rayan}
\vspace{1.0cm}

Intuitively, Convergence in Probability means that when n is very large, the
probability that $Z_n(w)$ differs from ${\mu}$ by a small amount is very
small; that is, $Z_n(w)$ gets more concentrated around ${\mu}$. But
{\sl ``very small''} is not zero and it may happen
that for some $k>n$ $Z_k$ differs from ${\mu}$ by more than ${\epsilon}$.
An stronger criteria of convergence is the {\sl Almost Sure Convergence}.

\subsection{Almost Sure Convergence}

A sequence $\{X_n(w)\}_{n=1}^{\infty}$ of random quantities converges 
{\sl almost sure} to $X(w)$ if, and only if:
\begin{eqnarray*}
     \lim_{n{\rightarrow}\infty}\,
     X_n(w)\,=\,X(w)
     \hspace{1.5cm}
  \end{eqnarray*}   
for all $w {\in}{\Omega}$ except at most on a set $W{\subset}{\Omega}$ of
zero measure ($P(W)=0$ so it is also referred to as {\sl convergence
almost everywhere}).
This means that for all $\epsilon >0$ and all
$w{\in}W^c={\Omega}-W$, ${\exists}n_0(\epsilon,w)>0$ such that
$|X_n(w)-X(w)|<\epsilon$ for all $n>n_0(\epsilon,w)$. Thus, we have the
equivalent forms:
\begin{eqnarray*}
     P\left[ \lim_{n{\rightarrow}\infty}\,
     |X_n(w)-X(w)|\,{\geq}\,{\epsilon} \right]\,=\,0
\hspace{0.5cm}{\rm or}\hspace{0.5cm}
 P\left[ \lim_{n{\rightarrow}\infty}\,
     |X_n(w)-X(w)|\,<\,{\epsilon} \right]\,=\,1
  \end{eqnarray*}       
for all ${\epsilon}>0$.
Needed less to say that the random quantities
$X_1,\,X_2\,...$ and $X$ are defined on the same probability space.
Again, Almost Sure Convergence implies Convergence in Probability but the
converse is not true. 
An important consequence of the Almost Sure Convergence is the:

\vspace{0.5cm}
\noindent
$\bullet$ {\bf Strong Law of Large Numbers} 
(E. Borel 1909, A.N. Kolmogorov,...){\bf :} 
Let $\{X_i(w)\}_{i=1}^{\infty}$ be a sequence of independent
random quantities all with the same Distribution
Function and first order moment $E[X_i(w)]={\mu}$. Then the
sequence $\{Z_n(w)\}_{n=1}^{\infty}$ with
\begin{eqnarray*}
  Z_n(w)\,=\, \frac{\textstyle 1}{\textstyle n} \sum_{i=i}^{n}X_i(w)
\end{eqnarray*}
converges almost sure to $\mu$; that is:
\begin{eqnarray*}
     P\left[ \lim_{n{\rightarrow}\infty}\,
     |Z_n(w)-{\mu}|\,{\geq}\,{\epsilon} \right]\,=\,0
     \hspace{1.5cm}\forall{\epsilon}>0
  \end{eqnarray*}

Intuitively, Almost Sure Convergence means that the probability that for
some $k>n$, $Z_k$ differs from ${\mu}$ by more than ${\epsilon}$ 
becomes smaller as n grows.

\subsection{Convergence in Distribution}
Consider the sequence of random quantities
$\{X_n(\omega)\}_{n=1}^{\infty}$ and of their corresponding Distribution
Functions $\{F_n(x)\}_{n=1}^{\infty}$. In the limit $n{\rightarrow}{\infty}$,
the random quantity $X_n(w)$ tends to be distributed as $X(w){\sim}F(x)$ iff
\begin{eqnarray*}
     \lim_{n{\rightarrow}\infty}\,F_n(x)\,=\,F(x)
     \hspace{0.5cm}{\Leftrightarrow}\hspace{0.5cm}
      \lim_{n{\rightarrow}\infty}\,
     P(X_n{\leq}x)\,=\,
     P(X{\leq}x) 
\hspace{0.3cm};\hspace{0.5cm}{\forall}x\,{\in}\,C(F)
\end{eqnarray*}       
with $C(F)$ the set of points of continuity of $F(x)$. Expressed in a 
different manner, the sequence $\{X_n(w)\}_{n=1}^{\infty}$ Converges in
Distribution to $X(w)$  if, and only if, for all
${\epsilon}>0$ and $x{\in}C(F)$, ${\exists}\,n_0({\epsilon},x)$
such that $|F_n(x)\,-\,F(x)|<{\epsilon}$,
${\forall}n>n_0({\epsilon},x)$. Note that, in general,
$n_0$ depends on $x$ so it is possible that, given an ${\epsilon}>0$, 
the value of $n_0$ for which the condition 
$|F_n(x)\,-\,F(x)|<{\epsilon}$ is satisfied for certain values of
$x$ may not be valid for others.
It is important to note also that we have not made any statement about the
statistical independence of the random quantities and that the Convergence
in Distribution is determined only by the Distribution Functions so
the corresponding random quantities do not have to be defined on the same
probability space. To study the Convergence in Distribution, the following 
theorem it is very useful:

\vspace{0.5cm}
\noindent
$\bullet$ {\bf Theorem} (L\'evy 1937; Cram\`er 1937){\bf :}
Consider a sequence of Distribution Functions
$\{F_n(x)\}_{n=1}^{\infty}$ and of the corresponding Characteristic
Functions $\{\Phi_n(t)\}_{n=1}^{\infty}$. Then

  \begin{itemize}
  \item[$\triangleright$] 
    if ${\rm lim}_{n{\rightarrow}\infty}\,F_n(x)=F(x)$, then
    ${\rm lim}_{n{\rightarrow}\infty}\,{\Phi}_n(t)={\Phi}(t)$
    for all $t{\in}{\Rcal}$ 
    with ${\Phi}(t)$ the Characteristic Function of $F(x)$. 
   \item[$\triangleright$] 
     Conversely, if
     ${\Phi}_n(t)\stackrel{n{\rightarrow}\infty}
      {\longrightarrow}{\Phi}(t)$
     ${\forall}t{\in}{\Rcal}$ and ${\Phi}(t)$ is continuous at $t=0$,
     then 
     $F_n(x)\stackrel{n{\rightarrow}\infty}
      {\longrightarrow}F(x)$
  \end{itemize}

This criteria of convergence is weak in the sense that
if there is convergence if probability or almost sure or in quadratic mean
then there is convergence in distribution but the converse is not
necessarily true. However, there is a very important consequence of the
Convergence in Distribution:

\vspace{0.5cm}
\noindent
$\bullet$ {\bf Central Limit Theorem} (Lindberg-Levy){\bf :} 
Let $\{X_i(w)\}_{i=1}^{\infty}$ be a sequence of independent
random quantities all with the same Distribution
Function and with second order moments so $E[X_i(w)]={\mu}$ and
$V[X_i(w)]={\sigma}^2$. Then the
sequence $\{Z_n(w)\}_{n=1}^{\infty}$ of random quantities
\begin{eqnarray*}
  Z_n(w)\,=\, \frac{\textstyle 1}{\textstyle n} \sum_{i=i}^{n}X_i(w)
\end{eqnarray*}
with
\begin{eqnarray*}
     E[Z_n]\,=\,\frac{\textstyle 1}{\textstyle n} \sum_{i=i}^{n}E[X_i]\,=\,\mu
\hspace{0.5cm}{\rm and}\hspace{0.5cm}
     V[Z_n]\,
=\,\frac{\textstyle 1}{\textstyle n^2} \sum_{i=i}^{n}V[X_i]\,=\,
\frac{\textstyle {\sigma}^2}{\textstyle n}
  \end{eqnarray*}
tends, in the limit $n{\rightarrow}\infty$, to be distributed as
$N(z|\mu,\sigma/\sqrt{n})$ or, what is the same, the {\sl standardized}
random quantity
\begin{eqnarray*}
 \stackrel{\sim}{Z}_n\,=\,\frac{Z_n-\mu}{\sqrt{V[Z_n]}}\,=\,
   \frac{\frac{\textstyle 1}{\textstyle n}\,
         \displaystyle{ \sum_{i=1}^{n}}X_i\,-\,\mu}
        {{\sigma}/\sqrt{n}}
\end{eqnarray*}
tends to be distributed as $N(x|0,1)$.

\vspace{0.5cm}
\noindent
{\rayan}                   
\vspace{0.35cm}
\footnotesize

\noindent
Consider, without loss of generality, the random quantity 
$W_i=X_i-{\mu}$ so that $E[W_i]=E[X_i]-\mu=0$ and
$V[W_i]=V[X_i]={\sigma}^2$. Then, 
\begin{eqnarray*}
 {\Phi}_{W}(t)\,=\,1\,-\,\frac{1}{2}t^2
       {\sigma}^2\,+\, {\Ocal}(t^k) 
\end{eqnarray*}
Since we require that the random quantities $X_i$ have at least moments of
order two, the remaining terms ${\Ocal}(t^k)$ are either zero or powers of 
$t$ larger than 2.
Then,
\begin{eqnarray*}
  Z_n\,=\, \frac{\textstyle 1}{\textstyle n} \sum_{i=i}^{n}X_i
  \,=\,\frac{\textstyle 1}{\textstyle n} \sum_{i=i}^{n}W_i+{\mu}
\hspace{0.5cm};\hspace{0.5cm}
 E[Z_n]\,=\,{\mu}
\hspace{0.5cm};\hspace{0.5cm}
 V[Z_n]\,=\,{\sigma}^2_{Z_n}\,=\,
\frac{\textstyle {\sigma}^2}{\textstyle n}
\end{eqnarray*}
so
\begin{eqnarray*}
 {\Phi}_{Z_n}(t)\,=\,e^{it\mu}\,
   \left[ {\Phi}_{W}(t/n) \right]^n
\hspace{0.5cm}{\longrightarrow}\hspace{0.5cm}
 \lim_{n{\rightarrow}\infty}\,{\Phi}_{Z_n}(t)\,=\,e^{it\mu}\,
 \lim_{n{\rightarrow}\infty}
   \left[ {\Phi}_{W}(t/n) \right]^n
\end{eqnarray*}
Now, since:
\begin{eqnarray*}
 {\Phi}_{W}(t/n)=
      1-\frac{1}{2}\,(\frac{t}{n})^2{\sigma}^2+{\Ocal}(t^k/n^k)=
      1\,-\,\frac{1}{2}\frac{t^2}{n}\,{\sigma}_{Z_n}^2+{\Ocal}(t^k/n^k)
\end{eqnarray*}
we have that:
\begin{eqnarray*}
 \lim_{n{\rightarrow}\infty}
   \left[ {\Phi}_{W}(t/n) \right]^n\,=\,
   {\rm lim}_{n{\rightarrow}\infty}
   \left[
   1\,-\,\frac{1}{2}\,
   \frac{t^2}{n}\,{\sigma}_{Z_n}^2\,+\,
   {\Ocal}(t^k/n^k) \right]^n\,=\,
   {\rm exp}\,\{-\,\frac{1}{2}t^2{\sigma}_{Z_n}^2\} 
\end{eqnarray*}
and therefore:
\begin{eqnarray*}
\lim_{n{\rightarrow}\infty}\,
 {\Phi}_{Z_n}(t)\,=\,
     e^{it\mu}\,
     e^{-\,\frac{1}{2}t^2{\sigma}^2/n}
\end{eqnarray*}
so,  $\lim_{n\rightarrow\infty}Z_n{\sim}N(x|{\mu},{\sigma}/\sqrt{n})$.

The first indications about the Central Limit Theorem are due to
A. De Moivre (1733). Later, C.F. Gauss and P.S. Laplace enunciated the
behavior in a general way and, in 1901, A. Lyapunov gave the first rigorous
demonstration under more restrictive conditions. The theorem in the
form we have presented here is due to Lindeberg and L\'evy and requires that
the random quantities $X_i$ are:
\begin{itemize}
\item[i)]    Statistically Independent;
\item[ii)]   have the same Distribution Function;
\item[iii)]  First and Second order moments exist 
             (i.e. they have mean value and variance).
\end{itemize}
In general, there is a set of Central Limit Theorems depending on which 
of the previous conditions are satisfied and justify the empirical fact
that many natural phenomena are adequately described by the Normal 
Distribution. To quote E. T. Whittaker and G. Robinson 
({\em Calculus of Observations}):

\vspace*{0.2cm}
  "Everybody believes in the exponential law of errors;\\
  \hspace*{0.6cm} The experimenters because they think that it can be
   proved by mathematics;\\
  \hspace*{0.6cm} and the mathematicians because they believe it has
   been established by
observation"

\smill
{\rayan}
\vspace{1.0cm}

\noindent
\footnotesize

\noindent
{\bf Example 1.29:}
From the limiting behavior of the Characteristic Function, show that:
\begin{itemize}
 \item[$\bullet$]
     If $X{\sim}Bi(r|n,p)$, in the limit $p \rightarrow 0$
     with $np$ constant tends to a Poisson Distribution $Po(r|{\mu}=np)$; 
 \item[$\bullet$]
     If $X{\sim}Bi(r|n,p)$, in the limit 
     $n \rightarrow \infty$ the standardized random quantity
   \begin{eqnarray*}
   Z\,=\,\frac{\textstyle X - {\mu}_{X}}
                   {\textstyle {\sigma}_{X}}\,=\,
              \frac{\textstyle X - n p}
                   {\textstyle \sqrt{n p q}}
      \hspace{0.5cm}\stackrel{n{\rightarrow}{\infty}}{\sim}\hspace{0.5cm}  
        N(x|0,1)
   \end{eqnarray*}
 \item[$\bullet$] If $X{\sim}Po(r|{\mu})$, then
   \begin{eqnarray*}
   Z\,=\,\frac{\textstyle X - {\mu}_{X}}
                   {\textstyle {\sigma}_{X}}\,=\,
              \frac{\textstyle X - {\mu}}
                   {\textstyle \sqrt{ \mu }}
                                   \nonumber 
      \hspace{0.5cm}\stackrel{\mu{\rightarrow}{\infty}}{\sim}\hspace{0.5cm}  
        N(x|0,1)
   \end{eqnarray*}
 \item[$\bullet$]
   $X{\sim}{\chi}^2(x|n)$, then
   $n{\rightarrow}\infty$ the standardized random quantity
   \begin{eqnarray*}
   Z\,=\,\frac{\textstyle X - {\mu}_{X}}
                   {\textstyle {\sigma}_{X}}\,=\,
              \frac{\textstyle X - {\nu}}
                   {\textstyle \sqrt{ 2 {\nu} }}
     \hspace{0.5cm}\stackrel{n{\rightarrow}{\infty}}{\sim}\hspace{0.5cm}   
                N(x|0,1)             
   \end{eqnarray*}
 \item[$\bullet$] The Student's Distribution $St(x|0,1,{\nu})$
    converges to $N(x|0,1)$ in the limit ${\nu}{\rightarrow}\infty$;
 \item[$\bullet$] The Snedecor's Distribution $Sn(x|{\nu}_1,{\nu}_2)$
    converges to ${\chi}^2(x|{\nu}_1)$ in the limit
    ${\nu}_2{\rightarrow}\infty$, to
    $St(x|0,1,{\nu}_2)$ in the limit ${\nu}_1{\rightarrow}\infty$ 
    and to $N(x|0,1)$ in the limit
    ${\nu}_1,{\nu}_2{\rightarrow}\infty$.
\end{itemize}
\vspace{0.35cm}
\noindent
{\bf Example 1.30:} It is interesting to see the Central Limit Theorem at work.
   For this, we have done a Monte Carlo sampling of the random quantity 
   $X{\sim}Un(x|0,1)$. The sampling distribution is shown in the figure
   1.2(1) and the following ones show the sample mean of 
   $n=2$ (fig. 1.2(2)), 5 (fig. 1.2(3)), 10 (fig. 1.2(4)),
   20 (fig. 1.2(5)) y 50 (fig. 1.2(6)) consecutive values. Each histogram
   has 500000 events and, as you can see, as $n$ grows the distribution
   ``looks'' more Normal. For
   $n=20$ and $n=50$ the Normal distribution is
   superimposed. 

   The same behavior is observed in figure 1.3 where we have generated
   a sequence of values from a parabolic distribution with minimum at $x=1$
   and support on ${\Omega}=[0,2]$.

   Last, figure 1.4 shows the results for a sampling from the  
   Cauchy Distribution $X{\sim}Ca(x|0,1)$.
   As you can see, the sampling averages
   follow a Cauchy Distribution regardless the value of $n$.
   For 
   $n=20$ and $n=50$ a Cauchy and a Normal
   distributions have been superimposed.
   In this case, since the Cauchy Distribution has no moments 
   the Central Limit Theorem does not apply.
\vspace{0.35cm}

\noindent
{\bf Example 1.31:} Let $\{X_i(w)\}_{i=1}^{\infty}$ be a sequence of 
independent random
quantities all with the same Distribution Function, mean value
$\mu$ and variance ${\sigma}^2$ and consider the random quantity
\begin{eqnarray*}
  Z(w)\,=\, \frac{\textstyle 1}{\textstyle n} \sum_{i=i}^{n}X_i(w)
\end{eqnarray*}
What is the value of $n$ such that the probability that $Z$ differs from
$\mu$ more than ${\epsilon}$ is less than ${\delta}=0.01$?

From the Central Limit Theorem we know that in the limit
${n{\rightarrow}\infty}$, $Z{\sim}N(x|{\mu},{\sigma}/\sqrt{n})$ so
we may consider that, for large $n$:
\begin{eqnarray*}
P(|Z-{\mu}|\geq {\epsilon})\,&=&\,
P({\mu}-{\epsilon} \geq Z \geq {\mu}+{\epsilon})\,{\simeq}\\
    &\simeq&\,
        \displaystyle {\int_{-\infty}^{{\mu}-{\epsilon}}\,
                        N(x|{\mu},{\sigma})\,dx}\,+\,
        \displaystyle {\int_{{\mu}+{\epsilon}}^{+\infty}\,
                        N(x|{\mu},{\sigma})\,dx}=
1-{\rm erf}\left[\frac{\sqrt{n}{\epsilon}}{{\sigma}\sqrt{2}}\right]<{\delta}
\end{eqnarray*}
For ${\delta}=0.01$ we have that 
\begin{eqnarray*}
\frac{\sqrt{n}{\epsilon}}{\sigma}\,\geq\,2.575  
\hspace{0.5cm}{\longrightarrow}\hspace{0.5cm}
 n\,\geq\,\frac{6.63\,{\sigma}^2}{{\epsilon}^2}
\end{eqnarray*}

\smill
{\raya}                   
\vspace{1.0cm}

\begin{figure}
\begin{center}
\mbox{\epsfig{file=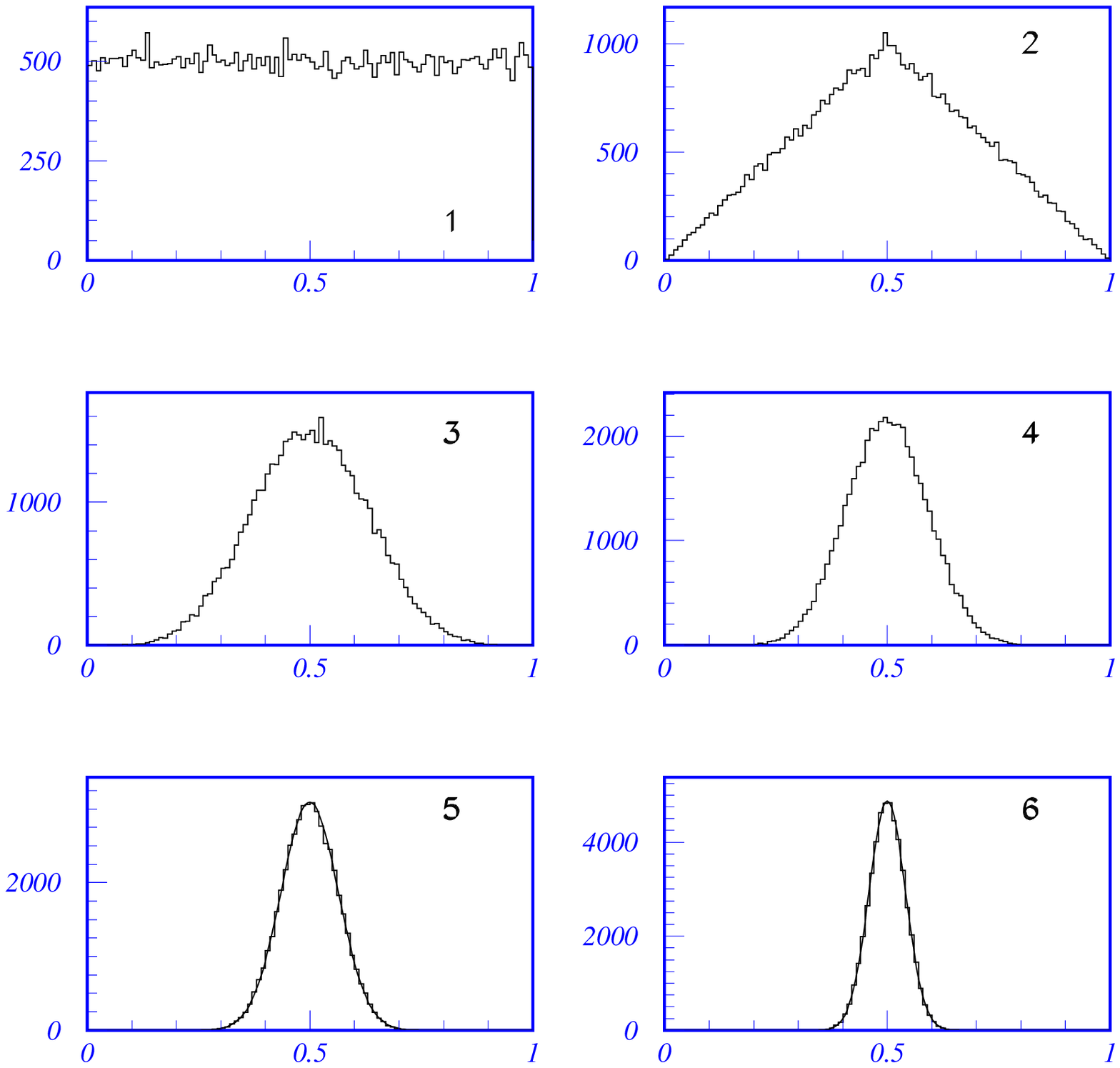,height=18cm,width=14.5cm}}

\footnotesize
{\bf Figure 1.2}.- Generated sample from $Un(x|0,1)$ 
(1) and sampling distribution of the mean of
2 (2), 5 (3), 10 (4), 20 (5) y 50 (6) generated values.

\end{center}
\end{figure}
\begin{figure}
\begin{center}

\mbox{\epsfig{file=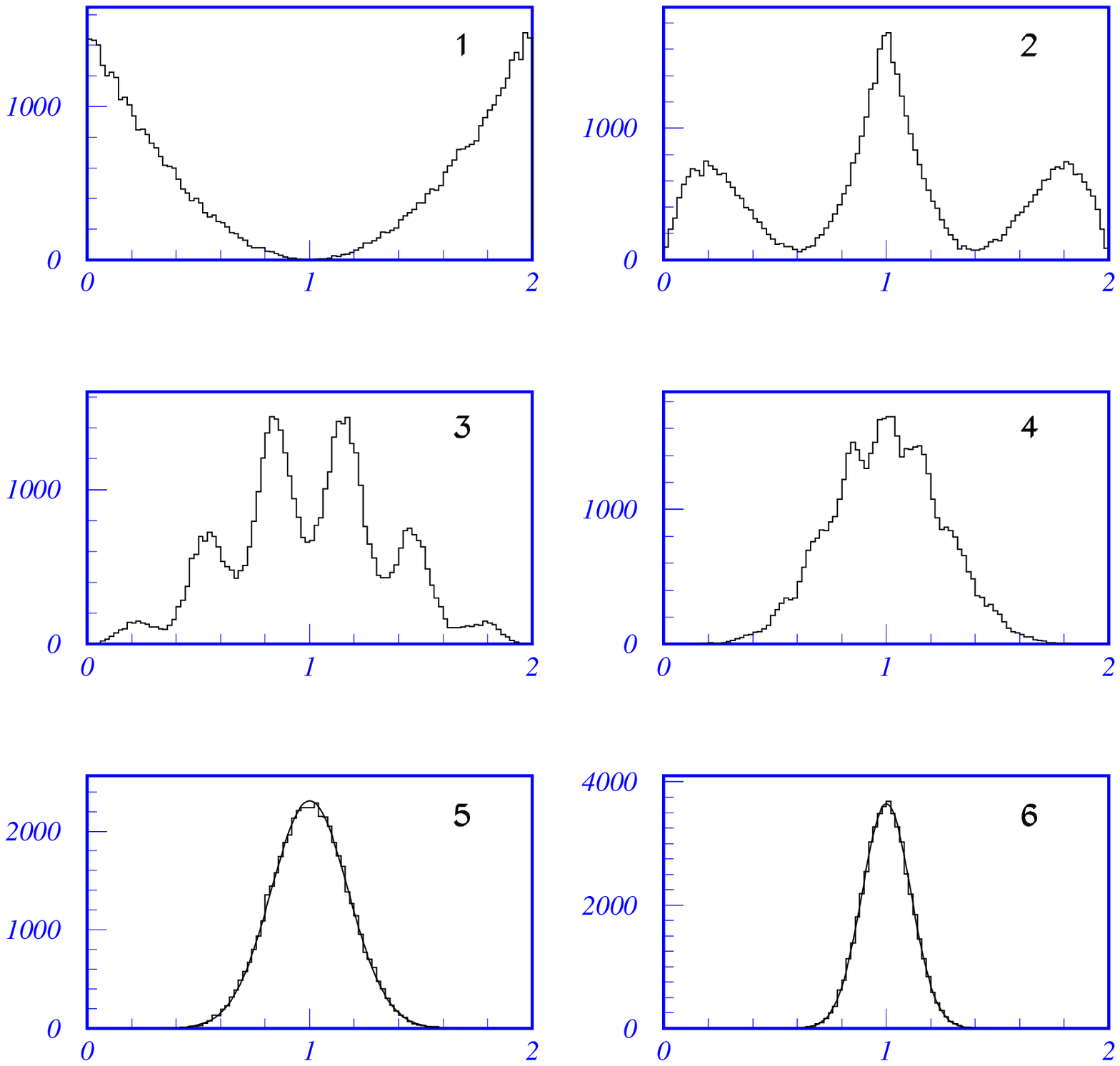,height=18cm,width=14.5cm}}

\footnotesize
{\bf Figure 1.3}.-  Generated sample from a parabolic distribution
with minimum at $x=1$ and support on ${\Omega}=[0,2]$
(1) and sampling distribution of the mean of
2 (2), 5 (3), 10 (4), 20 (5) y 50 (6) generated values.

\end{center}
\end{figure}
\begin{figure}
\begin{center}

\mbox{\epsfig{file=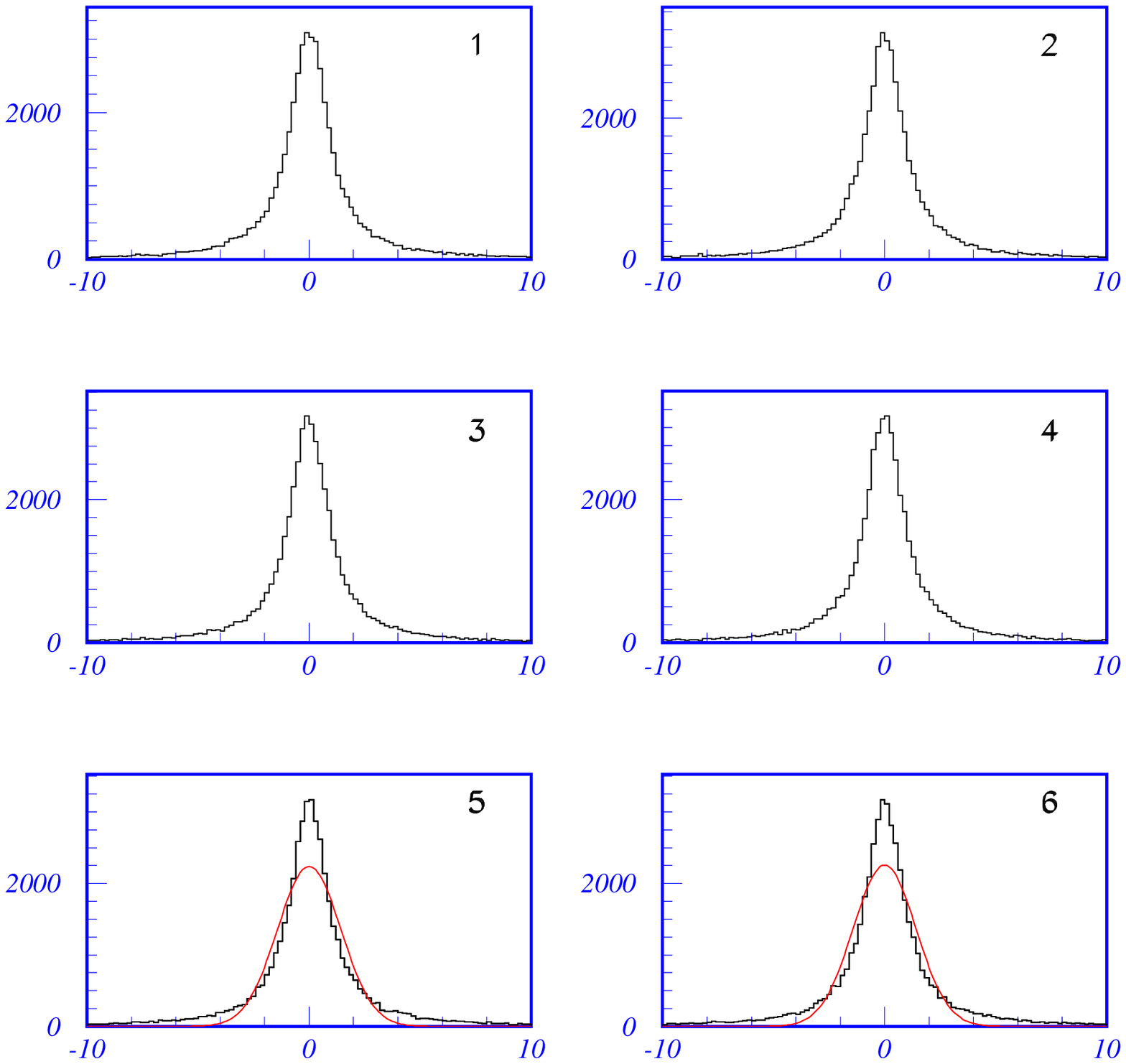,height=18cm,width=14.5cm}}

\footnotesize
{\bf Figure 1.4}.- Generated sample from a Cauchy distribution $Ca(x|0,1)$
(1) and sampling distribution of the mean of
2 (2), 5 (3), 10 (4), 20 (5) y 50 (6) generated values.

\end{center}
\end{figure}

\subsection{Convergence in $L_p$ Norm}
A sequence of random quantities
$\{X_n(w)\}_{n=1}^{\infty}$ converges to $X(w)$ in $L_p({\Rcal})$ 
$(p{\geq}1)$ norm iff,
\begin{eqnarray*}
X(w){\in}L_p({\Rcal}),\hspace{0.3cm} 
         X_n(w){\in}L_p({\Rcal})\,\,{\forall}n
\hspace{0.3cm}{\rm and} \hspace{0.3cm} 
     \lim_{n{\rightarrow}\infty}\,E[|X_n(w)-X(w)|^p]\,=\,0
\end{eqnarray*}       
that is, iff for any real ${\epsilon}>0$ there exists a natural 
$n_0({\epsilon})>0$ such that for all $n{\geq}n_0({\epsilon})$ it holds that
$E[|X_n(w)-X(w)|^p]<{\epsilon}$. In the particular case that $p=2$ it is
called {\sl Convergence in Quadratic Mean}.

From Chebyshev's Theorem
\begin{eqnarray*}
   P(|X_n(w)-X(w)| \geq {\alpha}^{1/p})\,\leq\,
        \frac{\textstyle E[(X_n(w)-X(w))^p]}
             {\textstyle \alpha }
\end{eqnarray*}
so, taking ${\alpha}={\epsilon}^p$, if there is convergence in 
$L_p({\Rcal})$ norm:
\begin{eqnarray}
   \lim_{n{\rightarrow}\infty}\,
   P(|{X_n(w)}-X(w)| \geq {\epsilon})\,\leq\,
   \lim_{n{\rightarrow}\infty}\,
        \frac{\textstyle E[(X_n(w)-X(w))^p]}
             {\textstyle {\epsilon}^p }\,=\,0
             \,\,\,\,\,{\forall}{\epsilon}>0
                                      \nonumber
\end{eqnarray}
and, in consequence, we have convergence in probability.

\subsection{Uniform Convergence}
In some cases, point-wise convergence of Distribution Functions
is not strong enough to guarantee
the desired behavior and we require a stronger type of convergence.
To some extent one may think that, more than a criteria of convergence, 
Uniform Convergence refers to the way in which it is achieved.
Point-wise convergence requires the existence
of an $n_0$ that may depend on ${\epsilon}$ and on $x$ so that
the condition $|f_n(x)-f(x)|<{\epsilon}$ for $n{\geq}n_0$ 
may be satisfied for some values of $x$ and not for others, for which a 
different value of $n_0$ is needed. The idea behind uniform convergence is
that we can find a value of $n_0$ for which the condition is satisfied 
regardless the value of $x$. Thus, we say that a sequence
$\{f_n(x)\}_{n=1}^{\infty}$ converges uniformly to $f(x)$ iff:
\begin{eqnarray*}
\forall{\epsilon}>0\,\,,\,\,\,{\exists}n_0{\in}{\Ncal}\,\,\,\,\,
{\rm such\,\,that}\,\,\,\,\,
  |f_n(x)-f(x)|<{\epsilon}\,\,\,\,\,\,\,\,\,\,
  \forall n>n_0\,\,\,\,\,{\rm and}\,\,\,\,\,
  \forall x
\end{eqnarray*}
or, in other words, iff:
\begin{eqnarray*}
{\rm sup}_x\,\,
  |f_n(x)-f(x)|\hspace{0.3cm}
\stackrel{n\rightarrow\infty}{\longrightarrow}\hspace{0.3cm}0
\end{eqnarray*}
Thus, it is a stronger type of convergence that implies point-wise convergence.
Intuitively, one may visualize the uniform convergence 
of $f_n(x)$ to $f(x)$ if one can draw a band $f(x){\pm}{\epsilon}$ 
that contains all $f_n(x)$ for any $n$ sufficiently large.
Look for instance at the sequence of functions
$f_n(x)=x(1+1/n)$ with $n=1,2,{\ldots}$ and $x{\in}{\Rcal}$. 
It is clear that converges point-wise to
$f(x)=x$ because ${\rm lim}_{n{\rightarrow}\infty}\,f_n(x)=f(x)$ for all
$x{\in}{\Rcal}$; that is, if we take
$n_0(x,{\epsilon})=x/{\epsilon}$, for all 
$n>n_0(x,{\epsilon})$ it is true that $|f_n(x)-f(x)|<{\epsilon}$ 
but for larger values of $x$ we need larger values of $n$. Thus, the
the convergence is not uniform because
\begin{eqnarray}
    {\rm sup}_x\,\,|f_n(x)-f(x)|\,=\,
    {\rm sup}_x\,\,|x/n|\,=\,{\infty}
    \,\,\,\,\,{\forall}n{\in}{\Ncal}
      \nonumber
\end{eqnarray}
Intuitively, for whatever small a given ${\epsilon}$ is, the band
$f(x){\pm}\epsilon=x{\pm}{\epsilon}$ does not contain
$f_n(x)$ for all $n$ sufficiently large. As a second example, take
$f_n(x)=x^n$ with $x{\in}(0,1)$. We have that 
${\lim}_{n{\rightarrow}{\infty}}f_n(x)=0$ but
${\rm sup}_x|g_n(x)|=1$ so the convergence is not uniform.
For the cases we shall be interested in, 
if a Distribution Function $F(x)$ is continuous and the sequence of
$\{F_n(x)\}_{n=1}^{\infty}$ {\sl converges in distribution} to $F(x)$
(i.e. point-wise) then it does {\sl uniformly} too.
An important case of uniform convergence is the (sometimes called 
{\sl Fundamental Theorem of Statistics}):

\vspace{0.5cm}
\noindent
$\bullet$ {\bf Glivenko-Cantelli Theorem} 
(V. Glivenko-F.P. Cantelli; 1933){\bf :} Consider the random quantity
 $X{\sim}F(x)$ and a statistically independent (essential point) 
 sampling of size $n$ 
 $\{x_1,x_2,{\ldots},x_n\}$. The {\sl empirical Distribution Function}
\begin{eqnarray*}
  F_n(x)\,=\,\frac{\textstyle 1}{\textstyle n}
           \,\sum_{i=1}^{n}\,{\mbox{\boldmath $1$}}_{(-{\infty},x]}(x_i)
\end{eqnarray*}
converges uniformly to $F(x)$; that is (Kolmogorov-Smirnov Statistic): 
 \begin{eqnarray*}
   {\rm lim}_{n{\rightarrow}\infty}\,
   {\rm sup}_x\, |F_n(x)\,-\,F(x)| \,=\,0
 \end{eqnarray*}

\vspace{0.5cm}
\noindent
{\rayan}                   
\vspace{0.35cm}
\footnotesize

Let's see the convergence in probability, in quadratic mean and,
in consequence, in distribution. 
For a fixed value $x=x_0$, $Y={\mbox{\boldmath $1$}}_{(-{\infty},x_0]}(X)$
is a random quantity that follows a Bernoulli
distribution with probability
 \begin{eqnarray*}
   p\,=\,P(Y\,=\,1)\,&=&\,
   P({\mbox{\boldmath $1$}}_{(-{\infty},x_0]}(x)\,=\,1)\,=\,
   P(X{\leq}x_0)\,=\,F(x_0) \\
   P(Y\,=\,0)\,&=&\,P({\mbox{\boldmath $1$}}_{(-{\infty},x_0]}(x)\,=\,0)\,=\,
   P(X>x_0)\,=\,
   1\,-\,F(x_0)
 \end{eqnarray*}
and Characteristic Function
 \begin{eqnarray*}
  {\Phi}_Y(t)\,=\,E[e^{itY}]\,=\,
               e^{it}\,p\,+\,(1\,-\,p)\,=\,
               e^{it}\,F(x_0)\,+\,(1\,-\,F(x_0))
 \end{eqnarray*}
Then, for a fixed value of $x$ we have for the random quantity
\begin{eqnarray*}
  Z_n(x)\,=\,\sum_{i=1}^{n}\,{\mbox{\boldmath $1$}}_{(-{\infty},x]}(x_i)
  \,=\,n\,F_n(x)
 \hspace{0.5cm}{\longrightarrow}\hspace{0.5cm}
{\Phi}_{Z_n}(t)\,=\,
         \left(  e^{it}\,F(x)\,+\,(1\,-\,F(x)) \right)^n
\end{eqnarray*}
and therefore $Z_n(x)\,{\sim}\,Bi(k|n,F(x))$ so, if
$W=n\,F_n(x)$, then
\begin{eqnarray*}
    P(W=k|n,F(x))\,=\,
       \left(  \begin{array}{c}
                 n \\ k
               \end{array}
     \right)\,F(x)^k\,(1\,-\,F(x))^{n-k}
\end{eqnarray*}
with
 \begin{eqnarray}
 && E[W]\,=\,n\,F(x)
  \hspace{2.2cm}{\longrightarrow}\hspace{0.5cm}
       E[F_n(x)]\,=\,F(x)   \nonumber \\
 && V[W]\,=\,n\,F(x)\,(1-F(x))
  \hspace{0.7cm}{\longrightarrow}\hspace{0.5cm}
       V[F_n(x)]\,=\,
       \frac{\textstyle 1}{\textstyle n}\,F(x)\,(1-F(x))
        \nonumber
 \end{eqnarray}
From Chebishev's Theorem
\begin{eqnarray*}
 P\left(|F_n(x)\,-\,F(x)|\,{\geq}\,{\epsilon}\right)\,{\leq}\,
       \frac{\textstyle 1}{\textstyle n\,{\epsilon}^2}\,F(x)\,(1-F(x))
 \end{eqnarray*}
and therefore
 \begin{eqnarray*}
 {\rm lim}_{n{\rightarrow}\infty}\,
  P[|F_n(x)\,-\,F(x)|\,{\geq}\,{\epsilon}]\,=\,0
 \,\,\,\,\, ;\,\,\, {\forall}  {\epsilon}>0
 \end{eqnarray*}
so the empirical Distribution Function $F_n(x)$ 
converges in probability to $F(x)$. In fact, since
 \begin{eqnarray*}
 {\rm lim}_{n{\rightarrow}\infty}\,
  E[|F_n(x)\,-\,F(x)|^2]\,=\,
 {\rm lim}_{n{\rightarrow}\infty}\,
       \frac{\textstyle F(x)\,(1-F(x))}{\textstyle n\,}\,=\,0
 \end{eqnarray*}
converges also in quadratic mean and therefore in distribution.

\vspace{0.35cm}
{\rayan}
\vspace{1.0cm}

\begin{figure}[t]
\begin{center}

\mbox{\epsfig{file=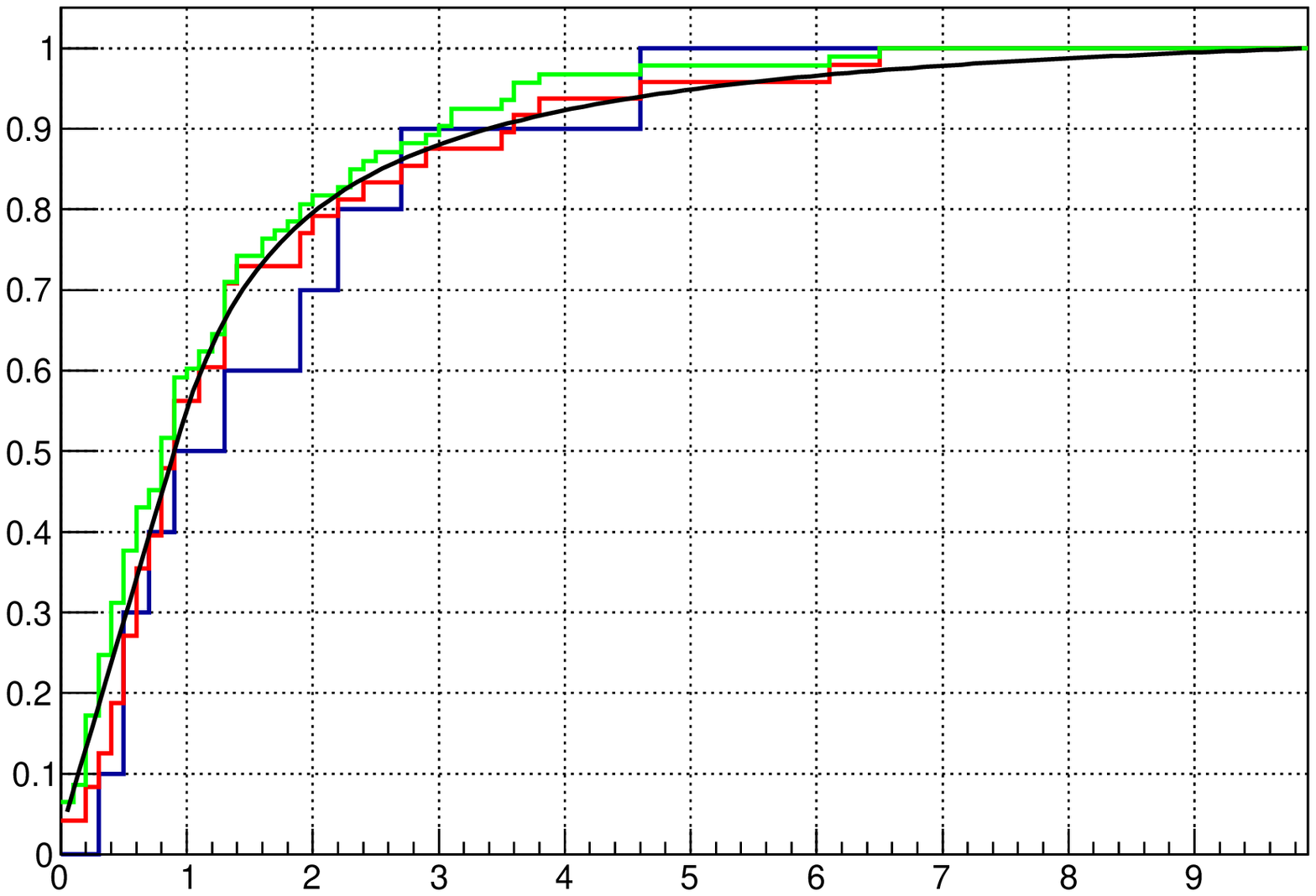,height=8cm,width=8cm}}

\footnotesize
{\bf Figure 1.5.}-  Empirical Distribution Function of example 1.32
for sample sizes 10 (blue), 50 (green) and 100 (red) together with the
Distribution Function (black).

\end{center}
\end{figure}
\vspace{0.35cm}

\noindent
{\bf Example 1.32:}
Let $X=X_1/X_2$ with $X_i{\sim}Un(x|0,1)$; $i=1,2$ and Distribution Function
\begin{eqnarray*}
  F(x)\,=\,\frac{\textstyle x}{\textstyle 2}
                {\mbox{\boldmath $1$}}_{(0,1]}(x)\,+\,
                  \left(1-\frac{x}{2}\right)
                {\mbox{\boldmath $1$}}_{(1,\infty)}(x)
\end{eqnarray*}
that you can get (exercise) from the Mellin Transform. This is
depicted in black in figure 1.5.
There are no moments for this distribution; that is $E[X^n]$ 
does not exist for $n{\geq}1$.  
We have done Monte Carlo samplings of size $n=10,\,50$ and 100 and
the corresponding empirical Distribution Functions
\begin{eqnarray*}
  F_n(x)\,=\,\frac{\textstyle 1}{\textstyle n}
           \,\sum_{i=1}^{n}\,{\mbox{\boldmath $1$}}_{(-{\infty},x]}(x_i)
\end{eqnarray*}
are shown in blue, read and green respectively. 

\smill

\vspace{0.5cm}
\noindent
{\rayan}                   
\vspace{0.35cm}
\footnotesize
\noindent
{\bf NOTE 4: Divergence of measures.} 

Consider the measurable space $(\Omega,{\Bcal}_{\Omega})$ and the
probability measures ${\lambda}$, ${\mu}<<{\lambda}$ and
${\nu}<<{\lambda}$. The Kullback's divergence between $\mu$ and $\nu$
is defined as (see Chapter 4)
\begin{eqnarray*}
  K(\mu,\nu)\,=\,\int_{\Omega}\frac{d\mu}{d\lambda}\log
\left(\frac{{d\mu}/{d\lambda}}{{d\nu}/{d\lambda}}\right)\,d\lambda
\end{eqnarray*}
and the Hellinger distance as
\begin{eqnarray*}
  d_H^2(\mu,\nu)\,=\,\frac{1}{2}\int_{\Omega}|\sqrt{{d\mu}/{d\lambda}}-
\sqrt{{d\nu}/{d\lambda}}|^2d\lambda
\end{eqnarray*}
For $\lambda$ Lebesgue measure, we can write
\begin{eqnarray*}
  K(p,q)\,=\,\int_{\Omega}p(x)\log
\left(\frac{p(x)}{q(x)}\right)\,dx
\hspace{1.cm}{\rm and}\hspace{1.cm}
 d_H^2(p,q)\,=\,1\,-\,\int_{\Omega}\sqrt{p(x)q(x)}dx
\end{eqnarray*}
The Kullback's divergence will be relevant for Chapter 2. It is left as an
exercise to show that the Normal density that best approximates in 
Kullback's sense a given density $p(x)$  is that with 
the same mean and variance (assuming they exist) and,
using the Calculus of Variations, that 
\begin{eqnarray*}
  p(x|\lambdabold)\,=\,f(x)\,{\rm exp}\left\{\sum_{i=0}^k\,\lambda_i\,
h_i(x)\right\}
\end{eqnarray*}
is the form (exponential family) that satisfies $k+1$ constraints
$\int h_i(x)p(x)dx=c_i<{\infty}$; $i=0,{\ldots},k$
with specified constants $\{c_j\}_{j=0}^k$ (
$c_0=1$ for $h_0(x)=1$) and best approximates a given density $q(x)$.
The Hellinger distance is a metric on the set of all probability 
measures on ${\Bcal}_{\Omega}$ and we shall make use of it, for instance, as 
a further check for convergence in Markov Chain Monte Carlo sampling.

{\rayan}                   
\vspace{1.0cm}
\smill


\newpage\null\thispagestyle{empty}\newpage

\rm

\footnotesize
\begin{tabular}{p{4.0cm}p{9.5cm}}
  &  "... some rule could be found, according to which we ought to
       estimate the chance that he probability for the happening of an
       event perfectly unknown, should lie between any two named
       degrees of probability, antecedently to any experiments made
       about it;..." 
\end{tabular}
\begin{flushright}
\emph{An Essay towards solving a Problem in the Doctrine of Chances.\\
      By the late Rev. Mr. Bayes ...}
\end{flushright}

\vspace*{1.cm}
\huge
\noindent
{\bf Lecture 2}

\vspace{0.5cm}
\noindent
\Huge
{\bf Bayesian Inference}
\vspace*{0.8cm}
\smill
\setcounter{section}{0}

\small

The goal of {\sl statistical inference} is to get information from
experimental observations about quantities (parameters, models,...) on which
we want to learn something, be them directly observable or not. Bayesian
inference \footnote{For a gentle reading on the subject see [Da03]} 
is based on the {\sl Bayes rule} and considers probability 
as a measure of the the degree of knowledge we have 
on the quantities of interest. Bayesian methods provide a framework
with enough freedom to analyze different models, as complex as needed,
using in a natural and conceptually simple way
all the information available from the experimental data within a
scheme that allows to understand the different steps
of the learning process:
\begin{itemize}
\item[1)] state the knowledge we have before we do the experiment;
\item[2)] how the knowledge is modified after the data is taken;
\item[3)] how to incorporate new experimental results.
\item[4)] predict what shall we expect in a future
          experiment from the knowledge acquired.
\end{itemize}

It was Sir R.A Fisher, one of the greatest statisticians ever, who said that
"The Theory of Inverse Probability (that is how Bayesianism was called at 
the beginning of the XX$^{th}$ century) is founded upon an error and must be 
wholly rejected"  although, as time went by, he became a little more
acquiescent with Bayesianism. You will see that
that Bayesianism is great, rational, coherent, conceptually simple,...
"even useful",... and worth to, at least, take a look at it and at the more 
detailed references on the subject given along the section. 
At the end, to quote Lindley, "Inside every non-Bayesian there is a 
Bayesian struggling to get out". 
For a  more classical approach to Statistical Inference see [Ja06] where
most of what you will need in Experimental Physics is covered in detail.

\section{\LARGE \bf Elements of Parametric Inference}

Consider an experiment designed to provide information about the set of 
parameters
$\thetabold=\{{\theta}_1,{\ldots},{\theta}_k\}$ 
${\in}\,{\Theta}\subseteq R^{k}$ and
whose realization results in the random sample
$\xbold=\{x_1,x_2,{\ldots},x_n\}$. The inferential process entails:
   \begin{itemize}
     \item[1)] Specification of the probabilistic model for the random
               quantities of interest; that is, state the joint density:
               \begin{eqnarray*}
                p({\thetabold},{\xbold})\,=\,
                p({\theta}_1,{\theta}_2,{\ldots},{\theta}_k,
                  x_1,x_2,{\ldots},x_n);
                  \hspace{0.5cm}
                  \thetabold={\in}\,{\Theta}\subseteq R^{k};
                  \hspace{0.5cm}
                  \xbold{\in}\,X
               \end{eqnarray*}
     \item[2)] Conditioning the observed data $(\xbold)$ to
               the parameters $(\thetabold)$ of the model:
               \begin{eqnarray*}
                 p({\thetabold},{\xbold})\,=\,
                 p({\xbold}|{\thetabold})\,p({\thetabold})
               \end{eqnarray*}
     \item[3)] Last, since $
                 p({\thetabold},{\xbold})\,=\,
                    p({\xbold}|{\thetabold})\,p({\thetabold})\,=\,
                    p({\thetabold}|{\xbold})\,p({\xbold}) $
               and
               \begin{eqnarray*}
                 p(\xbold)\,=\,
                 \int_{\Theta}   p(\xbold,\thetabold)\,d{\thetabold}\,=\,
                 \int_{\Theta}   p(\xbold|\thetabold)\,p(\thetabold)\,
                 d{\thetabold}
               \end{eqnarray*}
               we have ({\sl Bayes Rule}) that:

\begin{eqnarray*}
    p({\thetabold}|{\xbold})\,=\,
        \frac {\textstyle  p(\xbold|\thetabold)\,p(\thetabold) }
              {\textstyle
             \displaystyle  \int_{\Theta}
            p(\xbold|\thetabold)\,p(\thetabold) d\thetabold }
\end{eqnarray*}

\end{itemize}

This is the basic equation for parametric inference. The integral of the
denominator does not depend on the parameters $({\thetabold})$ of interest; 
is just a normalization factor so we can write in a general way;
\begin{eqnarray*}
       p({\thetabold}|{\xbold})\,\propto\,
       p(\xbold|\thetabold)\,p(\thetabold) 
\end{eqnarray*}

Let's see these elements in detail:

\vspace{0.5cm}
\begin{tabular}{p{1.5cm}p{11.5cm}}
  $p({\thetabold}|\xbold):$ & This is the {\sl Posterior Distribution} 
     that quantifies
     the knowledge we have on the parameters of interest $\thetabold$ 
     conditioned to the observed data $\xbold$ (that is, after the
     experiment has been done) and will allow
     to perform inferences about the parameters; \\
              &                         \\
\end{tabular}

\vspace{0.5cm}
\begin{tabular}{p{1.5cm}p{11.5cm}}
  $p(\xbold|{\thetabold}):$ & The {\sl Likelihood}; the
  sampling distribution considered
  as a function of the parameters $\thetabold$ for the 
  {\sl fixed} values (already observed) $\xbold$. Usually, it is written
  as
  ${\ell}({\thetabold};{\xbold})$
  to stress the fact that it is a function of the parameters.
              \\
\end{tabular}

\begin{tabular}{p{1.5cm}p{11.5cm}}
  & 
  The experimental results modify the prior knowledge we have on the parameters
  ${\thetabold}$ only through the likelihood so,
  for the inferential process, we can consider
  the likelihood function defined up to multiplicative factors provided they
  do not depend on the parameters.
   \\
              &                         \\
\end{tabular}

\vspace{0.5cm}
\begin{tabular}{p{1.5cm}p{11.5cm}}
  $p({\thetabold})\,:$ & This is {\sl a reference function}, independent of 
  the results of the experiment, that quantifies or expresses, in a sense
  to be discussed later, the knowledge we have on the parameters $\thetabold$
  {\sl before} the experiment is done. It is termed {\sl Prior Density} 
  although, in many cases, 
  it is an improper function and therefore not a probability density.
     \\
              &                         \\
\end{tabular}

\section{\LARGE \bf Exchangeable sequences}

The inferential process to obtain information about a set of parameters
 $\thetabold{\in}\,{\Theta}$ of a model $X{\sim}p(x|\thetabold)$
with $X{\in}\Omega_X$
is based on the realization of an experiment $e(1)$ that provides an
observation $\{x_1\}$. The $n-$fold repetition of the 
experiment under the same conditions, $e(n)$, will provide the random
sample $\xbold=\{x_1,x_2,{\ldots},x_n\}$ and this can be considered as
a draw of the $n$-dimensional random quantity $\Xbold=(X_1,X_2,\ldots,X_n)$
where each $X_i{\sim}p(x|\thetabold)$. 

In Classical Statistics, the inferential process makes extensive use of the 
idea that the observed sample is originated from a sequence of
{\sl independent and identically distributed} (iid) random quantities while
Bayesian Inference rests on the less restrictive idea 
of {\sl exchangeability} [Be96].
An infinite sequence of random quantities 
$\{X_i\}_{i=1}^{\infty}$ is said to be {\sl exchangeable} if {\sl any} 
finite sub-sequence $\{X_1,X_2,{\ldots},X_n\}$ is {\sl exchangeable}; 
that is, if the joint density $p(x_1,x_2,{\ldots},x_n)$ is invariant 
under {\sl any} permutation of the indices.

The hypothesis of {\sl exchangeability} assumes a symmetry of the experimental
observations $\{x_1,x_2,$ ${\ldots},x_n\}$ 
such that the subscripts which identify
a particular observation (for instance the order in which they appear) are 
irrelevant for the inferences. 
Clearly, if $\{X_1,X_2,{\ldots}X_n\}$  are iid then the
conditional joint density can be expressed as:
\begin{eqnarray*}
    p(x_1,x_2,{\ldots},x_n)\,=\,
    \prod_{i=1}^{n}\,p(x_i)
    \nonumber
\end{eqnarray*}
and therefore, since the product is invariant to reordering, is an 
{\sl exchangeable} sequence. 
The converse is not necessarily true
\footnote{It is easy to 
check for instance that if $X_0$ is a non-trivial random quantity independent 
of the $X_i$,
the sequence $\{X_0+X_1,X_0+X_2,{\ldots}X_0+X_n\}$ is exchangeable but not
iid.} so
the hypothesis of exchangeability is weaker than the hypothesis of 
independence.  
Now, if $\{X_i\}_{i=1}^{\infty}$ is an exchangeable sequence 
of real-valued random
quantities it can be shown that, for {\sl any} finite subset,
there exists a parameter $\thetabold{\in}{\Theta}$, a
parametric model $p(x|\thetabold)$ and measure $d{\mu}({\thetabold})$ such that
\footnote {This is referred as {\sl De Finetti's Theorem} after B. de Finetti
(1930s) and was generalized by E. Hewitt and L.J. Savage in the 1950s.
See [Be94].}:
\begin{eqnarray*}
    p(x_1,x_2,{\ldots},x_n)\,=\,\int_{\Theta}
    \prod_{i=1}^{n}\,p(x_i|\thetabold)\,d{\mu}(\thetabold)
\end{eqnarray*}
Thus, any finite sequence of exchangeable observations 
is described by a model $p(x|\thetabold)$ and, if 
$d{\mu}(\thetabold)=p(\thetabold)d\thetabold$, there is
a prior density 
$p(\thetabold)$ that we may consider as describing the available 
information on the parameter $\thetabold$ before the experiment is done.
This justifies and, in fact, leads to the
Bayesian approach in which, by formally applying Bayes Theorem
\begin{eqnarray*}
    p(\xbold,\thetabold)\,=\,    
    p(\xbold|\thetabold)\,p(\thetabold)\,=\,
    p(\thetabold|\xbold)\,p(\xbold)
\end{eqnarray*}
we obtain the
{\sl posterior density} $p(\thetabold|\xbold)$ that accounts for the 
degree of knowledge we have on the parameter after the experiment has been 
performed.
Note that the random quantities of the exchangeable sequence
$\{X_1,X_2,{\ldots},X_n\}$ are
{\sl conditionally independent given} $\thetabold$ 
{\sl but not iid} because
\begin{eqnarray*}
    p(x_j)\,=\,
   \int_{\Theta}\,p(x_i|\thetabold)\, d{\mu}(\thetabold)\,
   \left(\prod_{i(\neq j)=1}^{n}\int_{{\Omega}_X}p(x_i|\thetabold)\,dx_i\right)
\end{eqnarray*}
and
\begin{eqnarray*}
    p(x_1,x_2,{\ldots},x_n)\,{\neq}\,
    \prod_{i=1}^{n}\,p(x_i)
    \nonumber
\end{eqnarray*}

There are situations for which 
the hypothesis of exchangeability can not be assumed to hold.
That is the case, for instance, when the data collected by
an experiment depends on the running conditions that may be different
for different periods of time, for
data provided by two different experiments with different acceptances, 
selection criteria, efficiencies,... or the same medical treatment when applied
to individuals from different environments, sex, ethnic groups,...
In these cases, we shall have different {\sl units of observation}
and it may be more sound to assume {\sl partial exchangeability} within
each unit (data taking periods, detectors, hospitals,...) and
design a {\sl hierarchical structure} with 
parameters that account for the
relevant information from each unit analyzing all the data in a  more
global framework.

\newpage
\noindent
{\rayan}                   
\vspace{0.35cm}
\footnotesize
\noindent
{\bf NOTE 5:} Suppose that we have a 
parametric model $p_1(x|\thetabold)$ and the exchangeable sample 
$\xbold_1=\{x_1,x_2,$ ${\ldots},x_n\}$ provided by the experiment $e_1(n)$.
The inferences 
on the parameters $\thetabold$ will be drawn from the posterior density
  $p(\thetabold|\xbold_1){\propto}
  p_1(\xbold_1|\thetabold)p(\thetabold)$.
Now, we do a second experiment $e_2(m)$, statistically independent of the
first, that provides the exchangeable sample
$\xbold_2=\{x_{n+1},x_{n+2},{\ldots},x_{n+m}\}$ 
from the model $p_2(x|\thetabold)$.
It is sound to take as prior density
for this second experiment the posterior of the first including therefore
the information that we already have about $\thetabold$ so
\begin{eqnarray*}
  p({\thetabold}|\xbold_2){\propto}
  p_2(\xbold_2|\thetabold)
  p(\thetabold|\xbold_1){\propto}
  p_2(\xbold_2|\thetabold)
  p_1(\xbold_1|\thetabold)p(\thetabold).
\end{eqnarray*}
Being the two experiments statistically independent and their sequences
exchangeable, if they have the same sampling distribution
$p(x|\thetabold)$ we have that
$p_1(\xbold_1|\thetabold) p_2(\xbold_2|\thetabold)=p(\xbold|\thetabold)$
where $\xbold=\{\xbold_1,\xbold_2\}=\{x_1,\ldots,x_n,x_{n+1},\ldots,x_{n+m}\}$
and therefore
$p(\thetabold|\xbold_2)\,{\propto}\,p(\xbold|\thetabold)\, p(\thetabold)$.
Thus, 
the knowledge we have on $\thetabold$ including the
information provided by the experiments $e_1(n)$ 
and $e_2(m)$ is determined by the likelihood
function $p(\xbold|\thetabold)$ and,
in consequence, under the aforementioned conditions
the realization of $e_1(n)$ first and $e_2(m)$ after is equivalent,
from the inferential point of view,
to the realization of the experiment $e(n+m)$.

{\rayan}                   
\vspace{1.0cm}
\small

\section{\LARGE \bf Predictive Inference}

Consider the realization of the experiment $e_1(n)$ that provides the sample
$\xbold=\{x_1,x_2,{\ldots},x_n\}$ drawn from the model
$p(x|\thetabold)$. Inferences about 
$\thetabold{\in}{\Theta}$ are determined by the posterior density
\begin{eqnarray*}
  p(\mbox{\boldmath ${\theta}$}|\xbold)\,
  &{\propto}&\,
  p(\xbold|\mbox{\boldmath ${\theta}$})\,\pi(\mbox{\boldmath ${\theta}$})
\end{eqnarray*}
Now suppose that, under the same model and the same experimental conditions,
we think about doing a new independent experiment $e_2(m)$. 
What will be the distribution of the random sample  
$\ybold=\{y_{1},y_{2},{\ldots},y_{m}\}$ not yet observed?
Consider the experiment $e(n+m)$ and the sampling density
\begin{eqnarray*}
p(\thetabold,\xbold,\ybold)\,=\,
p(\xbold,\ybold|\thetabold)\,
\pi(\thetabold)
\end{eqnarray*}
Since both experiments are independent and iid, we have the joint density
\begin{eqnarray*}
p(\xbold,\ybold|\thetabold)\,=\,p(\xbold|\thetabold)\,p(\ybold|\thetabold)
\hspace{1.cm}{\longrightarrow}\hspace{1.cm}
p(\thetabold,\xbold,\ybold)\,=\,
p(\xbold|\thetabold)\,p(\ybold|\thetabold)\,
\pi(\thetabold)
\end{eqnarray*}
and integrating the parameter
$\thetabold\,{\in}\,{\Theta}$:
\begin{eqnarray*}
p(\ybold,\xbold)\,=\,p(\ybold|\xbold)\,p(\xbold)\,=\,
  \int_{\Theta}
p(\ybold|\thetabold)\,p(\xbold|\thetabold)\pi(\thetabold)
  d\thetabold\,=\,p(\xbold)\,
\int_{\Theta}
p(\ybold|\thetabold)\,p(\thetabold|\xbold)\,  d\thetabold
\end{eqnarray*}
Thus, we have that
\begin{eqnarray*}
    p(\ybold|\xbold)\,=\,
       \int_{\Theta}
       p(\ybold|\thetabold)\,p(\thetabold|\xbold)\,  d\thetabold
\end{eqnarray*}

This is the basic expression for the {\sl predictive inference} and allows
us to predict the results $\ybold$ of a future experiment from the 
results $\xbold$ observed in a previous experiment within the same parametric
model. Note that $p(\ybold|\xbold)$ is the density of the quantities not yet
observed conditioned to the observed sample. Thus, even though the experiments
$e(\ybold)$ and $e(\xbold)$ are statistically independent, the realization
of the first one $(e(\xbold))$ modifies the knowledge we have on the 
parameters $\thetabold$ of the model and therefore affect the prediction
on future experiments for, if we do not consider the results of the first
experiment or just don't do it, the predictive distribution for
$e(\ybold)$ would be
\begin{eqnarray*}
 p(\ybold)\,=\,
\int_{\Theta}
p(\ybold|\thetabold)\,
  \pi(\thetabold)\,d\thetabold
\end{eqnarray*}
It is then clear from the expression of {\sl predictive inference} that
in practice it is equivalent to consider as prior density for the second 
experiment the proper density
$\pi(\thetabold)=p(\thetabold|\xbold)$. If the first experiment provides
very little information on the parameters, then
$p(\thetabold|\xbold)\,{\simeq}\,\pi(\thetabold)$ and
\begin{eqnarray*}
p(\ybold|\xbold)\,\simeq\,
\int_{\Theta} p(\ybold|\thetabold)\,\pi(\thetabold)\,d\thetabold\,\simeq\,
p(\ybold)
\end{eqnarray*}
On the other hand, if after the first experiment we know the parameters
with high accuracy then, in distributional sense, 
$<p(\thetabold|\xbold),\cdot>\simeq<\delta(\thetabold_0),\cdot>$ and
\begin{eqnarray*}
p(\ybold|\xbold)\,\simeq\,
<\delta(\thetabold_0),p(\ybold|\thetabold)>\,=\,
p(\ybold|\theta_0)
\end{eqnarray*}

\section{\LARGE \bf Sufficient Statistics}

Consider $m$ random quantities
$\{\xbold_1,\xbold_2,{\ldots},\xbold_m\}$ 
that take values in
${\Omega}_1{\times}{\ldots}{\times}{\Omega}_m$. A random vector
\begin{eqnarray*}
\mbox{\boldmath ${t}$}\,:\,
 {\Omega}_1{\times}{\ldots}{\times}{\Omega}_m\,{\longrightarrow}\,
{\Rcal}^{k(m)}
    \nonumber
\end{eqnarray*}
whose $k(m){\leq}m$ components are functions of the random quantities
$\{\xbold_1,\xbold_2,{\ldots},\xbold_m\}$ is a 
$k(m)-${\sl di\-men\-sional statistic}. 
The practical interest is in the existence of statistics that contain all
the relevant information about the parameters so we don't
have to work with the whole sample and simplify considerably the expressions.
Thus, of special relevance are the {\sl sufficient statistics}.
Given the model $p(x_1,x_2,{\ldots},x_n|\thetabold)$, the set of statistics
$\tbold=\mbox{\boldmath ${t}$}(x_1,{\ldots},x_m)$ is {\sl sufficient} for
$\thetabold$ if, and only if, $\forall m{\geq}1$ and any prior distribution
${\pi}(\thetabold)$ it holds that
\begin{eqnarray*}
p(\mbox{\boldmath ${\theta}$}|x_1,x_2,{\ldots},x_m)\,=\,
p(\mbox{\boldmath ${\theta}$}|\mbox{\boldmath ${t}$})
    \nonumber
\end{eqnarray*}
Since the data act in the Bayes formula only through the likelihood,
it is clear that to specify the posterior density of
$\thetabold$ we can consider
\begin{eqnarray*}
p(\mbox{\boldmath ${\theta}$}|x_1,x_2,{\ldots},x_m)\,=\,
p(\mbox{\boldmath ${\theta}$}|\mbox{\boldmath ${t}$})\,
{\propto}\,p(\mbox{\boldmath ${t}$}|\mbox{\boldmath ${\theta}$})\,
{\pi}(\mbox{\boldmath ${\theta}$})
    \nonumber
\end{eqnarray*}
and all other aspects of the data but $\mbox{\boldmath ${t}$}$ are irrelevant.
It is obvious however that 
$\tbold=\{x_1,{\ldots},x_m\}$ 
is sufficient and gives no simplification in the modeling.
For this we should have 
$k(m)={\rm dim}(\tbold)<m$
({\sl minimal sufficient statistics})
and, in the ideal case, we would like that 
$k(m)=k$ does not depend on $m$. Except some irregular cases, 
the only distributions that admit a fixed number of sufficient statistics
independently of the sample size
(that is, $k(m)=k<m$ $\forall m$) are those that belong to the 
exponential family.

\vspace{0.5cm}
\noindent
{\raya}                   
\vspace{0.35cm}
\footnotesize

\noindent
{\bf Example 2.1:}
\vspace{0.3cm}

\noindent
{\bf 1)} Consider the exponential model $X{\sim}Ex(x|{\theta})$: 
   and the iid experiment $e(m)$ that provides the sample
$\xbold=\{x_1,{\ldots},x_m\}$. The likelihood function is:
   \begin{eqnarray*}
      p(\xbold|{\theta})\,=\,{\theta}^m\,
      e^{-{\theta}\,(x_1+{\ldots}+x_m)}\,=\,
         {\theta}^{t_1}\,
      e^{-{\theta}\,{t_2}}
                \nonumber 
   \end{eqnarray*}
and therefore we have the sufficient statistic
$\tbold=(m,\,\sum_{i=1}^{m}x_i):
    {\Omega}_1{\times}{\ldots}{\times}{\Omega}_m{\longrightarrow}
{\Rcal}^{k(m)=2}$
\vspace{0.3cm}

\noindent
{\bf 2)} Consider the Normal model $X{\sim}N(x|{\mu},{\sigma})$ and the iid
   experiment $e(m)$ again with $\xbold=\{x_1,{\ldots},x_m\}$.
The likelihood function is:
\begin{eqnarray*}
      p(\xbold|{\mu},{\sigma})\,&{\propto}&\,{\sigma}^{-m}\,
      {\rm exp}\left\{-\frac{1}{2\sigma^2}\sum_{i=1}^{m}(x_i\,-\,{\mu})^2
       \right\}\,=\,{\sigma}^{-t_1}\,
      {\rm exp}\left\{-\frac{1}{2\sigma^2}(t_3\,-\,2\,{\mu}\,t_2\,+\,
                      {\mu}^2\,t_1)\right\}
\end{eqnarray*}
and
$\tbold\,=\,(m,\,\sum_{i=1}^{m}\,x_{i},\,\sum_{i=1}^{m}\,x_{i}^2)\,:
    {\Omega}_1{\times}{\ldots}{\times}{\Omega}_m\,{\longrightarrow}\,
{\Rcal}^{k(m)=3}$
a sufficient statistic. Usually we shall consider
$\tbold=\{m,\,\overline{x},s^2\}$ with 
\begin{eqnarray*}
\overline{x}\,=\,\frac{1}{m}\sum_{i=1}^{m}x_i
\hspace{1.cm}{\rm and}\hspace{1.cm}
s^2\,=\,\frac{1}{m}\sum_{i=1}^{m}(x_i-\overline{x})^2
\end{eqnarray*}
the sample mean and the sample variance. 
Inferences on the parameters ${\mu}$ and 
${\sigma}$ will depend on  $\tbold$ and all other aspects of the data 
are irrelevant.
\vspace{0.3cm}

\noindent
{\bf 3)} Consider the Uniform model $X{\sim}Un(x|0,{\theta})$ and the iid
sampling $\{x_1,x_2,{\ldots},x_m\}$. Then \newline \noindent
$\tbold\,=\,(m,\,{\rm max}\{x_{i},\,i=1,\ldots m\})\,:
    {\Omega}_1{\times}{\ldots}{\times}{\Omega}_m\,{\longrightarrow}\,
{\Rcal}^{k(m)=2}$
is a sufficient statistic for $\theta$.

\small
{\raya}                   
\vspace{1.0cm}

\small
\section{\LARGE \bf Exponential Family}

A probability density $p(x|\thetabold)$, with $x{\in}{\Omega}_X$ and
$\thetabold{\in}{\Theta}\,{\subseteq}\,{\Rcal}^k$ 
belongs to the {\sl k-parameter exponential family} if it has the form:
\begin{eqnarray*}
  p(x|\thetabold)\,=\,f(x)\,g(\thetabold)\,
{\rm exp}\left\{\sum_{i=1}^k\,c_i\,{\phi}_i(\thetabold)\,h_i(x)\right\}
\end{eqnarray*}
with
\begin{eqnarray*}
g(\thetabold)^{-1}\,=\,\int_{{\Omega}_x}\,
  f(x)\,\prod_{i=1}^k
{\rm exp}\left\{c_i\,{\phi}_i(\thetabold)\,h_i(x)\right\}\,
dx\,{\leq}\,
{\infty}
\end{eqnarray*}
The family is called {\sl regular} if 
${\rm supp} \{X\}$ is independent of $\thetabold$; {\sl irregular} otherwise.

If $\xbold=\{ x_1,x_2,{\ldots},x_n \}$
is an exchangeable random sampling from the 
k-parameter regular exponential family, then 
\begin{eqnarray*}
  p(\xbold|\thetabold)\,=\,
\left[{\prod}_{i=1}^{n}\,f(x_i)\right]
\,\left[g(\mbox{\boldmath ${\theta}$})\right]^n\,
{\rm exp}\left\{\sum_{i=1}^k\,c_i\,{\phi}_i(\mbox{\boldmath ${\theta}$})\,
 ( \sum_{j=1}^n\,h_i(x_j) )\right\}
    \nonumber
\end{eqnarray*}
and therefore
 $ \tbold(\xbold)\,=\,\left\{n,
 \sum_{i=1}^n\,h_1(x_i),\,{\ldots}\,
 \sum_{i=1}^n\,h_k(x_i) \right\}$
will be a set of {\sl sufficient statistics}.

\vspace{0.5cm}
\noindent
{\raya}                   
\vspace{0.35cm}
\footnotesize

\noindent
{\bf Example 2.2:} Several distributions of interest, like Poisson and 
Binomial, belong to the exponential family:
\begin{itemize}
\item[1)] Poisson $Po(n|{\mu})$:
   $P(n|{\mu})\,=\,\frac{\textstyle e^{-{\mu}}\,{\mu}^n}
                         {\textstyle {\Gamma}(n+1)}\,=\,
                    \frac{\textstyle e^{-({\mu}-n{\rm ln}\,{\mu})}}
                         {\textstyle {\Gamma}(n+1)}$

\item[2)] Binomial $Bi(n|N,{\theta})$:
  $P(n|N,{\theta})\,=\,
       \left(  \begin{array}{c}
                 N \\ n
               \end{array}
     \right)\,{\theta}^n\,(1-{\theta})^{N-n}\,=\,
       \left(  \begin{array}{c}
                 N \\ n
               \end{array}
     \right)\,e^{n\,{\rm ln}{\theta}\,+\,(N-n)\,{\rm ln}\,(1-{\theta})}$

\end{itemize}
However, the Cauchy $Ca(x|{\alpha},{\beta})$ distribution, for instance,
does not because
   \begin{eqnarray*}
  p(x_1,{\ldots},x_m|{\alpha},{\beta})\,&{\propto}&\,
  \prod_{i=1}^{n}\,\left(1\,+\,{\beta}(x_i-{\alpha})^2\right)^{-1}\,=\,
   {\rm exp}\left\{ \sum_{i=1}^{m}{\rm log}(1\,+\,{\beta}(x_i-{\alpha})^2)
            \right\}
      \nonumber
   \end{eqnarray*}
can not be expressed as the exponential family form. In consequence,
there are no sufficient {\sl minimal} statistics
(in other words $\tbold=\{n,x_1,{\ldots},x_n\}$ is the sufficient statistic)
and we will have to work with the whole sample.

\small
{\raya}                   
\vspace{1.0cm}

\section{\LARGE \bf Prior functions}

In the {\sl Bayes rule}, 
$p({\thetabold}|{\xbold})\propto p(\xbold|\thetabold)\,p(\thetabold)$,
the {\sl prior function} $p(\thetabold)$ represents the knowledge
({\sl degree of credibility}) that we have about the parameters 
before the experiment is done and it is a necessary element to obtain
the {\sl posterior density} $p({\thetabold}|{\xbold})$ from which we
shall make inferences.
If we have faithful information on them before we do the experiment, it is
reasonable to incorporate that in the specification of the prior density
({\sl informative prior})
so the new data will provide additional information that will update 
and improve our knowledge. The specific form of the prior can be
motivated, for instance, by the results obtained in previous experiments.
However, it is usual that before we do the experiment, either we have a vague 
knowledge of the parameters compared to what we expect to get from the 
experiment or simply we do not want to include previous results to perform
an independent analysis. In this case, all the new information will be
contained in the likelihood function $p(\xbold|\thetabold)$ of the experiment
and the prior density ({\sl non-informative prior})
will be merely a mathematical element needed for the
inferential process. Being this the case, we expect that the whole weight
of the inferences rests on the likelihood and the prior function
has the smallest possible influence on them. To learn something from the
experiment it is then desirable to have a situation like the one shown in
fig. 1.1 where the posterior distribution $p(\thetabold|\xbold)$ is dominated
by the likelihood function. Otherwise, the experiment will provide little 
information compared to the one we had before and, unless our previous 
knowledge is based on suspicious observations, it will be wise to design 
a better experiment.

A considerable amount of effort has been put to obtain reasonable
{\sl non-informative priors} that can be used as a standard reference function
for the Bayes rule. Clearly, {\sl non-informative} is somewhat misleading 
because we are never in a state of absolute ignorance about the parameters and
the specification of a mathematical model for the process 
assumes some knowledge about them (masses and life-times take
non-negative real values, probabilities have support on $[0,1]$,...). On the
other hand, it doesn't make sense to think about a function that
represents ignorance in a formal and objective way so {\sl knowing little
a priory} is relative 
to what we may expect to learn from the experiment.
Whatever prior we use will certainly have some effect on the
posterior inferences and, in some cases, it would be wise to consider a
reasonable set of them to see what is the effect. 

The ultimate task of this section is to present the most usual approaches
to derive a non-informative prior function to be used as a standard reference 
that contains little information 
about the parameters compared to what we expect to get from the experiment
\footnote{For a comprehensive discussion see [Ka96]}.
In many cases, these priors will not be Lebesgue integrable 
({\sl improper functions}) and, obviously,
can not be considered as probability density functions that quantify
any knowledge on the parameters 
(although, with little rigor, sometimes 
we still talk about prior {\sl densities}).
If one is reluctant to use them
right the way one can, for instance, define them on a sufficiently large 
compact support that contains the region where the likelihood is dominant.
However, since
\begin{eqnarray*}
  p({\thetabold}|{\xbold})\,d\thetabold\,\propto\, 
  p(\xbold|\thetabold)\,p(\thetabold)\,d\thetabold\,=\,
  p(\xbold|\thetabold)\,d\mu(\thetabold)
\end{eqnarray*}
in most cases  it will be sufficient to consider them simply as what they 
really are: a measure.
In any case, what is mandatory is that the posterior is a 
well defined proper density.

\begin{figure}[t]
\begin{center}

\mbox{\epsfig{file=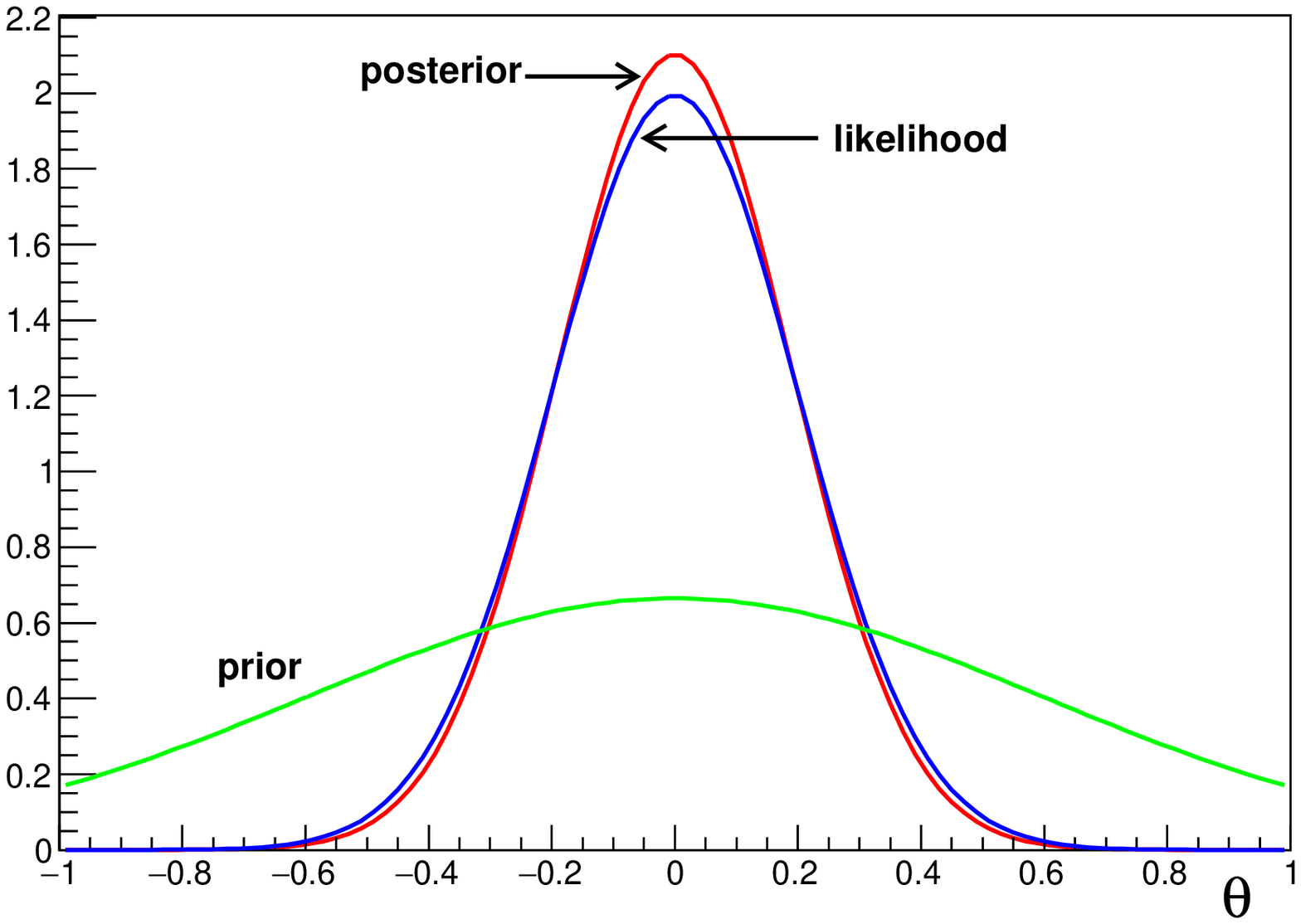,height=9.5cm,width=11cm}}

\footnotesize{
{\bf Figure 2.1}.- Prior, likelihood and posterior as function of the
the parameter $\theta$. In this case, the prior is a smooth function 
and the posterior is dominated by the likelihood.
             }
\end{center}
\end{figure}

\small
\subsection{Principle of Insufficient Reason}
The  {\sl Principle of Insufficient Reason}
\footnote{Apparently, {\sl ``Insufficient Reason''} was coined
by Laplace in reference to the Leibniz's {\sl Principle of Sufficient
Reason} stating essentially that 
every fact has a sufficient reason
for why it is the way it is and not other way.}
 dates back to
J. Bernoulli and P.S. Laplace and, originally, it states that
if we have $n$ exclusive and exhaustive hypothesis and there is no special 
reason to prefer one over the other, it is reasonable to consider them equally 
likely and assign a prior probability $1/n$ to each of them.
This certainly sounds reasonable and the idea was right the way
extended to parameters
taking countable possible values and to those with continuous support that,
in case of compact sets, becomes a uniform density.
It was extensively used by P.S. Laplace and T. Bayes, being he 
the first to use a uniform prior density for making 
inferences on the parameter of a Binomial distribution, and is is 
usually referred to as the {\sl ``Bayes-Laplace Postulate''}. 
However, a uniform prior density is
obviously not invariant under reparameterizations. If prior to the experiment
we have a very vague knowledge about the parameter
${\theta}{\in}[a,b]$,
we certainly have a vague knowledge about 
${\phi}=1/{\theta}$ or ${\zeta}={\rm log}{\theta}$
and a uniform distribution for $\theta$:
\begin{eqnarray*}
{\pi}({\theta})\,d{\theta}\,=\,
    \frac{\textstyle 1}
         {\textstyle b-a}\,d{\theta}
\nonumber
\end{eqnarray*}
implies that:
\begin{eqnarray*}
{\pi}({\phi})\,d{\phi}\,=\,
    \frac{\textstyle 1}
         {\textstyle {\phi}^2}\,d{\phi}
\hspace*{1.cm}\, {\rm and}\, \hspace*{1.cm}
{\pi}({\zeta})\,d{\zeta}\,=\,
    e^{\textstyle {\zeta}}  \,d{\zeta}
\nonumber
\end{eqnarray*}
Shouldn't we take as well a uniform density for  
 ${\phi}$ or ${\zeta}$?

Nevertheless, we shall see that a uniform density, that is 
far from representing ignorance on a parameter,
may be a reasonable choice in many cases even though,
if the support of the parameter is infinite, it is 
an improper function. 

\small
\subsection{Parameters of position and scale}
An important class of parameters we are interested in are those
of position and scale. Let's treat them separately and leave for a
forthcoming section the argument behind that.
Start with a random quantity $X{\sim}p(x|{\mu})$ with ${\mu}$ a
{\sl location parameter}. The density has the form
$ p(x|{\mu})\,=\,f(x-{\mu}) $
so, taking a prior function ${\pi}({\mu})$ we can write
\begin{eqnarray*}
 p(x,{\mu})\,dx\,d{\mu}\,=\,
\left[p(x|{\mu})\,dx\right]\,\left[{\pi}({\mu})\,d{\mu}\right]\,=\,
\left[f(x-{\mu})\,dx\right]\,\left[{\pi}({\mu})\,d{\mu}\right]
\nonumber
\end{eqnarray*}
Now, consider random quantity $X'\,=\,X\,+\,a$ with
$a{\in}{\Rcal}$ a known value. Defining the new parameter ${\mu}'={\mu}+a$ 
we have
\begin{eqnarray*}
p(x',{\mu}')\,dx'\,d{\mu}'\,=\,
\left[p(x'|{\mu}')\,dx'\right]\,\left[{\pi}'({\mu}')\,d{\mu}'\right]
\,=\,
\left[f(x'-{\mu}')\,dx'\right]\,\left[{\pi}({\mu}'-a)\,d{\mu}'\right]
\nonumber
\end{eqnarray*}
In both cases the models have the same structure so making
inferences on ${\mu}$ 
from the sample $\{x_1,x_1,{\ldots},x_n\}$ is formally equivalent to making 
inferences on ${\mu}'$ from the shifted sample 
$\{x'_1,x'_2,{\ldots},x'_n\}$. 
Since we have the same prior degree of knowledge on
${\mu}$ and ${\mu}'$, it is reasonable to take the same functional form
for ${\pi}({\cdot})$ and ${\pi}'({\cdot})$ so:
 \begin{eqnarray*}
{\pi}({\mu}'-a)\,d{\mu}'\,=\,
\,=\,{\pi}({\mu}')\,d{\mu}'
\hspace*{1.0cm} \forall a{\in}{\Rcal}
\nonumber
\end{eqnarray*}
and, in consequence:
\begin{eqnarray*}
 {\pi}({\mu})\,=\,{\rm constant}
\nonumber
\end{eqnarray*}

If $\theta$ is a {\sl scale parameter}, the model has the form
$p(x|{\theta})\,=\,{\theta}f(x{\theta})$ so taking a prior function
${\pi}({\theta})$ we have that
\begin{eqnarray*}
 p(x,{\theta})\,dx\,d{\theta}\,=\,
\left[p(x|{\theta})\,dx\right]\,\left[{\pi}({\theta})\,d{\theta}\right]\,=\,
\left[{\theta}f(x{\theta})\,dx\right]\,\left[{\pi}({\theta})\,d{\theta}\right]
\nonumber
\end{eqnarray*}
For the scaled random quantity $X'\,=\,a\,X$ with $a{\in}{\Rcal}^+$
known, we have that:
\begin{eqnarray*}
 p(x',{\theta}')\,dx'\,d{\theta}'\,=\,
\left[p(x'|{\theta}')\,dx'\right]\,
\left[{\pi}'({\theta}')\,d{\theta}'\right]\,=\,
\left[{\theta}'f(x'{\theta}')\,dx'\right]\,\left[{\pi}(a{\theta}')\,
ad{\theta}\right]
\nonumber
\end{eqnarray*}
where we have defined the new parameter ${\theta}'={\theta}/a$. Following
the same argument as before, it is sound to assume the same functional form
for ${\pi}({\cdot})$ and ${\pi}'({\cdot})$ so:
 \begin{eqnarray*}
{\pi}(a{\theta}')\,a\,d{\theta}'\,=\,{\pi}({\theta}')\,d{\theta}'
\hspace*{1.0cm} \forall a{\in}{\Rcal}
\nonumber
\end{eqnarray*}
and, in consequence:
\begin{eqnarray*}
{\pi}({\theta})\,=\,
\frac{\textstyle 1}{\textstyle \theta}
\nonumber
\end{eqnarray*}

Both prior functions are improper so they may be explicited as
\begin{eqnarray*}
{\pi}({\mu},{\theta})\,\propto\,
\frac{\textstyle 1}{\textstyle \theta}\,
{\mbox{\boldmath ${1}$}}_{\Theta}({\theta})\,
{\mbox{\boldmath ${1}$}}_{M}({\mu})\,
\nonumber
\end{eqnarray*}
with ${\Theta},M$ an appropriate sequence of compact sets
or considered as prior measures
provided that the posterior densities are well defined. 
Let's see some examples.

\vspace{0.5cm}
\noindent
{\raya}                   
\vspace{0.35cm}
\footnotesize
\noindent
{\bf Example 2.3: The Exponential Distribution.}
Consider the sequence of independent observations 
$\{x_1,x_2,{\ldots},x_n\}$ of the random quantity $X\sim Ex(x|{\theta})$ 
drawn under the same conditions. The joint density is
\begin{eqnarray*}
   p(x_1,x_2,{\ldots},x_n|\theta)\,=\,{\theta}^n\,
e^{\textstyle -\theta(x_1+x_2+{\cdots}\,x_n)}
\end{eqnarray*}
The statistic
$t=n^{-1}\sum_{i=1}^{n}x_i$  is sufficient for $\theta$ and is distributed as
\begin{eqnarray*}
   p(t|\theta)\,=\,\frac{\textstyle (n\theta)^n}
                             {\textstyle \Gamma(n)}\,t^{n-1}\,
{\rm exp}\{\textstyle -n\theta t\}\,
\end{eqnarray*}
It is clear that
$\theta$ is a scale parameter so we shall take the prior function
$\pi(\theta)=1/{\theta}$. 
Note that if we make the change
$z=\log t$ and ${\phi}=\log{\theta}$ we have that
\begin{eqnarray*}
   p(z|\phi)\,=\,\frac{\textstyle n^n}{\textstyle \Gamma(n)}\,
   {\rm exp}\{n\left((\phi+z)-e^{\phi+z}\right)\}
\end{eqnarray*}
In this parameterization,
${\phi}$ is a position parameter and therefore
$\pi({\phi})={\rm const}$ in consistency with $\pi(\theta)$.
Then, we have the proper posterior for inferences:
\begin{eqnarray*}
   p(\theta|t,n)\,=\,\frac{\textstyle (nt)^{n}}
                             {\textstyle \Gamma(n)}\,
{\rm exp}\{\textstyle -nt\theta \}\,\theta^{n-1}\,
			     \hspace{1.cm};\hspace{1.cm}\theta > 0
\end{eqnarray*}
Consider now the sequence of compact sets $C_k=\left[1/k,k\right]$ covering
$R^{+}$ as $k{\rightarrow}{\infty}$. Then, with support on $C_k$ we have
the proper prior density
\begin{eqnarray*}
\pi_k(\theta)=
\frac{\textstyle 1}{\textstyle 2\,\log k}
\frac{\textstyle 1}{\textstyle \theta}
{\mbox{\boldmath ${1}$}}_{C_k}({\theta})\,
\end{eqnarray*}
and the sequence of posteriors:
\begin{eqnarray*}
   p_k(\theta|t,n)\,=\,\frac{\textstyle (nt)^{n}}
                             {\textstyle \gamma(n,ntk)-\gamma(n,nt/k)}\,
{\rm exp}\{\textstyle -nt\theta\}\,\theta^{n-1}\,
{\mbox{\boldmath ${1}$}}_{C_k}({\theta})
\end{eqnarray*}
with $\gamma(a,x)$ the Incomplete Gamma Function. It is clear that
\begin{eqnarray*}
   \lim_{k{\rightarrow}\infty}\,p_k(\theta|t,n)\,=\,p(\theta|t,n)
\end{eqnarray*}

\vspace{0.35cm}

\noindent
{\bf Example 2.4: The Uniform Distribution.}
Consider the random quantity $X\sim Un(x|0,{\theta})$ and the independent 
sampling $\{x_1,x_2,{\ldots},x_n\}$. To draw inferences on $\theta$, the
statistics
$x_M=\max\{x_1,x_2,{\ldots},x_n\}$ is sufficient and is distributed as
(show that):
\begin{eqnarray*}
   p(x_M|\theta)\,=\,n\,\frac{\textstyle x_M^{n-1}}
                             {\textstyle \theta^n}
{\mbox{\boldmath ${1}$}}_{[0,{\theta}]}(x_M)\,
\end{eqnarray*}
As in the previous case, $\theta$ is a scale parameter and
with the change $t_M=\log x_M$, 
${\phi}=\log{\theta}$ is a position parameter. Then, we shall take
$\pi(\theta)\propto\theta^{-1}$ and get the posterior density (Pareto):
\begin{eqnarray*}
   p(\theta|x_M,n)\,=\,n\,\frac{\textstyle x_M^{n}}
                             {\textstyle \theta^{n+1}}
\,{\mbox{\boldmath ${1}$}}_{[x_M,\infty)}(\theta)
\end{eqnarray*}

\vspace{0.35cm}

\noindent
{\bf Example 2.5: The one-dimensional Normal Distribution.}
Consider the random quantity $X{\sim}N(x|{\mu},{\sigma})$ and
the experiment $e(n)$ that provides the independent and exchangeable sequence
${\xbold}=\{x_1,x_2,{\ldots},x_n\}$ of observations.
The likelihood function will then be:
\begin{eqnarray*}
    p({\xbold}|{\mu},{\sigma})\,=\,
  \prod_{i=1}^{n}\,p(x_i|{\mu},{\sigma})\,{\propto}\,
    \frac{\textstyle 1}{\textstyle {\sigma}^n}\,
       {\rm exp}\{ 
       { -\,\frac{1}{2{\sigma}^2}\,\sum_{i=1}^{n}(x_i-{\mu})^2}
                \}
                \nonumber 
\end{eqnarray*}
There is a three-dimensional sufficient statistic
${\tbold}\,=\, \{n,\,\overline{x},\,s^2\}$ where
\begin{eqnarray*}
     \overline{x}\,=\,
          \frac{\textstyle 1}{\textstyle n}\,
          \sum_{i=1}^{n}\,x_i\hspace{1.cm}{\rm and}\hspace{1.cm}
       s^2\,=\,
          \frac{\textstyle 1}{\textstyle n}\,
          \sum_{i=1}^{n}\,(x_i-\overline{x})^2
           \nonumber
\end{eqnarray*}
so we can write
\begin{eqnarray*}
    p({\xbold}|{\mu},{\sigma})\, {\propto}\,
    \frac{\textstyle 1}{\textstyle {\sigma}^n}\,
       {\rm exp}\{ 
                  -\,\frac{n}{2{\sigma}^2}\,
                  \left(  s^2\,+\,
                          (\overline{x}-{\mu})^2
                  \right)
                \}
                \nonumber 
\end{eqnarray*}
In this case we have both position 
and scale parameters so we take 
${\pi}({\mu},{\sigma})={\pi}({\mu}){\pi}({\sigma})={\sigma}^{-1}$  
and get the proper posterior
\begin{eqnarray*}
    p({\mu},{\sigma}|{\xbold})\, {\propto}\,
                  p({\xbold}|{\mu},{\sigma})\,
                  {\pi}({\mu},{\sigma})\,{\propto}\,
    \frac{\textstyle 1}{\textstyle {\sigma}^{n+1}}\,
       {\rm exp}\{ 
                  -\,\frac{n}{2{\sigma}^2}\,
                  \left[  s^2\,+\,
                          (\overline{x}-{\mu})^2 \right] \}
                \nonumber 
\end{eqnarray*}

\vspace{0.5cm}
\noindent
$\bullet$ {\bf Marginal posterior density of} $\sigma$: Integrating
the parameter ${\mu}{\in}\cal{R}$ we have that:
\begin{eqnarray*}
    p({\sigma}|{\xbold})\,=\,
    \int_{-\infty}^{+\infty}\,
         p({\mu},{\sigma}|{\xbold})\,d{\mu}\,
     {\propto}\,{\sigma}^{-n}\,
       {\rm exp}\left\{ 
                  -\,\frac{n\,s^2}{2{\sigma}^2}
                \right\}
                    {\mbox{\boldmath ${1}$}}_{(0,\infty)}({\sigma})
                \nonumber 
\end{eqnarray*}
and therefore, the random quantity
\begin{eqnarray*}
    Z\,=\,
    \frac{\textstyle n\,s^2}
         {\textstyle {\sigma}^2}\,{\sim}\,{\chi}^2(z|n-1)
                \nonumber 
\end{eqnarray*}

\vspace{0.5cm}
\noindent
$\bullet$ {\bf Marginal posterior density of} $\mu$: Integrating
the parameter ${\sigma}{\in}[0,{\infty})$ we have that:
\begin{eqnarray*}
    p({\mu}|{\xbold})\,=\,
    \int_{0}^{+\infty}\,
         p({\mu},{\sigma}|{\xbold})\,d{\sigma}\,
         {\propto}\, \left(1\,+\,
    \frac{\textstyle ({\mu}-\overline{x})^2}
         {\textstyle s^2}
    \right)^{-n/2}\,
                    {\mbox{\boldmath ${1}$}}_{(-\infty,\infty)}({\mu})
                \nonumber 
\end{eqnarray*}
so the random quantity
\begin{eqnarray*}
    T\,=\,\frac{\textstyle \sqrt{n-1}({\mu}-\overline{x})}
               {\textstyle s}\,{\sim}\,St(t|n-1)
                \nonumber 
\end{eqnarray*}
It is clear that 
$p({\mu},{\sigma}|{\xbold}){\neq}p({\mu}|{\xbold})\,p({\sigma}|{\xbold})$
and, in consequence, are not independent.

\vspace{0.5cm}
\noindent
$\bullet$ {\bf Distribution of } $\mu$ {\bf conditioned to} ${\sigma}$:
Since
$ p({\mu},{\sigma}|{\xbold})\,=\,
  p({\mu}|{\sigma},{\xbold})\,p({\sigma}|{\xbold})$
we have that                 
\begin{eqnarray*}
    p({\mu}|{\sigma},{\xbold})\,{\propto}\,
    \frac{\textstyle 1}{\textstyle {\sigma}}\,
       {\rm exp}\{ 
                  -\,\frac{n}{2{\sigma}^2}\,({\mu}-\overline{x})^2
                \}
                \nonumber 
\end{eqnarray*}
so $\mu|\sigma\,{\sim}N({\mu}|\overline{x},{\sigma}/\sqrt{n})$.
\vspace{0.35cm}

\noindent
{\bf Example 2.6: Contrast of parameters of Normal Densities}. 
Consider two independent random quantities
$X_1{\sim}N(x_1,|{\mu}_1,{\sigma}_1)$ and $X_2{\sim}N(x_2,|{\mu}_2,{\sigma}_2)$
and the random samplings
${\xbold}_1=\{x_{11},x_{12},{\ldots},x_{1n_1}\}$ and
${\xbold}_2=\{x_{21},x_{22},{\ldots},x_{2n_2}\}$ of sizes $n_1$ and $n_2$
under the usual conditions. 
From the considerations of the previous example, we can write
\begin{eqnarray*}
    p({\xbold}_i|{\mu}_i,{\sigma}_i)\, {\propto}\,
    \frac{\textstyle 1}{\textstyle {\sigma}_i^{n_i}}\,
       {\rm exp}\left\{ 
                  -\,\frac{n_i}{2{\sigma}_i^2}\,
                  \left(  s^2_i\,+\,(\overline{x_i}-{\mu}_i)^2
                  \right)                                   
                \right\}\hspace{0.3cm};\hspace{1.cm}i=1,2
\end{eqnarray*}
Clearly, $({\mu}_1,{\mu}_2)$ are position parameters and
$({\sigma}_1,{\sigma}_2)$ scale parameters so, in principle, we shall take
the improper prior function
\begin{eqnarray*}
    {\pi}({\mu}_1,{\sigma}_1,{\mu}_2,{\sigma}_2)\,=\,
    {\pi}({\mu}_1){\pi}({\mu}_2)
    {\pi}({\sigma}_1){\pi}({\sigma}_2)\,{\propto}\,
    \frac{\textstyle 1}{\textstyle {\sigma}_1\,{\sigma}_2}
                \nonumber 
\end{eqnarray*}
However, if we have know that both distributions have the same variance,
then we may set ${\sigma}={\sigma}_1={\sigma}_2$ and, in this case, the
prior function will be
\begin{eqnarray*}
    {\pi}({\mu}_1,{\mu}_2,{\sigma})\,=\,
    {\pi}({\mu}_1){\pi}({\mu}_2){\pi}({\sigma})
\,{\propto}\,\frac{\textstyle 1}{\textstyle {\sigma}}
                \nonumber 
\end{eqnarray*}
Let's analyze both cases.

\vspace{0.5cm}
\noindent
$\bullet$ {\bf Marginal Distribution of} ${\sigma}_1$ {\bf and}
          ${\sigma}_2$: In this case we assume that
${\sigma}_1\ne{\sigma}_2$ and we shall take the prior
${\pi}({\mu}_1,{\sigma}_1,{\mu}_2,{\sigma}_2)\,{\propto}\,
({\sigma}_1{\sigma}_2)^{-1}$.
Integrating ${\mu}_1$ and ${\mu}_2$ we get:
\begin{eqnarray*}
p({\sigma}_1,{\sigma}_2|{\xbold}_1,{\xbold}_2)\,=\,
p({\sigma}_1,|{\xbold}_1)p({\sigma}_2,|{\xbold}_2)
    {\propto}\,{\sigma}_1^{-n_1}\,{\sigma}_2^{-n_2}\,
       {\rm exp}\left\{ 
                  -\,\frac{1}{2}\,
                  \left(  
                   \frac{n_1 s^2_1}{{\sigma}_1^2}\,+\,
                   \frac{n_2 s^2_2}{{\sigma}_2^2}
                  \right)
                \right\}
\end{eqnarray*}
Now, if we define the new random quantities
\begin{eqnarray*}
Z\,=\,\frac{\textstyle s^2_2}{\textstyle w^2\,s^2_1}\,=\,
\frac{\textstyle \left({\sigma}_1/s_1\right)^2}
         {\textstyle \left({\sigma}_2/s_2\right)^2}
\hspace{1.cm}{\rm and}\hspace{1.cm}
W\,=\,\frac{\textstyle n_1 s^2_1}{\textstyle {\sigma_1}^2}
\nonumber
\end{eqnarray*}
both with support in $(0,\,+{\infty})$, and integrate the last we get
we get that $Z$ follows a Snedecor Distribution 
$Sn(z|n_2-1,n_1-1)$ whose density is
\begin{eqnarray*}
p(z|{\xbold}_1,{\xbold}_2)\,=\,
    \frac{\textstyle \left({\nu}_1/{\nu}_2\right)^{{\nu}_1/2}}
         {\textstyle {\rm Be}\left({\nu}_1/2,{\nu}_2/2\right)}\,
z^{({\nu}_1/2)-1}\,
                  \left(1\,+\,
                   \frac{{\nu}_1}{{\nu}_2}\,z
                  \right)^{-({\nu}_1+{\nu}_2)/2}\,
                    {\mbox{\boldmath ${1}$}}_{(0,\infty)}(z)
                \nonumber 
\end{eqnarray*}

\vspace{0.5cm}
\noindent
$\bullet$ {\bf Marginal Distribution of} ${\mu}_1$ {\bf and}
                                           ${\mu}_2$: In this case, it
is different whether we assume that, although unknown, 
the variances are the same or not.  
In the first case, we set 
${\sigma}_1={\sigma}_2={\sigma}$  and take the reference prior
${\pi}({\mu}_1,{\mu}_2,{\sigma})\,=\,{\sigma}^{-1}$. Defining
\begin{eqnarray*}
 A\,=\,n_1\,\left[s^2_1\,+\,(\overline{x}_1-{\mu}_1)^2\right]\,+\,
       n_2\,\left[s^2_2\,+\,(\overline{x}_2-{\mu}_2)^2\right]
\end{eqnarray*}
we can write
\begin{eqnarray*}
    p({\mu}_1,{\mu}_2,{\sigma}|
      {\xbold},{\ybold})\,&{\propto}&\,
        \frac{\textstyle 1}{\textstyle {\sigma}^{n_1+n_2+1}}
       {\rm exp}\{ 
                  -\,\frac{1}{2}\,
                  A/{\sigma}^2 \}
                \nonumber  
\end{eqnarray*}
It is left as an exercise to show that if we make the transformation
\begin{eqnarray*}
   w\,=\,{\mu}_1-{\mu}_2\,{\in}\,(-\infty,+{\infty}) \hspace{0.3cm};
\hspace{0.5cm}
   u\,=\,{\mu}_2\,{\in}\,(-\infty,+{\infty}) 
\hspace{0.5cm}{\rm and}\hspace{0.5cm}
   z\,=\,{\sigma}^{-2}\,{\in}\,(0,+{\infty})
\end{eqnarray*}
and integrate the last two, we get
\begin{eqnarray*}
    p(w|{\xbold}_1,{\xbold}_2)\,{\propto}\,
   \left(1\,+\,
        \frac{\textstyle n_1\,n_2}{\textstyle n_1+n_2}\,
        \frac{\textstyle [(\overline{x}_1-\overline{x}_2)-w]^2}
             {\textstyle n_1\,s^2_1\,+\,n_2\,s^2_2}
   \right)^{-(n_1+n_2-1)/2}
                \nonumber  
\end{eqnarray*}
Introducing the more usual terminology
\begin{eqnarray*}
    s^2\,=\,
        \frac{\textstyle n_1\,s^2_1\,+\,n_2\,s^2_2}
             {\textstyle n_1\,+\,n_2\,-\,2}
                \nonumber  
\end{eqnarray*}
we have that
\begin{eqnarray*}
    p(w|{\xbold}_1,{\xbold}_2)\,{\propto}\,
   \left(1\,+\,
        \frac{\textstyle n_1\,n_2}{\textstyle n_1+n_2}\,
        \frac{\textstyle [w-(\overline{x}_1-\overline{x}_2)]^2}
             {\textstyle s^2\,(n_1+n_2-2)}
   \right)^{-[(n_1+n_2-2)+1]/2}
                \nonumber  
\end{eqnarray*}
and therefore the random quantity
\begin{eqnarray*}
    T\,=\,
        \frac{\textstyle ({\mu}_1-{\mu}_2)-(\overline{x}_1-\overline{x}_2)}
             {\textstyle s\,(1/n_1+1/n_2)^{1/2}}
                \nonumber  
\end{eqnarray*}
follows a Student's Distribution $St(t|\nu)$ with 
$\nu=n_1+n_2-2$ degrees of freedom.

Let's see now the case where we can not assume that the variances are equal.
Taking the prior reference function
$ {\pi}({\mu}_1,{\mu}_2,{\sigma}_1,{\sigma}_2)\,=\,
    ({\sigma}_1\,{\sigma}_2)^{-1}$
we get
\begin{eqnarray*}
    p({\mu}_1,{\mu}_2,{\sigma}_1,{\sigma}_2|
      {\xbold}_1,{\xbold}_2)\,{\propto}\, 
{\sigma}_1^{-(n_1+1)}\,{\sigma}_2^{-(n_2+1)}
       {\rm exp}\left\{ 
                  -\,\frac{1}{2}\,\sum_{i=1}^2
    \frac{\textstyle s^2_i\,+\,(\overline{x}_i-{\mu}_i)^2}
         {\textstyle {\sigma}_i^2/n_i}
                \right\}
\end{eqnarray*}
After the appropriate integrations (left as exercise), defining 
$w=\mu_1-\mu_2$ and $u=\mu_2$ we end up with the density
\begin{eqnarray*}
    p(w,u|{\xbold}_1,{\xbold}_2)\,{\propto}\,
   \left(1\,+\,
     \frac{\textstyle (\overline{x}_1-w-u)^2}
          {\textstyle s^2_1}
   \right)^{-n_1/2} \,
   \left(1\,+\,
     \frac{\textstyle (\overline{x}_2-u)^2}
          {\textstyle \,s^2_2}
   \right)^{-n_2/2}
\end{eqnarray*}
where integral over $u{\in}\cal{R}$ can not be expressed in a 
simple way. The density
\begin{eqnarray*}
    p(w|{\xbold}_1,{\xbold}_2)\,{\propto}\,
   \int_{-\infty}^{+\infty}\,
    p(w,u|{\xbold}_1,{\xbold}_2)\,du
                \nonumber  
\end{eqnarray*}
is called the {\sl Behrens-Fisher Distribution}.
Thus, to make statements on the difference of Normal means, we should  
analyze first the sample variances and decide how shall we treat them.

\small
{\raya}                   
\vspace{1.0cm}

\small
\subsection{Covariance under reparameterizations}
The question of how to establish a reasonable criteria to obtain a prior 
for a given model $p(\mbox{\boldmath $x$}|{\thetabold})$
that can be used as a standard reference function 
was studied by Harold Jeffreys [Je39] in the mid ${\rm XX}^{th}$  century. 
The rationale behind
the argument is that if we have the model 
$p(\mbox{\boldmath $x$}|{\thetabold})$ with
${\thetabold}{\in}{\Omega}_{\theta}{\subseteq}R^n$ and make a reparameterizations
$\phibold=\phibold({\thetabold})$ with $\phibold({\cdot})$ 
a one-to-one differentiable function, the statements we make about 
$\thetabold$ should be consistent with 
those we make about ${\phibold}$ and, in consequence, 
priors should be related by
\begin{eqnarray*}
{\pi}_{\theta}({\thetabold})d{\thetabold}\,=\,
{\pi}_{\phi}({\phibold}({\thetabold}))\,
\left|\,{\rm det}\left[
\frac{\textstyle {\partial}{\phi}_i({\theta})}
     {\textstyle {\partial}{\theta}_j}
 \right] \right|\,d{\thetabold}
\nonumber
\end{eqnarray*}
Now, assume that the Fisher's matrix (see Note 6)
\begin{eqnarray*}  
{\Ibold}_{ij}({\thetabold})\,=\,E_{\Xbold}\left[
    \frac{\textstyle {\partial}\log\,p(\mbox{\boldmath $x$}|\thetabold)}
             {\textstyle {\partial} {\theta}_i}\,
    \frac{\textstyle {\partial}\log\,p(\mbox{\boldmath $x$}|\thetabold)}
             {\textstyle {\partial} {\theta}_j}\,\right]
\nonumber
\end{eqnarray*}
exists for this model. Under a differentiable one-to-one transformation  
$\phibold=\phibold({\thetabold})$ we have that 
\begin{eqnarray*}  
{\Ibold}_{ij}({\phibold})\,=\,
    \frac{\textstyle {\partial}{\theta}_k}{\textstyle {\partial} {\phi}_i}\,
    \frac{\textstyle {\partial}{\theta}_l}{\textstyle {\partial} {\phi}_j}\,
         {\Ibold}_{kl}({\thetabold})
\nonumber
\end{eqnarray*}
so it behaves as a covariant symmetric tensor of second order
(left as exercise). Then, since 
\begin{eqnarray*}  
{\rm det}\left[{\Ibold}({\phibold})\right]\,=\,
\left|\,{\rm det}\left[
\frac{\textstyle {\partial}{\theta}_i}
     {\textstyle {\partial}{\phi}_j}
 \right] \right|^{2}\,{\rm det}\left[{\Ibold}({\thetabold})\right]
\nonumber
\end{eqnarray*}
Jeffreys proposed to consider the prior
\begin{eqnarray*}  
       {\pi}({\thetabold})\,{\propto}\,
\left[ {\rm det}[{\Ibold}({\thetabold}] \right]^{1/2}
\end{eqnarray*}  
In fact, if we consider
the parameter space as a Riemannian manifold 
the Fisher's matrix is the metric tensor (Fisher-Rao metric) and this is
just the invariant volume element.
Intuitively, if we make a transformation such that at a particular value
${\phibold}_0={\phibold}({\thetabold}_0)$ the Fisher's tensor is constant and
diagonal, the metric in a neighborhood of ${\phibold}_0$ is Euclidean and
we have location parameters for which a constant prior is appropriate and
therefore
\begin{eqnarray*} 
{\pi}({\phibold})d{\phibold}\,\propto\, d{\phibold}\,=\,\left[
{\rm det}\left[{\Ibold}({\thetabold})\right]^{1/2}\right]\,d{\thetabold}\,=\,
{\pi}({\thetabold})d{\thetabold}
\end{eqnarray*}
It should be pointed out that there may be other priors that are also
invariant under reparameterizations and that, as usual, we
talk loosely about {\sl prior densities} although they usually are improper 
functions.

For one-dimensional parameter, the density function
expressed in terms of 
\begin{eqnarray*}  
{\phi}\,{\sim}\,\int\,\left[{\Ibold}({\theta}) \right]^{1/2}\,d{\theta}
\nonumber
\end{eqnarray*}
may be reasonably well approximated by a Normal density (at least in the
parametric region where the likelihood is dominant) because 
${\Ibold}({\phi})$ is constant and then, due to
translation invariance, a constant prior for $\phi$ is justified.
Let' see some examples.


\rayan
\footnotesize
\vspace{0.5cm}
\noindent
{\bf NOTE 6:} The {\bf Fisher's Matrix} is a non-negative symmetric matrix that
plays a very important role in statistical inference and is defined as:
\begin{eqnarray}  
{\Ibold}_{ij}({\phibold})\,=\,\int_{{\Omega}_X}\, 
p(\mbox{\boldmath $x$}|{\phibold})
             \left(
    \frac{\textstyle {\partial}\log\,p(\mbox{\boldmath $x$}|{\phibold})}
             {\textstyle {\partial} {\phi}_i}
    \frac{\textstyle {\partial}\log\,p(\mbox{\boldmath $x$}|{\phibold})}
             {\textstyle {\partial} {\phi}_j}
             \right)\,d{\mbox{\boldmath $x$}}=
\int_{{\Omega}_X}\, 
p(\mbox{\boldmath $x$}|{\phibold})
             \left(-\,
    \frac{\textstyle {\partial}^2\log\,p(\mbox{\boldmath $x$}|{\phibold})}
             {\textstyle {\partial} {\phi}_i{\partial} {\phi}_j}
             \right)\,d{\mbox{\boldmath $x$}}
\nonumber
\end{eqnarray}
provided it exists. This is the case for regular distributions where:
\begin{itemize}
\item[1)] ${\rm supp}_{\mbox{\boldmath $x$}}
          \{p(\mbox{\boldmath $x$}|{\phibold})\}$ 
          does not depend on $\phibold$; 
\item[2)] $p(\mbox{\boldmath $x$}|{\phibold}){\in}C_k({\phibold})$
          for $k{\geq}2$ and
\item[3)] The integrand is well-behaved so 
          $\frac{\textstyle {\partial}}{\textstyle {\partial} {\phi}}
           \int_{{\Omega}_X}(\bullet)d{\mbox{\boldmath $x$}}=
           \int_{{\Omega}_X}
           \frac{\textstyle {\partial}(\bullet)}{\textstyle {\partial} {\phi}}
           d{\mbox{\boldmath $x$}}$
\end{itemize}
Since
$\int_{\Omega_X}\,
p(\mbox{\boldmath $x$}|{\theta})\,d{\mbox{\boldmath $x$}}=1$ we have that
the diagonal elements can be expressed also as:
\begin{eqnarray}  
{\Ibold}_{ii}({\thetabold})\,=\,E_{\Xbold}\left[
    \frac{\textstyle {\partial}\log\,p(\mbox{\boldmath $x$}|\thetabold)}
             {\textstyle {\partial} {\theta}_i}\,
    \frac{\textstyle {\partial}\log\,p(\mbox{\boldmath $x$}|\thetabold)}
             {\textstyle {\partial} {\theta}_i}\,\right]\,=\,
E_{\Xbold}\left[-\,
    \frac{\textstyle {\partial}^2\log\,p(\mbox{\boldmath $x$}|\thetabold)}
             {\textstyle {\partial} {\theta}_i^2}\,\right]
\nonumber
\end{eqnarray}
If $\Xbold=(X_1,\ldots,X_n)$ is a n-dimensional random quantity and
$\{X_i\}_{i=1}^n$ are independent, then
$I_{\Xbold}(\thetabold)=\sum_i I_{X_i}(\thetabold)$. Obviously, if they are
iid then $I_{X_i}(\thetabold)=I_{X}(\thetabold)$ for all $i$ and
$I_{\Xbold}(\thetabold)=n I_{X}(\thetabold)$.

Suppose that the experiment $e(n)$ provides an independent and exchangeable
sequence of observations $\{x_1,x_2,{\ldots},x_n\}$ from the model 
$p(x|{\thetabold})$ with ${\rm dim}({\thetabold})=d$.
The information that the experiment provides about ${\thetabold}$ is contained  
in the likelihood function and,
being non-negative function, consider for simplicity its logarithm:
\begin{eqnarray}
w({\thetabold}|{\cdot})\,=\,
\sum_{i=1}^{n}\,\log\,p(x_i{\mid}{\thetabold})
\nonumber
\end{eqnarray}
and the Taylor expansion around the maximum $\widehat{\thetabold}$:
\begin{eqnarray}
w({\thetabold}|{\cdot})\,=\,
  w(\widehat{\thetabold}|{\cdot})\,-\,
  \frac{\textstyle n}{\textstyle 2}\,\sum_{k=1}^{d}\sum_{m=1}^{d}
\left[\frac{\textstyle 1}{\textstyle n}
\sum_{i=1}^{n}\,\left(-\,
  \frac{\textstyle {\partial}^2\,{\log}\,p(x_i|{\thetabold})}
                 {\textstyle {\partial}{\theta}_k{\partial}{\theta}_m} 
           \right)_{\widehat{\theta}}\,\right]
              \,({\theta}_k-\widehat{\theta}_k)({\theta}_m-\widehat{\theta}_m)
          +\,\ldots
\nonumber
\end{eqnarray}
where the second term has been multiplied and divided by $n$.
Under sufficiently regular conditions we have that
$\widehat{\thetabold}$ converges in probability to the true value
${\thetabold}_0$ so we can neglect higher order terms and,
by the Law of Large Numbers, approximate
\begin{eqnarray}
\lim_{n{\rightarrow}{\infty}}\, \left[ \frac{\textstyle 1}{\textstyle n}
\,{\sum}_{i=1}^{n}\,\left(
-\,\frac{\textstyle {\partial}^2\,{\log}\,p(x_i{\mid}{\thetabold})}
        {\textstyle {\partial}{\theta}_k{\partial}{\theta}_m}
\right)\right]_{\widehat{\theta}}\,{\simeq}\,
E_X
\left[
-\,\frac{\textstyle {\partial}^2\,{\log}\,p(x{\mid}{\thetabold})}
        {\textstyle  {\partial}{\theta}_k{\partial}{\theta}_m}
\right]_{\widehat{\theta}}\,\simeq\,
{\Ibold}_{km}(\widehat{\thetabold})
\nonumber
\end{eqnarray}
Therefore
\begin{eqnarray}
w({\thetabold}|{\cdot})\,=\,
  w(\widehat{\thetabold}|{\cdot})\,-\,
  \frac{\textstyle 1}{\textstyle 2}\,\sum_{k=1}^{d}\sum_{m=1}^{d}
    ({\theta}_k-\widehat{\theta}_k)
    \left[n{\Ibold}_{km}(\widehat{\thetabold})\right]
({\theta}_m-\widehat{\theta}_m)\,+\,{\ldots}
\nonumber
\end{eqnarray}
and, under regularity conditions, we can approximate the likelihood
function by a Normal density with mean $\widehat{\thetabold}$ and
covariance matrix ${\Sigmabold}^{-1}=n{\Ibold}(\widehat{\thetabold})$
\rayan
\small
\vspace{0.35cm}
\footnotesize
\noindent

{\bf Example 2.7: The Binomial Distribution}.
Consider the random quantity
$X{\sim}Bi(x|{\theta},n)$:
 \begin{eqnarray*}
   p(x|n,\theta)\,=\,
\left(\begin{array}{c}
  n \\
  x
\end{array}\right)\,
{\theta}^k\,(1-{\theta})^{n-k}
\hspace{0.5cm};\hspace{1.cm}
n,k{\in}N_0;\,k{\leq}n
\nonumber
\end{eqnarray*}
with $0<{\theta}<1$. Since  $E[X]=n{\theta}$ we have that:
\begin{eqnarray*}
   {\Ibold}({\theta})\,=\,
    E_X\left[         \left(-\,
    \frac{\textstyle {\partial}^2 \log  p(x|n,\theta)}
         {\textstyle {\partial} {\theta}^2}
             \right) \right] \,=\,
  \frac{\textstyle n}{\textstyle {\theta}\,(1-{\theta})}
\nonumber
\end{eqnarray*}
so the Jeffreys prior (proper in this case) for the parameter ${\theta}$ is
\begin{eqnarray*}
       {\pi}({\theta})\,{\propto}\,
             \left[
              {\theta}\,(1-{\theta})
           \right]^{-1/2}
\nonumber
\end{eqnarray*}
and the posterior density will therefore be
\begin{eqnarray*}
   p({\theta}|k,n)\,\,{\propto}\,\,
              {\theta}^{k-1/2}\,(1-{\theta})^{n-k-1/2}
\nonumber
\end{eqnarray*}
that is; a $Be(x|k+1/2, n-k+1/2)$ distribution. Since
\begin{eqnarray*}  
{\phi}\,=\,
    \int\,
    \frac{\textstyle d{\theta} }
         {\textstyle \sqrt{{\theta}\,(1-{\theta})}}\,=\,
       2\,{\rm asin}\, ({\theta}^{1/2})
\nonumber
\end{eqnarray*}
we have that ${\theta}={\rm sin}^2{\phi}/2$ and, parameterized in
terms of $\phi$,
${\Ibold}({\phi})$ is constant so the distribution {\sl ``looks''} more
Normal (see fig. 2.2.)

\begin{figure}[t]
\begin{center}

\mbox{\epsfig{file=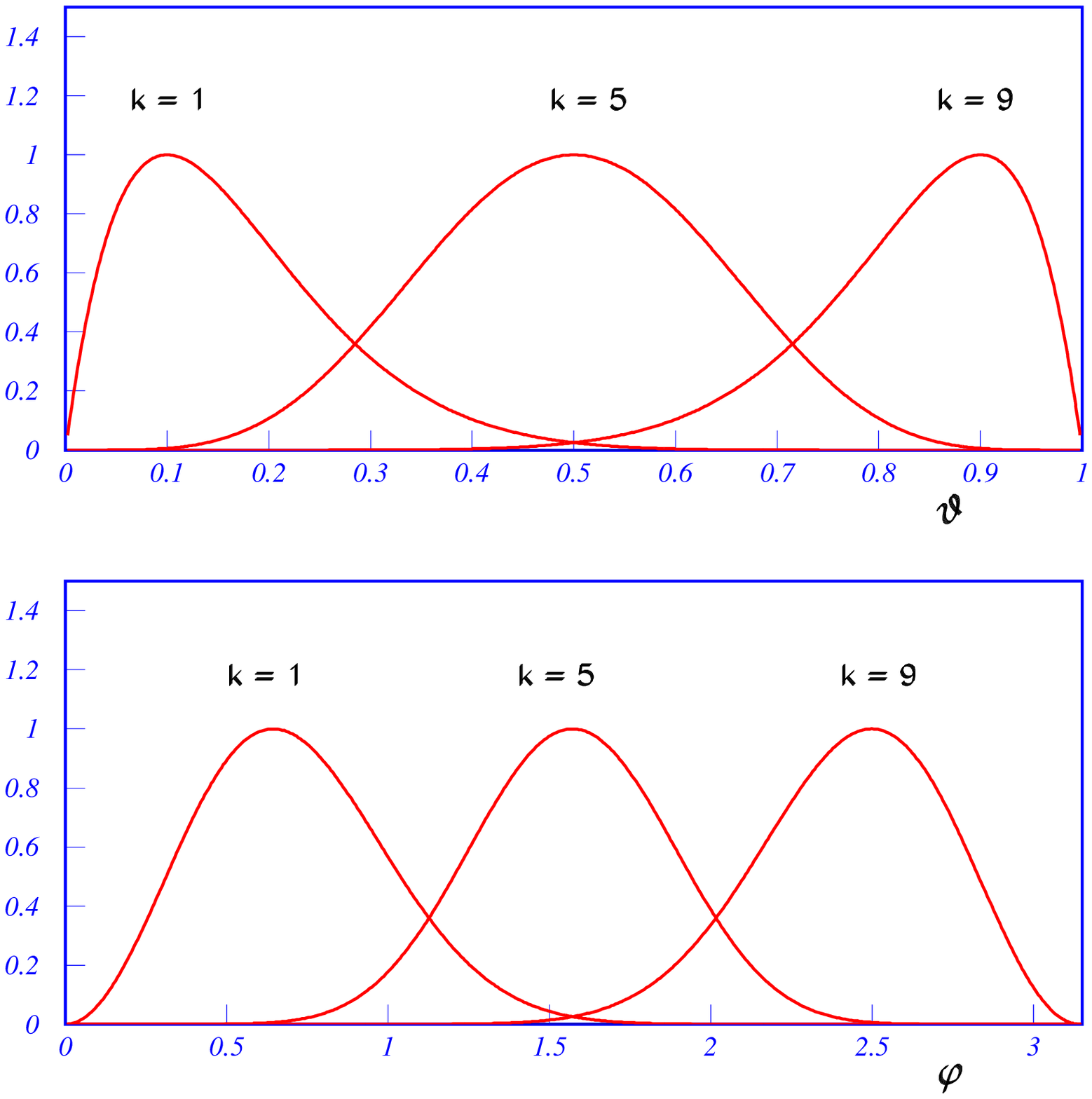,height=11cm,width=11cm}}

\footnotesize{
{\bf Figure 2.2}.- Dependence of the likelihood function with the parameter
${\theta}$ (upper) and with 
${\phi}=2\,{\rm asin}({\theta}^{1/2})$ (lower) for a Binomial process
with $n=10$ and $k=1,\,5$ and 9.  
             }
\end{center}
\end{figure}

\vspace{0.35cm}
\noindent
{\bf Example 2.8: The Poisson Distribution.} 
Consider the random quantity
$X{\sim}Po(x|{\mu})$:
   \begin{eqnarray*}
   p(x|{\mu})\,=\,e^{-{\mu}}\,\frac{\textstyle {\mu}^x}
                                  {\textstyle \Gamma(x+1)}
\hspace{0.5cm};\hspace{1.cm}
x{\in}N;\,{\mu}{\in}R^{+}
                \nonumber 
   \end{eqnarray*}
Then, since $E[X]={\mu}$ we have
\begin{eqnarray*}
   {\Ibold}({\mu})\,=\,
    E_X\left[         \left(-\,
    \frac{\textstyle {\partial}^2 \log p(x|{\mu})}
         {\textstyle {\partial} {\mu}^2}
             \right) \right] \,=\,
  \frac{\textstyle 1}{\textstyle {\mu}}
\nonumber
\end{eqnarray*}
so we shall take as {\sl prior} (improper):
\begin{eqnarray*}  
{\pi}({\mu})\,=\,\left[ {\Ibold}({\mu}) \right]^{1/2}\,=\,
           {\mu}^{-1/2}
\nonumber
\end{eqnarray*}
and make inferences on $\mu$ from the proper posterior density
   \begin{eqnarray*}
   p({\mu}|x)\,{\propto}\,e^{-{\mu}}\,{\mu}^{x-1/2}
                \nonumber 
   \end{eqnarray*}
that is, a $Ga(x|1,x+1/2)$ distribution.

\vspace{0.35cm}
\noindent
{\bf Example 2.9: The Pareto Distribution.} 
Consider the random quantity
$X{\sim}Pa(x|{\theta},x_0)$ with $x_0{\in}R^{+}$ known and density
\begin{eqnarray*}
   p(x|{\theta},x_0)\,=\,\frac{\textstyle \theta}
                              {\textstyle x_0}\,
\left(\frac{\textstyle x_0}{\textstyle x}\right)^{\theta+1}\,
{\mbox{\boldmath $1$}}_{(x_0,\infty)}(x)
\hspace{0.5cm};\hspace{1.cm}{\theta}{\in}R^{+}
                \nonumber 
   \end{eqnarray*}
Then, 
\begin{eqnarray*}
   {\Ibold}({\theta})\,=\,
    E_X\left[         \left(-\,
    \frac{\textstyle {\partial}^2 \log p(x|{\theta},x_0)}
         {\textstyle {\partial} {\theta}^2}
             \right) \right] \,=\,
  \frac{\textstyle 1}{\textstyle {\theta}^2}
\nonumber
\end{eqnarray*}
so we shall take as {\sl prior} (improper):
\begin{eqnarray*}  
{\pi}({\theta})\,{\propto}\,\left[ {\Ibold}({\mu}) \right]^{1/2}\,=\,
           {\theta}^{-1}
\nonumber
\end{eqnarray*}
and make inferences from the posterior density (proper)
   \begin{eqnarray*}
   p({\theta}|x,x_0)\,=\,x^{-\theta}\log x
                \nonumber 
   \end{eqnarray*}
Note that if we make the transformation $t=\log x$, the density becomes 
\begin{eqnarray*}
p(t|{\theta},x_0)={\theta}\,x_0^{\theta}\,e^{-{\theta}t} 
   {\mbox{\boldmath $1$}}_{(\log x_0,\infty)}(t)
                \nonumber 
   \end{eqnarray*}
for which
$\theta$ is a scale parameter and, from previous considerations, 
we should take ${\pi}({\theta})\,{\propto}\,{\theta}^{-1}$ in consistency with
Jeffreys's prior.

\vspace{0.35cm}
\noindent
{\bf Example 2.10: The Gamma Distribution.} 
Consider the random quantity
$X{\sim}Ga(x|{\alpha},{\beta})$ with ${\alpha},{\beta}{\in}R^{+}$ 
and density
\begin{eqnarray*}
   p(x|{\alpha},{\beta})\,=\,\frac{\textstyle {\alpha}^{\beta}}
                              {\textstyle \Gamma({\beta})}\,
e^{-{\alpha}x}\,x^{{\beta}-1}\,
{\mbox{\boldmath $1$}}_{(0,\infty)}(x)
                \nonumber 
   \end{eqnarray*}
Show that the Fisher's matrix is
\begin{eqnarray}
   {\Ibold}({\alpha},{\beta})\,=\,
\left(\begin{array}{cc}

  {\beta}{\alpha}^{-2} & -{\alpha}^{-1} \\
  -{\alpha}^{-1}   & {\Psi}'({\beta})
\end{array}\right)\
\nonumber
\end{eqnarray}
with ${\Psi}'(x)$ the first derivative of the Digamma Function and,
following Jeffreys' rule, we should take the prior 
\begin{eqnarray}  
{\pi}({\alpha},{\beta})\,{\propto}\,{\alpha}^{-1}\,\left[
             {\beta}{\Psi}'({\beta})-1 \right]^{1/2}
\nonumber
\end{eqnarray}
Note that ${\alpha}$ is a scale parameter so, from previous considerations, 
we should take ${\pi}({\alpha})\,{\propto}\,{\alpha}^{-1}$. Furthermore, if we
consider ${\alpha}$ and ${\beta}$ independently, we shall get
\begin{eqnarray}  
{\pi}({\alpha},{\beta})\,=\,{\pi}({\alpha}){\pi}({\beta})
\,{\propto}\,{\alpha}^{-1}\,\left[{\Psi}'({\beta})\right]^{1/2}
\nonumber
\end{eqnarray}

\vspace{0.35cm}
\noindent
{\bf Example 2.11: The Beta Distribution.}

Show that for the $Be(x|{\alpha},{\beta})$ 
distribution with density
\begin{eqnarray}
   p(x|{\alpha},{\beta})\,=\,\frac{\textstyle \Gamma({\alpha}+{\beta})}
                              {\textstyle \Gamma({\alpha})\Gamma({\beta})}\,
(x^{{\alpha}-1}\,(1-x)^{{\beta}-1}\,
{\mbox{\boldmath $1$}}_{[0,1]}(x)
\hspace{0.5cm};\hspace{1.cm}{\alpha},{\beta}{\in}R^{+}
                \nonumber 
   \end{eqnarray}
the Fisher's matrix is given by
\begin{eqnarray}
   {\Ibold}({\alpha},{\beta})\,=\,
\left(\begin{array}{cc}

  {\Psi}'({\alpha})-{\Psi}'({\alpha}+{\beta}) & - {\Psi}'({\alpha}+{\beta})\\
  -{\Psi}'({\alpha}+{\beta})   & {\Psi}'({\beta})-{\Psi}'({\alpha}+{\beta})
\end{array}\right)\
\nonumber
\end{eqnarray}
with ${\Psi}'(x)$ the first derivative of the Digamma Function.

\vspace{0.35cm}
\noindent
{\bf Example 2.12:} {\bf The Normal Distribution.}
\vspace{0.20cm}

\noindent
{\bf Univariate:}
The Fisher's matrix is given by
\begin{eqnarray*}
   {\Ibold}({\mu},{\sigma})\,=\,
\left(\begin{array}{cc}
  {\sigma}^{-2} & 0 \\
  0 & 2{\sigma}^{-2}
\end{array}\right)
\end{eqnarray*}
so 
\begin{eqnarray*}
   {\pi}_1({\mu},{\sigma})\,{\propto}\,
   \left[ {\rm det}[{\Ibold}({\mu},{\sigma}] \right]^{1/2}
   \,{\propto}\,\frac{\textstyle 1}{\textstyle {\sigma}^2}
\end{eqnarray*}
However, had we treated the two parameters independently, we should have
obtained
\begin{eqnarray*}
   {\pi}_2({\mu},{\sigma})\,=\,{\pi}({\mu})\,{\pi}({\sigma})
\,{\propto}\,\frac{\textstyle 1}{\textstyle {\sigma}}
\end{eqnarray*}
The prior ${\pi}_2{\propto}{\sigma}^{-1}$ is the one we had used
in example 2.5 where the problem was treated as two
one-dimensional independent problems and, as we saw: 
\begin{eqnarray*}
    T\,=\,\frac{\textstyle \sqrt{n-1}({\mu}-\overline{x})}
               {\textstyle s}\,{\sim}\,St(t|n-1)
\hspace{0.3cm}{\rm and}\hspace{0.3cm}               
    Z\,=\,
    \frac{\textstyle n\,s^2}
         {\textstyle {\sigma}^2}\,{\sim}\,{\chi}^2(z|n-1)
\end{eqnarray*}
with $E[Z]=n-1$.
Had we used prior ${\pi}_1{\propto}{\sigma}^{-2}$, we would have obtained that
$Z{\sim}{\chi}^2(z|n)$ and therefore $E[Z]=n$.
This is not reasonable. On the one hand, we know
from the sampling distribution $N(x|{\mu},{\sigma})$ that 
$E[ns^2{\sigma}^{-2}]=n-1$. On the other hand, we have two parameters
$({\mu},{\sigma})$ and integrate on one $(\sigma)$ so the number of degrees of
freedom should be $n-1$.
\vspace{0.20cm}

\noindent
{\bf Bivariate:}
The Fisher's matrix is given by
\begin{eqnarray*}
   {\Ibold}({\mu}_1,{\mu}_2)\,=\, (1-{\rho}^2)^{-1}\,
   \left(\begin{array}{cc}
     {\sigma}_1^{-2} & {-\rho}({\sigma}_1{\sigma}_2)^{-1} \\ 
     {-\rho}({\sigma}_1{\sigma}_2)^{-1} & {\sigma}_2^{-2} 
\end{array}\right)
\end{eqnarray*}
\begin{eqnarray*}
   {\Ibold}({\sigma}_1,{\sigma}_2,{\rho})\,=\,(1-{\rho}^2)^{-1}\,
   \left(\begin{array}{ccc}
     (2-{\rho}^2){\sigma}_1^{-2} & 
     -{\rho}^2({\sigma}_1{\sigma}_2)^{-1} & 
     -{\rho}{\sigma}_1^{-1} \\
     -{\rho}^2({\sigma}_1{\sigma}_2)^{-1} &
          (2-{\rho}^2){\sigma}_2^{-2} &
            -{\rho}{\sigma}_2^{-1} \\
     -{\rho}{\sigma}_1^{-1} &
     -{\rho}{\sigma}_2^{-1} &
     (1+{\rho}^2)(1-{\rho}^2)^{-1}
\end{array}\right)
\end{eqnarray*}

\begin{eqnarray*}
 {\Ibold}({\mu}_1,{\mu}_2,{\sigma}_1,{\sigma}_2,{\rho})\,=\,
\left(\begin{array}{cc}
     {\Ibold}({\mu}_1,{\mu}_2) & {\mbox{\boldmath $0$}} \\
     {\mbox{\boldmath $0$}} & {\Ibold}({\sigma}_1,{\sigma}_2,{\rho})
\end{array}\right)
\end{eqnarray*}
Form this,
\begin{eqnarray*}
{\pi}({\mu}_1,{\mu}_2,{\sigma}_1,{\sigma}_2,{\rho})\,{\propto}\,
|{\rm det}{\Ibold}({\mu}_1,{\mu}_2,{\sigma}_1,{\sigma}_2,{\rho})|^{1/2}\,=\,
\frac{\textstyle 1}{\textstyle {\sigma}_1^2{\sigma}_2^2(1-{\rho}^2)^2}
\end{eqnarray*}
while if we consider
${\pi}({\mu}_1,{\mu}_2,{\sigma}_1,{\sigma}_2,{\rho})=
{\pi}({\mu}_1,{\mu}_2)
{\pi}({\sigma}_1,{\sigma}_2,{\rho})$
we get
\begin{eqnarray*}
{\pi}({\mu}_1,{\mu}_2,{\sigma}_1,{\sigma}_2,{\rho})\,{\propto}\,
\frac{\textstyle 1}{\textstyle {\sigma}_1{\sigma}_2(1-{\rho}^2)^{3/2}}
\end{eqnarray*}

\vspace{0.35cm}
\noindent
{\bf Problem 2.1:} Show that for the density
$p(\mbox{\boldmath $x$}|\thetabold)$;
${\xbold}{\in}{\Omega}{\subseteq}R^n$,
the Fisher's matrix (if exists) 
\begin{eqnarray*}  
{\Ibold}_{ij}({\thetabold})\,=\,E_{\Xbold}\left[
    \frac{\textstyle {\partial}\log\,p(\xbold|\thetabold)}
             {\textstyle {\partial} {\theta}_i}\,
    \frac{\textstyle {\partial}\log\,p(\xbold|\thetabold)}
             {\textstyle {\partial} {\theta}_j}\,\right]
\nonumber
\end{eqnarray*}
transforms under a differentiable one-to-one transformation  
$\phibold=\phibold({\thetabold})$ as a covariant symmetric tensor of
second order; that is 
\begin{eqnarray*}  
{\Ibold}_{ij}({\phibold})\,=\,
    \frac{\textstyle {\partial}{\theta}_k}{\textstyle {\partial} {\phi}_i}\,
    \frac{\textstyle {\partial}{\theta}_l}{\textstyle {\partial} {\phi}_j}\,
         {\Ibold}_{kl}({\thetabold})
\nonumber
\end{eqnarray*}

\vspace{0.35cm}
\noindent
{\bf Problem 2.2:} Show that for
$X{\sim}Po(x|{\mu}+b)$ with $b{\in}R^{+}$ known
(Poisson model with known background),
we have that ${\Ibold}({\mu})=({\mu}+b)^{-1}$
and therefore the posterior (proper) is given by:
\begin{eqnarray}
   p({\mu}|x,b)\,{\propto}\,e^{-({\mu}+b)}\,({\mu}+b)^{x-1/2}
                \nonumber 
\end{eqnarray}

\vspace{0.35cm}
\noindent
{\bf Problem 2.3:} Show that for the one parameter mixture model
$p(x|\lambda)={\lambda}p_1(x)+(1-{\lambda})p_2(x)$ with $p_1(x){\neq}p_2(x)$
properly normalized and $\lambda\in(0,1)$,
\begin{eqnarray*}
   \Ibold({\lambda})\,=\,\frac{1}{\lambda(1-\lambda)}\,\left\{
   1-\int_{-\infty}^{\infty}\frac{p_1(x)p_2(x)}{p(x|\lambda)}dx\right\}
\end{eqnarray*}
When $p_1(x)$ and $p_2(x)$ are {\sl "well separated"}, the integral
is $<<1$ and therefore $\Ibold({\lambda})\sim
[\lambda(1-\lambda)]^{-1}$.
On the other hand, 
when they {\sl "get closer"} we can write $p_2(x)=p_1(x)+\eta(x)$ with
$\int_{-\infty}^{\infty}\eta(x)dx=0$ and, after a 
Taylor expansion for $|\eta(x)|<<1$ get to first order that
\begin{eqnarray*}
   \Ibold({\lambda})\,\simeq\,
   \int_{-\infty}^{\infty}\frac{(p_1(x)-p_2(x))^2}{p_1(x)}dx\,+\,...
\end{eqnarray*}
independent of $\lambda$. 
Thus, for this problem it will be sound to consider the prior
${\pi}(\lambda|a,b)= Be(\lambda|a,b)$ with parameters between 
$(1/2,1/2)$ and $(1,1)$.

\small
{\raya}                   
\vspace{1.0cm}
\subsection{Invariance under a Group of Transformations}
Some times, we may be interested to provide the prior with invariance
under some transformations of the parameters (or a subset of them)
considered of interest for the problem at hand. As we have stated,
from a formal point of view the prior can be treated as an
absolute continuous measure with respect to Lebesgue so
$   p({\thetabold}|{\xbold})\,d{\thetabold}\,{\propto}\,
    p({\xbold}|{\thetabold})\,{\pi}({\thetabold})\,d{\thetabold}\,=\,
    p({\xbold}|{\thetabold})\,d{\mu}({\thetabold})$.
Now, consider the probability space $({\Omega},B,{\mu})$ and a measurable 
homeomorphism $T:{\Omega}{\rightarrow}{\Omega}$. A measure ${\mu}$ on the 
Borel algebra $B$ would be invariant by the mapping $T$ if for 
any $A{\subset}B$, we have that ${\mu}(T^{-1}(A))={\mu}(A)$. 
We know, for instance, that there is a unique
measure $\lambda$ on $R^n$ that is invariant under 
translations and such that for the unit cube $\lambda([0,1]^n)=1$: 
the Lebesgue measure (in fact, it could have been defined that way). This
is consistent with the constant prior specified already for position 
parameters. The Lebesgue measure is also the unique measure in $R^n$
that is invariant under the rotation group $SO(n)$ (see problem 2.4).
Thus, when expressed in
spherical polar coordinates, it would be reasonable
for the spherical surface $S^{n-1}$ the rotation invariant prior
\begin{eqnarray}
    d{\mu}({\phibold})\,=\,\prod_{k=1}^{n-1}
    (\sin {\phi}_k)^{(n-1)-k}\,d{\phi}_k
     \nonumber  
\end{eqnarray}
with ${\phi}_{n-1}{\in}[0,2{\pi})$ and ${\phi}_{j}{\in}[0,{\pi}]$ for the rest.
We shall use this prior function in a later problem.

In other cases, the group of invariance is suggested by the model 
\begin{eqnarray*}
M:\{p({\xbold}|{\thetabold}),\,
{\xbold}{\in}{\Omega}_X,\,{\thetabold}{\in}{\Omega}_{\Theta}\}
\end{eqnarray*}
in the sense that
we can make a transformation of the random quantity 
$\Xbold{\rightarrow}{\Xbold}'$
and absorb the change
in a redefinition of the parameters ${\thetabold}{\rightarrow}{\thetabold}'$
such that the expression of the probability density remains unchanged. 
Consider a group of transformations 
\footnote{In this context, the use of Transformation Groups arguments
 was pioneered by E.T. Jaynes [Ja64].}
$G$ that acts

\begin{tabular}[h]{ll}  
   &    \\
       on the Sample Space: &
      $x{\rightarrow}x'=g{\circ}x\,;\hspace{0.5cm}g{\in}G;x,x'{\in}{\Omega}_X$
       \\   & \\
       on the Parametric Space: &
      ${\theta}{\rightarrow}{\theta}'=g{\circ}{\theta}\,;
      \hspace{0.5cm}g{\in}G;{\theta},{\theta}'{\in}{\Omega}_{\Theta}$
       \\
\end{tabular}

\vspace{0.5cm}
\noindent
The model $M$ is said to be invariant under $G$ if ${\forall}g{\in}G$ and 
${\forall}{\theta}{\in}\Omega_{\Theta}$  the random quantity
$X'=g{\circ}X$ is distributed as 
$p(x'|{\theta}'){\equiv}p(g{\circ}x|g{\circ}{\theta})$. Therefore,
transformations of data under $G$ will make no difference on the
inferences if we assign consistent ``prior beliefs'' to the original and 
transformed parameters.
Note that the action 
of the group on the sample and parameter spaces will, in general, be different.
The essential point is that, as Alfred Haar showed in 1933, for the action
of the group $G$ of transformations there is an invariant measure $\mu$
({\sl Haar measure}; [Bo06]) such that
\begin{eqnarray}
  \int_{\Omega_X}f(g{\circ}x)d{\mu}(x)\,=\,
  \int_{\Omega_X}f(x')d{\mu}(x')
     \nonumber  
\end{eqnarray}
for any Lebesgue integrable function $f(x)$ on $\Omega_X$. Shortly after,
it was shown (Von Neumann (1934); Weil and Cartan (1940)) that this measure is
unique up to a multiplicative constant.
In our case, the function will be 
$p({\cdot}|\theta){\mbox{\boldmath ${1}$}}_{\Theta}(\theta)$ and
the invariant measure we are looking for is
$d{\mu}(\theta){\propto}{\pi}(\theta)d{\theta}$. Furthermore,
since the group may be non-abelian, we shall consider the action on the
right and on the left of the parameter space. Thus, we shall have:
\begin{eqnarray}
  \int_{\Theta}p(\cdot|g{\circ}\theta)\,{\pi}_L(\theta)\,d\theta\,=\,
  \int_{\Theta}p(\cdot|{\theta}')\,{\pi}_L({\theta}')\,d{\theta}'
     \nonumber  
\end{eqnarray}
if the group acts on the left and
\begin{eqnarray}
  \int_{\Theta}p(\cdot|{\theta}{\circ}g)\,{\pi}_R(\theta)\,d\theta\,=\,
  \int_{\Theta}p(\cdot|{\theta}')\,{\pi}_R({\theta}')\,d{\theta}'
     \nonumber  
\end{eqnarray}
if the action is on the right.
Then, we should start by identifying the group of transformations under which
the model is invariant (if any; in many cases, either there is no invariance 
or at least not obvious) work in the parameter space.
The most interesting cases for us are:

\begin{tabular}[h]{ll}  
       \\
       Affine Transformations: & $x{\rightarrow}x'=g{\circ}x\,=\,a\,+\,b\,x$
       \\ & \\
       Matrix Transformations: & $x{\rightarrow}x'=g{\circ}x\,=\,R\,x$
       \\
\end{tabular}

\vspace{0.5cm}
\noindent
Translations and scale transformations are a particular case of the first and
rotations of the second. Let's start with the location and scale 
parameters; that is, a density 
\begin{eqnarray}
  p(x|{\mu},{\sigma})\,dx\,=\,\frac{\textstyle 1}{\textstyle \sigma}\,
 f\left( \frac{\textstyle x\,-\,{\mu}}{\textstyle \sigma}\right)\,dx
                \nonumber 
\end{eqnarray}
the Affine group $G=\{g\equiv(a,b);a{\in}R;b{\in}R^{+}\}$
so $x'=g{\circ}x=a+bx$ and the model will be invariant if
\begin{eqnarray*}
({\mu}',{\sigma}')\,=\,g{\circ}({\mu},{\sigma})\,=\,
(a,b){\circ}({\mu},{\sigma})\,=\,(a+b{\mu},b{\sigma})
\end{eqnarray*}
Now,
\begin{eqnarray*}
 \int p({\cdot}|{\mu}',{\sigma}')\,{\pi}_L({\mu}',{\sigma}')\,
       d{\mu}'\,d{\sigma}'\,=\,
  \int p({\cdot}|g{\circ}({\mu},{\sigma})\,{\pi}_L({\mu},{\sigma})\,
       d{\mu}\,d{\sigma}\,= \nonumber \\
=\,\
\int p({\cdot}|{\mu}',{\sigma}')\,\left\{{\pi}_L[g^{-1}({\mu}',{\sigma}')]\,
J({\mu}',{\sigma}';{\mu},{\sigma})\right\}\,d{\mu}'\,d{\sigma}'\,=
\nonumber \\
=\,
\int p({\cdot}|{\mu}',{\sigma}')\,\left\{
{\pi}_L\left(\frac{{\mu}'-a}{b},\frac{{\sigma}'}{b}\right)\,
\frac{1}{b^2}\right\}\,
d{\mu}'\,d{\sigma}'
\end{eqnarray*}
and this should hold for all $(a,b){\in}R{\times}R^{+}$ so, in consequence:
\begin{eqnarray*}
d{\mu}_L({\mu},{\sigma})\,=\,
{\pi}_L({\mu},{\sigma})\,d{\mu}\,d{\sigma}\,{\propto}\,
\frac{\textstyle 1}{\textstyle {\sigma}^2}\,d{\mu}\,d{\sigma}
\end{eqnarray*}
However, the group of Affine Transformations is non-abelian so if we 
study the the action on the left, 
there is no reason why we should not consider also the
action on the right. Since
\begin{eqnarray*}
({\mu}',{\sigma}')\,=\,({\mu},{\sigma}){\circ}g\,=\,
({\mu},{\sigma}){\circ}(a,b)\,=\,({\mu}+a{\sigma},b{\sigma})
\end{eqnarray*}
the same reasoning leads to (left as exercise):
\begin{eqnarray*}
d{\mu}_R({\mu},{\sigma})\,=\,
{\pi}_R({\mu},{\sigma})\,d{\mu}\,d{\sigma}\,{\propto}\,
\frac{\textstyle 1}{\textstyle {\sigma}}\,d{\mu}\,d{\sigma}
\end{eqnarray*}
The first one $({\pi}_L)$ is the one we obtain using Jeffrey's rule
in two dimensions while ${\pi}_R$ is the one we get for position
and scale parameters or Jeffrey's rule treating both parameters
independently; that is, as two one-dimensional problems instead a
one two-dimensional problem. Thus, although from the invariance point of view
there is no reason why one should prefer one over the other, the right
invariant Haar prior gives more consistent results. In fact ([St65],[St70]),
a necessary and sufficient condition for a sequence of posteriors based on 
proper priors to converge in probability to an invariant posterior is that
the prior is the right Haar measure.

\noindent 
\vspace{0.5cm}
\noindent
{\raya}                   
\vspace{0.35cm}
\footnotesize
\noindent
{\bf Problem 2.4:} Show that the measure
$d{\mu}({\theta})=\left[{\theta}(1-{\theta})\right]^{-1/2}d{\theta}$
is invariant under the mapping $T:[0,1]{\rightarrow}[0,1]$ such that
$ T:{\theta}{\rightarrow}{\theta}'=T{\circ}
{\theta}=4{\theta}(1-{\theta})$.
This is the Jeffrey's prior for the Binomial model $Bi(x|N,{\theta})$.

\vspace{0.35cm}
\noindent
{\bf Problem 2.5:} Consider the n-dimensional spherical surface $S_n$  
of unit radius, $\xbold{\in}S_n$ and the transformation
${\xbold}'={\Rbold}\xbold{\in}S_n$ where ${\Rbold}{\in}SO(n)$. 
Show that the Haar invariant measure is the Lebesgue measure on the sphere.

\noindent
Hint: Recall that ${\Rbold}$ is an orthogonal matrix so 
${\Rbold}^t={\Rbold}^{-1}$; that $|{\rm det}{\Rbold}|=1$ so
$J({\xbold}';{\xbold})=|{\partial}{\xbold}/{\partial}{\xbold}')|=
|{\partial}{\Rbold}^{-1}{\xbold}'/{\partial}{\xbold}')|=
|{\rm det}{\Rbold}|=1$ and that
${\xbold}'^{t}{\xbold}'={\xbold}^{t}{\xbold}=1$.

\vspace{0.35cm}
\noindent
{\bf Example 2.12: Bivariate Normal Distribution.}
Let $\Xbold=(X_1,X_2){\sim}
N(\xbold|\mbox{\boldmath ${0}$},{\phibold})$ 
with ${\phibold}=\{{\sigma}_1,{\sigma}_2,{\rho}\}$; that is:
\begin{eqnarray*}
    p({\xbold}|{\phibold})\,=\,(2{\pi})^{-1}\,|{\rm det}[{\Sigmabold}]|^{-1/2}\,
       {\rm exp}\left\{ 
     -\,\frac{1}{2}\left(\xbold^t\,{\Sigmabold}^{-1}\,\xbold\right)
                \right\}
\end{eqnarray*}
with the covariance matrix
\begin{eqnarray*}
   {\Sigmabold}\,=\,
   \left(\begin{array}{cc}
     {\sigma}_1^{2} & {\rho}{\sigma}_1{\sigma}_2 \\ 
     {\rho}{\sigma}_1{\sigma}_2 & {\sigma}_2^{2} 
\end{array}\right)
\hspace{0.3cm}{\rm and}\hspace{0.3cm}
{\det}[{\Sigmabold}]={\sigma}_1^{2} {\sigma}_2^{2}
(1-{\rho})^2
 \end{eqnarray*}
Using the Cholesky decomposition we can express 
${\Sigmabold}^{-1}$ 
as the product of two lower (or upper) triangular matrices:
\begin{eqnarray*}
 {\Sigmabold}^{-1}=\frac{1}{{\det}[{\Sigmabold}]}
   \left(\begin{array}{cc}
     {\sigma}_2^{2} & -{\rho}{\sigma}_1{\sigma}_2 \\ 
     -{\rho}{\sigma}_1{\sigma}_2 & {\sigma}_1^{2} 
\end{array}\right)
   ={\Abold}^t{\Abold}
   \hspace{0.3cm}{\rm with}\hspace{0.3cm}
   {\Abold}=\left(\begin{array}{cc}
     \frac{\textstyle 1}{\textstyle {\sigma}_1} & 0 \\
     \frac{\textstyle -{\rho}}{\textstyle {\sigma}_1\sqrt{1-{\rho}^2}} & 
     \frac{\textstyle 1}{\textstyle {\sigma}_2\sqrt{1-{\rho}^2}} 
\end{array}\right)
\end{eqnarray*}For the action on the left:
\begin{eqnarray*}
   {\Mbold}\,=\,{\Tbold}\,=\,
   \left(\begin{array}{cc}
     a & 0 \\
     b & c 
\end{array}\right)\hspace{0.3cm};{a,b>0}\hspace{0.3cm}
{\longrightarrow}\hspace{0.3cm}
J({\Abold}';{\Abold})\,=\,a^2c
 \end{eqnarray*}
and, in consequence
\begin{eqnarray*}
{\pi}(aa'_{11},aa'_{21}+ba'_{22},ca'_{22})\,ac^2\,=\,
{\pi}(a'_{11},a'_{21},a'_{22})\,\longrightarrow\,
{\pi}(a'_{11},a'_{21},a'_{22})\,\propto\,
\frac{\textstyle 1}{\textstyle {a'_{11}}^{2}a'_{22}}
\end{eqnarray*}
and
${\det}[{\Sigmabold}]=({\det}[{\Sigmabold}^{-1}])^{-1}=
({\det}[{\Abold}])^{-2}$.
Thus, in the new parameterization ${\thetabold}=\{a_{11},a_{21},a_{22}\}$
\begin{eqnarray*}
    p({\xbold}|{\thetabold})\,=\,(2{\pi})^{-1}\,
    |{\rm det}[{\Abold}]|\,
       {\rm exp}\left\{ 
     -\,\frac{1}{2}\left(\xbold^t\,{\Abold}^{t}{\Abold}\,\xbold\right)
                \right\}
\end{eqnarray*}
Consider now the group of lower triangular $2x2$ matrices
\begin{eqnarray*}
G_l\,=\,\{\Tbold\,{\in}\,LT_{2x2}\,;\hspace{0.3cm}T_{ii}>0\}
\end{eqnarray*}
Since $\Tbold^{-1}{\in}G_l$, inserting the identity matrix 
${\Ibold}={\Tbold}{\Tbold}^{-1}={\Tbold}^{-1}{\Tbold}$ we have:
action 
\begin{center}
\begin{tabular}[h]{cc}  
       \\
    \underline{On the Left} &  \underline{On the Right} \\ & \\
    ${\Tbold}{\circ}{\xbold}{\rightarrow}{\Tbold}{\xbold}={\xbold}'$ &
    ${\xbold}{\circ}{\Tbold}{\rightarrow}{\Tbold}^{-1}{\xbold}={\xbold}'$ 
    \\ & \\
$\left[{\xbold}^t\,({\Tbold}^t({\Tbold}^t)^{-1})\,{\Abold}^{t}{\Abold}\,
({\Tbold}^{-1}{\Tbold})\,{\xbold}\right]$ & 
$\left[{\xbold}^t\,(({\Tbold}^t)^{-1}{\Tbold}^t)\,{\Abold}^{t}{\Abold}\,
({\Tbold}{\Tbold}^{-1})\,{\xbold}\right]$ \\ & \\
${\Mbold}\,=\,{\Tbold}$ & ${\Mbold}\,=\,{\Tbold}^{-1}$
\end{tabular}
\end{center}
Then
\begin{eqnarray*}
  {\Mbold}{\xbold}={\xbold}'\,\hspace{0.3cm};\hspace{0.3cm}
  {\xbold}={\Mbold}^{-1}{\xbold}'\,\hspace{0.3cm};\hspace{0.3cm}
  {{\xbold}'}^{t}={\xbold}^{t}{\Mbold}^{t}\,\hspace{0.3cm}{\rm and}\hspace{0.3cm}
  d{\xbold}=\frac{1}{|{\rm det}[{\Mbold}]|}d{\xbold}'
\end{eqnarray*}
so
\begin{eqnarray*}
    p({\xbold}'|{\thetabold})\,=\,(2{\pi})^{-1}\,
    \frac{|{\rm det}[{\Abold}]|}{|{\rm det}[{\Mbold}]|}\,
       {\rm exp}\left\{ 
     -\,\frac{1}{2}\left({{\xbold}'}^t\,({\Abold}{\Mbold}^{-1})^{t}
({\Abold}{\Mbold}^{-1})\,
           {\xbold}'\right)
                \right\}
\end{eqnarray*}
and the model is invariant under $G_l$ if the action on the parameter space is
\begin{eqnarray*}
    G_l:{\Abold}{\longrightarrow}{\Abold}'\,=\,{\Abold}{\Mbold}^{-1}
\hspace{0.3cm};\hspace{0.3cm}
{\Abold}\,=\,{\Abold}'{\Mbold}
\hspace{0.3cm};\hspace{0.3cm}
{\rm det}[{\Abold}]\,=\,{\rm det}[{\Abold}']\,{\rm det}[{\Mbold}]
\end{eqnarray*}
so
\begin{eqnarray*}
    p({\xbold}'|{\thetabold}')\,=\,(2{\pi})^{-1}\,
    |{\rm det}[{\Abold}']|\,
       {\rm exp}\left\{ 
     -\,\frac{1}{2}\left({{\xbold}'}^t\,{{\Abold}'}^{t}
                          {\Abold}'\,{\xbold}'\right)
                \right\}
\end{eqnarray*}
Then, the Haar equation reads
\begin{eqnarray*}
  \int_{\Theta}p(\bullet|{\Abold}')\,{\pi}({\Abold}')\,d{\Abold}'
  \,=\,
\int_{\Theta}p(\bullet|g{\circ}\Abold)\,{\pi}(\Abold)\,d\Abold\,=\,
\int_{\Theta}p(\bullet|{\Abold}')\,{\pi}({\Abold}'{\Mbold})\,
J({\Abold}';{\Abold})\,d\Abold
     \nonumber  
\end{eqnarray*}
and, in consequence, ${\forall}{\Mbold}{\in}G$
\begin{eqnarray*}
{\pi}({\Abold}'{\Mbold})\,J({\Abold}';{\Abold})\,da'_{11}da'_{21}da'_{22}
\,=\,
{\pi}({\Abold}')\,da'_{11}da'_{21}da'_{22}
\end{eqnarray*}
For the action on the left:
\begin{eqnarray*}
   {\Mbold}\,=\,{\Tbold}\,=\,
   \left(\begin{array}{cc}
     a & 0 \\
     b & c 
\end{array}\right)\hspace{0.3cm};{a,b>0}\hspace{0.3cm}
{\longrightarrow}\hspace{0.3cm}
J({\Abold}';{\Abold})\,=\,a^2c
 \end{eqnarray*}
and, in consequence
\begin{eqnarray*}
{\pi}(aa'_{11},aa'_{21}+ba'_{22},ca'_{22})\,a^2c\,=\,
{\pi}(a'_{11},a'_{21},a'_{22})\,\longrightarrow\,
{\pi}(a'_{11},a'_{21},a'_{22})\,\propto\,
\frac{\textstyle 1}{\textstyle {a'_{11}}^{2}a'_{22}}
\end{eqnarray*}
For the action on the right:
\begin{eqnarray*}
   {\Mbold}\,=\,{\Tbold}^{-1}\,=\,
   \left(\begin{array}{cc}
     a^{-1} & 0 \\
     -b(ac)^{-1} & c^{-1} 
\end{array}\right)\hspace{0.3cm}
{\longrightarrow}\hspace{0.3cm}
J({\Abold}';{\Abold})\,=\,(ac^2)^{-1}
 \end{eqnarray*}
and, in consequence
\begin{eqnarray*}
{\pi}(\frac{a'_{11}}{a},\frac{ca'_{21}-ba'_{22}}{ac},
\frac{a'_{22}}{c})\,\frac{1}{ac^2}\,=\,
{\pi}(a'_{11},a'_{21},a'_{22})\,\longrightarrow\,
{\pi}(a'_{11},a'_{21},a'_{22})\,\propto\,
\frac{\textstyle 1}{\textstyle a'_{11}{a'_{22}}^2}
\end{eqnarray*}
In terms of the parameters of interest $\{{\sigma}_1,{\sigma}_2,{\rho}\}$,
since
\begin{eqnarray*}
     da_{11}da_{21}da_{22}=
     \frac{\textstyle 1}{\textstyle {\sigma}_1^2{\sigma}_2^2(1-{\rho}^2)^2}\,
     d{\sigma}_1d{\sigma}d{\rho}
\end{eqnarray*}
we have finally that for invariance under $G_l$:
\begin{eqnarray*}
  {\pi}_L^l({\sigma}_1,{\sigma}_2,{\rho})\,=\,
     \frac{\textstyle 1}{\textstyle {\sigma}_1{\sigma}_2(1-{\rho}^2)^{3/2}}
\hspace{0.5cm}{\rm and}\hspace{0.5cm}
  {\pi}_R^l({\sigma}_1,{\sigma}_2,{\rho})\,=\,
     \frac{\textstyle 1}{\textstyle {\sigma}_2^2(1-{\rho}^2)}
\end{eqnarray*}
The same analysis with decomposition in upper triangular matrices
leads to
\begin{eqnarray*}
  {\pi}_L^u({\sigma}_1,{\sigma}_2,{\rho})\,=\,
     \frac{\textstyle 1}{\textstyle {\sigma}_1{\sigma}_2(1-{\rho}^2)^{3/2}}
\hspace{0.5cm}{\rm and}\hspace{0.5cm}
  {\pi}_R^u({\sigma}_1,{\sigma}_2,{\rho})\,=\,
     \frac{\textstyle 1}{\textstyle {\sigma}_1^2(1-{\rho}^2)}
\end{eqnarray*}
As we see, in both cases the left Haar invariant prior coincides with
Jeffrey's prior when $\{{\mu}_1,{\mu}2\}$ and $\{{\sigma}_1,{\sigma}2,{\rho}\}$
are decoupled.

At this point, one may be tempted to use a right Haar invariant prior 
where the two
parameters ${\sigma}_1$ and ${\sigma}_2$ are treated on equal footing
\begin{eqnarray*}
  {\pi}({\sigma}_1,{\sigma}_2,{\rho})\,=\,
     \frac{\textstyle 1}{\textstyle {\sigma}_1{\sigma}_2(1-{\rho}^2)}
\end{eqnarray*}
Under this prior, since the sample correlation
\begin{eqnarray*}
  r\,=\,
     \frac{\textstyle \sum_i(x_{1i}-\overline{x}_1)(x_{2i}-\overline{x}_2)}
          {\textstyle \left( \sum_i(x_{1i}-\overline{x}_1)^2
                              \sum_i(x_{2i}-\overline{x}_2)^2\right)^{1/2}}
\end{eqnarray*}
is a sufficient statistics for $\rho$, we have that the posterior for
inferences on the correlation coefficient will be
\begin{eqnarray*}
  p({\rho}|\xbold)\,\propto\,(1-{\rho}^2)^{(n-3)/2}\,
  F(n-1,n-1,n-1/2;(1+r{\rho})/2)
\end{eqnarray*}
with $F(a,b,c;z)$ the Hypergeometric Function.

\vspace{0.35cm}
\noindent
{\bf Example 2.13:} If 
${\theta}{\in}{\Theta}\,{\longrightarrow}\,g{\circ}{\theta}={\phi}({\theta})=
{\theta}'{\in}{\Theta}$
with ${\phi}({\theta})$ is a one-to-one differentiable mapping, then
\begin{eqnarray*}
\int_{\Theta}p({\bullet}|{\theta}')d{\mu}({\theta})\,&=&\,
\int_{\Theta}p({\bullet}|{\theta}'){\pi}({\theta})d{\theta}\,=\,
\int_{\Theta}p({\bullet}|{\theta}'){\pi}({\phi}^{-1}({\theta}'))
\left|\frac{\partial {\phi}^{-1}({\theta}')}{\partial {\theta}'}\right|
d{\theta}'\,=\\
&=&\,\int_{\Theta}p({\bullet}|{\theta}'){\pi}({\theta}')d{\theta}'\,=\,
\int_{\Theta}p({\bullet}|{\theta}')d{\mu}({\theta}')
\end{eqnarray*}
and therefore,
Jeffreys' prior defines a Haar invariant measure.

\small
{\raya}                   
\vspace{1.0cm}

\subsection{Conjugated Distributions}
In as much as possible, we would like to consider reference priors
${\pi}({\theta}|a,b,...)$ versatile enough such that
by varying some of the parameters $a,b,{\ldots}$ we get diverse forms
to analyze the effect on the final results and, on the other hand,
to simplify the evaluation of integrals like:
\begin{eqnarray}
    p(x)\,=\,\int\,p(x|{\theta}){\cdot}p({\theta})\,d{\theta}
\,\,\,\,\,\,\,\,\,\,{\rm and}\,\,\,\,\,\,\,\,\,\,
    p(y|x)\,=\,\int\,p(y|{\theta}){\cdot}p({\theta}|x)\,d{\theta}
                \nonumber 
\end{eqnarray}
This leads us to consider as reference priors the 
{\sl Conjugated Distributions} [Ra61].

Let $\Scal$ be a class of sampling distributions 
$p(x|{\theta})$ and $\Pcal$ the class of prior densities for the
parameter ${\theta}$. If 
\begin{eqnarray}
    p({\theta}|x)\,{\in}\,{\Pcal}
\,\,\,\,\,\,\,\,\,\,{\rm for}\,\,{\rm all}\,\,\,\,\,\,\,\,\,\,
    p(x|{\theta})\,{\in}\,{\Scal}
\,\,\,\,\,{\rm and}\,\,\,\,\,
    p({\theta})\,{\in}\,{\Pcal}
                \nonumber 
\end{eqnarray}
we say that the class $\Pcal$ is conjugated to $\Scal$. We are mainly 
interested in  the class of priors $\Pcal$ that have the same functional 
form as the likelihood. In this case, since both the prior density and the
posterior belong to the same family of distributions, we say that they
are {\sl closed under sampling}. 
It should be stressed that the criteria for taking conjugated reference priors
is eminently practical and, in many cases, they do not exist.
In fact, only the {\sl exponential family} of distributions has
conjugated prior densities.
Thus, if
$\xbold=\{ x_1,x_2,{\ldots},x_n \}$
is an exchangeable random sampling from the 
k-parameter regular exponential family, then
\begin{eqnarray}
  p(\xbold|\thetabold)\,=\,
 f(\xbold)\,g(\thetabold)\,
  {\rm exp}\left\{\sum_{j=1}^k\,c_j\,{\phi}_j(\thetabold)\,
 \left( \sum_{i=1}^n h_j(x_i) \right)
      \right\}
    \nonumber
\end{eqnarray}
and the {\sl conjugated prior} will have the form: 
\begin{eqnarray}
  {\pi}({\thetabold}|{\taubold})\,=\,
 \frac{\textstyle 1}
      {\textstyle K(\taubold)}\,
      [g({\thetabold})]^{{\tau}_0}\,
  {\rm exp}\left\{\sum_{j=1}^k\,c_j\,
{\phi}_j({\thetabold})\,{\tau}_j \right\}
    \nonumber
\end{eqnarray}
where
$\mbox{\boldmath ${\theta}$}{\in}{\Theta}$,
${\taubold}=\{ {\tau}_0,{\tau}_1,{\ldots},{\tau}_k \}$ the 
{\sl hyperparameters} and $K(\taubold)<\infty$ the normalization factor so
$\int_{\Theta}{\pi}({\thetabold}|{\taubold})d{\thetabold}=1$.
Then, the general scheme will be
\footnote{We can go an step upwards and assign a prior to the
hyperparameters with hyper-hyperparameters,...}:
\begin{itemize}
\item[1)] Choose the class of priors ${\pi}({\thetabold}|{\taubold})$ 
that reflect the structure of the model;
\item[2)] Choose a prior function $\pi({\taubold})$ for the 
{\sl hyperparameters};
\item[3)] Express the posterior density as
$p(\thetabold,\taubold|\xbold){\propto}p(\xbold|\thetabold)
{\pi}({\thetabold}|{\taubold}){\pi}({\taubold})$;
\item[4)] Marginalize for the parameters of interest:
\begin{eqnarray*}
p(\thetabold|\xbold){\propto}\int_{\Phi}p(\xbold|\thetabold)
{\pi}({\thetabold}|{\taubold}){\pi}({\taubold})d{\taubold}
\end{eqnarray*}
or, if desired, get the conditional density 
\begin{eqnarray*}
  p(\thetabold|\xbold,\taubold)\,=\,\frac{p(\xbold,\thetabold,\taubold)}
                                       {p(\xbold,\taubold)}\,=\,
             \frac{p(\xbold|\thetabold){\pi}(\thetabold|\taubold)}
                          {p(\xbold|\taubold)}
\end{eqnarray*}
\end{itemize}

The obvious question that arises is how do we choose the prior
${\pi}({\phibold})$ for the hyperparameters. Besides {\sl reasonableness},
we may consider two approaches. Integrating the parameters $\thetabold$
of interest, we get
\begin{eqnarray*}
  p(\taubold,\xbold)\,=\,{\pi}(\taubold)\,\int_{\Theta}
  p(\xbold|\thetabold){\pi}(\thetabold|\taubold)\,d{\thetabold}\,=\,
  {\pi}(\taubold)\,p(\xbold|\taubold)
\end{eqnarray*}
so we may use any of the procedures under discussion to take
${\pi}(\taubold)$ as the prior for the model $p(\xbold|\taubold)$ and then
obtain
\begin{eqnarray*}
{\pi}(\thetabold)\,=\,\int_{\Omega_{\tau}}
  {\pi}(\thetabold|\taubold)\,
  {\pi}(\taubold)\,d{\taubold}
\end{eqnarray*}
The beauty of Bayes rule but not very practical in complicated 
situations. A second approach, more ugly and practical, is the so called
{\sl Empirical Method} where we assign numeric values to the hyperparameters
suggested by $p({\xbold}|{\taubold})$ (for instance, moments, 
maximum-likelihood estimation,...); that is, setting, in a distributional sense,
${\pi}(\taubold)={\delta}_{{\taubold}^{\star}}$ so
$<{\pi}(\taubold),p(\thetabold,\xbold,\taubold)>=
p(\thetabold,\xbold,{\taubold}^{\star})$. Thus, 
\begin{eqnarray*}
  p(\thetabold|\xbold,\taubold^{\star})
  \propto\,
           p(\xbold|\thetabold){\pi}(\thetabold|\taubold^{\star})
\end{eqnarray*}
Obviously, fixing
the hyperparameters assumes a perfect knowledge of them and does not
allow for variations but the procedure may be useful to guess at least were
to go.

Last, it may happen that a single conjugated prior does not represent 
sufficiently well our beliefs. In this case, we may consider
a k-mixture of conjugated priors
\begin{eqnarray*}
{\pi}(\thetabold|\taubold_1,{\ldots},\taubold_k)\,=\,
  \sum_{i=1}^{k}\,w_i\,{\pi}(\thetabold|\taubold_i)
\end{eqnarray*}
In fact [Da83], any prior density for 
a model that belongs to the exponential family can be approximated 
arbitrarily close by a mixture of conjugated priors.
\vspace{0.5cm}
\noindent
{\raya}                   
\vspace{0.35cm}
\footnotesize

\noindent
{\bf Example 2.14:} Let's see the conjugated prior 
distributions for some models:

\noindent
$\bullet$ {\bf Poisson model} $Po(n|{\mu})$: Writing
   \begin{eqnarray}
     p(n|{\mu})\,=\,\frac{\textstyle e^{-{\mu}}\,{\mu}^n}
                         {\textstyle {\Gamma}(n+1)}\,=\,
                    \frac{\textstyle e^{-({\mu}-n{\log}\,{\mu})}}
                         {\textstyle {\Gamma}(n+1)}
                \nonumber 
   \end{eqnarray}
it is clear that the Poisson distribution belongs to the exponential family
and the conjugated prior density for the parameter ${\mu}$ is
\begin{eqnarray*}
     {\pi}({\mu}|{\tau}_1,{\tau}_2)\,{\propto}\,
        e^{-{\tau}_1{\mu}+{\tau}_2{\log}\,{\mu}}\,{\propto}\,
Ga({\mu}|{\tau}_1,{\tau}_2)
\end{eqnarray*}
If we set a prior ${\pi}(\tau_1,\tau_2)$ for the hyperparameters
we can write
\begin{eqnarray*}
p(n,\mu,\tau_1,\tau_2)\,p(n|\mu)\,{\pi}(\mu|\tau_1,\tau_2)\,=\,
{\pi}(\tau_1,\tau_2)
\end{eqnarray*}
and integrating $\mu$:
\begin{eqnarray*}
p(n,\tau_1,\tau_2)\,=\,\left[
   \frac{\Gamma(n+\tau_2)}{\Gamma(\tau_1)}\,
   \frac{\tau_1^{\tau_2}}{(1+\tau_1)^{n+\tau_2}}\,
\right]
{\pi}(\tau_1,\tau_2)\,=\,p(n|\tau_1,\tau_2)\,{\pi}(\tau_1,\tau_2)
\end{eqnarray*}

\noindent
$\bullet$ {\bf Binomial model} $Bi(n|N,{\theta})$: Writing
   \begin{eqnarray}
    P(n|N,{\theta})\,=\,
       \left(  \begin{array}{c}
                 N \\ n
               \end{array}
     \right)\,{\theta}^n\,(1-{\theta})^{N-n}\,=\,
       \left(  \begin{array}{c}
                 N \\ n
               \end{array}
     \right)\,e^{n\,{\log}\,{\theta}\,+\,(N-n)\,{\log}\,(1-{\theta})}
                \nonumber 
   \end{eqnarray}
it is clear that it belong to the exponential family and
and the conjugated prior density for the parameter
${\theta}$ will be:
\begin{eqnarray}
    {\pi}({\theta}|{\tau}_1,{\tau}_2)\,=\,Be({\tau}|{\tau}_1,{\tau}_2)
               \nonumber
  \end{eqnarray}

\noindent
$\bullet$ {\bf Multinomial model} 
Let $\Xbold=(X_1,X_2,{\ldots},X_k){\sim}Mn(\xbold|\thetabold)$; that is:
\begin{eqnarray*}
    \Xbold{\sim}p(\xbold|{\thetabold})\,=\,\Gamma(n+1)\,\prod_{i=1}^{k}
    \frac{{\theta}_i^{x_i}}{\Gamma(x_i+1)}
\hspace{0.5cm}
\left\{
\begin{array}{l}
  X_i{\in}N\,,\hspace{0.5cm} \sum_{i=1}^k X_i=n \\ \\
   {\theta}_i{\in}[0,1]\,,\hspace{0.5cm} \sum_{i=1}^k{\theta}_i=1
               \end{array}
\right.
\end{eqnarray*}
The Dirichlet distribution $Di(\thetabold|\alphabold)$:
\begin{eqnarray*}
    {\pi}({\thetabold}|{\alphabold})\,=\,D(\alphabold)\,\prod_{i=1}^{k}
    {\theta}_i^{{\alpha}_i-1}
\hspace{0.5cm}
\left\{
\begin{array}{l}
  {\alphabold}=({\alpha}_1,{\alpha}_2,\ldots,{\alpha}_k),\hspace{0.3cm}
  {\alpha}_i>0\,,\hspace{0.3cm} \sum_{i=1}^k {\alpha}_i=\alpha_0 \\ \\
  D({\alphabold})={\Gamma(\alpha_0)}\,
    \left[ \prod_{i=1}^k\Gamma({\alpha}_i) \right]^{-1}
               \end{array}
\right.
\end{eqnarray*}
is the natural conjugated prior for this model. It is a degenerated distribution
in the sense that
\begin{eqnarray*}
    {\pi}({\thetabold}|{\alphabold})\,=\,D(\alphabold)\,
   \left[\prod_{i=1}^{k-1}{\theta}_i^{{\alpha}_i-1}\right]\,
   \left[1-\sum_{i=1}^{k-1}{\theta}_i\right]^{\alpha_k-1}
\end{eqnarray*}
The posterior density will then be 
${\thetabold}{\sim}Di(\thetabold|\xbold+\alphabold)$ with
\begin{eqnarray*}
    E[\theta_i]\,=\,\frac{x_i+\alpha_i}{n+\alpha_0}
    \hspace{0.5cm}{\rm and}\hspace{0.5cm}
    V[\theta_i,\theta_j]\,=\,\frac{E[\theta_i]({\delta}_{ij}-E[\theta_j])}
{n+\alpha_0+1}\,
\end{eqnarray*}

The parameters $\alphabold$ of 
the Dirichlet distribution $Di(\thetabold|\alphabold)$
determine the expected values
$E[\theta_i]\,=\,{\alpha}_i/{\alpha}_0$. In practice, it is more convenient
to control also the variances and use the {\sl Generalized Dirichlet
Distribution} $GDi(\thetabold|\alphabold,\betabold)$:
\begin{eqnarray*}
    {\pi}({\thetabold}|{\alphabold},\betabold)\,=\,
\prod_{i=1}^{k-1}\, \frac{\Gamma(\alpha_i+\beta_i)}
                      {\Gamma(\alpha_i)\Gamma(\beta_i)}\,
                      {\theta}_i^{{\alpha}_i-1}
                       \left[1-\sum_{j=1}^{i}{\theta}_j\right]^{\gamma_i}
\end{eqnarray*}
where:
\begin{eqnarray*}
&&0<{\theta}_i<1\,,\hspace{0.3cm} \sum_{i=1}^{k-1} {\theta}_i<1\,
   ,\hspace{0.3cm} {\theta}_n=1-\sum_{i=1}^{k-1} {\theta}_i \\ \\
&&{\alpha}_i>0\,,\hspace{0.3cm}{\beta}_i>0\,,
\hspace{0.3cm}{\rm and}\hspace{0.3cm}
{\gamma}_i\,\left\{
\begin{array}{l}
  {\beta}_i-{\alpha}_{i+1}-{\beta}_{i+1}\,;\hspace{0.3cm}i=1,2,{\ldots},k-2 \\
  {\beta}_{k-1}-1\,;\hspace{0.3cm}i=k-1 
               \end{array}
\right.
\end{eqnarray*}
When ${\beta}_i=\alpha_{i+1}+\beta_{i+1}$ it becomes the Dirichlet distribution.
For this prior we have that
\begin{eqnarray*}
    E[\theta_i]\,=\,\frac{\alpha_i}{\alpha_i+\beta_i}\,S_i
    \hspace{0.5cm}{\rm and}\hspace{0.5cm}
    V[\theta_i,\theta_j]\,=\,E[{\theta}_j]\left(
      \frac{\alpha_i+\delta_{ij}}{\alpha_i+\beta_i+1}T_i-E[\theta_i]\right)
\end{eqnarray*}
where
\begin{eqnarray*}
    S_i\,=\,\prod_{j=1}^{i-1}\frac{\beta_j}{\alpha_j+\beta_j}
    \hspace{0.5cm}{\rm and}\hspace{0.5cm}
    T_i\,=\,\prod_{j=1}^{i-1}\frac{\beta_j+1}{\alpha_j+\beta_j+1}
\end{eqnarray*}
with $S_1=T_1=1$ and we can have control over the prior means and variances.

\small
{\raya}                   
\vspace{1.0cm}
\subsection{Probability Matching Priors}
A {\sl probability matching prior} is a prior function such that
the one sided credible intervals derived from the posterior distribution
coincide, to a certain level of accuracy, with those derived by
the classical approach. This condition leads to a differential
equation for the prior distribution [We63],[Go84]. 
We shall illustrate in the following lines the rationale behind 
for the simple one parameter case assuming 
that the needed regularity conditions are satisfied. 

\vspace{0.5cm}
\noindent
{\rayan}                   
\vspace{0.35cm}
\footnotesize
\noindent

Consider then a random quantity $X{\sim}p(x|{\theta})$ and an iid sampling
$\mbox{\boldmath $x$}\,=\,\{x_1,x_2,{\ldots},x_n\}$ with
$\theta$ the parameter of interest. 
The classical approach for inferences is based on the the likelihood 
\begin{eqnarray*}
p(\mbox{\boldmath $x$}|{\theta})\,=\,
p(x_1,x_2,{\ldots}x_n|{\theta})\,=\,
\prod_{i=1}^n\,p(x_i|{\theta})
\end{eqnarray*}
and goes through the following reasoning:
\begin{itemize}
\item[1)] Assumes that the parameter $\theta$ has the {\sl true} but
          unknown  value $\theta_0$ so the sample is actually drawn
          from $p(x|\theta_0)$;
\item[2)] Find the estimator ${\theta}_m(\mbox{\boldmath $x$})$ 
          of $\theta_0$ as the value of $\theta$ 
	  that maximizes the likelihood; that is:
\begin{eqnarray*}
{\theta}_m\,=\,\max_{\theta}\{ p(\mbox{\boldmath $x$}|{\theta}) \}\,
\hspace{0.3cm}\longrightarrow\hspace{0.3cm}
\left(
\frac{\textstyle \partial \ln p(\mbox{\boldmath $x$}|{\theta})}
     {\textstyle \partial \theta}
\right)_{{\theta}_m}\,=\,0
\end{eqnarray*}
\item[3)] Given the model $X{\sim}p(x|\theta_0)$, after the appropriate change 
          change of variables get the distribution 
	  \begin{eqnarray*}
            p({\theta}_m|{\theta}_0)
	  \end{eqnarray*}
	  of the random quantity ${\theta}_m(X_1,X_2,{\ldots}X_n)$ and
	  draw inferences from it.
\end{itemize}
The Bayesian inferential process 
considers a prior distribution ${\pi}({\theta})$
and draws inferences on ${\theta}$ from the posterior
distribution of the quantity of interest
\begin{eqnarray*}
p({\theta}|\mbox{\boldmath $x$})\,\propto\,
p(\mbox{\boldmath $x$}|{\theta})\,{\pi}({\theta})
\end{eqnarray*}

Let's start with the Bayesian and expand the term
on the right around ${\theta}_m$. On the one hand:
\begin{eqnarray*}
\ln \frac{p(\mbox{\boldmath $x$}|{\theta})}
         {p(\mbox{\boldmath $x$}|{\theta}_m)}=
\frac{\textstyle 1}{\textstyle 2!}
\left(
\frac{\textstyle \partial^2 \ln p(\mbox{\boldmath $x$}|{\theta})}
     {\textstyle \partial \theta^2}
\right)_{{\theta}_m}(\theta-\theta_m)^2+ 
\frac{\textstyle 1}{\textstyle 3!}
\left(
\frac{\textstyle \partial^3 \ln p(\mbox{\boldmath $x$}|{\theta})}
     {\textstyle \partial \theta^3}
\right)_{{\theta}_m}(\theta-\theta_m)^3+{\ldots}
\end{eqnarray*}

\noindent
Now, 
\begin{eqnarray*}
-\frac{1}{n}
\frac{\textstyle \partial^2 \ln p(\mbox{\boldmath $x$}|{\theta})}
     {\textstyle \partial \theta^2}=
\frac{\textstyle 1}{\textstyle n}
      \sum_{i=1}^{n}
      \frac{\textstyle \partial^2 (-\ln p(x_i|{\theta}))}
     {\textstyle \partial \theta^2}\,\,\stackrel{n{\rightarrow}{\infty}}
    {\longrightarrow}\,\,
{\rm E}_X\left[
      \frac{\textstyle \partial^2 (-\ln p(x|{\theta}))}
           {\textstyle \partial \theta^2}\right]\,=\,I({\theta})
\end{eqnarray*}
so we can substitute:
\begin{eqnarray*}
\left(
\frac{\textstyle \partial^2 \ln p(\mbox{\boldmath $x$}|{\theta})}
     {\textstyle \partial \theta^2}
\right)_{{\theta}_m}\,=\,-n\,I({\theta}_m)
\hspace{0.5cm}{\rm and}\hspace{0.5cm}
\left(
\frac{\textstyle \partial^3 \ln p(\mbox{\boldmath $x$}|{\theta})}
     {\textstyle \partial \theta^3}
\right)_{{\theta}_m}\,=\,-n\,\left(
\frac{\textstyle \partial I({\theta})}
      {\textstyle \partial \theta}\right)_{\theta_m}
\end{eqnarray*}
to get
\begin{eqnarray*}
p(\mbox{\boldmath $x$}|{\theta})\,=\,
e^{\textstyle \ln p(\mbox{\boldmath $x$}|{\theta})}\,\propto\,
e^{ -\, 
\frac{\textstyle nI({\theta}_m)}{\textstyle 2}\,
\textstyle{({\theta}-{\theta}_m)^2}}\,
\left(1\,-\,
\frac{\textstyle n}{\textstyle 3!}
\left(
\frac{\textstyle \partial I({\theta})}
      {\textstyle \partial \theta}\right)_{{\theta}_m}\,
(\theta-\theta_m)^3\,+\,{\ldots}\right)
\end{eqnarray*}
On the other hand:
\begin{eqnarray*}
\pi({\theta})\,=\,\pi({\theta}_m)\,
\left(1+
\left(
\frac{\textstyle 1}{\textstyle {\pi}({\theta})}
\frac{\textstyle \partial \pi({\theta})}
      {\textstyle \partial \theta}\right)_{{\theta}_m}\,
(\theta-\theta_m)\,+\,{\ldots}\right)
\end{eqnarray*}
so If we define the random quantity
$T=\sqrt{nI({\theta}_m)}({\theta}-{\theta}_m)$
and consider that
\begin{eqnarray*}
I^{-3/2}({\theta})\,
       \frac{\textstyle \partial I({\theta})}
            {\textstyle \partial \theta}  \,=\,
-2\,       \frac{\textstyle \partial I^{-1/2}}
            {\textstyle \partial \theta}
\end{eqnarray*}
we get finally:
\begin{eqnarray*}
p(t|\mbox{\boldmath $x$})=
\frac{\textstyle {\rm exp}(-t^2/2)}{\textstyle \sqrt{2\pi}}\,
\left(1+
\frac{\textstyle 1}{\textstyle \sqrt{n}}
\left[
      \left(\frac{\textstyle I^{-1/2}({\theta})}{\textstyle {\pi}({\theta})}
            \frac{\textstyle \partial \pi({\theta})}
                 {\textstyle \partial \theta}\right)_{{\theta}_m}t+
\frac{\textstyle 1}{\textstyle 3}
\left(     \frac{\textstyle \partial I^{-1/2}}
            {\textstyle \partial \theta}
\right)_{{\theta}_m}
t^3\right]+O(\frac{\textstyle 1}{\textstyle n})\right)
\end{eqnarray*}
Let's now find 
\begin{eqnarray*}
P(T{\leq}z|\mbox{\boldmath $x$})\,=\,
\int_{-\infty}^{z}p(t|\mbox{\boldmath $x$})dt
\end{eqnarray*}
Defining
\begin{eqnarray*}
Z(x)\,=\,\frac{\textstyle 1}{\textstyle \sqrt{2\pi}}\,
e^{ -\, x^2/2} \hspace{0.3cm}{\rm and}\hspace{0.3cm}
P(x)\,=\,\int_{-\infty}^{x}Z(t)dt
\end{eqnarray*}
and considering that
\begin{eqnarray*}
\int_{-\infty}^{z}Z(t)\,t\,dt\,=\,-Z(z)
\hspace{0.3cm}{\rm and}\hspace{0.3cm}
\int_{-\infty}^{z}Z(t)\,t^3\,dt\,=\,-Z(z)\,(z^2+2)
\end{eqnarray*}
it is straight forward to get:
\begin{eqnarray*}
P(T{\leq}z|\mbox{\boldmath $x$})\,=\,
P(z)\,-\,
\frac{\textstyle Z(z)}{\textstyle \sqrt{n}}
\left[
      \left(\frac{\textstyle I^{-1/2}({\theta})}{\textstyle {\pi}({\theta})}
            \frac{\textstyle \partial \pi({\theta})}
                 {\textstyle \partial \theta}\right)_{{\theta}_m}\,+\,
\frac{\textstyle z^2+2}{\textstyle 3}
\left(     \frac{\textstyle \partial I^{-1/2}}
            {\textstyle \partial \theta}
\right)_{{\theta}_m}\,\right]\,+\,O(\frac{\textstyle 1}{\textstyle n})
\end{eqnarray*}

From this probability distribution, we can infer what the classical 
approach will get. Since he will draw inferences from
$p(\mbox{\boldmath $x$}|{\theta}_0)$, 
we can take a sequence of proper priors
${\pi}_k(\theta|\theta_0)$ for $k=1,2,{\ldots}$
that induce a sequence of distributions 
such that 
\begin{eqnarray*}
{\lim}_{k\rightarrow{\infty}}<{\pi}_k(\theta|\theta_0),
p(\mbox{\boldmath $x$}|{\theta})>=
p(\mbox{\boldmath $x$}|{\theta}_0)
\end{eqnarray*}
In Distributional sense, the sequence of distributions generated by
\begin{eqnarray*}
{\pi}_k(\theta|\theta_0)\,=\,
\frac{\textstyle k}{\textstyle 2}\,
{\mbox{\boldmath $1$}}_{[{\theta}_0-1/k,{\theta}_0+1/k]}
\hspace{0.5cm};\hspace{1.0cm}k=1,2,{\ldots}
\end{eqnarray*}
converge to the Delta distribution 
$\delta_{\theta_0}$ and, from distributional derivatives,
as $k{\rightarrow}{\infty}$,
\begin{eqnarray*}
<\frac{\textstyle d}{\textstyle d\theta}
{\pi}_k(\theta|\theta_0),I^{-1/2}({\theta})>\,=\,
-<{\pi}_k(\theta|\theta_0),\frac{\textstyle d}{\textstyle d\theta}
I^{-1/2}({\theta})>\,
{\simeq}\, -\left(
\frac{\textstyle \partial I^{-1/2}({\theta}))}
           {\textstyle \partial \theta}\right)_{\theta_0}
\end{eqnarray*}
But ${\theta}_0={\theta}_m+O(1/\sqrt{n})$
so, for a sequence of priors that shrink to 
${\theta}_0{\simeq}{\theta}_m$,
\begin{eqnarray*}
P(T{\leq}z|\mbox{\boldmath $x$})\,=\,
P(z)\,-\,
\frac{\textstyle Z(z)}{\textstyle \sqrt{n}}
\left[
\frac{\textstyle z^2+1}{\textstyle 3}
\left(     \frac{\textstyle \partial I^{-1/2}}
            {\textstyle \partial \theta}
\right)_{{\theta}_m}\,\right]\,+\,O(\frac{\textstyle 1}{\textstyle n})
\end{eqnarray*}
For terms of order $O(1/\sqrt{n})$ in the equations (1) and (2) to be
the same, we need that:
\begin{eqnarray*}
      \left(\frac{\textstyle 1}
                 {\textstyle \sqrt{I({\theta})}}
		 \frac{\textstyle 1}
		      {\textstyle {\pi}({\theta})}
		 \frac{\textstyle \partial \pi({\theta})}
                      {\textstyle \partial \theta}
       \right)_{{\theta}_m}\,=\,-
       \left(\frac{\textstyle \partial I^{-1/2}}
                  {\textstyle \partial \theta}
       \right)_{{\theta}_m}
\end{eqnarray*}
and therefore
\begin{eqnarray*}
\pi({\theta})\,=\,I^{1/2}(\theta)
\end{eqnarray*}
that is, Jeffrey's prior.
In the case of n-dimensional parameters, the reasoning goes along 
the same lines but the expressions and the development become much more
lengthy and messy and we refer to the literature.

{\rayan}                   
\vspace{1.0cm}
\small

\noindent
The procedure for
a first order {\sl probability matching prior} 
[Da96],[Da04] starts from the likelihood
\begin{eqnarray*}
p(x_1,x_2,{\ldots}x_n|{\theta}_1,{\theta}_2,{\ldots}{\theta}_p)\
\end{eqnarray*}
and then:
\begin{itemize}
\item[1)] Get the Fisher's matrix
${\Ibold}({\theta}_1,{\theta}_2,{\ldots}{\theta}_p)$
and the inverse
${\Ibold}^{-1}({\theta}_1,{\theta}_2,{\ldots}{\theta}_p)$;
\item[2)] Suppose
we are interested in the parameter 
$t=t({\theta}_1,{\theta}_2,{\ldots}{\theta}_p)$
a twice continuous and differentiable function of the parameters.
Define the column vector
\begin{eqnarray*}
{\nabla}_t\,=\,
\left(
\frac{\textstyle {\partial}t}
     {\textstyle {\partial}{\theta}_1},\,
\frac{\textstyle {\partial}t}
     {\textstyle {\partial}{\theta}_2},{\ldots},
\frac{\textstyle {\partial}t}
     {\textstyle {\partial}{\theta}_p}\right)^T
\end{eqnarray*}
\item[3)] Define the column vector
\begin{eqnarray*}
{\eta}\,=\,
 \frac{\textstyle {\Ibold}^{-1}\,{\nabla}_t}
      {\textstyle ({\nabla}_t^T{\Ibold}^{-1}\,{\nabla}_t)^{1/2}}
\hspace{1.5cm}{\rm so\,\,that}\hspace{1.5cm}
{\eta}^T{\Ibold}{\eta}=1
\end{eqnarray*}
\item[4)] The probability matching prior
for the parameter $t=t(\mbox{\boldmath $\theta$})$ in terms of
${\theta}_1,{\theta}_2,{\ldots}{\theta}_p$ is given by the equation:
\begin{eqnarray*}
\sum_{k=1}^p
\frac{\textstyle {\partial}}
     {\textstyle {\partial}{\theta}_k}\,
[{\eta}_k(\mbox{\boldmath $\theta$})\,
{\pi}(\mbox{\boldmath $\theta$})]\,=\,0
\end{eqnarray*}
Any solution 
${\pi}({\theta}_1,{\theta}_2,{\ldots}{\theta}_p)$
will do the job. 
\item[5)]
Introduce $t=t(\mbox{\boldmath $\theta$})$ in this expression,
say, for instance 
${\theta}_1={\theta}_1(t,{\theta}_2,{\ldots}{\theta}_p)$, and the
corresponding Jacobian
$J(t,{\theta}_2,{\ldots}{\theta}_p)$.
Then we get the prior for the parameter $t$ of interest and the nuisance
parameters ${\theta}_2,{\ldots}{\theta}_p$ that, eventually, will be 
integrated out.
\end{itemize}
\vspace{0.5cm}
\noindent
{\raya}                   
\vspace{0.35cm}
\footnotesize

\noindent
{\bf Example 2.15:}
Consider two independent random quantities $X_1$ and $X_2$ such
that 
\begin{eqnarray*}
P(X_i=n_k)=Po(n_k|{\mu}_i).
\end{eqnarray*}
We are interested in the parameter $t={\mu_1}/{\mu_2}$ so
setting $\mu=\mu_2$ we have the ordered parameterization
$\{t,\mu\}$.
The joint probability is
\begin{eqnarray*}
P(n_1,n_2|{\mu}_1,{\mu}_2)\,=\,
P(n_1|{\mu}_1)\,P(n_2|{\mu}_2)\,=\,
e^{\textstyle -(\mu_1+\mu_2)}
\frac{\textstyle {\mu}_1^{n_1}{\mu}_2^{n_2}}
     {\textstyle {\Gamma}(n_1+1){\Gamma}(n_2+1)}
\end{eqnarray*}
from which we get the Fisher's matrix
\begin{eqnarray*}
{\bf I}({\mu}_1,{\mu}_2)\,=\,
\left( \begin{array}{ccc}
                   1/{\mu}_1 & 0 \\
                   0 & 1/{\mu}_2
                   \end{array}
            \right)
\hspace{1.cm}{\rm and}\hspace{1.cm}
{\bf I}^{-1}({\mu}_1,{\mu}_2)\,=\,
\left( \begin{array}{ccc}
                   {\mu}_1 & 0 \\
                   0 & {\mu}_2
                   \end{array}
            \right)
\end{eqnarray*}

We are interested in the parameter $t={\mu}_1/{\mu}_2$,
a twice continuous and differentiable function of the parameters, so
\begin{eqnarray*}
{\nabla}_t({\mu}_1,{\mu}_2)\,=\,
\left(
\frac{\textstyle {\partial}t}
     {\textstyle {\partial}{\mu}_1},\,
\frac{\textstyle {\partial}t}
     {\textstyle {\partial}{\mu}_2}
\right)^T\,=\,
\left(
{\mu}_2^{-1},
\,-{\mu}_1{\mu}_2^{-2}
\right)^T\,=\,
\left( \begin{array}{c}
{\mu}_2^{-1} \\ 
-{\mu}_1{\mu}_2^{-2}
                   \end{array}
            \right)
\end{eqnarray*}
Therefore:
\begin{eqnarray*}
{\bf I}^{-1}\,{\nabla}_t\,=\,
\left( \begin{array}{c}
            {\mu}_1{\mu}_2^{-1} \\
            -{\mu}_1{\mu}_2^{-1}
       \end{array}
            \right)
\hspace{1.cm}
S\,=\,{\nabla}_t^{T}\,{\bf I}^{-1}\,{\nabla}_t\,=\,
\frac{\textstyle {\mu}_1({\mu}_1+{\mu}_2)}
     {\textstyle {\mu}_2^3}
\end{eqnarray*}
\begin{eqnarray*}
{\eta}\,=\,
 \frac{\textstyle {\bf I}^{-1}\,{\nabla}_t}
      {\textstyle ({\nabla}_t^T{\bf I}^{-1}\,{\nabla}_t)^{1/2}}\,=\,
\left( \begin{array}{c}
            ({\mu}_1{\mu}_2)^{1/2}({\mu}_1+{\mu}_2)^{-1/2} \\
           -({\mu}_1{\mu}_2)^{1/2}({\mu}_1+{\mu}_2)^{-1/2} 
       \end{array}
            \right)
\end{eqnarray*}
so that ${\eta}^T{\bf I}{\eta}=1$. The probability matching prior
for the parameter $t={\mu}_1/{\mu}_2$ in terms of
${\mu}_1$ and ${\mu}_2$ is given by the equation:
\begin{eqnarray*}
\sum_{k=1}^2
\frac{\textstyle {\partial}}
     {\textstyle {\partial}{\mu}_k}\,
[{\eta}_k({\mu})\,{\pi}({\mu})]\,=\,0
\end{eqnarray*}
so, if $f(\mu_1,\mu_2)=({\mu}_1{\mu}_2)^{1/2}({\mu}_1+{\mu}_2)^{-1/2}$,
we have to solve
\begin{eqnarray*}
\frac{\textstyle {\partial}}
     {\textstyle {\partial}{\mu}_1}\,f(\mu_1,\mu_2)\,\pi(\mu_1,\mu_2)\,=\,
\frac{\textstyle {\partial}}
     {\textstyle {\partial}{\mu}_2}\,f(\mu_1,\mu_2)\,\pi(\mu_1,\mu_2)
\end{eqnarray*}
Any solution will do so:
\begin{eqnarray*}
{\pi}({\mu}_1,{\mu}_2)\,{\propto}\,
f^{-1}({\mu}_1,{\mu}_2)\,=\,
  \frac{\textstyle \sqrt{{\mu}_1+{\mu}_2}}
       {\textstyle \sqrt{{\mu}_1{\mu}_2}}
\end{eqnarray*}
Substituting 
${\mu}_1=t{\mu}_2$ and including the Jacobian $J={\mu}_2$
we have finally:
\begin{eqnarray*}
{\pi}(t,{\mu}_2)\,{\propto}\,\sqrt{\mu_2}\,
  \sqrt{\frac{\textstyle 1+t}
             {\textstyle t}}
\end{eqnarray*}
The posterior density will be:
\begin{eqnarray*}
p(t,{\mu}_2|n_1,n_2)\,{\propto}\,
p(n_1,n_2|t,{\mu}_2){\pi}(t,{\mu}_2)\,{\propto}\,
e^{\textstyle -{\mu}_2(1+t)}
t^{n_1-1/2}\,(1+t)^{1/2}\,{\mu}_2^{n+3/2-1}
\end{eqnarray*}
and, integrating the nuisance parameter ${\mu}_2{\in}[0,{\infty})$,
we get the posterior density:
\begin{eqnarray*}
p(t|n_1,n_2)\,=\,N\,
\frac{\textstyle t^{n_1-1/2}}
     {\textstyle (1+t)^{n+1}}
\end{eqnarray*}
with $N^{-1}=B(n_1+1/2,n_2+1/2)$.

\vspace{0.35cm}
\noindent
{\bf Example 2.16: Gamma distribution.} Show that for $Ga(x|\alpha,\beta)$:
\begin{eqnarray}
   p(x|{\alpha},{\beta})\,=\,\frac{\textstyle {\alpha}^{\beta}}
                              {\textstyle \Gamma({\beta})}\,
e^{-{\alpha}x}\,x^{{\beta}-1}\,
{\mbox{\boldmath $1$}}_{(0,\infty)}(x)
                \nonumber 
   \end{eqnarray}
the probability matching prior for the ordering 
\begin{itemize}
\item[$\bullet$] $\{\beta,\alpha\}$ is 
     ${\pi}(\alpha,\beta)={\beta}^{-1/2}
\left[{\alpha}^{-1}\sqrt{{\beta}{\Psi}'(\beta)-1}\right]$
\item[$\bullet$] $\{\alpha,\beta\}$ is
  ${\pi}(\alpha,\beta)=\left[{\alpha}^{-1}\sqrt{{\Psi}'(\beta)}\right]
          \sqrt{{\beta}{\Psi}'(\beta)-1}$
\end{itemize}
to be compared with Jeffrey's prior
${\pi}^J_{2}(\alpha,\beta)={\alpha}^{-1}\sqrt{\beta{\Psi}'(\beta)-1}$
and Jeffrey's prior when both parameters are treated individually
${\pi}^J_{1+1}(\alpha,\beta)={\alpha}^{-1}\sqrt{{\Psi}'(\beta)}$

\vspace{0.35cm}
\noindent
{\bf Example 2.17: Bivariate Normal Distribution.}

For the ordered parameterization ${\rho,{\sigma}_1,{\sigma}_2}$:
the Fisher's matrix (see example 2.12) is:
\begin{eqnarray*}
   {\Ibold}({\rho},{\sigma}_1,{\sigma}_2)\,=\,(1-{\rho}^2)^{-1}\,
   \left(\begin{array}{ccc}
     (1+{\rho}^2)(1-{\rho}^2)^{-1} &
     -{\rho}{\sigma}_1^{-1} &
     -{\rho}{\sigma}_2^{-1} \\
     -{\rho}{\sigma}_1^{-1} &
     (2-{\rho}^2){\sigma}_1^{-2} &  
     -{\rho}^2({\sigma}_1{\sigma}_2)^{-1} \\ 
     -{\rho}{\sigma}_2^{-1} &
     -{\rho}^2({\sigma}_1{\sigma}_2)^{-1} &
     (2-{\rho}^2){\sigma}_2^{-2} 
\end{array}\right)
\end{eqnarray*}
and the inverse:
\begin{eqnarray*}
   {\Ibold}^{-1}({\rho},{\sigma}_1,{\sigma}_2)\,=\,\frac{1}{2}\,
   \left(\begin{array}{ccc}
     2(1-{\rho}^2)^2 &
     {\sigma}_1{\rho}(1-{\rho}^2) &
     {\sigma}_2{\rho}(1-{\rho}^2) \\
     {\sigma}_1{\rho}(1-{\rho}^2) &
     {\sigma}_1^{2} &  
     {\rho}^2{\sigma}_1{\sigma}_2 \\ 
     {\sigma}_2{\rho}(1-{\rho}^2) &
     {\rho}^2{\sigma}_1{\sigma}_2 & 
     {\sigma}_2^{2}  
\end{array}\right)
\end{eqnarray*}
Then
\begin{eqnarray*}
   \frac{2}{\rho}\frac{\partial}{\partial \rho}[{\pi}(1-{\rho}^2)]\,+\,
   \frac{\partial}{\partial {\sigma}_1}[{\pi}{\sigma}_1]\,+\,
   \frac{\partial}{\partial {\sigma}_2}[{\pi}{\sigma}_2]\,=\,0
\end{eqnarray*}
for which 
\begin{eqnarray*}
  {\pi}({\sigma}_1,{\sigma}_2,{\rho})\,=\,
     \frac{\textstyle 1}{\textstyle {\sigma}_1{\sigma}_2(1-{\rho}^2)}
\end{eqnarray*}
is a solution.

\vspace{0.35cm}
\noindent
{\bf Problem 2.6:} Consider
\begin{eqnarray*}
X\,{\sim}\,p(x|a,b,{\sigma})\,=\,
\frac{{\sinh}[{\sigma}(b-a)]}{2(b-a)}\,
\frac{1}{{\cosh}[{\sigma}(x-a)]{\cosh}[{\sigma}(b-x)]}
{\mbox{\boldmath $1$}}_{(-\infty,\infty)}(x)
\end{eqnarray*}
where $a<b{\in}{\Rcal}$ and ${\sigma}{\in}(0,\infty)$. Show that
\begin{eqnarray*}
E[X]\,=\,\frac{b+a}{2}\hspace{1.5cm}{\rm and}\hspace{1.5cm}
V[x]\,=\,\frac{(b-a)^2}{12}\,+\,\frac{{\pi}^2}{12{\sigma}^2}
\end{eqnarray*}
and that, for known ${\sigma}>>$, the probability matching prior for $a$ and
$b$ tends to
${\pi}_{pm}(a,b){\sim}(b-a)^{-1/2}$. Show also that, under the same limit,
${\pi}_{pm}({\theta}){\sim}{\theta}^{-1/2}$
for $(a,b)=(-{\theta},{\theta})$ and $(a,b)=(0,{\theta})$. Since 
$p(x|a,b,{\sigma})\stackrel{{\sigma}>>}{\rightarrow}Un(x|a,b)$
discuss in this last case what is the difference with the example 2.4. 

\small
{\raya}                   
\vspace{1.0cm}

\subsection{Reference Analysis}

The expected amount of information ({\sl Expected Mutual Information}) 
on the parameter $\theta$ provided by
$k$ independent observations of the model $p(\xbold|\theta)$ relative
to the prior knowledge on $\theta$ described by $\pi(\theta)$ is
\begin{eqnarray*}
  I\left[ e(k),{\pi}(\theta)\right]\,=\,
  \int_{\Theta}{\pi}(\theta)\,d{\theta}\,
  \int_{\Omega_{X}}p(\zbold_k|\theta)\,
   \log\,\frac{\textstyle p({\theta}|{\zbold}_k)} 
              {\textstyle {\pi}({\theta})} 
d{\zbold_k}
\end{eqnarray*}
where $\zbold_k=\{\xbold_1,\ldots,\xbold_k\}$.
If $\lim_{k{\rightarrow}{\infty}}I\left[e(k),{\pi}(\theta)\right]$
exists, it will quantify the maximum amount of information that we could  
obtain on $\theta$ from experiments described by this model relative to
the prior knowledge ${\pi}(\theta)$. The central idea of the 
{\sl reference analysis} [Be79,Be94]
is to take as {\sl reference prior for the the model}
 $p(\xbold|\theta)$ that which maximizes the maximum amount 
of information we may get so it will be the {\sl less informative} 
for this model. From Calculus of Variations, if we introduce the prior
${\pi}^{\star}(\theta)={\pi}(\theta)+{\epsilon}{\eta}(\theta)$ with
${\pi}(\theta)$ an extremal of the expected information
$I\left[e(k),{\pi}(\theta)\right]$ and ${\eta}(\theta)$ such that
\begin{eqnarray*}
  \int_{\Theta}{\pi}(\theta)d{\theta}\,=\,
  \int_{\Theta}{\pi}^{\star}(\theta)d{\theta}\,=\,1
  \hspace{0.5cm}{\longrightarrow}\hspace{0.5cm}
  \int_{\Theta}{\eta}(\theta)d{\theta}=0
\end{eqnarray*}
it is it is easy to see (left as exercise) that
\begin{eqnarray*}
  {\pi}(\theta)\,\propto\,
 {\rm exp}\left\{
  \int_{\Omega_{X}}p(\zbold_k|\theta)\,
   \log\,p({\theta}|{\zbold}_k)\, d{\zbold}_k \right\}\,=\,f_k({\theta})
\end{eqnarray*}
This is a nice but complicated implicit equation because, on the one hand,
$f_k({\theta})$ depends on ${\pi}(\theta)$ through the posterior 
$p({\theta}|{\zbold}_k)$ and,
on the other hand,  the limit $k{\rightarrow}{\infty}$ is usually
divergent (intuitively, the more precision we want for $\theta$, 
the more information
is needed and to know the actual value from the experiment requires an 
infinite amount of information). This can be circumvented regularizing
the expression as
\begin{eqnarray*}
  {\pi}(\theta)\,
  \propto\,{\pi}(\theta_0)\,\lim_{k{\rightarrow}{\infty}}
  \frac{\textstyle f_k(\theta)}{\textstyle f_k(\theta_0)}
\end{eqnarray*}
with $\theta_0$ any interior point of $\Theta$ (we are used to that in
particle physics!). Let's see some examples.

\vspace{0.5cm}
\noindent
{\raya}                   
\footnotesize
\vspace{0.35cm}

\noindent
{\bf Example 2.18:} Consider again the exponential model for which 
$t=n^{-1}\sum_{i=1}^{n}x_i$  is sufficient for $\theta$ and distributed as
\begin{eqnarray*}
   p(t|\theta)\,=\,\frac{\textstyle (n\theta)^n}
                             {\textstyle \Gamma(n)}\,t^{n-1}\,
{\rm exp}\{\textstyle -n\theta t\}\,
\end{eqnarray*}

Taking  ${\pi}(\theta)=\mbox{\boldmath $1$}_{(0,{\infty})}({\theta})$ 
we have the proper posterior 
\begin{eqnarray*}
   \pi^{\star}(\theta|t)\,=\,\frac{\textstyle (nt)^{n+1}}
                        {\textstyle \Gamma(n+1)}\,
\exp\left\{-n{\theta}t\right\}\,{\theta}^{n}
\end{eqnarray*}
Then $\log \pi^{\star}(\theta|t)\,=\,-(n{\theta})t\,+\,n\,\log {\theta}\,+\,
        (n+1) \log t \,+\,g_1(n)$ and
\begin{eqnarray*}
f_n({\theta})\,=\,
 {\rm exp}\left\{
  \int_{\Omega_{X}}p(t|\theta)\,
   \log\,{\pi}^{\star}({\theta}|t)\, dt \right\}\,=\,
\frac{\textstyle g_2(n)}{\textstyle \theta}
\hspace{0.5cm}{\longrightarrow}\hspace{0.5cm}
 {\pi}(\theta)\,
  \propto\,{\pi}(\theta_0)\,\lim_{n{\rightarrow}{\infty}}
  \frac{\textstyle f_n(\theta)}{\textstyle f_n(\theta_0)}\,\propto\,
  \frac{\textstyle 1}{\textstyle \theta}
\end{eqnarray*}

\vspace{0.35cm}
\noindent
{\bf Example 2.19:}
Prior functions depend on the particular model we are treating.
To learn about a parameter, we can do different experimental designs that
respond to different models and, even though the parameter is the same, they
may have different priors. For instance, we may be interested in the
{\sl acceptance}; the probability to accept an event under some conditions.
For this, we can generate for instance a sample of $N$ observed events and 
see how many $(x)$ pass the conditions. 
This experimental design corresponds to a Binomial distribution 
\begin{eqnarray*}
   p(x|N,\theta)\,=\,
   \left(\begin{array}{c}
      N \\
      x
\end{array}\right)\,\theta^{x}\,(1-\theta)^{N-x}
\end{eqnarray*}
with $x=\{0,1,{\ldots},N\}$. For this model, the reference prior (also
Jeffrey's and PM) is
${\pi}(\theta)={\theta}^{-1/2}(1-{\theta})^{-1/2}$ and the posterior
$\theta{\sim}Be({\theta}|x+1/2,N-x+1/2)$.
Conversely, we
can generate events until $r$ are accepted and see how many $(x)$
have we generated. This experimental design corresponds to a 
Negative Binomial distribution 
\begin{eqnarray*}
   p(x|r,\theta)\,=\,
   \left(\begin{array}{c}
      x-1 \\
      r-1
\end{array}\right)\,\theta^{r}\,(1-\theta)^{x-r}
\end{eqnarray*}
where $x=r,r+1,{\ldots}$ and $r\geq 1$. 
For this model, the reference prior (Jeffrey's and PM too) is
${\pi}(\theta)={\theta}^{-1}(1-{\theta})^{-1/2}$ and the posterior 
$\theta{\sim}Be({\theta}|r,x-r+1/2)$.

\vspace{0.35cm}
\noindent
{\bf Problem 2.7:} Consider 
\vspace{0.3cm}

\noindent
{\bf 1)}
$X\,{\sim}\,Po(x|{\theta})\,=\,\exp\{-\theta\}\frac{\textstyle \theta^x}
                                            {\textstyle \Gamma(x+1)}$
and the experiment $e(k)\stackrel{iid}{\rightarrow}\{x_1,x_2,\ldots x_k\}$.
Take ${\pi}^{\star}(\theta)=\mbox{\boldmath $1$}_{(0,{\infty})}({\theta})$,
and show that 
\begin{eqnarray*}
 {\pi}(\theta)\,
  \propto\,{\pi}(\theta_0)\,\lim_{k{\rightarrow}{\infty}}
  \frac{\textstyle f_k(\theta)}{\textstyle f_k(\theta_0)}\,\propto\,
  {\theta}^{-1/2}
\end{eqnarray*}
\vspace{0.3cm}

\noindent
{\bf 2)}
$  X\,{\sim}\,Bi(x|N,{\theta})\,=\,
       \left(  \begin{array}{c}
                 N \\ x
               \end{array}
     \right)\,{\theta}^x\,(1-{\theta})^{N-x}
$
and the experiment $e(k)\stackrel{iid}{\rightarrow}\{x_1,x_2,\ldots x_k\}$.
Take 
${\pi}^{\star}(\theta){\propto}{\theta}^{a-1}(1-{\theta})^{b-1}
\mbox{\boldmath $1$}_{(0,1)}({\theta})$ with $a,b>0$  and show that 
\begin{eqnarray*}
  {\pi}(\theta)\,
  \propto\,{\pi}(\theta_0)\,\lim_{k{\rightarrow}{\infty}}
  \frac{\textstyle f_k(\theta)}{\textstyle f_k(\theta_0)}\,\propto\,
  {\theta}^{-1/2}(1-{\theta})^{-1/2}
\end{eqnarray*}

\noindent
(Hint: For 1) and 2) consider the Taylor expansion of 
$\log \Gamma(z,\cdot)$ around
$E[z]$ and the asymptotic behavior of the Polygamma Function
${\Psi}^{\left.n\right)}=a_nz^{-n}+a_{n+1}z^{-(n+1)}+...$).
\vspace{0.3cm}

\noindent
{\bf 3)}
$X\,{\sim}\,Un(x|0,{\theta})$ and the iid sample
$\{x_1,x_2,\ldots x_k\}$. For inferences on $\theta$, show that
$f_k={\theta}^{-1}g(k)$ and in consequence the posterior is Pareto
$Pa(\theta|x_M,n)$ with $x_M=\max\{x_1,x_2,\ldots x_k\}$
the sufficient statistic.

\small
{\raya}                   
\vspace{1.0cm}

A very useful constructive theorem to obtain the 
{\sl reference prior} is given in [Be09].
First, a {\sl permissible prior} for the model $p(\xbold|\theta)$ is defined 
as a strictly positive function $\pi(\theta)$ such that it renders a proper
posterior; that is,
\begin{eqnarray*}
\forall x{\in}{\Omega}_X\hspace{1.cm}
\int_{\Theta}p(\xbold|\theta)\,{\pi}(\theta)\,d\theta\,<\,{\infty}
\end{eqnarray*}
and that for some approximating sequence 
${\Theta}_k{\subset}{\Theta}$;
$\lim_{k{\rightarrow}{\infty}}{\Theta}_k=\Theta$, the sequence of posteriors
$p_k(\theta|\xbold){\propto}p(\xbold|\theta){\pi}_k({\theta})$ converges
logarithmically to
$p(\theta|\xbold){\propto}p(\xbold|\theta){\pi}({\theta})$. 
Then, the
{\sl reference prior} is just a {\sl permissible prior} that maximizes
the maximum amount of information the experiment can provide for the 
parameter. The constructive procedure for a one-dimensional
parameter consists on:
\begin{itemize}
\item[1)] Take ${\pi}^{\star}(\theta)$ as a continuous strictly positive 
          function such that the corresponding posterior
          \begin{eqnarray*}
            {\pi}^{\star}(\theta|\zbold_k)\,=\,
            \frac{\textstyle p(\zbold_k|\theta)\,{\pi}^{\star}(\theta)}
                 {\textstyle \int_{\Theta}p(\zbold_k|\theta)\,{\pi}(\theta)\,
                     d\theta}
          \end{eqnarray*}
          is proper and asymptotically consistent.  ${\pi}^{\star}(\theta)$
          is arbitrary so it can be taken for convenience to simplify the 
          integrals.
\item[2)] Obtain 
  \begin{eqnarray*}
    f_k^{\star}({\theta})\,=\,
    {\rm exp}\left\{
    \int_{\Omega_{X}}p(\zbold_k|\theta)\,
    \log\,{\pi}^{\star}({\theta}|{\zbold}_k)\, d{\zbold}_k \right\}
    \hspace{0.5cm}{\rm and}\hspace{0.5cm}
     h_k(\theta;\theta_0)\,=\,
    \frac{\textstyle f_k^{\star}({\theta})}
         {\textstyle f_k^{\star}({\theta}_0)}
  \end{eqnarray*}
  for any interior point ${\theta}_0{\in}{\Theta}$;
\item[3)] If
  \begin{itemize}
    \item[3.1)] each $f_k^{\star}({\theta})$ is continuous;
    \item[3.2)] for any fixed $\theta$ and large $k$, is 
           $h_k(\theta;\theta_0)$ is either monotonic in $k$ or bounded
           from above by $h(\theta)$ that is integrable on any compact set;
    \item[3.3)] ${\pi}({\theta})=\lim_{k\rightarrow\infty}h_k({\theta};\theta_0)$
           is a {\sl permissible prior function}
  \end{itemize}
\end{itemize}
then ${\pi}({\theta})$ is a reference prior for the model $p(\xbold|\theta)$.
It is important to note that there is no requirement on the existence of the
Fisher's information $\Ibold(\theta)$. If it exists, a simple Taylor expansion
of the densities shows that for a one-dimensional parameter
${\pi}({\theta})=[\Ibold(\theta)]^{1/2}$ in
consistency with
Jeffrey's proposal. Usually, the last is easier to evaluate but not always 
as we shall see. 

In many cases ${\rm supp}(\theta)$  is unbounded and the prior  
$\pi(\theta)$ is not a propper density. As we have seen
this is not a problem as long as the posterior
$p(\theta|\zbold_k){\propto}p(\zbold_k|\theta){\pi}({\theta})$ is propper
although, in any case, one can proceed {\sl "more formally"} 
considering a sequence
of proper priors ${\pi}_m({\theta})$ defined on a sequence of compact sets 
${\Theta}_m{\subset}{\Theta}$ such that 
$\lim_{m{\rightarrow}{\infty}}{\Theta}_m=\Theta$ and taking the limit
of the corresponding sequence of posteriors
$p_m(\theta|\zbold_k){\propto}p(\zbold_k|\theta){\pi}_m({\theta})$. 
Usually simple sequences as for example 
$\Theta_m=[1/m,m]$;
$\lim_{m{\rightarrow}{\infty}}\Theta_m=(0,\infty)$, or
$\Theta_m=[-m,m]$; $\lim_{m{\rightarrow}{\infty}}\Theta_m=(-\infty,\infty)$ 
will suffice.

When the parameter $\thetabold$ is n-dimensional, the procedure is more 
laborious.
First, one starts [Be94] 
arranging the parameters in decreasing order of importance 
$\{\theta_1,\theta_2,\ldots,\theta_n\}$ (as we did for the Probability Matching
Priors) and then follow the previous scheme to obtain the 
conditional prior functions
\begin{eqnarray*}
{\pi}(\theta_n|\theta_1,\theta_2,\ldots,\theta_{n-1})\,
{\pi}(\theta_{n-1}|\theta_1,\theta_2,\ldots,\theta_{n-2})\,{\cdots}\,
{\pi}(\theta_2|\theta_1)\,{\pi}(\theta_1)
\end{eqnarray*}
For instance in the case of two parameters and
the ordered parameterization $\{\theta,\lambda\}$:
 \begin{itemize}
    \item[1)] Get the conditional ${\pi}(\lambda|\theta)$ as the reference
              prior for $\lambda$ keeping $\theta$ fixed;
    \item[2)] Find the marginal model
        \begin{eqnarray*}
    p(\xbold|{\theta})\,=\,
    \int_{\Phi}p(\xbold|\theta,\lambda)\,{\pi}(\lambda|\theta)\,d{\lambda}
  \end{eqnarray*}
    \item[3)] Get the reference prior ${\pi}(\theta)$ from the marginal
              model $p(\xbold|\theta)$
  \end{itemize}
Then ${\pi}(\theta,\lambda){\propto}{\pi}(\lambda|\theta){\pi}(\theta)$.
This is fine if ${\pi}(\lambda|\theta)$ and ${\pi}(\theta)$ are propper
functions; seldom the case. Otherwise one has to
define the appropriate sequence of compact sets observing,
among other things, that this has to be done for the full parameter space 
and  usually the limits depend on the parameters. Suppose that
we have the sequence 
$\Theta_i\times\Lambda_i\stackrel{i\rightarrow\infty}
{\longrightarrow}\Theta\times\Lambda$. Then:
 \begin{itemize}
    \item[1)] Obtain ${\pi}_i(\lambda|\theta)$:
        \begin{eqnarray*}
          {\pi}_i^{\star}(\lambda|\theta)
          \mbox{\boldmath $1$}_{\Lambda_i}({\lambda})
          \longrightarrow
            {\pi}_i^{\star}(\lambda|\theta,\zbold_k)=
            \frac{p(\zbold_k|\theta,\lambda){\pi}_i^{\star}(\lambda|\theta)}
                 {\int_{\Lambda_i}p(\zbold_k|\theta,\lambda)
                 {\pi}_i^{\star}(\lambda|\theta)\,d\lambda}
           \longrightarrow
 {\pi}_i(\lambda|\theta)=\lim_{k\rightarrow\infty}
            \frac{f^{\star}_k(\lambda|\Lambda_i,\theta,...)}
                 {f^{\star}_k(\lambda_0|\Lambda_i,\theta,...)}
        \end{eqnarray*}
    \item[2)] Get the marginal density $p_i(\xbold|\theta)$:
        \begin{eqnarray*}
    p_i(\xbold|{\theta})\,=\,
    \int_{\Lambda_i}p(\xbold|\theta,\lambda)\,{\pi}_i(\lambda|\theta)\,
       d{\lambda}
  \end{eqnarray*}
    \item[3)] Determine ${\pi}_i(\theta)$:
 \begin{eqnarray*}
          {\pi}_i^{\star}(\theta)
          \mbox{\boldmath $1$}_{\Theta_i}({\theta})
          \longrightarrow
            {\pi}_i^{\star}(\theta|\zbold_k)=
            \frac{p_i(\zbold_k|\theta){\pi}_i^{\star}(\theta)}
                 {\int_{\Theta_i}p_i(\zbold_k|\theta)
                 {\pi}_i^{\star}(\theta)\,d\theta}
           \longrightarrow
 {\pi}_i(\theta)=\lim_{k\rightarrow\infty}
            \frac{f^{\star}_k(\theta|\Theta_i,\Lambda_i,...)}
                 {f^{\star}_k(\theta_0|\Theta_i,\Lambda_i,...)}
        \end{eqnarray*}
\item[4)] The reference prior for the ordered parameterization 
          $\{\theta,\lambda\}$ will be:
 \begin{eqnarray*}
          {\pi}(\theta,\lambda)\,=\,
          \lim_{i\rightarrow\infty}
            \frac{{\pi}_i(\lambda|\theta)\,{\pi}_i(\theta)}
                 {{\pi}_i(\lambda_0|\theta_0)\,{\pi}_i(\theta_0)}
        \end{eqnarray*}
  \end{itemize}

In the case of two parameters, if $\Lambda$ is independent of $\theta$
the Fisher's matrix usually exists and, if
$\Ibold(\theta,\lambda)$ and 
$\Sbold(\theta,\lambda)=\Ibold^{-1}(\theta,\lambda)$ 
are such that:
 \begin{eqnarray*}
   \Ibold_{22}(\theta,\lambda)\,=\,
                   a_1^2(\theta)\,b_1^2(\lambda)
\hspace{0.5cm}{\rm and}\hspace{0.5cm}
   \Sbold_{11}(\theta,\lambda)\,=\,
      a_0^{-2}(\theta)\,b_0^{-2}(\lambda)
  \end{eqnarray*}
then [Be98]
${\pi}(\theta,\lambda)={\pi}(\lambda|\theta){\pi}(\theta)=
      a_0(\theta)\,b_1(\lambda)$
is a permissible prior
even if the conditional reference priors are not proper.
The reference priors are usually {\sl probability matching 
priors}.
\vspace{0.5cm}

\noindent
{\raya}                   
\vspace{0.35cm}
\footnotesize

\noindent
{\bf Example 2.20:}
A simple example is the Multinomial distribution 
${\Xbold}{\sim}Mn(\xbold|\thetabold)$ with ${\rm dim}{\Xbold}=k+1$ and
probability
\begin{eqnarray*}
    p(\xbold|{\theta})\,\propto\,
    {\theta_1}^{x_1}\,    {\theta_2}^{x_2}\,{\cdots}\,
    {\theta_k}^{x_k}\, (1-{\delta}_k)^{x_{k+1}}
    \hspace{0.5cm};    \hspace{1.5cm}
    \delta_k=\sum_{j=1}^k{\theta}_j
\end{eqnarray*}
Consider the ordered parameterization
$\{{\theta_1},{\theta_2},\ldots,{\theta_k}\}$. Then
\begin{eqnarray*}
    \pi({\theta_1},{\theta_2},\ldots,{\theta_k})\,=\,
    \pi({\theta_k}|\theta_{k-1},\theta_{k-2}\ldots \theta_{2},\theta_{1})\,
\pi({\theta_{k-1}}|\theta_{k-2}\ldots \theta_{2},\theta_{1})\,\cdots\,
\pi({\theta_2}|\theta_{1})\,\pi(\theta_{1})
\end{eqnarray*}
In this case, all the conditional densities are proper
\begin{eqnarray*}
    \pi({\theta_m}|\theta_{m-1},\ldots\,\theta_{1})\,{\propto}\,
        {\theta_{m}}^{-1/2}\,(1-{\delta_m})^{-1/2}
\end{eqnarray*}
and therefore
\begin{eqnarray*}
    \pi({\theta_1},\theta_{2},\ldots\,\theta_{k})\,{\propto}\,
    \prod_{i=1}^{k}
        {\theta_{i}}^{-1/2}\,(1-{\delta_i})^{-1/2}
\end{eqnarray*}
The posterior density will be then
\begin{eqnarray*}
    p(\thetabold|\xbold)\,{\propto}\,
    \left[
    \prod_{i=1}^{k}
        {\theta_{i}}^{x_i-1/2}\,(1-{\delta}_i)^{-1/2}
    \right]\,
    (1-{\delta}_k)^{x_{k+1}}
\end{eqnarray*}

\vspace{0.35cm}

\noindent
{\bf Example 2.21:}
Consider again the case of two independent Poisson distributed random 
quantities $X_1$ and $X_2$ with joint density density
\begin{eqnarray*}
P(n_1,n_2|{\mu}_1,{\mu}_2)\,=\,
P(n_1|{\mu}_1)\,P(n_2|{\mu}_2)\,=\,
e^{\textstyle -(\mu_1+\mu_2)}
\frac{\textstyle {\mu}_1^{n_1}{\mu}_2^{n_2}}
     {\textstyle {\Gamma}(n_1+1){\Gamma}(n_2+1)}
\end{eqnarray*}
We are interested in the parameter $\theta={\mu_1}/{\mu_2}$ so
setting $\mu=\mu_2$ we have the ordered parameterization
$\{\theta,\mu\}$ and:
\begin{eqnarray*}
P(n_1,n_2|\theta,{\mu})\,=\,
e^{\textstyle -{\mu}(1+\theta)}
\frac{\textstyle \theta^{n_1}{\mu}^n}
     {\textstyle {\Gamma}(n_1+1){\Gamma}(n_2+1)}
\end{eqnarray*}
where $n=n_1+n_2$. 
Since $E[X_1]={\mu}_1=\theta{\mu}$ and $E[X_2]={\mu}_2={\mu}$
the Fisher's matrix and its inverse will be
\begin{eqnarray*}
{\bf I}=
\left( \begin{array}{ccc}
                   {\mu}/\theta & 1 \\
                   1 & (1+\theta)/{\mu}
                   \end{array}
            \right)\,;
     \hspace{0.5cm}\det({\bf I})\,=\,{\theta}^{-1}
     \hspace{0.5cm}{\rm and}\hspace{0.5cm}
{\bf S}={\bf I}^{-1}=
\left( \begin{array}{ccc}
                   \theta(1+\theta)/{\mu} & -\theta \\
                   -\theta & {\mu}
                   \end{array}
            \right)
\end{eqnarray*}
Therefore
\begin{eqnarray*}
S_{11}\,=\,\theta(1+\theta)/{\mu} \hspace{1.cm}{\rm and}\hspace{1.cm}
F_{22}\,=\,(1+\theta)/{\mu} 
\end{eqnarray*}
and, in consequence:
\begin{eqnarray*}
{\pi}(\theta)\,f_1({\mu})\,{\propto}\,
S_{11}^{-1/2}\,=\,
\frac{\textstyle \sqrt{\mu}}
     {\textstyle \sqrt{\theta(1+\theta)}} \hspace{2.cm}
{\pi}({\mu}|\theta)\,f_2(\theta)\,{\propto}\,
F_{22}^{1/2}\,=\,
\frac{\textstyle \sqrt{1+\theta}}
     {\textstyle \sqrt{\mu}}
\end{eqnarray*}
Thus, we have for the ordered parameterization $\{\theta,\mu\}$ 
the reference prior:
\begin{eqnarray*}
{\pi}(\theta,{\mu})\,=\,{\pi}({\mu}|\theta)\,{\pi}(\theta)\,{\propto}\,
\frac{\textstyle 1}
     {\textstyle \sqrt{{\mu}\theta(1+\theta)}}
\end{eqnarray*}
and the posterior density will be:
\begin{eqnarray*}
p(\theta,{\mu}|n_1,n_2)\,&{\propto}&\,
{\rm exp}\left\{\textstyle -{\mu}(1+\theta)\right\}\,
\theta^{n_1-1/2}\,(1+\theta)^{-1/2}\,{\mu}^{n-1/2}
\end{eqnarray*}
and, integrating the nuisance parameter ${\mu}{\in}[0,{\infty})$
we get finally
\begin{eqnarray*}
p(\theta|n_1,n_2)\,=\,N\,
\frac{\textstyle \theta^{n_1-1/2}}
     {\textstyle (1+\theta)^{n+1}}
\end{eqnarray*}
with $\theta={\mu}_1/{\mu}_2$, $n=n_1+n_2$ and
$N^{-1}=B(n_1+1/2,n_2+1/2)$.
The distribution function will be:
\begin{eqnarray*}
P(\theta|n_1,n_2)\,=\,\int_{0}^{\theta}
p(\theta'|n_1,n_2)\,d\theta'=\,
I(\theta/(1+\theta);n_1+1/2,n_2+1/2)
\end{eqnarray*}
with $I(x;a,b)$ the Incomplete Beta Function and the moments,
when they exist;
\begin{eqnarray*}
E[\theta^m]\,=\,
\frac{\textstyle {\Gamma}(n_1+1/2+m)\,{\Gamma}(n_2+1/2-m)}
     {\textstyle {\Gamma}(n_1+1/2)\,{\Gamma}(n_2+1/2)}
\end{eqnarray*}

It is interesting to look at the problem from a different point of view.
Consider again the ordered parameterization $\{\theta,\lambda\}$ with
$\theta={\mu_1}/{\mu_2}$  but now, the nuisance parameter is
$\lambda={\mu_1}+{\mu_2}$. The likelihood will be:
\begin{eqnarray*}
P(n_1,n_2|\theta,\lambda)\,=\,
\frac{\textstyle 1}
     {\textstyle {\Gamma}(n_1+1){\Gamma}(n_2+1)}
e^{\textstyle -\lambda}\,{\lambda}^{n}\,
\frac{\textstyle {\theta}^{n_1}}
     {\textstyle (1+\theta)^n}
\end{eqnarray*}
The domains are ${\Theta}=(0,\infty)$ and 
${\Lambda}=(0,\infty)$, independent.
Thus, no need to specify the prior for $\lambda$ since
\begin{eqnarray*}
p({\theta}|n_1,n_2)\,{\propto}\,{\pi}(\theta)\,
\frac{\textstyle {\theta}^{n_1}}
     {\textstyle (1+\theta)^n}\,\int_{\Lambda}
     e^{\textstyle -\lambda}\,{\lambda}^{n}{\pi}(\lambda)d{\lambda}
 \,{\propto}\,
\frac{\textstyle {\theta}^{n_1}}
     {\textstyle (1+\theta)^n}\,{\pi}(\theta)
\end{eqnarray*}
In this case we have that 
\begin{eqnarray*}
I(\theta)\,{\propto}\,\frac{\textstyle 1}
                   {\textstyle \theta\,(1+\theta)^2}
\hspace{0.3cm}\longrightarrow\hspace{0.3cm}
\pi(\theta)\,=\,\frac{\textstyle 1}
                   {\textstyle \theta^{1/2}\,(1+\theta)}
\end{eqnarray*}
and, in consequence, 
\begin{eqnarray*}
 p(\theta|n_1,n_2)\,=\,N\,
    \frac{\textstyle \theta^{n_1-1/2}}
         {\textstyle (1+\theta)^{n+1}}
\end{eqnarray*}

\vspace{0.35cm}

\noindent
{\bf Problem 2.8:} Show that the reference prior for the Pareto distribution
$Pa(x|{\theta},x_0)$ (see example 2.9) 
is $\pi({\theta},x_0){\propto}({\theta}x_0)^{-1}$ and that for an iid sample
${\xbold}=\{x_1,\ldots,x_n\}$, if $x_{m}={\rm min}\{x_i\}_{i=1}^n$ and
$a=\sum_{i=1}^n\ln(x_i/x_m)$ the posterior
\begin{eqnarray*}
 p(\theta,x_0|{\xbold})\,=\,
    \frac{\textstyle na^{n-1}}{x_m\Gamma(n-1)}\,
    e^{-a{\theta}}{\theta}^{n-1}\left(\frac{x_0}{x_m}\right)^{n{\theta}-1}
{\mbox{\boldmath $1$}}_{(0,\infty)}(\theta)
{\mbox{\boldmath $1$}}_{(0,x_m)}(x_0)
\end{eqnarray*}
is proper for a sample size $n>1$. Obtain the marginal densities
\begin{eqnarray*}
 p(\theta|{\xbold})\,=\,
    \frac{\textstyle a^{n-1}}{\Gamma(n-1)}\,
    e^{-a{\theta}}{\theta}^{n-2}{\mbox{\boldmath $1$}}_{(0,\infty)}(\theta)
\hspace{0.5cm}{\rm and}\hspace{0.5cm}
p(x_0|{\xbold})\,=\,
    \frac{\textstyle n(n-1)}{a}\,x_0^{-1}
 \left[1+\frac{n}{a}{\ln}\left(\frac{x_m}{x_0}\right)\right]^{-n}
{\mbox{\boldmath $1$}}_{(0,x_m)}(x_0)
\end{eqnarray*}
and show that for large $n$ (see section 10.2)
$E[{\theta}]{\simeq}na^{-1}$ and $E[x_0]{\simeq}x_m$.

\vspace{0.35cm}

\noindent
{\bf Problem 2.9:} Show that for the shifted Pareto distribution
(Lomax distribution):
\begin{eqnarray*}
   p(x|{\theta},x_0)\,=\,\frac{\textstyle \theta}
                              {\textstyle x_0}\,
\left(\frac{\textstyle x_0}{\textstyle x+x_0}\right)^{\theta+1}\,
{\mbox{\boldmath $1$}}_{(0,\infty)}(x)
\hspace{0.5cm};\hspace{1.cm}{\theta},x_0{\in}R^{+}
                \nonumber 
\end{eqnarray*}
the reference prior for the ordered parameterization
$\{\theta,x_0\}$ is 
${\pi}_r({\theta},x_0)\,\propto\,(x_0{\theta}({\theta}+1))^{-1}$ and for
$\{x_0,\theta\}$ is 
${\pi}_r(x_0,{\theta})\,\propto\,(x_0{\theta})^{-1}$. Show that the first one
is a first order probability matching prior while the second is not. 
In fact, show that for $\{x_0,\theta\}$, 
${\pi}_{pm}(x_0,{\theta})\,\propto\,(x_0{\theta}^{3/2}\sqrt{\theta+2})^{-1}$ 
is a matching prior and
that for both orderings the Jeffrey's prior is
${\pi}_{J}({\theta},x_0)\,\propto\,
(x_0({\theta}+1)\sqrt{\theta(\theta+2)})^{-1}$.

\vspace{0.35cm}

\noindent
{\bf Problem 2.10:} Show that for the Weibull distribution
\begin{eqnarray*}
   p(x|\alpha,\beta)\,=\,\alpha \beta \,x^{\beta -1} \exp \left\{
           -\alpha x^{\beta}\right\}
{\mbox{\boldmath $1$}}_{(0,\infty)}(x)
\end{eqnarray*}
with $\alpha,\beta>0$, the reference prior functions are
\begin{eqnarray*}
  {\pi}_r(\beta,\alpha)= (\alpha \beta)^{-1}
  \hspace{1.cm}{\rm and}\hspace{1.cm}
  {\pi}_r(\alpha,\beta)= \left(\alpha \beta 
     \sqrt{\zeta(2)+(\psi(2)-\ln \alpha)^2}\right)^{-1}
\end{eqnarray*}
for the ordered parameterizations $\{\beta,\alpha\}$ and
$\{\alpha,\beta\}$ respectively 
being $\zeta(2)=\pi^2/6$ the Riemann Zeta Function and $\psi(2)=1-\gamma$
the Digamma Function.

\small
{\raya}                   
\vspace{1.0cm}

\section{\LARGE \bf Hierarchical Structures}

In many circumstances, even though the experimental observations
respond to the same phenomena it is not always possible to consider
the full set of observations as an exchangeable sequence but rather
exchangeability within subgroups of observations. As stated earlier, this may
be the case when the results come from different experiments or when, 
within the
same experiment, data taking conditions (acceptances, efficiencies,...) 
change from run to run. A similar situation holds, for instance, for the
results of responses under a drug performed at different hospitals when  
the underlying conditions of the population vary between zones, countries,...
In general, we shall have different groups of observations
\begin{eqnarray*}
   &\xbold_1&\,=\,\{x_{11},x_{21},{\ldots},x_{n_11}\} \\
   &{\vdots}& \\
   &\xbold_j&\,=\,\{x_{1j},x_{2j},{\ldots},x_{n_jj}\} \\
   &{\vdots}& \\
   &\xbold_J&\,=\,\{x_{1J},x_{2J},{\ldots},x_{n_JJ}\}
\end{eqnarray*}
from $J$ experiments $e_1(n_1),e_2(n_2),{\ldots},e_J(n_J)$.
Within each sample $\xbold_j$, we can consider that exchangeability holds
and also for the sets of observations $\{\xbold_1,\xbold_2,\ldots,\xbold_J\}$
In this case, it is appropriate to consider {\sl hierarchical structures}.

Let's suppose that for each experiment
$e(j)$ the observations are drawn from the model
\begin{eqnarray*}
   p({\xbold}_j|\thetabold_j)\,\hspace{0.5cm};
   \hspace{1.cm}
   j=1,\,2,\,{\ldots},\,J
\end{eqnarray*}
Since the experiments are independent we assume that the parameters of 
the sequence
$\{\thetabold_1,\thetabold_2,{\ldots},\thetabold_J\}$ 
are exchangeable and that, although different,
they can be assumed to
have a common origin since they respond to the same phenomena.
Thus,we can set
\begin{eqnarray*}
   p(\thetabold_1,\,\thetabold_2,\dots,\thetabold_J|\phibold)\,=\,
\prod_{i=1}^{J}\,p(\thetabold_i|\phibold)
\end{eqnarray*}
with $\phibold$ the {\sl hyperparameters} for which we take a prior 
${\pi}(\phibold)$. Then we have the structure (fig. 2.3.)
\begin{eqnarray*}
   p(\xbold_1,{\ldots},\xbold_J,
     \thetabold_1,{\ldots},\thetabold_J,\phibold)\,=\,{\pi}(\phibold)\,
 \prod_{i=1}^{J}\,p(\xbold_i|\thetabold_i)\,\pi(\thetabold_i|\phibold)
\end{eqnarray*}
This structure can be repeated sequentially if we consider appropriate to
assign a prior ${\pi}(\phibold|\taubold)$ to the hyperparameters 
$\phibold$ so that
\begin{eqnarray*}
   p(\mbox{\boldmath $x$},\mbox{\boldmath $\theta$},
     \mbox{\boldmath $\phi$},\taubold)\,=\,
     p(\mbox{\boldmath $x$}|\mbox{\boldmath $\theta$})\,
    \pi(\mbox{\boldmath $\theta$}|\mbox{\boldmath $\phi$})\,
    \pi(\mbox{\boldmath $\phi$}|\taubold)\,
    \pi(\taubold)
\end{eqnarray*}

\begin{figure}[t]
\begin{center}

\mbox{\epsfig{file=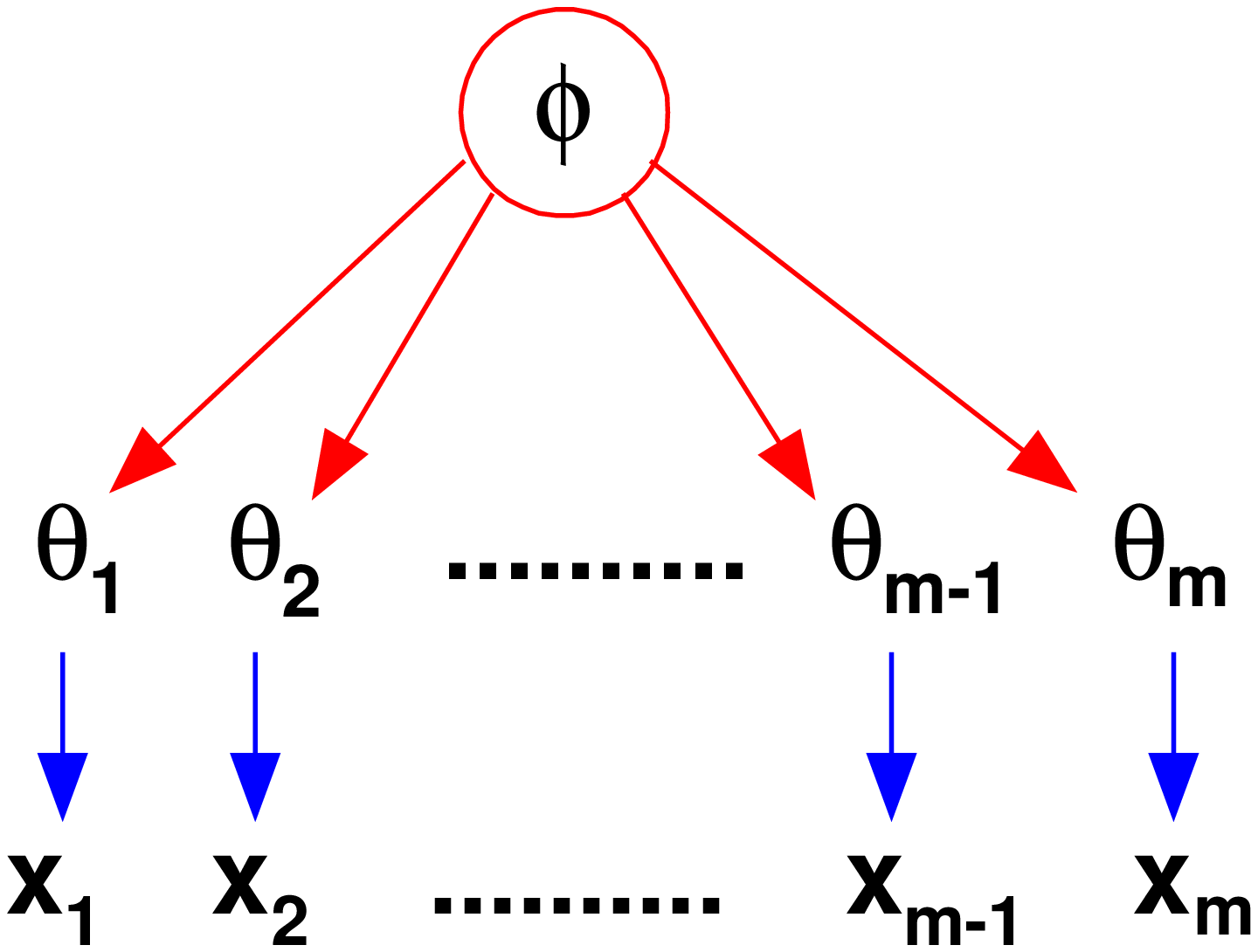,height=8cm,width=8cm}}

\vspace{-1.5cm}
\footnotesize{
{\bf Figure 2.3}.- Structure of the hierarchical model.
             }
\end{center}
\end{figure}

Now, consider the model $p(\xbold,\thetabold,\phibold)$. We may be interested
in $\thetabold$, in the hyperparameters $\phibold$ or in both. In general
we shall need the conditional densities:
\begin{eqnarray*}
&{\bullet}&\hspace{1.cm}
  p(\mbox{\boldmath ${\phi}$}|\mbox{\boldmath ${x}$})\,
{\propto}\,p(\mbox{\boldmath ${\phi}$})
    \int\,
    p(\mbox{\boldmath ${x}$}|\mbox{\boldmath ${\theta}$})\,
     p(\mbox{\boldmath ${\theta}$}|\mbox{\boldmath ${\phi}$})\,\,
     d\mbox{\boldmath ${\theta}$} \\
&{\bullet}&\hspace{1.cm}
  p(\mbox{\boldmath ${\theta}$}|
    \mbox{\boldmath ${x}$},\mbox{\boldmath ${\phi}$})\,=\,
    \frac{\textstyle 
      p(\mbox{\boldmath ${\theta}$},
    \mbox{\boldmath ${x}$},\mbox{\boldmath ${\phi}$})}
    {\textstyle p(\mbox{\boldmath ${x}$},\mbox{\boldmath ${\phi}$})}
    \hspace{2.cm}{\rm and} \\
&{\bullet}&\hspace{1.cm}
  p(\mbox{\boldmath ${\theta}$}|
    \mbox{\boldmath ${x}$})\,=\,
\frac{\textstyle p(\mbox{\boldmath ${x}$}|\mbox{\boldmath ${\theta}$})}
         {\textstyle p(\mbox{\boldmath ${x}$})}\,\,
	 p(\mbox{\boldmath ${\theta}$})
\,=\,
    \frac{\textstyle p(\mbox{\boldmath ${x}$}|\mbox{\boldmath ${\theta}$})}
         {\textstyle p(\mbox{\boldmath ${x}$})}\,\,
    \int\,
    p(\mbox{\boldmath ${\theta}$}|\mbox{\boldmath ${\phi}$})\,
     p(\mbox{\boldmath ${\phi}$})\,
     d\mbox{\boldmath ${\phi}$}
\end{eqnarray*}
that can be expressed as
\begin{eqnarray*}
  p(\mbox{\boldmath ${\theta}$}|
    \mbox{\boldmath ${x}$})\,=\,\int
    \frac{\textstyle p(\mbox{\boldmath ${x}$},\mbox{\boldmath ${\theta}$},
                     \mbox{\boldmath ${\phi}$})}
         {\textstyle p(\mbox{\boldmath ${x}$})}\,d\mbox{\boldmath ${\phi}$}
 \,=\,
    \int\,
    p(\mbox{\boldmath ${\theta}$}|\mbox{\boldmath ${x}$},
      \mbox{\boldmath ${\phi}$})\,
     p(\mbox{\boldmath ${\phi}$}|\mbox{\boldmath ${x}$})\,
     d\mbox{\boldmath ${\phi}$}
\end{eqnarray*}
and, since
\begin{eqnarray*}
  p(\mbox{\boldmath ${\theta}$}|
    \mbox{\boldmath ${x}$})\,=\,
     p(\mbox{\boldmath ${x}$}|\mbox{\boldmath ${\theta}$})\,
   \int
   p(\mbox{\boldmath ${\theta}$}|\mbox{\boldmath ${\phi}$})\,\,
    \frac{\textstyle p(\mbox{\boldmath ${\phi}$}|\mbox{\boldmath $x$})}
         {\textstyle p(\mbox{\boldmath ${x}$}|\mbox{\boldmath $\phi$})}
\,d\mbox{\boldmath ${\phi}$}
\end{eqnarray*}
we can finally write
\begin{eqnarray*}
\frac{\textstyle p(\mbox{\boldmath ${\theta}$},\mbox{\boldmath ${\phi}$})}
     {\textstyle p(\mbox{\boldmath ${x}$})}\,=\,
p(\mbox{\boldmath ${\theta}$}|\mbox{\boldmath ${\phi}$})\,
\frac{\textstyle p(\mbox{\boldmath ${\phi}$})}
     {\textstyle p(\mbox{\boldmath ${x}$})}\,=\,
p(\mbox{\boldmath ${\theta}$}|\mbox{\boldmath ${\phi}$})\,
\frac{\textstyle p(\mbox{\boldmath ${\phi}$}|\mbox{\boldmath ${x}$})}
         {\textstyle p(\mbox{\boldmath ${x}$}|\mbox{\boldmath ${\phi}$})}
\end{eqnarray*}

In general, this conditional densities have complicated expressions and
we shall use Monte Carlo methods to proceed 
(see Gibbs Sampling, example 3.15, in Lecture 3).

It is important to note that if the prior distributions are not proper
we can have improper marginal and posterior densities that obviously
have no meaning in the inferential process.
Usually, conditional densities are better behaved but, in any case, we have
to check that this is so. In general, the better behaved is the likelihood
the wildest behavior we can accept for the prior functions. We can also
used prior distributions that are a mixture of proper distributions:
\begin{eqnarray*}
p(\mbox{\boldmath ${\theta}$}|\mbox{\boldmath ${\phi}$})\,=\,
   \sum_i\,w_i\,
  p_i(\mbox{\boldmath ${\theta}$}|\mbox{\boldmath ${\phi}$})
\end{eqnarray*}
with $w_i{\geq}0$ and $\sum w_i=1$ so that the combination
is convex and we assure that it is proper density or, extending this to
a continuous mixture:
\begin{eqnarray*}
p(\mbox{\boldmath ${\theta}$}|\mbox{\boldmath ${\phi}$})\,=\,
   \int\,w(\mbox{\boldmath ${\sigma}$})\,
  p(\mbox{\boldmath ${\theta}$}|\mbox{\boldmath ${\phi}$},
\mbox{\boldmath ${\sigma}$})\,d\mbox{\boldmath ${\sigma}$}
\end{eqnarray*}

\section{\LARGE \bf Priors for discrete parameters}
So far we have discussed parameters with continuous support but in some
cases it is either finite or countable.
If the parameter of interest can take only a finite set of $n$ possible values,
the reasonable option for an {\sl uninformative prior} is 
a Discrete Uniform Probability $P(X=x_i)=1/n$.
In fact, it is left as an exercise to show that maximizing
the expected information provided by 
the experiment with the normalization constraint
(i.e. the probability distribution for which
the {\sl prior} knowledge is minimal)
drives to $P(X=x_i)=1/n$ in accordance with the {\sl Principle of 
Insufficient Reason}.

Even though finite discrete parameter spaces are either the most usual 
case we shall have to deal with or, at least, a sufficiently good 
approximation for the real situation, it may happen that 
a non-informative prior is not the most appropriate (see example 2.22).
On the other hand, if the
the parameter takes values on a countable set the problem is more involved.
A possible way out is to devise a hierarchical structure in which 
we assign the discrete parameter $\theta$ a prior
$\pi(\theta|\lambdabold)$ with $\lambdabold$ a set of continuous 
hyperparameters.
Then, since
 \begin{eqnarray*}
p(\xbold,\lambdabold)\,=\,
\sum_{\theta\in\Theta}p(\xbold|\theta)\pi(\theta|\lambdabold)\,
\pi(\lambdabold)\,=\,
p(\xbold|\lambdabold)\,\pi(\lambdabold)
\end{eqnarray*}
we get the prior $\pi(\lambdabold)$ by any of the previous procedures for 
continuous parameters with the model $p(\xbold|\lambdabold)$ and obtain
 \begin{eqnarray*}
\pi(\thetabold)\,\propto\,
\int_{\Lambda}\pi(\theta|\lambdabold)\,\pi(\lambdabold)\,d\lambdabold
\end{eqnarray*}
Different procedures are presented and discussed in [Be12].
\vspace{0.5cm}
\noindent
{\raya}                   
\vspace{0.35cm}
\footnotesize

\noindent
{\bf Example 2.22:}
The absolute value of the electric charge $(Z)$ of a particle is to be 
determined from the
number of photons observed by a Cherenkov Counter. We know from
test beam studies and Monte Carlo simulations that the number of observed
photons $n_{\gamma}$
produced by a particle of charge $Z$ is well described by a Poisson
distribution with parameter ${\mu}=n_0Z^2$; that is
\begin{eqnarray*}
    P(n_{\gamma}|n_0,Z)\,=\,e^{-n_0Z^2}\,\frac{(n_0Z^2)^{n_{\gamma}}}
                                           {\Gamma(n_{\gamma}+1)}
\end{eqnarray*}
so $E[n_{\gamma}|Z=1]=n_0$. First, by physics considerations $Z$ has a finite
support $\Omega_Z=\{1,2,{\ldots},n\}$.
Second, we know {\sl a priory} that not all incoming
nuclei are equally likely so a {\sl non-informative} prior may not be the best
choice. In any case, a discrete uniform prior will give the posterior:
\begin{eqnarray*}
    P(Z=k|n_{\gamma},n_0,n)\,=\,\frac{\textstyle e^{-n_0k^2}\,k^{2n_{\gamma}}}
                           {\textstyle \sum_{k=1}^n e^{-n_0k^2}\,k^{2n_{\gamma}}}
\end{eqnarray*}

\small
{\raya}                   
\vspace{1.0cm}
\section{\LARGE \bf Constrains on parameters and priors}

Consider a parametric model $p(\xbold|\thetabold)$ and the prior
$\pi_0(\thetabold)$. Now we have some information on the parameters that
we want to include in the prior. Typically we shall have say $k$ constraints of
the form
\begin{eqnarray*}
   \int_{\Theta}g_i(\thetabold)\,\pi({\thetabold})\,d\thetabold\,=\,a_i
   \hspace{0.5cm};\hspace{1.0cm}i=1,{\ldots},k
\end{eqnarray*}
Then, we have to find the prior $\pi(\thetabold)$ for which 
$\pi_0(\thetabold)$ is the best approximation, in the Kullback-Leibler sense,
including the constraints with the corresponding Lagrange multipliers
${\lambda}_i$; that is, the extremal of 
\begin{eqnarray*}
 {\Fcal}\,=\,
    \int_{\Theta}{\pi}(\thetabold)\,\log
         \frac{\textstyle \pi({\thetabold})}
              {\textstyle \pi_0({\thetabold})}\,d{\thetabold}\,+\,
    \sum_{i=1}^{k}\,{\lambda}_i\,\left(
   \int_{\Theta}g_i(\thetabold)\,\pi({\thetabold})\,d\thetabold\,-\,a_i\right)
\end{eqnarray*}
Again, it is left as an exercise to show that from Calculus of Variations we
have the well known solution
\begin{eqnarray*}
    {\pi}(\thetabold)\,\propto\,
    {\pi_0}(\thetabold)\,\exp\{\sum_{i=1}^{k}\,{\lambda}_i\,g_i(\thetabold)\}
   \hspace{0.8cm}{\rm where}\hspace{0.8cm}{\lambda}_i\, | \,
    \int_{\Theta}g_i(\thetabold)\,\pi({\thetabold})\,d\thetabold\,=\,a_i
\end{eqnarray*}

Quite frequently we are forced to include constraints on the 
support of the parameters: some are non-negative
(masses, energies, momentum, life-times,...), some are bounded in $(0,1)$
($\beta=v/c$, efficiencies, acceptances,...),...
At least from a formal point of view, 
to account for constraints on the support is a trivial problem.
Consider the model $p(\xbold|{\thetabold})$ with 
${\thetabold}{\in}{\Theta}_0$ and a reference prior
$\pi_0({\thetabold})$. Then, our inferences on $\thetabold$ shall be
based on the posterior
   \begin{eqnarray*}
     p({\thetabold}|\xbold)\,=\,
     \frac{\textstyle p(\xbold|{\thetabold})\,\pi_0({\thetabold})}
          {\textstyle \int_{{\Theta}_0}\,
           p(\xbold|{\thetabold})\,\pi_0({\thetabold})\,d{\thetabold} }
   \end{eqnarray*}
Now, if we require that
${\thetabold}{\in}{\Theta}{\subset}{\Theta}_0$ we define
\begin{eqnarray*}
     g_1({\thetabold})\,=\,\mbox{\boldmath $1$}_{\Theta}(\thetabold)
     &\longrightarrow&
     \int_{{\Theta}_0}\,g_1(\thetabold)\,\pi({\thetabold})\,d\thetabold\,=\,
     \int_{{\Theta}}\,\pi({\thetabold})\,d\thetabold\,=\,1-\epsilon \\
     g_2({\thetabold})\,=\,
     \mbox{\boldmath $1$}_{{\Theta}^c}(\thetabold)
     &\longrightarrow&
      \int_{{\Theta}_0}\,g_2(\thetabold)\,\pi({\thetabold})\,d\thetabold\,=\,
      \int_{{\Theta}^c}\,\pi({\thetabold})\,d\thetabold\,=\,\epsilon
\end{eqnarray*}
and in the limit $\epsilon\rightarrow 0$ we have the
{\sl restricted reference prior}
\begin{eqnarray*}
     \pi({\thetabold})\,=\,
     \frac{\textstyle \pi_0({\thetabold})}
          {\textstyle \int_{{\Theta}}\,
           \pi_0({\thetabold})\,d{\thetabold} }\,
          \mbox{\boldmath $1$}_{\Theta}(\thetabold)
   \end{eqnarray*}
as we have obviously expected. Therefore
   \begin{eqnarray*}
     p({\thetabold}|\xbold,{\thetabold}{\in}{\Theta})\,=\,
     \frac{\textstyle p(x|{\thetabold})\,\pi({\thetabold})}
          {\textstyle \int_{\Theta}\,
           p(x|{\thetabold})\,\pi({\thetabold})\,d{\thetabold} } \,=\,
     \frac{\textstyle p(x|{\theta})\,\pi_0({\thetabold})}
          {\textstyle \int_{{\Theta}}\,
           p(x|{\thetabold})\,\pi_0({\thetabold})\,d{\thetabold} }\,
      \mbox{\boldmath $1$}_{{\Theta}}(\thetabold)
   \end{eqnarray*}
that is, the same initial expression but normalized in the domain of interest
${\Theta}$.
\section{\LARGE \bf Decision Problems}

Even though all the information we have on the parameters of relevance is 
contained in the posterior density
it is interesting, as we saw in Lecture 1, to explicit some 
particular values that characterize the probability distribution.
This certainly entails a considerable and unnecessary reduction of the
available information but in the end, quoting Lord Kelvin,
{\sl ``... when you cannot express it in numbers, your knowledge is of a 
meager and unsatisfactory kind''}. In statistics, to specify a particular
value of the parameter is termed {\sl Point Estimation} and can be formulated
in the framework of {\sl Decision Theory}.

In general, {\sl Decision Theory} studies how to choose the 
{\sl optimal action} among several possible alternatives based on what has been
experimentally observed. Given a particular problem, we have to explicit the
set ${\Omega}_{\theta}$ of the possible {\sl "states of nature"}, the
set ${\Omega}_X$ of the possible experimental outcomes 
and the set ${\Omega}_A$ of the possible actions we can take.
Imagine, for instance, that we do a test on an individual suspected to
have some disease for which the medical treatment has some 
potentially dangerous collateral effects.
Then, we have:
\begin{eqnarray*}
{\Omega}_{\theta}\,&=&\,\{{\rm healthy},\,\,{\rm sic}\} \\
{\Omega}_{X}\,&=&\,\{{\rm test\,\, positive},\,\,{\rm test\,\, negative}\} \\
{\Omega}_{A}\,&=&\,\{{\rm apply\,\,treatment},\,\,{\rm do\,\,not\,\,apply\,\,
treatment}\}
\end{eqnarray*}
Or, for instance, a detector that provides within some accuracy
the momentum $(p)$ and the velocity
$(\beta)$ of charged particles. If we want to assign an hypothesis for the
mass of the particle we have that
${\Omega}_{\theta}={\Rcal}^+$ is the set of all possible states of nature (all
possible values of the mass),
${\Omega}_{X}$ the set of experimental observations (the momentum and the
velocity) and
${\Omega}_{A}$ the set of all possible actions that we can take 
(assign one or other value for the mass). In this case, we shall take a 
decision based on the probability density $p(m|p,\beta)$.

Obviously, unless we are in a state of absolute certainty we can not take
an action without potential losses. Based on the observed experimental
outcomes, we can for instance assign the particle a mass $m_1$ when the 
{\sl true state of nature} is $m_2{\neq}m_1$ or consider that the individual 
is healthy when is actually sic.  Thus,
the first element of Decision Theory is the {\sl Loss Function}:
\begin{eqnarray}
 l(a,{\theta})\,:\,({\theta},a)\,{\in}\,
      {\Omega}_{\theta}\,{\times}\,{\Omega}_A\,\,\,\,\,
      {\longrightarrow}\,\,\,\,\,{\Rcal}^++\{0\}
                \nonumber  
\end{eqnarray}
This is a non-negative function, defined for all
$\thetabold{\in}{\Omega}_{\theta}$ and the set of possible actions
$\mbox{\boldmath $a$}{\in}{\Omega}_A$, that
quantifies the {\sl loss} associated to take the action
$\abold$ (decide for $\abold$)
when the state of nature is $\thetabold$.

Obviously, we do not have a perfect knowledge of the {\sl state of nature}; 
what
we know comes from the observed data $\xbold$ and is contained in the posterior 
distribution $p(\thetabold|\xbold)$. Therefore, we define the 
{\sl Risk Function} ({\sl risk} associated to take the action
$\abold$, or decide for $\abold$ when we have observed the data $\xbold$)
as the expected value of the Loss Function:
   \begin{eqnarray}
   R(\mbox{\boldmath $a$}|\mbox{\boldmath $x$})\,=\,
   E_{\mbox{\boldmath ${\theta}$}}
   [l(\mbox{\boldmath $a$},\mbox{\boldmath ${\theta}$})]\,=\,
  \int_{{\Omega}_{\theta}}\,
     l(\mbox{\boldmath $a$},\mbox{\boldmath ${\theta}$})\,
     p(\mbox{\boldmath ${\theta}$}|\mbox{\boldmath $x$})\,
     d{\mbox{\boldmath ${\theta}$}}
                \nonumber 
   \end{eqnarray}

Sound enough, the Bayesian decision criteria consists on 
taking the action
$\abold(\xbold)$ ({\sl Bayesian action}) that minimizes the risk
$R(\abold|\xbold)$ ({\sl minimum risk}); that is, that minimizes the expected 
loss under the posterior density function
\footnote{The problems studied by {\sl Decision Theory}
can be addressed from the point of view of {\sl Game Theory}. 
In this case, instead of {\sl Loss Functions} one works with 
{\sl Utility Functions}
$u(\thetabold,\abold)$ that, in essence, are nothing else but
$u(\thetabold,\abold)=K-l(\thetabold,\abold){\geq}0$; 
it is just matter of personal optimism to work with
{\sl "utilities"} or  {\sl "losses"}. J. Von Neumann and
O. Morgenstern introduced in 1944 the idea of {\sl expected utility}
and the criteria to take as {\sl optimal action} hat which maximizes the
expected utility.}. 
Then, we shall encounter to kinds of problems:
\begin{itemize}
\item[$\bullet$] {\sl inferential problems}, where
                 ${\Omega}_A={\Rcal}$ y 
                 $\abold(\xbold)$ is a statistic that we shall take as
                 {\sl estimator} of the parameter $\thetabold$;
\item[$\bullet$] {\sl decision problems} (or {\sl hypothesis testing})
                 where ${\Omega}_A=\{{\rm accept},{\rm reject}\}$ or choose
                 one among a set of hypothesis.
\end{itemize}
Obviously, the actions depend on the loss function (that we have to specify)
and on the posterior density and, therefore, on the data through the model
$p(\xbold|\thetabold)$  and the prior function ${\pi}(\thetabold)$.
It is then possible that, for a particular model, two different loss functions
drive to the same decision or that the same loss function, depending
on the prior, take to different actions.

\subsection{Hypothesis Testing}
Consider the case where we have to choose between two 
exclusive and exhaustive hypothesis $H_1$ and $H_2(={H_1}^c)$. 
From the data sample and our prior beliefs
we have the posterior probabilities
   \begin{eqnarray}
     P(H_i|{\rm data})\,=\,
     \frac{\textstyle P({\rm data}|H_i)\,P(H_i)}
          {\textstyle P({\rm data})} \,\,\,\,\,\,\,\,\,\,;
          \,\,\,\,\,i=1,2
                \nonumber 
   \end{eqnarray}
and the actions to be taken are then:

\vspace{0.2cm}
\noindent
\hspace{3.cm}$a_1$: action to take if we decide upon $H_1$

\vspace{0.2cm}
\noindent
\hspace{3.cm}$a_2$: action to take if we decide upon $H_2$

\vspace{0.2cm}
\noindent
Then, we define the loss function $l(a_i,H_j)$; $i,j=1,2$ as:

\vspace{0.5cm}
\begin{tabular}{p{2.0cm}p{9.0cm}}
    \vspace{-2.9cm}
   \begin{eqnarray}
    l(a_i|H_j)\,=\,
       \left\{  \begin{array}{l}
        \\     \\     \\
        \\     \\     \\
        \\     \\     \\
        \\     \\   
               \end{array}
     \right.                  \nonumber
   \end{eqnarray}
 &
\begin{tabular}{p{2.0cm}p{6.cm}}
 $l_{11}=l_{22}=0$ & if we make the correct choice; that is, if we take 
       action $a_1$ when the state of nature is $H_1$ 
       or $a_2$ when it is $H_2$;
                 \\ & \\
$l_{12}>0$ &   if we take action $a_1$ (decide upon $H_1$)
                 when the state of nature is $H_2$
                 \\ & \\
$l_{21}>0$ &   if we take action $a_2$ (decide upon $H_2$)
                 when the state of nature is $H_1$
                 \\ & \\
\end{tabular}
\end{tabular}

\noindent
so the risk function will be:
   \begin{eqnarray*}
   R(a_i|{\rm data})\,=\,
   \sum_{j=1}^2\,l(a_i|H_j)\,P(H_j|{\rm data})
   \end{eqnarray*}
that is:
   \begin{eqnarray*}
   R(a_1|{\rm data})\,&=&\,l_{11}\,P(H_1|{\rm data})\,+\,
                          l_{12}\,P(H_2|{\rm data}) \\
   R(a_2|{\rm data})\,&=&\,l_{21}\,P(H_1|{\rm data})\,+\,
                          l_{22}\,P(H_2|{\rm data})
   \end{eqnarray*}
and, according to the minimum Bayesian risk, we shall choose the hypothesis
$H_1$ (action $a_1$) if
\begin{eqnarray*}
   R(a_1|{\rm data})\,<\,R(a_2|{\rm data})
\hspace{0.5cm}{\longrightarrow}\hspace{0.5cm}
 P(H_1|{\rm data})\,(l_{11}-l_{21})\,<\,
   P(H_2|{\rm data})\,(l_{22}-l_{12})
   \end{eqnarray*}
Since we have chosen $l_{11}=l_{22}=0$ in this particular case,
we shall take action $a_1$ (decide for hypothesis $H_1$) if: 
   \begin{eqnarray}
   \frac{\textstyle P(H_1|{\rm data})}
        {\textstyle P(H_2|{\rm data})}\,>\,
   \frac{\textstyle l_{12}}
        {\textstyle l_{21}}
                \nonumber 
   \end{eqnarray}
or action $a_2$
(decide in favor of hypothesis $H_2$) if:
   \begin{eqnarray}
   R(a_2,{\rm data})\,<\,R(a_1,{\rm data})
   \,\,\,\,\,{\longrightarrow}\,\,\,\,\,
   \frac{\textstyle P(H_2|{\rm data})}
        {\textstyle P(H_1|{\rm data})}\,>\,
   \frac{\textstyle l_{21}}
        {\textstyle l_{12}}
                \nonumber 
   \end{eqnarray}
that is, we take action $a_i$ $(i=1,2)$ if:
\begin{eqnarray}
   \frac{\textstyle P(H_i|{\rm data})}
        {\textstyle P(H_j|{\rm data})}\,=\,
   \left[ \frac{\textstyle P({\rm data}|H_i)}
               {\textstyle P({\rm data}|H_j)}\right]
   \left[ \frac{\textstyle P(H_i)}
               {\textstyle P(H_j)}\right]\,>\,
   \frac{\textstyle l_{ij}}
        {\textstyle l_{ji}}
                \nonumber 
   \end{eqnarray}
The ratio of likelihoods
   \begin{eqnarray}
   B_{ij}\,=\,
   \frac{\textstyle P({\rm data}|H_i)}
        {\textstyle P({\rm data}|H_j)}
                \nonumber 
   \end{eqnarray}
is called  {\bf Bayes Factor} $B_{ij}$ and
changes our prior beliefs on the two alternative hypothesis based on
the evidence we have from the data; that is, quantifies how strongly data 
favors one model over the other. Thus,
we shall decide in favor of hypothesis $H_i$ against $H_j$ $(i,j=1,2)$ if
   \begin{eqnarray}
 \frac{\textstyle P(H_i|{\rm data})}
        {\textstyle P(H_j|{\rm data})}\,>\,
   \frac{\textstyle l_{ij}}
        {\textstyle l_{ji}}
\hspace{1.cm}{\longrightarrow}\hspace{1.cm}
   B_{ij}\,>\,
   \frac{\textstyle P(H_j)}
        {\textstyle P(H_i)}\,
   \frac{\textstyle l_{ij}}
        {\textstyle l_{ji}}
                \nonumber 
   \end{eqnarray}
If we consider the same loss if we decide upon the wrong hypothesis
whatever it be, we have $l_{12}=l_{21}$ (Zero-One Loss Function).
In general, we shall be interested in testing: 
\begin{itemize}
\item[$1)$] {\bf Two simple hypothesis}, $H_1$ vs $H_2$, for which 
the models $M_i=\{X{\sim}p_i(x|{\theta}_i)\};\,i=1,2$ are fully specified
including the values of the parameters
(that is, ${\Theta}_i=\{\theta_i\}$). 
In this case, the Bayes Factor will be given by
the ratio of likelihoods 
\begin{eqnarray*}
   B_{12}\,=\,
       \frac{\textstyle p_1(\xbold|\theta_1)}
            {\textstyle p_2(\xbold|\theta_2)}
\hspace{2.cm}\left({\rm usually}\;\;
 \frac{\textstyle p(\xbold|\theta_1)}
            {\textstyle p(\xbold|\theta_2)}\right)
\end{eqnarray*}
The classical Bayes Factor is the ratio of the likelihoods for the two
competing models evaluated at their respective maximums.

\item[$2)$] {\bf A simple} $(H_1)$ {\bf vs a composite hypothesis}
$H_2$ for which the parameters of the model 
$M_2=\{X{\sim}p_2(x|{\theta}_2)\}$ have support on
${\Theta}_2$. Then we have to average the likelihood under $H_2$ and
\begin{eqnarray*}
B_{12}\,=\,
\frac{\textstyle p_1(\xbold|\theta_1)}
     {\textstyle \int_{\Theta_2}
            p_2(\xbold|\theta)\pi_2(\theta)d{\theta}}
\end{eqnarray*}

\item[$3)$] {\bf Two composite hypothesis:}
in which the models $M_1$ and $M_2$
have parameters that are not specified by the hypothesis so
\begin{eqnarray*}
B_{12}\,=\,
\frac{\textstyle \int_{\Theta_1}
            p_1(\xbold|\theta)\pi_1(\theta_1)d{\theta_1}}
     {\textstyle \int_{\Theta_2}
            p_2(\xbold|\theta_2)\pi_2(\theta_2)d{\theta_2}}
\end{eqnarray*}
\end{itemize}
and, since $P(H_1|{\rm data})+P(H_2|{\rm data})=1$, in the the posterior 
probability 
\begin{eqnarray*}
P(H_1|{\rm data})\,=\,
\frac{\textstyle B_{12}\,P(H_1)}
     {\textstyle P(H_2)\,+\,B_{12}\,P(H_2)}
\end{eqnarray*}
Usually, we consider equal prior probabilities for the two hypothesis 
$(P(H_1)=P(H_2)=1/2)$ but be aware that in some cases this may not be a 
realistic assumption.
 
Bayes Factors are independent of the prior beliefs on the hypothesis ($P(H_i)$) 
but, when we have composite hypothesis, we average the likelihood with a 
prior and if it is an improper function they are not well defined.
If we have prior knowledge
about the parameters, we may take informative priors that are proper but
this is not always the case. One possible way out is
to consider sufficiently general proper priors 
(conjugated priors for instance) so the Bayes factors are well defined 
and then study what is the sensitivity for different reasonable values
of the hyperparameters.
A more practical and interesting approach to avoid the
indeterminacy due to improper priors [OH95, BP96]
is to take a subset of the observed
sample to render a proper posterior (with, for instance,
reference priors) and use that as proper prior density to
compute the Bayes Factor with the remaining sample. Thus, if the 
sample $\xbold=\{x_1,{\ldots},x_n\}$ consists on iid observations, we may
consider  $\xbold=\{\xbold_1,\xbold_2\}$ and, with the
reference prior ${\pi}(\thetabold)$, obtain the proper posterior
\begin{eqnarray*}
{\pi}(\thetabold|\xbold_1)\,=\,
\frac{\textstyle p(\xbold_1|\thetabold)\,{\pi}(\thetabold)}
     {\textstyle \int_{\Theta}
            p(\xbold_1|\thetabold)\,{\pi}(\thetabold)\,d{\thetabold}}
\end{eqnarray*}
The remaining subsample $(\xbold_2)$ is then used to compute the partial
Bayes Factor
\footnote{Essentially, the ratio of the predictive inferences 
for $\xbold_2$ after $\xbold_1$ has been observed.}:
\begin{eqnarray*}
B_{12}(x_2|x_1)\,=\,
\frac{\textstyle 
       \int_{\Theta_1}p_1(\xbold_2|\thetabold_1)\,
            {\pi}_1(\thetabold_1|{\xbold}_1)\,d{\thetabold}_1}
{\textstyle 
       \int_{\Theta_2}p_2(\xbold_2|\thetabold_2)\,
            {\pi}_2(\thetabold_2|{\xbold}_1)\,d{\thetabold}_2}
\hspace{1.cm}\left(=
\frac{\textstyle BF(\xbold_1,\xbold_2)}{\textstyle BF(\xbold_1)}\right)
\end{eqnarray*}
for the hypothesis testing. 
Berger and Pericchi propose to use the 
minimal amount of data needed to specify a proper prior (usually
$\max\{{\rm dim}({\thetabold_i})\}$)
so as to leave most of the sample for the model testing and 
dilute the dependence on a particular election of
the training sample evaluating the
Bayes Factors with all possible minimal samples and choosing the truncated 
mean, the geometric mean or the 
median, less sensitive to outliers, as a characteristic value 
(see example 2.24).
A thorough analysis of Bayes Factors, with its caveats and advantages, 
is given in [Ka95].

A different alternative to quantify the evidence in favour of a particular
model that avoids the need of the prior specification and is easy to
evaluate
is the Schwarz criteria [Sc78] (or {\sl "Bayes Information Criterion (BIC)"}). 
The rationale is the following.
Consider a sample $\xbold=\{x_1,{\ldots},x_n\}$ and two alternative
hypothesis for the models 
$M_i=\{p_i(x|{\thetabold}_i);{\rm dim}(\thetabold_i)=d_i\};\,i=1,2$.
As we can see in Note 6,
under the appropriate conditions we can approximate the likelihood as
\begin{eqnarray*}
l({\thetabold}|{\xbold})\,{\simeq}\,
  l(\widehat{\thetabold}|{\xbold})\,{\rm exp}\left\{-
  \frac{\textstyle 1}{\textstyle 2}\,\sum_{k=1}^{d}\sum_{m=1}^{d}
    ({\theta}_k-\widehat{\theta}_k)
    \left[n{\Ibold}_{km}(\widehat{\thetabold})\right]
({\theta}_m-\widehat{\theta}_m)\right\}
\end{eqnarray*}
so taking a uniform prior for the parameters $\thetabold$, reasonable in
the region where the likelihood is dominant, we can approximate
\begin{eqnarray*}
J(\xbold)\,=\,
\int_{\Thetabold}p({\xbold}|{\thetabold})\,{\pi}(\thetabold)\,d{\thetabold}\,
{\simeq}\,
  p({\xbold}|\widehat{\thetabold})\,(2{\pi}/n)^{d/2}\,
|{\rm det}[{\Ibold}(\widehat{\thetabold})]|^{-1/2}
\end{eqnarray*}
and, ignoring terms that are bounded as $n{\rightarrow}{\infty}$, define 
the $BIC(M_i)$ for the model $M_i$ as
\begin{eqnarray*}
2\ln\,J_i(\xbold)\,{\simeq}\,BIC(M_i)\,\equiv\,
2\,\ln\,p_i({\xbold}|\widehat{\thetabold}_i)\,-\,d_i\,\ln\,n
\end{eqnarray*}
so:
\begin{eqnarray*}
B_{12}\,{\simeq}\,
\frac{p_1({\xbold}|\widehat{\thetabold}_1)}
     {p_2({\xbold}|\widehat{\thetabold}_2)}\,
   n^{(d_2-d_1)/2}
\hspace{0.5cm}{\longrightarrow}\hspace{0.5cm}
\Delta_{12}=
2\,{\ln}B_{12}\,{\simeq}\,
  2\,{\ln}\left(\frac{p_1({\xbold}|\widehat{\thetabold}_1)}
                    {p_2({\xbold}|\widehat{\thetabold}_2)}\right)\,-\,
   (d_1-d_2)\,{\ln}\,n
\end{eqnarray*}
and therefore, larger values of $\Delta_{12}=BIC(M_1)-BIC(M_2)$ 
indicate a preference for the hypothesis $H_1(M_1)$ against $H_2(M_2)$ 
being commonly accepted that for values grater than 6 the evidence is 
{\sl "strong"} 
\footnote{If $P(H_1)=P(H_2)=1/2$, then
$P(H_1|{\rm data})=0.95\,{\longrightarrow}\,B_{12}=19\,{\longrightarrow}\,
{\Delta}_{12}{\simeq}6$.}
although, in some cases, it is worth to study the behaviour
with a Monte Carlo sampling.
Note that
the last term penalises models with larger number of parameters and that
this quantification is sound when the sample
size $n$ is much larger than the dimensions $d_i$ of the parameters.  

\vspace{0.5cm}
\noindent
{\raya}                   
\vspace{0.35cm}
\footnotesize

\noindent
{\bf Example 2.23:} Suppose that from the information provided by a detector 
we estimate the mass of an incoming particle 
and we want to decide upon the two exclusive and alternative hypothesis
$H_1$ (particle of type 1) and $H_2(={H_1}^c)$ (particle of type 2).
We know from calibration data and Monte Carlo simulations that
the mass distributions for both hypothesis are, to a very
good approximation, Normal with means
$m_1$ and $m_2$ variances ${\sigma}_1^2$ and ${\sigma}_2^2$ respectively.
Then for an observed value of the mass $m_0$ we have:
 \begin{eqnarray*}
   B_{12}\,=\,
   \frac{\textstyle p(m_0|H_1)}
        {\textstyle p(m_0|H_2)}\,=\,
   \frac{\textstyle N(m_0|m_1,{\sigma}_1)}
        {\textstyle N(m_0|m_2,{\sigma}_2)}\,
  =\, \frac{\textstyle {\sigma}_2}
        {\textstyle {\sigma}_1}\,
        {\rm exp}\,
     \left\{
   \frac{\textstyle (m_0\,-\,m_2)^2}
        {\textstyle 2\,{\sigma}_2^2}\,-\,
   \frac{\textstyle (m_0\,-\,m_1)^2}
        {\textstyle 2\,{\sigma}_1^2}
     \right\}
   \end{eqnarray*}
Taking $(l_{12}=l_{21};l_{11}=l_{22}=0)$, the Bayesian decision criteria 
in favor of the hypothesis $H_1$ is:
   \begin{eqnarray}
   B_{12}\,>\,
   \frac{\textstyle P(H_2)}
        {\textstyle P(H_1)}
        \,\,\,\,\,{\longrightarrow}\,\,\,\,\,
 {\rm ln}\,  B_{12}\,>\, {\rm ln}\,
   \frac{\textstyle P(H_2)}
        {\textstyle P(H_1)}
                \nonumber 
   \end{eqnarray}
Thus,we have a critical value $m_c$ of the mass:
\begin{eqnarray}
   {{\sigma}_1}^2\,(m_c\,-\,m_2)^2\,-\,
   {{\sigma}_2}^2\,(m_c\,-\,m_1)^2\,=\,
   2\,{{\sigma}_1}^2\,{{\sigma}_2}^2\,{\rm ln}\,
 \left(  \frac{\textstyle P(H_2)\,{\sigma}_1}
        {\textstyle P(H_1)\,{\sigma}_2}     \right)
                \nonumber 
   \end{eqnarray}
such that, if $m_0<m_c$ we decide in favor of $H_1$ and for
$H_2$ otherwise. In the case that
${\sigma}_1={\sigma}_2$ and $P(H_1)=P(H_2)$, then $m_c=(m_1+m_2)/2$. This,
however, may be a quite unrealistic assumption for if $P(H_1)>P(H_2)$,
it may be more likely that the event is of type 1 being $B_{12}<1$.
\vspace{0.35cm}

\noindent
{\bf Example 2.24:} Suppose we have an iid 
sample $\xbold=\{x_1,\ldots,x_n\}$ of size $n$ with
$X{\sim}=N(x|\mu,1)$ 
and the two hypothesis $H_1=\{N(x|0,1)\}$ and 
$H_2=\{N(x|\mu,1);\mu{\neq}0\}$.
Let us take $\{x_i\}$ as
the minimum sample and, with the usual constant prior, consider the
proper posterior
\begin{eqnarray*}
   \pi(\mu|x_i)\,=\,\frac{1}{\sqrt{2\pi}}\,\exp\{-(\mu-x_i)^2/2\}
\end{eqnarray*}
that we use as a prior for the rest of the sample
$\xbold'=\{x_1,\ldots,x_{i-1},x_{i+1},\ldots,x_n\}$. Then
\begin{eqnarray*}
 \frac{P(H_1|\xbold',x_i)}{P(H_2|\xbold',x_i)}\,=\,
        B_{12}(i)\,  \frac{P(H_1)}{P(H_2)}
 \hspace{0.5cm}{\rm where}\hspace{0.5cm}
  B_{12}(i)\,=\,\frac{\textstyle p(\xbold'|0)}
     {\textstyle \int_{-\infty}^{\infty}
            p(\xbold'|\mu)\pi(\mu|x_i)d\mu}\,=\,n^{1/2}\,
    \exp\{-(n\,\overline{x}^2-x_i^2)/2\}
\end{eqnarray*}
and $\overline{x}=n^{-1}\sum_{k=1}^nx_k$. 
To avoid the effect that a particular choice of the minimal sample 
$(\{x_i\})$ may have, this is evaluated for all possible minimal samples
and the median (or the geometric mean) of 
all the $B_{12}(i)$ is taken. 
Since $P(H_1|\xbold)+P(H_2|\xbold)=1$, if we assign equal prior 
probabilities to the two hypothesis $(P(H_1)=P(H_2)=1/2)$
we have that
\begin{eqnarray*}
  P(H_1|\xbold)\,=\,\frac{\textstyle B_{12}}{\textstyle 1+B_{12}}\,=\,\left(
  1\,+\,n^{-1/2}\,\exp\{(n\,\overline{x}^2-{\rm med}\{x_i^2\})/2\}\right)^{-1}
\end{eqnarray*}
is the posterior probability that quantifies the evidence in favor of the 
hypothesis $H_1$. It is left as an exercise to compare the Bayes Factor
obtained from the geometric mean with what you would get if you were to take
a proper prior ${\pi}(\mu|\sigma)=N({\mu}|0,{\sigma})$.
\vspace{0.35cm}

\noindent
{\bf Problem 2.11:}
Suppose we have $n$ observations (independent, under the same experimental
conditions,...) of energies or decay time of particles above a certain known
threshold and we want to test the evidence of an exponential fall against a
power law. Consider then
a sample $\xbold=\{x_1,\ldots,x_n\}$ of observations with 
${\rm supp}(X)=(1,\infty)$ and the two models
\begin{eqnarray*}
M_1:\,p_1(x|\theta)\,=\,{\theta}\exp\{-{\theta}(x-1)\}
          \mbox{\boldmath $1$}_{(1,\infty)}(x)
          \hspace{1.cm}{\rm and}\hspace{1.cm}
M_2:\,p_2(x|\alpha)\,=\,
{\alpha}x^{-({\alpha}+1)}\mbox{\boldmath $1$}_{(1,\infty)}(x)
\end{eqnarray*}
that is, Exponential and Pareto
with unknown parameters $\theta$ and $\alpha$.
Show that for the minimal sample $\{x_i\}$ and reference priors, the
Bayes Factor $B_{12}(i)$ is given by
\begin{eqnarray*}
B_{12}(i)\,=\,\left(
        \frac{x_g\,\ln x_g}{\overline{x}-1}\right)^n
        \left(\frac{x_i-1}{x_i\,\ln x_i}\right)\,=\,
\frac{p_1({\xbold}|\widehat{\thetabold})}
     {p_2({\xbold}|\widehat{\alphabold})}\,
\left(\frac{x_i-1}{x_i\,\ln x_i}\right)
\end{eqnarray*}
where $(\overline{x},x_g)$ are the arithmetic and geometric sample means
and $(\widehat{\thetabold},\widehat{\alphabold})$ the values that maximize 
the likelihoods and therefore
\begin{eqnarray*}
{\rm med}\{B_{12}(i)\}_{i=1}^n\,=\,
 \left(\frac{x_g\,\ln x_g}{\overline{x}-1}\right)^n\,
{\rm med}\left\{\frac{x_i-1}{x_i\,\ln x_i}\right\}_{i=1}^n
\end{eqnarray*}
\vspace{0.35cm}

\noindent
{\bf Problem 2.12:}
Suppose we have two experiments $e_i(n_i);\,i=1,2$ in which, out of $n_i$
trials, $x_i$ successes have been observed and we are interested in testing
whether both treatments are different or not ({\sl contingency tables}). 
If we assume Binomial models 
$Bi(x_i|n_i,\theta_i)$ for both experiments and the two hypothesis
$H_1:\{\theta_1=\theta_2\}$ and $H_2:\{\theta_1{\neq}\theta_2\}$,
the Bayes Factor will be
\begin{eqnarray*}
B_{12}\,=\,
 \frac{\int_\Theta\,Bi(x_1|n_1,\theta)Bi(x_2|n_2,\theta){\pi}(\theta)d\theta }
      {\int_{\Theta_1}\,Bi(x_1|n_1,\theta_1){\pi}(\theta_1)d\theta_1\,
       \int_{\Theta_2}\,Bi(x_2|n_2,\theta){\pi}(\theta_2)d\theta_2}
\end{eqnarray*}
We may consider proper Beta prior densities $Be(\theta|a,b)$. 
In a specific pharmacological analysis, a sample of
$n_1=52$ individuals were administered a placebo and $n_2=61$ were treated with
an a priori beneficial drug. After the essay,
positive effects were observed in $x_1=22$ out of the 52 and $x_2=41$ out of
the 61 individuals.
It is left as an exercise to obtain the posterior probability 
$P(H_2|{\rm data})$ with Jeffreys' $(a=b=1/2)$ and Uniform $(a=b=1)$ priors
and to determine the BIC difference $\Delta_{12}$. 

\small

{\raya}                   
\vspace{1.0cm}

\subsection{Point Estimation}
When we have to face the problem to characterize the posterior density
by a single number, the most usual {\sl Loss Functions} are:

\vspace{0.5cm}
\noindent
$\bullet$ {\bf Quadratic Loss:} In the simple one-dimensional case, the
Loss Function is
   \begin{eqnarray}
     l({\theta},a)\,=\,({\theta}-a)^2
                \nonumber 
   \end{eqnarray}
so, minimizing the {\sl Risk}:
   \begin{eqnarray}
   {\rm min}\,
  \int_{{\Omega}_{\theta}}\,
     ({\theta}-a)^2\,
     p({\theta}|\mbox{\boldmath $x$})\,d{\theta}\;\;\;\;\;\;\;\;\;\;
     \hspace{0.5cm}{\longrightarrow}\hspace{0.5cm}
  \int_{{\Omega}_{\theta}}\,
     ({\theta}-a)\,
     p({\theta}|\mbox{\boldmath $x$})\,d{\theta}\,=\,0
                \nonumber 
   \end{eqnarray}
and therefore $a=E[{\theta}]$; that is, the posterior mean.

  In the $k-$dimensional case, if 
${\Acal}={\Omega}_{\theta}={\Rcal}^k$ we shall take as Loss Function
   \begin{eqnarray}
     l(\mbox{\boldmath ${\theta}$},
\mbox{\boldmath $a$})\,=\,
(\mbox{\boldmath $a$}-\mbox{\boldmath ${\theta}$})^T\,{\mathbf{H}}\,
                             (\mbox{\boldmath $a$}-\mbox{\boldmath ${\theta}$})
                \nonumber 
   \end{eqnarray}
where ${\mathbf{H}}$ is a positive defined symmetric matrix. It is clear that:
\begin{eqnarray*}
           {\rm min}\,\int_{{\Rcal}^k}\,
           (\mbox{\boldmath $a$}-\mbox{\boldmath ${\theta}$})^T\,
{\bf H}\,(\mbox{\boldmath $a$}-\mbox{\boldmath ${\theta}$})
  \,p(\mbox{\boldmath ${\theta}$}|\mbox{\boldmath $x$})\,
d{\mbox{\boldmath ${\theta}$}}
\hspace{0.5cm}{\longrightarrow}\hspace{0.5cm}
       {\mathbf{H}}\,\mbox{\boldmath $a$}\,=\,
       {\mathbf{H}}\,E[\mbox{\boldmath ${\theta}$}]
\end{eqnarray*}
so, if ${\mathbf{H}}^{-1}$ exists, then
$\abold=E[\thetabold]$. Thus, we have that the Bayesian estimate under a
quadratic loss function is the mean of
$p(\thetabold|\xbold)$ (... if exists!).

\vspace{0.5cm}
\noindent
$\bullet$ {\bf Linear Loss:} If
${\Acal}\,=\,{\Omega}_{\theta}\,=\,{\Rcal}$, 
we shall take the loss function:
   \begin{eqnarray}
 l({\theta},a)\,=\,c_1\,(a-{\theta})\,\mbox{\boldmath $1$}_
  {{\theta}{\leq}a}\,+\,
  c_2\,({\theta}-a)\,\mbox{\boldmath $1$}_{{\theta}>a}
                \nonumber 
   \end{eqnarray}
Then, the estimator will be such that
\begin{eqnarray*}
   {\rm min}\,
  \int_{{\Omega}_{\theta}}
     l(a,{\theta})
     p({\theta}|\mbox{\boldmath $x$})d{\theta}=    
   {\rm min}\left(c_1
  \int_{-{\infty}}^{a}
     (a-{\theta})
     p({\theta}|\mbox{\boldmath $x$})d{\theta}\,+\,
     c_2
       \int_{a}^{{\infty}}
        ({\theta}-a)
        p({\theta}|\mbox{\boldmath $x$})d{\theta} \right)
   \end{eqnarray*}
After derivative with respect to $a$ we have
   $(c_1\,+\,c_2)\,P({\theta}{\leq}a)\,-\,c_2\,=\,0$
and therefore the estimator will be the value of $a$ such that
\begin{eqnarray*}
   P({\theta}{\leq}a)\,=\,\frac{\textstyle c_2}
                               {\textstyle c_1\,+\,c_2}
\end{eqnarray*}
In particular, if $c_1=c_2$ then $P({\theta}{\leq}a)\,=\,1/2$ and we shall
have the median of the distribution
$p(\thetabold|\xbold)$. In this case, the Loss Function can be expressed
more simply as $l({\theta},a)=|{\theta}-a|$.

\vspace{0.5cm}
\noindent
$\bullet$ {\bf Zero-One Loss:} Si
${\Acal}\,=\,{\Omega}_{\theta}\,=\,{\Rcal}^k$, we shall take the Loss Function
\begin{eqnarray*}
  l(\mbox{\boldmath ${\theta}$},\mbox{\boldmath $a$})\,=\,1\,-\,
  \mbox{\boldmath $1$}_
  {{\Bcal}_{\epsilon}(\mbox{\boldmath $a$})}
\end{eqnarray*}
where ${\Bcal}_{\epsilon}(\mbox{\boldmath $a$}){\in}{\Omega}_{\theta}$ 
is an open ball of radius $\epsilon$ centered at $\abold$. The
corresponding point estimator will be:
\begin{eqnarray*}
    {\rm min}\,
  \int_{{\Omega}_{\theta}}\,
     (1\,-\,{\mathbf{1}}_{{\Bcal}_{\epsilon}(\mbox{\boldmath $a$})})\,
     p(\mbox{\boldmath ${\theta}$}|\mbox{\boldmath $x$})\,
d\mbox{\boldmath ${\theta}$}\,=\, 
   {\rm max}\,
  \int_{{\Bcal}_{\epsilon}(\mbox{\boldmath $a$})}\,
     p(\mbox{\boldmath ${\theta}$}|\mbox{\boldmath $x$})\,
d{\mbox{\boldmath ${\theta}$}}
\end{eqnarray*}
It is clear than, in the limit ${\epsilon}{\rightarrow}0$, 
the Bayesian estimator for the Zero-One Loss Function will be the mode
of $p(\thetabold|\xbold)$ if exists. 

As explained in Lecture 1, the mode, the median and the mean can be very
different if the distribution is not symmetric.
Which one should we take then?. Quadratic losses, for which 
large deviations from the {\sl true} value are penalized quadratically, 
are the most common option but, even if for 
unimodal symmetric the three statistics coincide, it may be misleading to
take this value as a characteristic number for the information we got
about the parameters or even be nonsense.
In the hypothetical case that the posterior is essentially the same
as the likelihood (that is the case for a sufficiently smooth prior),
the Zero-One Loss points to the classical estimate
of the {\sl Maximum Likelihood Method}. Other considerations of interest
in Classical Statistics (like bias, consistency, minimum variance,...)
have no special relevance in Bayesian inference. 

\vspace{0.5cm}
\noindent
{\raya}                   
\vspace{0.35cm}
\footnotesize

\noindent
{\bf Problem 2.13: The Uniform Distribution.}
Show that for the posterior density (see example 2.4)
\begin{eqnarray*}
   p(\theta|x_M,n)\,=\,n\,\frac{\textstyle x_M^{n}}
                             {\textstyle \theta^{n+1}}
\,{\mbox{\boldmath ${1}$}}_{[x_M,\infty)}(\theta)
\end{eqnarray*}
the point estimates under quadratic, linear and 0-1 loss functions are
\begin{eqnarray*}
  \theta_{QL}\,=\,x_M\,\frac{n}{n-1}
  \hspace{0.3cm};\hspace{0.8cm}
  \theta_{LL}\,=\,x_M\,2^{1/n}
  \hspace{0.3cm}{\rm and}\hspace{0.8cm}
  \theta_{01L}\,=\,x_M
\end{eqnarray*}
and discuss which one you consider more reasonable.

\small
{\raya}                   
\vspace{1.0cm}

\section{\LARGE \bf Credible Regions}

Let $p(\thetabold|\xbold)$, with
$\thetabold{\in}{\Omega}{\subseteq}{\Rcal}^n$ be a posterior density function. 
A credible region with probability content $1-\alpha$ is a region of
$V_{\alpha}{\subseteq}\Theta$ of the parametric space such that 
\begin{eqnarray*}
P(\thetabold{\in}V_{\alpha})\,=\,
\int_{V_{\alpha}}p(\thetabold|\xbold)\,d\thetabold\,=\,
1-\alpha 
\end{eqnarray*}
Obviously, for a given probability
content credible regions are not unique and a sound criteria
is to specify the one that the smallest possible volume. 
A region $C$ of the parametric space $\Omega$
is called {\sl Highest Probability Region} (HPD) 
with probability content $1-{\alpha}$ if:
\begin{itemize}
    \item[1)] $P({\thetabold}{\in}C)\,=\,1-{\alpha}$;
              $\,\,\,\,\,C{\subseteq}{\Omega}$;
    \item[2)] $p(\thetabold_1|\cdot)\,{\geq}\,p(\thetabold_2|\cdot)$ for all
              $\thetabold_1{\in}C$ and $\thetabold_2{\notin}C$ except, 
              at most, for a subset of $\Omega$ with zero probability measure.
\end{itemize}
It is left as an exercise to show that condition $2)$ implies that
the HPD region so defined is of minimum volume so both definitions are
equivalent. Further properties that are easy to demonstrate are:
\begin{itemize}
    \item[1)] If $p(\theta|\cdot)$ is {\sl not uniform},
              the HPD region
              with probability content $1-{\alpha}$ {\sl is unique};
    \item[2)] If $p(\thetabold_1|\cdot)=p(\thetabold_2|\cdot)$, then 
              $\thetabold_1$ and $\thetabold_2$ are both either included 
              or excluded of the HPD region;
    \item[3)] If $p(\thetabold_1|\cdot){\neq}p(\thetabold_2\cdot)$, 
              there is an HPD region for some value of $1-{\alpha}$ that
              contains one value of $\thetabold$ and not the other;
    \item[4)] $C=\{\thetabold{\in}{\Theta}|
               p(\thetabold|\xbold){\geq}k_{\alpha}\}$
               where $k_{\alpha}$ is the largest constant for which
               $P(\thetabold{\in}C){\geq}{\alpha}$;
    \item[5)] If $\phibold=f(\thetabold)$ is a one-to-one transformation, then
              \begin{itemize}
              \item[a)] any region with probability content 
              $1-{\alpha}$ for $\thetabold$ will have probability
              content $1-{\alpha}$ for $\phibold$ but...
              
              \item[b)] an HPD region for $\thetabold$ will not, in general,
              be an HPD region for $\phibold$ unless the transformation is 
              linear.
              \end{itemize}
\end{itemize}

In general, evaluation of credible regions is a bit messy task. A simple
way through is to do a Monte Carlo sampling of the posterior density and
use the $4^{\rm th}$ property.
For a one-dimensional parameter, the condition that the HPD region
with probability content $1-{\alpha}$ has the minimum length allows to
write a relation that may be useful to obtain those regions in an easier manner.
Let  $[{\theta}_1,{\theta}_2]$ be an interval such that
  \begin{eqnarray}
        \int_{{\theta}_1}^{{\theta}_2}\,
         p({\theta}|\cdot)\,d{\theta}\,=\,1\,-\,{\alpha}
        \nonumber
  \end{eqnarray}
For this to be an HPD region we have to find the extremal of
the function
  \begin{eqnarray}
        {\phi}({\theta}_1,{\theta}_2,{\lambda})\,=\,
         ({\theta}_2-{\theta}_1)\,+\,{\lambda}\,\left(
        \int_{{\theta}_1}^{{\theta}_2}\,
         p({\theta}|\cdot)\,d{\theta}\,-\,(1\,-\,{\alpha}) \right)
        \nonumber
  \end{eqnarray}
Taking derivatives we get:
  \begin{eqnarray*}
  \left(
  \frac{\textstyle {\partial}{\phi}({\theta}_1,{\theta}_2,{\lambda})}
       {\textstyle {\partial}{\theta}_i}\right)_{i=1,2}\,=\,0
       \hspace{1.cm}&{\longrightarrow}&\hspace{1.cm}
           p({\theta}_1|\cdot)=\,p({\theta}_2\cdot)\\
  \frac{\textstyle {\partial}{\phi}({\theta}_1,{\theta}_2,{\lambda})}
       {\textstyle {\partial}{\lambda}}\,=\,0
       \hspace{1.cm}&{\longrightarrow}&\hspace{1.cm}
        \int_{{\theta}_1}^{{\theta}_2}\,
         p({\theta})\,d{\theta}\,=\,1\,-\,{\alpha}
  \end{eqnarray*}
Thus, from the first two conditions we have that
$p({\theta}_1|\cdot)\,=\,p({\theta}_2|\cdot)$ 
and, from the third, we know that
${\theta}_1\,{\neq}{\theta}_2$. In the special case that the distribution
is unimodal and symmetric 
the only possible solution is ${\theta}_2=2E[{\theta}]-{\theta}_1$.

The HPD regions are useful to summarize the information on the parameters
contained in the posterior density $p(\thetabold|\xbold)$ 
but it should be clear that there is no justification
to reject a particular value $\thetabold_0$ just because is not included in
the HPD region (or, in fact, in whatever confidence region) and that in some
circumstances (distributions with more than one mode for instance) 
it may be the union of disconnected regions.

\section{\LARGE \bf Bayesian ($\Bcal$) vs Classical ($\Fcal$) Philosophy} 
The Bayesian philosophy aims at the right questions in a very intuitive and,
at least conceptually, simple manner. However
the {\sl "classical"} (frequentist) approach to statistics, that 
has been very useful in scientific reasoning over the last century, 
is at present more widespread in the Particle Physics community and
most of the stirred up controversies are originated by 
misinterpretations. 
It is worth to take a look for instance at [Ja06].  
Let's see
how a simple problem is attacked by the two schools. "We" are $\Bcal$, "they"
are $\Fcal$.

Suppose we want to estimate the life-time of a particle. We both {\sl "assume"} 
an exponential model $X{\sim}Ex(x|1/\tau)$ and do an experiment $e(n)$ that
provides an iid sample $\xbold=\{x_1,x_2,\ldots,x_n\}$. In this case there
is a sufficient statistic $\tbold=(n,\overline{x})$ 
with $\overline{x}$ the sample mean so let's define 
the random quantity 
\begin{eqnarray*}
X=\frac{1}{n}\sum_{i=1}^n X_i \hspace{0.5cm}{\sim}\hspace{0.5cm}
    p(x|n,\tau)\,=\,\left(\frac{n}{\tau}\right)^n\,
    \frac{1}{\Gamma(n)}\,\exp\left\{ -n{x}\tau^{-1}\right\}\,
    {x}^{n-1}\,{\mbox{\boldmath ${1}$}}_{(0,\infty)}({x})
\end{eqnarray*}
What can we say about the parameter of interest $\tau$?.

$\Fcal$ will start by finding the {\sl estimator} (statistic)
$\widehat{\tau}$ that maximizes the likelihood (MLE). 
In this case it is clear that
$\widehat{\tau}=\overline{x}$, the sample mean. 
We${\Bcal}$ may ask about the rationale behind 
because, apparently, there is no serious mathematical reasoning that justifies 
this procedure. $\Fcal$ will respond that,
in a certain sense, even for us this should be a
reasonable way because if we have a smooth prior function, the
posterior is dominated by the likelihood and one possible point estimator
is the mode of the posterior. Beside that, he will
argue that maximizing the likelihood renders an estimator that often has
``good'' properties like unbiasedness, 
invariance under monotonous one-to-one transformations,
consistency (convergence in probability),  
smallest variance within the class of unbiased estimators (Cram\`er-Rao bound),
approximately well known distribution,...
We$\Bcal$ may question some of them 
(unbiased estimators are not always the best option and invariance... well,
if the
transformation is not linear usually the MLE is biased),
argue that the others hold in 
the asymptotic limit,... Anyway; for this particular case one has that:
\begin{eqnarray*}
    E[\widehat{\tau}]=\tau \hspace{1.cm}{\rm and}\hspace{1.cm}
    V[\widehat{\tau}]=\frac{\tau^2}{n}
\end{eqnarray*}
and $\Fcal$ will claim that
{\sl ``if you repeat the experiment''} many times under the same
conditions, you will get a sequence of estimators
$\{\widehat{\tau}_1, \widehat{\tau}_2,...\}$ that eventually will
cluster around the life-time
$\tau$. Fine but we$\Bcal$ shall point out that, first, although desirable 
we usually 
do not repeat the experiments (and under the same conditions is even more rare)
so we have just one observed sample 
($\xbold \rightarrow \overline{x}=\widehat{\tau}$) from $e(n)$.
Second, {\sl ``if you repeat the experiment you will get''} is a free and
unnecessary hypothesis. You do not know what you will get, among other things,
because the model we are considering may not be the way
nature behaves. Besides that,
it is quite unpleasant that inferences on the life-time depend 
upon what you think you will get if you do what you know you are not 
going to do. And third, that
this is in any case a nice sampling property of the estimator $\hat{\tau}$
but eventually we are interested in $\tau$ so, What can we say about it?.

For us, the answer is clear.
Being $\tau$ a scale
parameter we write the posterior density function
\begin{eqnarray*}
    p(\tau|n,\overline{x})\,=\,\frac{(n\,\overline{x})^{n}}
                                    {\Gamma(n)}\,
    \exp\left\{ -n\,\overline{x}\tau^{-1}\right\}\,
    \tau^{-(n+1)}\,{\mbox{\boldmath ${1}$}}_{(0,\infty)}(\tau)
\end{eqnarray*}
for the {\sl degree of belief} we have on the parameter 
and easily get for instance:
\begin{eqnarray*}
    E[\tau^k]=(n\overline{x})^k\,\frac{\Gamma(n-k)}{\Gamma(n)}
    \hspace{0.5cm}\longrightarrow\hspace{0.5cm}
    E[\tau]=\overline{x}\frac{n}{n-1}
    \hspace{0.3cm};\hspace{0.3cm}
    V[\tau]=\overline{x}^2\frac{n^2}{(n-1)^2(n-2)}
    \hspace{0.3cm};\hspace{0.3cm}\ldots
\end{eqnarray*}
Cleaner and simpler impossible.

\begin{figure}[t]
\begin{center}

\mbox{\epsfig{file=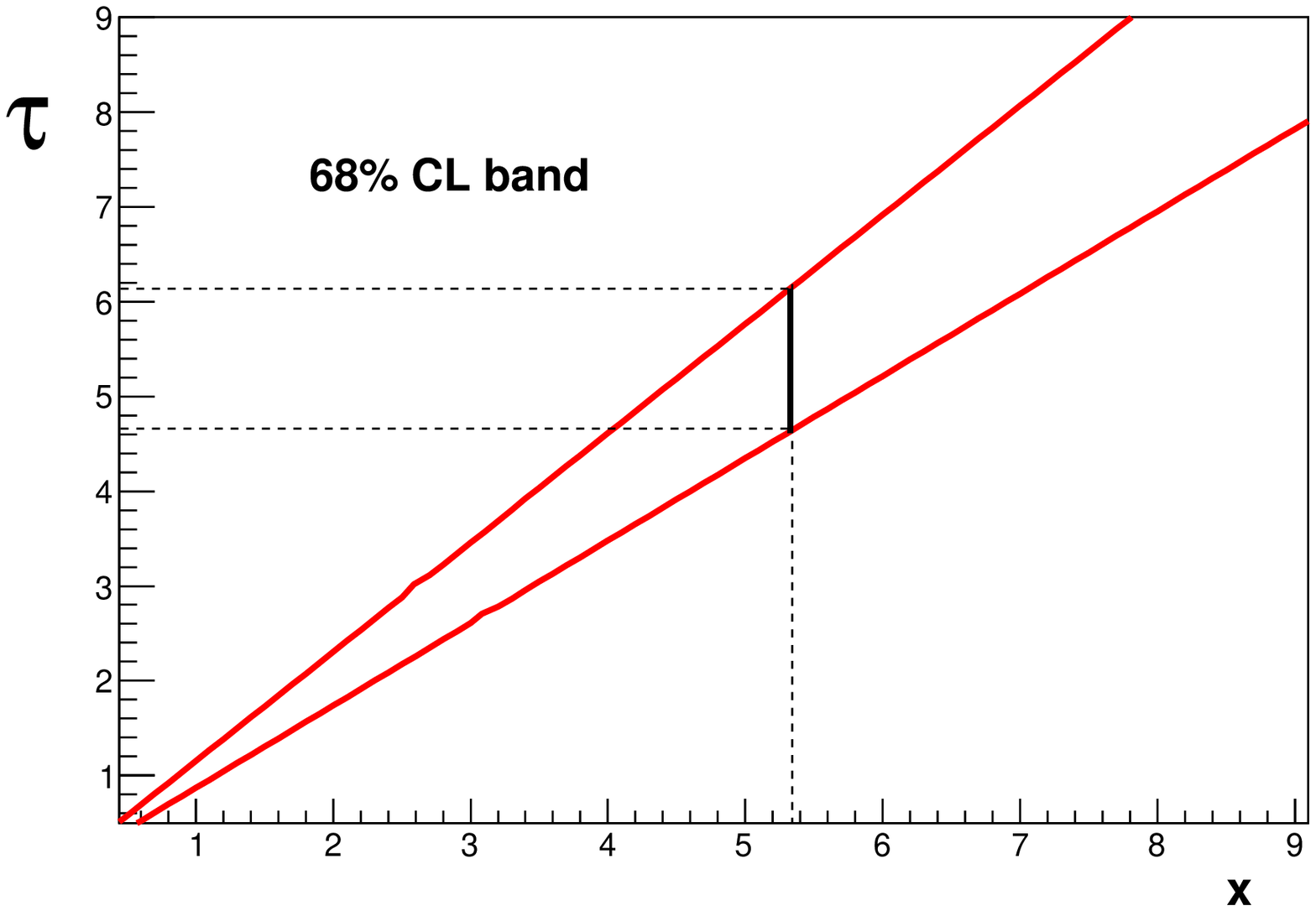,height=6cm,width=7cm}
      \epsfig{file=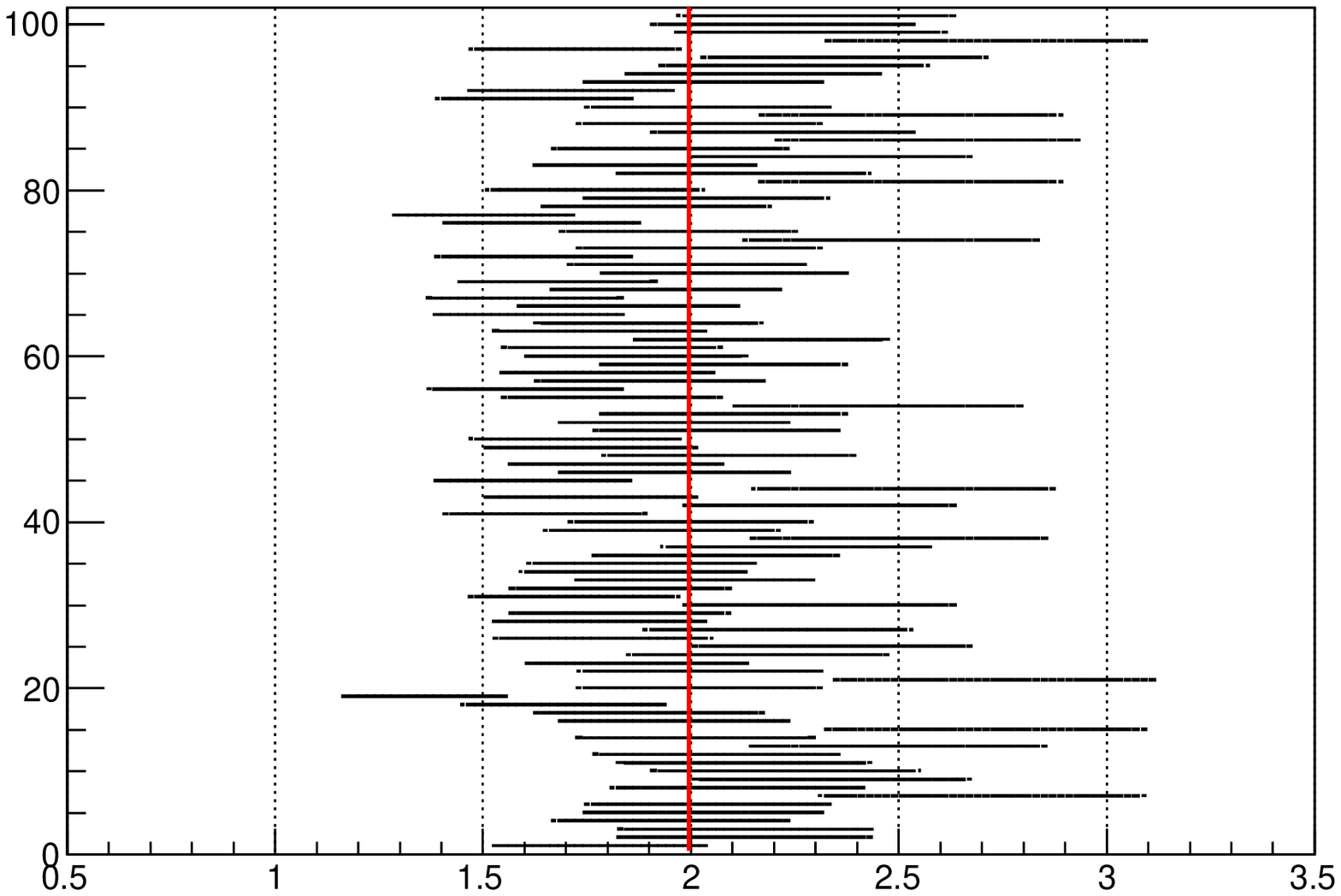,height=6cm,width=7cm}}

\vspace{0.5cm}
\footnotesize{
{\bf Figure 2.4}.- (1): 68\% confidence level bands in the $(\tau,X)$ plane.
                   (2): 68\% confidence intervals intervals obtained for 
                        100 repetitions of the experiment.
             }
\end{center}
\end{figure}

To bound the life-time, $\Fcal$ proceeds with the determination of the 
{\sl Confidence Intervals}. 
The classical procedure was introduced by J. Neyman in 1933 and
rests on establishing, for an specified 
probability content, the domain of the random quantity 
(usually a statistic) as function of the possible values the 
parameters may take.
Consider a one dimensional parameter $\theta$ and the
model $X{\sim}p(x|\theta)$.
Given a desired probability content
${\beta}{\in}[0,1]$, he determines the interval  
$[x_1,x_2]{\subset}{\Omega}_X$ such that
\begin{eqnarray*}
   P(X{\in}[x_1,x_2])\,=\,\int_{x_1}^{x_2}p(x|\theta)\,dx\,=\,\beta
\end{eqnarray*}
for a particular fixed value of $\theta$. Thus, for each possible
value of $\theta$ he has one interval 
$[x_1=f_1(\theta;\beta),x_1=f_2(\theta;\beta)]{\subset}{\Omega}_X$ 
and the sequence of those 
intervals gives a band in the $\Omega_{\theta}{\times}\Omega_{X}$ 
region of the real plane.
As for the {\sl Credible Regions}, these intervals
are not uniquely determined so one usually adds the condition:
\begin{eqnarray*}
1)\hspace{0.5cm}   \int_{\-\infty}^{x_1}p(x|\theta)\,dx\,=\,   
   \int_{x_2}^{\infty}p(x|\theta)\,dx\,=\,
\frac{1-\beta}{2}\hspace{1.5cm}{\rm or} \\
2)\hspace{0.5cm}   \int_{-\infty}^{x_1}p(x|\theta)\,dx\,=\,   
   \int_{x_1}^{\theta}p(x|\theta)\,dx\,=\,   
   \int_{\theta}^{x_2}p(x|\theta)\,dx\,=\,
\frac{\beta}{2}
\end{eqnarray*}
or, less often, (3) chooses the interval with smallest size.
Now, for an invertible mapping $x_i{\longrightarrow}f_i(\theta)$ one can write 
\begin{eqnarray*}
 \beta\,=\, P(f_1(\theta){\leq}X{\leq}f_2(\theta))\,=\,
P(f_2^{-1}(X){\leq}\theta{\leq}f_1^{-1}(X))
\end{eqnarray*}
and get the random interval $[f_2^{-1}(X),f_1^{-1}(X)]$ that contains
the given value of $\theta$ with probability $\beta$. Thus,
for each possible value that $X$ may take he will get an interval
$[f_2^{-1}(X),f_1^{-1}(X)]$ on the $\theta$ axis
and a particular experimental observation $\{x\}$
will single out one of them. This is the {\sl Confidence
Interval} that the frequentist analyst will quote. 
Let's continue with the life-time example and take, for illustration,
$n=50$ and $\beta=0.68$.
The bands $[x_1=f_1(\tau),x_2=f_2(\tau)]$ in the $(\tau,X)$ plane, in this
case obtained with the third prescription,
are shown in figure 2.4(1). They are essentially straight lines so
$P[X{\in}(0.847{\tau},1.126{\tau})]=0.68$. 
This is a correct statement, but doesn't say anything about $\tau$ 
so he inverts that and gets $0.89\,X<{\tau}<1.18\,X$ in such a way that 
an observed value $\{x\}$ singles out an interval in the vertical 
$\tau$ axis. We, Bayesian, will argue
this does not mean that $\tau$ has a 0.68 chance to lie in this interval
and the frequentist will certainly agree on that.
In fact, this is not an admissible question for him because
in the classical philosophy $\tau$ is a number,
unknown but a {\sl fixed} number. If he repeats the experiment $\tau$
will not change; it is the interval that will be different because
$x$ will change. They are {\sl random intervals} and
what the 68\% means is just that if he repeats the experiment a large number
$N$ of times, he will end up with $N$ intervals of which 
${\sim}68\%$ will contain the true value $\tau$ whatever it is. 
But the experiment is done only once so:
Does the interval derived from this observation contain $\tau$ or not?. 
We don't know, we have no idea if it does contain $\tau$, 
if it does not and
how far is the unknown true value. Figure 2.4(2) shows the 
68\% confidence intervals obtained after 100 repetitions of the experiment
for $\tau=2$ and 67 of them did contain the true value. But when the
experiment is done once, 
he picks up one of those intervals and has a 68\% chance that the
one chosen contains the true value. 
We{\Bcal} shall proceed in a different manner. After integration of
the posterior density we get
the HPD interval $P\left[{\tau}{\in}(0.85\,x,1.13\,x)\right]\,=\,0.68$;
almost the same but with a direct interpretation in terms of what we are 
interested in. 
Thus, both have an absolutely different philosophy:

\noindent
$\Fcal:$ "Given a particular value of the parameters of interest, 
           How likely is the observed data?"

\noindent
$\Bcal:$ "Having observed this data,
           What can we say about the parameters of interest?"

\noindent
... and the probability if the causes, as Poincare said, 
is the most important from the point of view of scientific applications.

\begin{figure}[t]
\begin{center}

\mbox{\epsfig{file=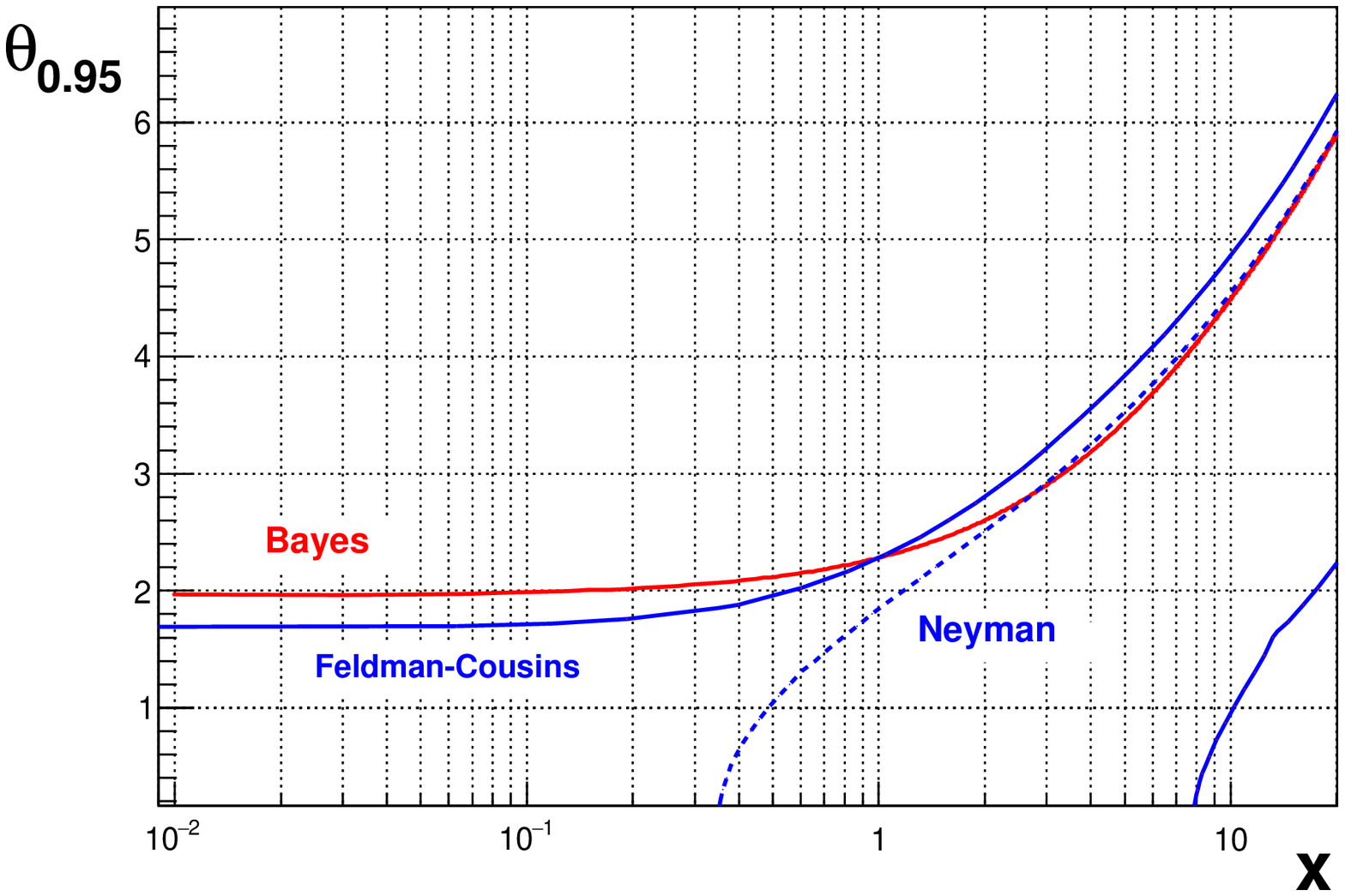,height=8cm,width=10cm}}

\vspace{0.5cm}
\footnotesize{
{\bf Figure 2.5}.- 95\% upper bounds on the parameter $\theta$ following
 the Bayesian approach (red), the Neyman approach (broken blue) and
 Feldman and Cousins (solid blue line).
             }
\end{center}
\end{figure}

In many circumstances we are also interested in one-sided intervals. That
is for instance the case when the data is consistent with the 
hypothesis $H:\{\theta=\theta_0\}$ and we want to give an upper bound
on $\theta$ so that $P(\theta{\in}(-\infty,\theta_{\beta}]={\beta}$.
The frequentist rationale is the same: obtain the interval
$[-\infty,x_2]{\subset}{\Omega}_X$ such that
\begin{eqnarray*}
 P(X{\leq}x_2)\,=\,  \int_{-\infty}^{x_2}p(x|\theta)\,dx\,=\,\beta
\end{eqnarray*} 
where $x_2=f_2(\theta)$; in this case without ambiguity. For the
the random interval $(-\infty,f_2^{-1}(X))$ $\Fcal$ has that
\begin{eqnarray*}
 P\left({\theta}<f_2^{-1}(X)\right)\,=\,
1-P\left({\theta}{\geq}f_2^{-1}(X)\right)\,=\,1-\beta
\end{eqnarray*}
so, for a probability content $\alpha$ (say 0.95), one should set
$\beta=1-\alpha$ (=0.05). Now, consider for instance the example of
the anisotropy is cosmic rays discussed in the last section 13.3. 
For a dipole moment (details are unimportant now) we have a statistic
\begin{eqnarray*}
X{\sim} p(x|{\theta},1/2)\,=\,\frac{\exp\{-\theta^2/2\}}
{\sqrt{2\pi}\theta}\,
\exp\{-x/2\}{\rm sinh}(\theta\sqrt{x})\,
  \mbox{\boldmath $1$}_{(0,\infty)}(x)
\end{eqnarray*}
where the parameter $\theta$ is the dipole coefficient multiplied by
a factor that is irrelevant for the example.
It is argued in section 13.3 that the reasonable prior for this model is
$\pi(\theta)={\rm constant}$ so we have the posterior
\begin{eqnarray*}
p(\theta|x,1/2)\,=\,\frac{\sqrt{2}}{\sqrt{{\pi}x}\,M(1/2,3/2,x/2)}\,
\exp\{-\theta^2/2\}\,{\theta}^{-1}\,
{\rm sinh}(\theta\sqrt{x})\,
  \mbox{\boldmath $1$}_{(0,\infty)}(\theta)
\end{eqnarray*}
with $M(a,b,z)$ the Kummer's Confluent Hypergeometric Function.
In fact, $\theta$ has a compact support but since the observed values 
of $X$ are consistent with $H_0:\{\theta=0\}$ and the sample size is very 
large [AMS13] 
\footnote{[AMS13]: Aguilar M. et al. (2013); Phys. Rev. Lett. 110, 141102.},
$p(\theta|x,1/2)$ is concentrated in a small interval $(0,\epsilon)$ and
it is easier for the evaluations to extend the domain to $\Rcal^{+}$ without
any effect on the results. Then we, Bayesians, shall derive the one-sided
upper credible region $[0,\theta_{0.95}(x)]$ with 
$\alpha=95\%$ probability content as simply as:
\begin{eqnarray*}
\int_{0}^{\theta_{0.95}}
p(\theta|x,1/2)\,d\theta\,=\,\alpha\,=\,0.95
\end{eqnarray*}
This upper bound shown as function of $x$ in figure 2.5 under {\sl "Bayes"}
(red line).
Neyman's construction is also straight forward. From
\begin{eqnarray*}
\int_{0}^{x_2}
p(x|\theta,1/2)\,dx\,=\,1\,-\,\alpha\,=\,0.05
\end{eqnarray*}
(essentially a $\chi^2$ probability for $\nu=3$), $\Fcal$
will get the upper bound shown in the same figure under {\sl "Neyman"}
(blue broken line). 
\begin{figure}[t]
\begin{center}

\mbox{\epsfig{file=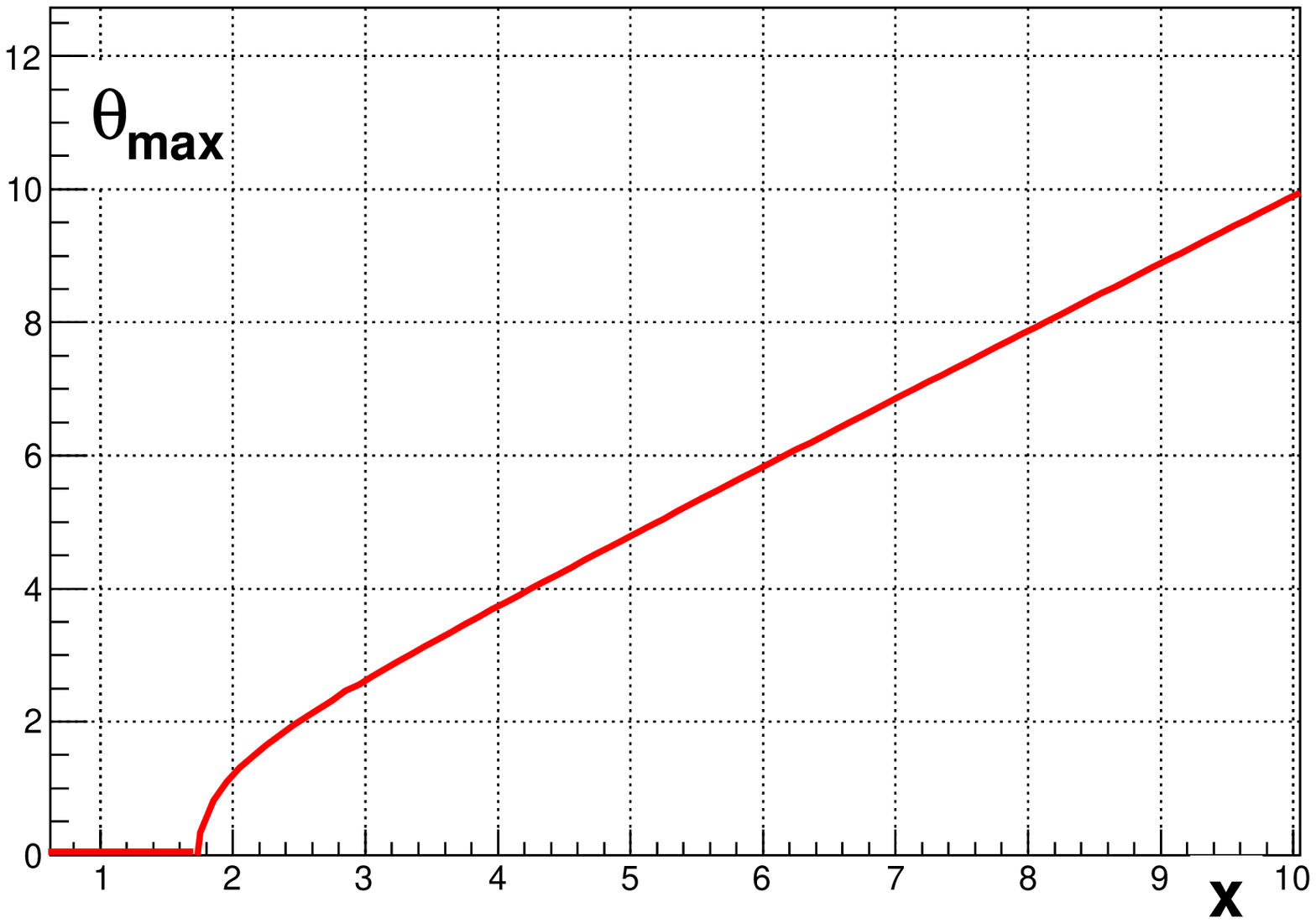,height=6.0cm,width=7cm}
\epsfig{file=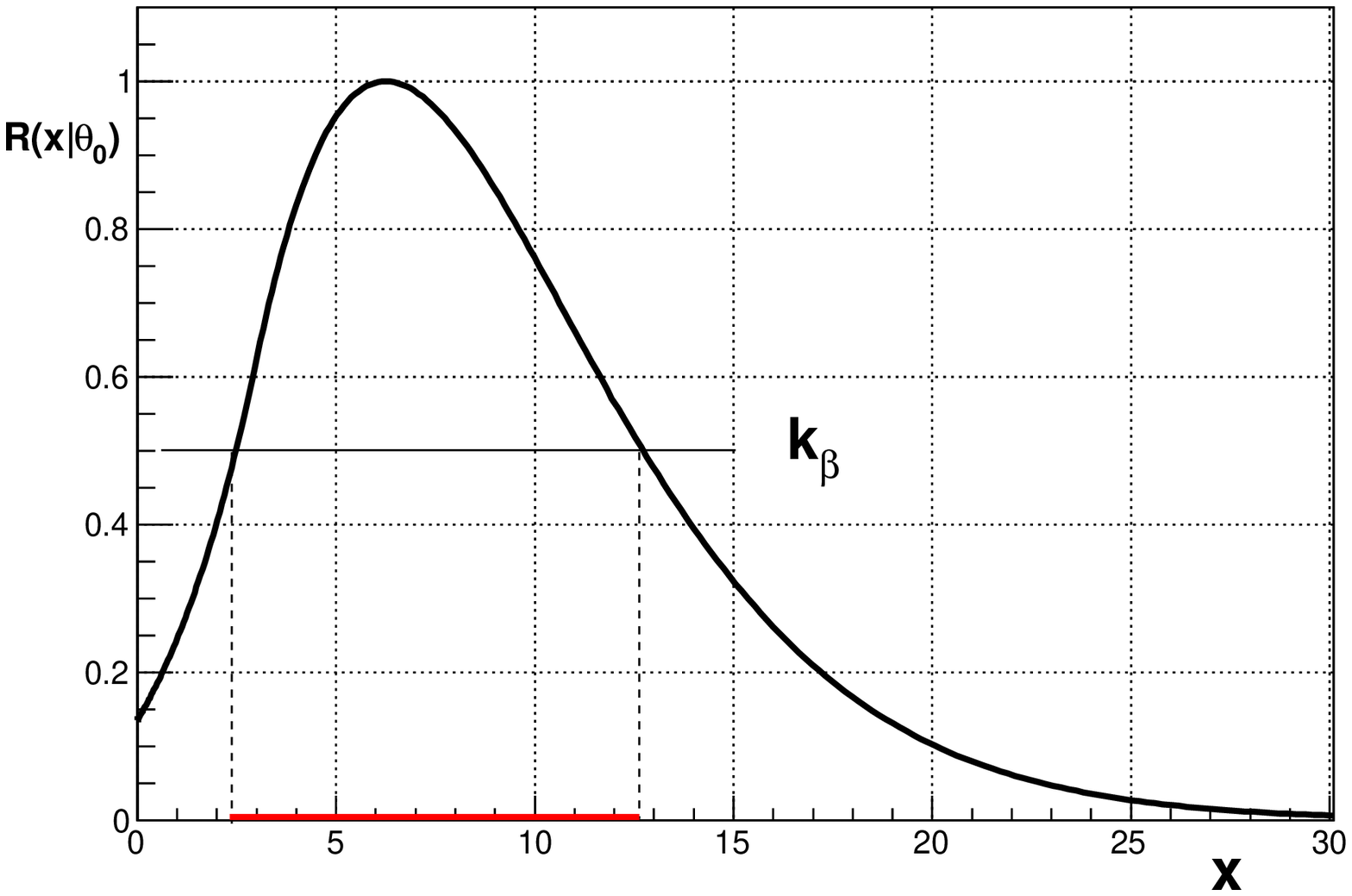,height=6.0cm,width=7cm}}

\vspace{0.5cm}
\footnotesize{
{\bf Figure 2.6}.- (1) Dependence of $\theta_m$ with $x$.
                   (2) Probability density ratio $R(x|\theta)$ for
                       $\theta=2$.
             }
\end{center}
\end{figure}
As you
can see, they get closer as $x$ grows but, first,
there is no solution for  $x{\leq}x_c=0.352$. In fact,
$E[X]={\theta}^2+3$ so if the dipole moment is $\delta=0$ ($\theta=0$),
$E[X]=3$ and observed
values below $x_c$ will be an unlikely fluctuation downwards
(assuming of course that the model is correct) but certainly a possible
experimental outcome.
In fact, you can see that
for values of $x$ less than 2, even though there is a solution 
Neyman's upper bound is underestimated.
To avoid this "little" problem, a different prescription has to be taken.

The most interesting solution is the one proposed by Feldman and Cousins
[Fe97] in which the region $\Delta_X{\subset}{\Omega}_X$
that is considered for the specified
probability content is determined by the ratio of probability densities.
Thus, for a given value $\theta_0$, the interval 
$\Delta_X$ is such that
\begin{eqnarray*}
\int_{\Delta_X}p(x|\theta_0)\,dx\,=\,{\beta}
\hspace{1.cm}{\rm with}\hspace{1.cm}
R(x|{\theta}_0)\,=\,\frac{p(x|\theta_0)}{p(x|{\theta}_b)}\,>\,k_{\beta}
\hspace{0.5cm};\,{\forall}x{\in}{\Delta_X}
\end{eqnarray*}
and where ${\theta}_b$ is the {\sl best} estimation of $\theta$
for a given $\{x\}$; usually
the one that maximizes the likelihood $(\theta_m)$. In our case,
it is given by:
\begin{eqnarray*}
{\theta}_m\,=\,\left\{
\begin{array}{ll}
  0 & \hspace{1.cm}{\rm if}\,\,x{\leq}\sqrt{3} \\ 
  {\theta}_m\,+\,{\theta}_m^{-1}\,-\,\sqrt{x}\,
  {\rm coth}({\theta}_m\sqrt{x})\,=\,0
  &\hspace{1.cm}{\rm if}\,\,x>\sqrt{3} 
\end{array}
\right.
\end{eqnarray*}
and the dependence with $x$ is shown in
figure 2.6(1) ($\theta_m{\simeq}x$ for $x>>$). As illustration,  
function $R(x|\theta)$ is shown in figure 2.6(2) for the 
particular value $\theta_0=2$. 
Following this procedure
\footnote{In most cases,a Monte Carlo simulation will simplify life.}, 
the $0.95$ probability content band is
shown in figure 2.5 under {\sl "Feldman-Cousins"} (blue line). Note that
for large values of $x$, the confidence region becomes an interval. It
is true that if we observe a large value of $X$, the hypothesis
$H_0:\{\delta=0\}$ will not be favoured by the data and a different analysis
will be more relevant although, by a simple modification of the ordering rule,
we still can get an upper bound if desired or use the standard Neyman's
procedure.

The Feldman and Cousins prescription allows to consider constrains on the 
parameters in a simpler way than Neyman's
procedure and,
as opposed to it, will always provide a region with the
specified probability content. However, on the one hand, they are
frequentist intervals and as such have to be interpreted. On the other hand,
for discrete random quantities with image in $\{x_1,\ldots,x_k,\ldots\}$
it may not be  possible to satisfy exactly the probability content equation 
since for the Distribution Function one has that 
$F(x_{k+1})=F(x_k)+P(X=x_{k+1})$. And last, it is not
straight forward to
deal with nuisance parameters. Therefore,
the best advice: "Be Bayesian!".
%
\small

\section{\LARGE \bf Some worked examples}
\subsection{Regression}
Consider the exchangeable sequence  
$\zbold=\{(x_1,y_1),(x_2,y_2),{\ldots},(x_n,y_n)\}$ 
of $n$ samplings from the two-dimensional model
$N(x_i,y_i|\cdot)=N(x_i|{\mu}_{x_i},{\sigma}_{x_i}^2)
N(y_i|{\mu}_{y_i},{\sigma}_{y_i}^2)$. Then
\begin{eqnarray*}
p(\zbold|\cdot)\,\propto\,
{\rm exp}\left\{-\frac{1}{2}
\sum_{i=1}^{n}\left[
   \frac{(y_i-{\mu}_{yi})^2}{\sigma_{yi}^2}\,+\,
   \frac{(x_i-{\mu}_{xi})^2}{\sigma_{xi}^2}\right]
      \right\}
\end{eqnarray*}
We shall assume that the precisions ${\sigma}_{xi}$ and ${\sigma}_{xi}$ are
known and that there is a functional relation
${\mu}_y=f({\mu}_x;\thetabold)$ with unknown parameters $\thetabold$. Then,
in terms of the new parameters of interest:
\begin{eqnarray*}
p(\ybold|\cdot)\,\propto\,
{\rm exp}\left\{-\frac{1}{2}
\sum_{i=1}^{n}\left[
   \frac{(y_i-f({\mu}_{x_i};\thetabold))^2}{\sigma_{y_i}^2}\,+\,
   \frac{(x_i-{\mu}_{x_i})^2}{\sigma_{x_i}^2}\right]
      \right\}
\end{eqnarray*}
Consider a linear relation $f({\mu}_x;a,b)=a+b{\mu}_x$ with ${a,b}$
the unknown parameters so:
\begin{eqnarray*}
p(\zbold|\cdot)\,\propto\,
{\rm exp}\left\{-\frac{1}{2}
\sum_{i=1}^{n}\left[
   \frac{(y_i-a-b{\mu}_{x_i})^2}{\sigma_{y_i}^2}\,+\,
   \frac{(x_i-{\mu}_{x_i})^2}{\sigma_{x_i}^2}\right]
      \right\}
\end{eqnarray*}
and assume, in first place, that ${\mu}_{x_i}=x_i$ without uncertainty. Then,
\begin{eqnarray*}
p(\ybold|a,b)\,\propto\,
{\rm exp}\left\{-\frac{1}{2}
\sum_{i=1}^{n}\left[
   \frac{(y_i-a-bx_i)^2}{\sigma_{y_i}^2}\right]
      \right\}
\end{eqnarray*}
There is a set of sufficient statistics for $(a,b)$:
\begin{eqnarray*}
\tbold\,=\,\{t_1,\,t_2,\,t_3,\,t_4,\,t_5\}\,=\,
   \left\{\sum_{i=1}^n\frac{1}{\sigma_i^2},\,
   \sum_{i=1}^n\frac{x_i^2}{\sigma_i^2},\,
   \sum_{i=1}^n\frac{x_i}{\sigma_i^2},\,
   \sum_{i=1}^n\frac{y_i}{\sigma_i^2},\,
   \sum_{i=1}^n\frac{y_ix_i}{\sigma_i^2}\right\}
\end{eqnarray*}
and, after a simple algebra, it is easy to write
\begin{eqnarray*}
p(\ybold|a,b)\,\propto\,
{\rm exp}\left\{-\frac{1}{2(1-{\rho}^2)}
\left[
  \frac{(a-a_0)^2}{\sigma_{a}^2}\,+\,
  \frac{(b-b_0)^2}{\sigma_{b}^2}\,-\,
  2\,{\rho}\frac{(a-a_0)}{\sigma_{a}}\frac{(b-b_0)}{\sigma_{b}}
\right]
      \right\}
\end{eqnarray*}
where the new statistics $\{a_0,b_0,{\sigma}_a,{\sigma}_b,\rho\}$ are defined
as:
\begin{eqnarray*}
a_0\,&=&\,\frac{t_2t_4-t_3t_5}{t_1t_2-t_3^2}\,,\hspace{0.5cm}
b_0\,=\,\frac{t_1t_5-t_3t_4}{t_1t_2-t_3^2} \\
{\sigma}_a^2\,&=&\,\frac{t_2}{t_1t_2-t_3^2}\,,\hspace{0.5cm}
{\sigma}_b^2\,=\,\frac{t_1}{t_1t_2-t_3^2}\,,\hspace{0.5cm}
{\rho}\,=\,-\frac{t_3}{\sqrt{t_1t_2}}
\end{eqnarray*}
Both $(a,b)$ are position parameters so we shall take a uniform prior and 
in consequence
\begin{eqnarray*}
p(a,b|\cdot)\,=\,\frac{1}{2{\pi}{\sigma}_a{\sigma}_b\sqrt{1-{\rho}^2}}
e^{\textstyle \left\{-\frac{1}{2(1-{\rho}^2)}
\left[
  \frac{(a-a_0)^2}{\sigma_{a}^2}\,+\,
  \frac{(b-b_0)^2}{\sigma_{b}^2}\,-\,
  2\,{\rho}\frac{(a-a_0)}{\sigma_{a}}\frac{(b-b_0)}{\sigma_{b}}
\right]
      \right\}}
\end{eqnarray*}
This was obviously expected. 

When ${\mu}_{x_i}$ are $n$ unknown parameters, if we take 
${\pi}({\mu}_{xi})={\mbox{\boldmath $1$}}_{(0,\infty)}({\mu}_{xi})$ and
marginalize for $(a,b)$ we have
\begin{eqnarray*}
p(a,b|\cdot){\propto}{\pi}(a,b)\,
 {\rm exp}\left\{ -  \frac{\textstyle 1}
                          {\textstyle 2}\,
  \sum_{i=1}^{n}\,
       \frac{\textstyle (y_i-a-b\,x_{i})^2 }
            {\textstyle {\sigma}_{yi}^2\,+\,b^2\,{\sigma}_{xi}^2}\right\}\,
\left\{\prod_{i=1}^{n}\,   
({\sigma}_{yi}^2\,+\,b^2\,{\sigma}_{xi}^2)\right\}^{-1/2}
\end{eqnarray*}
In general, the expressions one gets for non-linear regression problems 
are complicated and setting up priors is a non-trivial task but fairly vague 
priors easy to deal with are usually a reasonable choice. 
In this case, for instance, one may consider uniform priors or normal 
densities $N({\cdot}|0,{\sigma}>>)$ for both parameters $(a,b)$ and 
and sample the proper posterior with a Monte Carlo algorithm (Gibbs sampling
will be appropriate).

The same reasoning applies if we want to consider other models
or more involved relations with several explanatory variables like
$\theta_i=\sum_{j=1}^k{\alpha}_jx_{ij}^{b_j}$. In counting
experiments, for example, $y_i{\in}{\Ncal}$ so we may be interested in
a Poisson model $Po(y_i|{\mu}_i)$ where ${\mu}_i$ 
is parameterized as a simple log-linear form
$\ln({\mu}_i)={\alpha}_0+{\alpha}_1x_i$ (so ${\mu}_i>0$ for whatever
${\alpha}_0,{\alpha}_1{\in}{\Rcal}$). Suppose for instance that we have
the sample $\{(y_i,x_i)\}_{i=1}^n$. Then:
 \begin{eqnarray*}
p(\ybold|\alpha_1,\alpha_2,\xbold)\,\propto\,\prod_{i=1}^n
\exp\{-\mu_i\}{\mu_i}^{y_i}\,=\,
\exp\left\{{\alpha}_1s_1\,+\,{\alpha}_2s_2-e^{{\alpha}_1}\sum_{i=1}^ne^{\alpha_2x_i}
\right\}
\end{eqnarray*}
where $s_1=\sum_{i=1}^ny_i$ and $s_2=\sum_{i=1}^ny_ix_i$. 
In this case,
the Normal distribution $N(\alpha_i|a_i,\sigma_i)$ with $\sigma_i>>$ is
a reasonable smooth and easy to handle proper prior density for both 
parameters. Thus, we get the posterior conditional densities
 \begin{eqnarray*}
p(\alpha_i|\alpha_j,\ybold,\xbold)\,\propto\,
\exp\left\{-
\frac{{\alpha}_i^2}{2\sigma_i^2}\,+\,{\alpha}_i\left(
\frac{a_i}{\sigma_i^2}\,+\,s_i\right)\,-\,
e^{{\alpha}_1}\sum_{i=1}^ne^{\alpha_2x_i}\right\}\hspace{0.5cm};\,\,i=1,2
\end{eqnarray*} 
that are perfectly suited for the Gibbs sampling to be discussed in 
section 4.1 of lecture 3.

\begin{figure}[t]
\begin{center}

\mbox{\epsfig{file=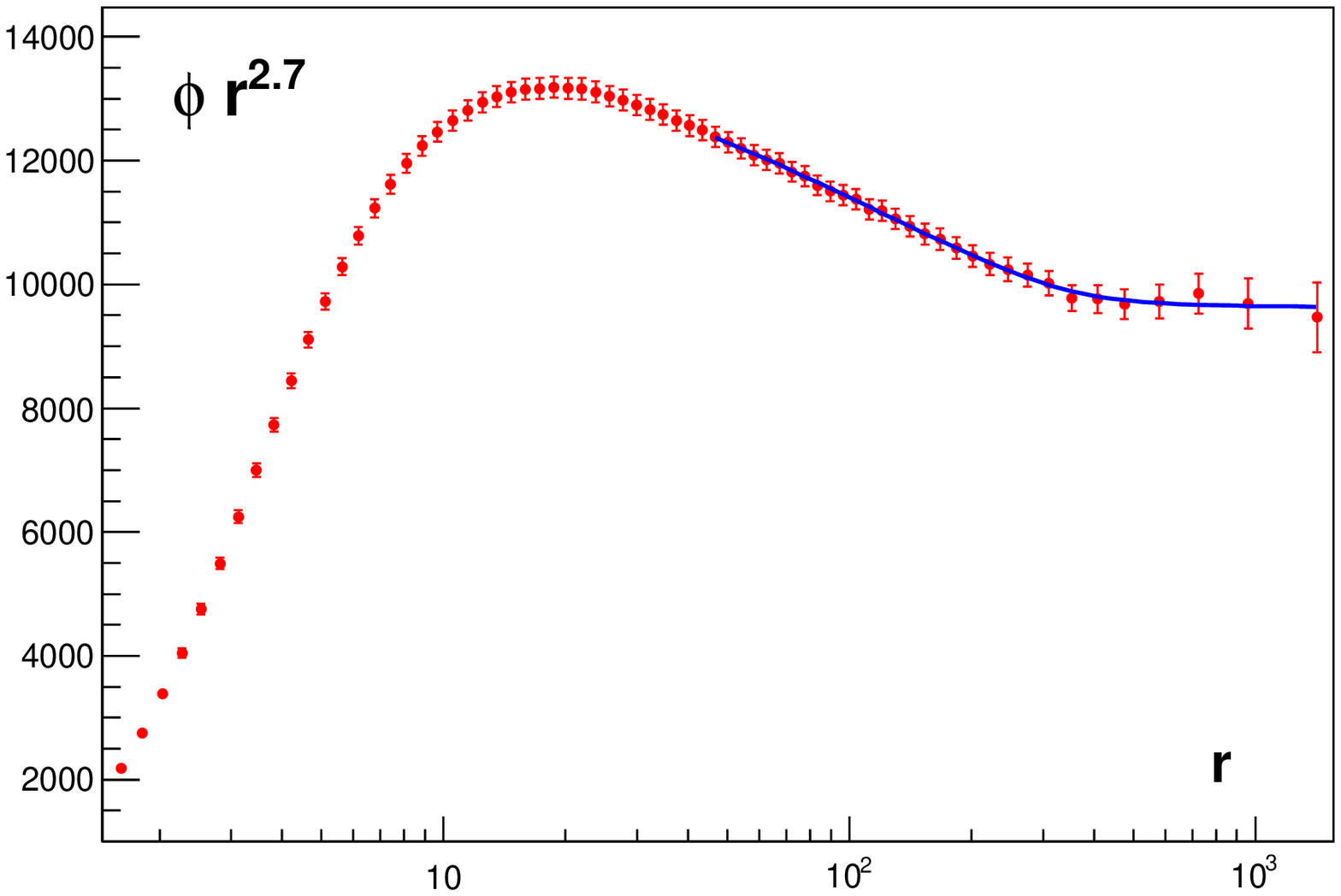,height=6.0cm,width=6.5cm}
  \epsfig{file=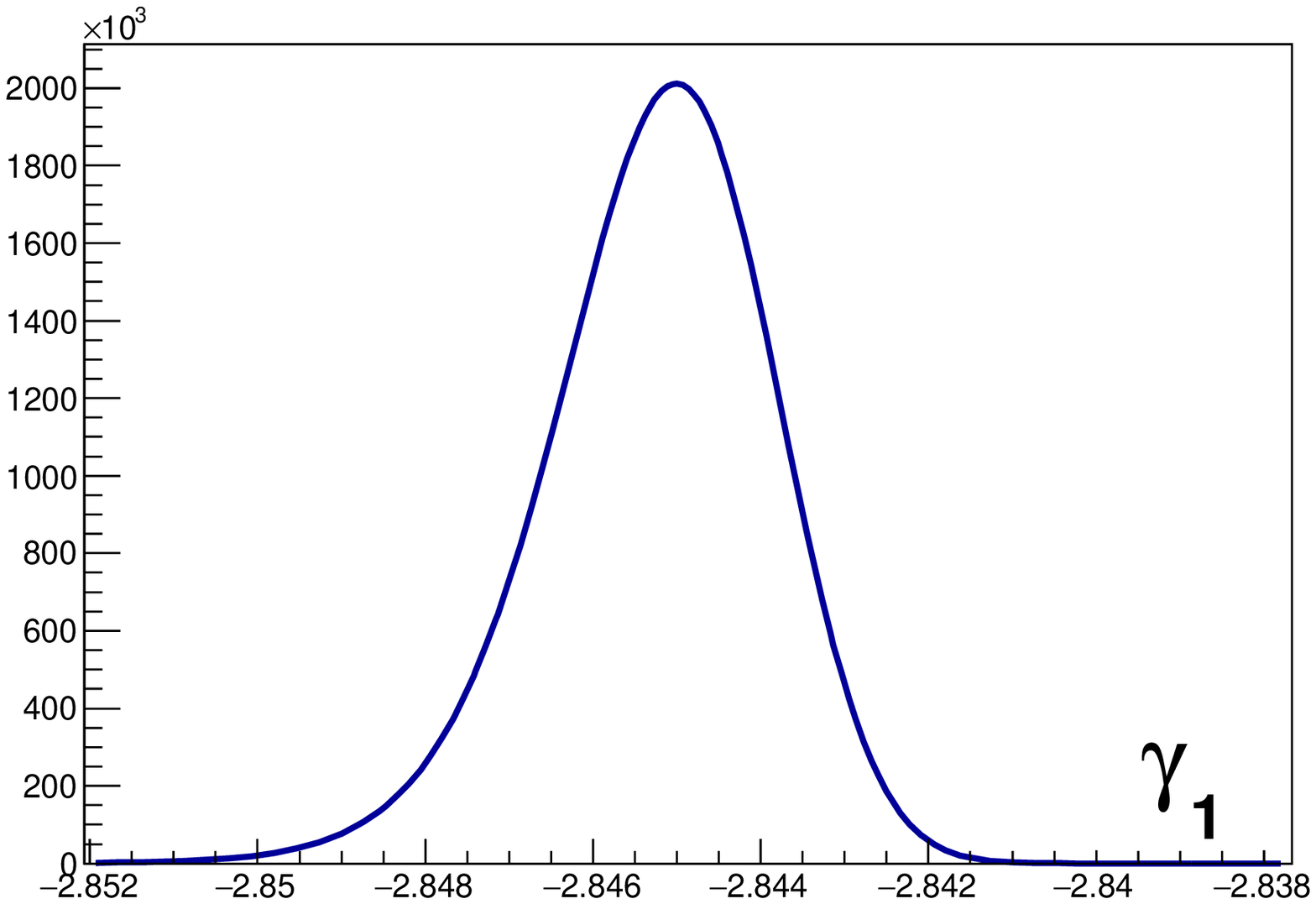,height=6.0cm,width=6.5cm}}
\mbox{\epsfig{file=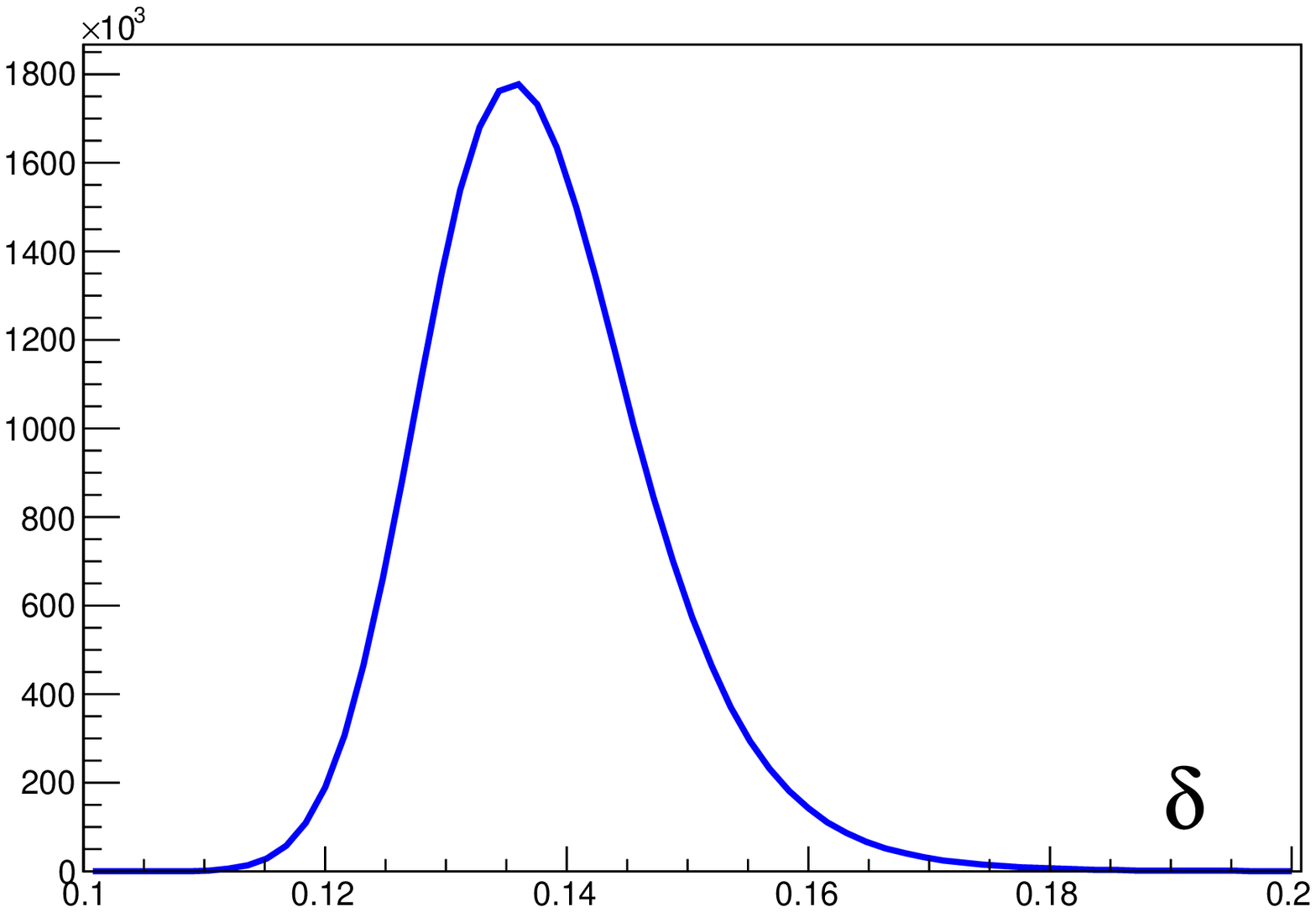,height=6.0cm,width=6.5cm}
  \epsfig{file=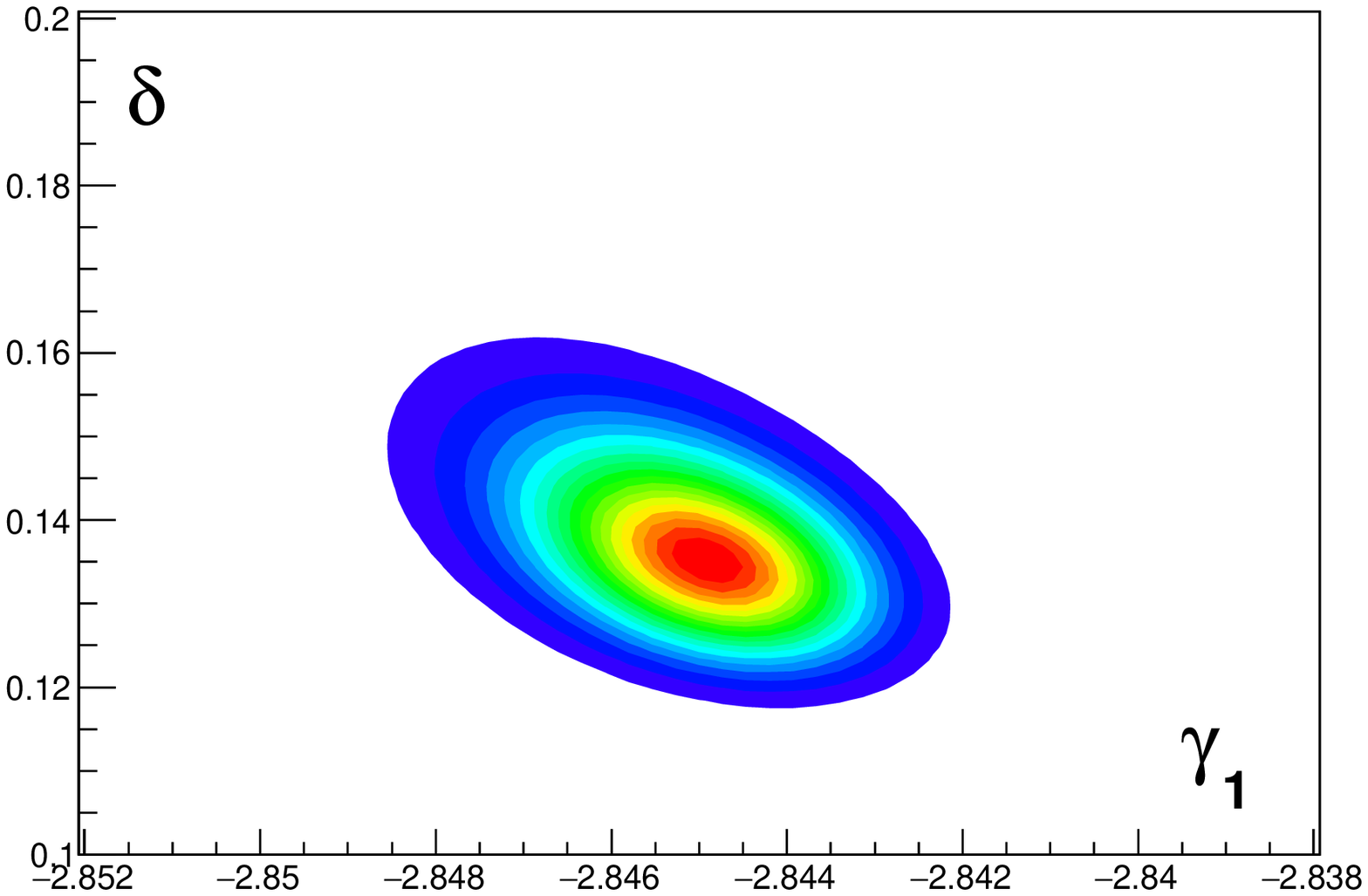,height=6.0cm,width=6.5cm}}

\footnotesize{
{\bf Figure 2.7}.- (1): Observed flux multiplied by $r^{2.7}$ in
$m^{-2}sr^{-1}sec^{-1}GV^{1.7}$ as given in [AMS15]; (2): Posterior density
of the parameter $\gamma_1$ (arbitrary vertical scale);
(3): Posterior density of the parameter $\delta={\gamma}_2-{\gamma}_1$
(arbitrary vertical scale); (4): Projection of the posterior density
$p(\gamma_1,\delta)$.}
\end{center}
\end{figure}
\vspace{0.5cm}
\noindent
{\raya}                   
\vspace{0.35cm}
\footnotesize

\noindent
{\bf Example 2.25: Proton Flux in Primary Cosmic Rays.}
For energies between ${\sim}20$ and ${\sim}200$ GeV, 
the flux of protons of the primary cosmic radiation 
is reasonably well described by a power law
$\phi(r)=c\,r^{\gamma}$ where $r$ is the {\sl rigidity}
\footnote{The {\sl rigidity} $(r)$ is defined as the momentum $(p)$ divided by 
the electric charge $(Z)$ so $r=p$ for protons.} 
and $\gamma=d{\ln}{\phi}/ds$, with $s=\ln r$, 
is the {\sl spectral index}. 
At lower energies, this dependence is significantly modified by the
geomagnetic cut-off and the solar wind but at higher energies, where
these effects are negligible,
the observations are not consistent with a single power law
(fig. 2.7(1)). One may
characterize this behaviour with a simple phenomenological model where the
spectral index is no longer constant but has a dependence
$\gamma(s)=\alpha+\beta\,{\rm tanh}[a(s-s_0)]$ such that 
$\lim_{s{\rightarrow}-\infty}\gamma(s)=\gamma_1$ ($r{\rightarrow}0$) and
$\lim_{s{\rightarrow}\infty}\gamma(s)=\gamma_2$ ($r{\rightarrow}+\infty$). After
integration, the flux can be expressed in terms of 5 parameters 
$\thetabold=\{
{\phi}_0,\gamma_1,\delta={\gamma}_2-{\gamma}_1,r_0,\sigma\}$ as:
\begin{eqnarray*}
\phi(r;\thetabold)\,=\,
{\phi}_0\,r^{\gamma_1}\,\left[1\,+\,
\left(\frac{r}{r_0}\right)^{\sigma}
\right]^{\delta/\sigma}
\end{eqnarray*}
For this example, I have used the data above 45 GeV
published by the AMS experiment 
\footnote{[AMS15]: Aguilar M. et al. (2015); PRL 114, 171103 and 
references therein.} 
and considered only the quoted {\sl statistical errors} (see fig. 2.7(1)).
Last, for a better description of the flux the previous expression 
has been modified to account for the effect of the solar wind with the
force-field approximation in consistency with [AMS15].
This is just a technical detail,
irrelevant for the purpose of the example. Then, 
assuming a Normal model for the observations we can write the posterior density
\begin{eqnarray*}
p(\thetabold|{\rm data})\,=\,\pi({\thetabold})\,
\prod_{i=1}^n{\rm exp}\left\{
-\frac{1}{2{\sigma_i}^2}\left(\phi_i\,-\,
\phi(r_i;\thetabold)\right)^2\right\}
\end{eqnarray*}
I have taken Normal priors with large variances $(\sigma_i>>)$
for the parameters ${\gamma}_1$ and ${\delta}$
and restricted the support to $\Rcal^+$ for $\{\phi_0,r_0,\sigma\}$. The
posterior densities for the parameters ${\gamma}_1$ and ${\delta}$ 
are shown in figure 2.7(2,3) together with the projection (2.7(4))
that gives an idea of correlation between them. For a visual inspection, 
the phenomenological form of the flux is shown in fig. 2.7(1) (blue line) 
overimposed to the data when the parameters are set to their expected
posterior values.

\small
{\raya}                   
\vspace{1.0cm}

\subsection{Characterization of a Possible Source of Events}
Suppose that we observe a particular region $\Omega$ of the sky during a
time $t$ and denote by $\lambda$ the rate at which events from this region
are produced. We take a Poisson model to describe the number of 
produced events: ${k}{\sim}Po(k|{\lambda}t)$. Now, denote by $\epsilon$
the probability to detect one event 
(detection area, efficiency of the detector,...). The number of observed
events $n$ from the region $\Omega$ after an exposure time $t$ and
detection probability $\epsilon$ will follow:
\begin{eqnarray*}
n\,{\sim}\,\sum_{k=n}^{\infty}\,Bi(k|n,\epsilon)\,Po(k|{\lambda}t)\,=\,
        Po(n|{\lambda}t{\epsilon})
\end{eqnarray*}

The approach to the problem will be the same for other counting process like,
for instance, events collected from a detector for a given integrated 
luminosity.
We suspect that the events observed in a particular region $\Omega_o$ of the 
sky are background events together with those from an emitting source. 
To determine the
significance of the potential source we analyze a nearby region, 
$\Omega_b$,
to infer
about the expected background. If after a time $t_b$ we observe $n_b$ events
from this region with detection probability $e_b$ then,
defining $\beta=\epsilon_bt_b$ we have that
\begin{eqnarray*}
n_b\,{\sim}\,Po(n_b|\lambda_b\,\beta)\,=\,
          {\rm exp}\left\{-\beta{\lambda}_b\right\}\,
      \frac{(\beta{\lambda}_b)^{n_b}}{\Gamma(n_b+1)}
\end{eqnarray*}
At $\Omega_o$ we observe $n_o$ events during a time $t_o$ with a 
detection probability $\epsilon_o$. Since $n_o=n_1+n_2$ with 
$n_1{\sim}Po(n_1|{\lambda}_s{\alpha})$ signal events 
(${\alpha}=\epsilon_ot_o$)  and
$n_2{\sim}Po(n_2|{\lambda}_b{\alpha})$ 
background events (assume reasonably that $e_s=e_b=e_o$ in the same region), 
we have that
\begin{eqnarray*}
n_o&{\sim}&\sum_{n_1=0}^{n_o}Po(n_1|{\lambda}_s{\alpha})
 Po(n_o-n_1|{\lambda}_b{\alpha})\,=\,Po(n_o|({\lambda}_s+{\lambda}_b){\alpha})
\end{eqnarray*}

Now, we can do several things. 
We can assume for instance that the overall rate form the region
$\Omega_o$ is $\lambda$, write $n_o{\sim}Po(n_o|{\alpha}{\lambda})$
and study the fraction ${\lambda}/{\lambda}_b$ of the rates from the 
information provided by the observations in the two different regions.
Then,
reparameterizing the model in terms of ${\theta}={\lambda}/{\lambda}_b$ 
and ${\phi}={\lambda}_b$ we have
\begin{eqnarray*}
p(n_o,n_b|\cdot)\,=\,Po(n_o|{\alpha}{\lambda})Po(n_b|{\beta}{\lambda}_b)
\,{\sim}\,
e^{-{\beta}{\phi}(1+{\gamma}{\theta})}{\theta}^{n_o}\,{\phi}^{n_o+n_b}
\end{eqnarray*}
where $\gamma={\alpha}/{\beta=}(\epsilon_st_s)/(\epsilon_bt_b)$.
For the ordering $\{\theta,\phi\}$ we have that
the Fisher's matrix and its inverse are
\begin{eqnarray*}
{\bf I}({\theta},{\phi})\,=\,
\left( \begin{array}{ccc}
    \frac{\gamma\beta\phi}{\theta} &  {\gamma}{\beta}  \\
        {\gamma}{\beta}  & \frac{\beta(1+{\gamma}{\theta})}{\phi}
                   \end{array}
            \right)
\hspace{0.5cm}{\rm and}\hspace{0.5cm}
{\bf I}^{-1}({\mu}_1,{\mu}_2)\,=\,
\left( \begin{array}{ccc}
   \frac{\theta(1+\gamma\theta)}{\phi\gamma\beta} & 
        -\frac{\theta}{\beta}   \\
        -\frac{\theta}{\beta}   & \frac{\phi}{\beta}
                   \end{array}
            \right)
\end{eqnarray*}
Then
\begin{eqnarray*}
{\pi}(\theta,\phi)\,=\,{\pi}({\phi}|\theta)\,{\pi}(\theta)\,\propto\,
\frac{{\phi}^{-1/2}}{\sqrt{\theta(1+\gamma\theta)}}
\end{eqnarray*}
and integrating the nuisance parameter ${\phi}$ we get finally:
\begin{eqnarray*}
p(\theta|n_o,n_b,\gamma)\,=\,
\frac{\gamma^{n_o+1/2}}{B(n_o+1/2,n_b+1/2)}\,
\frac{\theta^{n_o-1/2}}{(1+\gamma\theta)^{n_o+n_b+1}}
\end{eqnarray*}
From this:
\begin{eqnarray*}
E[\theta^m]\,=\,
\frac{1}{\gamma^m}\,
\frac{\Gamma(n_o+1/2+m)\,\Gamma(n_b+1/2-m)}
     {\Gamma(n_o+1/2)\,\Gamma(n_b+1/2)}\,\,\,{\longrightarrow}\,\,\,
E[\theta]\,=\,
\frac{1}{\gamma}\,
\frac{n_o+1/2}{n_b-1/2}
\end{eqnarray*}
and
\begin{eqnarray*}
P(\theta{\leq}{\theta}_0)\,=\,\int_0^{\theta_0}p(\theta|\cdot)\,d\theta\,=\,
1\,-\,IB(n_b+1/2,n_o+1/2;(1+\gamma\theta_0)^{-1})
\end{eqnarray*}
with $IB(x,y;z)$ the Incomplete Beta Function. 
Had we interest in ${\theta}={\lambda_s}/{\lambda_b}$, the
corresponding reference prior will be
\begin{eqnarray*}
{\pi}(\theta,\phi)\,\propto\,
\frac{{\phi}^{-1/2}}{\sqrt{(1+\theta)(\delta+\theta)}}
\hspace{1.cm}{\rm with}\hspace{1.cm}
\delta=\frac{1+\gamma}{\gamma}
\end{eqnarray*}

A different analysis can be performed to make inferences on $\lambda_s$.
In this case, we may consider as an informative prior for the nuisance 
parameter the posterior
what we had from the study of the background in the region  
$\Omega_b$; that is:
\begin{eqnarray*}
p(\lambda_b|n_b,\,{\beta})\,\propto\,
          {\rm exp}\left\{-{\beta}{\lambda}_b\right\}\,
          {\lambda_b}^{n_b-1/2}\,
\end{eqnarray*}
and therefore:
\begin{eqnarray*}
p(\lambda_s|\cdot)\,\propto\,\pi({\lambda_s})\,\int_{0}^{\infty}
p(n_o|{\alpha}(\lambda_s+\lambda_b))\,
                    p(\lambda_b|n_b,\,{\beta})\,d{\lambda}_b\,{\propto}\,
\pi({\lambda_s})\,e^{-{\alpha}{\lambda}_s}\,{\lambda}_s^{n_o}\,
\sum_{k=0}^{n_o}\,a_k\,{\lambda_s}^{-k}
\end{eqnarray*}
where
\begin{eqnarray*}
a_k\,=\,
\left(\begin{array}{c}
  n_o \\
  k
\end{array}\right)\,
\frac{\textstyle \Gamma(k+n_b+1/2)}{[(\alpha+\beta)]^k} 
\end{eqnarray*}
A reasonable choice for the prior will be a conjugated prior 
$\pi({\lambda_s})=Ga({\lambda_s}|a,b)$ that simplifies the calculations
and provides enough freedom analyze the effect of different shapes on the
inferences.
The same reasoning is valid if the knowledge on
$\lambda_b$ is represented by a different $p(\lambda_b|\cdot)$ from, say,
a Monte Carlo simulation. Usual distributions in this case are
the Gamma and the Normal with non-negative support.
Last, it
is clear that if the rate of
background events is known with high accuracy then, with
${\mu}_i=\alpha\lambda_i$ and $\pi({\mu}_s){\propto}(\mu_s+\mu_b)^{-1/2}$ we
have
\begin{eqnarray*}
p(\mu_s|\cdot)\,=\,\frac{1}{\Gamma(x+1/2,\mu_b)}\,
{\rm exp}\{-({\mu}_s+{\mu}_{b})\}\,
      ({\mu}_s+{\mu_{b}})^{x-1/2}
  \mbox{\boldmath $1$}_{(0,\infty)}(\mu_s)
\end{eqnarray*}
\begin{figure}[t]
\begin{center}

\mbox{\epsfig{file=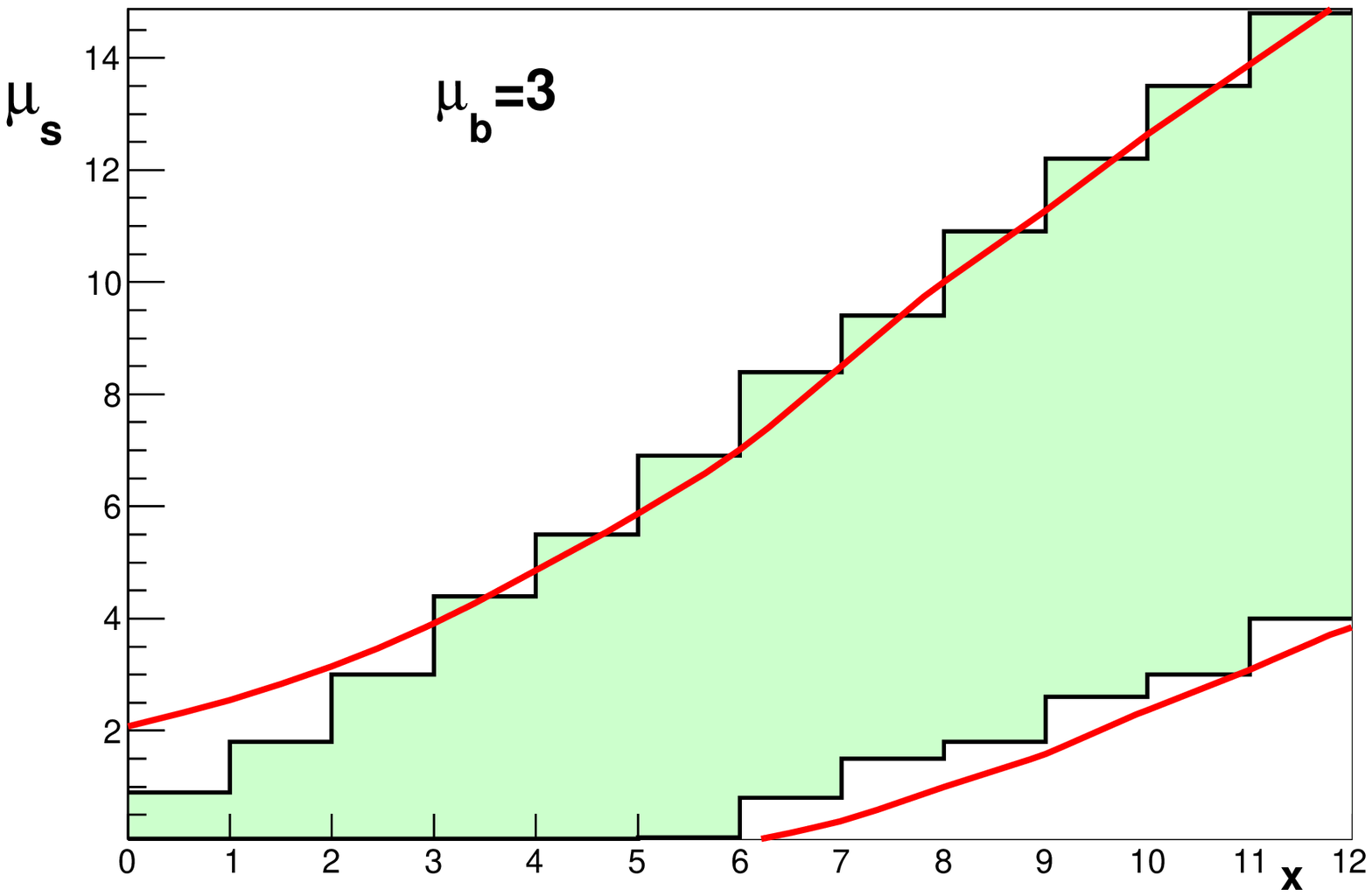,height=8.0cm,width=9cm}}

\vspace{0.5cm}
\footnotesize{
{\bf Figure 2.8}.-  $90\%$ Confidence Belt derived with Feldman and Cousins
                    (filled band) and the Bayesian HPD region (red lines)
                    for a background parameter ${\mu}_b=3$.
             }
\end{center}
\end{figure}
As an example, we show in figure 2.8 the $90\%$ 
HPD region obtained from the previous expression (red lines) as function 
of $x$ for $\mu_b=3$ (conditions as given in the example of [Fe97]) and the
Confidence Belt derived with the Feldman and Cousins  approach
(filled band). In this case, 
${\mu}_{s,m}={\rm max}\{0,x-{\mu}_b\}$ and therefore, for a given $\mu_s$:
\begin{eqnarray*}
\sum_{x_1}^{x_2}Po(x|\mu_s+\mu_b)\,=\,{\beta}
\hspace{0.5cm}{\rm with}\hspace{0.5cm}
R(x|{\mu}_s)\,=\,e^{({\mu}_{s,m}-{\mu}_s)}\,
\left(\frac{\mu_s+\mu_b}{\mu_{s,m}+\mu_b}\right)^x\,>\,k_{\beta}
\end{eqnarray*}
for all $x{\in}[x_1,x_2]$.
\vspace{0.5cm}
\noindent
{\raya}                   
\vspace{0.35cm}
\footnotesize

\noindent
{\bf Problem 2.14:}
In the search for a new particle, assume that the number of observed events
follows a Poisson distribution with $\mu_b=0.7$ known with enough
precision from extensive Monte Carlo simulations. Consider the hypothesis
$H_0:\{\mu_s=0\}$ and $H_1:\{\mu_s{\neq}0\}$. It is left as an
exercise to  obtain the Bayes Factor $BF_{01}$ with the proper prior
$\pi(\mu_s|\mu_b)={\mu_b}(\mu_s+\mu_b)^{-2}$
proposed in [BP04], $P(H_1|n)$ and the BIC difference
$\Delta_{01}$ as function of $n=1,{\ldots}7$ and decide when,
based on this results, will you consider that there is evidence for a signal.

\small
{\raya}                   
\vspace{1.0cm}

\subsection{Anisotropies of Cosmic Rays}
The angular distribution of cosmic rays in galactic coordinates is analyzed
searching for possible anisotropies. A well-behaved real function 
$f({\theta},{\phi}){\in}L_2({\Omega})$, 
with  
$({\theta},{\phi}){\in}{\Omega}=[0,{\pi}]{\times}[0,2{\pi}]$,
can be expressed in the real harmonics basis as:
\begin{eqnarray*}
f({\theta},{\phi})=\sum_{l=0}^{\infty}\,
  \sum_{m=-l}^{l}a_{lm}Y_{lm}({\theta},{\phi})
\hspace{0.5cm}{\rm where}\hspace{0.5cm}
a_{lm}=
\int_{\Omega}f({\theta},{\phi})
Y_{lm}({\theta},{\phi})\,d{\mu}\,\,\,;
\end{eqnarray*}
$a_{lm}{\in}R$ and $d{\mu}=\sin{\theta}d{\theta}d{\phi}$. 
The convention adopted for the spherical harmonic functions 
is such that ({\sl orthonormal basis}): 
\begin{eqnarray}
\int_{\Omega}\,
  Y_{lm}({\theta},{\phi})\,Y_{l'm'}({\theta},{\phi})\,d{\mu}\,=\,
  {\delta}_{ll'}\,\delta_{mm'}\hspace{1.cm}{\rm and}\hspace{1.cm}
\int_{\Omega}\,
  Y_{lm}({\theta},{\phi})\,d{\mu}\,=\,\sqrt{4{\pi}}\,{\delta}_{l0}
\nonumber
\end{eqnarray}
In consequence, a
probability density function $p({\theta},{\phi})$  with support in ${\Omega}$
can be expanded as
\begin{eqnarray}
p({\theta},{\phi})\,=\,c_{00}\,Y_{00}({\theta},{\phi})\,+\,
\sum_{l=1}^{\infty}\,
  \sum_{m=-l}^{l}\,c_{lm}\,Y_{lm}({\theta},{\phi})
\nonumber
\end{eqnarray}
The normalization imposes that $c_{00}=1/\sqrt{4{\pi}}$ 
so we can write
\begin{eqnarray*}
p({\theta},{\phi}|\mbox{{\boldmath ${a}$}})\,=\,
\frac{\textstyle 1}{\textstyle 4{\pi}}\,
\left(1\,+\,a_{lm}\,Y_{lm}({\theta},{\phi})\right) 
\end{eqnarray*}
where $l{\geq}1$,
\begin{eqnarray*}
a_{lm}\,=\,4{\pi}c_{lm}\,=\,
4{\pi}\int_{\Omega}p({\theta},{\phi})\,Y_{lm}({\theta},{\phi})\,d{\mu}
\,=\,4{\pi}\,E_{p;{\mu}}[Y_{lm}({\theta},{\phi})]
\end{eqnarray*}
and summation over repeated indices understood.
Obviously, for any
$({\theta},{\phi}){\in}{\Omega}$ we have that
$p({\theta},{\phi}|\mbox{{\boldmath ${a}$}})
{\geq}0$ so the set of parameters $\mbox{{\boldmath ${a}$}}$ are constrained 
on a compact support.

Even though we shall study the general case,
we are particularly interested in the expansion up to $l=1$
(dipole terms) so, to simplify the notation, we redefine the indices 
$(l,m)=\{(1,-1),(1,0),(1,1)\}$ as $i=\{1,2,3\}$ and, accordingly,
the coefficients $\mbox{{\boldmath ${a}$}}=(a_{1-1},a_{10},a_{11})$ as
$\mbox{{\boldmath ${a}$}}=(a_1,a_2,a_3)$. Thus:
\begin{eqnarray*}
p({\theta},{\phi}|\mbox{{\boldmath ${a}$}})
\,=\,\frac{1}{4{\pi}}\,\left( 1\,+\,
a_1Y_1\,+\,a_2Y_2\,+\,a_3Y_3 \right)
\end{eqnarray*}
In this case, the condition
$p({\theta},{\phi}|\mbox{{\boldmath ${a}$}}){\geq}0$
implies that the coefficients are bounded by the sphere
$a_1^2\,+\,a_2^2\,+\,a_3^2\,{\leq}\,4{\pi}/3$
and therefore, the coefficient of anisotropy 
\begin{eqnarray*}
{\delta}\,\stackrel{def.}=\,\sqrt{\frac{3}{4{\pi}}}\,
\left( a_1^2\,+\,a_2^2\,+\,a_3^2 \right)^{1/2}
\,{\leq}\,1
\end{eqnarray*}

There are no sufficient statistics for this model but the
Central Limit Theorem applies and, given the large amount of data, 
the experimental observations can be cast in the
statistic ${\abold}=(a_1,a_2,a_3)$ such that
\footnote{Essentially, 
          $a_{lm}\,=\,\frac{4\pi}{n}\sum_{i=1}^n Y_{lm}({\theta}_i,{\phi}_i)$
          for a sample of size $n$.}
\begin{eqnarray*}
p(\abold|\mubold)\,=\,\prod_{i=1}^3\,N(a_i|{\mu}_i,{\sigma}^2_i)
\end{eqnarray*}
with
$V({a}_i)=4\pi/n$ known and with
negligible correlations $({\rho}_{ij}{\simeq}0)$. 

Consider then  a $k$-dimensional random quantity 
${\Zbold}=\{Z_1,{\ldots},Z_k\}$
and the distribution
\begin{eqnarray*}
p(\mbox{{\boldmath ${z}$}}|{\mubold},{\sigmabold})\,=\,
\prod_{j=1}^{k}\,N({z}_j|{\mu}_j,{\sigma}_j^2)
\end{eqnarray*}
The interest is centered on the euclidean norm $||\mubold||$,
with $\dim \{{\mubold}\}=k$,
and its square; in particular, in 
\begin{eqnarray*}
{\delta}=\sqrt{\frac{\textstyle 3}{\textstyle 4{\pi}}}||{\mubold}||
\hspace{0.5cm}{\rm for\,\,}k=3\hspace{1.cm}{\rm and}\hspace{1.cm}
C_k=\frac{\textstyle ||{\mubold}||^2}{\textstyle k}
\end{eqnarray*} 

First, let us define
$X_j=Z_j/{\sigma}_j$ and ${\rho}_j={\mu}_j/{\sigma}_j$ so
$X_j{\sim}N({x}_j|{\rho}_j,1)$
and make a
transformation of the parameters ${\rho}_j$ to spherical coordinates:
\begin{eqnarray*}
{\rho}_1\,&=&\,{\rho}\,\cos{\phi}_1 \\
{\rho}_2\,&=&\,{\rho}\,\sin{\phi}_1 \cos{\phi}_2 \\
{\rho}_3\,&=&\,{\rho}\,\sin{\phi}_1 \sin{\phi}_2 \cos{\phi}_3 \\
&{\vdots}&\\
{\rho}_{k-1}\,&=&\,{\rho}\,\sin{\phi}_1 \sin{\phi}_2\ldots 
\sin{\phi}_{k-2}\cos{\phi}_{k-1} \\
{\rho}_{k}\,&=&\,{\rho}\,\sin{\phi}_1 \sin{\phi}_2\ldots 
\sin{\phi}_{k-2}\sin{\phi}_{k-1} 
\end{eqnarray*}
The Fisher's matrix is the Riemann metric tensor so
the square root of the 
determinant is the $k$-dimensional volume element:
\begin{eqnarray*}
dV^{k}\,=\,{\rho}^{k-1}d{\rho}\,dS^{k-1}
\end{eqnarray*}
with
\begin{eqnarray*}
dS^{k-1}=\sin^{k-2}{\phi}_1\,\sin^{k-3}{\phi}_2\,{\cdots}
\sin{\phi}_{k-2}\,d{\phi}_1\,d{\phi}_2\,{\cdots}d{\phi}_{k-1} 
=\prod_{j=1}^{k-1}\sin^{(k-1)-j}{\phi}_jd{\phi}_j
\end{eqnarray*}
the $k-1$ dimensional spherical surface element,
${\phi}_{k-1}\in[0,2{\pi})$ and ${\phi}_{1,{\ldots},k-2}\in[0,{\pi}]$.
The interest we have is
on the parameter $\rho$ so we should consider the ordered parameterization
$\{\rho;\,\phibold\}$ with
$\phibold=\{{\phi}_1,{\phi}_2,{\ldots},{\phi}_{k-1}\}$ nuisance parameters. 
Being
$\rho$ and ${\phi}_i$ independent for all $i$, we shall consider
the surface element (that is, the determinant of the submatrix obtained
for the angular part) as prior density (proper) for the nuisance parameters. 
As we have commented in Lecture 1, this is just the Lebesgue measure 
on the $k-1$ dimensional 
sphere (the Haar invariant measure under rotations) and therefore the natural 
choice for the prior; in other words, a uniform distribution on the $k-1$ 
dimensional sphere. 
Thus, we start integrating the the angular parameters. 
Under the assumption that the variances ${\sigma}_i^2$ are all the same
and considering that
\begin{eqnarray*}
\int_0^{\pi}\,e^{{\pm}{\beta}\cos{\theta}}\,{\sin}^{2{\nu}}{\theta}d{\theta}\,=\,
\sqrt{\pi}\,
\left(\frac{\textstyle 2}{\textstyle \beta}\right)^{\nu}\,
\Gamma({\nu}+\frac{1}{2})\,I_{\nu}({\beta})\hspace{0.5cm}{\rm for}
\hspace{0.5cm}Re({\nu})>-\frac{1}{2}
\end{eqnarray*}
one gets 
$p({\phi}|{\rm data})\,{\propto}\,p({\phi}_m|\phi)\pi({\phi})$
where
\begin{eqnarray*}
p({\phi}_m|{\phi},{\nu})\,=\,b\,e^{-b({\phi}+{\phi}_m)}\,
\left(\frac{\textstyle {\phi}_m}{\textstyle {\phi}}\right)^{{\nu}/2}
I_{\nu}(2b\sqrt{\phi_m}\sqrt{\phi})
\end{eqnarray*}
is properly normalized, 
\begin{eqnarray*}
{\nu}=k/2-1\;;\hspace{1.1cm}
{\phi}=||{\mubold}||^2\,;\hspace{1.1cm}
{\phi}_m=||\abold||^2\,;\hspace{1.1cm}b=\frac{\textstyle 1}
     {\textstyle 2{\sigma}^2}\,=\,\frac{\textstyle n}{\textstyle 8\pi}
\end{eqnarray*}
and $\dim \{{\mubold}\}=\dim \{{\abold}\}=k$.
This is nothing else but a non-central ${\chi}^2$ distribution.

From the series expansion of the Bessel functions it is easy to 
prove that this process is just a compound
Poisson-Gamma process
\begin{eqnarray*}
p({\phi}_m|{\phi},{\nu})\,=\,
\sum_{k=0}^{\infty}\,
Po(k|b{\phi})\,Ga({\phi}_m|b,{\nu}+k+1)
\end{eqnarray*}
and therefore the sampling distribution
is a Gamma-weighted Poisson distribution with the parameter
of interest that of the Poisson. From the Mellin Transform:
\begin{eqnarray*}
{\Mcal}(s)_{<-{\nu},{\infty}>}\,=\,\frac{\textstyle b\,e^{-b{\phi}}}
              {\textstyle \Gamma({\nu}+1)}\,
              \frac{\textstyle \Gamma(s+{\nu})}
              {\textstyle b^s}\,M(s+{\nu},{\nu}+1,b{\phi})
\end{eqnarray*}
with $M(a,b,z)$ the Kummer's function one can easily get
the moments ($E[{\phi}_m^n]=M(n+1)$); in particular
\begin{eqnarray*}
E[{\phi}_m]\,=\,{\phi}\,+\,b^{-1}({\nu}+1)\hspace{1.cm}{\rm and}\hspace{1.cm}
V[{\phi}_m]\,=\,2{\phi}b^{-1}\,+\,b^{-2}({\nu}+1)
\end{eqnarray*}

Now that we have the model $p({\phi}_m|{\phi})$,
let's go for the prior function ${\pi}({\phi})$ or ${\pi}({\delta})$.
One may guess already what shall we get.
The first element of the Fisher's matrix (diagonal)
corresponds to the norm and is constant so it would not be
surprising to get the Lebesgue measure for the norm 
$d{\lambda}({\delta})={\pi}({\delta})d{\delta}=c\,d{\delta}$. As a second 
argument, for large sample sizes ($n>>$) we have $b>>$ so 
${\phi}_m{\sim}N({\phi}_m|{\phi},{\sigma}^2=2{\phi}/b)$
and, to first order, Jeffreys' prior is ${\pi}({\phi}){\sim}{\phi}^{-1/2}$.
From the reference analysis, if we take for instance
\begin{eqnarray*}
{\pi}^{\star}({\phi})={\phi}^{({\nu}-1)/2}
\end{eqnarray*}
we end up, after some algebra, with
\begin{eqnarray*}
  {\pi}(\phi)\,\propto\,{\pi}(\phi_0)\,\lim_{k{\rightarrow}{\infty}}
  \frac{\textstyle f_k(\phi)}{\textstyle f_k(\phi_0)}
\,{\propto}\,
\left(\frac{\textstyle {\phi}_0}{\textstyle {\phi}}\right)^{1/2}\,
\lim_{b{\longrightarrow}{\infty}}\, 
e^{\textstyle -3b({\phi}-{\phi}_0)/2\,+\,[I({\phi},b)-I({\phi}_0,b)]}
\end{eqnarray*}
where
\begin{eqnarray*}
I({\phi},b)\,=\,\int_0^{\infty}\,p({\phi}_m|{\phi})\,\log
\frac{\textstyle I_{\nu}(2b\sqrt{{\phi}{\phi}_m})}
     {\textstyle I_{\nu/2}(b{\phi}_m/2)}
d{\phi}_m
\end{eqnarray*}
and ${\phi}_0$ any interior point of $\Lambda({\phi})=[0,\infty)$. From the
asymptotic behavior of the Bessel functions one gets
\begin{eqnarray*}
{\pi}({\phi})\,{\propto}\,{\phi}^{-1/2}
\end{eqnarray*}
and therefore, ${\pi}({\delta})\,=\,c$.
It is left as an exercise to get the same result with other priors like
${\pi}^{\star}({\phi})=c$ or ${\pi}^{\star}({\phi})={\phi}^{-1/2}$.

For this problem, it is easier to derive the prior from the reference analysis. 
Nevertheless,
the Fisher's information that can be expressed as:
\begin{eqnarray*}
F({\phi};{\nu})\,=\,b^2\left\{
-1\,+\,b \frac{\textstyle e^{-b{\phi}}}{\textstyle {\phi}^{{\nu}/2+1}}\,
\int_0^{\infty}e^{\textstyle -bz}\,z^{{\nu}/2+1}
\frac{\textstyle I_{{\nu}+1}^2(2b\sqrt{z{\phi}})}
     {\textstyle I_{\nu}(2b\sqrt{z{\phi}})}dz\right\}
\end{eqnarray*}
and, for large $b$ (large sample size), 
$F({\lambda};{\nu})\rightarrow{\phi}^{-1}$ {\sl regardless the number of degrees
of freedom} ${\nu}$. Thus, Jeffrey's prior is consistent with the result from 
reference analysis.
In fact, from the asymptotic behavior of the Bessel Function
in the corresponding expressions of the pdf, 
one can already see that
$F({\phi};{\nu})\,{\sim}\,{\phi}^{-1}$.
A cross check from a numeric integration is shown in fig. 2.9
where, for $k=3,5,7$ (${\nu}=1/2,3/2,5/2$), $F({\phi};{\nu})$ is depicted
as function of ${\phi}$ compared to $1/{\phi}$ in black for a sufficiently
large value of $b$.
Therefore we shall use ${\pi}({\phi})={\phi}^{-1/2}$ for the cases of interest
(dipole, quadrupole, ... any-pole). 

The posterior densities are
\vspace{0.5cm}

\noindent
$\bullet$ For  ${\phi}=||{\mubold}||^2:\,\,\,\,$
$p({\phi}|{\phi}_m,{\nu})\,=\,N\,e^{-b{\phi}}\,{\phi}^{-({\nu}+1)/2}\,
I_{\nu}(2b\sqrt{\phi_m}\sqrt{\phi})$
with
\begin{eqnarray*}
N\,=\,
\frac{\textstyle \Gamma({\nu}+1)\,b^{1/2-{\nu}}\,{\phi}_m^{-{\nu}/2}}
     {\textstyle \sqrt{\pi}\,M(1/2,{\nu}+1,b{\phi}_m)}
 \end{eqnarray*}
The Mellin Transform is
\begin{eqnarray*}
{\Mcal}_{\phi}(s)_{<1/2,{\infty}>}\,=\,\frac{\textstyle {\Gamma}(s-1/2)}
                    {\textstyle b^{s-1}\sqrt{\pi}}\,
                    \frac{\textstyle M(s-1/2,{\nu}+1,b{\phi}_m)}
                         {\textstyle M(1/2,{\nu}+1,b{\phi}_m)}
\end{eqnarray*}
and therefore the moments
\begin{eqnarray*}
E[{\phi}^n]\,=\,M(n+1)\,=\,
\frac{\textstyle \Gamma(n+1/2)\, M(n+1/2,{\nu}+1,b{\phi}_m)}
     {\textstyle \sqrt{\pi}\,b^n\,M(1/2,{\nu}+1,b{\phi}_m)}
\end{eqnarray*}
In the limit $|b{\phi}_m|{\rightarrow}{\infty}$, $E[{\phi}^n]={\phi}_m^n$.

\vspace{0.5cm}
\noindent
$\bullet$  For ${\rho}=||{\mubold}||:\,\,\,\,$
$p({\rho}|{\phi}_m,{\nu})\,=\,2\,N\,e^{-b{\rho}^2}\,{\rho}^{-{\nu}}\,
I_{\nu}(2b\sqrt{\phi_m}{\rho})$ and
\begin{eqnarray*}
{\Mcal}_{\rho}(s)={\Mcal}_{\phi}(s/2+1/2)\,\,\,{\longrightarrow}\,\,\,
E[{\rho}^n]\,=\,
\frac{\textstyle \Gamma(n/2+1/2)\, M(n/2+1/2,{\nu}+1,b{\phi}_m)}
     {\textstyle \sqrt{\pi}\,b^{n/2}\,M(1/2,{\nu}+1,b{\phi}_m)}
\end{eqnarray*}

In the particular case that $k=3$ (dipole; $\nu=1/2$), we have 
for ${\delta}=\sqrt{3/4{\pi}}{\rho}$ that 
the first two moments are:
\begin{eqnarray*}
E[{\delta}]\,=\,\frac{\textstyle 
{\rm erf}\left(z\right)}
{\textstyle a{\delta}_m M(1,3/2,-z^2)}
 \hspace{1.cm}
E[{\delta}^2]\,=\
  \frac{\textstyle 1}{\textstyle a\,M(1,3/2,-z^2)}    
\end{eqnarray*}
with $z=2{\delta}_m\sqrt{b{\pi}/3}$ and, when ${\delta}_m{\rightarrow}0$
we get
\begin{eqnarray*}
E[{\delta}]\,=\,  \sqrt{\frac{\textstyle 2}{\textstyle {\pi} a}}
\simeq \frac{\textstyle 1.38}{\textstyle \sqrt{n}} \hspace{1.cm}
E[{\delta}^2]\,=\,\frac{\textstyle 1}{\textstyle a}\hspace{1.cm}
{\sigma}_{\delta}{\simeq}\frac{\textstyle 1.04}{\textstyle \sqrt{n}}
\end{eqnarray*}
and a one sided $95\%$ upper credible region
(see section 12 for more details) 
of
${\delta}_{0.95}\,=\,\frac{\textstyle 3.38}{\textstyle \sqrt{n}}$.

\begin{figure}[t]
\begin{center}

\mbox{\epsfig{file=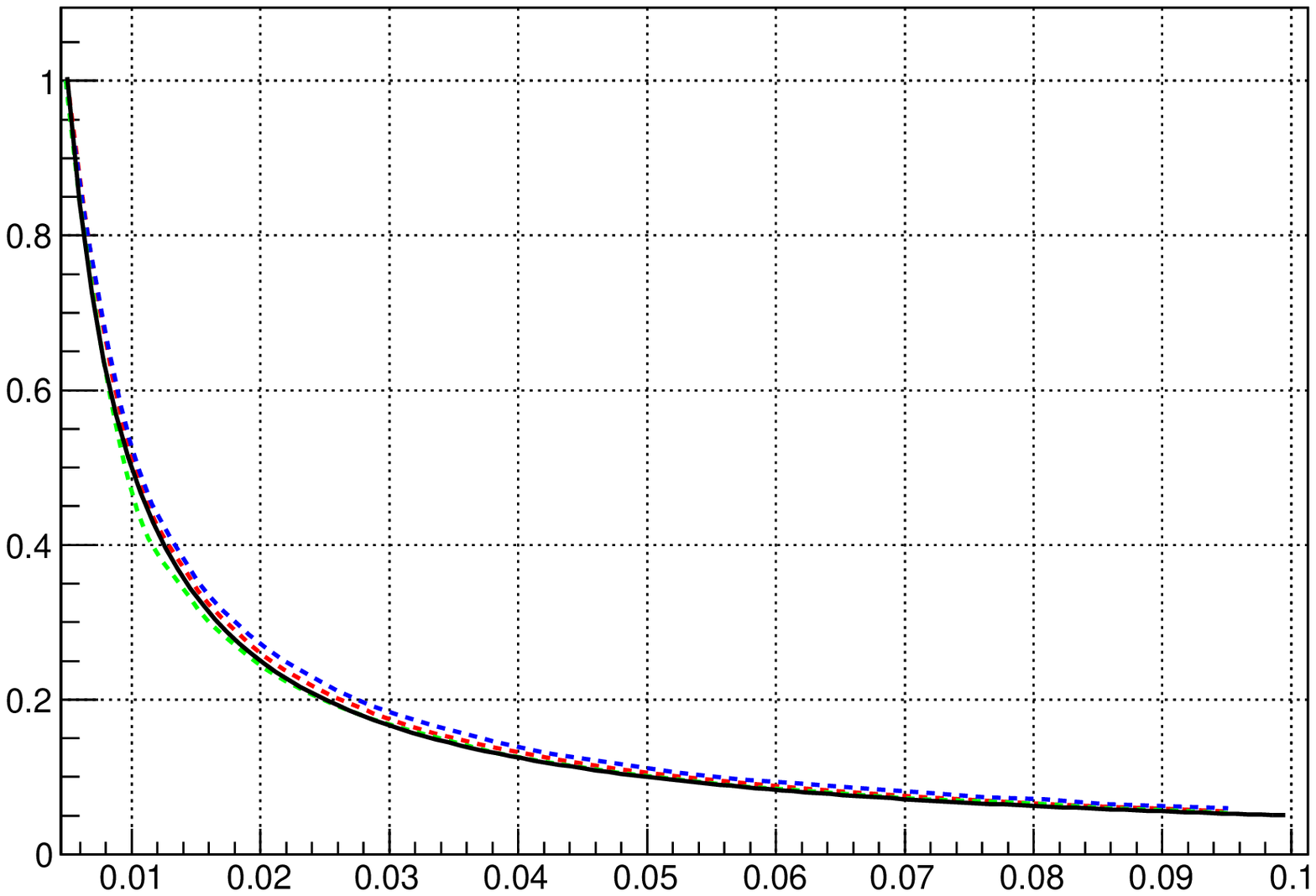,height=10.0cm,width=10.0cm}}

\footnotesize{
{\bf Figure 2.9}.- Fisher's information (numeric integration) 
as function of $\phi$ for $k=3,5,7$ (discontinuous lines)
and $f({\phi})={\phi}^{-1}$ (continuous line).
All are scaled so that $F({\phi}=0.005,{\nu})=1$.
             }
\end{center}
\end{figure}

So far, the analysis has been done assuming that the variances ${\sigma}^2_j$ 
are of the same size (equal in fact) and the correlations are small.
This is a very reasonable assumption but may not always be the case.
The easiest way 
to proceed then is to perform a transformation of the parameters of 
interest ($\mubold$) to polar coordinates 
$\mubold({\rho},{\Omega})$ and do a Monte Carlo sampling from the posterior:
\begin{eqnarray*}
p({\rho},{\Omega}|\mbox{{\boldmath ${z}$}},{\Sigmabold}^{-1})\,\propto\,
\left[\prod_{j=1}^{n}\,N({z}_j|{\mu}_j({\rho},{\Omega}),
{\Sigmabold}^{-1})\right]\,
{\pi}({\rho})d{\rho}\,dS^{n-1}
\end{eqnarray*}
with a constant prior for $\delta$ or 
$\pi(\phi){\propto}{\phi}^{-1/2}$ for ${\phi}$.

\newpage\null\thispagestyle{empty}\newpage
\newpage\null\thispagestyle{empty}\newpage

\rm

\footnotesize
\begin{tabular}{p{6.0cm}p{8.0cm}}
  &  ``Anyone who considers arithmetical methods 
       of producing random digits is, of course, in a state of sin''
\end{tabular}
\begin{flushright}
\emph{J. Von Neumann}
\end{flushright}

\vspace*{1.cm}
\huge
\noindent
{\bf Lecture 3}

\vspace{0.5cm}
\noindent
\Huge
{\bf Monte Carlo Methods}
\vspace{0.8cm}
\smill
\setcounter{section}{0}

\small

The Monte Carlo Method is a very useful and versatile numerical technique that 
allows to solve a large variety of problems difficult to tackle
by other procedures.
Even though the central idea is to simulate
experiments on a computer and make inferences from the {\sl ``observed''} 
sample, it is applicable to problems that do not have an explicit random
nature; it is enough if they have an adequate probabilistic approach. 
In fact, a frequent use of Monte Carlo techniques is the evaluation of 
definite integrals that at first sight have no statistical
nature but can be interpreted as expected values under some distribution. 

Detractors of the method used to argue that one uses Monte Carlo Methods 
because a manifest incapability to solve the problems by other 
{\sl more academic} means. 
Well, consider a {\sl ``simple''} process in particle physics:
$ee{\rightarrow} ee{\mu}{\mu}$. 
Just four particles in the final state;
the differential cross section in terms of eight variables that are not
independent due to kinematic constraints. To see what we expect for a
particular experiment, it has to be integrated within the acceptance region 
with dead zones between subdetectors, different materials and resolutions
that distort the momenta and energy, detection efficiencies,...
Yes. Admittedly we are not able to get nice expressions. Nobody in fact
and Monte Carlo comes to our help.
Last, it may be a truism but worth to mention that Monte Carlo is not a 
magic black box and will not give the answer to our problem out of nothing.
It will simply present the available information in a different and
more suitable manner after more or less complicated calculations are 
performed but all the needed information has to be put in to start with
in some way or another.

In this lecture we shall present and justify essentially all the procedures
that are commonly used in particle physics and statistics 
leaving aside subjects like Markov Chains
that deserve a whole lecture by themselves and for which
only the relevant properties will be stated without demonstration.
A general introduction to Monte Carlo techniques can be found in [Ja80].

\vspace{0.5cm}
\noindent
\section{\LARGE \bf Pseudo-Random Sequences}

\vspace{0.1cm}
Sequences of random numbers $\{x_1,x_2,{\ldots},x_n\}$ are the basis of 
Monte Carlo simulations and, in principle, their production 
is equivalent to perform an experiment $e(n)$ sampling $n$ times
the random quantity $X{\sim}p(x|\thetabold)$.
Several procedures have been developed for this purpose
(real experiments, dedicated machines, digits of transcendental numbers,...)
but, besides the lack of precise knowledge behind the generated sequences 
and the need of periodic checks, the complexity of the calculations we are
interested in  demands large sequences and fast generation procedures.
We are then forced to devise simple and efficient arithmetical algorithms to 
be implemented in a computer. Obviously neither the sequences produced are 
random nor we can produce truly random sequences by
arithmetical algorithms but we really do not need them.
It is enough for them to {\sl simulate} the relevant properties of truly
random sequences and be such that if I give you one of these
sequences and no additional information, 
you won't be able to tell after a bunch
of tests [Kn81] whether it is a truly random 
sequence or not 
(at least for the needs of the problem at hand). That's why they are
called {\sl pseudo-random} although, in what follows we shall call them
{\sl random}.
The most popular (and essentially the best) algorithms are based on 
congruential relations 
(used for this purpose as far as in the 1940s)
together with binary and/or shuffling operations with some free parameters 
that have to be fixed before the sequence is generated.
They are fast, easy to implement on any computer and, 
with the adequate initial setting of the parameters,
produce very long sequences with sufficiently good properties. 
And the easiest and fastest pseudo-random distribution to be 
generated on a computer is the {\sl Discrete Uniform}
\footnote{See [Ja88] and [Ec98] for a detailed review on random and
          quasi-random number generators.}.

Thus, let's assume that we have a {\sl good Discrete Uniform random number} 
generator
\footnote{For the examples in this lecture I have used
 {\sl RANMAR} [Ma87] that can be found, for instance, at 
 the CERN Computing Library.}
although, as Marsaglia said, {\sl ``A Random Number Generator is like sex:
When it is good it is wonderful; when it is bad... it is still pretty good''}.
Each call in a computer algorithm will produce an output ($x$) that we shall 
represent as $x{\Leftarrow}Un(0,1)$ and simulates a sampling of the 
random quantity $X{\sim}Un(x|0,1)$.
Certainly, we are not very much interested in the Uniform Distribution so the
task is to obtain a sampling of densities $p(\xbold|\thetabold)$ 
other than Uniform from
a {\sl Pseudo-Uniform Random Number Generator} for which there are several
procedures.

\vspace{0.5cm}
\noindent
{\raya}                   
\vspace{0.35cm}
\footnotesize

\noindent
{\bf Example 3.1: Estimate the value of $\pi$.} 
As a simple first example, let's see how we may estimate the 
value of $\pi$. Consider a circle of radius $r$ inscribed in a square 
with sides of length $2r$. Imagine now that we throw random {\sl points}
evenly distributed inside the square and count how many have fallen inside 
the circle. It is clear that since the area of the square is
$4r^2$ and the area enclosed by the circle is ${\pi}r^2$, the probability
that a throw falls inside the circle is $\theta=\pi/4$.

If we repeat the experiment $N$ times, the number $n$ of throws falling
inside the circle follows a Binomial law $Bi(n|N,p)$
and therefore, having observed $n$ out of $N$ trials we have that 
\begin{eqnarray*}
  p(\theta|n,N)\,\propto\,{\theta}^{n-1/2} (1-{\theta})^{N-n-1/2}
\end{eqnarray*}
Let's take ${\pi}^{\star}=4E[\theta]$ as point estimator and
${\sigma}^{\star}=4\sigma_{\theta}$ as a measure of the precision. 
The results obtained for samplings of different size are shown in the 
following table: 
\begin{center}
\begin{tabular}{|r|r|c|c|} \hline
 Throws $(N)$  &  Accepted $(n)$  & ${\pi}^{\star}$ & $\sigma^{\star}$
     \\    \hline
     $100$ &     $83$  & $3.3069$ & $0.1500$ \\
    $1000$ &    $770$  & $3.0789$ & $0.0532$  \\
   $10000$ &   $7789$  & $3.1156$ & $0.0166$  \\
  $100000$ &  $78408$  & $3.1363$ & $0.0052$  \\
 $1000000$ & $785241$  & $3.1410$ & $0.0016$  
              \\ \hline
\end{tabular}
\end{center}
It is interesting to see that the precision decreases with the sampling
size as $1/\sqrt{N}$. This dependence is a general feature of Monte Carlo 
estimations {\sl regardless} the number of dimensions of the problem.

A similar problem is that of Buffon's needle: {\sl A needle of
length $l$ is thrown at random on a horizontal plane with stripes of
width $d>l$. What is the probability that the needle intersects one of
the lines between the stripes?}. It is left as an exercise to shown that, 
as given already by Buffon in 1777, $P_{cut}=2l/{\pi}d$. Laplace pointed out,
in what may be the first use of the Monte Carlo method, that doing the
experiment one may estimate the value of ${\pi}$ {\sl ``... although with large
error''}.
\smill
{\raya}                   
\vspace{1.0cm}
\vspace*{0.3cm}
\section{\LARGE \bf Basic Algorithms}
\subsection{Inverse Transform}
\vspace*{0.1cm}
 This is, at least formally, the easiest procedure. Suppose we want a
sampling of the continuous one-dimensional random quantity 
$X{\sim}p(x)$
\footnote{Remember that if ${\rm supp}(X)=\Omega{\subseteq}{\Rcal}$,
it is assumed that the density is 
$p(x){\mbox{\boldmath $1$}_{\Omega}(x)}$.} so
\begin{eqnarray*}
P[X{\in}(-\infty,x]]\,=\,
\int_{-\infty}^{x}p(x')dx'\,=\,
\int_{-\infty}^{x}dF(x')\,=\,F(x)
\end{eqnarray*}
Now, we define the new random quantity
$U=F(X)$ with support in $[0,1]$. How is it distributed? Well,
\begin{eqnarray*}
F_U(u)\,\equiv\,P[U{\leq}u]\,=\,
P[F(X){\leq}u)]\,=\,P[X{\leq}F^{-1}(u)]\,=\,
\int_{-\infty}^{F^{-1}(u|\cdot)}dF(x')\,=\,u
\end{eqnarray*}
and therefore $U{\sim}Un(u|0,1)$. The algorithm is then clear; at step $i$:
\begin{itemize}
\item[$i_1$)] $u_i{\Leftarrow}Un(u|0,1)$
\item[$i_2$)] $x_i=F^{-1}(u_i)$
\end{itemize} 
After repeating the sequence $n$ times we end up with a sampling
$\{x_1,x_2,\ldots,x_n\}$ of $X{\sim}p(x)$.

\vspace{0.5cm}
\noindent
{\raya}                   
\vspace{0.35cm}
\footnotesize
\noindent
{\bf Example 3.2:} Let' see how we generate a sampling of the Laplace
distribution $X{\sim}La(x|{\alpha},{\beta})$ with $\alpha{\in}\Rcal$,
$\beta{\in}(0,\infty)$ and density
\begin{eqnarray*}
  p(x|\alpha,\beta)=\frac{1}{2\beta}\,e^{-|x-\alpha|/\beta}
 \mbox{\boldmath $1$}_{(-\infty,\infty)}(x)
\end{eqnarray*}
The distribution function is
\begin{eqnarray*}
   F(x)=\int_{-\infty}^{x}p(x'|\alpha,\beta)dx'=
 \left\{
     \begin{array}{l}
             \frac{1}{2}{\rm exp}\left(\frac{x-\alpha}{\beta}\right)
             \hspace*{0.2cm}{\rm if}\hspace*{0.2cm} x<\alpha \\
             \\
             1-\frac{1}{2}{\rm exp}\left(-\frac{x-\alpha}{\beta}\right)
             \hspace*{0.2cm}{\rm if}\hspace*{0.2cm} u{\geq}\alpha 
     \end{array}
   \right.
\end{eqnarray*}
Then, if $u{\Leftarrow}Un(0,1)$:
\begin{eqnarray*}
   x= \left\{
     \begin{array}{l}
             {\alpha}+{\beta}{\rm ln}(2u)
             \hspace*{0.2cm}{\rm if}\hspace*{0.2cm} u<1/2 \\
             \\
             {\alpha}-{\beta}{\rm ln}(2(1-u))
             \hspace*{0.2cm}{\rm if}\hspace*{0.2cm} u{\geq}1/2 
     \end{array}
   \right.
\end{eqnarray*}
\smill
{\raya}                   
\vspace{1.0cm}

The generalization of the {\sl Inverse Transform} method to n-dimensional
random quantities is trivial. We just have to consider the marginal and 
conditional distributions
 \begin{eqnarray}
    F(x_1,x_2,{\ldots},x_n)\,=\,
    F_n(x_n|x_{n-1},{\ldots},x_1)\,{\cdots}\,
    F_2(x_2|x_1)\,{\cdots}\,F_1(x_1)
              \nonumber
 \end{eqnarray}
or, for absolute continuous quantities, the probability densities
 \begin{eqnarray}
    p(x_1,x_2,{\ldots},x_n)\,=\,
    p_n(x_n|x_{n-1},{\ldots},x_1)\,{\cdots}\,
    p_2(x_2|x_1)\,{\cdots}\,p_1(x_1)
              \nonumber
 \end{eqnarray}
and then proceed sequentially; that is:
\begin{itemize}
\item[$i_{2,1}$)] $u_1{\Leftarrow}Un(u|0,1)$ and $x_1=F_1^{-1}(u_1)$;
\item[$i_{2,2}$)] $u_2{\Leftarrow}Un(u|0,1)$ and $x_2=F_2^{-1}(u_2|x_1)$;
\item[$i_{2,1}$)] $u_3{\Leftarrow}Un(u|0,1)$ and $x_3=F_3^{-1}(u_3|x_1,x_2)$;
\item[  ] \hspace*{3.cm} $\vdots$
\item[$i_{2,n}$)] $u_n{\Leftarrow}Un(u|0,1)$ and 
  $x_n=F_n^{-1}(u_3|x_{n-1},\ldots,x_1)$
\end{itemize}

If the random quantities are independent
there is a unique decomposition
\begin{eqnarray*}
    p(x_1,x_2,{\ldots},x_n)\,=\,\prod_{i=1}^{n}\,p_i(x_i)
    \hspace{0.5cm}{\rm and}\hspace{0.5cm}
    F(x_1,x_2,{\ldots},x_n)\,=\,\prod_{i=1}^{n}\,F_i(x_i)
 \end{eqnarray*}
but, if this is not the case, note that there are
$n!$ ways to do the decomposition and some may be
easier to handle than others (see example 3.3). 

\vspace{0.5cm}
\noindent
{\raya}                   
\vspace{0.35cm}
\footnotesize

\noindent
{\bf Example 3.3:} Consider the probability density
 \begin{eqnarray*}
    p(x,y)\,=\,2\,e^{-x/y}\,\,
      \mbox{\boldmath $1$}_{(0,\infty)}(x)
      \mbox{\boldmath $1$}_{(0,1]}(y)
 \end{eqnarray*}
We can express the Distribution Function as
$F(x,y)=F(x|y)F(y)$ where:
 \begin{eqnarray}
    p(y)\,&=&\,\int_{0}^{\infty}\,p(x,y)\,dx\,=\,2\,y
    \,\,\,\,\,\,\,\,\,\,{\longrightarrow}\,\,\,\,\,\,\,\,\,\,
    F(y)\,=\,y^2          \,\nonumber \\
    p(x|y)\,&=&\,\frac{\textstyle p(x,y)}
                     {\textstyle p(y)}\,=\,
                \frac{\textstyle 1}
                     {\textstyle y}\,e^{-x/y}
    \hspace*{0.6cm}\,\,\,\,\,{\longrightarrow}\,\,\,\,\,\,\,\,\,\,
    F(x|y)\,=\,1\,-\,e^{-x/y}
              \nonumber
 \end{eqnarray}
Both $F(y)$ and $F(x|y)$ are easy to invert so:
\begin{itemize}
\item[$i_1$)] $u{\Leftarrow}Un(0,1)$ and get $y\,=\,u^{1/2}$
\item[$i_2$)] $w{\Leftarrow}Un(0,1)$ and get $x\,=\,-\,y\,\ln w$
\end{itemize}
Repeating the algorithm $n$ times, we get the sequence
$\{(x_1,y_1),(x_2,y_2),{\ldots}\}$ that simulates a sampling from $p(x,y)$.

Obviously, we can also write $F(x,y)=F(y|x)F(x)$ and proceed in an 
analogous manner. However
\begin{eqnarray*}
    p(x)\,=\,\int_{0}^{1}\,p(x,y)\,dy\,=\,
    2\,x\,\int_{x}^{\infty}\,e^{-u}\,u^{-2}\,du
    \nonumber
 \end{eqnarray*}
is not so easy to sample.

\smill
{\raya}                   
\vspace{1.0cm}

Last, let's see how to use the {\sl Inverse Transform} procedure for
discrete random quantities.
If $X$ can take the values in
$\Omega_X=\{x_0,x_1,x_2,{\ldots}\}$ with probabilities 
$P({X}=x_k)=p_k$, the Distribution Function will be:
\begin{eqnarray}
       &&  F_0\,=\,P({X}{\leq}x_0)\,=\,p_0       \nonumber \\
       &&  F_1\,=\,P({X}{\leq}x_1)\,=\,p_0\,+\,p_1       \nonumber \\
       &&  F_2\,=\,P({X}{\leq}x_2)\,=\,p_0\,+\,p_1\,+\,p_2  \nonumber \\
       &&  {\ldots}
     \nonumber
\end{eqnarray}
Graphically, we can represent the sequence
$\{0,F_0,F_1,F_2,{\ldots},1\}$ as:

\newpage

\vspace{1.0cm}
{\hspace*{2.4cm}} $p_0$ {\hspace*{0.60cm}} $p_1$ {\hspace*{1.00cm}} $p_2$
{\hspace*{2.0cm}} ........

\hspace*{2.00cm} {\rule[-0.50mm]{0.50mm}{2.00mm}}
                 \hspace*{-1.50mm}{\rule{1.00cm}{0.50mm}}
                 {\hspace*{-1.50mm}\rule[-0.50mm]{0.50mm}{2.00mm}}
                 \hspace*{-1.50mm}{\rule{1.50cm}{0.50mm}}
                 {\hspace*{-1.50mm}\rule[-0.50mm]{0.50mm}{2.00mm}}
                 \hspace*{-1.50mm}{\rule{2.00cm}{0.50mm}}
                 {\hspace*{-1.50mm}\rule[-0.50mm]{0.50mm}{2.00mm}}
                 \hspace*{-1.50mm}{\rule{3.50cm}{0.50mm}}
                 {\hspace*{-1.50mm}\rule[-0.50mm]{0.50mm}{2.00mm}}

{\hspace*{1.95cm}} $0$ {\hspace*{0.58cm}} $F_0$ {\hspace*{0.80cm}} $F_1$
{\hspace*{1.30cm}} $F_2$ {\hspace*{0.80cm}} ........ {\hspace*{1.25cm}}$1$
\vspace{1.0cm}

\noindent
Then, it is clear that a random quantity
$u_i$ drawn from $U(x|0,1)$ will determine a point in the interval
$[0,1]$ and will belong to the subinterval
$[F_{k-1},F_k]$ with probability $p_k=F_k-F_{k-1}$ so 
we can set up the following algorithm:
\begin{itemize}
\item[$i_1$)] Get $u_i\,{\sim}\,Un(u|0,1)$;
\item[$i_2$)] Find the value $x_k$ such that $F_{k-1}\,{<}\,u_i\,{\leq}\,F_k$ 
\end{itemize}
The sequence $\{x_0,x_1,x_2,{\ldots}\}$ so generated will be a sampling
of the probability law $P({X}=x_k)=p_k$. Even though discrete
random quantities can be sampled in this way, some times there are specific
properties based on which faster algorithms can be developed. That is the 
case for instance for the Poisson Distribution as the following example shows.

\vspace{0.5cm}
\noindent
{\raya}                   
\vspace{0.35cm}

\footnotesize
\noindent
{\bf Example 3.4: Poisson Distribution} $Po(k|{\mu})$. From the recurrence
relation
\begin{eqnarray*}
    p_k\,=\,e^{-{\mu}}\,
    \frac{\textstyle {\mu}^k}{\textstyle \Gamma(k+1)}\,=\,
    \frac{\textstyle {\mu}}{\textstyle k}\,p_{k-1}
\end{eqnarray*}
\begin{itemize}
\item[$i_1$)] $u_i{\Leftarrow}Un(0,1)$ 
\item[$i_2$)] Find the value $k=0,1,...$ such that $F_{k-1}<u_i{\leq}F_k$ 
              and deliver  $x_k=k$ 
\end{itemize} 

For the Poisson Distribution, there is a faster procedure. Consider a
sequence of $n$ independent random quantities $\{X_1,X_2,{\ldots},X_n\}$,
each distributed as $X_i{\sim}Un(x|0,1)$, and introduce a new random
quantity 
\begin{eqnarray*}
    W_n\,=\,\prod_{k=1}^n X_k
\end{eqnarray*}
with ${\rm supp}\{W_n\}=[0,1]$. Then
\begin{eqnarray*}
    W_n\,{\sim}\,p(w_n|n)\,=\,
    \frac{\textstyle (-{\log}w_n)^{n-1}}{\textstyle \Gamma(n)}
     \hspace{1.cm}{\longrightarrow}\hspace{1.cm}
     P(W_n{\leq}a)\,=\,\frac{\textstyle 1}{\textstyle \Gamma(n)}
     \int_{-{\log}a}^{\infty}e^{-t}t^{n-1}dt
\end{eqnarray*}
and if we take $a=e^{-{\mu}}$ we have, in terms of the Incomplete Gamma 
Function $P(a,x)$:
\begin{eqnarray*}
  P(W_n{\leq}e^{-{\mu}})\,=\,1-P(n,\mu)\,=\,e^{-\mu}\,
    \sum_{k=0}^{n-1}
    \frac{\textstyle {\mu}^k}{\textstyle \Gamma(k+1)}\,=\,
    Po(X{\leq}n-1|\mu)
\end{eqnarray*}
Therefore, 
\begin{itemize}
\item[$i_0$)] Set $w_p=1$;
\item[$i_1$)] $u_i{\Leftarrow}Un(0,1)$  and set $w_p=w_pu$; 
\item[$i_2$)] Repeat step $i_1$ while $w_p{\leq}e^{-\mu}$, say $k$ times,
                and deliver  $x_k=k-1$ 
\end{itemize} 

\noindent
{\bf Example 3.5: Binomial Distribution} $Bn(k|N,\theta)$. From the recurrence
relation
\begin{eqnarray}
    p_k\,=\,
    \left( \begin{array}{c}
      N \\ k
    \end{array} \right)\,\theta^k\,(1-\theta)^{n-k}\,=\,
    \frac{\textstyle \theta}{\textstyle 1-\theta}\,
    \frac{\textstyle n-k+1}{\textstyle k}\,
     p_{k-1}
              \nonumber
 \end{eqnarray}
with $p_0=(1-\theta)^k$
\begin{itemize}
\item[$i_1$)] $u_i{\Leftarrow}Un(0,1)$ 
\item[$i_2$)] Find the value $k=0,1,{\ldots},N$ such that $F_{k-1}<u_i{\leq}F_k$ 
              and deliver  $x_k=k$ 
\end{itemize} 
\vspace{0.35cm}

\noindent
{\bf Example 3.6: Simulation of the response of a Photomultiplier tube.}

Photomultiplier tubes are widely used devices to detect electromagnetic
radiation by means of the external photoelectric effect. A typical 
photomultiplier consists of a vacuum tube with an input window, 
a photocathode, 
a focusing and a series of amplifying electrodes (dynodes) and
an electron collector (anode). Several materials are used for the input window 
(borosilicate glass, synthetic silica,...) which transmit radiation in 
different wavelength ranges and, due to absorptions (in particular in the UV
range) and external reflexions, the transmittance of the window is never 100\%.
Most photocathodes are compound semiconductors consisting of alkali metals
with a low work function. When the photons strike the photocathode the 
electrons in the valence band are excited and, if they get enough energy to 
overcome the vacuum level barrier, they are emitted into the vacuum tube as 
photoelectrons. The trajectory of the electrons inside the photomultiplier
is determined basically by the applied voltage and the geometry of the
focusing electrode and the first dynode. Usually, the photoelectron is 
driven towards the first dynode and originates an electron shower which is 
amplified in the following dynodes and collected at the anode.
However, a fraction of the incoming photons pass through the photocathode and
originates a smaller electron shower when it strikes the first dynode of the
amplification chain.

To study the response of a photomultiplier tube, an experimental set-up has 
been made with a LED as photon source. We are interested in the response
for isolated photons so we regulate the current and the frequency so as to
have a low intensity source.  
Under this conditions, the number of photons that arrive at the window of
the photomultiplier is well described by a Poisson law. When one of this
photons strikes on the photocathode, an electron is ejected and driven towards
the first dynode to start the electron shower. We shall assume that the
number of electrons so produced follows also a Poisson law
$n_{gen}{\sim}Po(n|{\mu})$. The parameter $\mu$ accounts for the
efficiency of this first process and depends on the characteristics of the
photocathode, the applied voltage and the geometry of the focusing
electrodes (essentially that of the first dynode). It has been estimated to
be $\mu=0.25$. Thus, we start our simulation with
\begin{itemize}
 \item[1)] $n_{gen}\,{\Leftarrow}\,Po(n|{\mu})$ 
\end{itemize}
electrons leaving the photocathode. They are driven towards the
the first dynode to start the electron shower but there is a chance 
that they miss the first and start the shower at the second.
Again, the analysis of the experimental data suggest that
this happens with probability $p_{d2}{\simeq}0.2$. Thus, we have to decide
how many of the $n_{gen}$ electrons start the shower at the second
dynode. A Binomial model is appropriate in this case:
\begin{itemize}
 \item[2)] $n_{d2}\,{\Leftarrow}\,Bi(n_{d2}|n_{gen},p_{d2})$ and therefore
           $n_{d1}=n_{gen}-n_{d2}$.
\end{itemize}
Obviously, we shall do this second step if $n_{gen}>0$.

Now we come to the amplification stage. Our photomultiplier has 12 dynodes
so let's see the response of each of them.
For each electron that strikes upon dynode $k$ $(k=1,...,12)$, $n_k$
electrons will be produced and directed towards the next element of the chain
(dynode $k+1$), the number of them again well described by a Poisson law
$Po(n_k|{\mu}_k)$.
If we denote by $V$ the total voltage applied between the photocathode and
the anode and by $R_k$ the resistance previous to dynode $k$ we have that
the current intensity through the chain will be
\begin{eqnarray}
 I\,=\,\frac{\textstyle V}{\textstyle \sum_{i=1}^{13}R_i}
\nonumber
\end{eqnarray}
where we have considered also the additional resistance between the last dynode
and the anode that collects the electron shower.
Therefore, the parameters $\mu_k$ are determined by the relation
\begin{eqnarray}
 {\mu}_k\,=\,a\,(I\,R_k)^b
\nonumber
\end{eqnarray}
where $a$ and $b$ are characteristic parameters of the photomultiplier. In
our case we have that
$N=12$, $a=0.16459$, $b=0.75$, a total applied voltage of $800$ V and
a resistance chain of
$\{2.4,\, 2.4,\, 2.4,\, 1.0,\, 1.0,\, 1.0,\, 1.0,$ $1.0,\,
 1.0,\,1.0,\,1.0,\,1.2,\,2.4\}$ Ohms. It is easy to see that if the response of
dynode $k$ to one electron is modeled as $Po(n_k|{\mu}_k)$, 
the response to $n_i$ incoming electrons is described by 
$Po(n_k|n_i{\mu}_k)$. Thus,
we simulate the amplification stage as:
\begin{itemize}
 \item[3.1)] If $n_{d1}>0:\hspace{1.cm}$
             {\rm do}\,\,{\rm from}\,\,k=1\,\,{\rm to}\,\,12: 
             \begin{eqnarray*}
               {\mu}\,=\,{\mu}_k\,n_{d1} \,\,\,
                     \longrightarrow\,\,\,n_{d1}\,{\Leftarrow}\,Po(n|{\mu})
             \end{eqnarray*}
\item[3.2)] If $n_{d2}>0:\hspace{1.cm}$
             {\rm do}\,\,{\rm from}\,\,k=2\,\,{\rm to}\,\,12: 
             \begin{eqnarray*}
               {\mu}\,=\,{\mu}_k\,n_{d2} \,\,\,
                     \longrightarrow\,\,\,n_{d2}\,{\Leftarrow}\,Po(n|{\mu})
             \end{eqnarray*}
\end{itemize}
Once this is done,
we have to convert the number of electrons at the anode in ADC counts.
The electron charge is $Q_e=1.602176\,10^{-19}$ C and in our set-up we have 
$f_{ADC}=2.1\,10^{14}$ ADC counts per Coulomb so
\begin{eqnarray*}
 ADC_{pm}\,=\,(n_{d1}\,+\,n_{d2})\,(Q_e\,f_{ADC})
\end{eqnarray*}

Last, we have to consider the noise ({\sl pedestal}). 
In our case, the number of pedestal ADC counts is well
described by a mixture model with two Normal densities
\begin{eqnarray*}
 p_{ped}(x|\cdot)\,=\,{\alpha}\,N_1(x|10.,1.)\,+\,
                    (1-{\alpha})\,N_1(x|10.,1.5)
\end{eqnarray*}
with $\alpha=0.8$. Thus,
with probability $\alpha$ we obtain
$ADC_{ped}{\Leftarrow}N_1(x|10,1.5)$, and with probability
$1-\alpha$, 
$ADC_{ped}{\Leftarrow}N_1(x|10,1)$ so
the total number of ADC counts will be
\begin{eqnarray*}
 ADC_{tot}\,=\,ADC_{ped}\,+\,ADC_{pm}
\nonumber
\end{eqnarray*}
Obviously, if in step $1)$ we get $n_{gen}=0$, then
$ ADC_{tot}\,=\,ADC_{ped}$.
Figure 1 shows the result of the simulation for a sampling size of $10^6$
together with the main contributions (1,2 or 3 initial photoelectrons) and
the pedestal. From these results, the parameters of the device can be
adjusted (voltage, resistance chain,...) to optimize the response for our
specific requirements.

\smill
{\raya}                   
\vspace{1.0cm}

\begin{figure}[t]
\begin{center}

\mbox{\epsfig{file=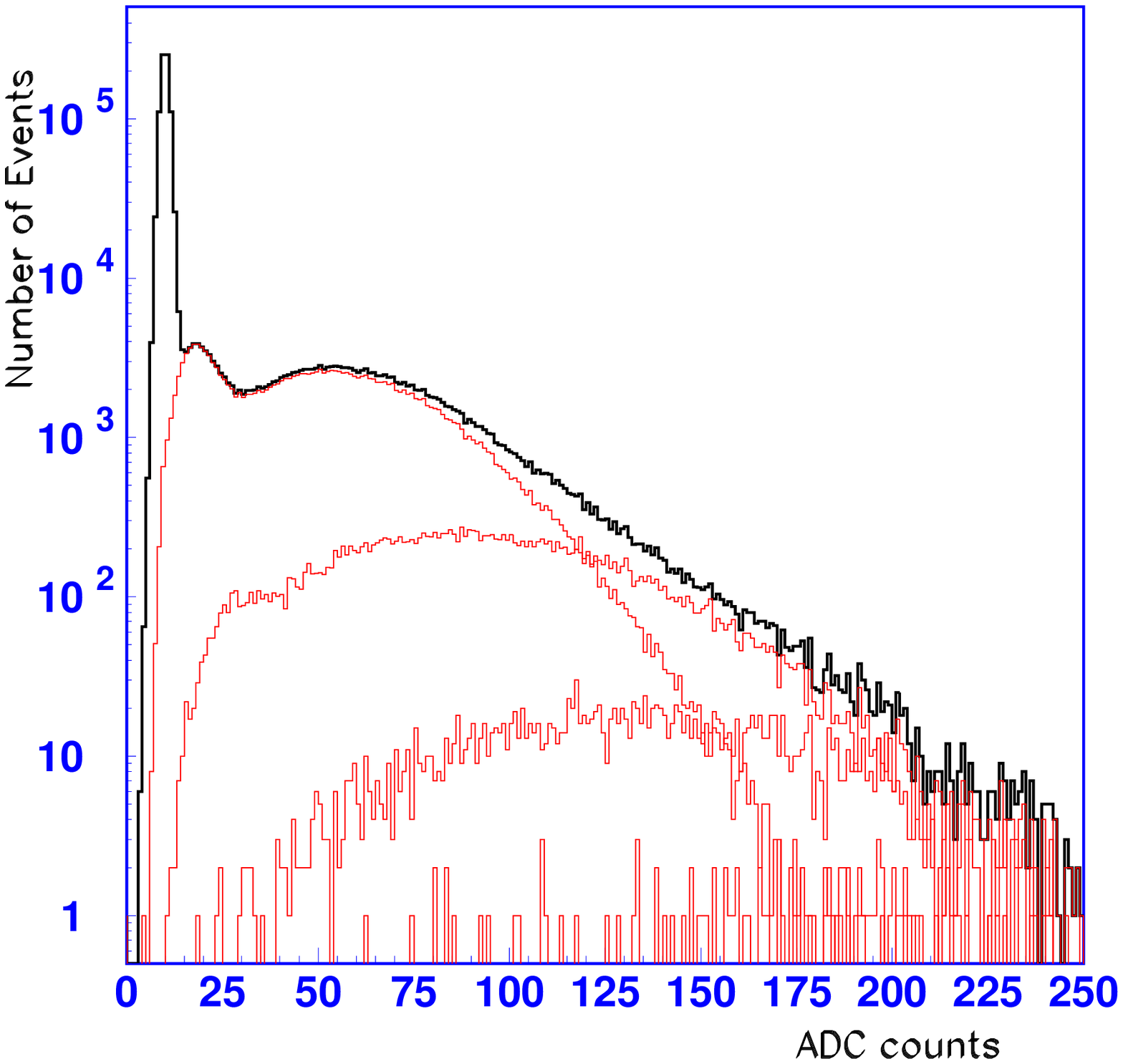,height=10cm,width=10cm}}

\footnotesize
{\bf Figure 3.1}.-
Result of the simulation  of the response of a photomultiplier tube.
The histogram contains $10^6$ events and shows the final ADC distribution
detailing the contribution of the pedestal and the response to 1, 2 and 3
incoming photons.
\end{center}
\end{figure}

The {\sl Inverse Transform}
method is conceptually simple and 
easy to implement for discrete distributions and many continuous distributions
of interest. Furthermore, is {\sl efficient} in the sense that
for each generated value $u_i$ as $Un(x|0,1)$ we get a value $x_i$ 
from $F(x)$. However, with the exception of easy distributions the 
inverse function $F^{-1}(x)$ has no simple expression in terms of 
elementary functions and may be difficult or time consuming to invert. 
This is, for instance, the case if you attempt to invert the Error Function
for the Normal Distribution. 
Thus, apart from simple cases, the {\sl Inverse Transform} method is used in
combination with other procedures to be described next.

\vspace{0.5cm}
\noindent
{\raya}                   
\vspace{0.35cm}
\footnotesize
\noindent
{\bf NOTE 7: Bootstrap.} Given the iid sample $\{x_1,x_2,\ldots,x_n\}$ 
of the random quantity $X{\sim}p(x|\theta)$ we know
(Glivenko-Cantelly theorem; see lecture 1 (7.6)) that:
\begin{eqnarray*}
F_n(x)=1/n\sum_{k=1}^{n}\mbox{\boldmath $1$}_{(-\infty,x]}(x_k)
\hspace{0.5cm}\stackrel{\rm unif.}{\longrightarrow}\hspace{0.5cm}F(x|\theta)
\end{eqnarray*}
Essentially the idea behind the {\sl bootstrap} is to sample from the
empirical Distribution Function $F_n(x)$ that, as we have seen for 
discrete random quantities, is equivalent to draw samplings
$\{x'_1,x'_2,\ldots,x'_n\}$ of size $n$ from the original sample with 
replacement. 
Obviously, increasing the number of resamplings does not provide more 
information than what is contained in the original data but,
used with good sense, each {\sl bootstrap} will lead to a 
posterior and can also be useful to give insight about the form of the 
underlying model $p(x|\theta)$ and the distribution of some statistics.
We refer to [Ru81] for further details. 

{\raya}                   
\vspace{1.0cm}
\small

\subsection{Acceptance-Rejection ({\sl Hit-Miss}; {\sl J. Von Neumann, 1951)}}

The {\sl Acceptance-Rejection} algorithm is easy to implement
and allows to sample a large variety of
n-dimensional probability densities with a less detailed
knowledge of the function. But nothing is for free; these advantages are in 
detriment of the generation efficiency. 

Let's start with the one dimensional case where
$X{\sim}p(x|\theta)$ is a continuous random quantity
with ${\rm supp}(X)=[a,b]$ and $p_m={\rm max}_xp(x|\theta)$. 
Consider now two independent random quantities 
$X_1{\sim}Un(x_1|\alpha,\beta)$ and $X_2{\sim}Un(x_2|0,\delta)$ where
$[a,b]\subseteq{\rm supp}(X_1)=[\alpha,\beta]$ and
$[0,p_m]\subseteq{\rm supp}(X_2)=[0,\delta]$. The covering does not
necessarily have to be a rectangle in ${\Rcal}^2$ 
(nor a hypercube ${\Rcal}^{n+1}$) and, in fact, in some cases it may be 
interesting to consider other coverings to improve the efficiency but 
the generalization is obvious. Then
\begin{eqnarray*}
p(x_1,x_2|\cdot)\,=\,\frac{\textstyle 1}{\textstyle \beta-\alpha}\,  
                     \frac{\textstyle 1}{\textstyle \delta}
\end{eqnarray*}
Now, let's find the distribution of $X_1$ conditioned to 
$X_2{\leq}p(X_1|\theta)$:
\begin{eqnarray*}
P(X_1{\leq}x|X_2{\leq}p(x|\theta))&=&
    \frac{\textstyle P(X_1{\leq}x,X_2{\leq}p(x|\theta))}
         {\textstyle P(X_2{\leq}p(x|\theta))}=
    \frac{\textstyle \int^x_{\alpha}dx_1\,\int^{p(x_1|\theta)}_{0}
                  p(x_1,x_2|\cdot)\,dx_2}
         {\textstyle \int^{\beta}_{\alpha}dx_1\,
                      \int_{0}^{p(x_1|\theta)} p(x_1,x_2|\cdot)\,dx_2}= \\
    &=&\,
    \frac{\textstyle \int^x_{\alpha}p(x_1|\theta)
                     \mbox{\boldmath $1$}_{[a,b]}(x_1)\,dx_1}
     {\textstyle \int^{\beta}_{\alpha}p(x_1|\theta)
                     \mbox{\boldmath $1$}_{[a,b]}(x_1)\,dx_1}
\,=\,\int^x_a p(x_1|\theta)\,dx_1 \,=\,F(x|\theta)
\end{eqnarray*}
so we set up the following algorithm:
\begin{itemize}
\item[$i_1$)] $u_i{\Leftarrow}Un(u|\alpha{\leq}a,\beta{\geq}b)$ and 
              $w_i{\Leftarrow}Un(w|0,\delta{\geq}p_m)$;
\item[$i_2$)] If $w_i{\leq}p(u_i|\theta)$ we {\sl accept} $x_i$;
              otherwise we {\sl reject} $x_i$ and start again from $i_1$
\end{itemize}
Repeating the algorithm $n$ times we get the sampling
$\{x_1,x_2,{\ldots},x_n\}$ from $p(x|\theta)$.

Besides its simplicity, the Acceptance-Rejection scheme does not require
to have normalized densities for it is enough to know an upper bound and 
in some cases, for instance when the support of the random quantity $X$ is 
determined by functional relations, it is easier to deal with a 
simpler covering of the support. 
However, the price to pay is a low {\sl generation efficiency}:
\begin{eqnarray*}
  {\epsilon}\,\stackrel{def.}{=}\,
                 \frac{\textstyle {\rm accepted\,\,trials}}
                 {\textstyle {\rm total\,\,trials}}\,=\,
                  \frac{\textstyle {\rm area\,\,under\,\,}p(x|\theta)}
                 {\textstyle {\rm area\,\,of\,\,the\,\,covering}}\,\leq\,1
\end{eqnarray*}
Note that
the efficiency so defined refers only to the fraction of accepted trials and,
obviously, the more adjusted the covering is the better but for the 
{\sl Inverse Transform} $\epsilon=1$ and it does not necessarily imply
that it is more efficient attending to other considerations. 
It is interesting to observe
that if we do not know the normalization factor of the density function, 
it can be estimated as
\begin{eqnarray*}
\int_X p(x|\theta)\,dx\,{\simeq}\,
            (\rm area\,\,of\,\,the\,\,covering)\,{\epsilon}
\end{eqnarray*}
Let's see some examples before we proceed.

\vspace{0.5cm}
\noindent
{\raya}                   
\vspace{0.35cm}
\footnotesize

\noindent
{\bf Example 3.7:} Consider $X{\sim}Be(x|{\alpha},{\beta})$. In this case,
   what follows is just for pedagogical purposes since other procedures
   to be discussed later are more efficient. Anyway, the density is
   \begin{eqnarray*}
     p(x|\alpha,\beta)\,{\propto}\,x^{{\alpha}-1}\,(1\,-\,x)^{{\beta}-1}
     \hspace{0.5cm};\hspace{1.5cm} x\,{\in}\,[0,1]
   \end{eqnarray*}
  Suppose that ${\alpha},{\beta}>1$ so the mode
$x_o=({\alpha}-1)({\alpha}+{\beta}-2)^{-1}$
  exists and is unique. Then
\begin{eqnarray*}
  p_m{\equiv}{\rm max}_x\{p(x|\alpha,\beta)\}\,=\,p(x_0|\alpha,\beta)\,=\,
 \frac{\textstyle ({\alpha}-1)^{{\alpha}-1}\,({\beta}-1)^{{\beta}-1}}
      {\textstyle ({\alpha}+{\beta}-2)^{{\alpha}+{\beta}-2}}
\end{eqnarray*}
so let's take then the domain 
$[{\alpha}=0,{\beta}=1]\,{\times}\,[0,p_m]$ and
\begin{itemize}
\item[$i_1$)] Get $x_i\,{\sim}\,Un(x|0,1)$ and 
                  $y_i\,{\sim}\,Un(y|0,p_m)$;
\item[$i_2$)] If
          \begin{itemize}
            \item[a)] $y_i{\leq}p(x_i|\alpha,\beta)$ we deliver 
                      ({\sl accept}) $x_i$
            \item[r)] $y_i>p(x_i|\alpha,\beta)$ we {\sl reject} $x_i$ and start
                      again from $i_1$
          \end{itemize}
\end{itemize}
Repeating the procedure $n$ times, we get a sampling $\{x_1,x_2,...,x_n\}$ 
from $Be(x|{\alpha},{\beta})$. In this case we know the normalization
so the area under $p(x|{\alpha},{\beta})$ is $Be(x|{\alpha},{\beta})$ 
and the {\sl generation efficiency} will be:
\begin{eqnarray*}
  {\rm {\epsilon}}\,=\,B({\alpha,\beta})\,
                 \frac{\textstyle ({\alpha}+{\beta}-2)^{{\alpha}+{\beta}-2}}
                 {\textstyle ({\alpha}-1)^{{\alpha}-1}\,({\beta}-1)^{{\beta}-1}}
\end{eqnarray*}

\vspace{0.35cm}

\noindent
{\bf Example 3.8:} Let's generate a sampling of the spatial distributions
of a bounded electron in a Hydrogen atom. In particular, as an example,
those for the principal quantum number $n=3$.
The wave-function is
${\psi}_{nlm}(r,{\theta},{\phi})=R_{nl}(r)Y_{lm}({\theta},{\phi})$ with:
\begin{eqnarray*}
       R_{30}\,{\propto}\,(1-2r/2-2r^2/27)e^{-r/3} \,\,;\hspace{1.cm}
       R_{31}\,{\propto}\,r(1-r/6)e^{-r/3} \hspace{1.cm}{\rm and}\hspace{1.cm}
       R_{32}\,{\propto}\,r^2e^{-r/3}
\end{eqnarray*}
the radial functions of  the 3s, 3p and 3d levels and
\begin{eqnarray*}
       |Y_{10}|^2\,{\propto}\,{\rm cos}^2{\theta}\,\,;\hspace{.3cm}
       |Y_{1{\pm}1}|^2\,{\propto}\,{\rm sin}^2{\theta}\,\,;\hspace{.3cm}
       |Y_{20}|^2\,{\propto}\,(3\,{\rm cos}^2{\theta}\,-\,1)^2
\,\,;\hspace{.3cm}
 |Y_{2{\pm}1}|^2\,{\propto}\,{\rm cos}^2{\theta}{\rm sin}^2{\theta}
\,\,;\hspace{.3cm}
|Y_{2{\pm}2}|^2\,{\propto}\,{\rm sin}^4{\theta}
\end{eqnarray*}
the angular dependence from the spherical harmonics.
Since $d{\mu}=r^2 {\rm sin}{\theta} dr d{\theta} d{\phi}$, the probability
density will be
\begin{eqnarray}
       p(r,{\theta},{\phi}|n,l,m)\,=\,R^2_{nl}(r)\,
       |Y_{lm}|^2\,r^2\,{\rm sin}{\theta}\,=\,
       p_{r}(r|n,l)\,
       p_{\theta}({\theta}|l,m)\,p_{\phi}({\phi})
                                       \nonumber
\end{eqnarray}
so we can sample independently $r,{\theta}$ and ${\phi}$. It is left
as an exercise to explicit a sampling procedure.
Note however that, for the marginal radial density, the mode is at
$r=13,12$ and 9 for $l=0,1$ and 2 and decreases exponentially so
even if the support is $r{\in}[0,{\infty})$ 
it will be a reasonable approximation to
take a covering
$r{\in}[0,r_{max})$ such that $P(r{\geq}r_{max})$ is small enough.
After $n=4000$ samplings 
for the quantum numbers $(n,l,m)=(3,1,0)$, $(3,2,0)$ and $(3,2,{\pm}1)$,
the projections on the planes $\pi_{xy}$, $\pi_{xz}$ and $\pi_{yz}$ are
shown in figure 3.2.

\smill
{\raya}                   
\vspace{1.0cm}       

\begin{figure}[t]
\begin{center}

\mbox{\epsfig{file=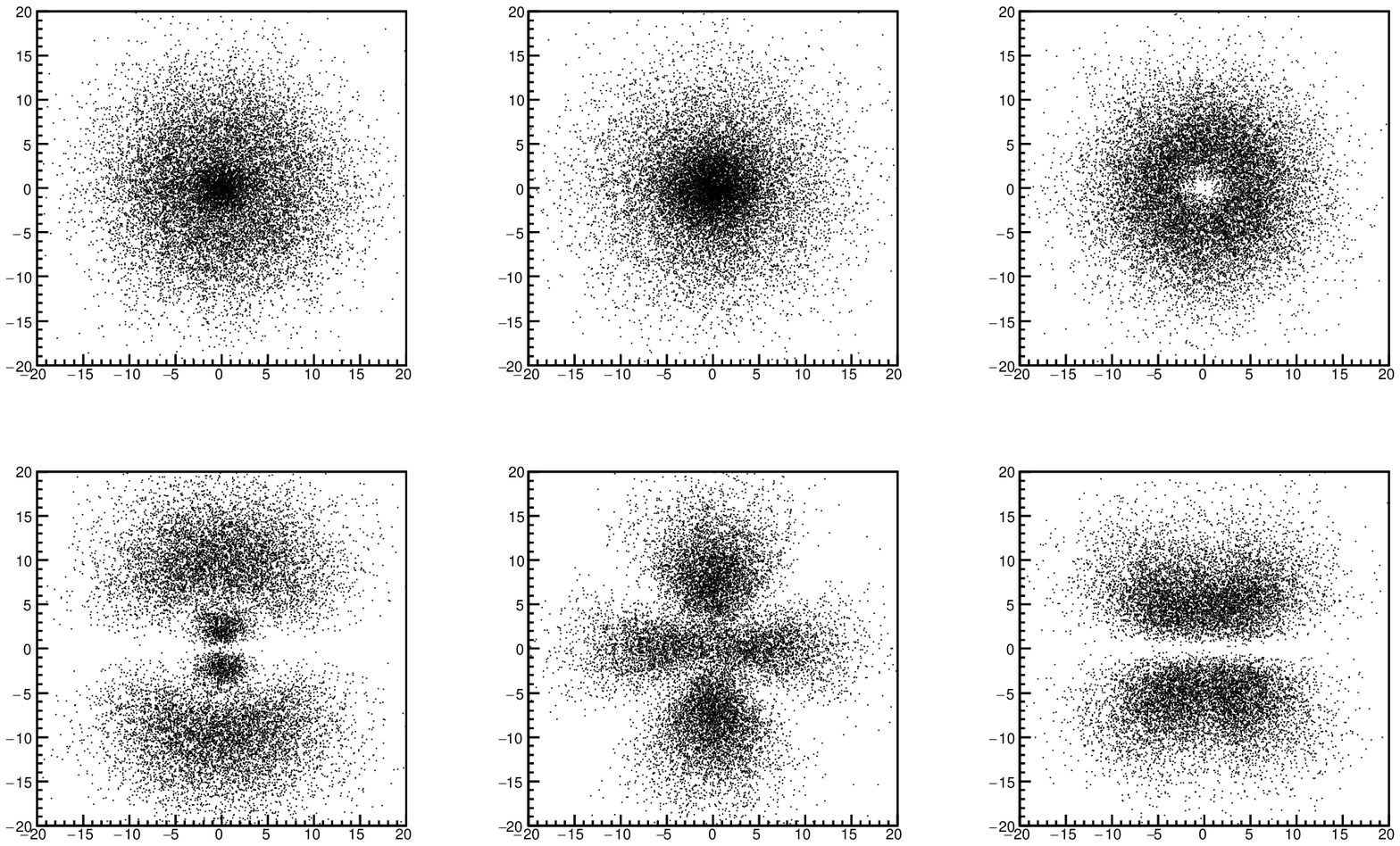,height=9cm,width=13cm}}

\footnotesize
{\bf Figure 3.2}.- Spatial probability distributions of an 
electron in a hydrogen atom corresponding to the quantum states 
$(n,l,m)=(3,1,0)$, $(3,2,0)$ and $(3,2,{\pm}1)$ (columns 1,2,3) and
projections $(x,y)$ and $(x,z)=(y,z)$ (rows 1 and 2) (see example 3.8).

\end{center}
\end{figure}

The generalization of the {\sl Acceptance-Rejection}
method to sample a n-dimensional density 
$\Xbold{\sim}p(\xbold|\thetabold)$ with 
${\rm dim}({\xbold})=n$ is straight forward. Covering with an
n+1 dimensional hypercube:
\begin{itemize}
\item[$i_1$)] Get a sampling $\{x_i^{1)},x_i^{2)},{\ldots},x_i^{n)};y_i\}$ 
              where 
          \begin{eqnarray*}
      \{x_i^{k)}\,{\Leftarrow}\,Un(x_i^{k)}|{\alpha}_k,{\beta}_k)\}_{k=1}^n\,;
       \hspace{0.3cm}y_i\,{\Leftarrow}\,Un(y|0,k)
       \hspace{0.3cm}{\rm and}\hspace{0.3cm}
          k{\geq}{\rm max}_{\xbold}p(\xbold|\thetabold)
          \end{eqnarray*}

\item[$i_2$)] Accept the n-tuple
              $\xbold_i=(x_i^{1)},x_i^{2)},{\ldots},x_i^{n)})$ if
              $ y_i{\leq}p(\xbold_i|\thetabold)$
              or reject it otherwise.
\end{itemize}

\subsubsection{Incorrect estimation of $\mathbf{max}_x\mathbf{\{p(x|\cdot)\}}$}
Usually, we know the support $[\alpha,\beta]$ of the random quantity but 
the pdf is complicated enough to know the maximum.
Then, we start the generation with our best guess for 
${\rm max}_xp(x|\cdot)$, say $k_1$, and after having generated
$N_1$ events ({\sl generated}, not {\sl accepted}) in 
$[{\alpha},{\beta}]{\times}[0,k_1]$,... wham!, we generate a value
$x_m$ such that $p(x_m)>k_1$. Certainly, our estimation of the maximum
was not correct. A possible solution is to forget about what has been
generated and start again with the new maximum $k_2=p(x_m)>k_1$ but,
obviously, this is not desirable among other things because we have no 
guarantee that this is not going to happen again. We better keep what
has been done and proceed in the following manner:
\begin{itemize}
  \item[1)] We have generated $N_1$ pairs $(x_1,x_2)$ in
            $[\alpha,\beta]\times[0,k_1]$ and, in particular, 
            $X_2$ uniformly in $[0,k_1]$. How many additional
            pairs $N_a$ do we have to generate?
            Since the density of pairs is constant in both domains
            $[\alpha,\beta]\times[0,k_1]$ and $[\alpha,\beta]\times[0,k_2]$
            we have that
\begin{eqnarray*}
   \frac{\textstyle N_1}
        {\textstyle ({\beta}-{\alpha})\,k_1}\,=\,
   \frac{\textstyle N_1\,+\,N_a}
        {\textstyle ({\beta}-{\alpha})\,k_2}
     \hspace{1.0cm}{\longrightarrow}\hspace{1.0cm} 
   N_a\,=\,N_1\,\left(
   \frac{\textstyle k_2}
        {\textstyle k_1}\,-\,1\right)
\end{eqnarray*}
  \item[2)] How do we generate them?
    Obviously in the domain $[{\alpha},{\beta}]{\times}[k_1,k_2]$ but from
    the truncated density
    \begin{eqnarray*}
      p_e(x|\cdot)\,=\,(p(x|\cdot)-k_1)\,
      \mbox{\boldmath $1$}_{p(x|\cdot)>k_1}(x)
    \end{eqnarray*}
  \item[3)] Once the $N_a$ additional events have been generated (out of which
            some have been hopefully accepted) we continue with the usual
            procedure but on the domain $[{\alpha},{\beta}]{\times}[0,k_2]$.
\end{itemize}
The whole process is repeated as many times as needed.

\vspace{0.5cm}
\noindent
{\rayan}                   
\vspace{0.35cm}
\footnotesize
\noindent
{\bf NOTE 8: Weighted events.}

The {\sl Acceptance-Rejection} algorithm just explained is equivalent to:
\begin{itemize}
\item[$i_1$)] Sample $x_i$ from $Un(x|{\alpha},{\beta})$ and
          $u_i$ from $Un(u|0,1)$;
\item[$i_2$)] Assign to each generated event $x_i$ a {\sl weight}:
$ w_i=p(x_i|\cdot)/p_m$; $0{\leq}w_i{\leq}1$
and accept the event if $u_i{\leq}w_i$ or reject it otherwise. 
\end{itemize}
It is clear that:
\begin{itemize}
\item[$\bullet$] Events with a higher weight will have a higher chance to be 
                 accepted;
\item[$\bullet$] After applying the {\sl acceptance-rejection} criteria at
                 step $i_2$, all events will have a weight either 1 if it
                 has been accepted or 0 if it was rejected.
\item[$\bullet$] The generation efficiency will be
  \begin{eqnarray*}
            {\rm {\epsilon}}\,=\,
            \frac{\textstyle {\rm accepted\,\,trials}}
                 {\textstyle {\rm total\,\,trials} (N)}\,=\,
     \frac{1}{N}\,\sum_{i=1}^{N}\,w_i\,=\,
                     \overline{w}
                 \end{eqnarray*}
\end{itemize}
In some cases it is interesting to keep all the events, accepted or not.

{\rayan}                   
\vspace{1.0cm}
\small

\noindent
\footnotesize

\noindent
{\bf Example 3.9:} Let's obtain a sampling 
$\{\xbold_1,\xbold_2,{\ldots}\}$, ${\rm dim}(\xbold)=n$, 
of points inside a n-dimensional sphere centered at $\xbold^c$ and
radius $r$. For a direct use of the {\sl Acceptance-Rejection} algorithm
we enclose the sphere in a n-dimensional hypercube   
\begin{eqnarray*}
  C_n\,=\, \prod_{i=1}^n\,[x_i^c-r,x_i^c+r]
\end{eqnarray*}
and:
\begin{itemize}
\item[1)] $x_i\,{\Leftarrow}\,Un(x|x_i^c-r,x_i^c+r)$ for $i=1,\ldots,n$
\item[2)] Accept $\xbold_i$ if ${\rho}_i=||\xbold_i-\xbold^c||\,{\leq}\,r$
          and reject otherwise.
\end{itemize}
The generation efficiency will be
\begin{eqnarray*}
  {\epsilon}(n)\,=\,
 \frac{\textstyle {\rm volume\,\,of\,\,the\,\,sphere}}
      {\textstyle {\rm volume\,\,of\,\,the\,\,covering}}\,=\,
 \frac{\textstyle 2\,{\pi}^{n/2}}
      {\textstyle n\,{\Gamma}(n/2)}\,
 \frac{\textstyle 1}
      {\textstyle 2^n}
\end{eqnarray*}
Note that the sequence $\{\xbold_i/\rho_i\}_{i=1}^n$ will be a sampling of
points uniformly distributed on the sphere of radius $r=1$. This we can
get also as:
\begin{itemize}
\item[1)] $z_i\,{\Leftarrow}\,N(z|0,1)$ for $i=1,\ldots,n$
\item[2)] ${\rho}=||\zbold_i||$ and $\xbold_i={\zbold_i}/\rho$ 
\end{itemize}

\smill
{\raya}                   
\vspace{1.0cm}

Except for simple densities, the efficiency of the {\sl Acceptance-Rejection}
algorithm is not very high and decreases quickly with the number of 
dimensions. For instance, we have seen in the previous example that
covering the n-dimensional sphere with a hypercube has a
generation efficiency
\begin{eqnarray*}
  {\epsilon}(n)\,=\,
 \frac{\textstyle 2\,{\pi}^{n/2}}
      {\textstyle n\,{\Gamma}(n/2)}\,
 \frac{\textstyle 1}
      {\textstyle 2^n}
\end{eqnarray*}
and ${\rm lim}_{n{\rightarrow}{\infty}}{\rm {\epsilon}}(n)=0$. Certainly,
some times we can refine the covering since there is no need other
than simplicity for a hypercube (see {\sl Stratified Sampling}) but, 
in general,
the origin of the problem will remain: when we generate points uniformly 
in whatever domain, we are sampling with constant density regions 
that have a very low probability content or even zero when they have null 
intersection with the support of the random quantity $\Xbold$.
This happens, for instance, when we want to sample from a differential
cross-section that has very sharp peaks (sometimes of several orders 
of magnitude as in the case of bremsstrahlung).
Then, the problem of having a low efficiency is not just
the time expend in the generation but the accuracy and convergence of the
evaluations. We need a more clever way to generate sequences and
the {\sl Importance Sampling} method comes to our help.

\subsection{Importance Sampling}

The {\sl Importance Sampling} generalizes the {\sl Acceptance-Rejection} 
method sampling the density function with higher frequency in regions 
of the domain where the probability of acceptance is larger (more 
{\sl important}). Let's see
the one-dimensional case since the extension to n-dimensions is 
straight forward.

Suppose that we want a sampling of 
$X{\sim}p(x)$ with support 
${\Omega}_X{\in}[a,b]$ and $F(x)$ the corresponding distribution function.
We can always express $p(x)$ as:
\begin{eqnarray*}
  p(x)\,=\,c\,g(x)\,h(x)
\end{eqnarray*}
where:
\begin{itemize}
\item[1)] $h(x)$ is a probability density function, i.e., non-negative
          and normalized in ${\Omega}_X$;
\item[2)] $g(x){\geq}0\,\,\,\,\,;\,\,\,{\forall}x{\in}\Omega_X$
          and has a finite maximum $g_m={\rm max}\{g(x);x{\in}\Omega_X\}$;
\item[3)] $c>0$ a constant normalization factor.
\end{itemize}
Now, consider a 
sampling $\{x_1,x_2,{\ldots},x_n\}$ drawn from the density $h(x)$. If we apply
the {\sl Acceptance-Rejection} criteria with $g(x)$, 
how are the accepted values distributed? It is clear that, if
$g_m={\rm max}(g(x))$ and $Y{\sim}Un(y|0,g_m)$
\begin{eqnarray*}
  P(X{\leq}x|Y{\leq}g(x))\,&=&\,
    \frac{\textstyle \int^{x}_{a}h(x)\,dx\,\int^{g(x)}_{0}dy}
         {\textstyle \int^{b}_{a}h(x)\,dx\,\int^{g(x)}_{0}dy}
         \,=\,
    \frac{\textstyle \int^{x}_{a}h(x)\,g(x)\,dx}
         {\textstyle \int^{b}_{a}h(x)\,g(x)\,dx}
     \,=\,
     F(x)
\end{eqnarray*}
and therefore, from a sampling of $h(x)$ we get a sampling of
$p(x)$ applying the {\sl Acceptance-Rejection} with the function $g(x)$.
There are infinite options for $h(x)$. First, the simpler the better for then
the Distribution Function can be easily inverted and the
{\sl Inverse Transform} applied efficiently. The Uniform Density 
$h(x)=Un(x|a,b)$ is the simplest one but then $g(x)=p(x)$ and this is just the
{\sl Acceptance-Rejection} over $p(x)$. The second consideration is that
$h(x)$ be a fairly good approximation to $p(x)$ so that
$g(x)=p(x)/h(x)$ is as smooth as possible and the {\sl Acceptance-Rejection}
efficient. Thus, if
$h(x)>0$ ${\forall}x{\in}[a,b]$:
\begin{eqnarray}
  p(x)\,dx\,=\,
    \frac{\textstyle p(x)}
         {\textstyle h(x)}\,h(x)\,dx\,=\,g(x)\,dH(x)
                                       \nonumber
\end{eqnarray}

\subsubsection{Stratified Sampling}
The {\sl Stratified Sampling} is a particular case of the {\sl Importance
Sampling} where the density  $p(x)$; $x{\in}\Omega_X$ is approximated by
a simple function over $\Omega_X$. Thus, in the one-dimensional case,
if ${\Omega}=[a,b)$ and we take the partition ({\sl stratification}) 
\begin{eqnarray*}
  \Omega\,=\,\cup_{i=1}^n \Omega_i\,=\,\cup_{i=1}^n [a_{i-1},\,a_i)
  \hspace{1.cm};\hspace{0.3cm}a_0=a\,,
  \hspace{0.3cm}a_{n}=b
\end{eqnarray*}
with measure $\lambda({\Omega}_i)=(a_i-a_{i-1})$, we have
\begin{eqnarray*}
  h(x)\,=\,\sum_{i=1}^n\,
  \frac{\textstyle \mbox{\boldmath $1$}_{[a_{i-1},\,a_i)}(x)}
       {\textstyle \lambda({\Omega}_i)}
  \hspace{1.cm}{\longrightarrow}\hspace{1.0cm}
 \int_{a_0}^{a_n}\,h(x)\,dx\,=\,1
\end{eqnarray*}
Denoting by $p_m(i)\,=\,{\rm max}_x\{p(x)|x{\in}\Omega_i\}$, we have that
for the {\sl Acceptance-Rejection} algorithm the
volume of each sampling domain is
$V_i=\lambda(\Omega_i)\,p_m(i)$. In consequence, for a partition of size $n$,
if $Z{\in}\{1,2,\ldots,n\}$ and define
\begin{eqnarray*}
P(Z=k)\,=\,\frac{\textstyle V_k}{\textstyle \sum_{i=1}^{n}\,V_i}
\hspace{1.0cm};\hspace{1.5cm}
F(k)=P(Z{\leq}k)\,=\,\sum_{j=1}^k\,P(Z=j)
\end{eqnarray*}
we get a sampling of $p(x)$ from the following algorithm:
\begin{itemize}
\item[$i_1$)] $u_i{\Leftarrow}Un(u|0,1)$ and select the partition
              $k\,=\,{\rm Int}[{\rm min}\{F_i\,|\,F_i\,>\,n{\cdot}u_i\}]$;
\item[$i_2$)] $x_i{\Leftarrow}Un(x|a_{k-1},a_k)$,
              $y_i{\Leftarrow}Un(y|0,p_m(k))$ and accept $x_i$ if
              $y_i{\leq}p(x_i)$ (reject otherwise).
\end{itemize}

\subsection{Decomposition of the probability density}
Some times it is possible to express in a simple manner the density function 
as a linear combination of densities; that is
\begin{eqnarray}
  p(x)\,=\,\sum_{j=1}^{k}\,a_j\,p_j(x)\,\,\,\,\,\,\,\,\,\, ; \,\,\,
        a_j\,>\,0\,\,\,\forall{j}\,=\,1,2,{\ldots},k
                                       \nonumber
\end{eqnarray}
that are easier to sample. Since normalization imposes that
\begin{eqnarray}
  \int_{-\infty}^{\infty}\,p(x)\,dx\,=\,\sum_{j=1}^{k}\,a_j\,
  \int_{-\infty}^{\infty}\,p_j(x)\,dx\,=\,
  \sum_{j=1}^{k}\,a_m\,=\,1
                                       \nonumber
\end{eqnarray}
we can sample from $p(x)$ selecting, at each step $i$, one of the 
k densities $p_i(x)$ with probability $p_i=a_i$ 
from which we shall obtain $x_i$ and therefore sampling with
higher frequency from those densities that have a higher relative weight.
Thus:
\begin{itemize}
\item[$i_1$)] Select which density $p_i(x)$ are we going to sample
              at step $i_2$ with probability $p_i=a_i$;
\item[$i_2$)] Get $x_i$ from $p_i(x)$ selected at $i_1$.
\end{itemize}

It may happen that some densities $p_j(x)$ can not be easily integrated so
we do not know a priory the relative weights. If this is the case,
we can sample from $f_j(x)\propto p_j(x)$ and estimate with the generated 
events from $f_i(x)$ the corresponding normalizations $I_i$ with, for instance,
from the sample mean
\begin{eqnarray*}
   I_i\,=\,\frac{1}{n}\sum_{k=1}^n\,f_i(x_k)
\end{eqnarray*}
Then, since $p_i(x)=f_i(x)/I_i$ we have that 
\begin{eqnarray*}
   p(x|\cdot)\,=\,\sum_{i=1}^K\,a_i\,f_i(x|\cdot)\,=\,
                  \sum_{i=1}^K\,a_i\,I_i\,\frac{f_i(x)}{I_i}\,=\,
                  \sum_{i=1}^K\,a_i\,I_i\,p_i(x)
\end{eqnarray*}
so each generated event from $f_i(x)$ has a weight $w_i=a_iI_i$.

\newpage
\noindent
{\raya}                   
\vspace{0.35cm}
\footnotesize

\noindent
{\bf Example 3.10:} Suppose we want to sample from the density
\begin{eqnarray}
   p(x)\,=\,\frac{3}{8}\,(1\,+\,x^2)\,\,\,\,\,\,\,\,\,\, ; \,\,\,
   x\,{\in}\,[-1,1]
                                       \nonumber
\end{eqnarray}
Then, we can take:
\begin{eqnarray}
 \left. 
      \begin{array}{l}
         p_1(x)\,{\propto}\,1 \\
         p_2(x)\,{\propto}\,x^2
      \end{array} 
      \right\}\,\,\,\,\,{\longrightarrow}\,\,\,\,\,
         {\rm normalization}
         \,\,\,\,\,{\longrightarrow}\,\,\,\,\, \left\{
      \begin{array}{l}
         p_1(x)\,=\,1/2 \\
         p_2(x)\,=\,3\,x^2/2
      \end{array} 
       \right.
                                       \nonumber
\end{eqnarray}
so:
\begin{eqnarray}
   p(x)\,=\,\frac{3}{4}\,p_1(x)\,+\,
            \frac{1}{4}\,p_2(x)
                                       \nonumber
\end{eqnarray}
Then:
\begin{itemize}
\item[$i_1$)] Get $u_i$ and $w_i$ as $Un(u|0,1)$;
\item[$i_2$)] Get $x_i$ as:
  \begin{itemize}
            \item[] if $u_i{\leq}3/4$ then $x_i\,=\,2\,w_i\,-\,1$
            \item[] if $u_i>3/4$ then  $x_i\,=\,(2\,w_i\,-\,1)^{1/3}$
          \end{itemize}
\end{itemize}
In this case,  $75\%$ of the times we sample from the trivial density 
$Un(x|-1,1)$.

\smill
{\raya}                   
\vspace{1.0cm}

\section{\LARGE \bf Everything at work}
\subsection{The Compton Scattering}
When problems start to get complicated, we have to combine
several of the aforementioned methods; in this case
{\sl Importance Sampling}, {\sl Acceptance-Rejection} and
{\sl Decomposition} of the probability density.

Compton Scattering is one of the main processes that occur in the interaction 
of photons with matter. When a photon interacts with one of the atomic 
electrons with an energy greater than the binding energy of the electron,
is suffers an inelastic scattering resulting in a photon of less energy
and different direction than the incoming one and an ejected free electron 
from the atom. If we make the simplifying assumptions that
the atomic electron initially at rest and neglect
the binding energy we have that if the incoming photon has an energy
$E_{\gamma}$ its energy after the interaction  $(E^{'}_{\gamma})$ is:
\begin{eqnarray*}
{\epsilon}\,=\,\frac{\textstyle E^{'}_{\gamma}}
                    {\textstyle E_{\gamma}}\,=\,
               \frac{\textstyle 1}
                    {\textstyle 1\,+\,a\,(1-{\rm cos}{\theta})}
\end{eqnarray*}
where ${\theta}{\in}[0,{\pi}]$ is the angle between the 
momentum of the outgoing photon and the incoming one and $a=E_{\gamma}/m_e$. 
It is clear that if the dispersed photon goes in the forward direction
(that is, with ${\theta}=0$), it will
have the maximum possible energy $({\epsilon}=1)$ and when it goes backwards
(that is, ${\theta}={\pi}$) the smallest possible
energy $({\epsilon}=(1+2a)^{-1})$. 
Being a two body final state, given the energy
(or the angle) of the outgoing photon the rest of the kinematic quantities
are determined uniquely:
\begin{eqnarray}
E^{'}_e\,=\,E_{\gamma}\,(1\,+\,
           \frac{\textstyle 1}
                {\textstyle a}\,-\,{\epsilon})
   \,\,\,\,\,\,\,\,\,\,{\rm and}\,\,\,\,\,\,\,\,\,\,
{\rm tan}{\theta}_e\,=\,
           \frac{\textstyle {\rm cot}{\theta}/2}
                {\textstyle 1\,+\,a}
  \nonumber
\end{eqnarray}
The cross-section for the Compton Scattering can be calculated perturbatively 
in Relativistic Quantum Mechanics resulting in the Klein-Nishina expression:
\begin{eqnarray}
           \frac{\textstyle d\,{\sigma}_0}
                {\textstyle d\,x}\,=\,
           \frac{\textstyle 3\,{\sigma}_T}
                {\textstyle 8}\,f(x)
  \nonumber
\end{eqnarray}
where $x={\rm cos}({\theta})$,
${\sigma}_T=0.665 \,{\rm barn}\,=\,0.665\,{\cdot}\,10^{-24}\,{\rm cm}^2$
is the {\sl Thomson cross-section} 
and
\begin{eqnarray}
f(x)\,=\,
           \frac{\textstyle 1}
                {\textstyle [1\,+\,a\,(1-x)]^2}\,
\left(
1\,+\,x^2\,+
           \frac{\textstyle a^2\,(1-x)^2}
                {\textstyle 1\,+\,a\,(1-x)}
\right)
  \nonumber
\end{eqnarray}
has all the angular dependence. Due to the azimuthal symmetry, there
is no explicit dependence with 
${\phi}{\in}[0,2{\pi}]$ and has been integrated out.
Last, integrating this expression for
$x{\in}[-1,1]$ we have the total cross-section of the process:
\begin{eqnarray}
{\sigma}_{0}(E_{\gamma})\,=\,
           \frac{\textstyle {\sigma}_T}
                {\textstyle 4}\,
\left[ 
      \left(
           \frac{\textstyle 1\,+\,a}
                {\textstyle a^2}\,
      \right)
      \left(
           \frac{\textstyle 2\,(1\,+\,a)}
                {\textstyle 1\,+\,2\,a}\,-\,
           \frac{\textstyle {\rm ln}\,(1+2a)}
                {\textstyle a}
      \right)\,+\,
           \frac{\textstyle {\rm ln}\,(1+2a)}
                {\textstyle 2a}\,-\,
           \frac{\textstyle 1\,+\,3a}
                {\textstyle (1\,+\,2\,a)^2}
\right]
  \nonumber
\end{eqnarray}
For a material with $Z$ electrons, the atomic cross-section
can be approximated by ${\sigma}=Z\,{\sigma}_0$ $cm^2$/atom.

\begin{figure}[!t]
\begin{center}

\mbox{\epsfig{file=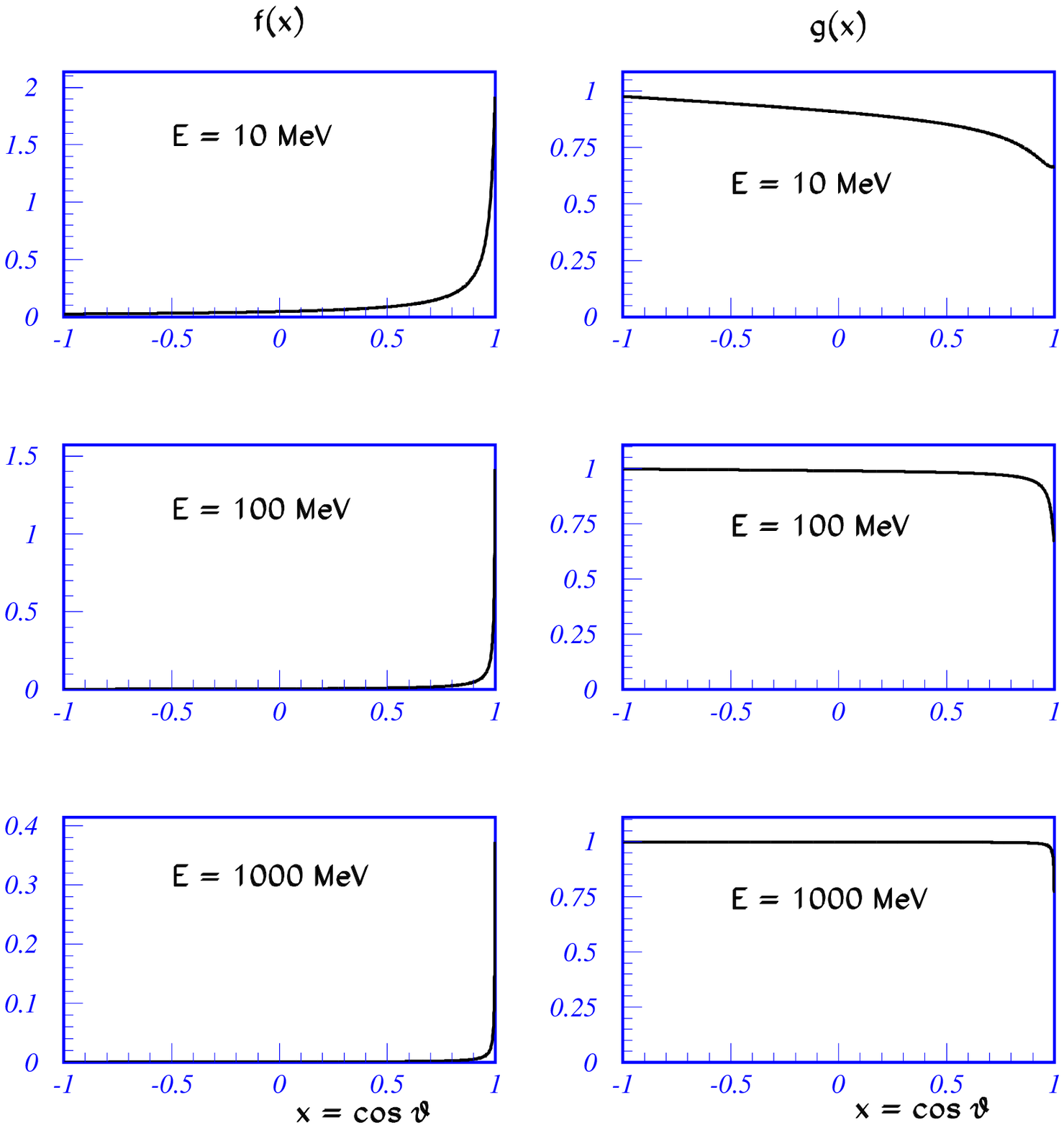,height=13cm,width=13cm}}

\footnotesize
{\bf Figure 3.3}.-
Functions $f(x)$ (left) and $g(x)$ (right) for different values of the
incoming photon energy.

\end{center}
\end{figure}

Let's see how to simulate this process sampling the
angular distribution $p(x){\sim}f(x)$. Figure 3.3 (left) shows this 
function for
incoming photon energies of 10, 100 and 1000 MeV. It is clear that it is
peaked at $x$ values close to 1 and gets sharper with the incoming energy;
that is, when the angle between the incoming and outgoing photon momentum
becomes smaller. In consequence, for high energy photons the 
{\sl Acceptance-Rejection} algorithm becomes very inefficient.
Let's then define the functions 
\begin{eqnarray*}
f_n(x)\,=\,\frac{\textstyle 1}
                {\textstyle [1\,+\,a\,(1-x)]^n}
\end{eqnarray*}
and express $f(x)$ as
$f(x)\,=\,\left(f_1(x)\,+\,f_2(x)\,+\,f_3(x)\right)\,{\cdot}\,g(x)$
where
\begin{eqnarray*}
g(x)\,=\,1\,-\,(2-x^2)\,
           \frac{\textstyle f_1(x)}
                {\textstyle 1\,+\,f_1(x)\,+\,f_2(x)}
\end{eqnarray*}
The functions $f_n(x)$ are easy enough to use the {\sl Inverse Transform}
method and apply afterward the {\sl Acceptance-Rejection} on $g(x)>0$
${\forall}x{\in}[-1,1]$. The shape of this function is shown in fig. 3.3
(right) for different values of the incoming photon energy and clearly is much
more smooth than $f(x)$ so the {\sl Acceptance-Rejection} will be 
significantly more efficient. Normalizing properly the densities
\begin{eqnarray}
p_i(x)\,=\,\frac{\textstyle 1}{\textstyle w_i}\,f_i(x)
\,\,\,\,\,\,\,\,\,\,{\rm such\,\,that}\,\,\,\,\,\,\,\,\,\,
\int_{-1}^{1}\,p_i(x)\,dx\,=\,1
\,\,\,\,\,;\,\,i=1,2,3
  \nonumber
\end{eqnarray}
we have that, with $b=1+2a$:
\begin{eqnarray*}
w_1\,=\,\frac{\textstyle 1}{\textstyle a}\,{\rm ln}b 
\hspace{1.cm}
w_2\,=\,\frac{\textstyle 2}{\textstyle a^2\,(b^2-1)} 
\hspace{1.cm}
w_3\,=\,\frac{\textstyle 2\,b}{\textstyle a^3\,(b^2-1)^2}
\end{eqnarray*}
and therefore
\begin{eqnarray}
f(x)\,&=&\,\left(f_1(x)\,+\,f_2(x)\,+\,f_3(x)\right)\,
{\cdot}\,g(x)\,= \,
     \left(w_1\,p_1(x)\,+\,w_2\,p_2(x)\,+\,w_3\,p_3(x)\right)\,
     {\cdot}\,g(x)\,= \nonumber \\
    &=&\,w_t\,\left({\alpha}_1\,p_1(x)\,+\,{\alpha}_2\,p_2(x)\,+\,
    {\alpha}_3\,p_3(x)\right)\,
         {\cdot}\,g(x) \nonumber \\
  \nonumber
\end{eqnarray}
where $w_t=w_1+w_2+w_3$,
\begin{eqnarray}
{\alpha}_i\,=\,\frac{\textstyle w_i}{\textstyle w_t}\,>\,0
\,\,\,\,\, ; \,\, i=1,2,3
\,\,\,\,\,\,\,\,\,\,{\rm and}\,\,\,\,\,\,\,\,\,\,
\sum_{i=1}^{i=3}\,{\alpha}_i\,=\,1
  \nonumber
\end{eqnarray}
Thus, we set up the following algorithm:

\begin{itemize}
\item[1)] Generate $u{\Leftarrow}Un(u|0,1)$,
          \begin{itemize}
             \item[1.1)] if $u{\leq}{\alpha}_1$ we sample
                         $x_g\,{\sim}\,p_1(x)$;
             \item[1.2)] if ${\alpha}_1<u{\leq}{\alpha}_1+{\alpha}_2$
                         we sample $x_g\,{\sim}\,p_2(x)$ and
             \item[1.3)] if ${\alpha}_1+{\alpha}_2<u$ we sample
                         $x_g\,{\sim}\,p_2(x)$;
          \end{itemize}
\item[2)] Generate $w{\Leftarrow}Un(w|0,g_M)$ where
          \begin{eqnarray}
             g_M\,{\equiv}\,{\rm max}[g(x)]\,=\,g(x=-1)\,=\,
            1\,-\,\frac{\textstyle b}{\textstyle 1\,+\,b\,+\,b^2}
               \nonumber
           \end{eqnarray}
          If $w{\leq}g(x_g)$ we accept $x_g$; otherwise we go back to step 1).
\end{itemize}
Let's see now to to sample from the densities $p_i(x)$. If 
$u{\Leftarrow}Un(u|0,1)$ and
\begin{eqnarray*}
       F_i(x)\,=\,\int_{-1}^{x}\,p_i(s)\,ds\hspace{1.0cm}i=1,2,3
\end{eqnarray*}
then:
\begin{itemize}
\item[$\bullet$] $x\,{\sim}\,p_1(x)$: 
$
       F_1(x)\,=\,1-\frac{\textstyle {\rm ln}(1+a(1-x))}
                         {\textstyle {\rm ln}(b)}
                         \hspace{1.5cm}{\longrightarrow}\hspace{0.5cm}
                         x_g=\frac{\textstyle 1\,+a\,-\,b^u}
                         {\textstyle a}
$
\item[$\bullet$] $x\,{\sim}\,p_2(x)$: 
$        F_2(x)=
        \frac{\textstyle b^2\,-\,1}
             {\textstyle 2\,(b-x)}\,-\,
        \frac{\textstyle 1}
             {\textstyle 2\,a}
             \hspace{2.3cm}{\longrightarrow}\hspace{0.5cm}
             x_g\,=\,b\,-\,\frac{\textstyle a\,(b^2-1)}
                          {\textstyle 1\,+\,2au}
$
\item[$\bullet$] $x\,{\sim}\,p_3(x)$:
$  F_3(x)=
        \frac{\textstyle 1}
             {\textstyle 4a(1+a)}\,\left(
        \frac{\textstyle (b+1)^2}
             {\textstyle (b-x)^2}\,-\,1 \right)
             \hspace{0.5cm}{\longrightarrow}\hspace{0.5cm}
             x_g\,=\,b\,-\,\frac{\textstyle b+1}
                          {\textstyle [1+4a(1+a)u]^{1/2}}$
\end{itemize}
Once we have
$x_g$ we can deduce the remaining quantities of interest from the 
kinematic relations. In particular, the energy of the outcoming photon will be
   \begin{eqnarray}
       {\epsilon}_g\,=\,\frac{\textstyle E^{'}_g}
                             {\textstyle E}\,=\,
                        \frac{\textstyle 1}
                             {\textstyle 1\,+\,a(1-x_g)}
   \nonumber
   \end{eqnarray}
Last, we sample the azimuthal outgoing photon angle as
${\phi}\,{\Leftarrow}\,Un(u|0,2{\pi})$. 

Even though in this example we are going to simulate only the Compton effect,
there are other processes by which the photon
interacts with matter. At low energies (essentially ionization energies:
${\leq}\,E_{\gamma}\,{\leq}\,100\,KeV$) the dominant interaction is the
photoelectric effect
\begin{eqnarray*}
    {\gamma}\,+\,{\rm atom}\,\,{\longrightarrow}\,\,
    {\rm atom}^+\,+\,e^- \,\,\,\,\, ;
\end{eqnarray*}
at intermediate energies
$(E_{\gamma}\,{\sim}\,1-10\,MeV)$ the Compton effect
   \begin{eqnarray}
   {\gamma}\,+\,{\rm atom}\,\,{\longrightarrow}\,\,
   {\gamma}\,+\,e^-\,+\,
    {\rm atom}^+
   \nonumber
   \end{eqnarray}
and at high energies $(E_{\gamma}\,{\geq}\,100\,MeV)$ the dominant one is
pair production
   \begin{eqnarray}
   {\gamma}\,+\,{\rm nucleus}\,\,{\longrightarrow}\,\,e^+\,+\,e^-\,+\,
    {\rm nucleus}
   \nonumber
   \end{eqnarray}
To first approximation, the contribution of other processes is negligible.
Then, at each step in the evolution of the photon along the material we have
to decide first which interaction is going to occur next.
The cross section is a measure of the interaction probability expressed in
${\rm cm}^2$ so, since the total interaction cross section will be in this 
case:
\begin{eqnarray}
  {\sigma}_t\,=\,{\sigma}_{\rm phot.}\,+\,
               {\sigma}_{\rm Compt.}\,+\,
               {\sigma}_{\rm pair}
   \nonumber
\end{eqnarray}
we decide upon the process $i$ that is going to happen next 
with probability $p_i={\sigma}_i/{\sigma}_t$; 
that is, $u{\Leftarrow}Un(0,1)$ and
\begin{itemize}
\item[1)] if $u{\leq}p_{\rm phot.}$ we simulate the
          photoelectric interaction;
\item[2)] if $p_{\rm phot.}<u{\leq}
         (p_{\rm phot.}+p_{\rm Compt.})$:
          we simulate the Compton effect and otherwise
\item[3)] we simulate the pair production
\end{itemize}

Once we have decided which interaction is going to happen next, we have to
decide where. The probability that the photon interacts after traversing
a distance $x$ (cm) in the material is given by
\begin{eqnarray*}
       F_{int}\,=\,1\,-\,e^{-x/{\lambda}}
\end{eqnarray*}
where ${\lambda}$ is the {\sl mean free path}. Being $A$
the atomic mass number of the material, $N_A$ the Avogadro's number,
${\rho}$ the density of the material in ${\rm g}/{\rm cm}^3$, and ${\sigma}$
the cross-section of the process under discussion, we have that
\begin{eqnarray*}
  {\lambda}\,=\,\frac{\textstyle A}
  {\textstyle {\rho}\,N_A\,{\sigma}}
  \,\,\,\,\,[{\rm cm}]
\end{eqnarray*}
Thus, if $u{\Leftarrow}Un(0,1)$, the next interaction is going to happen
at $x=-{\lambda}{\ln}u$ along the direction of the photon momentum.

\begin{figure}[!t]
\begin{center}

\vspace*{-1.0cm}
\mbox{\epsfig{file=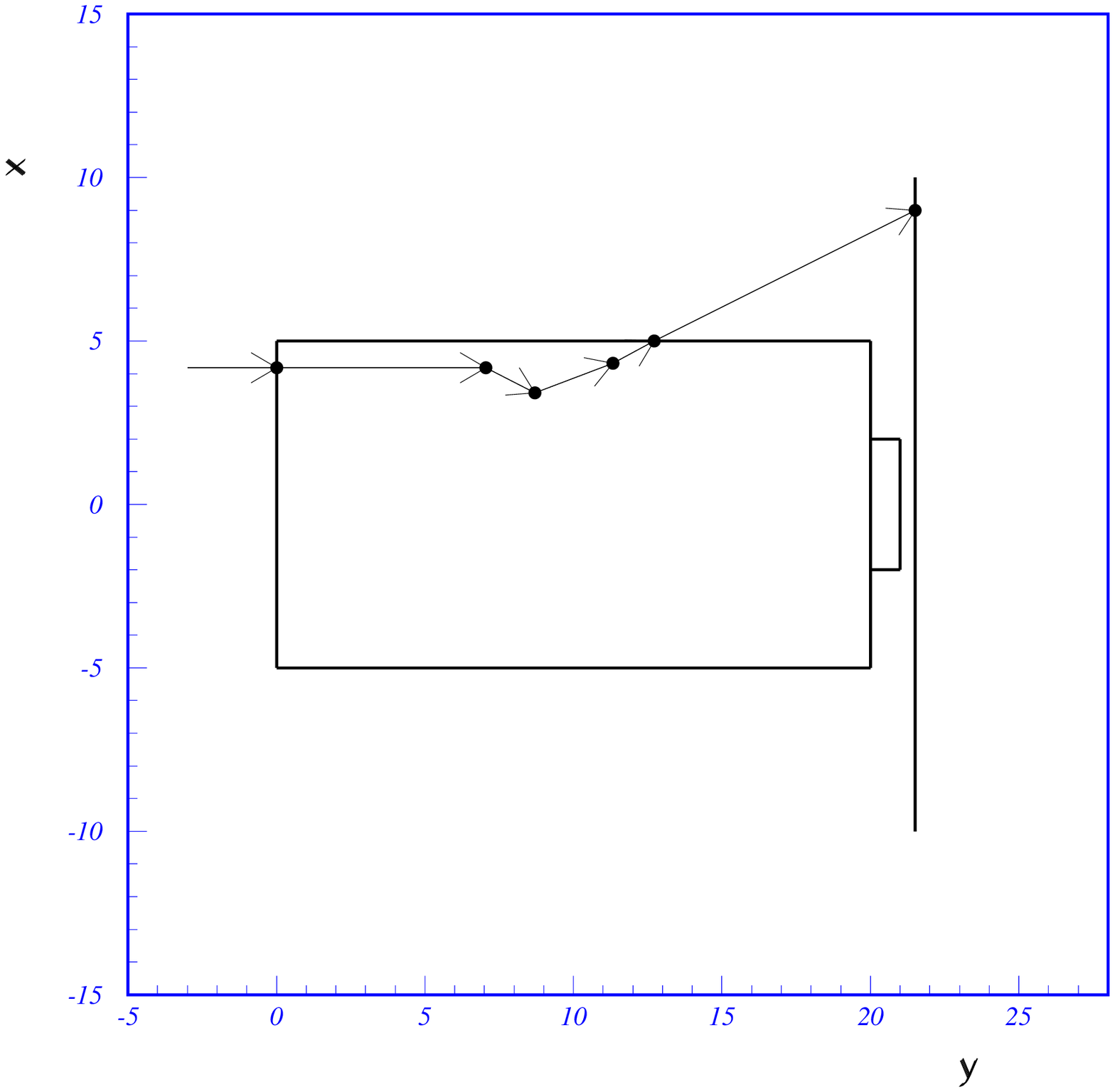,height=9.0cm,width=9.0cm}}
\vspace*{-1.5cm}
\mbox{\epsfig{file=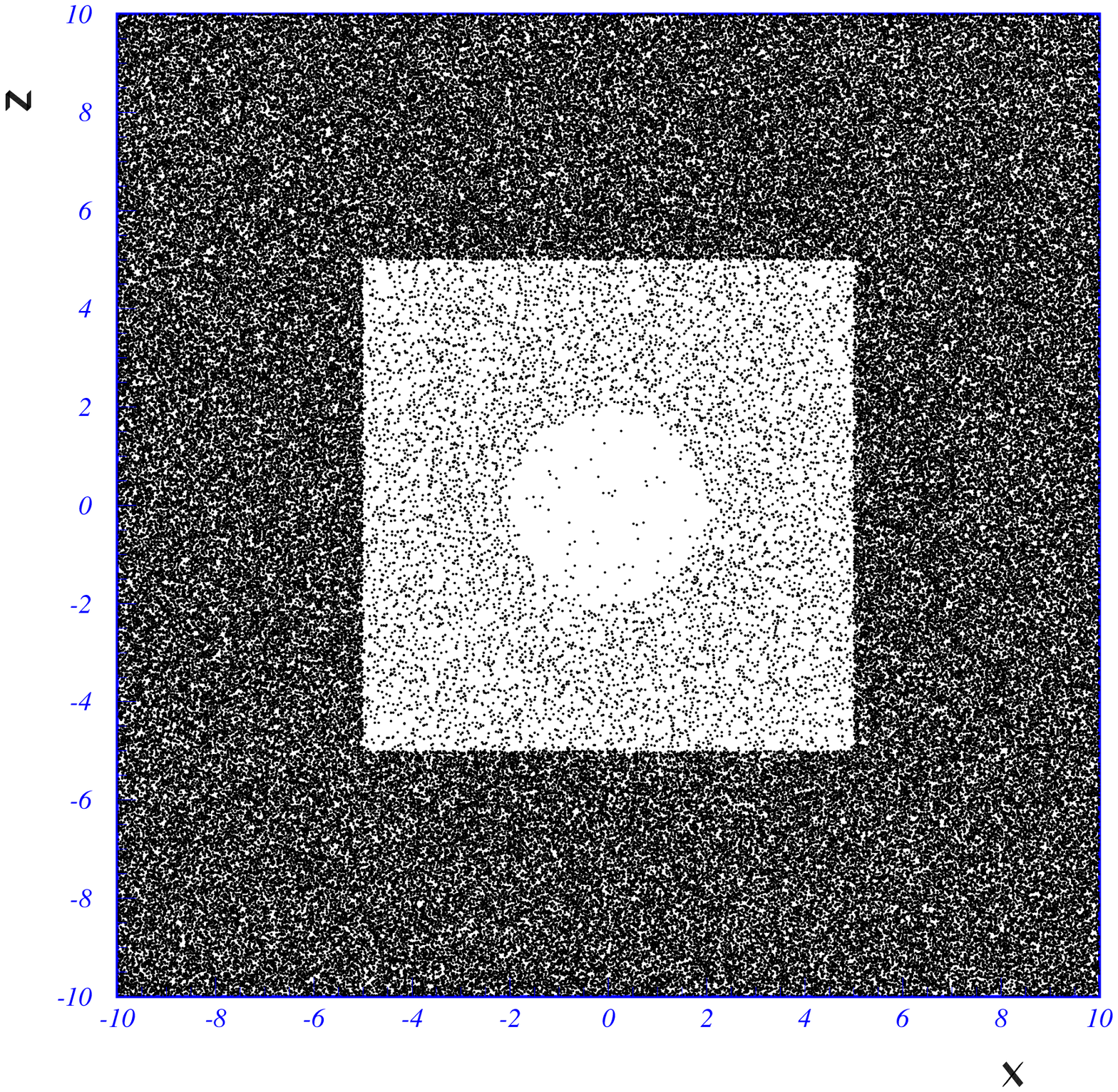,height=9.0cm,width=9.0cm}}
\vspace*{1.0cm}

\footnotesize
{\bf Figure 3.4}.-
The upper figure shows an sketch of the experimental set-up
and the trajectory of one of the simulated photons 
until it is detected on the screen. The lower figure shows the
density of photons collected at the screen for a initially 
generated sample $10^5$ events.

\end{center}
\end{figure}

As an example, we are going to simulate what happens when a beam of photons 
of energy $E_{\gamma}=1$ MeV (X rays) incide normally on the side of a 
rectangular block of carbon
$(Z=6,A=12.01,{\rho}=2.26)$ of $10\,{\times}\,10$ ${\rm cm}^2$ surface
y 20 cm depth. Behind the block, we have hidden an iron coin 
$(Z=26,A=55.85,{\rho}=7.87)$ centered on the surface and in contact with it
of 2 cm radius and 1 cm thickness. Last, at 0.5 cm from the coin there is
a photographic film that collects the incident photons.

The beam of photons is wider than the block of carbon so some of them
will go right the way without interacting and will burn the film.
We have assumed for simplicity that when the photon energy is
below 0.01 MeV, the photoelectric effect is dominant and the ejected
electron will be absorbed in the material. The photon will then be lost
and we shall start with the next one. 
Last, an irrelevant technical issue: the angular variables of the photon 
after the interaction are referred to the direction of the incident photon 
so in each case we have to do the appropriate rotation.

Figure 3.4 (up) shows the sketch of the experimental set-up and the trajectory
of one of the traced photons of the beam collected by the film.
The {\sl radiography} obtained after tracing 100,000 photons is shown in
figure 3.4 (down). 
The black zone corresponds to photons that either go straight to
to the screen or at some point leave the block before getting to the end.
The mid zone are those photons that cross the carbon block and the central
circle, with less interactions, those that cross the carbon block
and afterward the iron coin.

\subsection{An incoming flux of particles}
Suppose we have a detector and we want to simulate a flux of
isotropically distributed incoming particles. It is obvious that
generating them one by one in the space and tracing them backwards 
is extremely inefficient.
Consider a large cubic volume $V$ that encloses the detector, both
centered in the reference 
frame $S_0$. At time $t_0$, we have for particles uniformly 
distributed inside this volume that:
\begin{eqnarray}
  p(\rbold_0)\,d{\mu}_0\,=\,
     \frac{\textstyle 1}{\textstyle V}\,
     dx_0\,dy_0\,dz_0
\nonumber
\end{eqnarray}
Assume now that the velocities are isotropically distributed; that is:
\begin{eqnarray}
  p(\vbold)\,d{\mu}_v\,=\,
     \frac{\textstyle 1}{\textstyle 4{\pi}}\,
     \sin{\theta}\,d{\theta}\,d{\phi}\,f(v)\,dv
\nonumber
\end{eqnarray}
with $f(v)$ properly normalized. Under independence of positions and
velocities at $t_0$, we have that:
\begin{eqnarray}
  p(\rbold_0,\vbold)\,d{\mu}_0\,d{\mu}_v\,=\,
     \frac{\textstyle 1}{\textstyle V}\,
     dx_0\,dy_0\,dz_0\, 
    \frac{\textstyle 1}{\textstyle 4{\pi}}\,
     \sin{\theta}\,d{\theta}\,d{\phi}\,f(v)\,dv
\nonumber
\end{eqnarray}
Given a square of surface $S=(2l)^2$,
parallel to the $(x,y)$ plane, centered at $(0,0,z_c)$
and well inside the volume $V$,
we want to find the probability and distribution of particles that,
coming from the top, cross the surface $S$ in unit time.

For a particle having a coordinate $z_0$ at $t_0=0$, we have that
$z(t)=z_0+v_zt$. The surface $S$ is parallel to the plane $(x,y)$ at
$z=z_c$ so particles will cross this plane at time $t_c=(z_c\,-\,z_0)/v_z$
from above iff:
\begin{itemize}
\item[$0)$] $z_0{\geq}z_c$; obvious for otherwise they are below the
            plane $S$ at $t_0=0$;
\item[$1)$] ${\theta}{\in}[{\pi}/2,{\pi})$; also obvious because if they
             are above $S$ at $t_0=0$ and cut the plane at some $t>0$, 
             the only way is that 
	     $v_z=v\cos{\theta}<0\,\,{\rightarrow}\,\,
             \cos{\theta}<0\,\,{\rightarrow}\,\,
             {\theta}{\in}[{\pi}/2,{\pi})$.
\end{itemize}
But to cross the squared surface $S$ of side $2l$ we also need that
\begin{itemize}
\item[$2)$] $-l\,{\leq}\,x(t_c)=x_0+v_x\,t_c\,{\leq}\,l\,\,\,$ and
    $-l\,{\leq}\,y(t_c)=y_0+v_y\,t_c\,{\leq}\,l$
\end{itemize}
Last, we want particles crossing in unit time; that is $t_c{\in}[0,1]$ so
$0{\leq}t_c=(z_c\,-\,z_0)/v_z{\leq}1$ and therefore
\begin{itemize}
\item[$3)$] $z_0{\in}[z_c,z_c-v\cos{\theta}]$
\end{itemize}
Then, the desired subspace with conditions $1),2),3)$ is 
\begin{eqnarray*}
  {\Omega}_c\,=\,\{
  {\theta}{\in}[{\pi}/2,{\pi});\,
    z_0{\in}[z_c,z_c-v\cos{\theta}];\,
    x_0{\in}[-l-v_xt_c,l-v_xt_c];\,
    y_0{\in}[-l-v_yt_c,l-v_yt_c]
  \}
\end{eqnarray*}
After integration:
\begin{eqnarray}
 \int_{z_c}^{z_c-v\cos{\theta}}dz_0\,
 \int_{-l-v_xt_c}^{l-v_xt_c}dx_0\,
 \int_{-l-v_yt_c}^{l-v_yt_c}dy_0\,=\,-(2l)^2\,v\,\cos{\theta}
     \nonumber
\end{eqnarray}
Thus, we have that 
for the particles crossing the  surface $S=(2l)^2$ from above in unit time
\begin{eqnarray}
  p({\theta},{\phi},v)\,d{\theta}d{\phi}dv\,=\,-\,
     \frac{\textstyle (2l)^2}{\textstyle V}\,
    \frac{\textstyle 1}{\textstyle 4{\pi}}\,
     \sin{\theta}\,\cos{\theta}\,d{\theta}\,d{\phi}\,f(v)\,v\,dv
\nonumber
\end{eqnarray}
with ${\theta}{\in}[{\pi}/2,{\pi})$ and ${\phi}{\in}[0,2{\pi})$. 
If we define the {\sl average velocity}
\begin{eqnarray}
  E[v]\,=\,\int_{\Omega_{v}}\,v\,f(v)\,dv
\nonumber
\end{eqnarray}
the probability to have a cut per unit time is
\begin{eqnarray*}
  P_{cut}(t_c{\leq}1)\,=\,
  \int_{{\Omega}_c{\times}{\Omega}_v}
   p({\theta},{\phi},v)\,d{\theta}d{\phi}dv\,=\,
     \frac{\textstyle S\,E[v]}{\textstyle 4\,V}
\end{eqnarray*}
and the pdf for the angular distribution of velocities (direction of
crossing particles) is 
\begin{eqnarray}
  p({\theta},{\phi})\,d{\theta}d{\phi}\,=\,-
    \frac{\textstyle 1}{\textstyle {\pi}}\,
     \sin{\theta}\,\cos{\theta}\,d{\theta}\,d{\phi}\,=\,
    \frac{\textstyle 1}{\textstyle 2{\pi}}\,d({\cos}^2{\theta})\,d{\phi}
\nonumber
\end{eqnarray}
If we have a density of $n$ particles per unit volume, the expected
number of crossings per unit time
due to the $n_V=n\,V$ particles in the volume is
\begin{eqnarray}
  n_c\,=\,n_V\,P_{cut}(t_c{\leq}1)\,=\,
     \frac{\textstyle n\,E[v]}{\textstyle 4}\,S
\nonumber
\end{eqnarray}
so the {\sl flux}, number of particles crossing the surface from 
one side per unit time and unit surface is
\begin{eqnarray}
  {\Phi}_c^0\,=\,     \frac{\textstyle n_c}{\textstyle S}\,=\,
     \frac{\textstyle n\,E[v]}{\textstyle 4}
\nonumber
\end{eqnarray}

Note that the requirement that the particles cross the square
surface $S$ in a finite time ($t_c{\in}[0,1]$) modifies the
angular distribution of the direction of particles. Instead of
\begin{eqnarray}
  p_1({\theta},{\phi})\,{\propto}\,\sin{\theta}\hspace{1.cm};
 {\theta}{\in}[0,{\pi});\,\,\,{\phi}{\in}[0,2{\pi})
     \nonumber
\end{eqnarray}
we have
\begin{eqnarray}
\hspace{1.5cm}
  p_2({\theta},{\phi})\,{\propto}\,-
     \sin{\theta}\,\cos{\theta}\hspace{1.cm};
{\theta}{\in}[{\pi}/2,{\pi});\,\,\,{\phi}{\in}[0,2{\pi})
\nonumber
\end{eqnarray}
The first one spans a solid angle of
\begin{eqnarray}
 \int_{0}^{\pi}d{\theta}
 \int_{0}^{2\pi}d{\phi}
  p_1({\theta},{\phi})\,=\,4{\pi}
     \nonumber
\end{eqnarray}
while for the second one we have that
\begin{eqnarray}
 \int_{0}^{{\pi}/2}d{\theta}
 \int_{0}^{2\pi}d{\phi}
  p_2({\theta},{\phi})\,=\,{\pi}
     \nonumber
\end{eqnarray}
that is; one fourth the solid angle spanned by the sphere. 
Therefore, the flux expressed as number of particles crossing
from one side of the square surface $S$ per unit time and solid angle
is
\begin{eqnarray*}
  {\Phi}_c\,=\,     \frac{\textstyle {\Phi}_c^0}{\textstyle {\pi}}\,=\,
\frac{\textstyle n_c}{\textstyle {\pi}\,S}\,=\,
     \frac{\textstyle n\,E[v]}{\textstyle 4\,{\pi}}
\end{eqnarray*}
Thus, if we generate a total of $n_T=6n_c$ particles on the of the 
surface of a cube, each face of area $S$, 
with the angular distribution
\begin{eqnarray*}
  p({\theta},{\phi})\,d{\theta}d{\phi}\,=\,
    \frac{\textstyle 1}{\textstyle 2{\pi}}\,d({\cos}^2{\theta})\,d{\phi}
\end{eqnarray*}
for each surface
with ${\theta}{\in}[{\pi}/2,{\pi})$ and ${\phi}{\in}[0,2{\pi})$ 
defined with $\vec{k}$ normal to the surface,
the equivalent {\sl generated flux} per unit time, unit
surface and solid angle is ${\Phi}_T=n_T/6{\pi}S$
and corresponds to a density of 
$n=2n_T/3SE[v]$ particles per unit volume.

\vspace{0.5cm}
\noindent
{\rayan}                   
\vspace{0.35cm}
\footnotesize
\noindent
{\bf NOTE 9: Sampling some continuous distributions of interest.}

These are some procedures to sample from continuous distributions of
interest. There are several algorithms for each case with efficiency 
depending on the parameters but those outlined here have in general
high efficiency. In all cases, it is assumed that
$u{\Leftarrow}Un(0,1)$. 

\vspace*{0.3cm}
\noindent
$\bullet$ {\bf Beta:} $Be(x|{\alpha},{\beta})$;
${\alpha},{\beta}{\in}(0,\infty)$:
\begin{eqnarray*}
  p(x|{\cdot})\,=\,\frac{1}{B({\alpha},{\beta})}
        x^{\alpha-1}(1-x)^{\beta-1}
       \mbox{\boldmath $1$}_{(0,1)}(x)
  \hspace*{0.5cm}{\longrightarrow}\hspace*{0.5cm}
  x=\frac{\textstyle x_1}{\textstyle x_1\,+\,x_2}
\end{eqnarray*}
where $x_1\,{\Leftarrow}\,Ga(x|1/2,{\alpha})$ and
$x_2\,{\Leftarrow}\,Ga(x|1/2,{\beta})$

\vspace*{0.3cm}
\noindent
$\bullet$ {\bf Cauchy:} $Ca(x|{\alpha},{\beta})$;
${\alpha}{\in}{\Rcal};\,{\beta}{\in}(0,\infty)$:
\begin{eqnarray*}
  p(x|{\cdot})\,=\,\frac{\beta/\pi}
        {\left(1+{\beta}^2(x-\alpha)^2\right)}
       \mbox{\boldmath $1$}_{(-\infty,\infty)}(x)
  \hspace*{0.5cm}{\longrightarrow}\hspace*{0.5cm}
 x\,=\,{\alpha}\,+\,{\beta}^{-1}\,
     {\rm tan}({\pi}(u-1/2))
\end{eqnarray*}

\vspace*{0.3cm}
\noindent
$\bullet$ {\bf Chi-squared:} For ${\chi}^2(x|{\nu})$ see $Ga(x|1/2,{\nu}/2)$.

\vspace*{0.3cm}
\noindent
$\bullet$ {\bf Dirichlet} $Di({\xbold}|{\alphabold})$; 
$\rm dim (\xbold,\alphabold)=n$, 
$\alpha_j{\in}(0,\infty)$, $x_j{\in}(0,1)$ and $\sum_{j=1}^{n}x_j=1$
\begin{eqnarray*}
p({\xbold}|{\alphabold})\,=\,
\frac{\textstyle {\Gamma}({\alpha}_1+{\ldots}+{\alpha}_{n})}
{\textstyle {\Gamma}({\alpha}_1){\cdots}{\Gamma}({\alpha}_{n})}
\prod_{j=1}^{n}\,x_j^{{\alpha}_j-1}
       \mbox{\boldmath $1$}_{(0,1)}(x_j)
  \hspace*{0.5cm}{\longrightarrow}\hspace*{0.5cm} \{x_j=z_j/z_0\}_{j=1}^n
\end{eqnarray*}
where $z_j{\Leftarrow}Ga(z|1,\alpha_j)$ and $z_0=\sum_{j=1}^nz_j$.

\vspace*{0.3cm}
\noindent
\,\,\, {\bf Generalized Dirichlet}
$GDi({\xbold}|{\alphabold},{\betabold})$; $\rm dim (\betabold)=n$,
$\beta_j{\in}(0,\infty)$, $\sum_{j=1}^{n-1}x_j<1$
\begin{eqnarray*}
p(x_1,{\ldots},x_{n-1}|{\alphabold},{\betabold})\,=\,
\prod_{i=1}^{n-1}\,
\frac{\textstyle {\Gamma}({\alpha}_i+{\beta}_i)}
{\textstyle {\Gamma}({\alpha}_i){\Gamma}({\beta}_i)}\,
x_i^{{\alpha}_i-1}\left(1-\sum_{k=1}^{i}x_k\right)^{{\gamma}_i}
\end{eqnarray*}
with
\begin{eqnarray*}
    {\gamma}_i\,=\,
             \left\{ \begin{array}{l}
   {\beta}_i-{\alpha}_{i+1}-{\beta}_{i+1} \,\,\,\,\,{\rm for}\,\,
           i=1,2,{\ldots},n-2  \\
   {\beta}_{n-1}-1
             \,\,\,\,\,\,\,\hspace{1.1cm}{\rm for}\,\,i=n-1
                    \end{array}
               \right.
\end{eqnarray*}
When ${\beta}_i={\alpha}_{i+1}+{\beta}_{i+1}$ reduces to the
usual Dirichlet. If
$z_k{\Leftarrow}{\rm Be}(z|{\alpha}_k,{\beta}_k)$ then
$x_k\,=\,z_k(1-{\sum}_{j=1}^{k-1}x_j)$ for $k=1,\ldots ,n-1$ and
$x_n=1-{\sum}_{i=1}^{n-1}x_i$.

\vspace*{0.3cm}
\noindent
$\bullet$ {\bf Exponential:} $Ex(x|{\alpha})$; 
${\alpha}{\in}(0,\infty)$:
\begin{eqnarray*}
   p(x|\cdot)={\alpha}\exp\{-{\alpha}x\}
       \mbox{\boldmath $1$}_{[0,\infty)}(x)
  \hspace*{0.5cm}{\longrightarrow}\hspace*{0.5cm}
 x\,=\,-{\alpha}^{-1}\,{\ln}u
\end{eqnarray*}

\vspace{0.35cm}
\noindent
$\bullet$ {\bf Gamma Distribution $Ga(x|{\alpha},{\beta})$;
 ${\alpha},{\beta}{\in}(0,{\infty})$.}

\noindent
The probability density is
\begin{eqnarray*}
    p(x|\alpha,\beta)\,=\,\frac{\textstyle {\alpha}^{\beta}}
                  {\textstyle {\Gamma}({\beta})}\,
       e^{-\,{\alpha}x}\,x^{{\beta}-1}
       \,\mbox{\boldmath $1$}_{(0,\infty)}(x)
\end{eqnarray*}
Note that $Z={\alpha}X{\sim}Ga(z|1,{\beta})$ so let's see 
how to simulate a sampling of $Ga(z|1,{\beta})$ 
and, if ${\alpha}{\neq}1$, take $x=z/\alpha$.
Depending on the value of the parameter $\beta$ we have that 
\begin{itemize}
\item[$\rhd$] ${\beta}\,=\,1$: This is the Exponential
              distribution $Ex(x|1)$ already discussed;
\item[$\rhd$] ${\beta}\,=\,m\,{\in}{\Ncal}$: As we know, the sum 
              $X_s=X_1+\ldots+X_n$ of n independent random quantities 
              $X_i{\sim}Ga(x_i|\alpha,\beta_i)$; $i=1,{\ldots},n$ 
              is a random quantity distributed as
              $Ga(|x_s|{\alpha},{\beta}_1+ \cdots +{\beta}_n)$. 
              Thus, if we have $m$ independent samplings
              $x_i{\Leftarrow}Ga(x|1,1)=Ex(x|1)$, that is
              \begin{eqnarray*}
               x_1\,=\,-{\rm ln}u_1\,,\,\,\,\,\,{\ldots}\,\,\,\,\, ,
               x_m\,=\,-{\rm ln}u_m
               \nonumber
              \end{eqnarray*}
              with $u_i{\Leftarrow}Un(0,1)$, then
              \begin{eqnarray*}
               x_s\,=\,x_1\,+\,x_2\,{\ldots}\,+\,x_m\,=\,
               -{\rm ln}\,\prod_{i=1}^{m}u_i
              \end{eqnarray*}
              will be a sampling 
              of $Ga(x_s|1,{\beta}=m)$. 
\item[$\rhd$] ${\beta}\,>\,1\,{\in}{\Rcal}$: Defining $m=[{\beta}]$
              we have that ${\beta}=m+{\delta}$ with ${\delta}{\in}[0,1)$. 
              Then, if $u_i{\Leftarrow}Un(0,1);\,i=1,\ldots,m$ and 
              $w{\Leftarrow}Ga(w|1,{\delta})$,
              \begin{eqnarray}
               z\,=\,
               -{\rm ln}\,{\prod}_{i=1}^{m}u_i \,+\, w
               \nonumber
              \end{eqnarray}
              will be a sampling from $Ga(x|1,{\beta})$. The problem is
              reduced to get a sampling 
              $w{\Leftarrow}Ga(w|1,{\delta})$ with ${\delta}{\in}(0,1)$.
\item[$\rhd$] $0<{\beta}\,<\,1$: In this case, for small values of x the 
              density is dominated by $p(x)\,{\sim}\,x^{{\beta}-1}$
              and for large values by $p(x)\,{\sim}\,e^{-x}$. Let's then
              take the approximant
              \begin{eqnarray*}
                g(x)\,=\,
                  x^{{\beta}-1}\,\mbox{\boldmath ${1}$}_{(0,1)}(x)\,+\,
                  e^{-x}\,\mbox{\boldmath ${1}$}_{[1,\infty)}(x)
              \end{eqnarray*}
              Defining
               \begin{eqnarray*}
                p_1(x)\,&=&\,{\beta}x^{\beta-1}\mbox{\boldmath ${1}$}_{(0,1)}(x)
                \hspace{0.75cm}{\longrightarrow}\hspace{0.5cm}
                F_1(x)\,=\,x^{\beta}\\
                p_2(x)\,&=&\,e^{-(x-1)} \mbox{\boldmath ${1}$}_{[1,\infty)}(x)
                  \hspace{0.5cm}{\longrightarrow}\hspace{0.5cm}
                F_2(x)\,=\,1-e^{-(x-1)}
              \end{eqnarray*}
               $w_1=e/(e+\beta)$ and $w_2=\beta/(e+\beta)$ we have that
                  \begin{eqnarray*}
                    g(x)\,=\,w_1\,p_1(x)\,+\,w_2\,p_2(x)
                  \end{eqnarray*}
              and therefore:
             \begin{itemize}
             \item[1)] $u_i{\Leftarrow}Un(0,1)$; $i=1,2,,3$
             \item[2)] If $u_1\,{\leq}\,w_1$,  2et $x\,=\,u_2^{1/{\beta}}$ and 
                           accept $x$ if
                           $u_3{\leq}e^{-x}$; otherwise go to 1);

                       If $u_1\,>\,w_1$,
                       set $x\,=\,1\,-\,{\rm ln}u_2$ and accept $x$ if
                       $u_3{\leq}x^{{\beta}-1}$; otherwise go to 1);
             \end{itemize}
\end{itemize}
The sequence of accepted values will simulate a sampling from
$Ga(x|1,{\beta})$. It is easy to see that the generation efficiency is
\begin{eqnarray}
     {\epsilon}({\beta})\,=\,
     \frac{\textstyle e}
          {\textstyle e\,+\,{\beta}}\,{\Gamma}({\beta}+1)
           \nonumber
\end{eqnarray}
and ${\epsilon}_{min}(\beta\simeq 0.8)\,{\simeq}\,0.72$.

\vspace*{0.3cm}
\noindent
$\bullet$ {\bf Laplace}: $La(x|{\alpha},{\beta})$; $\alpha{\in}\Rcal$,
$\beta{\in}(0,\infty)$:
\begin{eqnarray*}
  p(x|\alpha,\beta)=\frac{1}{2\beta}\,e^{-|x-\alpha|/\beta}
 \mbox{\boldmath $1$}_{(-\infty,\infty)}(x)
 \hspace*{0.4cm}{\longrightarrow}\hspace*{0.4cm}
   x= \left\{
     \begin{array}{l}
             {\alpha}+{\beta}{\rm ln}(2u)
             \hspace*{0.2cm}{\rm if}\hspace*{0.2cm} u<1/2 \\
             \\
             {\alpha}-{\beta}{\rm ln}(2(1-u))
             \hspace*{0.2cm}{\rm if}\hspace*{0.2cm} u{\geq}1/2 
     \end{array}
   \right.
\end{eqnarray*}

\vspace*{0.3cm}
\noindent
$\bullet$ {\bf Logistic:} $Lg(x|{\alpha},{\beta})$;
${\alpha}{\in}{\Rcal};\,{\beta}{\in}(0,\infty)$:
\begin{eqnarray}
  p(x|{\cdot})\,=\,{\beta}\frac{\exp\{-{\beta}(x-{\alpha})\}}
  {(1+\exp\{-{\beta}(x-{\alpha})\})^2}
       \mbox{\boldmath $1$}_{(-\infty,\infty)}(x)
  \hspace*{0.5cm}{\longrightarrow}\hspace*{0.5cm}
 x\,=\,{\alpha}\,+\,{\beta}^{-1}\,
 {\rm ln}\left(
   \frac{\textstyle u}{\textstyle 1\,-\,u} \right)
         \nonumber
\end{eqnarray}

\vspace*{0.3cm}
\noindent
$\bullet$ {\bf Normal Distribution
$N(\mathbf{x}|\mbox{\boldmath ${\mu}$},\mathbf{V})$.}

\noindent
There are several procedures to generate samples from a Normal Distribution.
Let's start with the
one-dimensional case $X{\sim}N(x|{\mu},{\sigma})$ 
considering two independent standardized random quantities 
$X_i{\sim}N(x_i|0,1);\,i=1,2$  
[Bo58]
with joint density 
\begin{eqnarray*}
 p(x_1,x_2)\,=\,\frac{\textstyle 1}{\textstyle 2{\pi}}\,
    e^{- (x_1^2+x_2^2)/2}
\end{eqnarray*}
After the transformation
$X_1=R\,{\rm cos}{\Theta}$ and
$X_2=R\,{\rm sin}{\Theta}$ with
$R{\in}[0,{\infty});\, {\Theta}{\in}[0,2{\pi})$
we have
\begin{eqnarray*}
 p(r,{\theta})\,=\,\frac{\textstyle 1}{\textstyle 2{\pi}}\,
    e^{- r^2/2}\,r
         \nonumber
\end{eqnarray*}
Clearly, both quantities $R$ and $\Theta$ are independent and their 
distribution functions 
\begin{eqnarray*}
 F_{r}(r)\,=\,      1\,-\,e^{- r^2/2} 
\hspace{0.5cm}{\rm and}\hspace{0.5cm}
 F_{\theta}({\theta})\,=\,
     \frac{\textstyle {\theta}}{\textstyle 2{\pi}}
\end{eqnarray*}
are easy to invert so, using then the {\sl Inverse Transform} algorithm:
\begin{itemize}
\item[1)] $u_1{\Leftarrow}Un(0,1)$ and $u_2{\Leftarrow}Un(0,1)$;
\item[2)] $r=\sqrt{-2\,{\rm ln}u_1}$ and ${\theta}=2{\pi}u_2$;
\item[3)] $x_1=r{\rm cos}{\theta}$ and $x_2=r{\rm sin}{\theta}$.
\end{itemize}
Thus, we get two independent samplings 
$x_1$ and $x_2$ from $N(x|0,1)$ and
\begin{itemize}
\item[4)]  $z_1={\mu}_1+{\sigma}_1x_1$ and 
           $z_2={\mu}_2+{\sigma}_2x_2$
\end{itemize}
will be two independent samplings from $N(x|{\mu}_1,{\sigma}_1)$ and
$N(x|{\mu}_2,{\sigma}_2)$.

For the n-dimensional case, $\Xbold{\sim}N(\xbold|\mubold,\Vbold)$ and
$\Vbold$ the covariance matrix, we proceed from the conditional densities
\begin{eqnarray*}
   p(\xbold|\mubold,\Vbold)\,=\,
   p(x_n|x_{n-1},x_{n-2}\ldots,x_1;\cdot)\,
   p(x_{n-1}|x_{n-2}\ldots,x_1;\cdot)\,{\cdots}\,
   p(x_1|\cdot)
\end{eqnarray*}
For high dimensions this is a bit laborious and it is easier if we do first
a bit of algebra.
We know from {\sl Cholesky's Factorization Theorem} that if
${\mathbf{V}}{\in}{\Rcal}^{n{\times}n}$ is a symmetric positive defined matrix
there is a unique lower triangular matrix ${\mathbf{C}}$, with
positive diagonal elements, such that
${\mathbf{V}}={\mathbf{C}}{\mathbf{C}}^T$.
Let then
$\mbox{\boldmath ${Y}$}$ be an n-dimensional random quantity distributed
as $N(\mathbf{y}|{\mathbf{0}},{\mathbf{I}})$ and define a new random
quantity
\begin{eqnarray}
            \mbox{\boldmath ${X}$}\,=\,\mbox{\boldmath ${\mu}$}\,+\,
            {\mathbf{C}}\,\mbox{\boldmath ${Y}$}
      \nonumber
\end{eqnarray}
Then
${\mathbf{V}}^{-1}=[{\mathbf{C}}^{-1}]^T\,{\mathbf{C}}^{-1}$ and
\begin{eqnarray}
    \mbox{\boldmath ${Y}$}^T\mbox{\boldmath ${Y}$}\,=\,
            (\mbox{\boldmath ${X}$}-\mbox{\boldmath ${\mu}$})^T\,
            [{\mathbf{C}}^{-1}]^T\,[{\mathbf{C}}^{-1}]\,
            (\mbox{\boldmath ${X}$}-\mbox{\boldmath ${\mu}$})\,=\,
            (\mbox{\boldmath ${X}$}-\mbox{\boldmath ${\mu}$})^T\,
            [{\mathbf{V}}^{-1}]\,
            (\mbox{\boldmath ${X}$}-\mbox{\boldmath ${\mu}$})
      \nonumber
\end{eqnarray}
After some algebra, the elements if the matrix ${\mathbf{C}}$ can be easily
obtained as 
\begin{eqnarray*}
            {\mathbf{C}}_{i1}\,&=&\,
            \frac{\textstyle {\mathbf{V}}_{i1}}
                 {\textstyle \sqrt{{\mathbf{V}}_{11}} }\,\,\,\,\,\,\,\,\,\,
                 1{\leq}i{\leq}n   \nonumber \\
             {\mathbf{C}}_{ij}\,&=&\,
             \frac{\textstyle
              {\mathbf{V}}_{ij}\,-\,\sum_{k=1}^{j-1}\,C_{ik}C_{jk} }
              {\textstyle C_{jj}}
                  \,\,\,\,\,\,\,\,\,\,
                  1<j<i{\leq}n            \nonumber \\
             {\mathbf{C}}_{ii}\,&=&\,
             \left({\mathbf{V}}_{ii}\,-\,\sum_{k=1}^{i-1}\,C_{ik}^2\right)^{1/2}
                  \,\,\,\,\,\,\,\,\,\,
                  1{<}i{\leq}n 
\end{eqnarray*}
and, being lower triangular, $C_{ij}=0$ ${\forall}j>i$. 
Thus, we have the following algorithm:
\begin{itemize}
\item[1)] Get the matrix ${\mathbf{C}}$ from the covariance 
          matrix ${\mathbf{V}}$;
\item[2)] Get $n$ independent samplings $z_i{\Leftarrow}N(0,1)$ with
          $i=1,\ldots,n$;
\item[3)] Get $x_i={\mu}_i+\sum_{j=1}^{n}{\mathbf{C}}_{ij}z_j$
\end{itemize}
In particular, for a two-dimensional random quantity we have that
 \begin{eqnarray}
    {\mathbf{V}}\,=\,
                    \left(
                          \begin{array}{cc}
                             {\sigma}^2_1 & {\rho}{\sigma}_1{\sigma}_2 \\
                             {\rho}{\sigma}_1{\sigma}_2 & {\sigma}^2_2
                          \end{array}
                    \right)
              \nonumber
 \end{eqnarray}
and therefore:
\begin{eqnarray*}
            {\mathbf{C}}_{11}\,&=&\,
            \frac{\textstyle {\mathbf{V}}_{11}}
                 {\textstyle \sqrt{{\mathbf{V}}_{11}} }\,=\,{\sigma}_1
                           \hspace{0.7cm};\hspace{1.0cm}
            {\mathbf{C}}_{12}\,=\,0    \\                       
            {\mathbf{C}}_{21}\,&=&\,
            \frac{\textstyle  {\mathbf{V}}_{21}}
                 {\textstyle \sqrt{{\mathbf{V}}_{11}} }\,=\,
                 {\rho}\,{\sigma}_2  
               \hspace{0.5cm};\hspace{1.0cm}
            {\mathbf{C}}_{22}\,=\,
            ({\mathbf{V}}_{22}\,-\,{\mathbf{C}}_{21}^2)^{1/2}\,=\,{\sigma}_2\,
            \sqrt{1\,-\,{\rho}^2}
      \nonumber
\end{eqnarray*}
so:
 \begin{eqnarray}
    {\mathbf{C}}\,=\,
                    \left(
                          \begin{array}{cc}
                           {\sigma}_1 & 0  \\
                             {\rho}{\sigma}_2 & {\sigma}_2\sqrt{1\,-\,{\rho}^2}
                          \end{array}
                    \right)
              \nonumber
 \end{eqnarray}
Then, if $z_{1,2}{\Leftarrow}N(z|0,1)$ we have that:
 \begin{eqnarray}
                    \left(
                          \begin{array}{c}
                             x_1  \\
                             x_2
                          \end{array}
                    \right)\,=\,
                    \left(
                          \begin{array}{c}
                             {\mu}_1  \\
                             {\mu}_2
                          \end{array}
                    \right)\,+\,
                    \left(
                \begin{array}{c}
                    z_1\,{\sigma}_1   \\
                  {\sigma}_2(z_1\,{\rho}\,+\,z_2\,\sqrt{1\,-\,{\rho}^2})
                          \end{array}
                    \right)
              \nonumber
 \end{eqnarray}

\vspace*{0.3cm}
\noindent
$\bullet$ {\bf Pareto:} $Pa(x|\alpha,\beta)$;
${\alpha},{\beta}{\in}(0,\infty)$:
\begin{eqnarray*}
   p(x|\cdot)={\alpha}{\beta}^{\alpha}x^{-(\alpha+1)}
       \mbox{\boldmath $1$}_{(\beta,\infty)}(x)
  \hspace*{0.5cm}{\longrightarrow}\hspace*{0.5cm}
 x\,=\,\beta\,u^{-1/\alpha}
\end{eqnarray*}

\vspace*{0.3cm}
\noindent
$\bullet$ {\bf Snedecor:} $Sn(x|{\alpha},{\beta})$; 
${\alpha},{\beta}{\in}(0,\infty)$:
\begin{eqnarray*}
  p(x|{\cdot})\,\propto\,x^{\alpha/2-1}\,
    ({\beta}+{\alpha}x)^{-(\alpha+\beta)/2}
       \mbox{\boldmath $1$}_{(0,\infty)}(x)
  \hspace*{0.5cm}{\longrightarrow}\hspace*{0.5cm}
 x\,=\,\frac{\textstyle x_1/{\alpha}}{\textstyle x_2/{\beta}}
\end{eqnarray*}
where $x_1\,{\Leftarrow}\,Ga(x|1/2,{\alpha}/2)$ and
$x_2\,{\Leftarrow}\,Ga(x|1/2,{\beta}/2)$.

\vspace*{0.3cm}
\noindent
$\bullet$ {\bf Student} $St(x|{\nu})$; ${\nu}{\in}(0,\infty)$:
\begin{eqnarray*}
  p(x|{\cdot})\,\propto\,\frac{1}
                    {\left( 1\,+\,x^2/\nu \right)^{(\nu+1)/2}}    
       \mbox{\boldmath $1$}_{(-\infty,\infty)}(x)
  \hspace*{0.5cm}{\longrightarrow}\hspace*{0.5cm}
 x\,=\,\sqrt{\nu(u_1^{-2/\nu}-1)}\sin(2\pi u_2)
\end{eqnarray*}
where $u_{1,2}\,{\Leftarrow}\,Un(0,1)$.

\vspace*{0.3cm}
\noindent
$\bullet$ {\bf Uniform} $Un(x|a,b)$; $a<b\,{\in}{\Rcal}$
\begin{eqnarray*}
  p(x|\cdot)=(b-a)^{-1}\mbox{\boldmath $1$}_{[a,b]}(x)
  \hspace*{0.5cm}{\longrightarrow}\hspace*{0.5cm}
  x\,=\,(b-1)+a\,u
\end{eqnarray*}

\vspace*{0.3cm}
\noindent
$\bullet$ {\bf Weibull:} $We(x|{\alpha},{\beta})$;
${\alpha},{\beta}{\in}(0,\infty)$:
\begin{eqnarray}
  p(x|{\cdot}={\alpha}{\beta}^{\alpha}x^{\alpha-1}\exp\{-(x/{\beta})^{\alpha} \}
       \mbox{\boldmath $1$}_{(0,\infty)}(x)
  \hspace*{0.5cm}{\longrightarrow}\hspace*{0.5cm}
  x\,=\,{\beta}\,\left(-\,{\rm ln}\,u\right)^{1/{\alpha}}
         \nonumber
\end{eqnarray}

{\rayan}                   
\vspace{1.0cm}
\small


\noindent
\section{\LARGE \bf Markov Chain Monte Carlo}

With the methods we have used up to now we can simulate samples 
from distributions that are more or less easy to handle. 
Markov Chain Monte Carlo allows to sample from more complicated 
distributions. The basic idea is to consider each sampling as a state of 
a system that evolves in consecutive steps of a Markov Chain
converging (asymptotically) to the desired distribution. 
In the simplest version were introduced by Metropolis in the 1950s and
were generalized by Hastings in the 1970s.

Let's start for simplicity with a discrete distribution. Suppose that we
want a sampling of size $n$ from the distribution
\begin{eqnarray*}
    P({X}=k)\,=\,{\pi}_k\,\,\,\,\,{\rm with}
    \,\,\,\,\,\,\,\,\,\,k=1,2,\ldots ,N
\end{eqnarray*}
that is, from the {\sl probability vector}
\begin{eqnarray*}
\pibold\,=\,
   ({\pi}_1,{\pi}_2,{\ldots},{\pi}_N)
\hspace{0.5cm};\hspace{0.5cm} {\pi}_i{\in}[0,1]\,{\forall}i=1,\ldots,N
\hspace{0.5cm}{\rm and}\hspace{0.5cm} \sum_{i=1}^N{\pi}_i=1
\end{eqnarray*}
and assume that it is difficult to generate a sample from this distribution by
other procedures. Then, we may start from a sample of size $n$
generated from a simpler distribution; for instance, a  Discrete Uniform 
with
\begin{eqnarray*}
    P_0({X}=k)\,=\,\frac{\textstyle 1}{\textstyle N}\,\, ;
    \,\,\,\,\,\,\,\,\,\,{\forall}k
\end{eqnarray*}
and from the sample obtained $\{n_1,n_2,{\ldots},n_N\}$, where
   $n\,=\,\sum_{i=1}^{N}\,n_i$,
we form the {\sl initial sample probability vector}
\begin{eqnarray*}
{\mbox{\boldmath ${\pi}$}}^{(0)}\,=\,
   ({\pi}^{(0)}_1,{\pi}^{(0)}_2,{\ldots},
   {\pi}^{(0)}_N)\,=\,
   (n_1/n, n_2/n,{\dots},n_N/n)
\end{eqnarray*}
Once we have the $n$ events distributed in the $N$ classes
of the sample space $\Omega=\{1,2,{\ldots},N\}$ we just have to 
redistribute them according to some criteria in different steps 
so that eventually we
have a sample of size $n$ drawn from the desired 
distribution $P({X}=k)={\pi}_k$.

We can consider the process of redistribution as an evolving system 
such that, if at step $i$ the system is described by the probability vector
$\pibold^{(i)}$, the new state at step $i+1$, described by $\pibold^{(i+1)}$,
depends only on the present state of the system ($i$) and not
on the previous ones; that is, as a Markov Chain.
Thus, we start from the state
$\pibold^{(0)}$ and the aim is to find a 
{\sl Transition Matrix} ${\bf P}$, of dimension $N{\times}N$, such
that $\pibold^{(i+1)}=\pibold^{(i)}{\bf P}$ and allows 
us to reach the desired state $\pibold$.
The matrix ${\bf P}$ is 
\begin{eqnarray*}
  {\bf P}\,=\,
              \left( \begin{array}{cccc}
               P_{11} & P_{12} & \cdots & P_{1N} \\
               P_{21} & P_{22} & \cdots & P_{2N} \\
               \vdots  & \vdots  & \cdots & \vdots    \\
               P_{N1} & P_{N2} & \cdots & P_{NN} \\
                   \end{array}
            \right)
\end{eqnarray*}
where each element $({\bf P})_{ij}=P(i{\rightarrow}j){\in}[0,1]$ represents the 
probability for an event in class $i$ to move to class $j$ in one step. 
Clearly,
at any step in the evolution the probability that an event in class $i$
goes to any other class $j=1,{\ldots},N$ is 1 so
\begin{eqnarray*}
  \sum_{j=1}^{N}({\bf P})_{ij}\,=\,\sum_{j=1}^{N}P(i{\rightarrow}j)\,=\,1
\end{eqnarray*}
and therefore is a {\sl Probability Matrix}.
If the Markov Chain is:
\begin{itemize}
\item[$1)$] {\sl irreducible}; that is, all the states of the system 
  communicate among themselves;
\item[$2)$] {\sl ergodic}; that is, the states are:
  \begin{itemize}
  \item[$2.1)$] {\sl recurrent}: being at one state we shall return to it
                at some point in the evolution with probability 1;
  \item[$2.2)$] {\sl positive}: we shall return to it in a finite number of
                steps in the evolution;
  \item[$2.3)$] {\sl aperiodic}: the system is not trapped in cycles;
  \end{itemize}
\end{itemize}
then there is a {\sl stationary distribution} $\pibold$ such that:
\begin{itemize}
 \item[1)] $\pibold\,=\,\pibold\,{\bf P}$\,\,;
 \item[2)] Starting at any arbitrary state $\pibold^{(0)}$ of the system,
           the sequence
\begin{eqnarray*}
 && \hspace{-3.cm}\pibold^{(0)} \\
 && \hspace{-2.5cm}\pibold^{(1)}=\pibold^{(0)}\,{\bf P} \\
 &&\hspace{-2.0cm}\pibold^{(2)}=\pibold^{(1)}\,{\bf P}=
       \pibold^{(0)}\,{\bf P}^2\\
 &&  \hspace{-1.5cm}{\vdots}\\
 &&  \hspace{-1.0cm}\pibold^{(n)}=
       \pibold^{(0)}\,{\bf P}^n\\
 &&      \hspace{-0.5cm}{\vdots}
\end{eqnarray*}
converges asymptotically to the {\sl fix vector} $\pibold$;
 \item[3)] 
\begin{eqnarray*}
{\rm lim}_{n{\rightarrow}{\infty}}{\bf P}^n\,=\,
              \left( \begin{array}{cccc}
               {\pi}_1 & {\pi}_2 & \cdots & {\pi}_{N} \\
               {\pi}_1 & {\pi}_2 & \cdots & {\pi}_{N} \\
               \vdots  & \vdots  & \cdots & \vdots    \\
               {\pi}_1 & {\pi}_2 & \cdots & {\pi}_{N} \\
                   \end{array}
            \right)
\end{eqnarray*}
\end{itemize}

There are infinite ways to choose the transition matrix ${\bf P}$.
A {\sl sufficient} (although not necessary) condition for this matrix 
to describe a Markov Chain with fixed vector 
$\pibold$ is that the {\sl Detailed Balance} condition is satisfied
(i.e.; a reversible evolution); that is
\begin{eqnarray*}
     {\pi}_i\,({\bf P})_{ij} \,=\,
      {\pi}_j\,({\bf P})_{ji}
\hspace{1.cm}{\Longleftrightarrow}\hspace{1.cm}
{\pi}_i\,P(i{\rightarrow}j)\,=\,
      {\pi}_j\,P(j{\rightarrow}i)
\end{eqnarray*}
It is clear that if this condition is satisfied, then $\pibold$ is a
fixed vector since:
\begin{eqnarray*}
 {\mbox{\boldmath ${\pi}$}}\,{\bf P}\,=\,
\left(
\sum_{i=1}^{N}{\pi}_i\,({\bf P})_{i1},
\sum_{i=1}^{N}{\pi}_i\,({\bf P})_{i2},{\ldots} 
\sum_{i=1}^{N}{\pi}_i\,({\bf P})_{iN}\right)\,=\,
 {\mbox{\boldmath ${\pi}$}}
\end{eqnarray*}
due to the fact that
\begin{eqnarray*}
\sum_{i=1}^{N}{\pi}_i\,({\bf P})_{ik}\,=\,
\sum_{i=1}^{N}{\pi}_k\,({\bf P})_{ki}\,=\,{\pi}_k          
    \,\,\,\,\,{\rm for}                                   
    \,\,\,\,\,\,\,\,\,\,k=1,2,\ldots ,N
\end{eqnarray*}
Imposing the {\sl Detailed Balance} condition,
we have freedom to choose the elements $({\bf P})_{ij}$. 
We can obviously take 
$({\bf P})_{ij}={\pi}_j$ so that it is satisfied trivially 
(${\pi}_i{\pi}_j={\pi}_j{\pi}_i$)
but this means that being at class $i$ we shall select the new possible 
class $j$ with probability 
$P(i{\rightarrow}j)={\pi}_j$ and, therefore, to sample directly the desired
distribution that, in principle, we do not know how to do. 
The basic idea of Markov Chain Monte Carlo simulation is to
take
\begin{eqnarray}
({\bf P})_{ij}\,=\,q(j|i)\,{\cdot}\,a_{ij}
                                       \nonumber
\end{eqnarray}
where

\vspace*{0.5cm}
\begin{tabular}{p{1.5cm}p{9.8cm}}
 $q(j|i)$: & is a probability law to select the possible new class
           $j=1,{\ldots},N$ 
           for an event that is actually in class $i$;
           \\ & \\
 $a_{ij}$: & is the probability to accept the proposed new class
           $j$ for an event that is at $i$ taken such that the 
           {\sl Detailed Balance}  condition is satisfied for the
           desired distribution $\pibold$.

                 \\ & \\
\end{tabular}

Thus, at each step in the evolution, for an event that is in class $i$
we propose a new class $j$ to go
according to the probability $q(j|i)$ and accept the transition
with probability $a_{ij}$. Otherwise, we reject the transition and leave the
event in the class where it was. 
The Metropolis-Hastings [Ha70] algorithm consists in 
taking the acceptance function
\begin{eqnarray}
a_{ij}\,=\,{\rm min}\,\left\{1,\,
\frac{\textstyle {\pi}_j\,q(i|j)}
     {\textstyle {\pi}_i\,q(j|i)} \right\}
                                       \nonumber
\end{eqnarray}
It is clear that this election of $a_{ij}$ satisfies the {\sl Detailed Balance}
condition. Indeed, if ${\pi}_iq(j|i)>{\pi}_jq(i|j)$ we have that:
\begin{eqnarray*}
a_{ij}={\rm min}\,\left\{1,\,
\frac{\textstyle {\pi}_j\,q(i|j)}
     {\textstyle {\pi}_i\,q(j|i)} \right\}=
\frac{\textstyle {\pi}_j\,q(i|j)}
     {\textstyle {\pi}_i\,q(j|i)}
              \hspace{0.5cm}{\rm and}\hspace{0.5cm}
a_{ji}={\rm min}\,\left\{1,\,
\frac{\textstyle {\pi}_i\,q(j|i)}
     {\textstyle {\pi}_j\,q(i|j)} \right\}=1
\end{eqnarray*}
and therefore:
\begin{eqnarray*}
     {\pi}_i\,({\bf P})_{ij} \,&=&\,
     {\pi}_i\,q(j|i)\,a_{ij}\,=\,
     {\pi}_i\,q(j|i)\,
\frac{\textstyle {\pi}_j\,{\cdot}\,q(i|j)}
     {\textstyle {\pi}_i\,{\cdot}\,q(j|i)}\,=\,   \\
   &=&\,  {\pi}_j\,q(i|j)\,=\,
     {\pi}_j\,q(i|j)\,a_{ji}=\,
     {\pi}_j\,({\bf P})_{ji}
\end{eqnarray*}
The same holds if ${\pi}_iq(j|i)<{\pi}_jq(i|j)$ and is trivial if
both sides are equal. 
Clearly, if $q(i|j)={\pi}_i$ then $a_{ij}=1$ so the closer
$q(i|j)$ is to the desired distribution the better.

A particularly simple case is to choose a symmetric probability
$q(j|i)=q(i|j)$ [Me53]
\begin{eqnarray*}
a_{ij}\,=\,{\rm min}\,\left\{1,\,
\frac{\textstyle {\pi}_j}
     {\textstyle {\pi}_i} \right\}
                                       \nonumber
\end{eqnarray*}
In both cases, it is clear that since the acceptance of the proposed class
depends upon the ratio ${\pi}_j/{\pi}_i$, the normalization
of the desired probability is not important. 

The previous expressions are directly applicable in the case
we want to sample an absolute continuous random quantity $X{\sim}\pi(x)$.
If reversibility holds, $p(x'|x){\pi}(x)=p(x|x'){\pi}(x')$ and therefore
\begin{eqnarray*}
\int_X p(x'|x)\,dx'\,=\,1
\hspace{0.5cm}{\longrightarrow}\hspace{0.5cm}
\int_X p(x'|x){\pi}(x)\,dx\,=\,{\pi}(x')\int_X p(x|x')\,dx\,=\,
{\pi}(x')
\end{eqnarray*}
The transition kernel is expressed as
\begin{eqnarray*}
p(x'|x)\,\equiv\,
p(x{\rightarrow}x')\,=\,
q(x'|x)\,{\cdot}\,
a(x{\rightarrow}x')
\end{eqnarray*}
and the acceptance probability given by
\begin{eqnarray*}
a(x{\rightarrow}x')\,=\,{\rm min}\,\left\{1,\,
\frac{\textstyle \pi(x')\,q(x|x')}
     {\textstyle \pi(x)\,q(x'|x)} \right\}
\end{eqnarray*}
Let's see one example.

\vspace{0.5cm}
\noindent
{\raya}                   
\vspace{0.35cm}
\footnotesize

\noindent
{\bf Example 3.11: The Binomial Distribution}. 
Suppose we want a sampling of size $n$
of the random quantity $X{\sim}Bi(x|N,{\theta})$
Since $x=0,1,{\ldots},N$ we have 
$i=1,2,\ldots,N+1$ classes and the desired probability vector, of
dimension $N+1$, is
\begin{eqnarray*}
   {\pibold}\,=\,(p_0,\,p_1,\ldots,p_N)
\hspace{1.cm}{\rm where}\hspace{1.cm}
   p_k\,=\,P(X=k|\cdot)\,=\,
   \left(\begin{array}{c}
     N \\ k
   \end{array}\right)\,{\theta}^k\,(1-\theta)^{N-k}  
\end{eqnarray*}
Let's take for this example $N=10$ (that is, $11$ classes),
$\theta=0.45$ and $n=100,000$.
We start from a sampling of size $n$ from a uniform distribution (fig. 3.5-1).
At each step of the evolution we swap over the $n$ generated events. For
and event that is in bin $j$ we choose a new possible bin to go 
$j=1,\ldots,10$ with uniform probability $q(j|i)$. Suppose that we look
at an event in bin $i=7$ and choose $j$ with equal probability among
the 10 possible bins. If, for instance, $j=2$, then we accept the move
with probability
\begin{eqnarray*}
a_{72}\,=\,a(7{\rightarrow}2)\,{\rm min}\,\left(1,\,
\frac{\textstyle {\pi}_2=p_2}
     {\textstyle {\pi}_7=p_7} \right)\,=\,0.026
\end{eqnarray*}
if, on the other hand, we have $j=6$,
\begin{eqnarray*}
a_{76}\,=\,a(7{\rightarrow}6)\,{\rm min}\,\left(1,\,
\frac{\textstyle {\pi}_6=p_6}
     {\textstyle {\pi}_7=p_7} \right)\,=\,1.
\end{eqnarray*}
so we make the move of the event. After two swaps over all the sample
we have the distribution shown in fig. 3.5-2 
and after 100 swaps that shown in 
fig. 3.5-3, both compared to the desired distribution:
\begin{eqnarray*}
\pibold^{0)}&=&(
0.091,0.090,0.090,0.092,0.091,0.091,0.093,0.089,0.092,0.090,0.092)\\
\pibold^{2)}&=&(0.012,0.048,0.101,0.155,0.181,0.182,0.151,0.100,0.050,0.018,
0.002)\\
\pibold^{100)}&=&(0.002,0.020,0.077,0.167,0.238,0.235,0.159,0.074,0.022,0.004,
0.000)\\
\pibold&=&(0.000,0.021,0.076,0.166,0.238,0.234,0.160,0.075,0.023,0.004,
0.000)
\end{eqnarray*}

The evolution of the 
moments, in this case the mean value and the variance with the number of
steps is shown in fig. 3.6 together with the Kullback-Leibler
logarithmic discrepancy
between each state and the new one defined as
\begin{eqnarray*}
{\delta}_{KL}\{\pibold|{\pibold}^{n)}\}\,=\,
\sum_{k=1}^{10}\,\pi_k^{n)}\,\ln \frac{\pi_k^{n)}}{\pi_k}
\end{eqnarray*}

\begin{figure}[h]
\begin{center}
\mbox{\epsfig{file=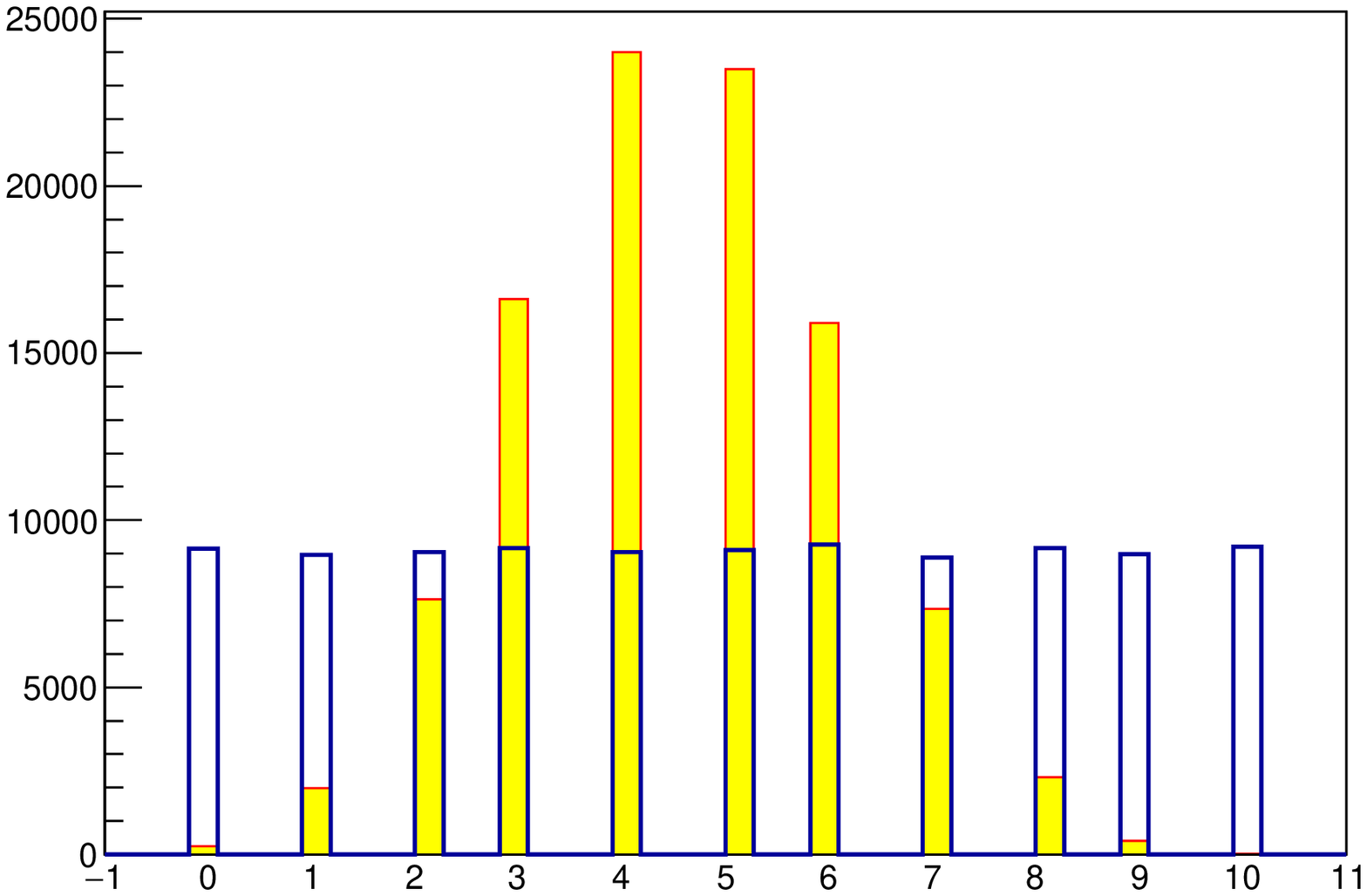,height=6.5cm,width=6.5cm}
      \epsfig{file=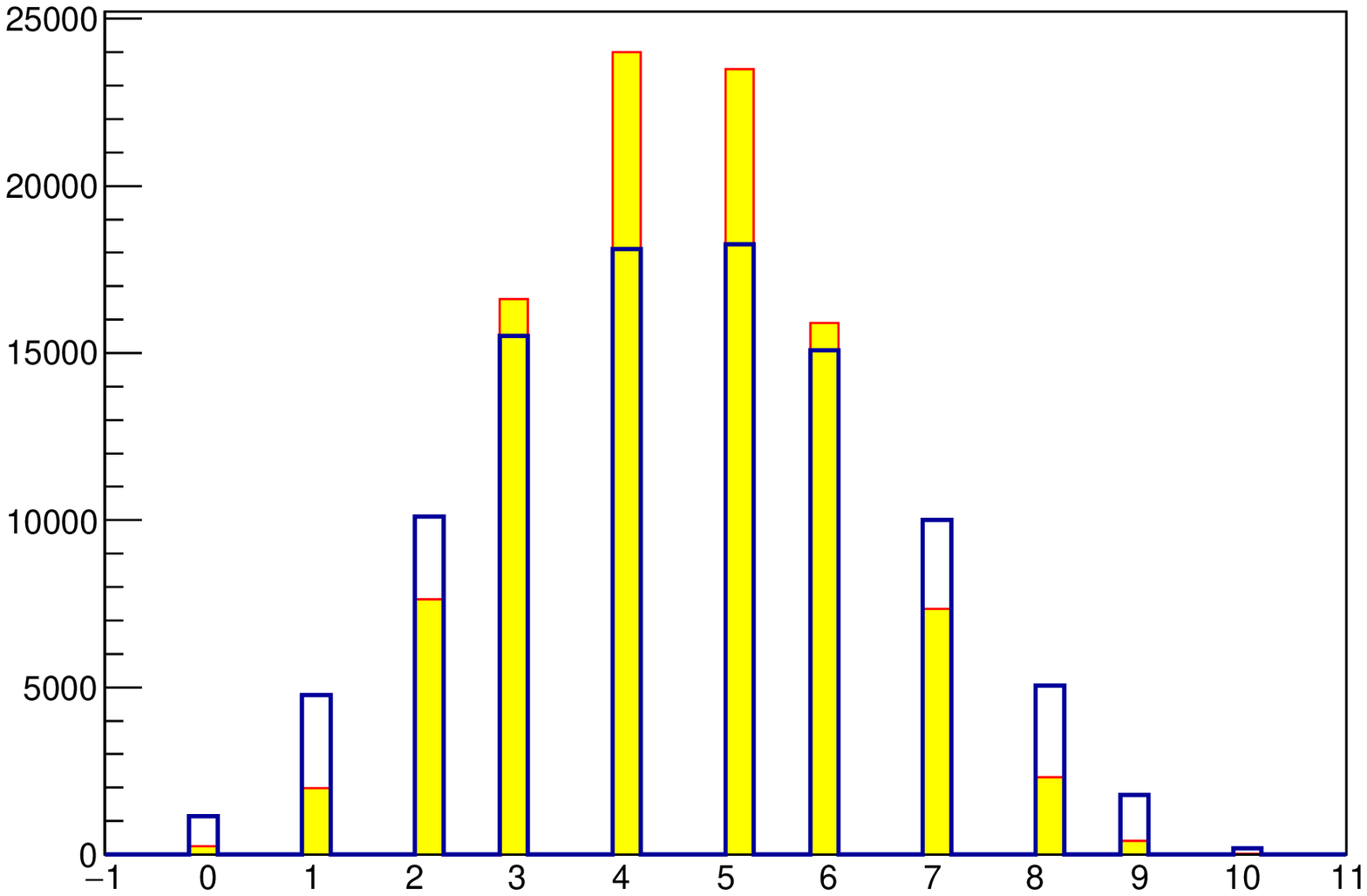,height=6.5cm,width=6.5cm}}
\mbox{\epsfig{file=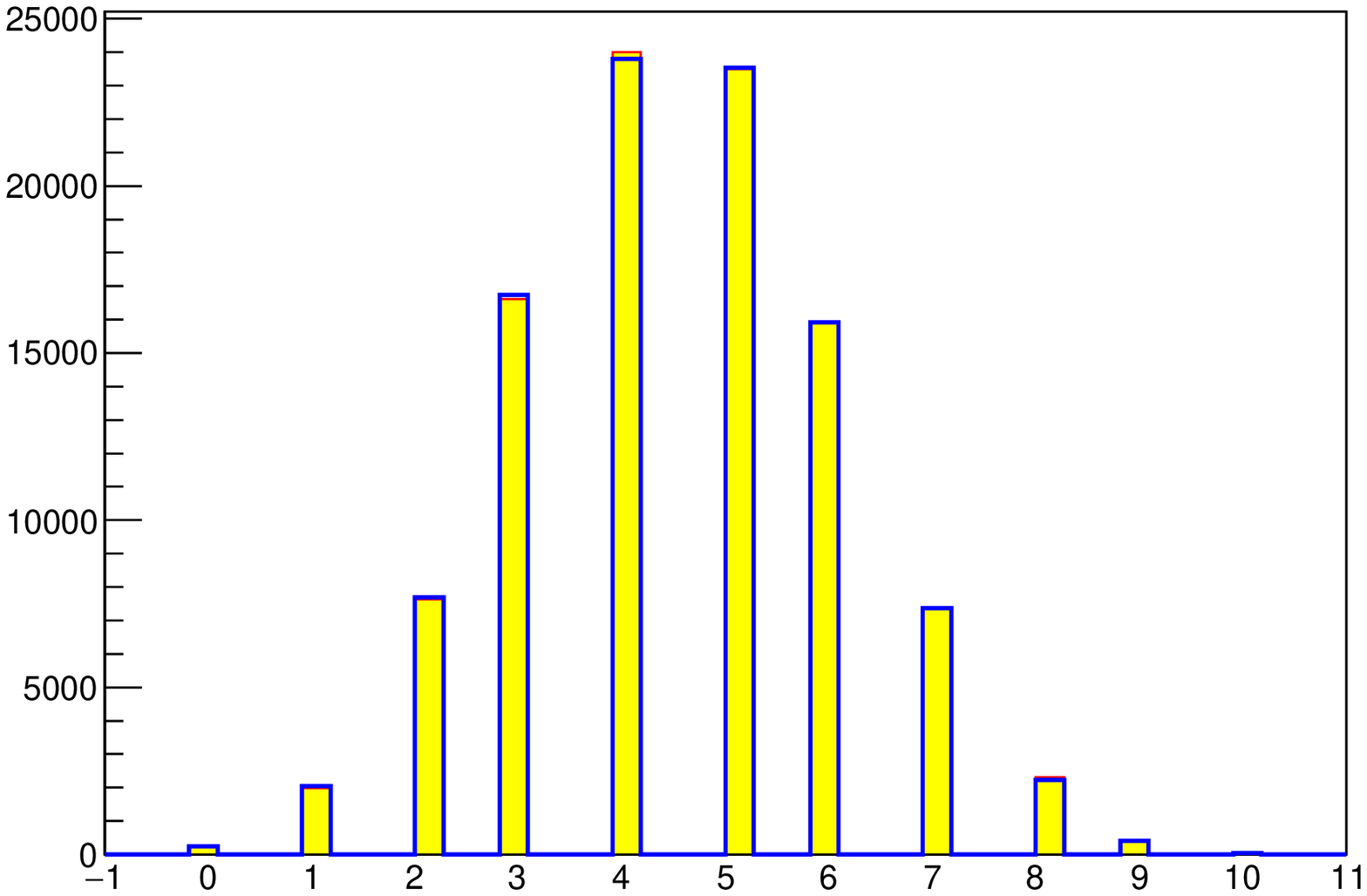,height=6.5cm,width=6.5cm}}

\footnotesize
{\bf Figure 3.5}.-
Distributions at steps 0, 2 and 100 (figs. 3.5-1,2,3; blue)
of the Markov Chain with the desired 
Binomial distribution superimposed in yellow.

\end{center}
\end{figure}
\begin{figure}[h]
\begin{center}

\mbox{\epsfig{file=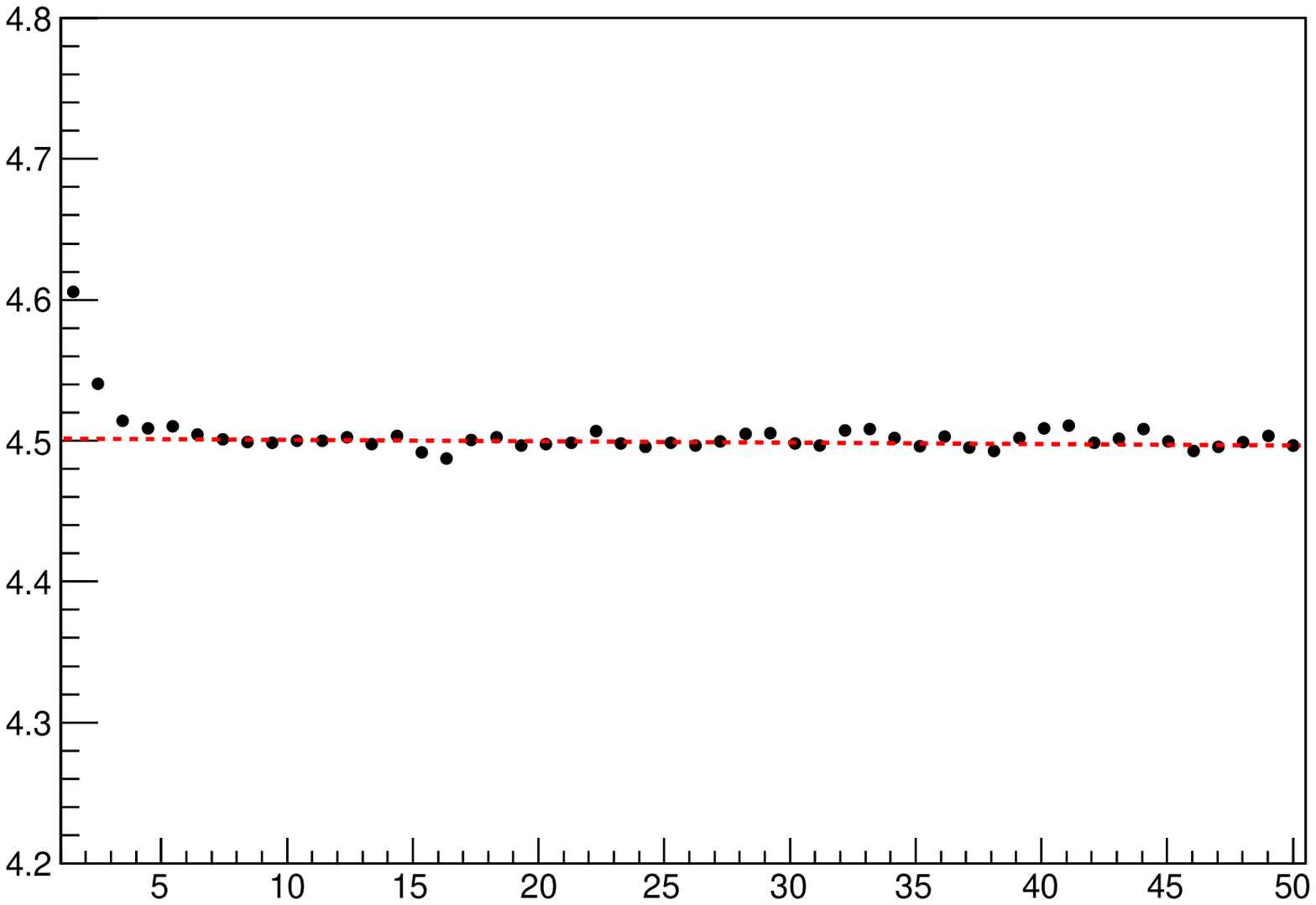,height=6.5cm,width=6.5cm}
      \epsfig{file=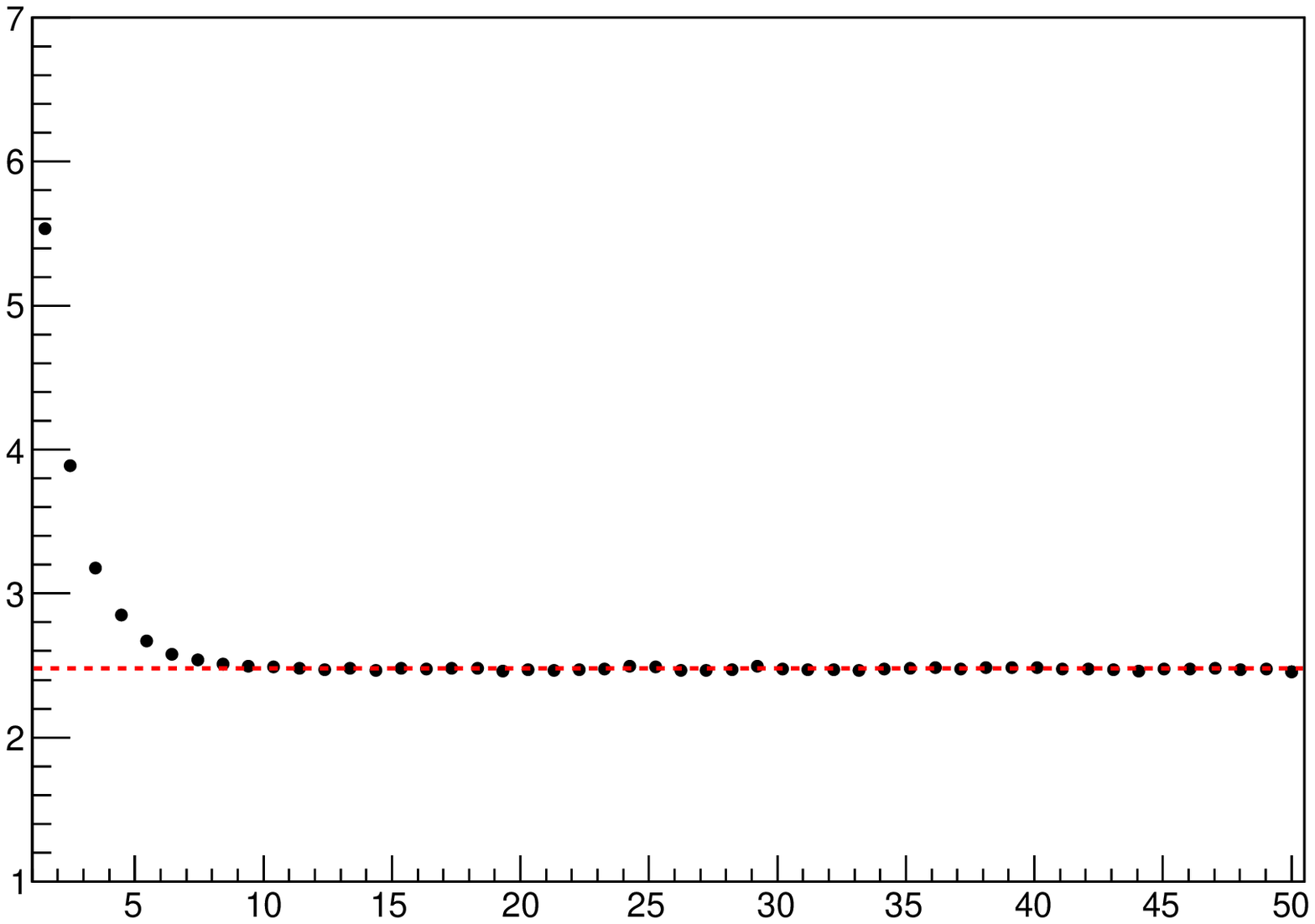,height=6.5cm,width=6.5cm}}
\mbox{\epsfig{file=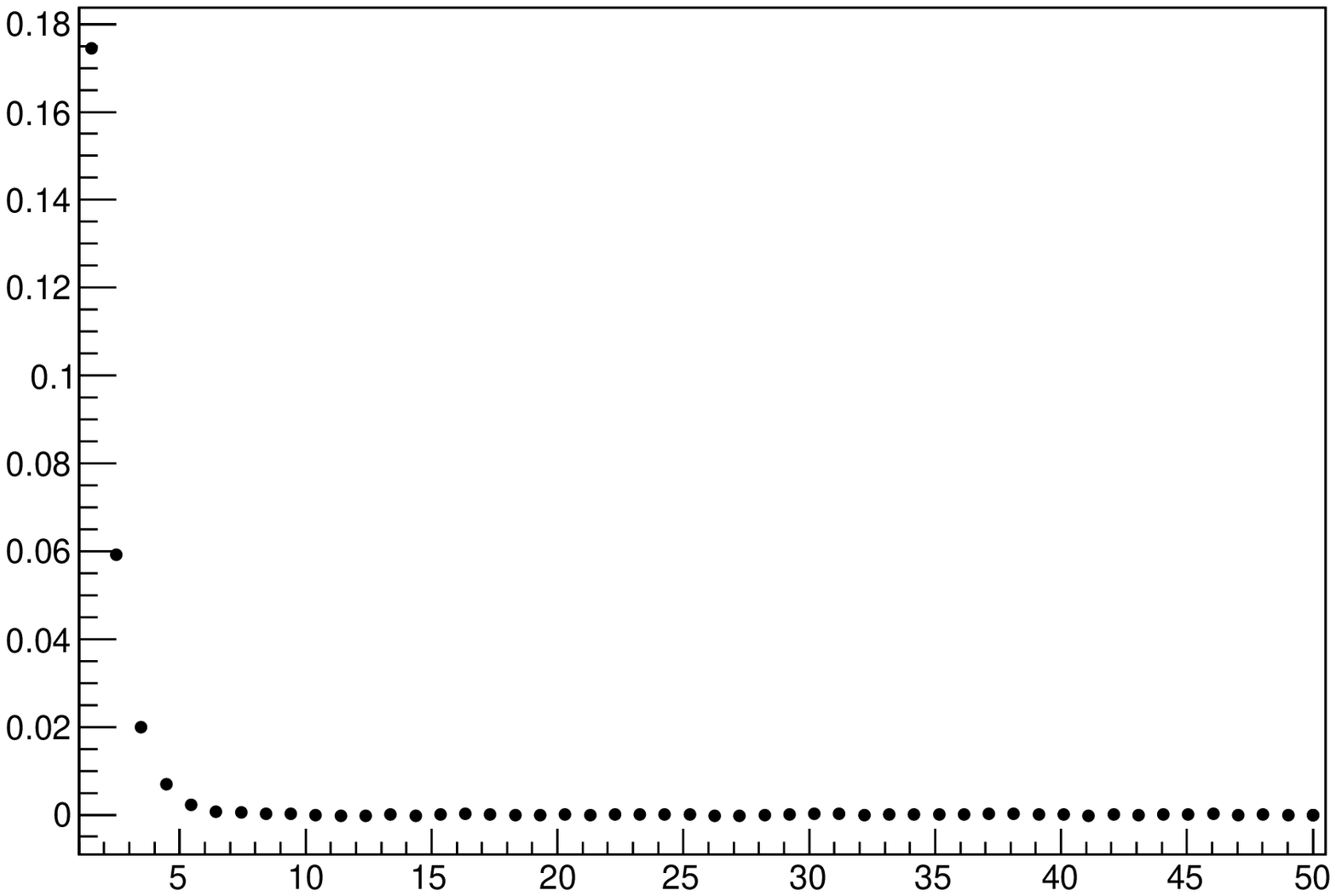,height=6.5cm,width=6.5cm}}

\footnotesize
{\bf Figure 3.6}.-
Distributions of the mean value, variance and logarithmic discrepancy
vs the number of steps. For the first two, the red line indicates what
is expected for the Binomial distribution.

\end{center}
\end{figure}

\smill
{\raya}                   
\vspace{1.0cm}

As the previous example shows, we have to let the system evolve some steps
(i.e.; some initial sweeps for {\sl ``burn-out'' or ``thermalization''}) 
to reach the stable running conditions and get close to the stationary
distribution after starting from an arbitrary state. Once this is achieved,
each step in the evolution will be a sampling from the desired distribution
so we do not have necessarily to generate a sample of the 
desired size to start with. In fact, we usually don't do that;
we choose one admissible state and let the system evolve. 
Thus, for instance if we want a sample of
$X{\sim}p(x|\cdot)$ with $x{\in}{\Omega}_X$, we may start with
a value $x_0{\in}{\Omega}_X$. At a given step $i$ the system will
be in the state $\{x\}$ and at the step $i+1$ the system will be
in a new state $\{x'\}$ if we accept the change $x\rightarrow x'$ or in
the state $\{x\}$ if we do not accept it. After thermalization, each
trial will simulate a sampling of $X{\sim}p(x|\cdot)$.
Obviously, the sequence of states of the system is not independent so, if
correlations are important for the evaluation of the quantities of interest,
it is a common practice to reduce them by taking for the evaluations
one out of few steps.

As for the {\sl thermalization} steps, 
there is no universal criteria to tell whether stable conditions
have been achieved. One may look, for instance, at the evolution of
the discrepancy between the desired probability distribution and the 
probability vector of the state of the system
and at the moments of the
distribution evaluated with a fraction of the last steps. More details about 
that are given in [Ge95].
It is interesting also to look at the acceptance
rate; i.e. the number of accepted new values over the number of trials.
If the rate is low, the proposed new values are rejected with high
probability (are far away from the more likely ones) and 
therefore the chain will mix slowly. 
On the contrary, a high rate indicates that the steps are short,
successive samplings move slowly around the space and therefore the 
convergence is slow. In both cases we should think about tuning the
parameters of the generation.

\begin{figure}[h]
\begin{center}

\mbox{\epsfig{file=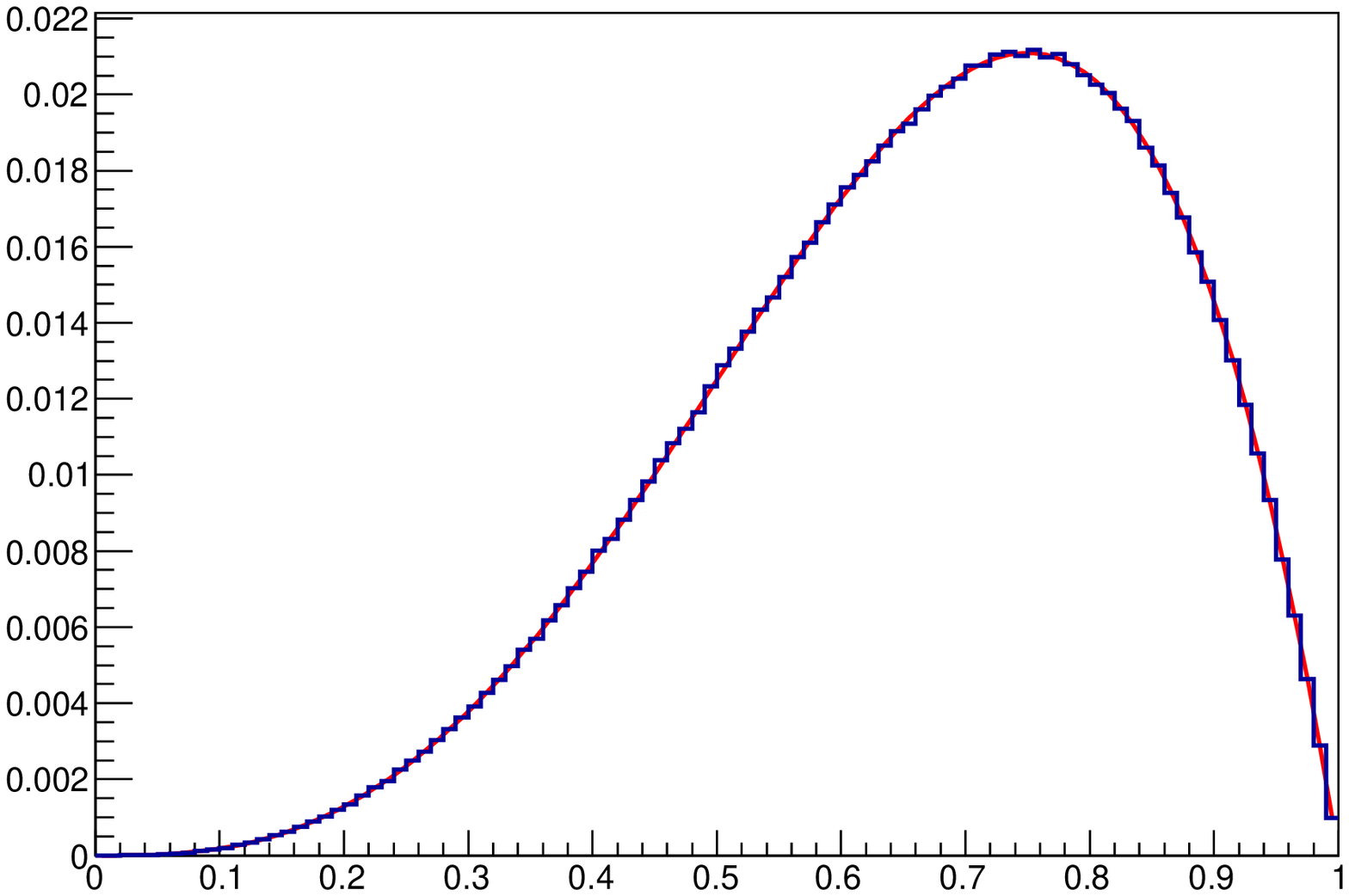,height=8cm,width=8cm}}

\footnotesize
{\bf Figure 3.7}.-
Sampling of the Beta distribution $Be(x|4,2)$ (blue) of the example 3.12 
compared to the desired distribution (continuous line).

\end{center}
\end{figure}
\vspace{0.5cm}
\noindent
{\raya}                   
\vspace{0.35cm}
\footnotesize

\noindent
{\bf Example 3.12: The Beta distribution.}

Let's simulate a sample of size $10^7$ from a Beta distribution
$Be(x|4,2)$; that is:
\begin{eqnarray*}
 {\pi}(x)\,{\propto}\,x^3\,(1-x)
\hspace{0.5cm}{\rm with}\hspace{0.5cm}
x{\in}[0,1]
\end{eqnarray*}
In this case, we start from the admissible state $\{x=0.3\}$ and select a
new possible state $x'$ from the density $q(x'|x)=2x'$; 
not symmetric and independent of $x$.
Thus we generate a new possible state as
 \begin{eqnarray*}
 F_q(x')=\int_0^{x'}q(s|x)ds=\int_0^{x'}2sds=x'^2
    \hspace{0.5cm}{\longrightarrow}\hspace{0.5cm}
    x'=u^{1/2}
\hspace{0.5cm}{\rm with}\hspace{0.5cm}
   u{\Leftarrow}Un(0,1)
\end{eqnarray*}
The acceptance function will then be
\begin{eqnarray*}
a(x{\rightarrow}x')\,=\,{\rm min}\,\left\{1,\,
\frac{\textstyle {\pi}(x')\,{\cdot}\,q(x|x')}
     {\textstyle {\pi}(x)\,{\cdot}\,q(x'|x)} \right\}\,=\,
{\rm min}\,\left\{1,\,
\frac{\textstyle x'^2(1-x')}
     {\textstyle x^2(1-x)}\right\}
\end{eqnarray*}
depending on which we set at the state $i+1$ the system in $x$ or $x'$.
After evolving the system for thermalization, the distribution is shown
in fig. 3.7 where we have taken one $x$ value out of 5 consecutive ones.
The red line shows the desired distribution $Be(x|4,2)$.

\vspace{0.35cm}

\noindent
{\bf Example 3.13: Path Integrals in Quantum Mechanics}.

In Feynman's formulation of non-relativistic Quantum Mechanics, the
probability amplitude to find a particle in $x_f$ at time $t_f$ when at
$t_i$ was at $x_i$ is given by
\begin{eqnarray}
 K(x_f,t_f|x_i,t_i)\,=\,
\int_{paths}\,e^{i/\hatch{h}\,S[x(t)]}\,D[x(t)]
                                       \nonumber
\end{eqnarray}
where the integral is performed over all possible trajectories
$x(t)$ that connect the initial state $x_i=x(t_i)$ with the final state 
$x_f=x(t_f)$,
$S[x(t)]$ is the classic action functional
\begin{eqnarray}
 S[x(t)]\,=\,\int_{t_i}^{t_f}\,L(\dot{x},x,t)\,dt
                                       \nonumber
\end{eqnarray}
that corresponds to each trajectory and
$L(\dot{x},x,t)$ is the Lagrangian of the particle.
All trajectories contribute to the amplitude with the same weight but
different phase. In principle, small differences in the trajectories
cause big changes in the action compared to $\hatch{h}$ and, due to
the oscillatory nature of the phase, their contributions cancel.
However, the action does not change, to first order, for the trajectories
in a neighborhood of the one for which the action is extremal and, since 
they have similar phases (compared to $\hatch{h}$) their contributions will
not cancel. The set of trajectories around the extremal one that produce
changes in the action of the order of $\hatch{h}$ define the limits of
classical mechanics and allow to recover is laws expressed as the
{\sl Extremal Action Principle}.

The transition amplitude ({\sl propagator}) allows to get the
wave-function ${\Psi}(x_f,t_f)$ from ${\Psi}(x_i,t_i)$ as:
\begin{eqnarray}
 {\Psi}(x_f,t_f)\,=\,\int\,
 K(x_f,t_f|x_i,t_i)\,{\Psi}(x_i,t_i)\,dx_i
    \,\,\,\,\,\,\,\,{\rm for}                                   
    \,\,\,\,\,\,\,\,\,\,t_f>t_i
                                       \nonumber
\end{eqnarray}
In non-relativistic Quantum Mechanics there are no trajectories evolving
backwards in time so in the definition of the propagator a Heaviside
step function ${\theta}(t_f-t_i)$ is implicit. Is this clear from this 
equation that $K(x_f,t|x_i,t)={\delta}(x_f-x_i)$.

For a local Lagrangian (additive actions), it holds that:
\begin{eqnarray}
 K(x_f,t_f|x_i,t_i)\,=\,\int\,
 K(x_f,t_f|x,t)\,
 K(x,t|x_i,t_i)\,dx
                                       \nonumber
\end{eqnarray}
analogous expression to the Chapman-Kolmogorov equations that are satisfied
by the conditional probabilities of a Markov process. If the Lagrangian
is not local, the evolution of the system will depend on the intermediate 
states and this equation will not be true.
On the other hand, if the Classical Lagrangian has no explicit time dependence
the propagator admits an expansion
({\sl Feynman-Kac Expansion Theorem}) in terms of a compete set of 
eigenfunctions $\{{\phi}_n\}$ of the Hamiltonian as:
\begin{eqnarray}
 K(x_f,t_f|x_i,t_i)\,=\,\sum_{n}\,
e^{-i/\hatch{h}\,E_n\,(t_f-t_i)}\,{\phi}_n(x_f)\,
                   {\phi}_n^{\star}(x_i)
                                       \nonumber
\end{eqnarray}
where the sum is understood as a sum for discrete eigenvalues and as
an integral for continuous eigenvalues.
Last, remember that expected value of an operator 
$A(x)$ is given by:
\begin{eqnarray}
 <A>\,=\,
\int \,A[x(t)]\,e^{i/\hatch{h}\,S[x(t)]}\,D[x(t)]   \,/\,
\int          \,e^{i/\hatch{h}\,S[x(t)]}\,D[x(t)]
                                       \nonumber
\end{eqnarray}

Let's see how to do the integral over paths to get the propagator in a
one-dimensional problem.
For a particle that follows a trajectory
$x(t)$ between $x_i=x(t_i)$ and $x_f=x(t_f)$ under the action of a potential
$V(x(t))$, the Lagrangian is:
\begin{eqnarray}
L(\dot{x},x,t)\,=\,\frac{\textstyle 1}
                        {\textstyle 2}\,m\,\dot{x}(t)^2\,-\,V(x(t))
                                       \nonumber
\end{eqnarray}
and the corresponding action:
\begin{eqnarray}
S[x(t)]\,=\,\int_{t_i}^{t_f}\,\left(
           \frac{\textstyle 1}
                {\textstyle 2}\,m\,\dot{x}(t)^2\,-\,V(x(t))
            \right)\,dt
                                       \nonumber
\end{eqnarray}
so we have for the propagator:
\begin{eqnarray}
 K(x_f,t_f|x_i,t_i)\,&=&\,
\int_{Tr}\,e^{i/\hatch{h}\,S[x(t)]}\,D[x(t)]\,=\, \nonumber \\
&=&\,\int_{Tr}\,\exp\left\{
   \frac{\textstyle i}
        {\textstyle \hatch{h}}\,
\int_{t_i}^{t_f}\,\left(
           \frac{\textstyle 1}
                {\textstyle 2}\,m\,\dot{x}(t)^2\,-\,V(x(t))
            \right)\,dt
\right\}\,D[x(t)]
                                       \nonumber
\end{eqnarray}
where the integral is performed over the set
$Tr$ of all possible trajectories that start at $x_i=x(t_i)$ and end at
$x_f=x(t_f)$. Following Feynman, a way to perform this integrals is to
make a partition of the interval 
$(t_i,t_f)$ in $N$ subintervals of equal length $\epsilon$ (fig. 3.8); that is,
with
\begin{eqnarray}
{\epsilon}\,=\,
           \frac{\textstyle t_f\,-\,t_i}
                {\textstyle N}\,\,\,\,\,\,\,\,\,\,{\rm so\,\,that}
         \,\,\,\,\,\,\,\,\,\,t_j\,-\,t_{j-1}\,=\,{\epsilon}\;\; ;
         \,\,\,\,\,j=1,2,{\ldots},N
                                       \nonumber
\end{eqnarray}
Thus, if we identify $t_0=t_i$ and $t_N=t_f$ we have that 
\begin{eqnarray*}
[t_i,t_f)\,=\,\cup_{j=0}^{N-1}\,[t_j,t_{j+1})
\end{eqnarray*}
On each interval $[t_j,t_{j+1})$, the possible trajectories 
$x(t)$ are approximated by straight segments so they are defined by the 
sequence
\begin{eqnarray}
\{x_0=x_i=x(t_i),\,x_1=x(t_1),\,x_2=x(t_2),{\ldots}\,
                       x_{N-1}=x(t_{N-1}),\,x_N=x_f=x(t_f)\}
                                       \nonumber
\end{eqnarray}
Obviously, the trajectories so defined are continuous but not differentiable
so we have to redefine the velocity. An appropriate prescription is to
substitute $\dot{x}(t_j)$ by
\begin{eqnarray*}
\dot{x}(t_j)\hspace{0.5cm}{\longrightarrow}\hspace{0.5cm}
           \frac{\textstyle x_j\,-\,x_{j-1}}
                {\textstyle {\epsilon}}
\end{eqnarray*}
so the action is finally expressed as:
\begin{eqnarray*}
S_{N}[x(t)]\,=\,{\epsilon}\,\sum_{j=1}^{N}\,
      \left[\frac{\textstyle 1}{\textstyle 2}\,m\,
           \left(\frac{\textstyle x_j\,-\,x_{j-1}}
                 {\textstyle {\epsilon}}\right)^2
      \,-\,V(x_j) \right]
\end{eqnarray*}

\begin{figure}[!t]
\begin{center}

\mbox{\epsfig{file=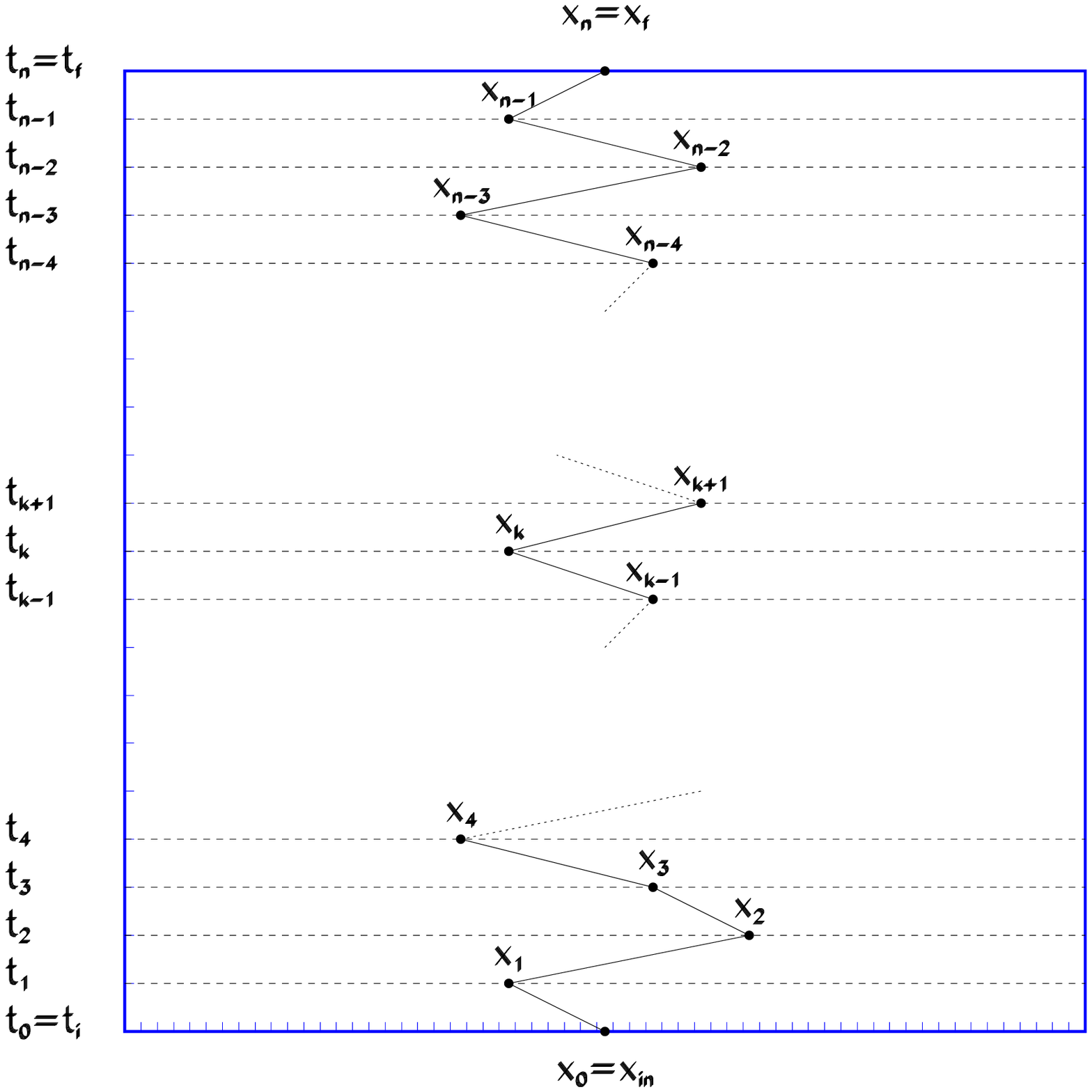,height=11cm,width=11cm}}

\footnotesize
{\bf Figure 3.8}.- Trajectory in a space with discretized time.

\end{center}
\end{figure}

Last, the integral over all possible trajectories that start at
$x_0$ and end at $x_N$ is translated in this space with discretized
time axis as an integral over the quantities
$x_1,x_2,{\ldots},x_{N-1}$ so the differential measure
for the trajectories $D[x(t)]$ is substituted by
\begin{eqnarray}
D[x(t)]\,\,{\longrightarrow}\,\,A_N\,{\prod}_{j=1}^{j=N-1}\,dx_j
                                       \nonumber
\end{eqnarray}
with $A_N$ a normalization factor. The propagator is finally expressed as:
\begin{eqnarray}
K_{N}(x_f,t_f|x_i,t_i)\,= \hspace{7.cm}\nonumber \\
=\, A_N\,
\int_{x_1}dx_1\,{\cdots}\,\int_{x_{N-1}}dx_{N-1}\,\,
\exp\left\{
     \frac{\textstyle i}{\textstyle \hatch{h}}\,
\sum_{j=1}^{N}\,
      \left[\frac{\textstyle 1}{\textstyle 2}\,m\,
           \left(\frac{\textstyle x_j\,-\,x_{j-1}}
                 {\textstyle {\epsilon}}\right)^2
      \,-\,V(x_j) \right]\,{\cdot}\,{\epsilon}
   \right\}
                                       \nonumber
\end{eqnarray}
After doing the integrals, taking the limit
${\epsilon}{\rightarrow}0$ (or $N{\rightarrow}{\infty}$ since the product
$N{\epsilon}=(t_f-t_i)$ is fixed) we get the expression of the propagator.
Last, note that the interpretation of the integral over trajectories
as the limit of a multiple Riemann integral is valid only in Cartesian
coordinates.

To derive the propagator from path integrals is a complex problem
and there are few potentials that can be treated exactly (the {\sl ``simple''}
Coulomb potential for instance was solved in 1982).
The Monte Carlo method allows to attack satisfactorily this type of 
problems but before we have first to convert the complex integral in
a positive real function.
Since the propagator is an analytic function of time, it can be 
extended to the whole complex plane of $t$ 
and then perform a rotation of the time axis
({\sl Wick's rotation}) integrating along
\begin{eqnarray}
 {\tau}\,=\,e^{i{\pi}/2}\,t\,=\,i\,t\,\,\,\,\,;{\rm that\,\,is,}\,\,\,\,\,
  t\,{\longrightarrow}\,-i\,{\tau}
                                       \nonumber
\end{eqnarray}
Taking as prescription the analytical extension over the imaginary
time axis, the oscillatory exponentials are converted to decreasing
exponentials, the results are consistent with those derived by other 
formulations
(Schrodinger or Heisenberg for instance) and it is manifest the
analogy with the partition function of Statistical Mechanics.
Then, the action is expressed as:
\begin{eqnarray}
S[x(t)]\,\,{\longrightarrow}\,\,i\,\int_{{\tau}_i}^{{\tau}_f}\,\left(
           \frac{\textstyle 1}
                {\textstyle 2}\,m\,\dot{x}(t)^2\,+\,V(x(t))
            \right)\,dt
                                       \nonumber
\end{eqnarray}
Note that the integration limits are real as corresponds to integrate
along the imaginary axis and not to just a simple change of variables.
After partitioning the time interval, the propagator is expressed as:
\begin{eqnarray*}
K_{N}(x_f,t_f|x_i,t_i)\,
=\, A_N\,
\int_{x_1}dx_1\,{\cdots}\,\int_{x_{N-1}}dx_{N-1}\,\,
\exp\left\{ -
     \frac{\textstyle 1}{\textstyle \hatch{h}}\,
S_{N}(x_0,x_1,{\ldots},x_N)\,
   \right\}
\end{eqnarray*}
where
\begin{eqnarray}
S_{N}(x_0,x_1,{\ldots},x_N)\,=
\sum_{j=1}^{N}\,
      \left[\frac{\textstyle 1}{\textstyle 2}\,m\,
           \left(\frac{\textstyle x_j\,-\,x_{j-1}}
                 {\textstyle {\epsilon}}\right)^2
      \,+\,V(x_j) \right]\,{\cdot}\,{\epsilon}
                                       \nonumber
\end{eqnarray}
and the expected value of an operator $A(x)$ will be given by:
\begin{eqnarray}
 <A>\,=\, \frac{\textstyle
\int \,
{\prod}_{j=1}^{j=N-1}\,dx_j\,
A(x_0,x_1,{\ldots},x_N)\,
\exp\left\{ -
     \frac{\textstyle 1}{\textstyle \hatch{h}}\,
S_{N}(x_0,x_1,{\ldots},x_N)\,
   \right\} }
            {\textstyle
\int \,
{\prod}_{j=1}^{j=N-1}\,dx_j\,
\exp\left\{ -
     \frac{\textstyle 1}{\textstyle \hatch{h}}\,
S_{N}(x_0,x_1,{\ldots},x_N)
   \right\} }
                                       \nonumber
\end{eqnarray}
Our goal is to generate $N_{gen}$ trajectories with the
Metropolis criteria according to
\begin{eqnarray}
p(x_0,x_1,{\ldots},x_N)\,{\propto}\,
\exp\left\{ -
     \frac{\textstyle 1}{\textstyle \hatch{h}}\,
S_{N}(x_0,x_1,{\ldots},x_N)\,
   \right\} 
                                       \nonumber
\end{eqnarray}
Then, over these trajectories we shall evaluate the expected value of
the operators of interest
$A(x)$
\begin{eqnarray}
 <A>\,=\, \frac{\textstyle 1}{\textstyle N_{gen}}\,
 \sum_{k=1}^{N_{gen}}\,
A(x_0,x^{(k)}_1,{\ldots},x^{(k)}_{N-1},x_N)
                                       \nonumber
\end{eqnarray}
Last, note that if we take $({\tau}_f,x_f)=({\tau},x)$ and
$({\tau}_i,x_i)=(0,x)$ in the Feynman-Kac expansion
we have that
\begin{eqnarray}
 K(x,{\tau}|x,0)\,=\,\sum_{n}\,
e^{-1/\hatch{h}\,E_n\,{\tau}}\,{\phi}_n(x)\,
                   {\phi}_n^{\star}(x)
                                       \nonumber
\end{eqnarray}
and therefore, for sufficiently large times
\begin{eqnarray}
 K(x,{\tau}|x,0)\,{\approx}\,
e^{-1/\hatch{h}\,E_0\,{\tau}}\,{\phi}_0(x)\,
                   {\phi}_0^{\star}(x)\,+\,{\cdots}
                                       \nonumber
\end{eqnarray}
so basically only the fundamental state will contribute.

Well, now we have everything we need. Let's apply all that 
first to an harmonic potential
\begin{eqnarray}
 V(x)\,=\,
     \frac{\textstyle 1}{\textstyle 2}\,k\,x^2
                                       \nonumber
\end{eqnarray}
so the discretized action will be:
\begin{eqnarray}
S_{N}(x_0,x_1,{\ldots},x_N)\,=
\sum_{j=1}^{N}\,
      \left[\frac{\textstyle 1}{\textstyle 2}\,m\,
           \left(\frac{\textstyle x_j\,-\,x_{j-1}}
                 {\textstyle {\epsilon}}\right)^2
      \,+\,
     \frac{\textstyle 1}{\textstyle 2}\,k\,x_j^2
       \right]\,{\cdot}\,{\epsilon}
                                       \nonumber
\end{eqnarray}
To estimate the energy of the fundamental state we use the 
{\sl Virial Theorem}. Since:
\begin{eqnarray}
 <T>_{\Psi}\,=\,\frac{\textstyle 1}{\textstyle 2}\,
<\vec{x}{\cdot}\vec{\nabla}V(\vec{x})>_{\Psi}
                                       \nonumber
\end{eqnarray}
we have that $<T>_{\Psi}=<V>_{\Psi}$ and therefore
\begin{eqnarray}
 <E>_{\Psi}\,=\,<T>_{\Psi}\,+\,<V>_{\Psi}\,=\,k\,<x^2>_{\Psi}
                                       \nonumber
\end{eqnarray}
In this example we shall take $m=k=1$.

\begin{figure}[!t]
\begin{center}

\vspace*{-1.0cm}
\mbox{\epsfig{file=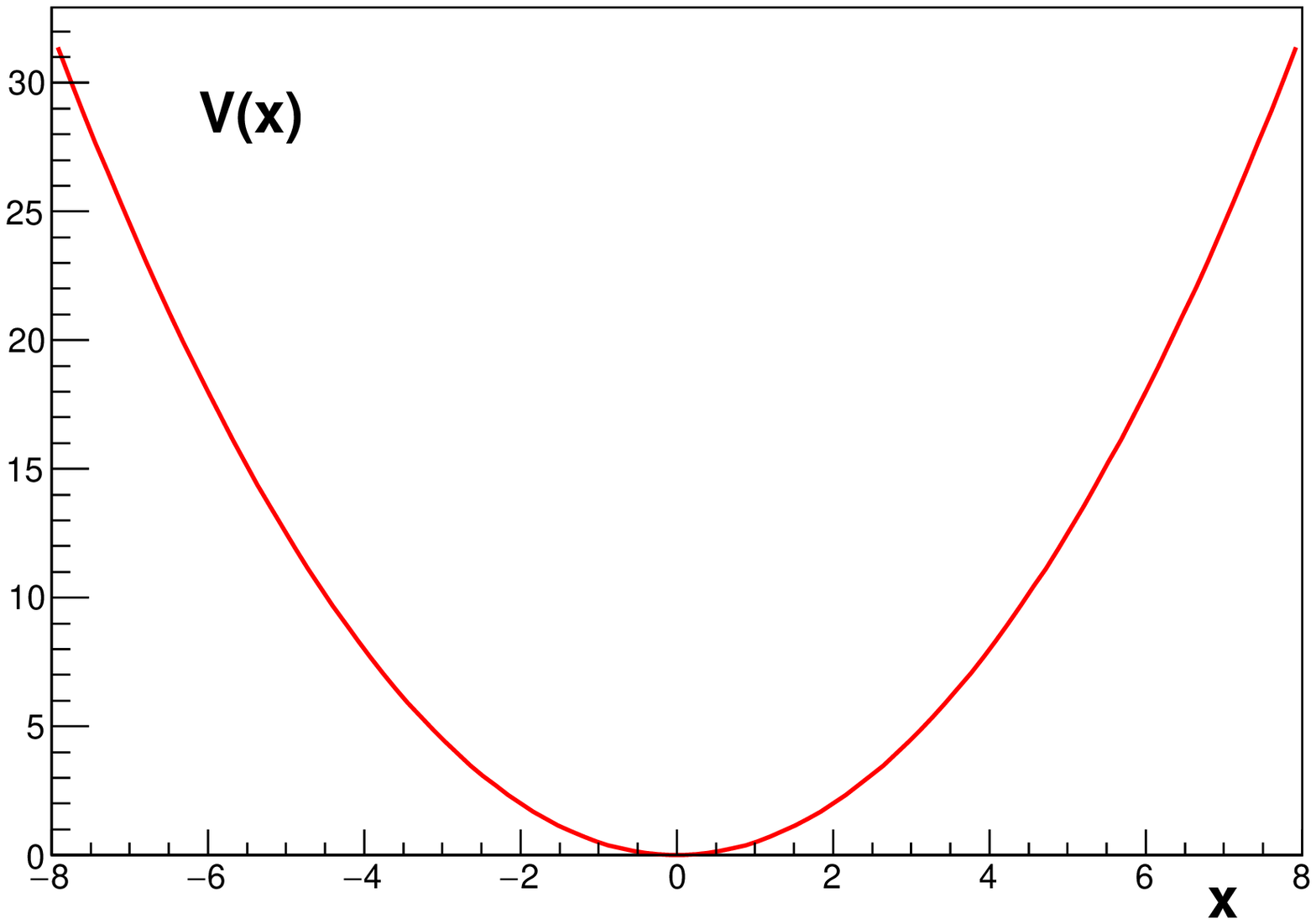,height=5.5cm,width=7.0cm}
      \epsfig{file=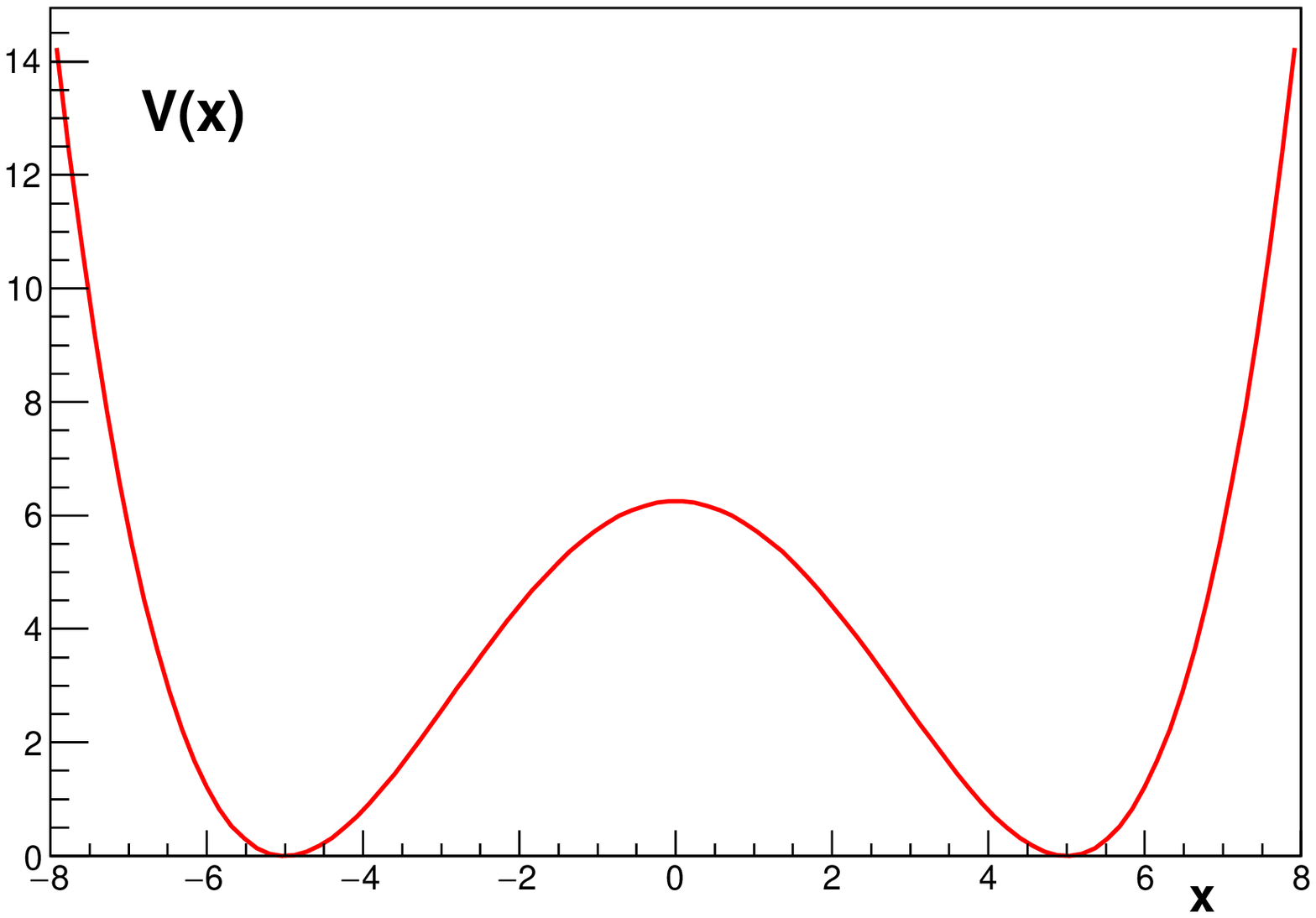,height=5.5cm,width=7.0cm}}
\vspace*{1.0cm}

\footnotesize
{\bf Figure 3.9}.- Potential wells studied in the example 3.13.

\end{center}
\end{figure}

We start with an initial trajectory from $x_0=x(t_i)=0$ to
$x_f=x(t_f)=0$ and the intermediate values
$x_1,x_2,{\ldots},x_{N-1}$ drawn from $Un(-10.,10.)$, sufficiently large
in this case since their support is $(-{\infty},{\infty})$. 
For the parameters of the grid we took
${\epsilon}=0.25$ and $N=2000$. The parameter $\epsilon$ has to be small enough
so that the results we obtain are close to what we should have for a continuous
time and $N$ sufficiently large so that
${\tau}=N{\epsilon}$ is large enough to isolate the contribution of the
fundamental state. With this election we have that
${\tau}=2000{\cdot}0.25=500$. Obviously, we have to check the stability
of the result varying both parameters.
Once the grid is fixed, we sweep over all the points
$x_1,x_2,{\ldots},x_{N-1}$ of the trajectory and  for each $x_j$,
$j=1,{\ldots},N-1$ we propose a new candidate
$x^{'}_j$ with support ${\Delta}$. Then, taking $\hatch{h}=1$ we have that:
\begin{eqnarray}
P(x_j{\longrightarrow}x^{'}_j)\,=
\exp\left\{ -
S_{N}(x_0,x_1,{\ldots}x^{'}_j,{\ldots},x_N)
   \right\} 
                                       \nonumber
\end{eqnarray}
and
\begin{eqnarray}
P(x_j{\longrightarrow}x_j)\,=
\exp\left\{ -
S_{N}(x_0,x_1,{\ldots}x_j,{\ldots},x_N)
   \right\} 
                                       \nonumber
\end{eqnarray}
so the acceptance function will be:
\begin{eqnarray}
a(x_j{\longrightarrow}x^{'}_j)\,=\,{\rm min}\,\left\{1,
\frac{\textstyle P(x_j{\longrightarrow}x^{'}_j)}
     {\textstyle P(x_j{\longrightarrow}x_j)    }\right\}
                                       \nonumber
\end{eqnarray}
Obviously we do not have to evaluate the sum over all the nodes because
when dealing with node $j$, only the intervals
$(x_{j-1},x_j)$ and $(x_j,x_{j+1})$ contribute to the sum. Thus, at node
$j$ we have to evaluate
\begin{eqnarray}
a(x_j{\longrightarrow}x^{'}_j)\,=\,
{\rm min}\,\left\{1,
\exp\left\{ -
                   S_{N}(x_{j-1},x^{'}_j,x_{j+1})\,+\,
                   S_{N}(x_{j-1},x_j,x_{j+1}) 
   \right\} \right\}
                                       \nonumber
\end{eqnarray}
Last, the trajectories obtained with the Metropolis algorithm will follow
eventually the desired distribution
$p(x_0,x_1,{\ldots},x_N)$ in the asymptotic limit. To have a reasonable
approximation to  that we shall not use the first
$N_{term}$ trajectories ({\sl thermalization}).
In this case we have taken $N_{term}=1000$ and again, we should check the
the stability of the result. After this, we have generated
$N_{gen}=3000$ and, to reduce correlations, we took one out of three
for the evaluations; that is $N_{used}=1000$ trajectories, each one 
determined by $N=2000$ nodes.
The distribution of the accepted values $x_j$will be an approximation
to the probability to find the particle at position $x$ for the fundamental
state; that is, $|{\Psi}_0(x)|^2$. 
Figure 3.10 shows the results of the simulation compared to 
\begin{eqnarray}
  |{\Psi}_0(x)|^2\,{\propto}\,e^{-x^2}
                                       \nonumber
\end{eqnarray}
together with one of the many trajectories generated.
The sampling average $<x^2>=0.486$ is a good approximation to the
energy of the fundamental state $E_0=0.5$.

\begin{figure}[!t]
\begin{center}

\vspace*{-1.0cm}
\mbox{\epsfig{file=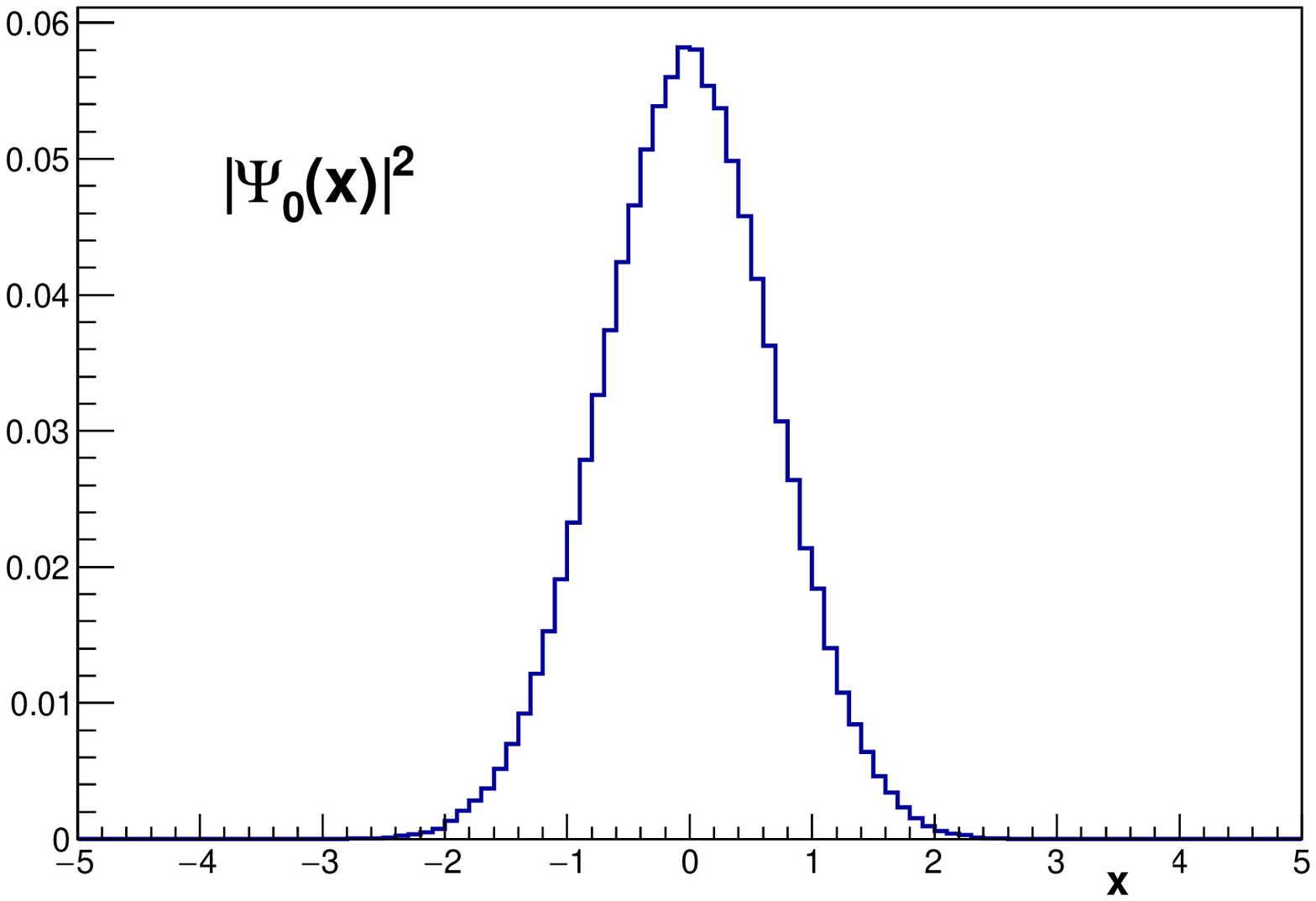,height=5.5cm,width=7.0cm}
      \epsfig{file=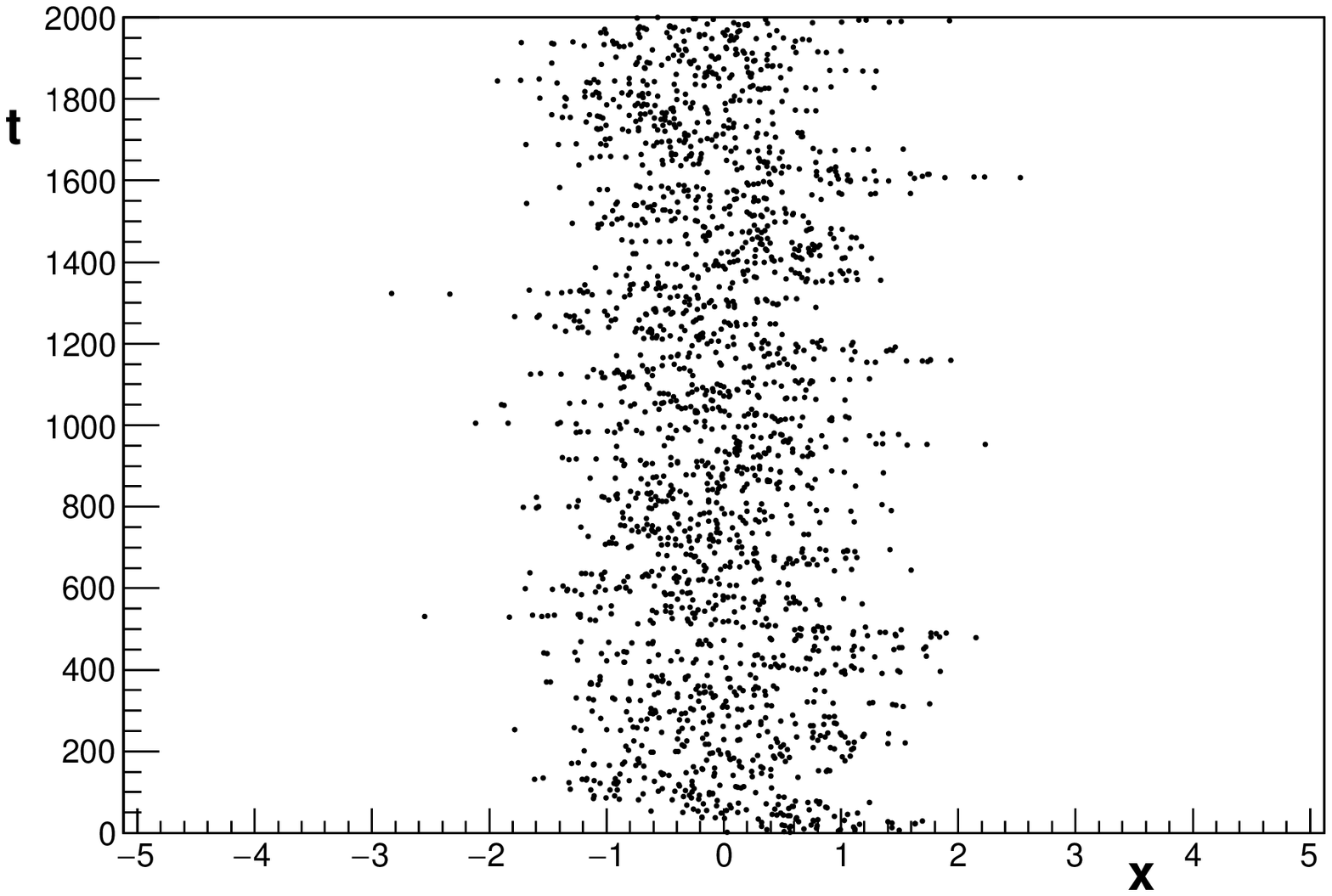,height=5.5cm,width=7.0cm}}
\vspace*{1.0cm}

\footnotesize
{\bf Figure 3.10}.- Squared norm of the fundamental state
wave-function for the harmonic potential and one of the 
simulated trajectories.

\end{center}
\end{figure}

As a second example, we have considered the potential well
\begin{eqnarray}
 V(x)\,=\,
 \frac{\textstyle a^2}{\textstyle 4}\,\left[
 \left( x/a \right)^2\,-\,1 \right]^2
                                       \nonumber
\end{eqnarray}
and, again from the Virial Theorem:
\begin{eqnarray}
 <E>_{\Psi}\,=\,
 \frac{\textstyle 3}{\textstyle 4\,a^2}\,<x^4>_{\Psi}\,-\,
 <x^2>_{\Psi}\,+\,
 \frac{\textstyle a^2}{\textstyle 4}
                                       \nonumber
\end{eqnarray}
We took $a=5$, a grid of $N=9000$ nodes and ${\epsilon}=0.25$ 
(so ${\tau}=2250$),  and as before
$N_{term}=1000$, $N_{gen}=3000$ and $N_{used}=1000$. 
From the generated trajectories we have the sample moments
$<x^2>=16.4264$ and $<x^4>=361.4756$ so the estimated
fundamental state energy is $<E_0>=0.668$ to be compared with the exact
result $E_0=0.697$. The norm of the wave-function for the fundamental
state is shown in figure 3.11 together with one of the simulated trajectories
showing clearly the tunneling between the two wells.

\begin{figure}[!t]
\begin{center}

\vspace*{-1.0cm}
\mbox{\epsfig{file=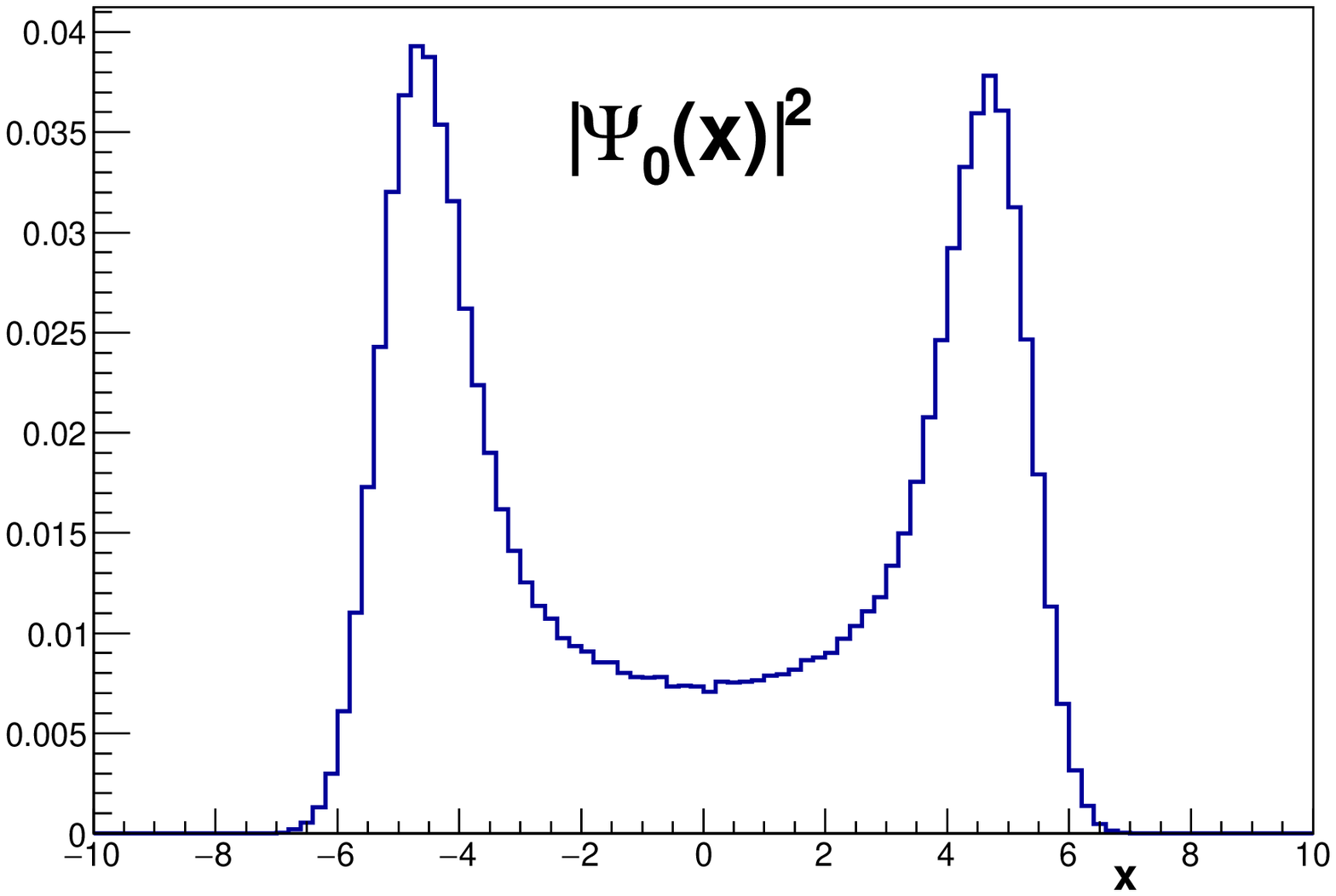,height=5.5cm,width=7.0cm}
      \epsfig{file=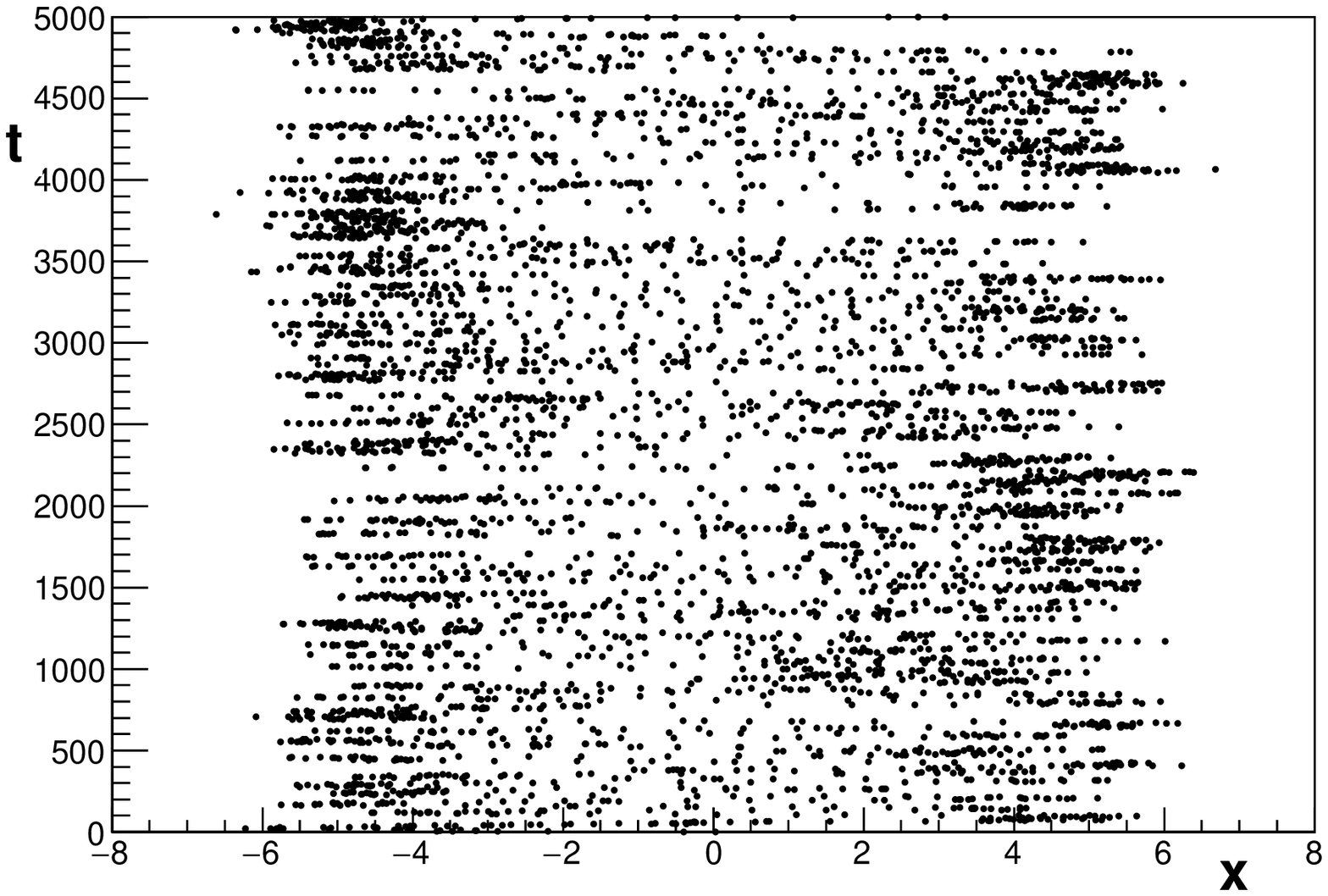,height=5.5cm,width=7.0cm}}
\vspace*{1.0cm}

\footnotesize
{\bf Figure 3.11}.- 
Squared norm of the fundamental state
wave-function for the quadratic potential and one of the 
simulated trajectories.

\end{center}
\end{figure}

\smill
{\raya}                   
\vspace{1.0cm}

\subsection{Sampling from Conditionals and Gibbs Sampling}
In many cases, the distribution of an n-dimensional random quantity
is either not known explicitly or difficult to sample directly but
sampling the conditionals is easier. In fact, sometimes it may help to
introduce an additional random quantity an consider the conditional densities
(see the example 3.14). 
Consider then the n-dimensional random quantity
${\bf X}=(X_1,{\ldots},X_n)$ with density $p(x_1,{\ldots},x_n)$,
the usually simpler conditional densities
\begin{eqnarray*}
 &&  p(x_1|x_2,x_3{\ldots},x_n)   \\
 &&  p(x_2|x_1,x_3{\ldots},x_n) \\
 &&    \hspace{1.cm} {\vdots}  \\
 &&  p(x_n|x_1,x_2,{\ldots},x_{n-1})
\end{eqnarray*}
and an arbitrary initial value
$\xbold^{0)}=\{x_1^{0)},x_2^{0)},{\ldots},x_n^{0)}\}\,{\in}\,{\Omega}_{\bf X}$.
If we take the approximating density $q(x_1,x_2,{\ldots},x_n)$ and the
conditional densities
\begin{eqnarray*}
 &&  q(x_1|x_2,x_3,{\ldots},x_n) \\
 &&  q(x_2|x_1,x_3,{\ldots},x_n) \\
 &&    \hspace{1.cm} {\vdots}   \\
 &&  q(x_n|x_1,x_2,{\ldots},x_{n-1}) 
\end{eqnarray*}
we generate for $x_1$ a proposed new value $x_1^{1)}$ from
$q(x_1|x_2^{0)},x_3^{0)},{\ldots},x_n^{0)})$
and accept the change with probability
\begin{eqnarray*}
a(x_1^{0)}\rightarrow x_1^{1)})\,&=&\,
\min\left\{1,\,
\frac{p(x_1^{1)},x_2^{0)},x_3^{0)},{\ldots},x_n^{0)})
      q(x_1^{0)},x_2^{0)},x_3^{0)},{\ldots},x_n^{0)})}
     {p(x_1^{0)},x_2^{0)},x_3^{0)},{\ldots},x_n^{0)})
      q(x_1^{1)},x_2^{0)},x_3^{0)},{\ldots},x_n^{0)})}\right\}\,=\\
&=&\,\min\left\{1,\,
\frac{p(x_1^{1)}|x_2^{0)},x_3^{0)},{\ldots},x_n^{0)})
      q(x_1^{0)}|x_2^{0)},x_3^{0)},{\ldots},x_n^{0)})}
     {p(x_1^{0)}|x_2^{0)},x_3^{0)},{\ldots},x_n^{0)})
      q(x_1^{1)}|x_2^{0)},x_3^{0)},{\ldots},x_n^{0)})}\right\}
\end{eqnarray*}
After this step, let's denote the value of $x_1$ by $x_1'$ (that is,
$x_1'=x_1^{1)}$ or  $x_1'=x_1^{0)}$ if it was not accepted). Then, we
proceed with $x_2$. We generate a proposed new value $x_2^{1)}$ from
$q(x_2|x_1',x_3^{0)},{\ldots},x_n^{0)})$
and accept the change with probability
\begin{eqnarray*}
a(x_2^{0)}\rightarrow x_2^{1)})\,&=&\,
\min\left\{1,\,
\frac{p(x_1',x_2^{1)},x_3^{0)},{\ldots},x_n^{0)})
      q(x_1',x_2^{0)},x_3^{0)},{\ldots},x_n^{0)})}
     {p(x_1',x_2^{0)},x_3^{0)},{\ldots},x_n^{0)})
      q(x_1',x_2^{1)},x_3^{0)},{\ldots},x_n^{0)})}\right\}\,=\\
&=&\,\min\left\{1,\,
\frac{p(x_2^{1)}|x_1',x_3^{0)},{\ldots},x_n^{0)})
      q(x_2^{0)}|x_1',x_3^{0)},{\ldots},x_n^{0)})}
     {p(x_2^{0)}|x_1',x_3^{0)},{\ldots},x_n^{0)})
      q(x_2^{1)}|x_1'x_3^{0)},{\ldots},x_n^{0)})}\right\}
\end{eqnarray*}
After we run over all the variables, we are in a new state
$\{x_1',x_2',{\ldots},x_n'\}$ and repeat the whole procedure until
we consider that stability has been reached so that we are sufficiently close
to sample the desired density. The same procedure can be applied if we
consider more convenient to express the density 
\begin{eqnarray*}
p(x_1,x_2,x_3{\ldots},x_n)\,=\,p(x_n|x_{n-1},\ldots,x_2{\ldots},x_1)\,
       \cdots\,p(x_2|x_1)\,p(x_1)
\end{eqnarray*}
Obviously, we need only one one admissible starting value $x_1^{0)}$.

Gibbs sampling is a particular case of this approach and
consists on sampling sequentially all the random quantities directly 
from the conditional densities; that is:
\begin{eqnarray*}
   q(x_i|x_1,\ldots,x_{i-1},x_{i+1},\ldots,x_n)\,=\,
   p(x_i|x_1,\ldots,x_{i-1},x_{i+1},\ldots,x_n)
\end{eqnarray*}
so the acceptance factor $a(x{\rightarrow}x')=1$.
This is particularly 
useful for Bayesian inference since, in more than one dimension,
densities are usually specified in conditional form after the ordering of
parameters.

\vspace{0.5cm}
\noindent
{\raya}                   
\vspace{0.35cm}
\footnotesize

\noindent
{\bf Example 3.14:} Sometimes it may be convenient to introduce additional
random quantities in the problem to ease the treatment.
Look for instance at the Student's distribution 
$X{\sim}St(x|\nu)$ with
\begin{eqnarray*}
   p(x|\nu)\,\propto\,\left(
   1\,+\,x^2/\nu\right)^{-(\nu+1)/2}
\end{eqnarray*}
Since
\begin{eqnarray*}
 \int_0^{\infty}e^{-au}u^{b-1}du\,=\,\Gamma(b)\,a^{-b}
\end{eqnarray*}
we can introduce an additional random quantity $U{\sim}Ga(u|a,b)$ in the
problem with $a=1+x^2/\nu$ and $b=(\nu+1)/2$ so that
\begin{eqnarray*}
p(x,u|\nu)\,{\propto}\,e^{-au}u^{b-1}
\hspace{0.5cm}{\rm and}\hspace{0.5cm}
p(x)\,{\propto}\,\int_0^{\infty}p(x,u|\nu)du\,\propto\,a^{-b}=
      \left(1+x^2/\nu\right)^{-(\nu+1)/2}
\end{eqnarray*}
The conditional densities are
\begin{eqnarray*}
p(x|u,\nu)\,&=&\,\frac{p(x,u|\nu)}{p(u|\nu)}\,=\,
                 N(x|0,{\sigma})
  \hspace{0.5cm};\hspace{0.5cm}{\sigma}^2={\nu}(2u)^{-1}\\
p(u|x,\nu)\,&=&\,\frac{p(x,u|\nu)}{p(x|\nu)}\,=\,
                 Ga(u|a,b)
\hspace{0.5cm};\hspace{0.5cm}a=1+x^2/\nu\,,\,\,b=(\nu+1)/2
\end{eqnarray*}
so, if we start with an arbitrary initial value $x{\in}{\Rcal}$, we can
\begin{itemize}
\item[$1)$] Sample $U|X$: $u{\Leftarrow}Ga(u|a,b)$ with
            $a=1+x^2/\nu$ and $b=(\nu+1)/2$
\item[$2)$] Sample $X|U$: $x{\Leftarrow}N(x|0,\sigma)$ with
            ${\sigma}^2={\nu}(2u)^{-1}$
\end{itemize}
and repeat the procedure so that, after equilibrium,
$X{\sim}St(x|\nu)$. We can obviously start from $u{\in}\Rcal$ and
reverse the steps 1) and 2).
Thus, instead of sampling from the Student's distribution
we may sample the conditional densities: Normal and a Gamma distributions. 
Following this approach for $\nu=2$ and $10^3$ thermalization sweeps 
(far beyond the needs),
the results of $10^6$ draws are shown in figure 3.12 together with the 
Student's distribution $St(x|2)$.

\begin{figure}[!t]
\begin{center}

\vspace*{-1.0cm}
\mbox{\epsfig{file=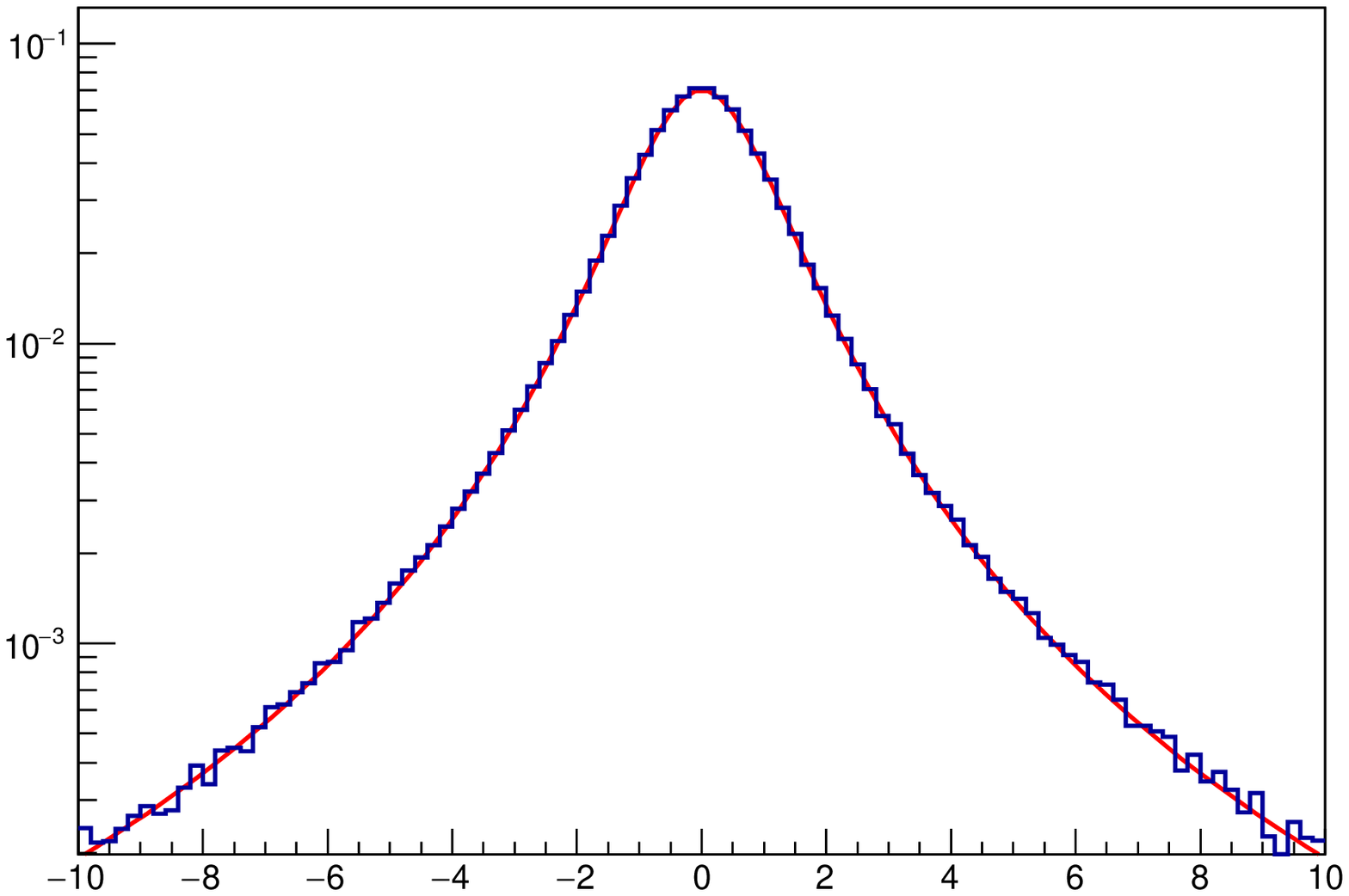,height=8.0cm,width=8.0cm}}
\vspace*{1.0cm}

\footnotesize
{\bf Figure 3.12}.- Sampling of the Student's Distribution
$St(x|2)$ (blue) compared to the desired distribution (red).

\end{center}
\end{figure}

\vspace{0.35cm}

\noindent
{\bf Example 3.15:}:
We have $j=1,\ldots,J$ groups of observations
each with a sample of size $n_j$; that is
$\xbold_j=\{x_{1j},x_{2j},\ldots,,x_{n_jj}\}$. Within each of the J groups,
observations are considered an exchangeable sequence and assumed to be
drawn from a distribution $x_{i,j}{\sim}N(x|{\mu}_j,\sigma^2)$ where
$i=1,\ldots,n_j$. Then:
\begin{eqnarray*}
p(\xbold_j|{\mu}_j,{\sigma})\,=\,\prod_{i=1}^{n_j}N(x_{ij}|{\mu}_j,\sigma^2)
\,{\propto}\,{\sigma}^{-n_j}\,{\exp}\left\{- \sum_{i=1}^{n_j}
\frac{\textstyle (x_{ij}-{\mu}_j)^2}
     {\textstyle 2\,{\sigma}^2}
      \right\}
\end{eqnarray*}
Then, for the $J$ groups we have the parameters
${\mubold}=\{{\mu}_1,{\ldots}{\mu}_J\}$ that, in turn, are also considered
as an exchangeable sequence drawn from a parent distribution
${\mu}_j{\sim}N({\mu}_j|{\mu},\sigma_{\mu}^2)$. 
We reparameterize the model in terms of
${\eta}={\sigma}_{\mu}^{-2}$ and  ${\phi}={\sigma}^{-2}$ and consider
conjugated priors for the parameters considered independent; that is
\begin{eqnarray*}
{\pi}(\mu,\eta,\phi)\,=\,
N(\mu|{\mu}_0,{\sigma}_0^2)\,Ga(\eta|c,d)\,Ga(\phi|a,b)
\end{eqnarray*}
Introducing the sample means
$\overline{x}_j=n_j^{-1}\sum_{i=1}^{n_j}x_{ij}$,
$\overline{x}=J^{-1}\sum_{j=1}^J\overline{x}_j$ and defining
$\overline{\mu}=J^{-1}\sum_{j=1}^J{\mu}_j$ and defining
After some simple algebra it is easy to see that the marginal densities are:
\begin{eqnarray*}
{\mu}_j\,&{\sim}&\,N\left(
\frac{\textstyle n_j{\sigma}_{\mu}^2\overline{x}_j+{\mu}{\sigma}^2}
     {\textstyle n_j{\sigma}_{\mu}^2+{\sigma}^2},
     \frac{\textstyle {\sigma}_{\mu}^2{\sigma}^2}
     {\textstyle n_j{\sigma}_{\mu}^2+{\sigma}^2}\right) \\
{\mu}\,&{\sim}&\,N\left(
\frac{\textstyle {\sigma}_{\mu}^2{\mu}_0+{\sigma}_0^2J\overline{\mu}}
     {\textstyle {\sigma}_{\mu}^2+J{\sigma}_0^2},
     \frac{\textstyle {\sigma}_{\mu}^2{\sigma}_0^2}
     {\textstyle {\sigma}_{\mu}^2+J{\sigma}_0^2}\right) \\
{\eta}={\sigma}_{\mu}^{-2}\,&{\sim}&\,Ga\left(
\frac{1}{2}\sum_{j=1}^J(\mu_j-{\mu})^2+c,\,
\frac{J}{2}+d\right) \\
{\phi}={\sigma}^{-2}\,&{\sim}&\,Ga\left(
\frac{1}{2}\sum_{j=1}^J\sum_{i=1}^{n_j}(x_{ij}-{\mu}_j)^2+a,\,
\frac{1}{2}\sum_{j=1}^Jn_j+b\right) 
\end{eqnarray*}

Thus, we set initially the parameters $\{{\mu}_0,{\sigma}_0,a,,b,c,d\}$
and then, at each step
\begin{itemize}
\item[1] Get $\{{\mu}_1,\ldots,{\mu}_J\}$ each as ${\mu}_j{\sim}N(\cdot,\cdot)$
\item[2] Get ${\mu}{\sim}N(\cdot,\cdot)$
\item[3] Get $\sigma_{\mu}={\eta}^{-1/2}$ with ${\eta}{\sim}Ga(\cdot,\cdot)$
\item[4] Get $\sigma={\phi}^{-1/2}$ with ${\phi}{\sim}Ga(\cdot,\cdot)$
\end{itemize}
and repeat the sequence until equilibrium is reached and 
samplings for evaluations can be done.

\smill
{\raya}               
\vspace{1.0cm}

\section{\LARGE \bf Evaluation of definite integrals}
\vspace*{0.1cm}
\noindent

A frequent use of Monte Carlo sampling is the evaluation of
definite integrals. Certainly, there are many numerical methods for 
this purpose and for low dimensions they usually give a better precision 
when fairly compared. In those cases one rarely uses Monte Carlo... 
although sometimes the domain of integration has a very complicated 
expression and the Monte Carlo implementation is far easier.
However, as we have seen the uncertainty of Monte Carlo estimations 
decreases with the sampling size $N$ as $1/\sqrt{N}$ regardless the number 
of dimensions so, at some point, it becomes superior. And,
besides that, it is fairly easy to estimate the accuracy of the evaluation. 
Let's see in this section the main ideas.

Suppose we have the n-dimensional definite integral
\begin{eqnarray*}
   I\,=\,{\int}_{\Omega}\,f(x_1,x_2,{\ldots},x_n)\,
   dx_1\,dx_2\,{\ldots}\,dx_n
\end{eqnarray*}
where $(x_1,x_2,{\ldots},x_n){\in}{\Omega}$ and  
$f(x_1,x_2,{\ldots},x_n)$ is a Riemann integrable function. If we consider a
random quantity $\Xbold=(X_1,X_2,\ldots,X_n)$ with distribution
function $P(\xbold)$ and support in $\Omega$, the
mathematical expectation of $\Ybold=g(\Xbold){\equiv}f(\Xbold)/p(\Xbold)$ 
is given by
\begin{eqnarray*}
   E[{\Ybold}]\,=\,{\int}_{\Omega}\,g(\xbold)\,dP(\xbold)\,=\,
                     {\int}_{{\Omega}}\,\frac{f(\xbold)}
                                            {p(\xbold)}\,dP(\xbold)\,=\,
   {\int}_{\Omega}\,f(\xbold)\,d\xbold\,=\,I
\end{eqnarray*}
Thus, if we have a sampling $\{\xbold_1,\xbold_2,\ldots,\xbold_N\}$, of 
size $N$,  of the random quantity $\Xbold$ under $P(\xbold)$ 
we know, by the Law of Large Numbers that, as $N\rightarrow{\infty}$,
the sample means
\begin{eqnarray*}
   I^{(1)}_N\,=\,
   \frac{\textstyle 1}{\textstyle N}\,
      \sum_{i=1}^{N}\,g(\xbold_i)
   \hspace{1.cm}{\rm and}\hspace{1.cm}
I^{(2)}_N\,=\,
   \frac{\textstyle 1}{\textstyle N}\,
      \sum_{i=1}^{N}\,g^2(\xbold_i)
\end{eqnarray*}
converge respectively to $E[Y]$ (and therefore to I) and to 
$E[Y^2]$ (as for the rest, all needed conditions for existence are assumed 
to hold). Furthermore, if we define 
\begin{eqnarray*}
   SI^2\,=\,
    \frac{\textstyle 1}{\textstyle N}\,\left(
    I^{(2)}_K\,-\,(I^{(1)}_K)^2 \right)
\end{eqnarray*}
we know by the Central Limit Theorem that the random quantity
\begin{eqnarray*}
    \Zbold\,=\,
    \frac{\textstyle I_K^{(1)}\,-\,I}{\textstyle SI}
\end{eqnarray*}
is, in the limit $N{\rightarrow}\infty$, distributed as $N(x|0,1)$. 
Thus, Monte Carlo integration provides a simple way to estimate the
integral $I$ and a quantify the accuracy. Depending on the problem
at hand, you can envisage several tricks to further improve the accuracy. For
instance, if $g(\xbold)$ is a function {\sl ``close''} to 
$f(\xbold)$ with the same support and known
integral $I_g$ one can write 
\begin{eqnarray*}
   I\,=\,{\int}_{\Omega}\,f(\xbold)\,d\xbold\,=\,
   {\int}_{\Omega}\,(f(\xbold)-g(\xbold))\,d\xbold\,+\,I_g
\end{eqnarray*}
and in consequence estimate the value of the integral as
\begin{eqnarray*}
\stackrel{\sim}{I}\,=\,
   \frac{\textstyle 1}{\textstyle N}\,
      \sum_{i=1}^{N}\,(f(\xbold_i)-g(\xbold_i))\,+\,I_g
\end{eqnarray*}
reducing the uncertainty.

\newpage\null\thispagestyle{empty}\newpage

\noindent
{\large {\bf References}}
\vspace{0.5cm}

\footnotesize

\noindent
\begin{tabular}{p{1.0cm}p{13.0cm}}
  $[Be79]$ & Bernardo J.M. (1979); J. Roy. Stat. Soc. Ser. B 41, 113-147. \\
  $[Be94]$ & Bernardo J.M., Smith A.F.M (1994); Bayesian Theory;
             John Wiley \& Sons;\\
  $[Be96]$ & Bernardo J.M. (1996); The Concept of Exchangeability and its
            Applications; Far East Journal of Mathematical Sciences 4, 111-121
            (www.uv.es/$\sim$bernardo/Exchangeability.pdf). \\
  $[Be98]$ & Bernardo J.M., Ram\'on J.M. (1998); The Statistician 47; 1-35. \\ 
  $[Be09]$ & Berger J.O., Bernardo J.M., Sun D. (2009); Ann. Stat; 37,  
            No. 2; 905-938. \\
  $[Be12]$ & Berger J.O., Bernardo J.M., Sun D. (2012); Objective Priors
            for Discrete Parameter Spaces; J. Am. Stat. Assoc. 107, No 498,
            636-648. \\
  $[Bo58]$ & Box G.E.P.,  M\"uller M.E. (1958); 
             {\sl A Note on the Generation of Random Normal Deviates}, 
             Ann. Math. Stat. 29, No 2, 610-611. \\
  $[Bo06]$ & Bogachev, V.I. (2006); {\sl Measure Theory}; Springer. \\
  $[BP04]$ & Berger, J.O., Pericchi L.R. (2004); 
             Ann. Stat. V32, No 3, 841-869. \\
  $[BP96]$ & Berger, J.O., Pericchi L.R. (1996); 
             J. Am. Stat. Assoc. V91, No 433, 109-122. \\
$[Co97]$ & Cowan, G. (1997);
           {\sl Statistical Data Analysis};
           Oxford Science Publications.\\
  $[Da03]$ & D'Agostini G. (2003); {\sl Bayesian Reasoning in Data Analysis};
             World Scientific Publishing.\\
  $[Da83]$ & Dalal S.R., Hall, W.J. (1983); 
           J. Roy. Stat. Soc. Ser. B, 45, 278-286. \\
  $[Da96]$ & Datta G.S., Ghosh M. (1996); Ann. Stat. 24, No 1, 141-159. \\
  $[Da04]$ & Datta G.S., Mukerjee R. (2004); {\sl Probability Matching Priors
             and Higher Order Asymptotics}, Springer, New York. \\
  $[Ec98]$ & L'Ecuyer P. (1998); {\sl Handbook of Simulations},
             Chap. 4, 93-137; Wiley.\\
  $[Fe97]$ & Feldman G.J. and Cousins R.D. (1997);
             arXiV:physics/9711021 v2.\\
  $[Ge95]$ & Gelman A.B., Carlin J.S., Stern H.S. and Rubin D.B. (1995);
             {\sl Bayesian Data Analysis}; Chapman $\&$ Hall \\
  $[Go84]$ & Gosh M., Mukerjee R. (1984) Biometrika 84, 970-975. \\
  $[Gu13]$ & Gut, A.(2013); {\sl Probability: A Graduate Course};
             Springer Texts in Statistics.\\ 
  $[Ha70]$ & Hastings W.K. (1970); Biometrika 57, 97-109. \\
  $[Ja80]$ & James F. (1980); {\sl Monte Carlo theory and practice},
             Rep. Prog. Phys. 43, 1145-1189.\\
  $[Ja99]$ & James F., Hoogland J., Kleiss R. (1999); 
             Comput. Phys. Comm 2-3, 180-220.\\
  $[Ja06]$ & James F. (2006); 
            {\sl Statistical Methods in Experimental Physics};              
            World Scientific Pub. Co. \\
  $[Ja64]$ & Jaynes E.T. (1964); {\sl Prior 
             Probabilities and Transformation Groups},NSF G23778. \\
  $[Je39]$ & Jeffreys H. (1939); {\sl Theory of Probability};
             Oxford Univ. Press.\\
  $[Ka95]$ &  Kass R.E., Raftery A.E. (1995); 
            J. Am. Stat. Assoc. V90, No 430, 773-795. \\
  $[Ka96]$ &  Kass R.E., Wasserman L. (1996); 
            {\sl The Selection of Prior Distributions by Formal Rules};
            J. Am. Stat. Assoc. V91, No 453, 1343-1370. \\
  $[Kn81]$ & Knuth D.E. (1981); 
             {\sl The Art of Computer Programming}, vol 2.; Addison-Wesley.\\
 $[Ly89]$ & Lyons, L. (1989);
             {\sl Statistics for Nuclear and Particle Physicists};
             Cambridge University Press. \\
  $[Ma87]$ & Marsaglia G., Zaman A. (1987); 
             {\sl Toward a Universal Random Number Generator}, 
             Florida State University Report FSU-SCRI-87-50. \\
  $[Me53]$ & Metropolis N., Rosenbluth A.W., Rosenbluth M.W., Teller A.H., 
             Teller E. (1953); Journal of Chemical Physics 21, 1087-1092. \\
  $[OH95]$ & O'Hagan A. (1995); J. Roy. Stat. Soc. B57, 99-138. \\
  $[Ra61]$ & Raiffa, H. and Schlaifer R. (1961); 
            {\sl Applied Statistical Decision Theory}; 
            Cambridge MA: Harvard University Press. \\
  $[Ru81]$ & Rubin D.B. (1981); Ann. Stat. 9, 130-134. \\
  $[Sc78]$ & Schwarz, G (1978); Ann. Stat. 6, 461-464. \\
  $[St65]$  & Stone M. (1965); 
            {\sl Right Haar Measures for Convergence in Probability to 
                 Invariant Posterior Distributions}; 
            Annals of Math. Stat. 36, 440-453. \\
  $[St70]$ &  Stone M. (1970); 
            {\sl Necessary and Sufficient Conditions for Convergence in 
                 Probability to Invariant Posterior Distributions}; 
            Annals of Math. Stat. 41, 1349-1353. \\
  $[We63]$ & Welch, B.; Pears H. (1963) J. Roy. Stat. Soc. B25, 318-329. 
\end{tabular}
\end{document}